\documentclass{amsbook}
\usepackage{graphics}
\usepackage{epsfig}
\makeindex
\usepackage{amssymb}
\usepackage[mathscr]{eucal}
\usepackage{youngtab}
\newtheorem{theorem}{Theorem}[chapter]
\newtheorem{lemma}{Lemma}[theorem]
\newtheorem{cor}{Corollary}[theorem]
\newtheorem{prop}[theorem]{Proposition}
\newtheorem{df}{Definition}[section]
\theoremstyle{definition}
\newtheorem{definition}{Definition}[chapter]
\newtheorem{example}{Example}[chapter]
\newtheorem{thm}{Theorem}[section]
\newtheorem{lem}[thm]{Lemma}
\theoremstyle{definition}
\newtheorem{defn}[thm]{Definition}
\newtheorem{pro}{Problem}

\theoremstyle{remark}
\newtheorem{remark}{Remark}[chapter]
\newcommand{\CL}{\mathcal{C}(\la)}
\newcommand{\CU}{\mathcal{C}(u,\la)}
\newcommand{\D }{\mathrm{d}}

\newcommand{\upo}[1]{\overset{\circ}{#1}}
\DeclareMathOperator{\Der}{Der}

\numberwithin{equation}{chapter}
\numberwithin{section}{chapter}

\def\ma#1#2#3#4{\left[{}^{#1}_{#3}{}^{#2}_{#4}\right]_2}

\newcommand{\fOA}{\theta^2\left[{}_1^0{}_1^1\right]_2}
\newcommand{\fOO}{\theta^2\left[{}_1^0{}_1^0\right]_2}

\newcommand{\ffCC}{\theta^2\left[{}_0^0{}_0^0\right]_2}
\newcommand{\ffCB}{\theta^2\left[{}_0^0{}_0^1\right]_2}
\newcommand{\ffCO}{\theta^2\left[{}_0^0{}_1^0\right]_2}
\newcommand{\ffBC}{\theta^2\left[{}_0^1{}_0^0\right]_2}
\newcommand{\ffOC}{\theta^2\left[{}_1^0{}_0^0\right]_2}
\newcommand{\ffAC}{\theta^2\left[{}_1^1{}_0^0\right]_2}
\newcommand{\ffCA}{\theta^2\left[{}_0^0{}_1^1\right]_2}
\newcommand{\ffBB}{\theta^2\left[{}_0^1{}_0^1\right]_2}
\newcommand{\ffBA}{\theta^2\left[{}_0^1{}_1^1\right]_2}
\newcommand{\ffAB}{\theta^2\left[{}_1^1{}_0^1\right]_2}
\newcommand{\ffBO}{\theta^2\left[{}_0^1{}_1^0\right]_2}
\newcommand{\ffOB}{\theta^2\left[{}_1^0{}_0^1\right]_2}
\newcommand{\ffAA}{\theta^2\left[{}_1^1{}_1^1\right]_2}
\newcommand{\ffAO}{\theta^2\left[{}_1^1{}_1^0\right]_2}

\newcommand{\ffOO}{\theta^2\left[{}_1^0{}_1^0\right]_2}

\newcommand{\kI}{\sqrt{1-\kappa^2}}
\newcommand{\lI}{\sqrt{1-\lambda^2}}
\newcommand{\mI}{\sqrt{1-\mu^2}}

\newcommand{\ml}{\sqrt{\lambda^2-\mu^2}}
\newcommand{\mk}{\sqrt{\kappa^2-\mu^2}}
\newcommand{\lk}{\sqrt{\kappa^2-\lambda^2}}

\newcommand{\kkI}{{1-\kappa^2}}
\newcommand{\llI}{{1-\lambda^2}}
\newcommand{\mmI}{{1-\mu^2}}

\newcommand{\mml}{{\lambda^2-\mu^2}}
\newcommand{\mmk}{{\kappa^2-\mu^2}}
\newcommand{\llk}{{\kappa^2-\lambda^2}}

\newcommand{\vek}[1]{\boldsymbol{#1}}

\newcommand{\M}{\phantom{-}}
\newcommand{\weglassen}[1]{}

\renewcommand{\imath}{\mathrm{i}}
\renewcommand{\d}{\mathrm{d}}

\def\C{\mathbb{C}}

\def\wt{\widetilde}
\def\wh{\widehat}
\def\pa{\partial}
\def\le{\leqslant}
\def\ge{\geqslant}

\def\a{\alpha}
\def\ga{\gamma}
\def\dt{\delta}
\def\la{\lambda}
\def\ph{\varphi}
\def\si{\sigma}
\def\Dt{\Delta}
\def\La{\Lambda}

\def\D{\mathrm{d}}

\def\Bga{\boldsymbol{\gamma}}
\def\Bla{\boldsymbol{\lambda}}
\def\BOm{\boldsymbol{\Omega}}
\def\Bmu{\boldsymbol{\mu}}
\def\Bnu{\boldsymbol{\nu}}
\def\Bp{\boldsymbol{p}}
\def\Bq{\boldsymbol{q}}
\def\Br{\boldsymbol{r}}
\def\Bt{\boldsymbol{t}}
\def\Bu{\boldsymbol{u}}
\def\Bv{\boldsymbol{v}}
\def\Bz{\boldsymbol{z}}
\def\BL{\boldsymbol{L}}
\def\BU{\boldsymbol{U}}
\def\BW{\boldsymbol{W}}
\def\BZ{\boldsymbol{Z}}
\def\Bwp{\boldsymbol{\wp}}

\def\cA{\mathcal{A}}
\def\cB{\mathcal{B}}
\def\cF{\mathcal{F}}
\def\cH{\mathcal{H}}
\def\cK{\mathcal{K}}
\def\cL{\mathcal{L}}
\def\cT{\mathcal{T}}

\def\sJ{\mathsf{J}}
\def\sK{\mathsf{K}}
\def\sM{\mathsf{M}}
\def\sU{\mathsf{U}}

\DeclareMathOperator\rank{rank}
\DeclareMathOperator\Jac{Jac}
\DeclareMathOperator\Kum{Kum}
\DeclareMathOperator\Sym{Sym}

\DeclareMathOperator{\inv}{inv}
\DeclareMathOperator{\diag}{diag}
\DeclareMathOperator{\id}{id}
\DeclareMathOperator{\odd}{od}
\DeclareMathOperator{\even}{ev}
\newcommand{\bwp}{\boldsymbol{\wp}}

\def\bW{\boldsymbol{W}}

\begin{document}
\title{Multi-Dimensional Sigma-Functions}
\author{V M Buchstaber}

\address{Steklov Institute of Mathematics, Russian Academy of Sciences, ul.~Gubkina~8,
Moscow, 119991 Russia} \email{buchstab@mi.ras.ru}

\author{V Z Enolski}
\address{Theoretical Physics Division\\
NASU Institute of Magnetism\\
36--b Vernadsky str.\\
Kiev-680\\
252142\\
Ukraine}
\email{venolski@googlemail.com}

\author{D V Leykin}
\address{Theoretical Physics Division\\
NASU Institute of Magnetism\\
36--b Vernadsky str.\\
Kiev-680\\
252142\\
Ukraine}
\email{dile@imag.kiev.ua}
\date{\empty}

\thanks{}

\maketitle

\chapter*{Preface}
In 1997 the present authors published a review \cite{bel97b} that recapitulated and developed classical theory of Abelian functions realized in terms of multi-dimensional sigma-functions.  This approach originated by K.Weierstrass and F.Klein was aimed to extend to higher genera Weierstrass theory of elliptic functions based on the Weierstrass $\sigma$-functions.
Our development was motivated by the recent achievements of mathematical physics and theory of integrable systems that were based of the results of classical theory of multi-dimensional theta functions. Both theta and sigma-functions are integer and quasi-periodic functions, but  worth to remark the fundamental difference between them. While theta-function are defined in the terms of the Riemann period matrix, the sigma-function can be constructed by coefficients of polynomial defining the curve. Note that the relation between periods and coefficients of polynomials defining the curve is transcendental.

Since the publication of our 1997-review a lot of new results in this area appeared (see below the list of Recent References), that promoted us to submit this draft to ArXiv without waiting publication a well-prepared book. We complemented the review by the list of articles that were published after 1997 year to develop the theory of $\sigma$-functions presented here. Although the main body of this review is devoted to hyperelliptic functions the method can be extended to an arbitrary algebraic curve and new material that we added in the cases when the opposite is not stated does not suppose hyperellipticity  of the curve considered.

 We already thankful to  readers for comments relevant to the text presented, in particular we thank to   A.Nakayashiki \cite{nakaya08}, \cite{nakayashiki09} who noticed an error in the formula illustrating algebraic representability of the $\sigma$-function in our short note \cite{bel99} and corrected it in his publications.

We are grateful to our colleges discussion with whom influenced on the content of our manuscript. They are: Ch.Athorne, V.Bazhanov, E.Belokolos, H.Braden, B.Dubrovin, Ch. Eilbeck, J.Elgin, J.Harnad, H.Holden, E.Hackmann, B.Hartmann, A.Hone,  A.Fordy, A.Its, F. Gesztesy, J.Gibbons,  T.Grava, V.Kagramanova, A.Kokotov, D.Korotkin, I.Krichever, J.Kunz, C.L\"ammerzahl, Sh.Matsutani, S.Natanson, A.Na\-ka\-ya\-shi\-ki, F.Nijhoff, S.Novikov, A.Mikhailov, Yo.$\hat{\rm O}$nishi, M.Pavlov, E.Previato, P.Richter, S.Novikov, V.Shramchenko, S.Shorina, A.Veselov.

The authors are grateful to ZARM, Bremen University, in particular to  Professor L\"ammerzahl for funding research/teaching visit of one from us (VZE) to Bremen and Oldenburg Universities in April-July 2012,  they also grateful to the Department of Physics of University of Oldenburg and personally to Prof. J.Kunz for the organizing  in Oldenburg   the research meeting at May 2012  of two from the authors (VMB and VZE) when this version of the manuscript was prepared.

\tableofcontents

\chapter*{Introduction}
Starting point for this monograph was the approach developed in papers
of the authors \cite{bel97b,bel97c,bel99} where the Kleinian program of construction of hypelliptic Abelian functions was modified and realized. Our approach to the modified Kleinian program (see below) can be extended to wider classes of algebraic curve. To demonstrate key ideas of the method that's enough to consider the non-degenerate hyperelliptic curve $V$,
\begin{equation}V= \{ (y, x)  \in  \mathbb{ C}^2:y^2- \sum_{i=0}^{2g+2} \lambda_{i}x^i=0 \}\label{hyperellipintroduc}
\end{equation} that yields the co-compact lattice $\mathbb{Z}^{2g} \simeq\Gamma_V\subset\mathbb{C}^g$
and the compact variety
\[   \mathbb{C}^g/\Gamma_V = \mathrm{Jac}(V)  \]
 which is called Jacobian of the curve $V$. The varieties $\mathrm{Jac}(V)$
 associated to hyperelliptic curve represent important class of Abelian tori $T^{g}_{\Lambda}$ that is associated to a hyperelliptic curve.

 Every Abelian variety represents compact variety $T_{\Gamma}^g=\mathbb{C}^g/\Gamma$ where
 $\mathbb{Z}^{2g} \simeq\Gamma\subset\mathbb{C}^g$ and $\Gamma$ is the lattice satisfying Riemann conditions. Let complex $g\times2g$ matrix $\Omega$ is the matrix of rank $g$ defining a basis in the lattice $\Gamma$ through standard basis $\mathbb{Z}^{2g}\subset \mathbb{C}^g$.  The matrix $\Omega$ called Riemann matrix if there exists a skew-symmetric integral matrix $J$ of rank $2g$, called a principal matrix, such that
 \begin{equation}
 \Omega J^T \Omega=0, \qquad \frac{1}{2\imath}\Omega J^T \overline{\Omega}>0
 \label{riemanncondition}
 \end{equation}
the second condition means that the hermitian matrix  $\frac{1}{2\imath}\Omega J^T \overline{\Omega}$ is positive-definite. The problem elimination of Jacobians $\mathrm{Jac}(V)$ among Abelian varieties $T^g_{\Gamma}$ is the well known as the Riemann-Schottky problem. On every Abelian variety $T^g_{\Gamma}$ defined $\theta$-function $\Theta(\boldsymbol{u};\Gamma)$. It is an integer function on $\mathbb{C}^g$ such that the functions
\begin{align*}
\mathfrak{P}_{j_1,\ldots,j_k} ( \boldsymbol{ u},\Gamma) &=- \frac{ \partial^k}{ \partial
u_{j_1}, \ldots, \partial u_{j_k}}  \mathrm{ln} \; \Theta ( \boldsymbol{ u}, \Gamma) ,\quad k\geq 2
\end{align*}
are meromorphic functions on $T^g_{\Gamma}$ and generate the whole field of meromorphic functions on this variety. A natural problem arises -
to describe properties of $\Theta(\boldsymbol{u},\Gamma)$ in the case $\Gamma=\Gamma_V$, the program to solve this problem was suggested by S.P.Novikov (1979) that was realized by Shiota \cite{shi86}, see also
\cite{bea87}.

Our approach to the problem of $\sigma$-function for hyperelliptic curves
(\ref{hyperellipintroduc}) can be formulated as follows:

{\em To built integer function $\sigma(\boldsymbol{u};V)$, $\boldsymbol{u}\in \mathbb{C}^g$ such that:

 {\bf (i)} In the expansion of $\sigma(\boldsymbol{u};V)$ into series in $\boldsymbol{u}$ the coefficients of monomials $u_1^{i_1}\cdots u_g^{i_g}$ are polynomials in coefficients $\lambda_k$, $k=0..2g$ of the polynomial defining the curve $V$

  {\bf (ii)} Functions $\mathfrak{P}_{\boldsymbol{J}}(\boldsymbol{u};V)$, where $\boldsymbol{J}=(j_1,\ldots,j_k)$, $k\geq 2$ generate the whole field of Abelian functions on $\mathrm{Jac}(V)$}

When such the function $\sigma(\boldsymbol{u};V)$ is built then according to the property  {\bf (ii)} it can be expressed as
\begin{equation}
\sigma(\boldsymbol{u};V)=C \mathrm{e}^{B(\boldsymbol{u},\Gamma_V)}\theta(A(\Gamma_V)\boldsymbol{u},\Gamma_V)
\label{sigmaintroduc}
\end{equation}
where $B(\boldsymbol{u};\Gamma_V)$ is a quadratic form with respect to the vector $\boldsymbol{u}=(u_1,\ldots,u_g)^T$ and non-degenerate matrix $A(\Gamma_V)$ is non-degenerate matrix built by the period lattice $\Gamma_V$ and $V$ is a constant. But according to the property {\bf{(ii)}} the expression (\ref{sigmaintroduc}) distinguishes the functions $\Theta(\boldsymbol{u};\Gamma_V)$ and   $\Theta(\boldsymbol{u};\Gamma)$, where $\Gamma$ is a general lattice of Abelian torus and $\Gamma_V$ is a lattice generated by the periods of the curve $V$.

That follows from {\bf (i)} that the function $\sigma(\boldsymbol{u};V)$ is independent on the basis choice. When the problem demands knowledge of solution dependence from coefficients of the curve an answer in terms of
$\sigma(\boldsymbol{u};V)$ has advantages in comparison with $\Theta(\boldsymbol{u};\Gamma_V)$.


We already mentioned that the problem of construction of multi-dimensional $\sigma$-functions is a classical one. In 1886 F.Klein suggested the following program:

{\em Modify multi-dimensional function $\Theta(\boldsymbol{u};\Gamma_V)$ to obtain an entire function which is

{\bf (1)} independent on a basis in $\Gamma_V$

{\bf (2)} a covariant of M\"obius transformation of a curve $V$}

Klein wrote at this subject papers  \cite{kl86,kl88,kl90} and in the foreword to the corresponding section of the 3-volumed collection of works \cite{kl923} Klein estimated state of art with his program and resumed there that it was not completely executed.

The claim {\bf (2)}
restricts realization of Klein's program only to the case of hyperelliptic curves and even in this case appear artificial complications in the realization of the key condition which we denoted above as {\bf (i)}. Klein called the cycle of his works as ``Ueber hyperelliptische Sigmafunctionen". Basing on that we suggested to call these functions as {\em Klein functions} in our preceding review \cite{bel97b}.

H.F.Baker abandoned {\bf (2)} and demonstrated that for $g=2$ a theory of $\sigma$-function can be constructed without any reference to $\theta$-function \cite{ba03}.

Development of the theory of hyperelliptic $\sigma$-functions is motivated both by the problem that left classical science as well by requests of modern theories. These are

-Problems of the theory of hyperelliptic integrals, solving Jacobi inversion problem, differentiation of classical integrals and Abelian functions in branch points, derivation of differential equations for Abelian functions, see in particular, Burkhardt  \cite{bur88},
Wiltheiss  \cite{wi88},  Bolza  \cite{bo95},  Baker  \cite{ba98,ba03} and
others; the detailed bibliography may be found in  \cite{kw15}.

- Problems of the modern theory of integrable systems, algebro-geomeric solutions of integrable equations of mathematical physics, various problems of algebraic geometry, theory of singularities of differential equations and singularity theory of the algebraic maps.
Various aspect these problems were discussed in the given above list of Recent References.



The book is organized as follows

In the Chapter \ref{chap:genc} we present the approach which
leads to the general notion of $\sigma$-function. The important
definitions of universal space and fiber bundle of the Jacobians
of Riemann surfaces of plane algebraic curves are given.  We
discuss the general outline of the construction which is
carried out in the next chapters.

The Chapter \ref{chap:hypp} is devoted to explicit realization of
the fundamental hyperelliptic $\sigma$-function. We provide
necessary details from the theory of hyperelliptic curves and
$\theta$-functions and fix notation, which is extensively used in
the sequel. We give here the exposition
of the classical  Theorems \ref{the-S}, \ref{J-s} and
\ref{the-Z} which constitute the background of the theory of
Kleinian functions.

In the Chapter \ref{chap:wpfun} we  derive the basic
relations connecting the functions $\wp_{gi}$ and their
derivatives and  find a basis set of functions closed
with respect to differentiations over the canonical fields
$\partial/\partial u_i$. We  use these results
to construct solutions of the KdV system and matrix families
satisfying to the zero curvature condition.  Next, we find the
fundamental  cubic and quartic relations connecting the odd
functions $\wp_{ggi}$ and even functions $\wp_{ij}$.  We use these
results to give the explicit solution of the ``Sine-Gordon".

We start the Chapter \ref{chap:hjac}  with an analysis of the
fundamental cubic and quartic relations. The results of this
analysis lead to the explicit matrix realization  of hyperelliptic
Jacobians $ \mathrm{ Jac} (V) $ and Kummer varieties $  \mathrm{
Kum} (V) $ of the curves $V$ with the fixed branching point
$e_{2g+2}=a= \infty$. Next we describe a dynamical system defined
on the universal space of the Jacobians of the Riemann surfaces of
the canonical hyperelliptic curves of genus $g$ such that
trajectories of its evolution lay completely  in the fibers of the
universal bundle of canonical hyperelliptic Jacobians.  We apply
the theory developed to construct systems of linear differential
operators for which the hyperelliptic curve $V(y,x)$ is their
common spectral variety.

The Chapter \ref{chap:add}
is based on some of our most recent results and contains the
explicit expression of the ratio
$\frac{\sigma(\boldsymbol{u}+\boldsymbol{v})\sigma(\boldsymbol{u}-
\boldsymbol{v})}{\sigma(\boldsymbol{u})^2\sigma(\boldsymbol{v})^2}$ as
a polynomial on $\wp_{i,j}(\boldsymbol{u})$
and $\wp_{i,j}(\boldsymbol{v})$  for
the cases of arbitrary genus. We briefly discuss an application of
this result to the addition theorems for Kleinian functions.

The Chapter \ref{chap:red} contains a short introduction to the theory of reduction of
theta-functions and Abelian integrals to lower genera. The Weierstrass-Poincar\'e theorem
on the complete reducibility is formulated there. The case of genus two curves and reduction
of associated theta-function to Jacobian theta's and holomorphic integrals to elliptic integrals is
considered in mere details. In particular, we are discussing Humbert variety and relevant
reductions of Abelian functions to elliptic functions.

The Chapter \ref{chap:rat} discussed the class
polynomials that satisfy an analog of Riemann vanishing theorem
It is shown there that these polynomials are completely characterized by this
property. By rational analogs of Abelian functions we mean logarithmic
derivatives of orders $\geq 2$ of these polynomials. We call the polynomials
thus obtained the \textit{Schur--Weierstrass polynomials}
because they are constructed from classical Schur polynomials,
which, however, correspond to special partitions related to
Weierstrass sequences. Since a Schur polynomial corresponding to
an arbitrary partition leads to a rational solution of the
Kadomtsev--Petviashvili hierarchy, the problem of connecting the above
solutions with those defined in terms of Abelian functions on
Jacobians naturally arose.Our results open the way to solve this problem
on the basis of the Riemann vanishing theorem.

The Chapter \ref{chap:div} considers subvarieties of the Jacobian - theta (sigma)-divisor
and its lower dimension strata. We describe analytically embedding of hyperelliptic curve
into Jacobian in terms of derivatives of sigma-functions. Using these formulae we construct
inversion of one hyperelliptic integral. Restriction of KdV flow on the strata of theta-divisor
are also considered here and the cases of hyperelliptic curves of genera two and three
are discussed in more details.   As am example of application of the method developed
the integration of double pendulum dynamics is considerd.

In the Chater \ref{chap:ns} a wide class of models of plane algebraic curves, so-called $(n,s)$-curves. The case $(2,3)$ is the classical Weierstrass model of an elliptic curve. On the basis of the theory of multivariative $\sigma$ function, for every pair of coprime $n$ and $s$ we obtain an effective description of the Lie algebra of derivation of the field of fiberwise Abelian functions defined on the total space of the bundle whose base is the parameter space of the family of nondegenerate $(n,s)$-curves and whose fibers are the Jacobi varieties of these curves. 
The essence of the method is demonstrated by the example of Weierstrass elliptic functions. Details are given for the case of a family of genus 2 curves. It is also considered in this chapter the system of heat equations in a nonholonomic frame and the solution in terms of $\sigma$-functions of Abelian tori is found. As a corollary the generators of a ring differential operators annihilating the $\sigma$-function of plane algebraic curves are described.

The Chapter \ref{chap:tau} considers the  algebro-geometric $\tau$ function and presents its expression
into  correspondence to a member of integrable hierarchy a Young diagram. The differential equations in this
approach follows from the Pl\"ucker relation associated to the given Young diagram. As examples
hyperelliptic genust two curve and trigonal genus three curve are considered. The $\tau$-functional
method is compared with residual derivation of integrable equations as well be means of Hirota bilinear
relations.

In the Chapter \ref{chap:abbs} Abelian Bloch solutions of the $2d$
Schr{\"o}dinger equations are studied using the Kleinian
functions of genus $2$.  The associated spectral problem leads to the 
fixed energy level which geometric sence we are clarifying. The main result of this chapter is the
addition theorem for Baker function on Jacobian. Spectral problems
on reducible and degenerate Jacobians are discussed.

In the Chapter \ref{chap:baf} we are considering the Baker-Akhiezer function within Krichever theory. We first introduce so-called muster function as a solution of the Schr\"odinger equation with finite-gap potential and then construct 
degenerate Baker-Akhiezer function. Baker-Akhiezer-Krichever function is introduce as the quotient of  of the two 
last ones.

In the Chapter \ref{chap:trig} we are considering the trigonal curve that belongs to the family of $(3,s)$-curves. We are showing that the most part of theory that was developed for hyperelliptic curves can be   extended to this case. To this end we introduce trigonal $\sigma$-function and correspondinf multi-dimensional $\wp$-functions. The Jacobi inversion problem is solved in these coordinate, also embedding of Jacobi variety into projective space is described as an algebraic variety that coordinates are trigonal $\wp$-functions. As an application show that partial differential equation from the Boussinesq hierarchy naturaly arise as differential relations between $\wp$-functions.

The Appendices  contain elements of handbooks for the Abelian
functions of genera two and three.

\chapter[General Construction]{General construction}\label{chap:genc}
In this chapter we present the general outline of the
construction which leads to $\sigma$-function. Necessary details
about $\theta$-functions may be found in e.g.
\cite{mu83,ig72,co56,ma79,kr03,ba97}, the bibliography of classical
literature may be found in \cite{kw15}.

\section{Universal bundle of Abelian varieties}
Let $\mathbb{H}^n$ be a $n$-di\-m\-en\-si\-o\-nal linear space
over the quaternions $\mathbb{H}$. Fixing an isomorphism
$\mathbb{H}^n\cong \mathbb{C}^n\times\mathbb{C}^n$ let us write
vectors from $\mathbb{H}^n$ in the form of row vectors
$\mathbf{v}=(\mathbf{v}_1,\mathbf{v}_2)$, where
$\mathbf{v}_1,\mathbf{v}_2\in\mathbb{C}^n$. For
$g\in\mathrm{Sp}(n,\mathbb{Z})$ let us put $g= \begin{pmatrix}
a^T&b^T\\c^T&d^T \end{pmatrix}$ where $a,b,c,d\in
\mathrm{GL}(n,\mathbb{Z})$, then
\[ \det g=1,
\quad\text{and}\quad g \begin{pmatrix} 0&-1_n\\1_n&0
\end{pmatrix}g^T=\begin{pmatrix} 0&-1_n\\1_n&0 \end{pmatrix}. \]
Let us fix the following {\em right} action of the group
\index{group!$\mathrm{Sp}(n,\mathbb{Z})$}
$\mathrm{Sp}(n,\mathbb{Z})$ on $\mathbb{H}^n$
\begin{equation} \label{H-act} \mathbf{v}\cdot
g=(\mathbf{v}_1d+\mathbf{v}_2c,\mathbf{v}_1b+\mathbf{v}_2a),
\end{equation}
in matrix notation
\begin{equation}
\mathbf{v}\cdot
g=(\mathbf{v}_1,\mathbf{v}_2)\begin{pmatrix}
0&1_n\\1_n&0
\end{pmatrix}g^T\begin{pmatrix}
0&1_n\\1_n&0
\end{pmatrix}.\label{mat-H-act}
\end{equation}
The {\em right} action of the translations  group
$\mathbb{Z}^n\times\mathbb{Z}^n$ we write in the form
\begin{equation}
\mathbf{v}\cdot\begin{pmatrix}\boldsymbol{m}\\
\boldsymbol{m}'
\end{pmatrix}=
(\mathbf{v}_1+\boldsymbol{m}^T,\mathbf{v}_2+{\boldsymbol{m}'}^T).
\label{H-transl}
\end{equation}
Let us denote  by $A_{\mathbb{H}}(n,\mathbb{Z})$ the subgroup  in the motion
group of $\mathbb{H}^n$  generated by  $\mathrm{Sp}(n,\mathbb{Z})$
and $\mathbb{Z}^n\times\mathbb{Z}^n$. This group,
$A_{\mathbb{H}}(n,\mathbb{Z})$  is an analogue of the affine group. It is the
extension $\mathrm{Sp}(n,\mathbb{Z})$
by $\mathbb{Z}^n\times\mathbb{Z}^n$, i.e.
there is an exact sequence:
\begin{equation*}
0\to\mathbb{Z}^n \times\mathbb{Z}^n\overset{i}{\rightarrow}
A_{\mathbb{H}}(n,\mathbb{Z})\overset{\pi}\rightarrow
\mathrm{Sp}(n,\mathbb{Z})\to 0,
\end{equation*}
where $i$ is the canonical embedding, and the projection $\pi$
projects an affine transformation $\gamma\in A_{\mathbb{H}}(n,\mathbb{Z})$ to
the corresponding rotation $\pi(\gamma)\in \mathrm{Sp}(n,\mathbb{Z})$, i.e.
\begin{equation}
\label{pi-proj}\pi(\gamma):\mathbf{v}\cdot\pi(\gamma)=
\mathbf{v}\cdot\gamma-\mathbf{0}\cdot\gamma.
\end{equation}
This extension corresponds to the following action of the group
$\mathrm{Sp}(n,\mathbb{Z})$ by automorphisms of the group
$\mathbb{Z}^n \times\mathbb{Z}^n$
\begin{equation*}
\begin{pmatrix}\boldsymbol{m}\\
\boldsymbol{m}'
\end{pmatrix}\cdot g=\begin{pmatrix}\widehat{\boldsymbol{m}}\\
\widehat{\boldsymbol{m}}'
\end{pmatrix} =\begin{pmatrix}d^T\boldsymbol{m}+c^T\boldsymbol{m}'
\\
b^T\boldsymbol{m}+a^T\boldsymbol{m}'
\end{pmatrix}.
\end{equation*}

Let us recall the notion of the {\em Siegel upper
half-space}\index{Siegel half-space}. It is the space of matrices
$\mathcal{ S}_n$ \[ \mathcal{ S}_n=\{\tau\in
\mathrm{GL}(n,\mathbb{C})\,|\,\tau^T
=\tau,\quad\mathrm{Im}(\tau)\text{ --- positively defined}\}.
\] \index{matrix!of periods}
We introduce an extension of the $\mathcal{ S}_n$, which we  denote
$\mathcal{ S}'_n$. It is the space of $n\times 2n$-matrices
$(2\omega, 2\omega')$, such that
their columns are independent over $\mathbb{R}$,
matrices $\omega$ and  $\omega'$ satisfy the equation
\begin{equation} \omega'{\omega}^T-\omega{\omega'}^T=0, \label
{Riem-eq} \end{equation}  $\det(\omega)\neq
0$ and  $\tau=\omega^{-1}\omega'
\in\mathcal{ S}_n$. Such matrices $(2\omega, 2\omega')\in\mathcal{ S}'_n$
generate so called {\em principally polarized lattices} in
$\mathbb{C}^n$. There is a natural projection to the Siegel
half-space $\pi_s:\mathcal{S}'_n\to\mathcal{
S}_n:\pi_s(\omega,\omega')=\omega^{-1}\omega'$ and embedding
$i_s:\mathcal{S}_n\to\mathcal{
S}'_n:i_s(\tau)=(1_n,\tau)$.

Action \eqref{H-act} induces the right action of the group
$\mathrm{Sp}(n,\mathbb{Z})$ on $\mathcal{S}'_n$:
\begin{equation}
(\omega,\omega')\cdot
g=(\widehat{\omega},\widehat{\omega}')= (\omega d+
\omega' c, \omega b+ \omega'a),\label{S'-act} \end{equation} which
under the projection $\pi_s:\mathcal{S}'_n\to\mathcal{ S}_n$ comes
to the canonical {\em right} action of the
$\mathrm{Sp}(n,\mathbb{Z})$ on the upper Siegel half-space:
\index{group!$\mathrm{Sp}(n,\mathbb{Z})$} \[ \pi_s((\omega,\omega')\cdot
g)=\pi_s(\widehat{\omega},\widehat{\omega}')=
(\omega d+ \omega' c)^{-1}(\omega b+ \omega'a)=
( d+ \tau c)^{-1}(b+ \tau a).
\]
So the canonical action of the
$\mathrm{Sp}(n,\mathbb{Z})$ on $\mathcal{S}_n$ is decomposed to
the composition $\tau\cdot g=\pi_s(i_s(\tau)\cdot g).$

Let us consider  spaces $\mathcal{U}_n=\mathbb{C}^n \times
\mathcal{ S}_n$ and
$\mathcal{U}'_n=\mathbb{C}^n \times
\mathcal{ S}'_n$ . The action \eqref{S'-act} of the group
$\mathrm{Sp}(n,\mathbb{Z})$ on $\mathcal{S}'_n$  extends to the
right action of the group
$A_{\mathbb{H}}(n,\mathbb{Z})$ on $\mathcal{U}'_n$ in the following way.

 Let
$(\boldsymbol{z},\omega,\omega')\in \mathcal{U}'_n$ and
$\gamma=\big(g,({}^{\boldsymbol{m}}_{
\boldsymbol{m}'})\big)\in A_{\mathbb{H}}(n,\mathbb{Z})$ then
\begin{equation}
(\boldsymbol{z},\omega,\omega')\cdot\gamma=(
\boldsymbol{z}+2\widehat{\omega}\boldsymbol{m}+2\widehat{\omega}'
\boldsymbol{m}',\widehat{\omega},\widehat{\omega}'),\label{G-acts}
\end{equation}
that is, the subgroup $\mathbb{Z}^n \times\mathbb{Z}^n $ acts on
 $\mathcal{U}'_n$ by transformations
$\big(1_{2n},({}^{\boldsymbol{m}}_{ \boldsymbol{m}'})\big)$:
 \begin{equation}
 (\boldsymbol{z},\omega,\omega')\mapsto(\boldsymbol{z}+2\omega
  \boldsymbol{m}+2\omega'
 \boldsymbol{m}',\omega,\omega')
\label{translat}
\end{equation}
and the modular subgroup $\mathrm{Sp}(n,\mathbb{Z})$ as
$\big(g,({}^{\boldsymbol{0}}_{0})\big)$: \begin{equation}
(\boldsymbol{z},\omega,\omega')\mapsto(
\boldsymbol{z},\widehat{\omega},\widehat{\omega}')=(\boldsymbol{z},
\omega d+ \omega' c,
\omega b+ \omega'a).
\label{modultransf}
\end{equation}

The projection $\pi_s:\mathcal{S}'_n\to\mathcal{
S}_n$ extends to
the projection $\tilde{\pi}_s:\mathcal{U}'_n\to\mathcal{
U}_n$  by the formula
\[
\tilde{\pi}_s(\boldsymbol{z},\omega,\omega')=(\omega^{-1}\boldsymbol{z},
\omega^{-1}\omega')
\]
and the embedding
$i_s:\mathcal{S}_n\to\mathcal{ S}'_n$ extends
to
 the embedding $\tilde{\imath}_s:\mathcal{U}_n\to\mathcal{
U}'_n$  according to the formula
\[
\tilde{\imath}_s(\boldsymbol{z},\tau)=(\boldsymbol{z},1,\tau).
\]
It is clear that $\tilde{\pi}_s\tilde{\imath}_s=\mathrm{id}$.

\begin{definition}
Factor-space
$\mathcal{UT}'_n=\mathcal{U}'_n/A_{\mathbb{H}}(n,\mathbb{Z})$ is
called the {\em universal space of $n$-dimensional principally
polarized Abelian varieties}.

Factor-space
$\mathcal{UM}'_n=\mathcal{S}'_n/A_{\mathbb{H}}(n,\mathbb{Z})$ is
called the {\em  space of moduli of $n$-dimensional principally
polarized Abelian varieties}.

Canonical projection $\mathcal{U}_n'\to\mathcal{S}_n'$ induces the
projection of factor-spaces $p:\mathcal{UT}_n'\to\mathcal{UM}_n'$.
The triple $(\mathcal{UT}_n',\mathcal{UM}_n',p)$ is called the
{\em universal bundle of $n$-dimensional principally polarized
Abelian varieties}.
\end{definition}

By construction, the fiber $p^{-1}(m)$ over a point $m\in
\mathcal{UM}'_n$ is an $n$-dimensional principally
polarized Abelian variety, which is a factor of the space
$\mathbb{C}^n$ over the lattice generated by the columns of a
matrix $(\omega,\omega')$ belonging to the equivalence class of
$m$.

\section{Construction of $\sigma$-functions}
For a point
$t=(\boldsymbol{z},\omega ,\omega')$ let us   define so called {\em
characteristic}\index{characteristic} by the mapping\[
\mathit{char}:\mathcal{U}'_n \to
\mathbb{R}^n\times\mathbb{R}^n:(\boldsymbol{z},\omega,\omega')\mapsto
[\epsilon]\]
according to equation \begin{equation*} \boldsymbol{z}=2
(\omega,\omega')[\epsilon],
\end{equation*}
which  uniquely defines $[\epsilon]$ due to the linear
independence of the columns of matrices $\omega$ and $\omega'$
over $\mathbb{R}$.

The characteristics $[\epsilon]=
\begin{bmatrix} \boldsymbol{\epsilon}\\ \boldsymbol{\epsilon}'
\end{bmatrix}
\in\frac{1}{2}\mathbb{Z}^{n}\times\frac{1}{2}\mathbb{Z}^{n}
\subset \mathbb{R}^n\times\mathbb{R}^n$ are
called {\em half-integer}\index{characteristic!half-integer}. Half-integer
characteristic is {\em even}\index{characteristic!even} or {\em
odd}\index{characteristic!odd} whenever $4 \boldsymbol{\epsilon}^T
\boldsymbol{\epsilon}'$ is even or odd.  Set of half-integer
characteristics is divided into classes modulo
$\mathbb{Z}^n\times\mathbb{Z}^n$.  Among $4^n$ classes of
half-integer  characteristics there are $ 2^{n-1}(2^n+1) $ even
classes and $ 2^{n-1}(2^n-1) $ odd classes.

Define on $\mathcal{U'}_n$ the
{\em $\theta$--function with
characteristic}\index{$\theta$--function!with characteristic}, as
a function \[\theta[\varepsilon]:\mathcal{U}'_n\to \mathbb{C}\]
given by the convergent Fourier series
\begin{multline}
\theta[\varepsilon] (\boldsymbol{z}, \omega,\omega') = \sum_{
  \boldsymbol{ m}  \in \mathbb{ Z}^n} \mathrm{exp}\pi\imath
  \left  \{  ( \boldsymbol{ m}+ \boldsymbol{ \varepsilon}') ^T
\omega^{-1}\omega' ( \boldsymbol{ m}+
\boldsymbol{\varepsilon}')\right.\\
+\left.(\omega^{-1}\boldsymbol{z}+ 2\boldsymbol{\varepsilon})
^T ( \boldsymbol{ m}+ \boldsymbol{\varepsilon}')   \right \},
\label{thetachar}
\end{multline}
for any vector
$[\varepsilon]=(\boldsymbol{\varepsilon},\boldsymbol{\varepsilon}')^T \in
\mathbb{R}^{n}\times\mathbb{R}^{n}$.

This vector $[\varepsilon]$
is called the {\em characteristic of
$\theta$--function}\index{characteristic!of $\theta$--function}.
As a function of ${\boldsymbol{z}}$ the $\theta$-function with odd
characteristic is odd and even with even characteristic.

Under translations
\eqref{translat}\index{$\theta$--function!transformation rules} the
function $\theta[\varepsilon]$ transforms according to
\begin{align} \theta&[\varepsilon]({\boldsymbol
z}+2\omega\boldsymbol{m}+2\omega'
\boldsymbol{m}',\omega,\omega')\notag=\\ &\mathrm{ exp}\bigl\{
  -\pi \imath [
{\boldsymbol{m}'}^T(\omega^{-1}{\boldsymbol z} +
\boldsymbol{m}')\label{thetachar22}
- \boldsymbol{m}^T\boldsymbol{\varepsilon}'
+{\boldsymbol{m}'}^T\boldsymbol{\varepsilon}
 ]\bigr\}
\theta[\varepsilon](\boldsymbol{z},\omega,\omega').
\end{align}

Under the  action \eqref{modultransf}  of
$\mathrm{Sp}(n,\mathbb{Z})$ the function $\theta[\varepsilon]$ is
taken to $\theta[\hat{\varepsilon}]$, the transformed
characteristic is given by
\begin{equation}
\left[\hat\varepsilon\right]= \begin{pmatrix}d&-c\\
-b&a\end{pmatrix}
\left[\varepsilon\right]
+\frac{1}{2}\left[\begin{array}{c}\mathrm{diag}(c^Td)\\
\mathrm{diag}(a^Tb)\end{array}\right].\label{chartransf}
\end{equation}
The complete rule of transformation is
\begin{equation}
\theta[\hat\varepsilon]
({\boldsymbol{z}},\widehat{\omega},\widehat{\omega}') =(\mathrm{
det}(\omega^{-1}\widehat{\omega}))^{1/2} \mathrm{
exp}\left\{\frac{\pi \imath}{4}  {\boldsymbol z}^T
({\omega^T})^{-1} c\,\widehat{\omega}^{-1}
\boldsymbol{z}\right\}
\theta[\varepsilon](\boldsymbol{z},\omega,\omega'),
\label{theta-mod}
\end{equation}

Now we are going to construct such a modification of
$\theta[\varepsilon]$ that it has as simple transformation rules
under action  \eqref{modultransf} as possible.
Indeed, by \eqref{theta-mod} the
$(\det(\omega))^{-1/2}\theta[\varepsilon]$ under
\eqref{modultransf} goes to
$(\det(\omega))^{-1/2}\theta[\varepsilon]$ and acquires a factor
$$\mathrm{
exp}\left\{\frac{\pi \imath}{4}  {\boldsymbol z}^T
({\omega^T})^{-1} c\,\widehat{\omega}^{-1}
\boldsymbol{z}\right\}
.$$
To get rid of this factor we need to find a $n\times n$  matrix
$\kappa$ such that it transforms under \eqref{modultransf} as
$\kappa\to\widehat{\kappa}$ and satisfies
$\kappa-\widehat{\kappa}= \frac{\pi \imath}{4} ({\omega^T})^{-1}
c\,\widehat{\omega}^{-1}$, then  $(\det(\omega))^{-1/2}
\mathrm{
exp}\left\{ {\boldsymbol z}^T\kappa
\boldsymbol{z}\right\}\theta[\varepsilon]$  will be taken by
\eqref{modultransf} just to
$$(\det(\widehat{\omega}))^{-1/2}
\mathrm{
exp}\left\{ {\boldsymbol z}^T\widehat{\kappa}
\boldsymbol{z}\right\}\theta[\widehat{\varepsilon}].$$ To do this
we need to introduce so called {\em associated
matrices}\index{associated matrices}.
\index{matrix!associated}

A matrix $(\eta,\,\eta')$ which for the given $(\omega,\omega')\in
\mathcal{S}'_n$ satisfies
\begin{equation}
\begin{pmatrix} \omega&
\omega'\\ \eta&\eta'\end{pmatrix} \begin{pmatrix} 0&-1_n\\1_n&0
\end{pmatrix} \begin{pmatrix} \omega& \omega'\\
\eta&\eta'\end{pmatrix}^T=-\frac{1}{2}\pi \imath
\begin{pmatrix} 0&-1_n\\1_n&0\end{pmatrix},\label{SSimp}
\end{equation}
is called
{\em associated to $(\omega,\omega')$}. An associated matrix is not
unique. In fact, \eqref{SSimp} is equivalent to equations
\begin{equation*}
\omega'\omega^T-\omega{\omega'}^T =0,\quad
\eta'\omega^T-\eta{\omega'}^T = -\frac{\pi \imath }{2}1_n,
\quad
\eta'\eta^T-\eta{\eta'}^T =0
\end{equation*}
first of these equations is satisfied due to \eqref{Riem-eq}, to
solve the rest take arbitrary matrix $\varkappa\neq0_n$ and put
$\eta(\varkappa)=2\varkappa\omega$ and $\eta'(\varkappa)
=2\varkappa\omega'-\frac{\pi \imath }{2}(\omega^T)^{-1}$, then
the rest of equations are equivalent to \eqref{Riem-eq}.
So for any nonzero $\varkappa_1$ and $\varkappa_2$  both
matrices $(\eta(\varkappa_1),\eta'(\varkappa_1))$ and
$(\eta(\varkappa_2),\eta'(\varkappa_2))$  are associated to
$(\omega,\omega')$.

Under the action of \eqref{modultransf} an associated matrix is
taken to associated,
i.e.  $(\widehat{\eta},\widehat{\eta}')$ and
$(\widehat{\omega},\widehat{\omega}')$ satisfy to \eqref{SSimp}.
An associated matrix $(\eta,\eta')$ transforms as
\begin{equation*}
(\widehat{\eta},\widehat{\eta}')=(\eta d+\eta'
c, \eta b+ \eta'a).
\end{equation*}

\begin{lemma}
For any  associated  matrix $(\eta,\,\eta')$  the matrix
$\varkappa=\frac12 \eta \omega^{-1}$ transforms by
\begin{equation}
\widehat{\varkappa}=\varkappa-\frac{\pi \imath}{4}
({\omega^T})^{-1} c\,\widehat{\omega}^{-1}.\label{kappa-trans}
\end{equation}
under the action of $\mathrm{Sp}(n,\mathbb{Z})$.
\end{lemma}
\begin{proof}
 Taking into account the relation
$\eta'=2\varkappa\omega'-\frac{\pi \imath }{2}(\omega^T)^{-1}$ we
have \begin{align*}
\widehat{\varkappa}&=\frac{1}{2}\widehat{\eta}(\widehat{\omega})^{-1}=
\frac{1}{2}(\eta d+\eta' c)(\widehat{\omega})^{-1}=
(\varkappa\omega d+\varkappa\omega'\,c-\frac{\pi \imath
}{4}(\omega^T)^{-1}\,c)(\widehat{\omega})^{-1}\\=
&\varkappa (\omega d+\omega'\,c)(\widehat{\omega})^{-1}-
\frac{\pi \imath
}{4}(\omega^T)^{-1}\,c(\widehat{\omega})^{-1}=
\varkappa-
\frac{\pi \imath
}{4}(\omega^T)^{-1}\,c(\widehat{\omega})^{-1}.
\end{align*}
\end{proof}

Now we can give
\begin{definition} Fix some  matrix $(\eta,\eta')$ associated to
$(\omega,\omega')\in \mathcal{S}'_n$, then {\em
$\sigma[\varepsilon]$--function with
characteristic}\index{$\sigma$--function!with characteristic} is
the function \[\sigma[\varepsilon]:\mathcal{U}'_n\to\mathbb{C}\]
defined by formula \begin{equation*}
\sigma[\varepsilon](\boldsymbol{z},\omega,\omega')=
\frac{C(\omega,\omega')}{\sqrt{\mathrm{det}(\omega)}}
\mathrm{exp}\{\frac{1}{2}
\boldsymbol{z}^T\eta\omega^{-1}{\boldsymbol
z}\}\;\theta[\varepsilon](\boldsymbol{z},\omega,\omega'),
\end{equation*}
where $C(\omega,\omega')$ is an invariant with  respect to
$\mathrm{Sp}(n,\mathbb{Z})$.
\end{definition}

From  \eqref{thetachar22} and \eqref{theta-mod} we deduce
\begin{theorem}
\label{IMPORTANT}
The $\sigma[\varepsilon]$--functions has the following
transformation
properties\index{$\sigma$--function!transformation rules}:
\begin{itemize} \item Translations:  put \[
\boldsymbol{E}(\boldsymbol{ m},\boldsymbol{ m}')=\eta\boldsymbol{
m} +\eta'\boldsymbol{ m}',\quad\text{and}\quad
\boldsymbol{\Omega}(\boldsymbol{ m},\boldsymbol{  m }')
  =\omega\boldsymbol{ m} +\omega' \boldsymbol{ m}', \]  where
$\boldsymbol{ m},\boldsymbol{ m}'\in \mathbb{Z}^{n}$, then
\begin{align*} &\sigma[\varepsilon](\boldsymbol{z}+
2{\boldsymbol\Omega} (\boldsymbol{ m }, \boldsymbol{  m
}'),\omega,\omega') =\mathrm{exp} \big\{ 2\boldsymbol{E}^T
(\boldsymbol{ m},\boldsymbol{ m}') \big({\boldsymbol z}+
\boldsymbol{\Omega}(\boldsymbol{  m }, \boldsymbol{  m
}')\big)\big\}\\
&\times \mathrm{exp} \{ -\pi \imath  {\boldsymbol m
}^T{\boldsymbol m}' -2\pi \imath  {\boldsymbol\varepsilon
}^T{\boldsymbol m}' \}
\sigma[\varepsilon]( {\boldsymbol z},\omega,\omega')
\end{align*}
\item Modular transformations:
\begin{equation*}
\sigma[\widehat{\varepsilon}](\boldsymbol{z},\widehat{\omega},
\widehat{\omega}')
=\sigma[{\varepsilon}]({\boldsymbol
z},\omega,\omega'),
\end{equation*}
where
\begin{eqnarray*}
\widehat{\omega}&=&\omega
d+\omega'c,\quad\widehat{\omega}'=\omega b+ \omega' a\\
\widehat{\eta}&=&\eta d+ \eta'
c,\quad\widehat{\eta}'=\eta b+\eta' a \end{eqnarray*} and
$[\varepsilon]$ is transforming to $[\widehat{\varepsilon}]$
according to \eqref{chartransf}.
\end{itemize}
\end{theorem}

As the transformation \eqref{chartransf} takes half-integer
characteristics to half-integer, we obtain
\begin{cor} The set
$\big\{\sigma[\varepsilon]\,| \,[\varepsilon]
\in\frac{1}{2}\mathbb{Z}^{n}\times\frac{1}{2}\mathbb{Z}^{n}
\big\}$ is taken by \eqref{modultransf}  to itself.
\end{cor}

\section{$\sigma$-functions on Jacobians}
There is an important particular case when
matrices $(2\omega,2\omega')$ appear as period matrices of
holomorphic differentials on the Riemann surface
$V$\index{algebraic curve!Riemann surface of} of a plane algebraic
curve $V(x,y)$\index{algebraic curve} of genus $n$:
\begin{equation*} V(x,y)=\{(x,y)\in\mathbb{C}^2\,|\,f(x,y)=0\},
\end{equation*}
where $f(x,y)$ is a polynomial of two variables over $\mathbb{C}$.
The basis in the $1$-dimensional cohomology
group $\mathrm{H}^1(V,\mathbb{C})$ consists
of $n$ holomorphic differentials\index{differential!holomorphic}
\begin{equation*} \mathrm{d} \mathbf{u}^T=(\mathrm{d}u_1,\ldots
\mathrm{d}u_n).
\end{equation*}
The differentials $\mathrm{d}u_k$ may be always chosen in the
form
\begin{equation}
\mathrm{d}u_k=
\frac{\phi_k(x,y) \mathrm{d}x}{\frac{\partial}{\partial
y}f(x,y)},\label{uuu}
\end{equation} where $\phi_k(x,y)$ are polynomials which are
defined by $f(x,y)$.

Let $(\boldsymbol{\mathfrak{
a}},\boldsymbol{\mathfrak{ b}})$ be a basis of cycles in
$\mathrm{H}_1(V,\mathbb{Z})$ with intersections $\mathfrak{
a}_i\circ\mathfrak{ a}_k=0$, $\mathfrak{ b}_i\circ\mathfrak{
b}_k=0$ and $\mathfrak{ a}_i\circ\mathfrak{ b}_k=-\mathfrak{
b}_k\circ\mathfrak{a
}_i=1$.\index{differential!holomorphic!period matrices of}
Canonical coupling
$\mathrm{H}^1(V,\mathbb{C})\times\mathrm{H}_1(V,\mathbb{Z})\to
\mathcal{S}'_n$ defines a matrix
\begin{equation*}
\left(\{\oint_{\mathfrak{a}_i}
\mathrm{d}u_j\},\{\oint_{\mathfrak{ b}_i}\mathrm{d}u_j\}
\right)_{i,j=1,\ldots,n}=(2\omega, 2\omega'),
\end{equation*}
and in such a way gives rise to a special subspace of
$\mathcal{S}'_n$ which we will denote $\mathcal{J}'_n$.

Now let us restrict  our construction to the subspace
$\mathcal{UJ}'_n=\mathbb{C}^n
\times\mathcal{J}'_n
\subset\mathcal{U}'_n$. The subspace
$\mathcal{J}'_n\in\mathcal{S}'_n$ is closed with respect to the
action of the group $A_{\mathbb{H}}(n,\mathbb{Z})$.
\begin{definition} \label{uni-jac}
Factor-space
$\mathcal{JM}'_n=\mathcal{J}'_n/A_{\mathbb{H}}(n,\mathbb{Z})$ is
called the {\em space of moduli of the Jacobians
of the Riemann surfaces of the plane algebraic curves of genus
$n$}. \index{Jacobians!moduli space of} \index{genus of plane
algebraic curve}

Bundle $p_{\mathcal{J}}:\mathcal{JT}'_n\to\mathcal{JM}'_n$ induced
by the embedding  $\mathcal{JM}'_n\subset\mathcal{UM}'_n$ is
called the  {\em universal bundle of the Jacobians of
the Riemann surfaces of the plane algebraic
curves of genus $n$}\index{Jacobians!universal bundle of}.

The space $\mathcal{JT}'_n$  of this bundle is called
{\em universal space of the Jacobians of
the Riemann surfaces of the plane algebraic
curves of genus $n$}\index{Jacobians!universal space of}.
\end{definition}

Remark, that on $V$, for a given set of basis holomorphic
differentials \eqref{uuu}, it always possible to construct the
associated set of basis meromorphic differentials,
\index{differential!meromorphic}
i.e. to find
such polynomials $g_k(x,y)$ that differentials
\begin{equation} \mathrm{d}
\mathbf{r}^T=(\mathrm{d}r_1,\ldots \mathrm{d}r_n),\quad
\mathrm{d}r_k=
\frac{g_k(x,y) \mathrm{d}x}{\frac{\partial}{\partial
y}f(x,y)},
\label{rrr}
\end{equation}
have the period matrix
\begin{equation*}
(2\eta, 2\eta')=\left(\{-\oint_{\mathfrak{a}_i}
\mathrm{d}r_j\},\{-\oint_{\mathfrak{ b}_i}\mathrm{d}r_j\}
\right)_{i,j=1,\ldots,n},
\end{equation*}
which differs only by factor $2$ from the matrix
$(\eta,\eta')$\index{associated matrices} associated to $(\omega,\omega')$.

There is another circumstance to be taken into account,
if we fix   $(\omega,\omega')$, then in $\mathcal{UJ}'_n$   there
is a distinguished class of half-integer characteristics, the
characteristics of the {\em vector of Riemann constants}
(relatively $(\omega,\omega')$), which under modular
transformations goes to the  vector of Riemann constants
\index{vector of Riemann constants}
(relatively $(\widehat{\omega},\widehat{\omega}')$). If we
claim that the origin of the lattice generated by \eqref{translat}
is shifted by the vector of Riemann constants, then there will
appear an ``unmovable'' with respect to \eqref{modultransf}
  class of
characteristics --- the origin.  Summarizing, we have

\begin{definition}\label{fund1} The {\em
fundamental Abelian $\sigma$--function} is a function
\index{$\sigma$--function!fundamental Abelian}
\[\sigma:\mathcal{UJ}'_n\to \mathbb{C}\]  defined by the formula
\begin{equation} \sigma(\boldsymbol{z},\omega,\omega')=
\frac{1}{\sqrt[4]{D(V)}}\sqrt{\frac{\pi^g}{\mathrm{det}(\omega)}}
 \mathrm{exp}\{\tfrac12\boldsymbol{z}^T\eta\omega^{-1}{\boldsymbol
z}\}\;\theta[\varepsilon_{R}](\boldsymbol{z},\omega,\omega'),
\label{sigmaabel}
\end{equation}
where the characteristic $[\varepsilon_{R}]$
is the characteristic of the vector of Riemann
\index{vector of Riemann constants!characteristic of}
constants, the function $D(V)$ is the discriminant of the
defining equation $f(x,y)=0$ of the curve $V$.
\end{definition}

For the fundamental Abelian $\sigma$--function we have:
\index{$\sigma$--function!transformation rules} \begin{equation*}
  \sigma(\boldsymbol{z},\omega,\omega')=
  \sigma(\boldsymbol{z},\widehat{\omega},\widehat{\omega}'),
\end{equation*}
moreover,  any second logarithmic derivative of $\sigma$-function
is an automorphic function with respect to the action
\eqref{G-acts}, namely, let
\begin{align*}
\wp_{ij}(\boldsymbol{z},\omega,\omega')&=
-\frac{\partial^2}{\partial z_i \partial z_j}
\mathrm{ln}\sigma(\boldsymbol{z},\omega,\omega'),\\
\intertext{then $\forall\, \gamma \in A_{\mathbb{H}}(n,\mathbb{Z})$}
\wp_{ij}((\boldsymbol{z},\omega,\omega')\cdot \gamma)&=
\wp_{ij}(\boldsymbol{z},\omega,\omega').
\end{align*}
\begin{prop}
Functions $\wp_{i,k}$ where $i,k=1,\dots,n$
  \index{$\wp$--function} as functions of
$(\boldsymbol{z},\omega,\omega')$ define the functions on
the universal space of the Jacobians $\mathcal{JT}'_n$,
as functions of  $\boldsymbol{z}$ at fixed $(\omega,\omega')$
they define Abelian, i.e. $2n$-periodic meromorphic, functions on
the Jacobian of the Riemann surface of the underlying plane
algebraic curve $V(x,y)$.  \end{prop}

We aim to study properties of the $\wp$-functions, in order to
benefit on use of them in applications. For this we need such a
realization of the fundamental $\sigma$-function, which
relates the objects involved to modular
invariants, i.e. constants in the equation $f(x,y)=0$ defining
the curve.  The polynomials $\{\phi_k\}_{k=1,\dots,n}$ and
$\{g_k\}_{k=1,\dots,n}$ in \eqref{uuu} and \eqref{rrr}, in turn,
substantially depend on the form of the equation $f(x,y)=0$.
A certain ambiguity, still remaining in
the construction of the polynomials $g_k$, is removed once we fix
a very important object --- the global $2$-differential of second
kind. The required shift of origin of lattices, discussed above,
is achieved by special choice of the base divisor of the Abel map,
so that the vector of Riemann constants has effectively zero
coordinates. The $\sigma$-function thus obtained is related rather
to the algebraic curve $V(x,y)$ than to the lattice
$(\omega,\omega')$.

In following sections we explicitly carry out
all the stages of this construction for the case of the {\em
hyperelliptic} defining equation.

In the hyperelliptic case the subspace
$\mathcal{H}'_n\subset\mathcal{J}_n'\subset\mathcal{U}_n'$  of
hyperelliptic period matrices $(\omega,\omega')$  is closed with
respect of the action of the group $A_{\mathbb{H}}(n,\mathbb{Z})$.
The moduli space of hyperelliptic Jacobians
$\mathcal{HM}'_n\subset\mathcal{JM}'_n\subset\mathcal{UM}'_n$,
the universal space of hyperelliptic Jacobians
$\mathcal{HT}'_n\subset\mathcal{JT}'_n\subset\mathcal{UT}'_n$ and
the universal bundle of hyperelliptic Jacobians are defined
analogously to the Definition \ref{uni-jac}.

\section{Remarks}
For the first time, the equation (\ref{SSimp}) in the context of
the theory of Abelian integrals appeared in the
seminal papers by Weierstrass \cite{w49,w54} and is the
generalization of Legendre relation for periods of complete
elliptic integrals\footnote{Here we abide by the standard notation
of the theory of elliptic function \cite{ba55}}, \begin{equation*}
\eta'\omega-\omega'\eta=-\frac{\pi \imath }{2}.  \end{equation*}

To conclude this chapter, consider the classical example.
The fundamental $\sigma$--function in
the elliptic case $n=1$ is the Weierstrass
$\sigma$--function\index{$\sigma$--function!Weierstrass'}
\begin{equation*} \sigma(z,\omega,\omega')=
 \sqrt{\frac{\pi}{\omega}}\frac{1}{\sqrt[8]\Delta} \mathrm{exp}\left(
 \frac{\eta
}{2\omega}z^2\right)\vartheta_1(v|\tau),\end{equation*}
 where \[
 v=\frac{z}{2\omega},\quad\;
 \tau=\frac{\omega'}{\omega}\]
The defining equation of the elliptic curve $V$ in this case is
 \[f(x,y)=y^2-4(x-e_1)(x-e_2)(x-e_3),\quad e_1+e_2+e_3=0,\]
and its discriminant $D(V)=\sqrt{\Delta}$, where
\index{algebraic curve!elliptic} \[\Delta=16(e_1-e_2)^2(e_2-e_3)^2(e_1-e_3)^2. \]
Due to the condition $e_1+e_2+e_3=0$ the expansion of $\sigma(z)$
has the form:  \[ \sigma(z)=z+ O(z^5). \]

The Weierstrass $\wp(z,\omega,\omega')$--function is an
\index{$\wp$--function!Weierstrass'} automorphic function of the
group $A_{\mathbb{H}}(1,\mathbb{Z})$.

\chapter[Hyperelliptic $\sigma$--functions]{Hyperelliptic
$\sigma$--functions} \label{chap:hypp} For the detailed exposition
of the material concerning curves and $\theta$-functions on the
Jacobians see e.g. \cite{gh78,mu75,ba97,fk80,fa73}; on the
classical background of the $\sigma$-functions see in \cite{kl88};
for the detailed account of the genus $2$ $\sigma$-functions and
related topics see Part 1 of the monograph \cite{ba07}.

\section{Hyperelliptic curves}
The set of points $V (y,
x) $ satisfying the \begin{eqnarray} y^2= \sum_{i=0}^{2g+2}
\lambda_{i}x^{i} = \lambda_{2g+2} \prod_{k=1}^{2g+2} (x-e_{k}) =f
(x) \label{curve} \end{eqnarray} is a model of a plane { \em
hyperelliptic curve} of genus $g$, realized as a $2$--sheeted
covering over Riemann sphere with the { \em branching
points}\index{branching points} $e_1,  \ldots, e_{2g+2}$.
\index{curve!hyperelliptic} The form of the defining equation such
that $\lambda_{2g+2}=0$ (this means that e.g.  $e_{2g+2}=\infty$),
$\lambda_{2g+1}=4$ and $\lambda_{2g}=0$ is called a {\em canonical
form}. Any defining equation can be reduced to the canonical form
by a rational transformation.
\index{curve!hyperelliptic!canonical form of}

Any pair  $ (y, x) $ in $V (y, x) $ is     \index{point}
\index{point!analytic} called an { \em analytic point}; an
analytic point, which is not a  branching point is called a { \em
regular point}.  The { \em hyperelliptic
involution}\index{involution!hyperelliptic} $ \phi ( \;) $  (the
swap of the sheets of covering) acts as $ (y, x) \mapsto (-y, x)
$, leaving the branching points fixed.    \index{point!branching}

To make  $y$ the single valued function of $x$ it suffices to
draw $g+1$ cuts,  connecting pairs of branching points
$e_i$---$e_{i'}$ for some partition of $ \{1,  \ldots,  2g+2 \}$ into
the set of $g+1$ disjoint pairs ${i,  i'}$.  Those of $e_j$,  at
which the cuts start we will denote ${a_i}$,  ending points of the
cuts we will denote $b_i$,  respectively; except for one of the
cuts which is denoted by starting point $a$ and ending point $b$.
In the case $ \lambda_{2g+2}  \mapsto 0$ this point $a  \mapsto  \infty$.
We  assume the branching points to be numbered so that
\begin{gather*}
e_{2n}=a_n,\quad e_{2n-1}=b_n,\quad n=1,\ldots g;\\
 e_{2g+1}=b,\quad
e_{2g+2}=a
 \end{gather*}
with no loss of generality.

The
equation of the curve,  in case $ \lambda_{2g+2}=0$ and
$ \lambda_{2g+1}=4$  can be rewritten as  \begin{eqnarray} &y^2=4
P (x) Q (x) ,  \label{alt-curve}  \\ &P (x) = \prod_{i=1}^{g}
  (x-a_i) ,
  \quad Q (x) = (x-b)  \prod_{i=1}^{g} (x-b_i) .
\nonumber \end{eqnarray}

The local parametrization of the point $ (y, x) $ in the  vicinity
of a point $ (w, z) $:   \[ x=z+ \left \{ \begin{array}{ll}  \xi,
\quad & \text{near regular point} \,  ( \pm w, z) ;  \\  \xi^2, &
\text{near branching point} \,  (0, e_i) ;  \\  \frac{1}{ \xi}, &
\text{near regular point} \,   ( \pm \infty,  \infty) ; \\
\frac{1}{ \xi^2}, & \text{near branching point} \,   ( \infty,
\infty)    \end{array} \right.   \] provides the structure of the
{ \em hyperelliptic Riemann surface} --- a one-dimensional compact
complex manifold. We will employ the same notation for the plane
curve and the Riemann surface --- $V (y, x) $ or $V$. All the
curves and Riemann surfaces through the paper are assumed to be
hyperelliptic,  if the converse not stated.
\index{Riemann!hyperelliptic surface} \index{manifold!compact
complex}

A { \em marking} on $V (y, x) $ is given by the base
\index{marking} point $x_0$ and the canonical basis of cycles
$\index{basis!of cycles! canonical} (\mathfrak{ a}_1,  \ldots,
\mathfrak{ a}_g;\mathfrak{ b}_1, \ldots, \mathfrak{ b}_g) $ ---
the basis in the group\index{group!of one-dimensional homologies}
of one-dimensional homologies $H_1 (V (y, x) , \mathbb{ Z}) $ on
the surface $V (y, x) $ with the symplectic intersection matrix
\index{matrix!symplectic} $\mathfrak{ I}= \left (
\begin{array}{cc}0&- 1_g \\ 1_g&0 \end{array} \right) $, here $
1_g$ denotes the unit $g\times g$--matrix.

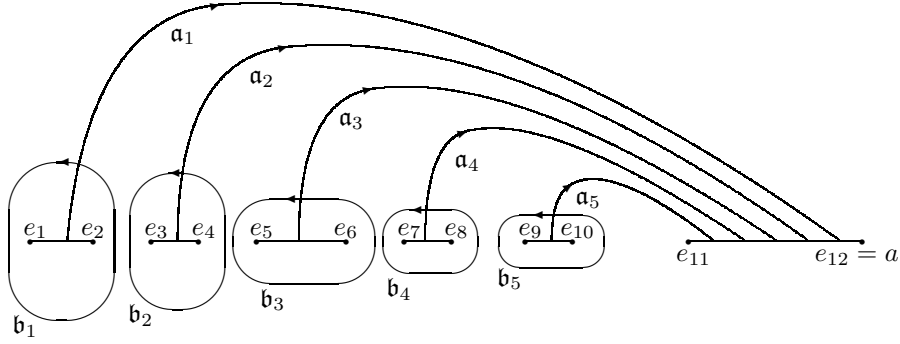
\begin{figure}
\begin{center}
\unitlength 0.7mm
\linethickness{0.4pt}
\begin{picture}(150.00,80.00)
\put(-11.,33.){\line(1,0){12.}}
\put(-11.,33.){\circle*{1}}
\put(1.,33.){\circle*{1}}
\put(-10.,35.){\makebox(0,0)[cc]{$e_1$}}
\put(1.,35.){\makebox(0,0)[cc]{$e_2$}}
\put(-5.,33.){\oval(20,30.)}
\put(-12.,17.){\makebox(0,0)[cc]{$\mathfrak{ b}_1$}}
\put(-5.,48.){\vector(-1,0){.7}}
\put(12.,33.){\line(1,0){9.}}
\put(12.,33.){\circle*{1}}
\put(21.,33.){\circle*{1}}
\put(13.,35.){\makebox(0,0)[cc]{$e_3$}}
\put(22.,35.){\makebox(0,0)[cc]{$e_4$}}
\put(17.,33.){\oval(18.,26.)}
\put(10.,19.){\makebox(0,0)[cc]{$\mathfrak{ b}_2$}}
\put(16.,46.){\vector(-1,0){.7}}
\put(32.,33.){\line(1,0){17.}}
\put(32.,33.){\circle*{1}}
\put(49.,33.){\circle*{1}}
\put(33.,35.){\makebox(0,0)[cc]{$e_5$}}
\put(49.,35.){\makebox(0,0)[cc]{$e_6$}}
\put(41.,33.){\oval(27,16.0)}
\put(35.,22.){\makebox(0,0)[cc]{$\mathfrak{ b}_3$}}
\put(38.,41.){\vector(-1,0){.7}}
\put(60.,33.){\line(1,0){9.}}
\put(60.,33.){\circle*{1}}
\put(69.,33.){\circle*{1}}
\put(61.,35.){\makebox(0,0)[cc]{$e_7$}}
\put(70.,35.){\makebox(0,0)[cc]{$e_8$}}
\put(65.,33.){\oval(18,12.0)}
\put(59.,24.){\makebox(0,0)[cc]{$\mathfrak{ b}_4$}}
\put(62.,39.){\vector(-1,0){.7}}
\put(83.,33.){\line(1,0){9.}}
\put(83.,33.){\circle*{1}}
\put(92.,33.){\circle*{1}}
\put(84.,35.){\makebox(0,0)[cc]{$e_9$}}
\put(93.,35.){\makebox(0,0)[cc]{$e_{10}$}}
\put(88.,33.){\oval(20,10.)}
\put(80.00,26.){\makebox(0,0)[cc]{$\mathfrak{ b}_5$}}
\put(85.,38.){\vector(-1,0){.7}}
\put(114.,33.00){\line(1,0){33.}}
\put(114.,33.){\circle*{1}}
\put(147.,33.){\circle*{1}}
\put(115.,30.){\makebox(0,0)[cc]{$e_{11}$}}
\put(146.,30.){\makebox(0,0)[cc]{$e_{12}=a$}}
\put(18.,72.){\makebox(0,0)[cc]{$\mathfrak{ a}_1$}}
\put(25.,78.){\vector(4,1){0.2}}
\bezier{484}(-4.,33.00)(0.,76.)(25.,78.)
\bezier{816}(25.00,78.)(75.00,82.00)(143.00,33.00)
\put(33.,64.){\makebox(0,0)[cc]{$\mathfrak{ a}_2$}}
\put(38.00,70.){\vector(4,1){0.2}}
\bezier{384}(17.,33.00)(17.,68.)(38.00,70.)
\bezier{516}(38.00,70.)(80.00,74.00)(137.00,33.00)
\put(50.,56.){\makebox(0,0)[cc]{$\mathfrak{ a}_3$}}
\put(54.,62.00){\vector(4,1){0.2}}
\bezier{284}(40.,33.)(40.,60.)(54.,62.)
\bezier{416}(54.,62.00)(82.,66.)(131.,33.00)
\put(72.,48.){\makebox(0,0)[cc]{$\mathfrak{ a}_4$}}
\put(72.,54.){\vector(3,1){0.2}}
\bezier{226}(64.00,33.)(64.00,52.00)(72.,54.)
\bezier{324}(72.,54.)(88.,58.00)(125.00,33.00)
\put(95.,41.){\makebox(0,0)[cc]{$\mathfrak{ a}_5$}}
\put(92.,44.){\vector(3,2){0.2}}
\bezier{186}(88.00,33.)(88.00,42.00)(92.,44.00)
\bezier{266}(92.,44.00)(100.,48)(119.,33.00)
\end{picture}
\end{center}
\caption{A homology basis on a Riemann surface of the
hyperelliptic curve of genus $5$ with the real branching points
$e_1,\ldots,e_{12}=a$ (upper sheet).  The cuts are drawn from
$e_{2i-1}$ to $e_{2i}$ for $i=1,\dots,6$.  The $\mathfrak
a$--cycles are completed on the lower sheet (the picture on lower
sheet is just flipped horizontally).} \label{figure-1}\end{figure}

\section{Differentials}
Traditionally three kinds of differential $1$--forms are
distinguished on a Riemann surface. They are {\em
holomorphic differentials} or differentials of the first kind,
{\em differentials of the second kind}, which have only poles
with zero residues and {\em differentials of the third kind}
which have only first order singularities (locally) and zero total
residue.  Any algebraic differential on $V$ can be presented as
a linear combination of the differentials of three kinds.
\index{differential!of the first kind}
\index{differential!of second kind}\index{differential!of the
third kind}

Let  $\mathrm {d}\Omega_1(x)$ and
$\mathrm {d}\Omega_2(x)$ be arbitrary differentials on $V$, then
the {\em Riemann bilinear relation} holds\index{Riemann!bilinear relation}
\begin{align} \int_{\partial V}\mathrm
{d}\Omega_1(x)\int_{x_0}^x\mathrm
{d}\Omega_2(x')&=\sum_{i=1}^g\left[
\oint_{\mathfrak{ a}_i}\mathrm {d}\Omega_1(x)
\oint_{\mathfrak{ b}_i}\mathrm
{d}\Omega_2(x')\right.\notag\\&-\left.  \oint_{\mathfrak{
b}_i}\mathrm {d}\Omega_1(x) \oint_{\mathfrak{ a}_i}\mathrm
{d}\Omega_2(x')\right], \label{brl} \end{align}
where $\partial V$  \index{domain!fundamental}
is the boundary of the {\em fundamental domain}, the one-connected
domain which is obtained by cutting the surface along all the
cycles of the homology basis.

Below we give explicit construction for differentials on $V$.

\subsection{Holomorphic differentials}
\index{differential!holomorphic} Holomorphic differentials or the
differentials of the first kind, are the differential $1$--forms $
\mathrm{ d}u$,  which can be locally given as $ \mathrm{
d}u=(\sum_{i=0}^ \infty \alpha_i \xi^i) d \xi$ in the vicinity of
any point $ (y, x) $ with some constants $ \alpha_i \in \mathbb{
C}$.  It can be checked directly,  that forms satisfying such a
condition are all of the form $ \sum_{i=0}^{g-1}  \beta_i x^i
\frac{ \mathrm{ d}x}{y}$.  Forms $ \{ \mathrm{ d}u_i \}_{i=1}^g$,
 \begin{equation} \mathrm{ d}u_i= \frac{x^{i-1}
 \mathrm{ d}x}{ y},  \quad i = 1,  \ldots, g\label{du}
 \end{equation} are the set of { \em canonical holomorphic
 differentials} \index{differential!holomorphic! canonical} in
 $H^1 (V, \mathbb{ C}) $.
The $g \times g$--matrices of their
 $\mathfrak{a}$ and $\mathfrak{b}$--periods, \[2 \omega= \left (
\oint_{\mathfrak{ a}_k} \mathrm{ d}u_l \right)_{k,l=1,\ldots,g} ,
 \quad 2 \omega'= \left ( \oint_{\mathfrak{ b}_k} \mathrm{ d}u_l
 \right)_{k,l=1,\ldots,g} \] are non-degenerate.  Under the action
of the transformation $ (2 \omega) ^{-1}$  the vector $ \mathrm{
 d} \mathbf{ u}= ( \mathrm{ d}u_1,  \ldots, \mathrm{ d}u_g)^T$
maps to the vector of normalized holomorphic differentials
\index{differential!holomorphic!normalized}
 $ \mathrm{ d} \mathbf{ v}= ( \mathrm{ d}v_1, \ldots, \mathrm{ d}v_g)
^T$ --- the vector in $ H^1 (V,  \mathbb{ C}) $ to satisfy the
conditions $ \oint_{\mathfrak{ a}_k} \mathrm{ d}v_k= \delta_{kl},
k, l=1 \ldots, g$.  \index{differential!holomorphic!period matrix
of}

Let us denote by $ \mathrm{ Jac} (V) $ the { \em Jacobian} of the curve
$V$,  i.e.  the factor $ \mathbb{ C}^g/ \Gamma$,  where
$ \Gamma=2 \omega \oplus 2 \omega'$ is the lattice generated by the
periods of canonical holomorphic differentials.  \index{Jacobian}

{ \em Divisor} \index{divisor} $ \mathcal{D}$ is a formal sum of
 subvarieties of codimension $1$ with coefficients from $ \mathbb{
 Z}$.  Divisors on Riemann surfaces are given by formal sums of
analytic points $ \mathcal{ D}= \sum_i^n m_i (y_i, x_i) $,  and $
\mathrm{ deg} \mathcal{ D}= \sum_i^n m_i$. The { \em effective
divisor} is such that $m_i>0 \forall i$. \index{divisor!effective}

Let $\mathcal{D}$ be a divisor of degree $0$,
$ \mathcal{D}= \mathcal{ X}- \mathcal{ Z}$,
 \index{divisor!degree of } with $ \mathcal{ X}$  and $\mathcal{
 Z}$ --- the effective divisors $ \mathrm{  deg } \;\mathcal{ X}=
\mathrm{  deg } \;\mathcal{ Z}=n$ presented by $\mathcal{ X}= \{
(y_1,  x_1) , \ldots,   (y_n,  x_n)  \}$ and $\mathcal{ Z}= \{
  (w_1,  z_1) , \ldots, (w_n,  z_n)  \} \in (V)^{n}$, where $
(V)^n$ is the $n$--th symmetric power of $V$.
\index{symmetric!power of $V$}

The {  \em Abel map}   \index{Abel map}
\[  \mathfrak{ A}: (V)^n \rightarrow   \mathrm{ Jac}(V) \]
puts into correspondence the divisor $ \mathcal{ D}$,  with
fixed $  \mathcal{ Z}$,  and the point
$  \boldsymbol{ u} \in   \mathrm{ Jac} (V)
$ according to the
\begin{equation} \boldsymbol{ u}=  \int_{  \mathcal{ Z}}^
\mathcal{ X}
\mathrm{ d}  \mathbf{ u}, \quad  \text{or}  \quad
  u_i=  \sum_{k=1}^n  \int_{z_k}^{x_k}
 \mathrm{ d}u_i, \quad
i=1,   \ldots, g.  \label{jip}
\end{equation}

The {  \em Abel's theorem} says that the points of the
\index{theorem!Abel's} divisors $  \mathcal{ Z}$ and $  \mathcal{
X}$ are respectively the  poles and zeros of a meromorphic
function on $V (y, x) $ if and only if  $  \int^  \mathcal{ X}_
\mathcal{ Z} \mathrm{ d} \mathbf{ u}=0 \mod    \Gamma$.  The { \em
Jacobi inversion problem} \index{Jacobi!inversion problem} is
formulated as the problem of inversion of the map $ \mathfrak{
A}$, when $n=g$ the $  \mathfrak{ A}$ is $1 \to1$, except for so
called { \em special divisors}. In our case special divisors  of
degree $g$ are such that  at least for one pair $j$ and $k   \in 1
\ldots g$ the point $ (y_j, x_j) $ is the image of the
hyperelliptic involution of the  point $ (y_k, x_k) $.
\index{divisor!special} \index{involution!hyperelliptic}

  \subsection{Meromorphic differentials} Meromorphic differentials
or the differentials of the second kind,  are the differential
\index{differential!meromorphic}
$1$-forms $  \mathrm{ d}r$ which can be locally given as $
\mathrm{ d}r= (  \sum_{i=-k}^ \infty  \alpha_i  \xi^i) \mathrm{ d}
\xi$ in the vicinity of any point $ (y, x) $ with some constants $
\alpha_i$,  and $  \alpha_{ (-1) }=0$.  It can be also checked
directly,  that forms satisfying such a condition are all of the
form $ \sum_{i=0}^{g-1} \beta_i x^{i+g}  \frac{ \mathrm{ d}x}{y}$
($  \mod$ holomorphic differential) .  Let us introduce the
  following {  \em canonical Abelian differentials of the second
kind} \index{differential!Abelian!of second kind}
 \begin{equation}
\mathrm{ d}r_j= \sum_{k=j}^{2g+1-j} (k+1-j)
   \lambda_{k+1+j} \frac{x^k \mathrm{ d}x }{ 4y},   \quad j=1,
  \ldots, g.  \label{rrj}
 \end{equation} We denote their matrices
  of $\mathfrak a$ and $\mathfrak b$--periods, \[2  \eta= \left (-
\oint_{\mathfrak{ a}_k} \mathrm{ d}r_l  \right)_{k,l=1,\ldots,g} ,
\quad 2  \eta' = \left (- \oint_{\mathfrak{b}_k} \mathrm{ d}r_l
\right)_{k,l=1,\ldots,g}  .  \]

\subsection{Fundamental $2$--differential of the
  second kind}  \index{2-differential!fundamental!of the second
kind} For any pair of analytic points $ \{ (y_1, x_1) ,  (y_2,
 x_2)   \} \in  (V) ^2$ we introduce  function  $F (x_1, x_2) $
 defined by the conditions \begin{eqnarray} (  \text{i}) .  &&F
 (x_1, x_2) =F (x_2, x_1) ,    \nonumber   \\ (  \text{ii}) .  &&F
 (x_1, x_1) =2f (x_1) ,   \nonumber   \\ (  \text{iii}) .  &&
 \frac{ \partial F (x_1, x_2) }{   \partial x_2}   \big|_{x_2=x_1}
 =
\frac{  \mathrm{ d}f (x_1) }{   \mathrm{ d}x_1}.  \label{propf}
  \end{eqnarray}

Such $F (x_1, x_2) $ can be presented in the following
equivalent forms
  \begin{eqnarray}
F (x_1, x_2) &=&2y_2^2+2 (x_1-x_2) y_2  \frac{  \mathrm{ d} y_2}{
  \mathrm{ d}x_2}  \nonumber  \\
&+& (x_1-x_2) ^2  \sum_{j=1}^gx_1^{j-1}  \sum_{k=j}^{2g+1-j} (k-j+1)
   \lambda_{k+j+1}x_2^k,
  \label{Formf-1}  \\
F (x_1, x_2) &=&2  \lambda_{2g+2}x_1^{g+1}x_2^{g+1}\cr&+&
\sum_{i=0}^{g}x_1^ix_2^i (2  \lambda_{2i}+  \lambda_{2i+1}
(x_1+x_2) ) .  \label{Formf-2}   \end{eqnarray}

Properties   \eqref{propf} of  $F (x_1, x_2) $ permit to construct
the {\em global Abelian $2$--dif\-f\-er\-en\-ti\-al of the second
kind} with the unique  pole of order $2$  along $x_1=x_2$ :
\index{2-differential!global!of second kind}
  \begin{equation}
\mathrm{d} \widehat \omega (x_1, x_2) =  \frac{2y_1 y_2+F (x_1,
  x_2) }{ 4 (x_1-x_2) ^2}  \frac{  \mathrm{ d}x_1}{ y_1}  \frac{
\mathrm{ d}x_2}{y_2}, \label{omega-1}   \end{equation} which
  expands in the vicinity of the pole as   \[
\mathrm{d} \widehat  \omega (x_1, x_2) =
  \left (  \frac{1}{(  \xi- \zeta) ^2}+O (1)    \right)
\mathrm{ d}  \xi \mathrm{ d}  \zeta, \] where $ \xi$ and $  \zeta$
are the local coordinates at the points $x_1$ and $x_2$
correspondingly.

Using the   \eqref{Formf-1},  rewrite the   \eqref{omega-1} in the form
\begin{equation}
\mathrm{d} \widehat{  \omega} (x_1, x_2) =  \frac{  \partial}{
\partial x_2}   \left (  \frac{y_1+y_2}{ 2y_1 (x_1-x_2) }
\right)   \mathrm{ d}x_1 \mathrm{ d}x_2+ \mathrm{ d}  \mathbf{
 u}^T (x_1)   \mathrm{ d} \mathbf{ r} (x_2) ,   \label{omega-2}
 \end{equation} where the
differentials $  \mathrm{ d}  \mathbf{ u},   \mathrm{ d}  \mathbf{
r}$ are as above. So,  the  periods of this $2$-form  (the double
integrals $  \oint  \oint \widehat \omega (x_1, x_2) $)  are
 expressible in terms of $ (2  \omega, 2  \omega') $ and $ (-2
 \eta, -2 \eta') $,  e.g.,  we have for $\mathfrak a$--periods:  \[
\left \{ \oint_{\mathfrak{ a}_i}  \oint_{\mathfrak{ a}_k}\widehat{
  \omega} (x_1, x_2) \right  \}_{i, k=1,   \ldots, g}= -4
\omega^T \eta.  \]

\begin{lemma} $2g  \times
2g$--matrix $  \mathfrak{ m}=  \left (  \begin{array} {cc}  \omega&
\omega'  \\ \eta&  \eta'  \end{array}  \right) $
satisfies to
\begin{equation}
\mathfrak{ m}
\begin{pmatrix} 0&-1_g\\1_g&0
\end{pmatrix} \mathfrak{ m}^T
=
-\frac{1}{2}\pi \imath
\begin{pmatrix} 0&-1_g\\1_g&0\end{pmatrix}.\label{wei1}
\end{equation}
\label{LPSR} \end{lemma}

\begin{proof}Let
$\varkappa$ be such a symmetric $g\times g$--matrix that the second
kind differential \begin{equation}\mathrm{d}\omega (x_1, x_2)
=\mathrm{d}\widehat\omega (x_1, x_2)
+2\mathrm {d}\mathbf{u}^T(x_1)\varkappa\mathrm {d}
\mathbf{ u}(x_2)\label{Eq}\end{equation} is  normalized by the
condition \begin{equation} \oint_{\mathfrak{ a}_j} \mathrm{d}\omega
(x_1, x_2)=0,\quad j=1,\ldots, g \quad \forall
x_1\in V.
\end{equation} Taking into the account (\ref{omega-2})
compute $\mathfrak a$ and $\mathfrak b$--periods over the variable $x_2$
from both the sides of (\ref{Eq}). Because of the
equalities $$\oint_{\mathfrak{ b}_j}\mathrm{d}\omega(x_1,x_2)=2\pi \imath
v_j(x_1),\quad j=1,\ldots,g,$$ which follows from the bilinear
Riemann relation (\ref{brl}) we obtain \begin{equation}
\eta=2\varkappa\omega,\quad \eta'=2\varkappa\omega'- \frac{\pi
i}{2 }(\omega^{-1})^T. \label{PSP1} \end{equation}
Hence (\ref{wei1}) holds. \end{proof}

\subsection{Differentials of the third kind}
\index{differential!of third kind}
Differentials of the third kind
are the differential 1-forms $  \mathrm{ d}  \Omega$ to have only
poles of order $1$ and $0$ total residue,  and so are locally
given in the vicinity of any of the poles as
$  \mathrm{ d}  \Omega= (  \sum_{i=-1}^  \infty  \alpha_i  \xi^i) d  \xi$
with some constants $  \alpha_i$,  $  \alpha_{-1}$ being nonzero. Such
forms  ($  \mod$  holomorphic differential)  may be presented as:
  \[
  \sum_{i=0}^{n}   \beta_i
  \left (  \frac{y+y^+_i}{x-x^+_i}-  \frac{y+y^-_i}{x-x^-_i}  \right)
  \frac{  \mathrm{ d}x}{y},
  \]
where $ (y^  \pm_i, x^  \pm_i) $ are the analytic points of the poles of
positive  (respectively,  negative)  residue.

Let us introduce the canonical differential of the third kind
  \begin{equation}
  \mathrm{ d}   \Omega (x_1, x_2) =
  \left (  \frac{y+y_1}{x-x_1}-  \frac{y+y_2}{x-x_2}  \right)
  \frac{  \mathrm{ d} x}{2y}
+\int\limits_{x_1}^{x_2}\mathrm{d}\mathbf{r}(z)^T\mathrm{d}\mathbf{u}(x),
   \label{third}
  \end{equation}
for this differential we have

 $$  \int_{x_3}^{x_4}  \mathrm{ d}
  \Omega (x_1, x_2) =  \int_{x_1}^{x_2}  \mathrm{ d}   \Omega (x_3,
x_4). $$.

\section{Riemann $\theta$--function} \index{Riemann!$\theta$--function}
The standard $  \theta$--function $   \theta
( \boldsymbol{ v}| \tau) $ on $  \mathbb{ C}^g  \times   \mathcal{
H}_g$ is defined by its Fourier series,  \begin{equation} \theta (
  \boldsymbol{ v}|   \tau) =  \sum_{  \boldsymbol{ m} \in
\mathbb{ Z}^g}  \mathrm{ exp}  \;  \pi i  \left  \{ \boldsymbol{
m}^T  \tau   \boldsymbol{ m}+2  \boldsymbol{ v}^T \boldsymbol{ m}
  \right  \}\label{thecan} \end{equation} The $ \theta$--function
possesses the periodicity properties $  \forall k   \in 1,   \ldots, g$
  \begin{eqnarray*} &&  \theta (v_1,    \ldots, v_k+1,   \ldots,
  v_g|  \tau) =  \theta (  \boldsymbol{ v}|  \tau) ,   \\ &&
\theta (v_1+  \tau_{1k},   \ldots, v_k+  \tau_{kk},   \ldots, v_g+
\tau_{gk}|  \tau) =  \mathrm{ e}^{-i  \pi   \tau_{kk}-2  \pi i v_k}
\theta (  \boldsymbol{ v}|  \tau) .  \end{eqnarray*}
\index{$\theta$--function}

$  \theta$--functions with characteristics $[  \varepsilon]=
  \left[   \begin{array}{c}  \varepsilon'  \\  \varepsilon   \end{array}  \right]
=  \left[   \begin{array}{ccc}  \varepsilon_1'&  \ldots&  \varepsilon_g'
  \\  \varepsilon_1&  \ldots&  \varepsilon_g  \end{array}  \right]   \in   \mathbb{
R}^{2g}$
\[
  \theta[  \varepsilon] (  \boldsymbol{ v}|   \tau) =  \sum_{
  \boldsymbol{ m}  \in   \mathbb{ Z}^g}  \mathrm{ exp}  \;  \pi i
\left  \{  (  \boldsymbol{ m}+  \varepsilon') ^T  \tau (
 \boldsymbol{ m}+  \varepsilon') +2 (  \boldsymbol{ v}+
 \varepsilon) ^T (  \boldsymbol{ m}+  \varepsilon')   \right  \},
  \] for which
the periodicity properties are
  \begin{eqnarray}
 && \theta[  \varepsilon] (v_1,   \ldots, v_k+1,   \ldots, v_g|  \tau)
=  \mathrm{ e}^{2  \pi i  \varepsilon_k'}  \theta[  \varepsilon]
 (  \boldsymbol{
v}|  \tau) ,  \label{thetachar2} \\
 &&\theta[  \varepsilon] (v_1+  \tau_{1k},    \ldots, v_k+
\tau_{kk},   \ldots, v_g +  \tau_{gk}|  \tau)\cr
&&\qquad  =  \mathrm{ e}^{-i
\pi   \tau_{kk}-2  \pi i v_k-2  \pi i \varepsilon_k}  \theta [  \varepsilon ]  (
  \boldsymbol{ v}|  \tau) \notag.  \end{eqnarray}

Let $ \boldsymbol{
w}^T= (w_1, \ldots, w_g)   \in \mathrm{ Jac} (V) $ be some fixed
vector, the function, \[   \mathcal{R} (x) =  \theta \left (
\int_{x_0}^x \mathrm{ d} \mathbf{ v} -   \boldsymbol{ w}| \tau
\right) , \quad x  \in V, \]
where $\theta$ is canonical theta
function of the first order \eqref{thecan}, is called { \em
Riemann $ \theta$--function}.
\index{$\theta$--function!periodicity property of}

The Riemann $\theta$--function $\mathcal{ R} (x)$ is either identically
$0$,  or it has exactly $g$ zeros $x_1,   \ldots, x_g  \in V$,  for which
the {  \em Riemann  vanishing theorem} says that
   \index{Riemann!vanishing theorem}
\[\sum_{k=1}^g\int_{x_0}^{x_i}\mathrm{
d} \mathbf{ v}= \boldsymbol{ w}+  \boldsymbol{ K}_{x_0}, \] where
$ \boldsymbol{ K}^T_{x_0}= (K_1,   \ldots, K_g) $ is the vector of
Riemann constants with  respect to the base point $x_0$ and is
defined by the formula \begin{equation} K_j=  \frac{1+
\tau_{jj}}{ 2}- \sum_{l  \neq j} \oint_{\mathfrak{ a}_l} \mathrm{
d}v_l (x)   \int_{x_0}^x \mathrm{ d}v_j, \quad j=1,   \ldots,  g.
\label{Rconst}
\end{equation}

\index{vector of Riemann constants}
\subsection{$\theta$--functions with half--integer characteristics}

\index{$\theta$--function!with characteristic!half--integer}
It follows from (\ref{thetachar2}) that
\begin{equation}
\theta[\varepsilon]({-\boldsymbol v}|\tau)= \mathrm {e}^
{-4\pi \imath  \boldsymbol{\varepsilon}^T\boldsymbol{\varepsilon}'}
\theta[\varepsilon](\boldsymbol{v}|\tau),
\end{equation}
thus the function $\theta[\varepsilon](\boldsymbol{v}|\tau)$,
with the characteristic  $[\varepsilon]$ consisting only of
half-integers, is either even  when $4\pi \imath
\boldsymbol{\varepsilon}^T\boldsymbol{\varepsilon}'$ is an even
integer and odd otherwise. We recall that half-integer
characteristic is called even or odd whenever $4\pi \imath
\boldsymbol{\varepsilon}^T\boldsymbol{\varepsilon}'$ is even or
odd, and among $4^g$ half integer characteristics there are $
\frac{1}{2} (4^g+2^g) $ even characteristics and $ \frac{1}{2}
(4^g-2^g) $ odd characteristics.
\index{$\theta$--function! even, odd}

The half--integer characteristics are
  \index{characteristic!of $\theta$--function} connected with
branching points of the $V(x,y)$ as follows.  Identify the every
branching point $e_j$ with a vector \index{characteristic! of
$\theta$--function!half-integer}

$$
\mathfrak{A}_j=\mathfrak{A}({e}_j,a)=\int_{a}^{e_j}
\mathrm{d}\mathbf{u} =\boldsymbol{\varepsilon}_j +\tau
\boldsymbol{\varepsilon}'_j\in \mathrm{Jac}(V),
$$
and choose the path of integration, such that all the non
zero
components of the vectors $\boldsymbol{\varepsilon}$ and
$\boldsymbol{\varepsilon}'$ are equal to $\frac12$, which is always
possible.

The $2\times g$ matrices
$$
[{\mathfrak A}_j]= \left[\begin{array}{c} {\boldsymbol{
\varepsilon}'}^T_j \\
 \boldsymbol{ \varepsilon}^T_j
\end{array}  \right]  =\left[\begin{array}{ccc}
\varepsilon'_{j1}&\ldots& \varepsilon'_{jg} \\
 \varepsilon_{j1}&\ldots& \varepsilon_{jg}
\end{array}  \right]
$$
will serve as characteristics for the $\theta$-functions.

Let us identify the half periods   $\boldsymbol{\mathfrak A}_{i}$,
$i=1,\ldots,2g+1$ (see e.g. \cite{fk80}). Evidently, $[{\mathfrak
A}_{2g+2}]=[0]$. Further
\begin{align*}
\boldsymbol{\mathfrak A}_{2g+1}&=\boldsymbol{\mathfrak A}_{2g+2}-
\sum_{k=1}^g\int\limits_{e_{2k-1}}^{e_{2k}}\mathrm{d}\mathbf{v}
=\sum_{k=1}^g\boldsymbol{f}_k,\\
 [{\mathfrak A}_{2g+1}] &=
\frac12\left[\begin{array}{cccc}1&1&\ldots&1\\0&0&\ldots&0
\end{array}\right],\\
\boldsymbol{\mathfrak A}_{2g} &=\boldsymbol{\mathfrak A}_{2g+1} -
\int\limits_{e_{2g+1}}^{e_{2g}}\mathrm{d}\mathbf{v}
=\sum_{k=1}^g\boldsymbol{f}_k+\boldsymbol{\tau}_g,\\
[ {\mathfrak A}_{2g} ]&= \frac12 \left[\begin{array}{ccccc}
1&1&\ldots&1&1\\
0&0&\ldots&0&1
\end{array}\right],\\
\boldsymbol{\mathfrak A}_{2g-1} &=\boldsymbol{\mathfrak A}_{2g} -
\int\limits_{e_{2g}}^{e_{2g-1}}\mathrm{d}\mathbf{v}
=\sum_{k=1}^{g-1}\boldsymbol{f}_k+\boldsymbol{\tau}_g,\\
[ {\mathfrak A}_{2g-1} ]&= \frac12 \left[\begin{array}{ccccc}
1&1&\ldots&1&0\\
0&0&\ldots&0&1
\end{array}\right],\\\end{align*}
where here and below $\boldsymbol{f}_k = \frac{1}{2}(\delta_{1k},
\ldots, \delta_{gk})^T$ and $\boldsymbol{\tau}_k$ is the $k$-th
column vector of the matrix $\tau$.

We have for arbitrary $k>1$
\begin{eqnarray*}
[{\mathfrak A}_{2k+1}]=
\frac12\left[\overbrace{\begin{array}{cccc}
                   1&1&\ldots&1\\
                   0&0&\ldots&0 \end{array}}^{k}
\begin{array}{cccc}0&0&\ldots&0\\
                   1&0&\ldots&0\end{array} \right],
\end{eqnarray*}
\begin{eqnarray*}
[{\mathfrak A}_{2k+2}]=
\frac12\left[\overbrace{\begin{array}{cccc}
                   1&1&\ldots&1\\
                   0&0&\ldots&0 \end{array}}^{k}
\begin{array}{cccc}1&0&\ldots&0\\
                   1&0&\ldots&0\end{array} \right]
\end{eqnarray*}
and  finally,
\[
[{\mathfrak
A}_2]=\frac12\left[\begin{array}{cccc}1&0&\ldots&0\\
                                      1&0&\ldots&0
\end{array}\right],\quad
[{\mathfrak A}_1]=\frac12\left[\begin{array}{cccc}0&0&\ldots&0\\
                                      1&0&\ldots&0\end{array}\right].\]

 \index{characteristic!of $\theta$--function!connection
with branching points}
 For example, we have for the homology basis  drawn on the Figure \ref{figure-1}  we have:
\begin{align*}
&[{\mathfrak
A}_{1\phantom{0}}]=\tfrac12\left[{}_1^0{}_0^0{}_0^0{}_0^0{}_0^0\right],\quad
[{\mathfrak
A}_{2\phantom{0}}]=\tfrac12\left[{}_1^1{}_0^0{}_0^0{}_0^0{}_0^0\right],\quad
[{\mathfrak
A}_{3\phantom{0}}]=\tfrac12\left[{}_0^1{}_1^0{}_0^0{}_0^0{}_0^0\right],\\
&[{\mathfrak
A}_{4\phantom{0}}]=\tfrac12\left[{}_0^1{}_1^1{}_0^0{}_0^0{}_0^0\right],\quad
[{\mathfrak
A}_{5\phantom{0}}]=\tfrac12\left[{}_0^1{}_0^1{}_1^0{}_0^0{}_0^0\right],\quad
[{\mathfrak
A}_{6\phantom{0}}]=\tfrac12\left[{}_0^1{}_0^1{}_1^1{}_0^0{}_0^0\right],\\
&[{\mathfrak
A}_{7\phantom{0}}]=\tfrac12\left[{}_0^1{}_0^1{}_0^1{}_1^0{}_0^0\right],\quad
[{\mathfrak
A}_{8\phantom{0}}]
=\tfrac12\left[{}_0^1{}_0^1{}_0^1{}_1^1{}_0^0\right],\quad
[{\mathfrak
A}_{9\phantom{0}}]=\tfrac12\left[{}_0^1{}_0^1{}_0^1{}_0^1{}_1^0\right],\\
&[{\mathfrak
A}_{10}]=\tfrac12\left[{}_0^1{}_0^1{}_0^1{}_0^1{}_1^1\right],\quad
[{\mathfrak
A}_{11}]=\tfrac12\left[{}_0^1{}_0^1{}_0^1{}_0^1{}_0^1\right],\quad
[{\mathfrak
A}_{12}]=\tfrac12\left[{}_0^0{}_0^0{}_0^0{}_0^0{}_0^0\right].
\end{align*}
Remark, that the characteristics with even indices (beside the
last one which is zero) are odd. These characteristics correspond
to the branching points $e_{2n}=a_n$, $n=1,\ldots,g$.

Let us choose $a$ as the base point $x_0$. Then the vector of
Riemann constants has the form
\begin{equation} \boldsymbol{ K}_{a}=
\sum_{k=1}^g \int_{a}^{a_{i}} \mathrm{ d}  \mathbf{ v}.
\label{rvector}
\end{equation}

Generally, $4^g$ half-periods are in $1\to 1$
\index{periods!half--periods} correspondence with $4^g$ partitions
\begin{eqnarray*} {\bf I}_m = {\bf J}_m \cup {\bf \tilde J}_m ,
\qquad {\bf J}_m &=&\{ i_1, \ldots, i_{g+1-2m}\},\\ {\bf \tilde
J}_m &=& \{ j_1, \ldots, j_{g+1+2m}\} \nonumber\end{eqnarray*} of
$\{1,\ldots,2g+2\}$ for integers  $\left[\frac{g+1}{2}\right] \ge m \ge
0$.  We will further use the correspondence of partitions to
characteristics $[\varepsilon]$ defined by \begin{equation*}
\boldsymbol{\varepsilon} +\tau
\boldsymbol{\varepsilon}'=\sum_{k=1}^{g+1-2m}\mathfrak{A}_{i_k}
- \boldsymbol{K}_a,
\end{equation*}
where  $\boldsymbol{K}_a$  is the vector of Riemann constants.
This means that we translate the origin from which we
calculate the characteristics by the vector of Riemann constants.
\index{vector of Riemann constants}
Let us denote by $[{\mathfrak A}_1],\ldots,[{\mathfrak A}_{2g+2}]$
the characteristics put into the correspondence to the
branching points of the curve $V(x,y)$ according to the rules
assumed above.
Denote the characteristics  of the point
$\sum_{k=1}^n{\mathfrak A}_{i_k}$ by $[\sum_{k=1}^n{\mathfrak
A}_{i_k}],\quad k=2,\ldots,2g+2$.

The non-vanishing values of $\theta[\epsilon]$--functions with
characteristics (which will be supposed further to be only
half-integer) and their derivatives at the zero argument are
called {\em $\theta$--constants}.\index{$\theta$--constants}

The {\em Thomae formulae for the even $\theta$ constants}
\index{Thomae formulae! for even $\theta$ constants}
\cite{tho870} give the
expressions of $\theta$-constants in terms of branching points
$e_1,\ldots e_{2g+2}$ of the curve $V$ as follows.  Let
\begin{eqnarray*} \mathbf{I}_0 = \mathbf{J}_0 \cup \mathbf{\tilde
J}_0 , \qquad \mathbf{J}_0 &=&\{ i_1, \ldots, i_{g+1}\},\\
\mathbf{\tilde J}_0 &=& \{ j_1, \ldots, j_{g+1}\}
\nonumber\end{eqnarray*} be such a partition to which the
characteristic of the point
\begin{equation*}
\sum_{k=1}^{g+1}\mathfrak{A}_{i_k} - {\boldsymbol
K}_a\end{equation*} corresponds. Then
\begin{equation}
\theta^2[\varepsilon](\boldsymbol{0}|\tau)
=\epsilon\frac{\mathrm{det} \;(2\omega)}{(\imath\pi)^g}
\sqrt{\prod_{1\leq i_k<i_l\leq g+1}(e_{i_k}-e_{i_l}) \prod_{1\leq
j_k<j_l\leq g+1}(e_{j_k}-e_{j_l})}, \label{th}\end{equation}
where $\epsilon^4=1$.  When
one of the branching points is moved to infinity the corresponding
multiplier in the formula (\ref{th}) is to be omitted.

The quotients of $\theta$--functions or {\em $\theta$-quotients}
 with half-integer characteristics are rational
\index{$\theta$--quotients} functions on the symmetric power
\index{symmetric!power of the curve} $(V)^g$ of the curve. In
particular, if $[{\mathfrak A}_i],[{\mathfrak A}_j], [{\mathfrak
A}_k]$, $i,j,k=1,\ldots,g$ be the characteristics of the branching
points $a_i$, $a_j$, $a_k$, then
\index{characteristic!of branching point}
\begin{align}
&\dfrac{\theta^2[{\mathfrak
A}_k](\boldsymbol{v})}{\theta^2(\boldsymbol{v})}
=-\frac{q_k^2}{P'(a_k)}
\prod_{l=1}^g(a_k-x_l),\quad l=1\ldots,g,\label{tq1}\\
&\dfrac{\theta^2[{\mathfrak A}_i+{\mathfrak A}_j]
(\boldsymbol{v})\theta^2(\boldsymbol{v})}
{\theta^2[{\mathfrak A}_i](\boldsymbol{v})\theta^2
[{\mathfrak A}_j](\boldsymbol{v})}
=
(a_i-a_j)\left\{\frac12
\sum_{l=1}^g\frac{y_l}{(x_l-a_i)(x_l-a_j)
\mathcal{P}'(x_l)}\right\}^2,\label{tq2}\\
&\dfrac{\theta^2[{\mathfrak A}_i+{\mathfrak A}_j+{\mathfrak A}_k]
(\boldsymbol{v})\theta^2(\boldsymbol{v})}
{\theta^2[{\mathfrak A}_i](\boldsymbol{v})
\theta^2[{\mathfrak A}_j](\boldsymbol{v})\theta^2[{\mathfrak A}_k](\boldsymbol{v})}
=
(a_i-a_j)(a_j-a_k)(a_k-a_i)\notag\\ &\hskip3cm \times\left\{\frac12
\sum_{l=1}^g\frac{y_l}{(x_l-a_i)(x_l-a_j)(x_l-a_k)
\mathcal{P}'(x_l)}\right\}^2,\label{tq3}\\
i\neq j\neq k\in \{1\ldots,g\};\notag
\end{align} 
where
\begin{equation}
q_k=\epsilon\sqrt[4]{\frac{P'(a_k)}{Q(a_k)}},\quad
k=1,\ldots, g \label{q_k}\end{equation}
and
\begin{equation*}
\mathcal{P}(x)=\prod_{i=1}^g(x-x_i),
\end{equation*}
and $\epsilon^8=1$.

In general, if the $\theta$-quotients with the characteristics
$[{\mathfrak A}_i]$ and $[{\mathfrak A}_i+{\mathfrak A}_j]$ are
known then the $\theta$-quotients
$\theta[\sum_{n=1}^k\mathfrak{A}_{i_n}
](\boldsymbol{v})/\theta(\boldsymbol{v})$, it is possible to
express in terms of them the quotients for $k>2$. More precisely,
let $k=2n$; let us divide the set of characteristics
$\{[{\mathfrak A}_{i_l}]\}|_{1}^{2n}$ in two subsets
\begin{equation} \{[{\mathfrak
A}_{i_l}]\}|_{1}^{2n}=\{[{\mathfrak A}_{r_1}],\ldots, [{\mathfrak
A}_{r_n}]\}\cup \{[{\mathfrak A}_{s_1}],\ldots,[{\mathfrak
A}_{s_n}]\}.  \label{par1}\end{equation}
Then we have \cite{ba98}, p.358
\begin{align}
{\theta }&^{n-1}(\boldsymbol v){\theta }[{\mathfrak
A}_1+\ldots+{\mathfrak A}_{k}](\boldsymbol v)= \label{evendec}\\&
\epsilon\sqrt{ W_{r_1\ldots r_n} W_{s_1\ldots s_n}} \det
\begin{pmatrix} {\theta }[{\mathfrak A}_{r_1}+{\mathfrak
A}_{s_1}](\boldsymbol{v})&\ldots& {\theta }[{\mathfrak
A}_{r_1}+{\mathfrak A}_{s_n}](\boldsymbol{v})\\ \vdots&{}&\vdots\\
{\theta}[{\mathfrak A}_{r_n}+{\mathfrak
A}_{s_1}](\boldsymbol{v})&\ldots& {\theta}[{\mathfrak
A}_{r_n}+{\mathfrak A}_{s_n}]({\boldsymbol
v}).\end{pmatrix}\notag
\end{align}
Next, let
$k=2n+1$; let us divide the set of characteristics $\{[{\mathfrak
 A}_{i_l}]\}|_{1}^{2n+1}$ in the subsets
\begin{equation}
\{[{\mathfrak A}_{i_l}]\}|_{1}^{2n}=\{[{\mathfrak
A}_{r_1}],\ldots, [{\mathfrak A}_{r_{n}}]\}\cup \{[{\mathfrak
A}_{s_1}],\ldots,[{\mathfrak A}_{s_{n+1}}]\}.
\label{par2}\end{equation}
And we have
\begin{align}
{\theta}&^{n}(\boldsymbol v){\theta}[{\mathfrak
A}_1+\ldots+{\mathfrak A}_{k}](\boldsymbol v)=\label{odddec}\\
&
\epsilon\sqrt{ W_{r_1\ldots r_n} W_{s_1\ldots s_{n+1}}}
\det
\begin{pmatrix} {\theta}[{\mathfrak
A}_{r_1}+{\mathfrak A}_{s_1}](\boldsymbol{v})&\ldots&
{\theta}[{\mathfrak A}_{r_1}+{\mathfrak
A}_{s_{n+1}}](\boldsymbol{v})\\ \vdots&{}&\vdots\\
{\theta}[{\mathfrak A}_{r_n}+{\mathfrak
A}_{s_1}](\boldsymbol{v})&\ldots& {\theta}[{\mathfrak
A}_{r_n}+{\mathfrak A}_{s_{n+1}}](\boldsymbol{v})\\
{\theta}[{\mathfrak A}_{s_1}](\boldsymbol{v})&\ldots&
{\theta}[{\mathfrak
A}_{s_{n+1}}](\boldsymbol{v})
\end{pmatrix}.
\notag
\end{align}
In these formulae $W_{i_1\ldots i_m}$ is the determinant of
Vandermond matrix of the elements $e_{i_1},\ldots,e_{i_m} $ and
$\epsilon^8=1$. Both formulae are
independent on the particular choice of partitions \eqref{par1}
and \eqref{par2}.                 \index{matrix!Vandermond}

\subsection{Vanishing properties of $\theta$--functions}
\index{$\theta$--function!vanishing properties} To complete the
construction of $\sigma$--functions we have to investigate the
vanishing properties of $\theta$--functions.

First we introduce special determinants. Let $k=1,\ldots,g$,
$l=1,\ldots,\left[\frac{g+1}{2}\right]$. Let $l<k$ and
\[\{i_1,\ldots,i_{k-l}\}\cup\{i_{k-l+1},\ldots,i_{g}\}=\{1,2,\ldots,g\}
\]
is the partition of $\{1,2,\ldots,g\}$ with $i_1<i_2\ldots<i_{k-l}$
and   $i_{k-l+1}<i_{k-l+2}\ldots<i_{g}$.

Introduce for any pair $(k,l)$ the functions
\begin{equation}
\mathfrak
{B}_{k;l}^{i_{k-l+1},\ldots,i_{g}}(\boldsymbol{u};\boldsymbol{a})
=\frac{1}{ W_{ i_1\ldots  i_{k-l} }(\boldsymbol{a})}
\det\left(\begin{array}{cccc}
u_1&u_2&\ldots&u_k\\
u_2&u_3&\ldots&u_{k+1}\\
\vdots&\vdots&\ldots&\vdots\\
u_l&u_{l+1}&\ldots&u_{k+l-1}\\
1&a_{i_1}&\ldots&a_{i_1}^{k-1}\\
\vdots&\vdots&\ldots&\vdots\\
1&a_{i_{k-l}}&\ldots&a_{i_{k-l}}^{k-1}
\end{array}\right),\label{newmatrix}
\end{equation}
where $W_{ i_1\ldots  i_{k-l} }(\boldsymbol{a})$ is the Vandermond
determinant of the elements $a_{i_1},\ldots,a_{i_{k-l}}$, i.e. $k$
the dimension of the determinant, $l$ is the number of lines
involving variables $u_k$, $i_{k-l+1},\ldots,i_g$ are the indices
of omitted branching points.

In particular, at $k=l=[\frac{g+1}{2}]$ the function
$\mathfrak{ B}_{k;l}^{1,\ldots,g}(\boldsymbol{u})$ is
the determinant of $[\frac{g+1}{2}]\times [\frac{g+1}{2}]$
the Hankel matrix of the elements
$u_1,\ldots,u_g$,\index{matrix!Hankel} \begin{equation}
\mathcal{H}_{\left[\frac{g+1}{2}\right]}\;=\; \begin{pmatrix}
u_1&u_2&\ldots&u_{\left[\frac{g+1}{2}\right]}\\
u_2&u_3&\ldots&u_{\left[\frac{g+1}{2}\right]+1}\\
\vdots&\vdots&\ldots&\vdots\\
u_{\left[\frac{g+1}{2}\right]}
&u_{\left[\frac{g+1}{2}\right]+1}&\ldots
&u_{2\left[\frac{g+1}{2}\right]-1}\end{pmatrix}.
\label{hankel}\end{equation}
which is frequently used.

At $k=[\frac{g+1}{2}], l=[\frac{g+1}{2}]-1$ the function
$ \mathfrak{ B}_{k;l}^{1,\ldots,i-1,i+1,\ldots,g}
(\boldsymbol{u};a_i)$
arises, in particular, in the theory of infinite continued
fractions and the orthogonal polynomials \cite{ak65}.  The
generalization (\ref{newmatrix}) of these determinants naturally
arises  when the $\theta$--function vanishing properties are
discussed and to the authors knowledge these special determinants
have not been treated \cite{mu28}.

The functions $\mathfrak
{B}_{k;l}^{i_1,\ldots,i_{k-l}} (\boldsymbol{u};\boldsymbol{a})$
(below we shall omit the argument
$(\boldsymbol{u};\boldsymbol{a})$)  satisfy to the
following equalities.

Let
\[\{i_1,\ldots,i_{n}\}\cup\{j_{1},\ldots,j_{n}\}=\{1,2,\ldots,2n\}
\subset\{1,\ldots,g\}
,\]
where $n\leq[\frac{g+1}{2}]$ is the partition with
$i_1<i_2\ldots<i_{n}$ and   $j_{1}<j_{2}\ldots<j_{n}$.
Then
\begin{eqnarray}
\det(\mathfrak{B}^{i_k,j_l}_{g-1;1})_{k,l=1,\ldots,n}=
\det(\mathfrak{B})^{i_1,\ldots, i_n, j_1,\ldots, j_n}_{g-n,n}
W_{i_1\ldots i_n}W_{j_1\ldots j_n}.\label{matmat}
\end{eqnarray}

Let
\[\{i_1,\ldots,i_{n}\}\cup\{j_{1},\ldots,j_{n},j_{n+1}\}=\{1,2,\ldots,2n+1\}
\subset\{1,\ldots,g\}\]
is the partition with
$i_1<i_2\ldots<i_{n}$ and   $j_{1}<j_{2}\ldots<j_{n+1}$.
Then
\begin{eqnarray}
&&\det\left(\begin{array}{ccl}
{}&(\mathfrak{B}^{i_k,j_l}_{g-1;1})_{k,l=1,\ldots,n+1}&
\\ \mathfrak{ B}^{j_1}_{g,1}&\ldots&\mathfrak{
B}^{j_{n+1}}_{g,1}\end{array}\right)
\label{mat}\\&&=
\det(\mathfrak{ B})^{i_1\ldots i_n, j_1\ldots
j_n,j_{n+1}}_{g-n,n+1} W_{i_1\ldots i_n}W_{j_1\ldots j_{n+1}}
\nonumber\end{eqnarray}

In particular, when $2n+1=g$, then  $(\mathfrak{B})^{i_1,\ldots,
i_n, j_1,\ldots, j_n}_{g-n,n+1}$ is $(n+1)\times (n+1)$ Hankel
matrix of elements $u_1,\ldots, u_g$.

Consider as the examples the cases $g=3, n=1$  and $g=4, n=2$.
For $g=3, n=1$ define the partition $\{1,2,3\}=\{1\}\cup\{2,3\}$.
The equality (\ref{mat}) then reads

\begin{eqnarray*} &&\left|\begin{array} {cc}
{\left|\begin{array}{cc}
u_1&u_2\\
1  &a_3\end{array}\right|}&
{\left|\begin{array}{cc}
u_1&u_2\\
1  &a_2\end{array}\right|}\\
{}&{}\\
\frac{1}{a_1-a_3}{\left|\begin{array}{ccc}
u_1&u_2&u_3\\
1  &a_1&a_1^2\\
1  &a_3&a_3^2\end{array}\right|}&
\frac{1}{a_1-a_2}{\left|\begin{array}{ccc}
u_1&u_2&u_3\\
1  &a_1&a_1^2\\
1  &a_2&a_2^2\end{array}\right|}
\end{array}\right|\\
&&=(a_2-a_3)\left|\begin{array}{cc}u_1&u_2\\u_2&u_3\end{array}
\right|.
    \end{eqnarray*}

In the case $g=4, n=2$ and partition
$\{1,2,3,4\}=\{1,2\}\cup\{3,4\}$ the equality (\ref{matmat}) reads
\begin{eqnarray*} &&\left|\begin{array} {cc}
\frac{1}{a_2-a_4}{\left|\begin{array}{ccc}
u_1&u_2&u_3\\
1  &a_2&a_2^2\\
1  &a_4&a_4^2\end{array}\right|}&
\frac{1}{a_2-a_3}{\left|\begin{array}{ccc}
u_1&u_2&u_3\\
1  &a_2&a_2^2\\
1  &a_3&a_3^2\end{array}\right|}\\
{}&{}\\
\frac{1}{a_1-a_4}{\left|\begin{array}{ccc}
u_1&u_2&u_3\\
1  &a_1&a_1^2\\
1  &a_4&a_4^2\end{array}\right|}&
\frac{1}{a_1-a_3}{\left|\begin{array}{ccc}
u_1&u_2&u_3\\
1  &a_1&a_1^2\\
1  &a_3&a_3^2\end{array}\right|}
\end{array}\right|\\
&&=(a_1-a_2)(a_3-a_4)\left|\begin{array}{cc}u_1&u_2\\u_2&u_3\end{array}
\right|.
\end{eqnarray*}

\begin{prop}Let $\mathfrak{A}_{i}$, $i=1,\ldots,g$ are
characteristics associated with the branching points $a_i$,
$i=1,\ldots,g$. Then the lowes term of the expansion of
$\theta[\mathfrak{A}_{i_1}+\ldots+\mathfrak{A}_{i_k}] (\boldsymbol
{u})$ is described by the formula
\begin{equation}
\theta[\mathfrak{A}_{i_1}+\ldots+\mathfrak{A}_{i_{2n+\upsilon}}]
(\boldsymbol {u})
=\frac{1}{C_{i_1\ldots i_{i_{2n+\upsilon}} }}
{\mathfrak B}^{i_{2n+\upsilon+1},\ldots,
i_{g}}_{g-n+\upsilon;n}(\boldsymbol{ u};\boldsymbol{ a} ),
\label{vanish} \end{equation}
where
\begin{equation} C_{i_1\ldots
i_k}=\frac{\epsilon\prod_{m=1}^kq_{i_m}}
{\theta(\boldsymbol{  0}|\tau)\sqrt{ \prod_{1\leq m<n\leq
k}(a_{i_m}-a_{i_n})}}\label{ccc}
\end{equation} $\upsilon$ is equal $0$ or
$1$  and $\epsilon^8=1$.
\end{prop}
\begin{proof}
Put $x_i=a_i+t_i^2$, $i=1,\ldots,g$, where $\boldsymbol{
t}^T=(t_1,\ldots,t_g)$ is small. Then we get from (\ref{jip})
\begin{eqnarray}
t_i&=&\sqrt{\frac{Q(a_i)}{P'(a_i)}}
\frac{1}{W_{1\ldots\widehat{i}\ldots g}(\boldsymbol{a})}
\det\left( \begin{array}{cccc}u_1&u_2&\ldots&u_g \\ 1
&a_1&\ldots&a_1^{g-1}\\ \vdots &\vdots&\ldots&\vdots\\ 1
                    &a_{i-1}&\ldots&a_{i-1}^{g-1}\\ 1
                    &a_{i+1}&\ldots&a_{i+1}^{g-1}\\ \vdots
                    &\vdots&\ldots&\vdots\\ 1
                    &a_g&\ldots&a_g^{g-1}\end{array}
\right)+o(\boldsymbol{u}^3)\cr
&=&\sqrt{\frac{Q(a_i)}{P'(a_i)}} \mathfrak{
B}^i_{g;1}+o(\boldsymbol{u}^3) \label{jip1} \end{eqnarray}

Let $k=1$, then we derive from the (\ref{tq1}) the expansion
\begin{equation}
\frac{\theta[\mathfrak{ A}_k](\boldsymbol{u})}{\theta(\boldsymbol{u})}=
q_kt_k+ o(\boldsymbol{t}^3), \quad
k=1,\ldots, g, \end{equation}
where $q_k$ are given in \eqref{q_k}, which due to
\eqref{jip1} comes to \eqref{vanish}.

At $k=2$, it follows from the (\ref{tq1}),
\begin{eqnarray}
&&\frac{\theta[\mathfrak{ A}_i+\mathfrak{ A}_j](\boldsymbol{v})}
{\theta(\boldsymbol{v})}=\epsilon\sqrt{a_i-a_j}
\frac{\theta[\mathfrak{
A}_i](\boldsymbol{v})}{\theta(\boldsymbol{v})}
\frac{\theta[\mathfrak{
A}_j](\boldsymbol{v})}{\theta(\boldsymbol{v})}\cr
&&\times\left(
\sqrt{\frac{Q(a_i)}{P'(a_i)}}\frac{1}{a_i-a_j}\frac{1}{t_i}
+\sqrt{\frac{Q(a_j)}{P'(a_j)}}\frac{1}{a_j-a_i}\frac{1}{t_j}
\right).\end{eqnarray}
Taking into the account the expansion which we have for $k=1$ and
the formula \[ {\mathfrak B}^i_{g;1}-{\mathfrak
B}^j_{g;1}={\mathfrak B}^{i,j}_{g-1;1}(a_j-a_i) \] we
obtain (\ref{vanish}) for $k=2$.  To complete the proof it remains
to use (\ref{evendec},\ref{odddec}) and the
determinant identities (\ref{matmat}) and  (\ref{mat}).
\end{proof}

For example, at $g=4$ we have
\begin{eqnarray*}
&&C_{1234}\theta\left[\sum_{i=1}^4\mathfrak{A}_i\right](\boldsymbol{u}|\tau)=
\left|\begin{array}{cc}u_1&u_2\\u_2&u_3\end{array}\right|
+O(\boldsymbol{ u}^4), \cr
&&C_{123}\theta\left[\sum_{i=1}^3\mathfrak{A}_i\right](\boldsymbol{u}|\tau)=
\left|\begin{array}{ccc}u_1&u_2&u_3\\u_2&u_3&u_4\\1&a_4&a_4^2
\end{array}\right|+O(\boldsymbol{ u}^4),\cr
&&C_{12}\theta\left[\sum_{i=1}^2\mathfrak{A}_i\right](\boldsymbol{u}|\tau)
=\frac{1}{W_{34}}
\left|\begin{array}{ccc}u_1&u_2&u_3\\1&a_3&a_3^2\\1&a_4&a_4^2
\end{array}\right|+O(\boldsymbol{ u}^3),\cr
&&C_{1}\theta\left[\mathfrak{A}_1\right]\boldsymbol{u}|\tau)
=\frac{1}{W_{234}}
\left|\begin{array}{cccc}u_1&u_2&u_3&u_4\\1&a_2&a_2^2&a_2^3
\\1&a_3&a_3^2&a_3^3\\1&a_4&a_4^2&a_4^3
\end{array}\right|+O(\boldsymbol{ u}^3).
\end{eqnarray*}

The formulae for the first term of the expansion
are equivalent to that of given in \cite{kl88,bur88} and
\cite{ba98}, where in contrast with our exposition the
determinantial character of the expansions was not clarified.

We remark that the knowledge of explicit form of the first term of
$\theta$-expansion  permit to obtain various relations between
$\theta$--constants, from which we mention as the most important
{\em Jacobi derivative formula} \cite{kw15, fa79, ig82}.
\index{Jacobi!derivative formula}

\section{Construction of the hyperelliptic $\sigma$--function}
Let $2\omega$ and $-2\eta$ are $g\times g$ period matrices of the
differentials (\ref{du}) and (\ref{rrj}) correspondingly.
\index{$\sigma$--function!hyperelliptic!construction of}

\begin{definition}
The {\em hyperelliptic functions $\sigma[\varepsilon]$} with the
characteristics $[\varepsilon]=[\mathfrak{ A}_{i_1}+\ldots
\mathfrak{ A}_{i_k}]$
are defined by the formula
\begin{equation*}
\sigma[\varepsilon](\boldsymbol{u})=C_{i_1\ldots
i_k}\mathrm{exp}\{{\boldsymbol
u}^T\kappa\boldsymbol{u}\}\theta[\varepsilon]
(\boldsymbol{v}|\tau), \end{equation*} where
\begin{equation*} \varkappa=\eta(2\omega)^{-1}\quad\text{and}\quad
\boldsymbol{v}=(2\omega)^{-1}\boldsymbol{u}
\end{equation*}
and $C_{i_1\ldots i_k}$ is given in (\ref{ccc}).
\index{$\sigma$--function!hyperelliptic!with characteristic}

The {\em hyperelliptic fundamental $\sigma$--function}
is defined by the formula
\begin{equation*}
\sigma(\boldsymbol{u})=C\mathrm{exp}\{{\boldsymbol
u}^T\kappa\boldsymbol{u}\}\theta[\varepsilon]
((2\omega)^{-1}\boldsymbol{u}|\tau), \end{equation*}
where $\varkappa=\eta(2\omega)^{-1}$ the characteristic
$[\varepsilon]$ equals to $[\mathfrak{ A}_1+\ldots+\mathfrak{
A}_g]$ and is the characteristic of the vector of Riemann
constants $\boldsymbol{K}_a$. The constant $V$ is given as follows
\begin{equation} C=\sqrt{\frac {\pi^g}{\mathrm{det}
(2\omega)}}\frac{\epsilon}{\sqrt[4]{\prod_{1\leq i<j\leq N}(a_i-a_j)}},
\label{cfun} \end{equation} where $\epsilon^4=1$, $N=2g+2$
for a curve without
a branching point at infinity and  $N=2g+1$ otherwise.
\label{fund2} \end{definition}
\index{$\sigma$--function!hyperelliptic!fundamental}

\begin{prop}\label{expansion_1}
In the vicinity of $\boldsymbol{u}=0$ the lowest term of the
expansion of fundamental hyperelliptic function
$\sigma(\boldsymbol{u})$ is the determinant of Hankel matrix
of order $\left[\frac{g+1}{2}\right]$ (\ref{hankel}).
In particular, for small genera we have,
\begin{align*}
\sigma(\boldsymbol{u})&=u_1+ o(\boldsymbol{u}^3)\quad
\text{for}\; g=1\; \text{and}\; 2 \\
\sigma(\boldsymbol{u})&=u_1u_3-u_2^2+ o(\boldsymbol{u}^4)
\quad
\text{for} \; g=3\; \text{and} \;4 \\
\sigma({\boldsymbol
u})&=-u_3^3+2u_2u_3u_4-u_1u_4^2-u_2^2u_5+u_1u_3u_5+
o(\boldsymbol{u}^5)\;\;  \text{for}\; g=5
\;\text{and}\; 6.\end{align*}\end{prop}
\index{matrix!Hankel}
\begin{proof}
Note first, that $C=C_{1\ldots g}$,
where the constants $V$ and $C_{1\ldots g}$ are given by
(\ref{cfun}) and (\ref{ccc}) at $k=g$ correspondingly. This can be
seen from the Thomae formula (\ref{th}) written for the
characteristic $[0]$ which corresponds to the partition to of the
branching points into groups $$\{a_1,\ldots,a_g, a\}\cup
\{b_1,\ldots,b_g, b\},$$  \begin{equation}
\theta[0](\boldsymbol{0}|\tau)=\epsilon
\sqrt{ \frac {\mathrm{det}\;(2\omega) }   {(\imath \pi) ^g}   }
 \sqrt[4]{  \prod_{1\leq i<j\leq g+1} (a_i-a_j)
 \prod_{1\leq i<j\leq g+1}  (b_i-b_j)        },
 \label{th0}\end{equation}
where $\epsilon^8=1$ and we denoted $b=b_{g+1}$.
The substitution of (\ref{th0}) to (\ref{ccc}), where $k=g$ is
put, comes to (\ref{cfun}).

The fact that the first term of the expansion on the fundamental
$\sigma$-function is the determinant of the Hankel matrix
directly follows from the vanishing properties of the
hyperelliptic $\theta$ functions given above.  \end{proof}

\index{$\sigma$--function!first term of expansion }

Let us give examples of the leading term of expansion of
                    $\sigma$-function with characteristics for the
                    case $g=5$ \begin{eqnarray*}
\sigma\left[\mathfrak{A}_1\right](\boldsymbol{u})&=&\frac{1}{W_{2345}}\left|
\begin{array}{ccccc}u_1& u_2 &u_3    &u_4   &u_5\\
                     1 & a_2 &a_2^2  &a_2^3 &a_2^4\\
                     1 & a_3 &a_3^2  &a_3^3 &a_3^4\\
                     1 & a_4 &a_4^2  &a_4^3 &a_4^4\\
                     1 & a_5 &a_4^2  &a_5^3 &a_5^4
\end{array}\right|+O(\boldsymbol{ u}^3). \\\\\\
\sigma\left[\sum_{l=1}^2\mathfrak{A}_l\right](\boldsymbol{u})&=&\frac{1}{W_{345}}\left|
\begin{array}{cccc}u_1& u_2 &u_3    &u_4   \\
                     1 & a_3 &a_3^2  &a_3^3 \\
                     1 & a_4 &a_4^2  &a_4^3\\
                     1 & a_5 &a_4^2  &a_4^3
\end{array}\right|+O(\boldsymbol{ u}^3). \\\\\\
\sigma\left[\sum_{l=1}^3\mathfrak{A}_l\right]
(\boldsymbol{u})&=&\frac{1}{a_4-a_5}\left|
\begin{array}{cccc}u_1& u_2 &u_3    &u_4   \\
                   u_2&  u_3 &u_4    &u_5   \\
                     1 & a_4 &a_4^2  &a_4^3\\
                     1 & a_5 &a_4^2  &a_4^3
\end{array}\right|+O(\boldsymbol{ u}^4). \\\\\\
\sigma\left[\sum_{l=1}^4\mathfrak{A}_l\right]
(\boldsymbol{u})&=&\left|
\begin{array}{ccc}u_1& u_2 &u_3      \\
                   u_2&  u_3 &u_4    \\
                     1 & a_5 &a_4^2
\end{array}\right|+O(\boldsymbol{ u}^4).
\end{eqnarray*}
and  finally,
\[
[{\mathfrak
A}_2]=\frac12\left[\begin{array}{cccc}1&0&\ldots&0\\
                                      1&0&\ldots&0
\end{array}\right],\quad
[{\mathfrak
A}_1]=\frac12\left[\begin{array}{cccc}0&0&\ldots&0\\
                                      1&0&\ldots&0
\end{array}\right].
\]


\section[Fundamental $2$--differential]{Realization of the fundamental $2$--differential of the
second kind}
\index{2-differential!of second kind}

Let us introduce the following notations.
\index{$\wp$--function!Kleinian} Kleinian $\zeta$ and
\index{$\zeta$--function!Kleinian}$ \wp$--functions are defined as
logarithmic derivatives of the fundamental $\sigma$--function
\begin{gather*} \zeta_i ( \boldsymbol{  u}) = \frac{  \partial
\mathrm{ ln}  \; \sigma ( \boldsymbol{  u}) }{   \partial u_i},
\quad i  = 1, \ldots, g;  \\  \wp_{ij} ( \boldsymbol{  u}) = -
  \frac{ \partial^2 \mathrm{  ln}  \; \sigma (  \boldsymbol{ u}) }
{ \partial u_i \partial u_j}, \;   \wp_{ijk} (  \boldsymbol{ u})
=- \frac{ \partial^3 \mathrm{ ln}  \;  \sigma (  \boldsymbol{  u})
}{ \partial u_i \partial u_i  \partial  u_k}  \ldots, \\ i, j, k,
\ldots  = 1, \ldots, g.  \end{gather*}

The functions $  \zeta_i (  \boldsymbol{  u}) $ and $
 \wp_{ij} (  \boldsymbol{  u}) $ have the
following periodicity properties
  \begin{eqnarray*}   \zeta_i (  \boldsymbol{
u}+  2\boldsymbol{   \Omega} (  \boldsymbol{  m},
    \boldsymbol{
m}') ) &= &  \zeta_i (  \boldsymbol{  u}) +2E_i (  \boldsymbol{
m},   \boldsymbol{  m}') ,
    \quad i  = 1,   \ldots, g, \\
  \wp_{ij} (  \boldsymbol{  u}+2 \boldsymbol{\Omega} (
  \boldsymbol{  m},
   \boldsymbol{  m}') ) &=&  \wp_{ij} (
\boldsymbol{  u}) ,    \quad i, j   = 1,   \ldots, g,
\end{eqnarray*}
where $E_i (\boldsymbol{m},\boldsymbol{m}') $ is the $i$-th
component of the vector
$E(\boldsymbol{m},\boldsymbol{m}')=\eta\boldsymbol{m}+\eta'\boldsymbol{m}'$
and $\boldsymbol{\Omega}(\boldsymbol{m},\boldsymbol{m}')=
\omega\boldsymbol{m}+\omega'\boldsymbol{m}'$.

The construction is based on
the following \begin{theorem} \label{the-S} Let $ (y (a_0),  a_0)
$,  $ (y, x) $ and $ (  \nu, \mu) $  be arbitrary distinct  points
on $V$ and let $  \{ (y_1, x_1),   \ldots,  (y_g, x_g)   \}$  and
  $  \{ ( \nu_1, \mu_1) , \ldots, (  \nu_g, \mu_g)   \}$ be
arbitrary sets of distinct points  $  \in (V) ^g$.  Then the
following relation is valid \begin{eqnarray} &&\qquad \lefteqn{
 \int_{  \mu}^x \sum_{i=1}^g \int_{ \mu_i}^{x_i} \frac{2yy_i+F (x,
 x_i) }{4 (x-x_i) ^2} \frac{ \mathrm{ d}x}{y} \frac{ \mathrm{
 d}x_i}{y_i}} \label{r11} \\ &=& \mathrm{ ln }   \; \left  \{
  \frac{  \sigma \left ( \int_{a_0}^x  \mathrm{ d} \mathbf{ u} -
 \sum_{i=1}^g \int_{a_i}^{x_i}  \mathrm{ d} \mathbf{ u}  \right)
  }{  \sigma \left (  \int_{a_0}^x  \mathrm{ d} \mathbf{ u}-
\sum_{i=1}^g \int_{a_i}^{  \mu_i}  \mathrm{ d} \mathbf{ u}
  \right) } \right \}-  \mathrm{  ln }   \;   \left  \{ \frac{
\sigma \left ( \int_{a_0}^{  \mu}  \mathrm{ d}  \mathbf{ u}-
  \sum_{i=1}^g \int_{a_i}^{x_{i}}  \mathrm{ d}  \mathbf{ u}
\right) }{  \sigma \left (  \int_{a_0}^{  \mu}  \mathrm{ d}
  \mathbf{ u}-  \sum_{i=1}^g  \int_{a_i}^{  \mu_i}  \mathrm{ d}
\mathbf{ u}  \right) }  \right  \} , \nonumber \end{eqnarray}
 where the function $F (x, z) $ is given by   \eqref{Formf-2}.
\end{theorem}

  \begin{proof} Let us consider the sum
  \begin{equation}
  \sum_{i=1}^g  \int_{  \mu}^x  \int_{  \mu_i}^{x_i}   \left[
 \mathrm{d} \hat{\omega} (x, x_i)  +2
  \mathrm{ d}  \mathbf{ u}^T (x)
  \varkappa
  \mathrm{ d}  \mathbf{ u} (x_i)   \right],
  \label{r10}
  \end{equation}
with $ \hat{ \omega} (  \cdot,   \cdot) $ given by
\eqref{omega-2}.  It is the normalized Abelian integral of the
third kind with the logarithmic residues in the points $x_i$ and $
\mu_i$. By Riemann vanishing theorem\index{Riemann!vanishing
theorem} we can express \eqref{r10} in terms of Riemann $
 \theta$--functions as \begin{equation} \mathrm{ ln} \left  \{
\tfrac{{  \displaystyle{ \theta}}   \left (  \int_{a_0}^x \mathrm{
d}  \mathbf{ v}- ( \sum_{i=1}^g \int_{a_0}^{x_i} \mathrm{ d}
\mathbf{ v}- \mathbf{ K}_{a_0}) \right) } {{ \displaystyle{
\theta}}   \left ( \int_{a_0}^{x} \mathrm{ d} \mathbf{ v}- (
\sum_{i=1}^g \int_{a_0}^{  \mu_i} \mathrm{ d} \mathbf{ v}-
\mathbf{ K}_{a_0}) \right) } \right  \} - \mathrm{ ln} \left  \{
\tfrac{{ \displaystyle{  \theta}} \left ( \int_{a_0}^{  \mu}
 \mathrm{ d} \mathbf{ v}- (  \sum_{i=1}^g \int_{a_0}^{x_i}
\mathrm{ d} \mathbf{ v}-  \mathbf{ K}_{a_0}) \right) }{{
\displaystyle{  \theta}} \left (  \int_{a_0}^{  \mu} \mathrm{ d}
  \mathbf{ v}- (  \sum_{i=1}^g  \int_{a_0}^{  \mu_i} \mathrm{ d}
\mathbf{ v}-  \mathbf{ K}_{a_0})   \right) }  \right \},
\label{in-depends} \end{equation} and to obtain right hand side of
\eqref{r11} we have to combine the \eqref{fund2}, expression of
  the vector $  \mathbf{ K}_{a_0}$ \eqref{rvector}, matrix $
  \varkappa = \eta(2  \omega) ^{-1}$ and Lemma \ref{LPSR}. Left
hand side  of \eqref{r11} is obtained using \eqref{omega-1}.
 \end{proof}

 The fact,  that right hand side of the
  \eqref{r11} is independent on the arbitrary point $a_0$,  to be
 employed further,  has its origin in the properties of the vector
 of Riemann constants. Consider the difference $  \mathbf{ K}_{a_0}-
   \mathbf{ K}_{a_0'}$ of vectors of Riemann constants with arbitrary
 base points $a_0$ and $a_0'$ by   \eqref{Rconst} we find
   \[
   \mathbf{ K}_{a_0}-
   \mathbf{ K}_{a_0'} = (g-1)   \int_{a_0'}^{a_0}  \mathrm{ d}  \mathbf{ v},
   \]
this property provides that
   \[
  \int_{a_0}^{x}  \mathrm{ d}  \mathbf{
 v}- (  \sum_{i=1}^g  \int_{a_0}^{x_i}  \mathrm{ d}  \mathbf{ v}-  \mathbf{
K}_{a_0}) =
   \int_{a_0'}^{x}  \mathrm{ d}  \mathbf{
 v}- (  \sum_{i=1}^g  \int_{a_0'}^{x_i}  \mathrm{ d}  \mathbf{ v}-  \mathbf{
K}_{a_0'})
   \]
for arbitrary $x_i, $ with $i  \in 0,   \ldots, g$  on $V$,  so the
arguments of $  \sigma$'s in   \eqref{r11} which are linear
transformations by $2  \omega$ of the arguments of $  \theta$'s in
\eqref{in-depends},  do not depend on $a_0$.

  \begin{cor}  \label{cor-P}  From Theorem   \ref{the-S} for arbitrary
distinct $ (y (a_0) , a_0) $ and $ (y, x) $ on $V$ and  arbitrary set of
distinct points $  \{ (y_1, x_1)   \ldots,  (y_g, x_g)   \}
  \in  (V) ^g$ follows:
  \begin{eqnarray}  && \sum_{i, j=1}^g  \wp_{ij}   \left (
\int_{a_0}^x \mathrm{ d}  \mathbf{ u}+   \sum_{k=1}^g
  \int_{a_k}^{x_k} \mathrm{ d}  \mathbf{ u}  \right)
x^{i-1}x_r^{j-1}\cr&&=  \frac{F (x, x_r) -2yy_r}{ 4 (x-x_r) ^2},
\; r=1,   \ldots, g.  \label{principal} \end{eqnarray} \end{cor}
\index{$\wp$--function!principal relation}
 \begin{proof}
Taking the partial derivative $  \partial^2/  \partial x_r  \partial x$
from the both sides of   \eqref{r11} and using the hyperelliptic
involution \index{involution!hyperelliptic} $  \phi (y, x) = (-y,
 x) $ and $  \phi (y (a_0) , a_0) = (-y (a_0) , a_0) ) $ we obtain
\eqref{principal}.  \end{proof}

In the case $g=1$ the formula   \eqref{principal} is
actually the
addition theorem for the Weierstrass elliptic functions,
\index{addition theorem!for Weierstrass $\wp$--functions}
  \[
  \wp (u+v) =-  \wp (u) -  \wp (v) +  \frac{1}{4}  \left[  \frac{  \wp' (u) -  \wp' (v) }{
  \wp (u) -  \wp (v)  }  \right]^2
  \]
on the  elliptic curve  $y^2=f (x) =4x^3-g_2
x-g_3$.

Now we can give the expression for $ \mathrm{d} \omega (x, x_r) $
in terms of Kleinian functions.  We send the base point $a_0$ to
the branch place $a$,  and  for $r  \in 1,   \ldots, g $ the
 fundamental $2$--differential of the second kind is given by   \[
 \mathrm{d}\omega (x, x_r) = \sum_{i, j=1}^g  \wp_{ij}   \left (  \int_a^x
  \mathrm{ d}  \mathbf{ u}- \sum_{k=1}^g  \int_{a_k}^{x_k}
  \mathrm{ d}  \mathbf{ u}  \right)   \frac{x^{i-1}  \mathrm{
d}x}{y}  \frac{x_r^{j-1} \mathrm{ d}x_r}{y_r}.  \] \begin{cor} $
  \forall r   \neq s   \in 1,   \ldots, g$ \begin{eqnarray}
  \sum_{i, j=1}^g  \wp_{ij}
  \left (  \sum_{k=1}^g  \int_{a_k}^{x_k}  \mathrm{ d}  \mathbf{ u}  \right)
x_s^{i-1}x_r^{j-1} =  \frac{F (x_s, x_r) -2y_sy_r}{ 4 (x_s-x_r) ^2}
.  \label{principal11}
  \end{eqnarray}
  \end{cor}
  \begin{proof}
In   \eqref{principal} we have for $s  \neq r$
  \begin{eqnarray*}
&  \int  \limits_{  \phi (a_0) }^x
  \mathrm{ d}  \mathbf{ u}+   \sum  \limits_{k=1}^g
  \int  \limits_{a_k}^{x_k}  \mathrm{ d}  \mathbf{
u}
 =-2  \omega (  \int  \limits_{a_0}^{  \phi (x) }  \mathrm{ d}  \mathbf{
 v}- (  \sum  \limits_{i=1}^g  \int  \limits_{a_0}^{x_i}
   \mathrm{ d}  \mathbf{ v}-  \mathbf{
K}_{a_0}) ) =  \\
&-2  \omega (  \int  \limits_{x_s}^{  \phi (x) }  \mathrm{ d}  \mathbf{
 v}- (  \sum  \limits_{  \begin{subarray}{l}i=1  \\
i  \neq s  \end{subarray}}^g  \int  \limits_{x_s}^{x_i}
   \mathrm{ d}  \mathbf{
v}-  \mathbf{ K}_{x_s}) ) =  \int  \limits_{a_s}^{x_s}  \mathrm{
d} \mathbf{ u}+  \sum  \limits_{  \begin{subarray}{l}i=1  \\ i
\neq s  \end{subarray}}^g  \int  \limits_{a_i}^{x_i} \mathrm{ d}
 \mathbf{u} \end{eqnarray*} and the change of notation $x   \to
x_s$ gives \eqref{principal11}.    \end{proof}

  \section{Solution of the Jacobi inversion problem}
\index{Jacobi!inversion problem!solution of}
The equations of Abel map in conditions of Jacobi
inversion problem
  \begin{equation}
u_i=  \sum_{k=1}^g  \int_{a_k}^{x_k}   \frac{x^{i-1}  \mathrm{ d}x}{y},
  \label{Abel-map}
  \end{equation}
are invertible if the points $ (y_k,  x_k) $ are distinct and
$  \forall j, k   \in 1,   \ldots, g  \;   \phi (y_k, x_k)   \neq (y_j, x_j)  $.  Using
  \eqref{principal} we find the solution of Jacobi inversion problem
on the curves with $a=  \infty$ in a very effective form.

  \begin{theorem}\label{J-s} The Abel pre-image of the point $
\boldsymbol{ u}  \in  \mathrm{ Jac} (V) $ is given  by the set $
\{ (y_1, x_1) ,   \ldots,  (y_g, x_g)   \}  \in  (V) ^g$,  where $
\{x_1,   \ldots, x_g  \}$ are the zeros of the polynomial
  \begin{equation}   \mathcal{ P} (x;  \boldsymbol{  u}) =0  \label{x},
     \end{equation}
where   \begin{equation}   \mathcal{ P} (x;  \boldsymbol{  u}) =x^g-x^{g-1}
   \wp_{g, g}
 (  \boldsymbol{ u}) -x^{g-2}  \wp_{g, g-1} (  \boldsymbol{ u}) -
    \ldots-  \wp_{g, 1} (  \boldsymbol{
u}) ,   \label{p}
  \end{equation}
and  $  \{y_1,   \ldots, y_g  \}$ are given by
  \begin{equation} y_k=-  \frac{  \partial   \mathcal{
P} (x;  \boldsymbol{  u}) }{   \partial u_g}  \Bigl\lvert_{x=x_k},
  \; \label{y} \end{equation} \end{theorem} \begin{proof} We tend
in \eqref{principal} $a_0  \to a=  \infty$.
Then we take
  \begin{equation}
  \lim_{x   \to  \infty}
  \frac{F (x, x_r) }{4x^{g-1} (x-x_r) ^2}=
  \sum_{i=1}^g  \wp_{gi} (  \boldsymbol{ u}) x_r^{i-1}.  \label{limit}
  \end{equation} The limit in the left hand side of  \eqref{limit} is
equal to $x_r^g$,  and we obtain   \eqref{x}.

We find from   \eqref{Abel-map},
  \[  \sum_{i=1}^g  \frac{x^{k-1}_i}{ y_i}
  \frac{  \partial x_i}{   \partial u_j}=  \delta_{jk},   \qquad
\frac{  \partial x_k}{   \partial u_g}=  \frac{y_k
}{  \prod_{i  \neq
k} (x_k-x_i) }.  \]
On the other hand we have
  \[
  \frac{  \partial   \mathcal{ P}}{   \partial u_g}
\Bigl\lvert_{x=x_k}= -  \frac{  \partial x_k}{   \partial u_g}
\prod_{i  \neq k} (x_i-x_k) ,   \] and we obtain   \eqref{y}.
\end{proof}

Let us denote by $  \boldsymbol{  \wp}$,  $  \boldsymbol{  \wp}'$ the
$g$--dimensional vectors,
  \[
  \boldsymbol{   \wp}=  \left (  \wp_{g1},   \ldots,   \wp_{gg}
  \right) ^T,
  \quad
  \boldsymbol{   \wp}'=  \frac{  \partial   \boldsymbol{   \wp}}
  {   \partial u_g}
  \] and the companion matrix
\index{matrix!companion}   \cite{hj86} of the polynomial
$  \mathcal{ P} (z;  \boldsymbol{ u}) $,  given by   \eqref{p}
  \[  \mathcal{ C}=  \mathcal{ B}_g+   \boldsymbol{   \wp}
\mathbf{ e}_g^T, \quad  \text{where}  \quad  \mathcal{ B}_g=
  \sum_{k=1}^g \mathbf{ e}_k  \mathbf{ e}^T_{k-1}.  \]   The
companion matrix $  \mathcal{ C}$ has the property
\begin{equation} x_k^n=  \mathbf{ X}^T_k  \mathcal{ C}^{n-g+1}
\mathbf{ e}_g=  \mathbf{ X}^T_k  \mathcal{ C}^{n-g} \boldsymbol{
  \wp},    \quad   \forall n  \in   \mathbb{ Z},    \label{prop-c}
\end{equation} with the vector $  \mathbf{ X}_k^T= (1, x_k,
\ldots, x_k^{g-1}) $,  where $x_k$ is one of the roots of
\eqref{x}.  From   \eqref{principal11} we find $-2y_ry_s=4
(x_r-x_s) ^2  \sum_{i=1}^{g}  \wp_{ij} (  \boldsymbol{ u})
x_r^{i-1}x_s^{j-1}-F (x_r, x_s) $.  Introducing matrices $  \Pi= (
\wp_{ij}) $, $  \Lambda_0=  \mathrm{ diag} (  \lambda_{2g-2},
\ldots,   \lambda_0) $ and $  \Lambda_1=  \mathrm{ diag}  \; (
\lambda_{2g-1},   \ldots,   \lambda_1) $, we have,  taking into
 account   \eqref{prop-c}, \begin{eqnarray*} \lefteqn{2  \mathbf{
  X}^T_r  \boldsymbol{ \wp}'{  \boldsymbol{   \wp}'}^T  \mathbf{
  X}_s=-4  \mathbf{ X}^T_r ({  \mathcal{ C}}^2  \Pi-2   \mathcal{
C}  \Pi   \mathcal{ C}^T+  \Pi {  \mathcal{ C}^T}^2 )    \mathbf{
X}_s }   \nonumber  \\&&+ 4  \mathbf{ X}^T_r (  \mathcal{ C}
\boldsymbol{ \wp}  \boldsymbol{   \wp}^T+  \boldsymbol{   \wp}
  \boldsymbol{   \wp}^T  \mathcal{ C}^T)   \mathbf{ X}_s +2
\mathbf{ X}^T_r  \Lambda_0  \mathbf{ X}_s+  \mathbf{ X}^T_r (
\mathcal{ C}  \Lambda_1+  \Lambda_1  \mathcal{ C}^T)   \mathbf{
X}_s.  \end{eqnarray*} Whence,   (see   \cite{be96b}) :
  \begin{cor}
The relation
  \begin{eqnarray}
2  \boldsymbol{
  \wp}'{  \boldsymbol{   \wp}'}^T&=&-4 ({  \mathcal{
C}}^2  \Pi-2   \mathcal{ C}  \Pi   \mathcal{ C}^T+
   \Pi {  \mathcal{ C}^T}^2 )
+ 4 (  \mathcal{ C}  \boldsymbol{
  \wp}  \boldsymbol{   \wp}^T\cr
  &+&  \boldsymbol{   \wp}
  \boldsymbol{ \wp}^T  \mathcal{ C}^T) +  \mathcal{ C}  \Lambda_1+
  \Lambda_1 \mathcal{ C}^T+2  \Lambda_0.  \label{principium}
\end{eqnarray} connects odd functions $  \wp_{ggi}$ with poles
order $3$ and  even functions $  \wp_{jk}$ with poles of order $2$
in the field of meromorphic functions on $  \mathrm{ Jac} (V) $.
  \end{cor}

  \begin{definition}  \label{umbral_D}
The {  \em umbral derivative}
 \index{umbral derivative}    \cite{ro84}
  $D_s (p (z) ) $ of a polynomial    $$p (z) =  \sum_{k=0}^n p_k z^k$$
is given by   \[ D_s p (z) =
  \left (  \frac{p (z) }{z^s}  \right) _+=
\sum_{k=s}^{n}p_k z^{k-s}, \] where $ (  \cdot) _+$ means taking
the purely polynomial part.  \end{definition}

Considering polynomials $p=  \prod_{k=1}^{n} (z-z_k) $
and ${  \tilde p}= (z-z_0) p$,  the elementary properties of $D_s$ are
immediately deduced:
  \begin{eqnarray}
&D_s (p) =z D_{s+1} (p) +p_s=z D_{s+1} (p)  +
S_{n-s} (z_1,   \ldots, z_n) ,   \nonumber  \\
&D_s ({  \tilde
p})\nonumber\\
= &(z-z_0) D_s (p) +p_{s-1}\nonumber\\=& (z-z_0) D_s (p) +S_{n+1-s}
  (z_1, \ldots, z_n) ,   \label{umbral_2} \end{eqnarray} where
  $S_l ( \cdots) $ is the $l$--th order elementary symmetric
function of its variables  times $ (-1) ^l$  (we assume $S_0 (
\cdots) =1$) .

From   \eqref{umbral_2} we see that $S_{n-s} (z_0,   \ldots,   \hat
z_l,   \ldots, z_n) =   \left (D_{s+1} ({  \tilde
p}) |_{z=z_l}  \right) $. This is particularly useful to  write
down the inversion of   \eqref{y}
  \begin{equation}
  \wp_{ggk} (  \boldsymbol{ u}) =  \sum_{l=1}^g y_l
  \left (  \frac{D_k (P (z) ) }
{  \frac{  \partial}{  \partial z}P (z) }  \Bigg|_{z=x_l}  \right) ,
  \label{inversiony}
  \end{equation}
where $P (z) =  \prod_{k=1}^{g} (z-x_k) $.

It is of importance to describe the set of common zeros of the
functions $  \wp_{ggk} (  \boldsymbol{ u}) $.

  \begin{cor}
The vector function $  \boldsymbol{   \wp}' (  \boldsymbol{ u}) $
  vanishes iff
$  \boldsymbol{ u}$ is a half period.   \end{cor}

  \begin{proof} The equations $  \wp_{ggk} (  \boldsymbol{ u}) =0,
  \;k  \in 1,   \ldots,  g$ yield due to   \eqref{inversiony} the equalities
$y_i=0,   \forall i  \in 1,   \ldots, g$. The latter is possible if and
only if the points $x_1,   \ldots,  x_g$ coincide with any $g$ points
$e_{i_1},   \ldots, e_{i_g}$ from the set  branching points
$e_1,   \ldots, e_{2g+2}$. So the point
  \[
  \boldsymbol{ u}=  \sum_{l=1}^{g}  \int_{a_l}^{e_{i_l}}  \mathrm{ d}
    \mathbf{
u}  \in   \mathrm{ Jac} (V)    \] is of the second order in
Jacobian and hence is a half-period.  \end{proof}

\section[$\zeta$-functions and second kind
differentials]{$\zeta$-functions and
differentials of the second kind} Imposing the conditions $
\lambda_{2g+2}=0, \, \lambda_{2g+1}=4$ we have the following
Theorem,  which will be the starting point for derivation of the
basic relations in the next chapter.

\index{$\zeta$--function!Kleinian}

  \begin{theorem}  \label{the-Z}
 Let $(y_0, x_0)\in V$ be an arbitrary point and  $$  \{ (y_1, x_1)
,   \ldots,  (y_g, x_g)   \}  \in  (V)^g$$ be the Abel pre-image
of the point $  \boldsymbol{ u}  \in \mathrm{ Jac} (V) $. Then
\begin{eqnarray} -  \zeta_j \left ( \int_{a}^{x_0} \mathrm{ d}
\mathbf{ u} +  \boldsymbol{ u}  \right) &=& \int_{a}^{x_0}
  \mathrm{ d}r_j + \sum_{k=1}^g  \int_{a_k}^{x_k} \mathrm{ d}r_j
\cr&-& \frac{1}{2}  \sum_{k=0}^g y_k  \left ( \frac{D_j (R' (z) )
-jD_{j+1} (R(z) )
  }{R' (z) }  \Bigg|_{z=x_k}  \right) , \label{principalz}
\end{eqnarray} where $ R (z) =  \prod_0^g (z-x_j) $ and $ R' (z) =
\frac{  \partial}{  \partial z} R (z) $.

And
  \begin{eqnarray}
-  \zeta_j (  \boldsymbol{ u}) =
  \sum_{k=1}^g  \int_{a_k}^{x_k}  \mathrm{ d}
r_j-  \frac12 {\mathfrak Z}_j(  \boldsymbol{ u}),
  \label{principalz1} \end{eqnarray}
where ${\mathfrak  Z}_j(\boldsymbol{u})$ is the determinant of the
lower Hessenberg matrix, \begin{eqnarray} {\mathfrak
Z}_j(\boldsymbol{u})= \mathrm{det}\left(\begin{array}{ccccc}
\phantom{1}&1&\phantom{1}&\phantom{1}&\\
\phantom{1}&\phantom{1}&1&{ 0}&\\
{T}&\phantom{1}&\phantom{1}&\ddots&\phantom{1}\\
{}&{}&{}&{}&1
\end{array}\right),\label{zzz0}
\end{eqnarray}
where $T=(T_{i,k})$ is $(g-j)\times(g-j)$ lower triangle matrix,
which entries are given by the formula
\[ T_{i,k}=\begin{cases}
(-1)^{i+k}i^{\delta_{1,k}}\wp_{g-i+k,g}({\boldsymbol
u})&\text{if}\quad i<g-j,\\
(-1)^{g-j+k}(g-j+k)^{\delta_{1,k}}\wp_{j+k,g,g}({\boldsymbol
u})&\text{if}\quad i=g-j.
\end{cases}\]

In particular, the following expressions are valid for
${\mathfrak Z}_j(\boldsymbol{u})$ with few
higher indices $j$
\begin{eqnarray*}
{\mathfrak Z}_g(\boldsymbol{u})&=&0,\cr
{\mathfrak Z}_{g-1}(\boldsymbol{u})&=&\wp_{ggg}(\boldsymbol{u}),\cr
{\mathfrak Z}_{g-2}(\boldsymbol{u})&=&
\wp_{gg}(\boldsymbol{u})\wp_{ggg}(\boldsymbol{u})
+2\wp_{g-1,g,g}(\boldsymbol{u}).
\end{eqnarray*}

\end{theorem}

\begin{proof} Putting in   \eqref{r11} $  \mu_i=a_i$ we have

  \begin{eqnarray}  && \mathrm{ ln}
  \left  \{  \frac{  \sigma (  \int_{a_0}^{x}  \mathrm{ d}
     \mathbf{ u}-  \boldsymbol{ u}) }
{  \sigma (  \int_{a_0}^{x}  \mathrm{ d}  \mathbf{ u}) }   \right  \}
- \mathrm{ln} \left  \{  \frac{  \sigma (  \int_{a_0}^{  \mu}
\mathrm{ d} \mathbf{ u}-  \boldsymbol{ u}) } {  \sigma (
\int_{a_0}^{ \mu}  \mathrm{ d}  \mathbf{ u}) } \right  \}\cr&& =
\int_{ \mu}^{x}  \mathrm{ d}  \mathbf{ r}^T  \, \boldsymbol{ u}+
  \sum_{k=1}^g  \int_{a_k}^{x_k}  \mathrm{ d} \Omega (x,   \mu) ,
\label{r20}  \end{eqnarray} where $ \mathrm{ d}  \Omega$ is as in
\eqref{third}. Taking derivative over $u_j$ from the both sides of
the equality   \eqref{r20}, after that letting $a_0  \to  \mu$ and
applying $  \phi (y, x) = (-y, x) $ and $  \phi (  \nu,   \mu) =
(-  \nu,   \mu) $,  we have \begin{eqnarray*} &&\zeta_j  \left (
  \int_{  \mu}^x  \mathrm{ d}  \mathbf{ u}+  \boldsymbol{ u}
  \right)  +  \int_{  \mu}^{x}   \mathrm{ d}  r_j   -
  \frac12  \sum_{k=1}^g  \frac{1}{y_k}  \frac{  \partial x_k}{  \partial
u_j}   \frac{y_k-y}{x_k-x}
\cr&&=
  \zeta_j  \left (
  \boldsymbol{ u}
  \right) -  \frac12  \sum_{k=1}^g  \frac{1}{y_k}
   \frac{  \partial x_k}{  \partial
u_j}   \frac{y_k-  \nu}{x_k-  \mu}.
  \end{eqnarray*}
Put $x=x_0$. Denoting $P (z) =  \prod_1^g (z-x_j) $ we find
  \begin{eqnarray*}
&&  \sum_{k=1}^g  \frac{1}{y_k}  \frac{  \partial x_k}{  \partial
u_j}   \frac{y_k-y}{x_k-x}=
  \sum_{k=1}^g  \left (  \frac{D_{j} (P (z) ) }{P' (z) }
    \Bigg|_{z=x_k}  \right)
  \frac{y_k-y}{x_k-x}  \\=
&&  \sum_{k=0}^g
y_k  \left (  \frac{D_j (R' (z) )-jD_{j+1} (R (z) ) }{R' (z) }
\Bigg|_{z=x_k} \right)\cr&& -  \sum_{k=1}^g y_k  \left (
\frac{D_{j} (P' (z) )-jD_{j+1} (P (z) ) }{P' (z) }
\Bigg|_{z=x_k}  \right) .  \end{eqnarray*} Hence,  using
\eqref{inversiony} and adding to both sides $  \sum_{k=1}^g
\int_{a_k}^{x_k}  \mathrm{ d}r_j$,  we deduce
\begin{eqnarray} &&
\zeta_j  \left (   \int_{  \mu}^{x_0} \mathrm{ d}  \mathbf{ u}+
\boldsymbol{ u}   \right)  +  \int_{ \mu}^{x_0}   \mathrm{ d}  r_j
+  \sum_{k=1}^g  \int_{a_k}^{x_k} \mathrm{ d}r_j\cr
&&  -
\frac12
\sum_{k=0}^g y_k  \left ( \frac{D_j (R' (z) )-jD_{j+1} (R(z) )
}{R' (z) } \Bigg|_{z=x_k}  \right) \nonumber
   \\ &&= \zeta_j  \left
( \boldsymbol{ u} \right) +  \sum_{k=1}^g \int_{a_k}^{x_k}
  \mathrm{ d}r_j -  \frac12  \sum_{k=1}^g \frac{1}{y_k}  \frac{
\partial x_k}{  \partial u_j}   \frac{y_k- \nu}{x_k-  \mu}\cr&&
-\frac12
\sum_{k=0}^g y_k  \left ( \frac{D_j (P' (z) )-jD_{j+1} (P(z) )
}{R' (z) } \Bigg|_{z=x_k}  \right). \nonumber
 \label{deduce}
\end{eqnarray} Now see, that the left hand side of the
\eqref{deduce}  is symmetrical in $x_0, x_1,    \ldots, x_g$,
while the right hand side does not depend on $x_0$. So,  it does
not depend on any of $x_i$.  We conclude, that it is a constant
depending only on $  \mu$.  Tending $  \mu \to a$ and applying the
hyperelliptic involution to the whole aggregate,  we find this
constant to be $0$. Using the solution of the Jacobi inversion
  problem we find

\begin{eqnarray*}
&&\sum_{k=0}^g y_k  \left ( \frac{D_j (P' (z) )-jD_{j+1} (P(z) )
}{P' (z) } \Bigg|_{z=x_k}  \right)=\cr
&&\mathrm{det}\left(\begin{array}{cccccc}
\wp_{gg}&-1&0&0&\ldots&0\\
2\wp_{g-1,g}&\wp_{gg}&-1&0&\ldots&0\\
\ldots&\ldots&\ldots&\ldots&\ldots&\\
(g-k)\wp_{k,g}&\wp_{k+1,g}&\ldots&\ldots&\ldots&\ldots\\
\ldots&\ldots&\ldots&\ldots&\ldots&\\
(g-j-1)\wp_{j+2,g}&\wp_{j+3,g}&\ldots&\ldots&\wp_{gg}&-1\\
(g-j)\wp_{j+1,g,g}&\wp_{j+2,g,g}&\ldots&\ldots&\wp_{g-1,g,g}&\wp_{ggg}
\end{array}\right)
\end{eqnarray*}
\end{proof}

Remark, that the expression for the function ${\mathfrak
Z}_j(\boldsymbol u)$ given in the monograph \cite{ba97} on the
page 323, Ex. vi,   is correct only in the particular cases $j=g$
and
  $j=g-1$ and is wrong at $j<g-1$. Alternatively the expression
  for ${\mathfrak Z}_j(\boldsymbol u)$ in terms of the divisor can
  be given as
  \begin{equation}
  {\mathfrak Z}_j(\boldsymbol
  u)=-\frac12\sum_{k,l=1}^gx_k^l\frac{\partial x_k}{\partial
  u_{j+1}}. \label{formulameum}
  \end{equation}

  \begin{cor} For $ (y, x)   \in V$ and $  \boldsymbol{
  \alpha}=  \int_a^{x}  \mathrm{ d}  \mathbf{ u}$ :
  \begin{equation}
  \zeta_{j} (  \boldsymbol{ u}+  \boldsymbol{   \alpha})
-  \zeta_{j} (  \boldsymbol{ u}) -  \int_{a}^x\mathrm{d}\mathbf{r}_j =  \frac{ (-y D_j+  \partial_j)    \mathcal{ P} (x;  \boldsymbol{
u}) } {2  \mathcal{ P} (x;  \boldsymbol{ u}) }
,   \label{stickelberger}   \end{equation}
where $  \partial_j=  \partial/  \partial u_j$ and $D_j$ is the umbral derivative of the order $j$.
  \end{cor}
\index{$\zeta$--function!principal relation}

  \begin{proof} To find $  \zeta_{j} (  \boldsymbol{
  \alpha}) $ take the limit
$  \{x_1,   \ldots, x_g  \}  \to  \{a_1,   \ldots, a_g  \}$ in
  \eqref{principalz}. The right hand side of
  \eqref{stickelberger} is obtained by rearranging $
  \frac12  \sum_{k=1}^g  \frac{1}{y_k}  \frac{  \partial x_k}{  \partial
u_j}   \frac{y_k-y}{x_k-x}$.  \end{proof}




\section{Moduli of the sigma-function}

\subsection{Thomae formulae}
\label{secThomae}

In this section we exhibit classical results of
Tho\-mae~\cite{tho870}. The Thomae formula which links branch
points with nonsingular even $\theta$--constants, i.e.
$\theta$--constants of the first kind, is well known and widely
used. We shall also implement another Thomae formula, written for
$\theta$-constants of the second kind. We refer to these formulae
as first and second Thomae theorems. The important Riemann-Jacobi
derivative relation which generalizes to higher genera Jacobi's
relation
\begin{equation}  \vartheta_1'(0)=\pi \vartheta_2(0)\vartheta_3(0)\vartheta_4(0)\label{jacobirel}\end{equation}
 follows from the second Thomae theorem.

The $\theta$-constants of the first kind are expressed in terms of
branch points and periods of holomorphic integrals as follows:

\begin{theorem}[{\bf First  Thomae theorem}]
Let ${\mathcal I}_0\cup{\mathcal  J}_0 $ be a partition of the set
${\mathcal G} = \{1,\ldots,2g+1\}$ of indices of the finite
branch points of the hyperelliptic curve $V$. Then the
following formula is valid
\begin{eqnarray}&&\theta^4\{{\mathcal I}_0\}=
\pm \frac{(\det 2\omega)^2}{\pi^{2g}} \Delta({\mathcal I}_0).
\label{thomae1}
\end{eqnarray}
\end{theorem}
The proof can be found in many places, see e.\,g.~
Thomae (1870), Bolza (1899), Fay (1973), Mumford (1983).
There are ${2g+1 \choose g}$ different
possibilities to choose the set ${\mathcal I}_0$.

Among various corollaries of the Thomae formula we shall single
out the following two:

\begin{cor} \label{cor1} Let $\mathcal S=\{i_1,\ldots,i_{g-1}\}$ and
$\mathcal T=\{j_1,\ldots,j_{g-1}\}$ be two disjoint sets of
non-coinciding integers taken from the set ${\mathcal G}$ of
indices of the finite branch points. Then for any two $k\neq l$
from the set ${\mathcal G} \backslash ({\mathcal S} \cup {\mathcal
T})$ the following formula is valid
\begin{equation}
\frac{e_l-e_m}{e_k-e_m}=\epsilon \frac{\theta^2\{k,{\mathcal
S}\}\theta^2\{k,{\mathcal T}\}} {\theta^2\{l,{\mathcal
S}\}\theta^2\{l,{\mathcal T}\}}, \label{fractions}
\end{equation}
where $m$ is the remaining number when $\mathcal S$, $\mathcal T$,
$k, l$ are taken away from~${\mathcal G}$, and $\epsilon^4 = 1$.
\end{cor}

\begin{cor} \label{cor2}
Let ${\mathcal I}_0= \{i_1,\ldots,i_g\}$ and ${{\mathcal J}}_0 =
\{j_1,\ldots j_{g+1}\} $ be the partition. Choose $k, n\in
{\mathcal I}_0$ and $i, j \in {\mathcal J}_0$. Define the sets
${\mathcal S}_{k}={\mathcal I}_0 \backslash \{k\}$, ${\mathcal
S}_{k,n}={\mathcal I}_0 \backslash \{k,n\}$, ${\mathcal T}_{i,j}
={\mathcal J}_0\backslash \{i,j\}$. Then
\begin{equation}
\frac{\prod\limits_{j_l\in{\mathcal
J}_0}(e_k-e_{j_l})}{\prod\limits_{i_l\in{\mathcal I}_0, i_l\neq k
}(e_k-e_{i_l})(e_k-e_n)^2} =\frac{\pm\theta^4\{i,{\mathcal
S}_{k}\}\theta^4\{j,{\mathcal S}_{k}\} \theta^4\{n,{\mathcal
T}_{i,j}\}}{  \theta^4\{i,j,{\mathcal S}_{k,n} \}
\theta^4\{i,{\mathcal T}_{i,j}\} \theta^4\{j,{\mathcal T}_{i,j}
\}}. \label{fractions1}
\end{equation}
\end{cor}
The signs $\pm$ and the values of $\epsilon$ should be determined
in each particular case by some limiting procedure (see e.g. Fay(1973))

The Thomae paper~(1870) , see also Krazer\&Wirtinger (1915), contains
another set of formulae expressing the nonsingular
$\theta$-constants of the second kind in terms of branch points
and periods of Abelian differentials:

 \begin{theorem}[{\bf Second  Thomae theorem}]
Let ${\mathcal I}_1\cup{\mathcal J}_1 $ be a partition of the set
${\mathcal G}$ of indices of the finite branch points, and
$v_1,\ldots, v_g$ the normalized holomorphic integrals. Then the
$\theta$-constants of the second kind are given by the formula

\begin{equation}
\frac{\partial }{\partial v_j}\theta\{ {\mathcal I}_1 \}
(\vek{v}|\tau)\big|_{\vek{v}=0} =: \theta_j\{ {\mathcal I}_1 \} =
2\epsilon\sqrt{\frac{\det 2\omega}{\pi^{g}}} \Delta({\mathcal
I}_1)^{\frac14} \sum_{i=1 }^{g }\omega_{ij} s_{g-i}({\mathcal
I}_1),\quad j=1,\ldots,g, \label{thomae2}\end{equation}
 where
$s_l({\mathcal I}_1)$ is the elementary symmetric function of
degree~l associated with the set~${\mathcal I}_1$ of indices of
the branch points.

\end{theorem}

It is convenient to rewrite this Thomae theorem in matrix form. To
do that we introduce for any set of nonsingular odd
characteristics  $[\delta_1],\ldots,[\delta_g]$ the Jacobi matrix
\begin{equation}
  D[\delta_1,\ldots,\delta_g]=\left(\begin{array}{cccc}
 \theta_1[\delta_1]& \theta_1[\delta_2]&\ldots& \theta_1[\delta_g]\\
 \vdots&\vdots&\ldots&\vdots\\
  \theta_g[\delta_1]& \theta_g[\delta_2]&\ldots& \theta_g[\delta_g]
    \end{array}\right) .
  \label{jacobimatrix}
\end{equation}

\begin{theorem}\label{thomaematrix}
Let ${\mathcal I}_0= \{i_1,\ldots,i_g\}$ and ${{\mathcal J}}_0 =
\{j_1,\ldots j_{g+1}\} $ be the sets of a partition ${\mathcal
I}_0 \cup {\mathcal J}_0 = {\mathcal G}$. Define the  $g$ sets
${\mathcal S}_{k}={\mathcal I}_0 \backslash \{i_k\}$ and use the
correspondence $[\delta_k] \Leftrightarrow \{{\mathcal S}_k\}$,
$k=1,\ldots,g$, for nonsingular odd characteristics. Then

\begin{equation}
  D[\delta_1,\ldots,\delta_g]=\epsilon
  \sqrt{\frac{\det\,2\omega}{\pi^g}}\,2\omega^T \,S\,M  ,
  \label{thomae2m}
\end{equation}
where $\epsilon^8=1$; the matrices $S$ and $M$ are given as

\begin{equation}
 \label{smmatrix}
 \begin{split}
   S &= \bigl(s_{g-i}({\mathcal S}_k)\bigr)_{k,i=1,\ldots,g}, \\
   M &= \mathrm{diag}\, \Bigl(\sqrt[4]{\smash[b]{\Delta({\mathcal S}_1)}},\ldots,
                \sqrt[4]{\smash[b]{\Delta({\mathcal S}_g)}}\Bigr),
 \end{split}
\end{equation}
where the $s_{g-i}({\mathcal S}_k)$ are the symmetric functions
of order $g-i$ built on the set of branch points $e_i$ with $i\in \mathcal{S}_k$.

Moreover, by choosing any $n\in {\mathcal S}_k$ and $i, j \in
{\mathcal J}_0$, the formula $\mathrm{({\ref{thomae2m}})}$ is
transformed to
 \begin{equation}
  D[\delta_1,\ldots,\delta_g]=\epsilon\,2\omega^T \,S\,N ,
  \label{thomae22m}
\end{equation}
with
\begin{equation}
 N = \theta\{{\mathcal I}_0\} \notag
     \times \mathrm{diag}\Bigl(\ldots,\sqrt{e_k-e_n}
\frac{\theta\{i,{\mathcal S}_{k}\}\theta\{j,{\mathcal S}_{k}\}
\theta\{n,{\mathcal T}_{i,j}\}}{  \theta\{i,j,{\mathcal S}_{k,n}\}
\theta\{i,{\mathcal T}_{i,j} \} \theta\{j,{\mathcal T}_{i,j}\}
},\ldots\Bigr)_{k=1,\ldots,g} ,
\end{equation}
where we defined the sets ${\mathcal S}_{k,n}:={\mathcal I}_0
\backslash \{k,n\}$,  ${\mathcal T}_{i,j} :={\mathcal
J}_0\backslash \{i,j\}$.
\end{theorem}

\begin{proof}  The formula (\ref{thomae2m}) is the matrix version
of  formula (\ref{thomae2}). To write (\ref{thomae2m})  in the
form (\ref{thomae22m}), we use (\ref{thomae1}) to obtain for every
$k=1,\ldots,g$

\[ \sqrt{\frac{\det\,2\omega}{\pi^g}}
 \sqrt[4]{\Delta ({\mathcal S}_k)} = \epsilon
 \theta\{{\mathcal I}_0\}
 \sqrt[4]{
  \frac{\Delta ({\mathcal S}_k)}{\Delta ({\mathcal I}_0)}}
  . \]
The quotient under the sign of the fourth root is exactly the left
hand side of the equality (\ref{fractions1}).
\end{proof}

As an immediate corollary of the second Thomae theorem we obtain

\begin{theorem}[{\bf Riemann-Jacobi formula}]
   \label{RiemannJacobi}
Fix $g$ different positive integers $ \{i_1,\ldots,i_g\} =:
{\mathcal I}_0$ of the set ${\mathcal G}$, and let ${\mathcal J}_0
= \{ j_1, \ldots, j_{g+1}\} $ be the complementary set. Define the
$g$ sets ${\mathcal S}_{k}={\mathcal I}_0 \backslash \{i_k\}$ and
use the correspondence $[\delta_k] \Leftrightarrow \{{\mathcal
S}_k\}$, $k=1,\ldots,g$, for nonsingular odd characteristics.
Similarly, define $g+2$ sets ${\mathcal T}_l$ where ${\mathcal
T}_0 = {\mathcal I}_0$ and ${\mathcal T}_l = {\mathcal
J}_0\backslash \{j_l\} $ $l=1,\ldots,g+1$, and use the
correspondence $[\varepsilon_l] \Leftrightarrow \{{\mathcal
T}_l\}$ with the $g+2$ nonsingular even characteristics. Then the
following formula is valid
\begin{equation}
\det \,D[\delta_1,\ldots,\delta_g]=\pm
\pi^g\theta[\varepsilon_{0}]
\theta[\varepsilon_{1}]\cdots\theta[\varepsilon_{g+1}].
\label{Riemann}
\end{equation}
\end{theorem}
\begin{proof}
Compute the determinant of both sides of the matrix equality
(\ref{thomae2m})
\[
\det \,D[\delta_1,\ldots,\delta_g]= \epsilon \left(\det \,2\omega
\right)^{\frac{g+2}{2}}\pi^{\frac{-g^2}{2}}
 \det\,M \det\, S.
 \]
One can see that the product
\begin{equation}
\mathrm{det}\,M^4\mathrm{det}\,S^4 = \prod_{l=0}^{g+1} \Delta(
{\mathcal T}_l ),
\end{equation}
where the partitions   ${\mathcal T}_l$ are given in the
formulation of the theorem. The final formula follows immediately
after expressing each $\Delta({\mathcal T}_l)$ in terms of
$\theta$-constants $\theta\{ {\mathcal T}_l \}$ by the formula
(\ref{thomae1}). Our analysis enables us to give the exact value
of $\epsilon$ from which it follows that the only remaining
ambiguity in (\ref{Riemann}) is the $\pm$ sign, corresponding to
the antisymmetry of the determinant.
\end{proof}

Formula (\ref{Riemann}) was called generalized Riemann-Jacobi
formula by Fay (1979). Its general theory, including
non-hyperelliptic curves, was developed in
the series of works by Igusa~ (1979, 1980, 1982). In
the elliptic case $g=1$ it reduces to~(\ref{jacobirel}).

By inverting Eq.~(\ref{thomae22m}) we obtain the periods of the
first kind and their inverse matrix $\rho$, i.\,e., the
normalizing constants for the holomorphic differentials in terms of $\theta$-constants:
\begin{equation}
 \label{finomega}
  \begin{split}
    2\omega &= \epsilon (S^T)^{-1} N^{-1}D^T[\delta_1,\ldots,\delta_g], \\
    \rho    &:= (2\omega)^{-1}= \epsilon D[\delta_1,\ldots,\delta_g]^{-1}NS^T .
  \end{split}
\end{equation}


\subsection{Differentiation over branch points}
\label{secPicFu}

Periods of the second kind can be obtained from the Picard-Fuchs
equations for the derivatives with respect to the branch points
$e_i$ of the two sets of periods $\omega$ and $\eta$. In the case of $g=1$ these equations read
\begin{align}
\frac{\partial \omega}{\partial e_i} =-\frac12 \frac{\eta+e_i\omega}{(e_i-e_j)(e_i-e_k)}\label{omegade}\\
\frac{\partial \eta}{\partial e_i} =\frac12 \frac{e_i\eta-(e_i^2+e_je_k)\omega}{(e_i-e_j)(e_i-e_k)}
\label{etade}
\end{align}
Bolza (1899)
described how to derive them by a variation procedure
which goes back to Riemann, Thomae, and Fuchs; it was generalized in
terms of Rauch's formula (see e.g. Rauch (1959), Fay(1992) ) which in the
case of the hyperelliptic curve~(\ref{curve}) reads
\begin{equation}
 \frac{\partial\ }{\partial e_k}\d \vek{v}(x,y) = -
 \left. 2\mathrm{Res}\right|_{z=e_k} \d\omega_{\text{norm}}(z,w;x,y)
\d\vek{v}(z,w)
\end{equation}
where $\d \vek{v}(x,y)$ is the vector of normalized holomorphic
differentials and $\d\omega_{\text{norm}}(z,w;x,y)$ the normalized
Kleinian bi-differential
\begin{equation}
 \begin{split}
   \d\omega_{\text{norm}}(z,w;x,y) &= \d\omega(z,w;x,y) +
   2\d\vek{u}^T(z,w)\,\varkappa\,\d\vek{u}(x,y), \\
   \varkappa^T = \varkappa = \eta(2\omega)^{-1}, \quad & \quad
\oint_{\mathfrak{a}_l}\d\omega_{\text{norm}}(z,w;x,y)
   =0, \quad l=1,\ldots, g.
 \end{split}
\end{equation}
In the proof
below we use the same set of ideas using the Klein
bi-differential~(\ref{omega-1}) because our aim is to derive
differential equations, with respect to the branch points, in
the space of periods $\omega, \omega', \eta, \eta'$ of the
non-normalized differentials.

\begin{theorem} For an arbitrary
branch point $e_l$ the following equations are valid

\begin{equation}
\frac{\partial}{\partial e_l}
\begin{pmatrix}
    \omega & \omega'\\ \eta & \eta'
\end{pmatrix}
= \begin{pmatrix}
     \alpha_l & \M\beta_l \\ \gamma_l & -\alpha_l^ t
 \end{pmatrix}
 \begin{pmatrix}
      \omega & \omega' \\ \eta & \eta'
 \end{pmatrix} ,
 \label{pf1}
\end{equation}
where
\begin{eqnarray}
\alpha_l &=&-\frac12\Bigl\{\frac{1}{R'(e_l)} \boldsymbol{\mathcal
U}(e_l)\boldsymbol{\mathcal R}^T(e_l)- M_l
\Bigr\}\label{ma},\\
\beta_l &=&-2\Bigl\{\frac{1}{R'(e_l)} \boldsymbol{\mathcal
U}(e_l)\boldsymbol{\mathcal U}^T(e_l) \Bigr\}\label{mb}, \\
\gamma_l &=&\M\frac18\Bigl\{\frac{1}{R'(e_l)} \boldsymbol{\mathcal
R}(e_l)\boldsymbol{\mathcal R}^T(e_l)- N_l \Bigr\}\label{mc},
\end{eqnarray}
with
\begin{equation}
M_l=\left(\begin{array}{cccccc}
0&0&0&\ldots &0&0\\
1&0&0&\ldots&0&0 \\
e_l&1&0&\ldots&0&0\\
e_l^2&e_l&1&\ldots&0&0\\
\vdots&\vdots&\ddots&\ddots&\ddots&\vdots\\
e_l^{g-2}&e_l^{g-3}&\ldots&e_l&1&0\end{array}\right)
\end{equation}
and
\begin{equation}
N_l = e_l(M_l Q_l + Q_l M_l^ t) + Q_l, \qquad Q_l =
\mathrm{diag}\Bigl(\ldots,\frac{{\mathcal R}_k(e_l)} {{\mathcal
U}_{k+1}(e_l)}, \ldots \Bigr) .
\end{equation}


\end{theorem}

\begin{proof} First we consider the equation for $\partial\omega/\partial e_l$.
It is obtained by integrating the following equivalence over
cycles $\mathfrak{a}_l$:
\begin{multline}
R'(e_i)\frac{\partial  }{\partial e_i} \frac{{\mathcal U}_m(x)}{y}
= \frac{{\mathcal U}_m(e_i)}{2y} \bigl\{\boldsymbol{\mathcal
U}^T(e_i)\boldsymbol{\mathcal R}(x)- \boldsymbol{\mathcal
R}^T(e_i)\boldsymbol{\mathcal U}(x) \bigr\} \\
-{\mathcal U}_m(e_i)\frac{\partial }{\partial x} \frac{y }{ x-e_i}
+\frac{R'(e_i)}{2y}\sum_{j=1}^{m-1}{\mathcal U}_{m-j}(x)
\mathcal{U}_{j}(e_i)\label{de}.
\end{multline}
To prove this, substitute $(z,w)=(e_i,0)$ into
the equality
\[    \frac{\partial}{\partial z} \frac12 \frac{y+w}{y(x-z)}\mathrm{d}x\mathrm{d}z+\mathrm{d}\boldsymbol{r}^T(z,w)
\mathrm{d}\boldsymbol{u}(x,y)=\frac{F(x,z)+2yw}{4(x-z)^2}\frac{\mathrm{d}x}{y}\frac{\mathrm{d}z}{w} \]
which leads to the relation
\begin{equation}
-\frac{\partial }{\partial x}\frac{y}{(x-e_i)}
+\frac{\boldsymbol{\mathcal U}^T(e_i)\boldsymbol{\mathcal
R}(x)}{2y} =\frac{F(x,e_i)}{2(x-e_i)^2y };
\end{equation}
Further insert
\begin{equation}F(x,e_i)=(x-e_i)R'(e_i)+(x-e_i)^2\;
\boldsymbol{\mathcal U}^T(x)\boldsymbol{\mathcal R}(e_i)
\label{fxe}
\end{equation}
to obtain the equality
\begin{equation}
-\frac{\partial }{\partial x}\frac{y}{(x-e_i)}
+\frac{\boldsymbol{\mathcal U}^T(e_i) \boldsymbol{\mathcal
R}(x)}{2y} -\frac{\boldsymbol{\mathcal
R}^T(e_i)\boldsymbol{\mathcal U}(x)}{2y}
=\frac{R'(e_i)}{2(x-e_i)y} \equiv R'(e_i)\frac{\partial}{\partial
e_i}\frac{1}{y},
\end{equation}
which is (\ref{de}) for $m=1$. The validity of (\ref{de}) for
$m=2,\ldots$ follows from the equivalence
\begin{equation}\frac{\partial }{\partial e_i}\frac{x^{m-1} }{y}=e_i^{m-1}
\frac{\partial }{\partial e_i}\frac{1 }{y}
+\frac12\sum_{j=1}^{m-1}e_i^{j-1}\frac{x^{m-j-1}}{y}\label{dem}
,\end{equation} which can be proved inductively.

Compute now the periods $2\omega_{m,l} =
\oint_{\mathfrak{a}_l}({\mathcal U}_m(x)/y)\,\d x$
 from (\ref{de}):
\begin{equation}
\frac{\partial\omega_{m,l}} {\partial e_i} = -\frac{2{\mathcal
U}_m(e_i)} {R'(e_i)} \sum_{k=1}^g\left\{{\mathcal
U}_k(e_i)\eta_{k,l} +\frac14{\mathcal R}_k(e_i)\omega_{k,l}
\right\} + \frac12\sum_{j=1}^{m-1}\omega_{m-j,l}{\mathcal
U}_{j}(e_i) .\label{rel}
\end{equation}
The upper left block of~(\ref{pf1}) is nothing but this formula
written in matrix form.

Next we derive the equation for $\partial \eta/\partial e_l$ in an
analogous way, using the equivalence which may be checked by
direct computing,
\begin{multline}
\label{deR} R'(e_i)\frac{\partial  }{\partial e_i} \frac{{\mathcal
R}_m(x)}{y} = \frac{{\mathcal R}_m(e_i)}{2y}
\bigl\{\boldsymbol{\mathcal U}^T(e_i)\boldsymbol{\mathcal R}(x)-
\boldsymbol{\mathcal R}^T(e_i)\boldsymbol{\mathcal U}(x) \bigr\}
\\
-{\mathcal R}_m(e_i)\frac{\partial }{\partial x} \frac{y }{x-e_i}
-\frac{R'(e_i)}{2y}\Bigl(\frac{{\mathcal R}_m(x) -{\mathcal
R}_m(e_i)}{x-e_i} +2\frac{\partial }{\partial e_i}{\mathcal
R}_m(x)\Bigr).
\end{multline}
The expression in the bracket of the last term can be written as
\begin{equation*}
\frac{-1}{{\mathcal U}_{m+1}(e_i)}\Bigl( \sum_{k=1}^m{\mathcal
U}_{m-k+1}(x){\mathcal U}_{k}(e_i)+ \sum_{k=m+1}^g({\mathcal
U}_k(x){\mathcal R}_k(e_i) -{\mathcal R}_k(x){\mathcal U}_k(e_i))
\Bigr).
\end{equation*}

The periods $2\eta_{m,l} = - \oint_{\mathfrak{a}_l}({\mathcal
R}_m(x)/y)\,\d x$ are obtained by integration:
\begin{multline}
\frac{\partial\eta_{m,l}}{\partial e_i}= \frac{{\mathcal
R}_m(e_i)}{2R'(e_i)}\sum_{k=1}^g\Bigl( {\mathcal
U}_k(e_i)\eta_{k,l}+\frac14{\mathcal R}_k(e_i)\omega_{k,l} \Bigr)
-\frac{1}{2}\sum_{k=m+1}^g{\mathcal
 U}_{k-m}(e_i)\eta_{k,l}\\
 -\frac18\Bigl(\sum_{k=m+1}^g
 \omega_{k,l}
 \frac{{\mathcal R}_k(e_i)}{{\mathcal
 U}_{m}(e_i)}+ \sum_{k=1}^m\omega_{m-k+1,l}
 {\mathcal U}_{k-m}(e_i)
 \Bigr).
\end{multline}
Written in matrix form, this is the lower left block of
Eq.~(\ref{pf1}). The equations for the derivatives of $\omega'$
and $\eta'$ are obtained in the same way.
\end{proof}

\begin{cor}
The following variation formula is valid:
\begin{eqnarray}
\frac{\partial \tau}{\partial e_l} &=& \frac{\imath\pi}{2}
\omega^{-1}\beta_l (\omega^T)^{-1}, \qquad e=1,\ldots,2g+1 ,
\label{detau}
\end{eqnarray}
where $\beta_l$ is given in eq.~$\mathrm{(\ref{mb})}$.
\end{cor}
This is a consequence of eqs.~(\ref{pf1}) and~(\ref{ma}) for
$\tau=\omega^{-1}\omega'$; it was derived in~ Thomae (1870) and has
been proved again in many places. For the case of genus one the
explicit formula reeads:
\begin{equation}
\frac{\partial \tau}{\partial e_i}=\frac{1}{4\omega^2}\frac{\imath\pi}{(e_i-e_j)(e_i-e_k)},\quad i\neq j \neq k\in \{1,2,3\}\label{taubye}
\end{equation}

We are now in the position to give expressions for the second kind
periods in terms of $\theta$-constants.
\begin{theorem}
Choose any $g$ different positive integers $ \{i_1,\ldots,i_g\} =:
{\mathcal I}_0$ from the set ${\mathcal G}$, and let
$e_{i_1},\ldots,e_{i_g}$ be the corresponding branch points.
Then the period matrix $\eta$ is given as
\begin{equation}
    B({\mathcal I}_0)\eta = \sum_{i_l\in{\mathcal I}_0}\frac{\partial \omega}{\partial e_{i_l}}
       - A({\mathcal I}_0) \omega , \label{etaperiods}
\end{equation}
where
\[
B({\mathcal I}_0) =\sum_{i_l\in{\mathcal I}_0}\beta(e_{i_l}),
\qquad A({\mathcal I}_0)=\sum_{i_l\in{\mathcal
I}_0}\alpha(e_{i_l}),
\]
and the matrix $B({\mathcal I}_0)$ is invertible.
\end{theorem}
\begin{proof}
Eq.~(\ref{etaperiods}) follows from (\ref{pf1}), and it is
straightforward to check that
\[\det\, B({\mathcal I}_0) = (-2)^g \frac{\prod\limits_{i_l<i_k\in{\mathcal
I}_0} (e_{i_l}-e_{i_k})^2}{ \prod\limits_{i_n\in{\mathcal I}_0}
R'(e_{i_n})}\neq 0.\]
\end{proof}

The derivative $\partial\omega/\partial e_l$ can be calculated
with the help of formula~(\ref{detau}), (\ref{finomega})  and the
heat equation.

Formula~(\ref{etaperiods}) can be applied as follows. Let
$\lambda_{2g}=\sum_k e_k =0$. Define the matrices
\begin{equation}
  C({\mathcal I}_0) = \sum_{i_k\in{\mathcal I}_0} e_{i_k} A({\mathcal I}_0)^{-1}B({\mathcal
  I}_0), \qquad
  D({\mathcal I}_0) = \sum_{i_k\in{\mathcal I}_0} e_{i_k} A({\mathcal
  I}_0)^{-1}.
\end{equation}
Then by taking a suitable sum of the $ {2g+1 \choose g}$ equations
(\ref{etaperiods}), we obtain
\begin{equation}
 \eta = \Bigl(\sum_{{\mathcal I}_0} C({\mathcal I}_0)\Bigr)^{-1}
 \sum_{{\mathcal I}_0} D({\mathcal I}_0)\sum_{i_k\in{\mathcal
 I}_0}\frac{\partial \omega}{\partial e_{i_k}} ,
\end{equation}
where the summation is over all subsets ${\mathcal I}_0$ of the
set ${\mathcal G}$ of indices. For genus one this formula reduces
to (\ref{omegade})

\subsection{Formula for the matrix $\varkappa$}
We present here the formula that permits to express the second period matrices $2\eta,2\eta',\varkappa$ in terms of $2\omega,2\omega'$, branching points and theta-constants
\begin{prop} \label{kappaprop}
Let $\boldsymbol{\mathfrak A}_{\mathcal{I}_0}+2\omega \boldsymbol{K}_{\infty}$ be an arbitrary even nonsingular half-period corresponding to the $g$ branch points of the set of indices $\mathcal{I}_0=\{i_1,\ldots,i_g\}$.
We define the symmetric $g\times g$ matrices
\begin{equation}
\mathfrak{P}(\boldsymbol{\mathfrak A}_{\mathcal{I}_0}) := \left(\wp_{ij}(\boldsymbol{\mathfrak A}_{\mathcal{I}_0})   \right)_{i,j=1,\ldots,g}
\label{matrixP}
\end{equation}
where
\begin{equation}
\theta_{i,j}[\varepsilon] =\left.\frac{\partial^2}{\partial z_i\partial z_j} \theta[\varepsilon](\boldsymbol{z})\right|_{\boldsymbol{z}=0}, i,j \in \{1,\ldots, g\}
\end{equation}
and the $g\times g$ matrix $H$ is expressible in terms of even non-singular theta-constants

\begin{equation}
H(\boldsymbol{\mathfrak A}_{\mathcal{I}_0}) =  \frac{1}{\theta[\varepsilon]} \left( \theta_{ij}[\varepsilon]
 \right)_{i,j=1,\ldots,g}, \quad\text{where}\quad \quad [\varepsilon]=[(2\omega)^{-1}\boldsymbol{\mathfrak A}_{\mathcal{I}_0}+\boldsymbol{K}_{\infty}].
\end{equation}
Then the $\varkappa$-matrix is given by
\begin{equation}
\varkappa = - \frac12\mathfrak{P}(\boldsymbol{\mathfrak A}_{\mathcal{I}_0})-\frac12
((2\omega)^{-1})^T
H(\boldsymbol{\mathfrak A}_{\mathcal{I}_0}) (2\omega)^{-1} \label{kappa}
\end{equation}
and the half-periods $\eta$ and $\eta'$ of the meromorphic differentials can be represented as
\begin{equation}
\eta = 2 \varkappa \omega, \qquad \eta' = 2 \varkappa \omega' - \frac{\imath\pi}{2}(\omega^{-1})^T \,.
\end{equation}
\end{prop}

We remark that~\eqref{kappa} represents the natural generalization of the Weierstra{\ss} formulae
\begin{equation}
2\eta\omega=-2e_1\omega^2-\frac12 \frac{\vartheta_2''(0)}{\vartheta_2(0)}\,,\quad
2\eta\omega=-2e_2\omega^2-\frac12 \frac{\vartheta_3''(0)}{\vartheta_3(0)}\,,\quad
2\eta\omega=-2e_3\omega^2-\frac12 \frac{\vartheta_4''(0)}{\vartheta_4(0)}\, ,
\end{equation}
see e.g. the Weierstra{\ss}--Schwarz lectures, \cite{w-lect} p. 44.
Therefore Proposition \ref{kappaprop} allows the reduction of the variety of moduli necessary for the calculation of the $\sigma$-- and $\wp$--functions to the first period matrix. More information about can be found in \cite{ehkklp12} and \cite{ksh12}.

\chapter[Multidimensional $\zeta$ and $\wp$-functions]
{Multi-dimensional $\zeta$ and $\wp$-functions}\label{chap:wpfun}

\section{Basic relations}

In this section we are going to derive the basic
relations connecting the functions $\wp_{gi}$ and their
derivatives. Further we will find a basis set of functions closed
with respect to differentiations over the canonical fields
$\partial/\partial u_i$. We give applications of these results
to the modern theory of the integrable equations: to construction
 of the explicit solutions of the KdV system in terms of Kleinian
functions and to the problem of the matrix families satisfying to
the zero curvature condition.

\index{basic relation!for $\wp_{gggk}$--functions}
\begin{prop}  The functions
$  \wp_{gggk}$,  for $k=1,   \ldots, g$ are given by   \begin{eqnarray}
  \wp_{gggi}= (6  \wp_{gg}+  \lambda_{2g})   \wp_{gi}
  +6  \wp_{g, i-1}-2  \wp_{g-1, i}
+  \frac{1}{2}  \delta_{gi}  \lambda_{2g-1}.  \label{wpgggi}
  \end{eqnarray}
  \end{prop}

  \begin{proof} Consider the relation
  \eqref{principalz1}. The differentials $  \mathrm{ d}  \zeta_i$,
$i=1,   \ldots, g$ can be presented in the following forms
  \[
-  \mathrm{ d}  \zeta_i=  \sum_{k=1}^g  \wp_{ik}  \mathrm{ d}u_k=
  \sum_{k=1}^g  \mathrm{ d}r_i (x_k) -   \frac12
  \sum_{k=1}^g  \wp_{gg, i+1, k}  \mathrm{ d}u_k.
  \]
Put $i=g-1$.
We obtain for each of the $x_k$,  $k=1,   \ldots, g$
  \begin{eqnarray*}
&&  \left (12x_k^{g+1}+2  \lambda_{2g}x_k^g+
  \lambda_{2g-1}x_k^{g-1}
-4  \sum_{j=1}^g  \wp_{g-1, j}x_k^{j-1}  \right)
   \frac{  \mathrm{ d}x_k}{ y_k}
\cr&&=2  \sum_{j=1}^g  \wp_{gggj}x_k^{j-1}  \frac{  \mathrm{ d}x_k}{
y_k}.  \end{eqnarray*}
Applying the formula   \eqref{p} to eliminate the
powers of $x_k$ greater than $g-1$,  and taking into account,  that
the differentials $  \mathrm{ d}x_k$ are independent,  we come to
  \begin{eqnarray*}
&& \sum_{i=1}^g  \left[ (6  \wp_{gg}+
\lambda_{2g}) \wp_{gi}+6  \wp_{g, i-1} -2  \wp_{g-1, i} +
\frac{1}{2} \delta_{gi}  \lambda_{2g-1}  \right]x_k^{i-1}\cr=&&
\sum_{j=1}^g \wp_{gggj}x_k^{j-1}.  \end{eqnarray*}  \end{proof}

Let us
calculate the difference
$  \frac{  \partial  \wp_{gggk}}{  \partial
u_i}-  \frac{  \partial  \wp_{gggi}}{  \partial u_k}$ according to the
  \eqref{wpgggi}. We obtain
  \begin{cor}
  \begin{equation}
  \wp_{ggk}  \wp_{gi}-  \wp_{ggi}  \wp_{gk}+  \wp_{g, i-1, k}-
 \wp_{gi, k-1}=0.
  \label{wp3}
  \end{equation}
  \end{cor}
This means that the $1$--form \index{differential!1--form!closed}
$  \sum_{i=1}^g (  \wp_{gg}  \wp_{gi}+  \wp_{g, i-1})   \mathrm{ d} u_i$ is
closed. We can rewrite this as $  \mathrm{ d}  \boldsymbol{ u}^T  \mathcal{
C}  \boldsymbol{  \wp}$.

Differentiation of   \eqref{wp3} by
$u_g$ yields
\begin{cor}
  \begin{equation}
  \wp_{gggk}  \wp_{gi}-  \wp_{gggi}  \wp_{gk}+  \wp_{gg, i-1, k}
  -  \wp_{ggi, k-1}=0.
  \label{i-1,k-1}
  \end{equation}
\end{cor}
And the corresponding closed $1$--form is
$  \mathrm{ d}   \boldsymbol{ u}^T  \mathcal{ C}  \boldsymbol{   \wp}'$.

\subsection{Vector form of the basic relations.} We are
going to find such a set of Kleinian functions which is
algebraically closed with respect to the differentiation over the
canonical variables $u_1,\dots,u_g$.  It is the most convenient to
carry out the deduction in the vector notation.

\index{basic relation!for $\wp_{gggk}$--functions!vector form of}

Introducing the vector $
{\boldsymbol{\wp}}''=\frac{\partial^2 {\boldsymbol{\wp}}}{
\partial u_g^2}$ we can rewrite (\ref{wpgggi}) in the form
\begin{eqnarray*} {\boldsymbol{\wp}}''=(6\mathcal{C}
+\lambda_{2g}){\boldsymbol{\wp}}
+\frac12\lambda_{2g-1}e_g-2\mathcal{P}_{g-1}, \end{eqnarray*} where
$e_i=(\delta_{i1},\ldots,\delta_{ig})$, $i=1,\ldots,g$ and
\[\mathcal{C}=\mathcal{B}_g+ {\boldsymbol{\wp}} \boldsymbol{e}_g^T,
 \quad  \mathcal{B}_g=
\sum_{k=1}^g \boldsymbol{e}_ke^T_{k-1}\] is the
companion matrix  of the polynomial $\mathcal P(z;u)$
defined by (\ref{p}) and
$\mathcal{P}_{g-1}=(\wp_{1,g-1},\dots,\wp_{g,g-1})^T$.
\begin{cor}
The following equation is valid \begin{eqnarray*} 2\frac{\partial
{\boldsymbol{\wp}}}{ \partial u_{g-1}}=\left[-\frac{\partial^3}{
\partial u_g^3}\right.  +\{\left.
6(\mathcal{B}_g+{\boldsymbol{\wp}}
 \boldsymbol{e}_g^T+{\boldsymbol{\wp}}^T \boldsymbol{e}_g)+\lambda_{2g}
\}  \frac{\partial}{ \partial
 u_g}\right]{\boldsymbol{\wp}}
\end{eqnarray*}
\end{cor}
\begin{cor}
The vectors ${\boldsymbol{\wp}}'$
  and ${\boldsymbol{\wp}}''$ satisfy the
following relations
\begin{eqnarray}
\frac{\partial}{\partial
 u_k}{\boldsymbol{\wp}}=\mathsf{A}_k{\boldsymbol{\wp}}'
,\label{wpgik}\\ \frac{\partial }{ \partial
u_k}{\boldsymbol{\wp}}'=\mathsf{A}_k
{\boldsymbol{\wp}}'',\label{wpggik} \end{eqnarray} where the
$g\times g$--matrices $\mathsf{A}_k$ are defined by
\begin{equation*}
\mathsf{A}_{k}=\sum_{i=k+1}^g\mathcal{B}_g^{g-i}
({\boldsymbol{\wp}} \boldsymbol{e}_i^T-\boldsymbol{e}_i^T{\boldsymbol{\wp}})
+\mathcal{B}_g^{g-k}
\end{equation*}
\end{cor}
\begin{proof} Let us write according to \eqref{wp3}
\begin{equation*}
\wp_{g,k-1,i}=\boldsymbol{e}_{i-1}^T\mathsf{A}_k{\boldsymbol{\wp}}',
\end{equation*}
where $\mathsf{A}_k$ some matrix and
$\wp_{ggk}=\boldsymbol{e}_k^T{\boldsymbol{\wp}}'$,
$\wp_{gk}=\boldsymbol{e}_k^T{\boldsymbol{\wp}}$.
Then,  from (\ref{wp3})
\begin{equation*}
\boldsymbol{e}_i^T\mathsf{A}_{k-1}{\boldsymbol{\wp}}'
={\boldsymbol{\wp}}^T(\boldsymbol{e}_i\boldsymbol{e}_k^T
-\boldsymbol{e}_k\boldsymbol{e}_i^T){\boldsymbol{\wp}}'
  +\boldsymbol{e}_{i-1}\mathsf{A}_k{\boldsymbol{\wp}}'.
\end{equation*}
Multiplying  the last equality by $\boldsymbol{e}_i$ we obtain the
recursion relation
\begin{equation*}
\mathsf{A}_{k-1}
=\sum_{i=1}^g \boldsymbol{e}_i{\boldsymbol{\wp}}^T(\boldsymbol{e}_i
\boldsymbol{e}_k^T
-\boldsymbol{e}_k\boldsymbol{e}_i^T)
+\boldsymbol{e}_i\boldsymbol{e}_{i-1}^T\mathsf{A}_k,
\end{equation*}
which takes  after evident simplifications the form
\begin{equation} \mathsf{A}_{k-1}=\mathcal{B}_g\mathsf{A}_k
+{\boldsymbol{\wp}} \boldsymbol{e}_k^T-{\wp}_{gk}\label{ak-1}
\end{equation}
where $\mathcal{B}_g$ is the $g\times g$ backward step matrix,
\index{matrix!step backward}
and the condition $\mathsf{A}_g=1$ is imposed to start the
recursion.

The (\ref{wpgik}) is an immediate consequence of (\ref{ak-1}).

By the same argument we obtain (\ref{wpggik}) from
\eqref{i-1,k-1}.  \end{proof}
\begin{cor} The following equality holds
\begin{equation*}\frac{\partial^2}{ \partial
u_g^2}{\boldsymbol{\wp}}'=\mathcal{
M}{\boldsymbol{\wp}}'\end{equation*}
with the $g\times g$ matrix
\begin{eqnarray*}
\mathcal{M}=4(\mathcal{B}_g+{\boldsymbol{\wp}} \boldsymbol{e}_g^T)
+8\wp_{gg}+\lambda_{2g}
\end{eqnarray*}\end{cor}
\begin{proof} Differentiating  \eqref{wpgggi} by $u_g$ and
applying \eqref{wp3} we obtain
\begin{equation*}\frac{\partial^2}{ \partial
u_g^2}{\boldsymbol{\wp}}'=
\left[6(\mathcal{B}_g+{\boldsymbol{\wp}}
\boldsymbol{e}_g^T)-2\mathsf{A}_{g-1}+6\wp_{gg}+\lambda_{2g}\right]
{\boldsymbol{\wp}}'.\end{equation*} Hence we obtain the corollary.
\end{proof}
\begin{cor}  Vectors ${\boldsymbol\wp}'$ and
  ${\boldsymbol\wp}''$ satisfy the following system
  \begin{eqnarray*}
[\mathsf{A}_i,\mathsf{A}_k]\mathcal{M}{\boldsymbol{\wp}}'=\left(
\frac{\partial \mathsf{A}_k }{ \partial u_i}-\frac{\partial
\mathsf{A}_i }{ \partial u_k}\right){\boldsymbol{\wp}}''
\\{}
 [\mathsf{A}_i,\mathsf{A}_k]{\boldsymbol{\wp}}''=\left(
\frac{\partial \mathsf{A}_k }{ \partial u_i}-\frac{\partial
 \mathsf{A}_i }{ \partial
u_k}\right){\boldsymbol{\wp}}'\end{eqnarray*}
\end{cor}
The proof is straightforward.

Summarizing the above results we have for the canonical case
$\lambda_{2g}=0$
\begin{prop}\label{b_s} The set of Kleinian functions
$\boldsymbol{\wp},\boldsymbol{\wp}'$ and $\boldsymbol{\wp}''$  is
algebraically closed with respect to differentiations over the
canonical variables $\boldsymbol{u}$, that is their derivatives
are expressed as polynomials on the basis set
$(\boldsymbol{\wp},\boldsymbol{\wp}',\boldsymbol{\wp}'')$ with
rational coefficients.
\end{prop}

We would like to pay attention to the complete analogy of this
Proposition to the well-known fact from elliptic theory, that any
derivative $\wp^{(n)}$ is a polynomial of $\wp,\wp'$  and $\wp''$.

Another aspect, which we would like to underline, is the actual
universality of the relations implied by the Proposition
\ref{b_s}. In fact, these relations are valid for any underlying
hyperelliptic curve, provided it is presented by a canonical
equation that is, in  the form with $\lambda_{2g+2}=0$,
$\lambda_{2g+1}=4$ and $\lambda_{2g}=0$.

\subsection{Zero curvature
condition and a generalized shift}
\index{zero curvature condition}

The theory of Kleinian functions developed above permits to
construct explicitly the family of operators satisfying to the zero
curvature condition (see. theorem \ref{t1}). In this section we
give a new explicit construction, which permits to built a
parametric family of operators possessing the same property.
The approach of operators of general shift lies in the ground of
the construction.

We introduce the family of matrices
\[
\mathcal{A}_k=\begin{pmatrix}
B_k&&A_k\\
C_k&&-B_k
\end{pmatrix},
\]
where
\begin{gather*}
A_k=\delta_{k,g}-\wp_{g,k+1}(\boldsymbol{u}),\\
B_k=-\frac{1}{2}\frac{\partial}{\partial
u_g} A_k=\frac{1}{2}\wp_{gg,k+1}({\boldsymbol
u}),\\
C_k=-\frac{1}{2}\frac{\partial^2}{\partial
u_g^2} A_k + (2\wp_{gg}({\boldsymbol
u})+\frac{\lambda_{2g}}{4})A_k. \end{gather*}

\begin{theorem}\label{t1} Let
$\partial_k=\frac{\partial}{\partial u_k}$, then the family of
matrices    $\{\mathcal{A}_j\}$ satisfies to zero curvature
condition:         $$
[\mathcal{A}_k,\mathcal{A}_i]=\partial_k\mathcal{A}_i
-\partial_i\mathcal{A}_k
$$ \end{theorem} \begin{proof} The required conditions are checked
directly by applying of the equalities  \eqref{wp3} and
\eqref{i-1,k-1}.  For example:
\begin{multline*}
\partial_k A_i-\partial_i A_k
-2(A_iB_k-A_kB_i)=
\wp_{gk,i+1}-\wp_{gi,k+1}-\\
\wp_{gg,k+1}(\delta_{g,i}-\wp_{g,i+1})
+ \wp_{gg,i+1}(\delta_{g,k}-\wp_{g,k+1})\equiv 0
\end{multline*}
due to \eqref{wp3}.
\end{proof}
\begin{cor}\label{ccc1}
Let $L(\xi)=\sum_{k\geq0}\xi^k \mathcal{A}_k$ and
$\partial^{\xi}=\sum_{k=1}^g \xi^k \partial_k$, then
$$
[L(\xi_1),L(\xi_2)]=\partial^{\xi_1}L(\xi_2)-\partial^{\xi_2}L(\xi_1).
$$
\end{cor}

Let us introduce the shift operator $\mathcal{D}_x^{\xi}$ by the
formula $$ \mathcal{D}_x^{\xi} G(x)=\frac{x G(x)-\xi
 G(\xi)}{x-\xi} $$where the lower index shows the argument which
is shifted and the upper index shows the argument at which the
aforementioned one is shifted.

 \begin{lemma} The operator
 $\mathcal{D}_x^{\xi}$ defines the commutative generalized shift,
i.e. satisfies to the associativity equation:  \index{equation!of
associativity}
$$\mathcal{D}_{\xi_1}^{\xi_2}\mathcal{D}_x^{\xi_1}=\mathcal{
D}_x^{\xi_1}\mathcal{D}_x^{\xi_2}.$$ \end{lemma} Proof follows
directly from the definition.  \begin{prop} The action
$\mathcal{D}_x^{\xi} G(x)$ on the space of  functions regular at
$x=0$ is defined by the formula $$ \mathcal{D}_x^{\xi} G(x)=
\sum_{k \geq 0}\xi^k D_k G(x), $$ where operators $D_k$ are
invariant with respect to shift $\mathcal{ D}_x^{\xi} $ and
coincide with those given in Definition \ref{umbral_D}.
\end{prop}
Remark, that $D_k=D_1^k$.  The action
$\mathcal{D}_x^{\xi} $ is extended to matrices, elements of which
are functions on $x$, and is defined by the same formula.

\index{operator!commutative generalized shift}
Let us introduce the matrix $\mathcal{L}(\xi,x)$ of the following
form $$\mathcal{L}(\xi,x)=\mathcal{
D}_x^{\xi}\left(L(x)+G(\xi,x)\right)
, $$ where
 $G(\xi,x)=\big[\sum_{i>0}(x^i-\xi^i)A_i\big]\begin{pmatrix}
0&0 \\1&0 \end{pmatrix}$.  Coefficients of the
expansion $\mathcal{ L}(\xi,x)$ define operators     $L_k(x)$:
\begin{equation} \mathcal{
L}(\xi,x)=\sum_{k \geq 0}\xi^k L_k(x),\label{oper-L}
\end{equation} Remark, that $L_k(x)$ vanishes at           $k>g$.

\begin{theorem}
For such the matrix $\mathcal{L}(\xi,x)$ and vector field
     $\partial^{\xi}=\sum_{k=1}^g \xi^k \frac{\partial}{\partial
u_k} $ the following relation is valid
$$
\partial^{\xi_1}\mathcal{L}(\xi_2,x)-\partial^{\xi_2}\mathcal{
L}(\xi_1,x)=[\mathcal{L}(\xi_1,x),\mathcal{L}(\xi_2,x)].
$$
\end{theorem}
\begin{proof}
We obtain directly from the definition of shift operator, that
$$
[\mathcal{L}(\xi_1,x_1),\mathcal{L}(\xi_2,x_2)]=
\mathcal{D}_{x_1}^{\xi_1}\mathcal{D}_{x_2}^{\xi_2}
[L(x_1)+G(\xi_1,x_1) , L(x_2)+G(\xi_2,x_2)]
$$
and
\begin{gather*}
\partial^{\xi_1}\mathcal{L}(\xi_2,x_2) =
\mathcal{D}_{x_1}^{\xi_1}\mathcal{D}_{x_2}^{\xi_2}
\partial^{\xi_1}(L(x_2)+G(\xi_2,x_2)),\\
\partial^{\xi_2}\mathcal{L}(\xi_1,x_1) =
\mathcal{D}_{x_1}^{\xi_1}\mathcal{D}_{x_2}^{\xi_2}
\partial^{\xi_2}(L(x_1)+G(\xi_1,x_1)).
\end{gather*}
Set
\begin{gather*}
F(\xi_1,x_1,\xi_2,x_2)=
[L(x_1)+G(\xi_1,x_1), L(x_2)+G(\xi_2,x_2)]\\
-
\partial^{\xi_1}(L(x_2)+G(\xi_2,x_2))
+
\partial^{\xi_2}(L(x_1)+G(\xi_1,x_1)).
\end{gather*}
Then
\begin{gather*}
\mathcal{D}_{x_1}^{\xi_1}\mathcal{D}_{x_2}^{\xi_2}
F(\xi_1,x_1,\xi_2,x_2)=\frac{1}{(x_1-\xi_2)(x_2-\xi_1)}
\bigg[
x_1 x_2 F(\xi_1,x_1,\xi_2,x_2)-\\
\xi_1 x_2F(\xi_1,\xi_1,\xi_2,x_2)-
x_1\xi_2 F(\xi_1,x_1,\xi_2,\xi_2)+
\xi_1\xi_2F(\xi_1,\xi_1,\xi_2,\xi_2)
\bigg].
\end{gather*}
Evidently the following statement is valid
\begin{lemma}\label{ll2}
$$
\mathcal{D}_{x_1}^{\xi_1}\mathcal{D}_{x_2}^{\xi_2}
F(\xi_1,x_1,\xi_2,x_2) \mid_{x_1=x_2=x} =0
$$
iff,  when
$$
F(\xi_1,\xi_1,\xi_2,\xi_2)=0
$$
and simultaneously
$$
x F(\xi_1,x,\xi_2,x)-
\xi_1 F(\xi_1,\xi_1,\xi_2,x)-
\xi_2 F(\xi_1,x,\xi_2,\xi_2)=0.
$$
\end{lemma}
The proof of the theorem results in the direct checking of
validity of conditions of lemma \ref{ll2}. The
condition $ F(\xi_1,\xi_1,\xi_2,\xi_2)=0$ is satisfied because of
the corollary \ref{ccc1}.  The second condition of the lemma is
equivalent to the equation

$$ \sum
x\bigg\{(\partial^{\xi_2}-\partial^{\xi_1})(L(x)+G(\xi_2,\xi_1))+
[L(x),G(\xi_2,\xi_1)]\bigg\}=0,
$$
where the summation run over all the cyclic permutations of
       $\{x,\xi_1,\xi_2 \}$. Matrix elements    $(i,j)$ of the
function situated over the sign of sum is reduced to the form
\begin{align*} (1,1):&\\ &x^{g+1}(\xi_1^{g+1}-\xi_2^{g+1})+
(x^{g+1}(\partial^{\xi_2}-\partial^{\xi_1})+(\xi_1^{g+1}-
\xi_2^{g+1})\partial^{x})\zeta_g+\\
&((\partial^{\xi_2}-\partial^{\xi_1})\zeta_g)\partial^{x}\zeta_g
-\frac{1}{2}(\partial^{\xi_2}-\partial^{\xi_1})\partial^{x}\wp_{gg};\\
(1,2):&\\
&(\partial^{\xi_2}-\partial^{\xi_1})\partial^{x}\zeta_g;\\
(2,1):&\\
&(\partial^{\xi_2}-\partial^{\xi_1})\big[
x(2\wp_{gg}+\frac{\lambda_{2g}}{4})
+\frac{1}{2}\partial_g\partial^{x}(\wp_{gg}) \big]+\\
&\big( x(\partial^{\xi_2}-\partial^{\xi_1})-(\partial^{x}\wp_{gg})
\big)
[\xi_1^{g+1}-\xi_2^{g+1}+(\partial^{\xi_2}-\partial^{\xi_1})\zeta_g];
\\
(2,2)&=-(1,1),
\end{align*}
where we used the equality
$A_i=-\partial_{i+1}\zeta_g(\boldsymbol{u})$,  and by performing
the summation we find that the condition of the lemma is
satisfied. The theorem is
proved.
\end{proof}
\begin{cor}\label{zero-L}
The parametric family of matrices
$\{ L_j(x)\}$ satisfies also to the zero curvature condition:

$$
[L_k(x),L_i(x)]=\partial_kL_i(x)-\partial_i L_k(x).
$$
\end{cor}
Thus, applying the generalized shift  $\mathcal{
D}_x^\xi$ to the generating function  $L(x)$ of the matrix family
$\mathcal{ A}_1,\ldots, \mathcal{A}_g$, corrected by the gauge
summand $G(\xi,x)$, we obtain the generating function of the
matrix function of the matrix  $L_0(x),\ldots,L_g(x)$ that depends
on a parameter, and the family thus obtained also satisfies the
zero curvature condition.\index{zero curvature condition}

The above result also solves the following problem: for a given
family of operators satisfying the zero curvature condition,
construct a generalized shift operator, which (after a gauge
correction) takes this family to a new family satisfying the same
condition for all  values of parameter.

\subsection{Solution of KdV hierarchy by $\wp$-functions}
The KdV system is the infinite hierarchy of
differential equations
\[
\mathcal{U}_{t_{k}} =  \mathcal{ X}_{k}[\mathcal{U}],
\]
for the function  $\mathcal{U}(t_1,t_2,t_3,\ldots)$. Set, as
usual,          $t_1=z$ and  $t_2=t$.  The first two equations from
the hierarchy have the form:  \[\mathcal{U}_{t_{1}} =\mathcal{U}_z
, \quad \text{and } \quad \mathcal{U}_{t_{2}} =\tfrac14
  (\mathcal{U}_{zzz}-6\mathcal{U} \mathcal{ U}_{z}), \] the second
equation is the Korteweg de Vries equation.
Higher           KdV equations are defined by the relation
\index{KdV equation!solution in Kleinian functions}
\[
\mathcal{ X}_{k+1}[\mathcal{U}]=\mathcal{ R} \mathcal{ X}_{k}[\mathcal{U}], \]
where  $  \mathcal{ R}= \frac14 \partial_z^2-(\mathcal{
  U}+c)-\frac12 \mathcal{U}_z \partial_z^{-1}$ ---is the Lenard
recursion operator and  $V$ --- is a constant.
\index{recursion!operator of Lenard}

The KdV hierarchy is a widely known object in the theory of
integrable systems. Large amount of papers being originated by the
pioneer paper of Novikov \cite{no74} is devoted to the
construction of algebraic geometric solutions for this system.

The theory of Kleinian hyperelliptic functions, being constructed
above, permits to give explicit solution for KdV hierarchy, which
depends directly on the canonical coordinates of the
hyperelliptic Jacobian.  Identifying time variables
$ (t_1, t_2, \ldots,  t_g) \to (u_{g}, u_{g-1}, \ldots, u_1)$ and
the constant $c=\frac{1}{12}\lambda_{2g},$ we have

\begin{theorem}\label{KDV}
The function $\mathcal{U}=2\wp_{gg}(\boldsymbol{
u})+\frac16\lambda_{2g} $ is a $g$--gap solution of the KdV
system.
\end{theorem}

\begin{proof}
Indeed,  we have $\mathcal{U}_z= \partial_g 2\wp_{gg}$  and
by  \eqref{wp3}
 \[\mathcal{U}_{t_{2}}= \partial_{g-1} 2 \wp_{gg}=
\frac12\big(\wp_{ggggg}-(12\wp_{gg}+\lambda_{2g})\wp_{ggg}\big).
\]
The action of $  \mathcal{ R}$
\[ \partial_{g-i-1} 2 \wp_{gg}=
  \left[\frac14 \partial_{g}^2-(2\wp_{gg}+\frac14\lambda_{2g})
  \right] 2\wp_{gg, g-i} -2 \wp_{ggg}\wp_{g, g-i}    \] is verified
  by \eqref{wp3} and \eqref{i-1,k-1}.

On the $g$--th step of recursion, the ``times''  $u_i$ are
exhausted and the stationary equation $\mathcal{ X}_{g+1}[\mathcal{U}]=0$
appears.  A periodic solution of $g+1$ higher stationary
equation is a $g$--gap potential  (see \cite{dmn76}) .
\end{proof}
\index{KdV equation!stationary flows}

\section{Fundamental cubic and quartic
relations}  \label{hyper-Jac} We are going to find the relations
connecting the odd functions $  \wp_{ggi}$ and even functions
$  \wp_{ij}$. These relations take in hyperelliptic theory the place
of the Weierstrass cubic relation   \[ {  \wp'}^2=4  \wp^3-g_2  \wp-g_3,
  \]
for elliptic functions,  which establishes the meromorphic map
between the elliptic Jacobian $  \mathbb{ C}/ (2  \omega, 2
\omega') $ and the plane cubic. We use these results to give the
explicit solution of the ``Sine-Gordon" equation by Kleinian
function. Another application is the theory of Kleinian functions
itself, in the case of genus $3$ we give the complete lists of the
expressions of first and second derivatives of
$\wp_{ik}$-functions by the basis set  and use them to calculate
the second nontrivial term of the expansion of $\sigma$-function
\index{``Sine-Gordon" equation!solution in Kleinian functions}

The theorem below is based
on the property of an Abelian function to be constant if  any
gradient  of it is identically $0$,  or,  if for Abelian
functions $G (  \boldsymbol{ u}) $ and
$F (  \boldsymbol{ u}) $ there exist such a
nonzero vector $  \boldsymbol{
  \alpha}  \in   \mathbb{ C}^g$,  that
$  \sum_{i=1}^g  \alpha_i  \frac{  \partial}{  \partial u_i} (G (  \mathbf{
u}) -F (  \boldsymbol{ u}) ) $  vanishes,  then
$G (  \boldsymbol{ u}) -F (  \boldsymbol{ u}) $ is a constant.

  \begin{theorem} The  functions
$  \wp_{ggi}$ and $  \wp_{ik}$ are related by
  \begin{eqnarray}
  \lefteqn{  \wp_{ggi}  \wp_{ggk}=4  \wp_{gg}  \wp_{gi}  \wp_{gk}-
2 (  \wp_{gi}  \wp_{g-1, k}+  \wp_{g, k}  \wp_{g-1, i}) }  \nonumber  \\
&&+
4 (  \wp_{gk}  \wp_{g, i-1}+  \wp_{gi}  \wp_{g, k-1}) +
4  \wp_{k-1, i-1}-2 (  \wp_{k, i-2}+  \wp_{i, k-2})    \nonumber  \\
&&+
  \lambda_{2g}  \wp_{gk}  \wp_{gi}+  \frac{  \lambda_{2g-1}}2
 (  \delta_{ig}  \wp_{kg}+  \delta_{kg}  \wp_{ig}) +c_{ (i, k) },
  \label{product3}
  \end{eqnarray}
where
  \begin{equation}c_{ (i, k) }=  \lambda_{2i-2}  \delta_{ik}
+  \frac12 (  \lambda_{2i-1}  \delta_{k, i+1}
+  \lambda_{2k-1}  \delta_{i, k+1}) .
  \label{cij}
  \end{equation}
  \end{theorem}
\index{Fundamental cubic relation}

  \begin{proof}
We are looking for such a function $G (  \boldsymbol{ u}) $ that
$  \frac{  \partial}{  \partial u_g} (  \wp_{ggi}  \wp_{ggk}-G) $ is $0$.
Direct check using   \eqref{wp3} shows that
  \begin{multline*}
  \frac{  \partial}{  \partial u_g} (  \wp_{ggi}  \wp_{ggk}-
 (4  \wp_{gg}  \wp_{gi}  \wp_{gk}-
2 (  \wp_{gi}  \wp_{g-1, k}+  \wp_{g, k}  \wp_{g-1, i})   \\
+
4 (  \wp_{gk}  \wp_{g, i-1}+  \wp_{gi}  \wp_{g, k-1}) +
4  \wp_{k-1, i-1}-2 (  \wp_{k, i-2}+  \wp_{i, k-2})    \\
+
  \lambda_{2g}  \wp_{gk}  \wp_{gi}+  \frac{  \lambda_{2g-1}}2
 (  \delta_{ig}  \wp_{kg}+  \delta_{kg}  \wp_{ig}) ) ) =0.
  \end{multline*}
It remains to determine $c_{ij}$.  From   \eqref{principium},  we
conclude,  that $c_{ (i, k) }$  for $k=i$ is equal to
$  \lambda_{2i-2}$,  for $k=i+1$ to $  \frac12{  \lambda_{2i-1}}$,
otherwise $0$. So $c_{ij}$ is given by   \eqref{cij}.
\end{proof}
Consider $  \mathbb{ C}^{g+ \frac{g (g+1) }{2}}$ with
coordinates $ (\boldsymbol{ z}, p=\{ p_{i, j}  \}_{i, j=1
\ldots g}) $ with $  \boldsymbol{ z}^T= (z_1, \ldots, z_g)$ and
 $p_{ij}=p_{ji}$,  then we
have
  \begin{cor} The map   \[ \varphi:
  \mathrm{ Jac} (V)   \backslash (  \sigma)   \to   \mathbb{
C}^{g+  \frac{g (g+1) }{2}},   \quad   \varphi (  \boldsymbol{
u}) = (  \boldsymbol{   \wp}' (  \boldsymbol{ u}) ,
      \Pi (  \boldsymbol{ u}) ) ,
  \]
where $  \Pi=  \{  \wp_{ij}  \}_{i, j=1,   \ldots, g}$,  is  meromorphic
embedding.

The image
$  \varphi (  \mathrm{ Jac} (V)   \backslash (  \sigma) )   \subset
   \mathbb{ C}^{g+  \frac{g (g+1) }{2}}$ is the intersection of
$  \frac{g (g+1) }{2}$ cubics,  induced by   \eqref{product3}.
  \end{cor}
$ (  \sigma) $ denotes the divisor of $0$'s of $  \sigma$.
\index{Fundamental quartic relation}

\begin{definition} Factor $K^g=T^g/\pm$ of a torus
$T^g=\mathbb{C}^g/(2 \omega\oplus 2\omega')$ over the involution
$T^g\to T^g:\boldsymbol{u}\mapsto(-\boldsymbol{u})$ is  called
{\em Kummer variety of the torus} $T^g$.
\index{Kummer!variety}

Kummer variety $\mathrm{Kum}(V)=\mathrm{Jac}(V)/\pm$ of the
Jacobian $\mathrm{Jac}(V)$ of the Riemann surface of an algebraic
curve $V$ is called {\em Kummer variety  of the Riemann surface of
an algebraic curve $V$}.
\end{definition}

Consider projection   \[
  \pi:  \mathbb{ C}^{  \frac{g+g (g+1) }{2}}  \to   \mathbb{
C}^{  \frac{g (g+1) }{2}},   \quad   \pi (  \boldsymbol{ z}, p) =p.
  \]
  \begin{cor}
The restriction  $  \pi  \circ  \varphi$ is the meromorphic embedding of
the Kummer variety
$  \mathrm{ Kum} (V) =
(  \mathrm{ Jac} (V)   \backslash (  \sigma) ) /  \pm$  into
$  \mathbb{ C}^{  \frac{g (g+1) }{2}}$. The image
$  \pi (  \varphi (  \mathrm{ Jac} (V)   \backslash (  \sigma) ) )
\linebreak[3]  \subset   \mathbb{ C}^{  \frac{g (g+1) }{2}}$ is
the intersection of quartics,  induced by \begin{equation} (
 \wp_{ggi}  \wp_{ggj})  (  \wp_{ggk}  \wp_{ggl}) - (  \wp_{ggi}
\wp_{ggk})  (  \wp_{ggj}  \wp_{ggl}) =0, \label{Kijkl}
  \end{equation} where the parentheses mean,  that substitutions
by   \eqref{product3} are made before expanding.  \end{cor} The
quartics   \eqref{Kijkl}  have no analogue in the elliptic theory.
The first example is given by genus $2$, where the celebrated
Kummer surface   \cite{hu05} appears.

\index{SG equation!solution in Kleinian functions}

\subsection{Solution of the sine--Gordon equation}
Consider the following system of equations
\begin{equation}
\begin{cases}
\quad\dfrac{\partial^2 }{\partial z_1 \partial
t}\varphi(z_1,z_2,t)&= 2\beta \sin\varphi(z_1,z_2,t) -\alpha \,
\psi(z_1,z_2,t) \mathrm{e}^{\{-2i\varphi(z_1,z_2,t)\}}\\ \\
\quad\dfrac{\partial}{\partial z_1
}\psi(z_1,z_2,t)&=\frac{1}2\beta \mathrm{e}^{\{i\varphi(z_1,z_2,t)
\}} \dfrac{\partial }{\partial z_2} \varphi(z_1,z_2,t) \end{cases}
\label{SG-2}\end{equation}
with respect two functions $\varphi(z_1,z_2,t)$ and
$\psi(z_1,z_2,t)$. At $\alpha=0$ the system splits to two
independent equations, from which the first represents itself the
known                      ``sine-Gordon" (SG) equation in the
light cone coordinates. In general case            ($\alpha\neq
0$) the system be a two dimensional generalization of     SG
equation.

The theory of Kleinian function permits to construct explicit
solutions for this system.

Set the correspondence of variables $ (z_1,z_2,t)\to
(u_{1}, u_{2}, u_g)$.
\begin{theorem}\label{SG}
Let $\alpha=\dfrac{\lambda_0}{\lambda_1}$ and
$\beta=\sqrt{\lambda_1}$, then the pair of functions
$\varphi=-i\mathrm{ln}\big({2\wp_{1,g}(\boldsymbol{
u})}/{\sqrt{\lambda_1}}\big)$ and $\psi=i\wp_{2,g}(\boldsymbol{u})$
is   $2g$-periodic finite-gap solution of the system  \eqref{SG-2}
for any genus   $g\geq2$.  \end{theorem} \begin{proof} In virtue
of \eqref{wp3},\eqref{i-1,k-1} and \eqref{product3} we have \[
\partial_1\partial_g \mathrm{ln}
\wp_{1,g}=2\wp_{1,g}-\frac{\lambda_1}{2\wp_{1,g}}+\frac{\lambda_0}
{\wp_{1,g}^2} \wp_{2,g}, \]
what provides the validity of the first equation from
\eqref{SG-2}; the second equation is equivalent to the equality
$\partial_1\wp_{2,g}= \partial_2\wp_{1,g}$.
\end{proof}

The results of the  Theorems \ref{KDV} and \ref{SG}
demonstrate the applicability of the Kleinian functions to the
modern theory of integrable systems.  We emphasize that the natural
identification of the independent variables    with the canonical
coordinates on Jacobian come to explicit solutions in terms of
Kleinian functions for the known system in the form being
available to the investigation of the solutions and applications.
This fact is one of the  important incentives for the further
development of the Kleinian functions theory.

\subsection{Veselov-Novikov equation}
The Veselov-Novikov equation
is given as
\begin{eqnarray}
&&\alpha {\mathcal U}_{xxx}+\beta
{\mathcal U}_{yyy}-2\alpha({\mathcal
UV})_x-3\beta({\mathcal UW})_y=u_t,\nonumber\\
&&{\mathcal W}_x={\mathcal U}_y,\quad {\mathcal V}_y={\mathcal
U}_x,
\label{vne} \end{eqnarray}
\begin{prop}
The Veselov-Novikov equation is satisfied
by the hyperelliptic Kleinian functions of the hyperelliptic curve
\begin{equation}
y^2=4x^{2g+1}+\sum_{k=1}^{2g}\lambda_{k}x^k,
\end{equation}
as follows
\begin{eqnarray}
{\mathcal U}(x,y)&=&2\wp_{1,g}(y,u_2,\ldots,u_{g-1},x),\cr
{\mathcal
V}(x,y)&=&2\wp_{g,g}(y,u_2,\ldots,u_{g-1},x)+V,\label{vn}\\
{\mathcal W}(x,y)&=&2\wp_{1,1}(y,u_2,\ldots,u_{g-1},x)+W,\nonumber
\end{eqnarray}
where the variables $x,y,t$ are identified with the Jacobian
variables $u_1,\ldots,u_g$ as
 \begin{equation} x=u_g,\quad y=u_1,\quad
t=\sum_{i=1}^{g}u_i\gamma_i \end{equation}
and
\[
\gamma_i=\begin{cases} -\alpha(3V-\lambda_{2g})&\text{if}\; i=g\\
-\frac12\alpha\lambda_{2g+1}&\text{if}\; i=g-1\\
-\alpha\lambda_{2g+2}&\text{if}\; i=g-2\\
0 &\text{if}\; g-3<i<3\\
-\beta\lambda_0 &\text{if}\; i=3\\
-\frac12\beta\lambda_1 &\text{if}\; i=2\\
-\beta(3W-\lambda_2)&\text{if}\; i=1
\end{cases}
\]
\end{prop}

\chapter[Hyperelliptic Jacobians]{Hyperelliptic Jacobians}\label{chap:hjac}
\section{Fundamental relations}  \label{hyper-Kum}
Let us take a second look at the fundamental cubics
  \eqref{product3} and quartics   \eqref{Kijkl}.
\index{Jacobian!hyperelliptic}
  \subsection{Sylvester's identity}\index{Sylvester's identity}
For any matrix $K$ of entries $k_{ij}$ with $i, j=1,   \ldots, N$
we introduce the symbol $K[{}^{i_1}_{j_1}  \cdots{}^{i_m}_{j_n}]$
to denote the $m  \times n$ sub-matrix:
  \[
K[{}^{i_1}_{j_1}  \cdots{}^{i_m}_{j_n}]
=  \{k_{i_k, j_l}  \}_{k=1,   \ldots, m;  \, l=1,   \ldots, n}
  \]
for subsets of rows $i_k$ and columns $j_l$.

We will need here the {  \em Sylvester's
identity}  (see,  for instance~\cite{hj86}) . Let us fix a subset of
indices $  \boldsymbol{  \alpha}=  \{i_1,   \ldots, i_k  \}$,  and make
up the
$N-k  \times N-k$ matrix $S (K,   \boldsymbol{  \alpha}) $ assuming that
  \[
S (K,   \boldsymbol{  \alpha}) _{  \mu,   \nu}=  \det
K[{}^{  \mu,   \boldsymbol{  \alpha}}_{  \nu,   \boldsymbol{
  \alpha}}]   \] and
$  \mu,   \nu$ are not in $  \boldsymbol{  \alpha}$,  then
  \begin{equation}
  \det S (K,   \boldsymbol{  \alpha}) =  \det
K[{}^{  \boldsymbol{  \alpha}}_{  \boldsymbol{  \alpha}}]^{ (N-k-1) }
 \det K .
  \label{Sylvester}   \end{equation}

  \subsection{Determinantal form}

We introduce  (cf.   \cite{le95})  new functions  $h_{ik}$ defined by
the formula   \begin{eqnarray} h_{ik}&=&4  \wp_{i-1, k-1} -2
  \wp_{k, i-2} -2  \wp_{i, k-2}   \nonumber  \\ &+&
  \frac{1}{2}  \left (  \delta_{ik} (  \lambda_{2i-2}+  \lambda_{2k-2})
+  \delta_{k, i+1}  \lambda_{2i-1}
+  \delta_{i, k+1}  \lambda_{2k-1}  \right) ,
  \label{variables}
  \end{eqnarray}
where the indices
$i, k  \in 1,   \ldots, g+2$.
We assume that $  \wp_{nm}=0$ if $n$ or $m$ is $<1$ and
$  \wp_{nm}=0$ if $n$ or $m$ is $>g$. It is evident that
$h_{ij}=h_{ji}$. We shall denote the matrix of $h_{ik}$ by $H$.

The map   \eqref{variables} from $  \wp$'s and $  \lambda$'s to $h$'s
respects the grading
  \[  \mathrm{ deg}  \;h_{ij}=i+j,   \qquad
  \mathrm{ deg}  \;  \wp_{ij}=i+j+2,   \qquad
  \mathrm{ deg}  \;  \lambda_{i}=i+2,   \] and on a fixed level $L$
  \eqref{variables} is
linear and invertible. From the definition follows
  \[
  \sum_{i=1}^{L-1}h_{i, L-i}=  \lambda_{L-2}   \Rightarrow   \boldsymbol{
X}^T H  \boldsymbol{ X}=  \sum_{i=0}^{2g+2}  \lambda_{i}x^{i}
  \]
for $  \boldsymbol{ X}^T= (1, x,   \ldots, x^{g+1}) $ with
arbitrary $x  \in  \mathbb{ C}$. Moreover, for any roots $x_r$ and
$x_s$ of the equation $  \sum_{j=1}^{g+2} h_{g+2, j}x^{j-1}=0$ we
have  (cf.   \eqref{principal11})  $ y_ry_s =  \boldsymbol{ X}_r^T
H  \boldsymbol{ X}_s$.

From    \eqref{variables} we have
\begin{gather}
-2  \wp_{ggi}=  \tfrac{  \partial}{  \partial
u_g}h_{g+2, i}=  \tfrac{  \partial}{  \partial
u_i}h_{g+2, g}=-  \tfrac12  \tfrac{  \partial}{  \partial
u_i}h_{g+1, g+1},   \notag  \\
2 ( \wp_{gi, k-1}- \wp_{g, i-1, k}) =  \tfrac{  \partial}{
  \partial u_k}h_{g+2, i-1}- \tfrac{  \partial}{  \partial
u_i}h_{g+2, k-1}= \tfrac12 \tfrac{  \partial}{  \partial
u_k}h_{g+1, k}
-\tfrac12  \tfrac{  \partial}{  \partial
u_i}h_{g+2, i}, \notag  \\
  \intertext{$\ldots$ etc.,  and  (cf. \eqref{wpgggi}):}  -2
\wp_{gggi}= \tfrac{ \partial^2}{\partial u_g^2}h_{g+2, i}= -\det
H[{}^{i, }_{g+1 }{}^{g+1 }_{g+2}] + \det H[{}^{i,
}_{g,}{}^{g+2}_{g+2}]- \det H[{}^{i-1, }_{g+1,}{}^{g+2 }_{g+2}].
\label{dg_H}
\end{gather}
Using \eqref{variables},  let us rewrite  \eqref{product3} in
more effective form:
\begin{eqnarray}
4  \wp_{ggi}  \wp_{ggk}=  \tfrac{  \partial}{  \partial
u_g}h_{g+2, i}  \tfrac{  \partial}{  \partial
u_g}h_{g+2, k}=- \det
H[{}^{i, }_{k, }{}^{g+1, }_{g+1, }{}^{g+2}_{g+2}]
  \label{wpggiwpggk}
  \end{eqnarray}
\index{Fundamental cubic relation!in determinant form}

\index{Fundamental quartic relation!in determinant form}

Consider,  as an example,  the case of genus $1$.

Let us define on the
Jacobian of a curve
\[ y^2=  \lambda_4 x^4+  \lambda_3 x^3 +  \lambda_2
x^2+  \lambda_1 x+  \lambda_0    \]
the Kleinian functions:
$  \sigma_{K} (u_1) $ with expansion $u_1+  \ldots$,  its second  and
third logarithmic derivatives $-  \wp_{11}$ and $-  \wp_{111}$. By
  \eqref{wpggiwpggk}
and following the definition
\eqref{variables} we have:
  \[
  -4\wp_{111}^2=\det H
  \left[{}_{1, 2, 3}^{1, 2, 3}  \right]= \det  \left (
  \begin{array}{ccc}   \lambda_0&   \tfrac{1}{2}  \lambda_1&-2
  \wp_{11}  \\ \tfrac12 \lambda_1&4  \wp_{11}+  \lambda_2&
  \tfrac12  \lambda_3  \\ -2  \wp_{11}&  \tfrac12
\lambda_3&  \lambda_4 \end{array} \right) ; \]
  expanding  the determinant we obtain:
\[ \wp_{111}^2=4  \wp_{11}^3+  \lambda_2
  \wp_{11}^2+ \wp_{11}  \frac{  \lambda_1   \lambda_3-4  \lambda_4
  \lambda_0}{4}+ \frac{  \lambda_0   \lambda_3^2+  \lambda_4 (
 \lambda_1^2-4  \lambda_2  \lambda_0) }{16}, \]
and the  \eqref{dg_H},  in complete accordance, gives
\[ \wp_{1111}=6
  \wp_{11}^2 +  \lambda_2   \wp_{11}+ \frac{  \lambda_1
  \lambda_3-4  \lambda_4  \lambda_0}{8}.
\]

These equations show
that $  \sigma_K$ differs only by a factor $  \mathrm{ exp} (-
\frac1{12} \lambda_2 u_1^2) $ from standard Weierstrass $
\sigma$-function built by the invariants $g_2=  \lambda_4
\lambda_0+ \frac1{12} \lambda_2^2-  \frac14  \lambda_3  \lambda_1$
and $g_3= \det \left (  \begin{smallmatrix} \lambda_0&  \frac14
\lambda_1& \frac16  \lambda_2  \\ \frac14  \lambda_1&  \frac16
  \lambda_2& \frac14  \lambda_3  \\ \frac16  \lambda_2&  \frac14
  \lambda_3& \lambda_4 \end{smallmatrix}  \right) $  (see,  e.g.
  \cite{ba55, ww73}) .

Further,
we find,  that $  \mathrm{ rank}  \; H=3$ in generic point of Jacobian,
$  \mathrm{ rank}  \;  H=2$ in half-periods.  At $u_1=0$,  where
$  \sigma_K$ has is $0$ of order $1$,  we have $  \mathrm{ rank}  \;
  \sigma_K^2 H=3$.

Concerning the general case,  on the ground of   \eqref{wpggiwpggk},
we prove the following:
  \begin{theorem}
$  \mathrm{ rank}  \;H=3$ in generic point $  \in  \mathrm{ Jac} (V) $  and
$  \mathrm{ rank}  \;H=2$ in the half-periods.
$  \mathrm{ rank}  \;  \sigma (  \boldsymbol{ u}) ^2 H=3$ in generic
point $  \in  (  \sigma) $ and
$  \mathrm{ rank}  \;  \sigma (  \boldsymbol{ u}) ^2 H=0$ in the
points of $ (  \sigma) _{  \mathrm{ sing}}$.  \end{theorem} Here $
(  \sigma)   \subset   \mathrm{ Jac} (V) $ denotes the divisor of
$0$'s of $  \sigma (  \boldsymbol{ u}) $. The
$ (  \sigma) _{  \mathrm{ sing}}  \subset  (  \sigma) $ is the so-called
singular set of $ (  \sigma) $. $ (  \sigma) _{  \mathrm{ sing}}$ is the set
of points where $  \sigma$ vanishes and all its first partial
derivatives vanish.  $ (  \sigma) _{  \mathrm{ sing}}$ is known  (see
  \cite{fa73} and references therein)  to be a subset of dimension
$g-3$ in hyperelliptic Jacobians of $g>3$,  for  genus $2$ it is
empty and consists of single point for $g=3$. Generally,  the
points of $ (  \sigma) _{  \mathrm{ sing}}$ are presented by
$  \{ (y_1, x_1) ,   \ldots,  (y_{g-3}  \, , x_{g-3}  \,  )   \}
\in (V) ^{g-3}$, and generic points of $ (  \sigma) _{  \mathrm{
sing}}$ are such that for all $i  \neq j \in 1, \ldots, g-3, $ $
\phi (y_i, x_i)   \neq (y_j, x_j) $.

\begin{proof} Consider the Sylvester's
matrix $$S=S   \left (H
 [{}^{i, j, g+1, g+2}_{k, l, g+1, g+2}],   \{g+1, g+2  \}
  \right). $$
According to  \eqref{wpggiwpggk} we have
$S=-4  \left (   \begin{array}{ll}
  \wp_{ggi}  \wp_{ggk}&  \wp_{ggi}  \wp_{ggl}  \\
  \wp_{ggj}  \wp_{ggk}&  \wp_{ggj}  \wp_{ggl}
  \end{array}  \right) $
and $  \det S=0$,  so applying   \eqref{Sylvester} we see,  that
$  \det H  [{}^{i, j, g+1, g+2}_{k, l, g+1, g+2} ]
  \det H  [{}^{g+1, g+2}_{g+1, g+2} ]$ vanishes
identically. As  $  \det H  [{}^{g+1, g+2}_{g+1, g+2} ]=
  \lambda_{2g+2} (4  \wp_{gg}+  \lambda_{2g}) -  \frac14  \lambda_{2g+1}^2$
is not an identical $0$,  we infer that
  \begin{equation}
  \det H  \left[{}^{i, j, g+1, g+2}_{k, l, g+1, g+2}  \right]=0.
  \label{subHdet}
  \end{equation}
\index{matrix!Sylvester's}

Remark,  that this equation is actually the   \eqref{Kijkl} rewritten
in terms of $h$'s.

Now from the   \eqref{subHdet},  putting $j=l=g$,
we obtain for any $i, k$,  except for such $  \boldsymbol{ u}$,  that
$H  \left[{}^{g, g+1, g+2}_{g, g+1, g+2}  \right]$ becomes degenerate,  and
those  where the entries become singular i.e.
$  \boldsymbol{u}  \in (  \sigma) $,    \begin{equation}
h_{ik}= (h_{i, g}, h_{i, g+1}, h_{i, g+2})
  \left (H  \left[{}^{g, g+1, g+2}_{g, g+1, g+2}  \right]  \right) ^{-1}
  \left (  \begin{array}{l}
h_{k, g}  \\
h_{k, g+1}  \\
h_{k, g+2}
  \end{array}
  \right) .  \label{hik}
  \end{equation}
This leads to the skeleton decomposition of the matrix $H$
  \begin{equation}
H=H  \left[{}^{1,   \, {  \displaystyle   \ldots}  \, , g+2}_{g, g+1, g+2}  \right]
  \left (H  \left[{}^{g, g+1, g+2}_{g, g+1, g+2}  \right]  \right) ^{-1}
H  \left[{}^{g, g+1, g+2}_{1,   \, {  \displaystyle   \ldots}  \, , g+2}  \right],
  \label{skeleton}
  \end{equation}
which shows,  that in generic point of $  \mathrm{ Jac} (V) $ rank
of $H$ equals $3$.

Consider the case $  \det
H  [{}^{g, g+1, g+2}_{g, g+1, g+2}]=0$. As by
  \eqref{wpggiwpggk} we have $$   \det
H [{}^{g, g+1, g+2}_{g, g+1, g+2}]= -4 \wp_{ggg}^2,$$  this may
happen only if and only if $  \boldsymbol{u}$ is a half-period.
And therefore
  we have instead of   \eqref{subHdet} the equalities $H
[{}^{i, g+1, g+2}_{k, g+1, g+2}]=0$ and consequently
in half-periods matrix $H$ is decomposed as
  \[
H=H  \left[{}^{1,   \, {  \displaystyle   \ldots}  \, , g+2}_{g+1, g+2}  \right]
  \left (H  \left[{}^{g+1, g+2}_{g+1, g+2}  \right]  \right) ^{-1}
H  \left[{}^{g+1, g+2}_{1,   \, {  \displaystyle   \ldots}  \, , g+2}  \right],
  \]
having the rank $2$.

Next,  consider $  \sigma (  \boldsymbol{ u}) ^2 H$ at the
$  \boldsymbol{ u}  \in (  \sigma) $. We have
$  \sigma (  \boldsymbol{ u}) ^2
h_{i, k}=4  \sigma_{i-1}  \sigma_{k-1}-2  \sigma_{i}  \sigma_{k-2}
-2  \sigma_{i-2}  \sigma_{k}, $  where
$  \sigma_{i}=  \frac{  \partial}{  \partial u_i}  \sigma (  \boldsymbol{
u}) $,  and,  consequently,  the
decomposition
  \[
   \sigma (  \boldsymbol{ u}) ^2 H|_{  \boldsymbol{
u}  \in (  \sigma) }=2 (  \boldsymbol{ s}_1,    \boldsymbol{ s}_2,
  \boldsymbol{ s}_3)    \left (   \begin{array}{rrr} 0&0&-1  \\ 0&2&0  \\
-1&0&0   \end{array}   \right)    \left (   \begin{array}{r}   \boldsymbol{
s}_1^T  \\   \boldsymbol{ s}_2^T  \\   \boldsymbol{ s}_3^T   \end{array}
  \right) ,    \] where
  \begin{align*}  \boldsymbol{
s}_1&= (  \sigma_1,   \ldots,   \sigma_g, 0, 0) ^T,\\
  \boldsymbol{ s}_2&= (0,   \sigma_1,   \ldots,   \sigma_g, 0)^T \\
\intertext{and} \boldsymbol{ s}_3&= (0, 0,   \sigma_1,   \ldots,
  \sigma_g) ^T.\end{align*}

We conclude,  that $  \mathrm{ rank} (  \sigma (  \boldsymbol{ u})
^2 H) $ is $3$ in generic point of $ (  \sigma) $,  and becomes
$0$ only when $  \sigma_1=  \ldots=  \sigma_g=0$,   is in the
points $  \in  (  \sigma) _{  \mathrm{ sing}}$,  while no other values
are possible.
 \end{proof}

{  \em Conclusion}. The map   \begin{align*}
h:  \boldsymbol{ u}  \mapsto&
  \{4  \sigma_{i-1}  \,   \sigma_{k-1}-2  \sigma_{i}  \,   \sigma_{k-2}
-2  \sigma_{i-2}  \,   \sigma_{k}  \\&-  \sigma
 (4  \sigma_{i-1, k-1}  \, -2  \sigma_{i, k-2}  \,
-2  \sigma_{i-2, k}) +
  \tfrac{1}{2}  \sigma^2 (  \delta_{ik} (  \lambda_{2i-2}+  \lambda_{2k-2})   \\
&+  \delta_{k, i+1}  \lambda_{2i-1}
+  \delta_{i, k+1}  \lambda_{2k-1})   \}_{i, k  \in 1,   \ldots,  g+2},
   \end{align*}  induced by
  \eqref{variables} establishes a meromorphic mapping of the
$  \big (  \mathrm{ Jac} (V)   \backslash (  \sigma) _{  \mathrm{ sing}}  \big) /  \pm
$ into the space $Q_3$
of complex symmetric $ (g+2)   \times (g+2) $ matrices of
$  \mathrm{ rank}$ not greater than $3$.

We give the example of genus   $2$ with $  \lambda_6=0$ and
$  \lambda_5=4$:

  \begin{equation}
H=  \left (  \begin{array}{cccc}  \lambda_0&  \frac{1}{2}
 \lambda_1&-2  \wp_{11}&-2  \wp_{12}  \\
  \frac{1}{2}  \lambda_1&  \lambda_2+4  \wp_{11}&  \frac{1}{2}
\lambda_3+ 2  \wp_{12}&-2  \wp_{22}  \\-2  \wp_{11}&  \frac{1}{2}
 \lambda_3+2  \wp_{12}&  \lambda_4+4  \wp_{22}&2  \\
-2  \wp_{12}&-2  \wp_{22}&2&0  \end{array}  \right) .
  \label{kum}  \end{equation}
In this case $ (  \sigma) _{  \mathrm{ sing}}=  \{  \varnothing  \}$,
so the Kummer surface
\index{Kummer!surface}
in $  \mathbb{ C}  \mathbb{ P}^3$ with coordinates  \newline
$ (X_0,  X_1,  X_2,  X _3)  =  (  \sigma^2,   \sigma^2  \wp_{11},   \sigma^2  \wp_{12},   \sigma^2  \wp_{22}) $ is
defined by the equation $  \det  \sigma^2 H=0$.

  \subsection{Extended cubic relation}
\index{extended cubic relation}
The extension   \cite{le95} of   \eqref{wpggiwpggk} is
described by \begin{theorem} \begin{eqnarray} \mathbf{ R}^T
  \boldsymbol{   \pi}_{jl}  \boldsymbol{   \pi}_{ik}^T  \mathbf{
S}=  \frac{1}{4}  \det   \left (  \begin{array}{cc} H
  \left[{}^i_j{}^k_l{}^{g+1}_{g+1}{}^{g+2}_{g+2}  \right]
&  \mathbf{ S}  \\  \mathbf{ R}^T&0
  \end{array}  \right) ,   \label{bakergen}
  \end{eqnarray}
where $  \mathbf{ R},   \,    \mathbf{ S}  \in   \mathbb{  C}^4$ are arbitrary
vectors and
  \[  \boldsymbol{   \pi}_{ik}=  \left (
  \begin{array}{c} -  \wp_{ggk}  \\
  \wp_{ggi}  \\
  \wp_{g, i, k-1}-  \wp_{g, i-1, k}  \\
  \wp_{g-1, i, k-1}-
  \wp_{g-1, k, i-1}+
  \wp_{g, k, i-2}-
  \wp_{g, i, k-2}
  \end{array}
  \right)   \]
  \end{theorem}

  \begin{proof}
Let us show, that vectors $  \boldsymbol{ {  \tilde  \pi}}=
  \boldsymbol{ \pi}_{ik}$ and $  \boldsymbol{   \pi}=
\boldsymbol{ \pi}_{jl}$ solve the equations    \[ H
\left[{}^i_j{}^k_l{}^{g+1}_{g+1}{}^{g+2}_{g+2}  \right]
\boldsymbol{ \pi}=0;   \quad   \boldsymbol{ {  \tilde  \pi}}^T{H}
\left[{}^i_j{}^k_l{}^{g+1}_{g+1}{}^{g+2}_{g+2} \right]=0.  \]
Consider a  system of linear equations:
\begin{equation}
H\left[{}^i_j{}^k_l{}^{g+1}_{g+1}{}^{g+2}_{g+2}\right]\boldsymbol{\pi}=0;
\quad
\boldsymbol{\tilde
\pi}^TH\left[{}^i_j{}^k_l{}^{g+1}_{g+1}{}^{g+2}_{g+2} \right]=0.
\label{eqq}
\end{equation}
Denote by $H_{\alpha}$, $\alpha=1,\ldots,4$ the columns
of the matrix
$H\left[{}^i_j{}^k_l{}^{g+1}_{g+1}{}^{g+2}_{g+2}\right]$, i.e.
$H\left[
{}^i_j{}^k_l{}^{g+1}_{g+1}{}^{g+2}_{g+2}\right]=\sum_{i=1}^4
H_{\alpha}{\bf e}_{\alpha}^T$ and by $\pi_i$ the corresponding
components of a vector $\pi$. To prove that vectors
$\boldsymbol{\pi}=\boldsymbol{\pi}_{jk}$ and
$\boldsymbol{\tilde\pi}=\pi_{ik}$ solve the system (\ref{eqq}),
firstly, we find that the equality $\pi_{ik}^T H_4=0$ is
exactly (\ref{wp3}) if we take into account that
$\lambda_{2g+1}=4$ and $\lambda_{2g+2}=0$ in our case.

Further, to show that  $\pi_{ik}^T H_3=0$ we have to
prove, that \begin{eqnarray*}
&&-\wp_{ggk}h_{i,g+1}+\wp_{ggi}h_{k,g+1}
-h_{g+1,g+1}(\wp_{g,i-1,k}
-\wp_{g,i,k-1})\\
&&h_{g+2,g+2}(\wp_{g-1,i,k-1}-\wp_{g-1,k,i-1}
+\wp_{g,k,i-2}-\wp_{g,i,k-2}) =0
\end{eqnarray*}
or taking into the account the definition of $h_{ij}$,
\begin{eqnarray*}
&-&\wp_{ggk}[2\wp_{g,i-1}-\wp_{g-1,i}+\frac{1}{4}\delta_{gi}\lambda_{2g-1}]\\
&+&\wp_{ggi}[2\wp_{g,k-1}-\wp_{g-1,k}+\frac{1}{4}\delta_{gk}\lambda_{2g-1}]\\
&-&2(\wp_{gg}+\frac{1}{4}\lambda_{2g})(\wp_{g,i-1,k}-\wp_{g,i,k-1})\\
&+&\wp_{g-1,i,k-1}-\wp_{g-1,k,i-1}+\wp_{g,k,i-2}-\wp_{g,i,k-2}=0.
 \end{eqnarray*}
Let us compute the difference $\partial
\wp_{gggi}/\partial u_{k-1} -\partial \wp_{gggk}/\partial u_{i-1}$
by (\ref{wpgggi}) and compare it with the derivative of
\eqref{i-1,k-1} on $u_g$.  After evident simplifications we have
\begin{eqnarray*}
&-&\wp_{ggk}[6\wp_{g,i-1}-2\wp_{g-1,i}+\frac{1}{2}\delta_{gi}\lambda_{2g-1}]
+\wp_{ggi}[6\wp_{g,k-1}-2\wp_{g-1,k}+\frac{1}{2}\delta_{gk}\lambda_{2g-1}]\\
&-&(6\wp_{gg}+\frac{1}{4}\lambda_{2g})(\wp_{g,i-1,k}-\wp_{g,i,k-1})
+2(\wp_{g-1,k-1,i}-\wp_{g-1,i-1,k})\\
&-&2(\wp_{gi}\wp_{g,g-1,k}-\wp_{gk}\wp_{g,g-1,i}=0.
\end{eqnarray*}
To complete the calculation we use the equalities
\begin{eqnarray*}
&-&\wp_{ggi}\wp_{g,k-1}+\wp_{ggk}\wp_{g,i-1}\\
&+&\wp_{g,g,k-1}\wp_{gi}-\wp_{g,g,i-1}\wp_{gk}
-\wp_{g,i,k-2}+\wp_{g,k,i-2}=0
\end{eqnarray*}
and
\begin{eqnarray*}
&&\wp_{gg}(\wp_{gi}\wp_{ggk}-\wp_{gk}\wp_{ggi})\\
&+&\wp_{gk}\wp_{g,g,i-1}-\wp_{gi}\wp_{g,g,k-1}
-\wp_{gk}\wp_{g,g-1,i}-\wp_{gi}\wp_{g,g-1,k}=0
\end{eqnarray*}
The first of these equations is deduced from (\ref{wp3}) by
substituting $i\rightarrow i-1$ and   $k\rightarrow k-1$;
to obtain the second  substitute $i=g$ in \eqref{wp3} and multiply
the result by $\wp_{gi}$ and then add
\eqref{wp3} with $k=g$, multiplied by $\wp_{gk}$.
We find, after some obvious
manipulations that the required equality holds.

Next, the equality  $\pi_{ik}^T H_2=0$ reads as
\begin{eqnarray}
h_{il}\pi_1-h_{kl}\pi_2+h_{l,g+1}\pi_3-h_{l,g+2}\pi_4=0;
\label{eq2} \end{eqnarray}
eliminating $\pi_3$ and $\pi_4$ with the use of
$\pi_{ik}^T H_4=0$ and $\pi_{ik}^T H_3=0$ we
obtain
\[
H\left[{}^i_l{}^{g+1}_{g+1}{}^{g+2}_{g+2}\right] \pi_1-
H\left[{}^k_l{}^{g+1}_{g+1}{}^{g+2}_{g+2}\right] \pi_2=0.
\]
Using
(\ref{wpggiwpggk}) we turn the last expression to
$-\pi_1\wp_{ggi}\wp_{ggj}+\pi_2\wp_{ggk}\wp_{ggj}$ which vanishes
by the definition of $\pi$.

Finally, notice, that $\pi_{ik}^T H_1=0$ differs from
(\ref{eq2}) only by the change of sign and  replacement of
indices $l \to j$.

To comlete the proof we need the following
\begin{lemma} Let ${ \sf  k}$ be such a degenerate $n\times
n$--matrix, that all its $(n-1)\times(n-1)$--minors are nonzero,
and $\boldsymbol{\pi}^T=(\pi_1,\ldots,\pi_n)$,
$\boldsymbol{\tilde\pi}^T=(\tilde\pi_1,\ldots,\tilde\pi_n)$,
satisfy the $${ \sf  k} \boldsymbol{\pi}=0,\quad
\boldsymbol{\tilde\pi}^T{ \sf k}=0.$$

Then, for $j=1,\ldots,n$:
\begin{eqnarray*}
\frac{\pi_1}{{ \sf  k}_{1j}}=\frac{\pi_2}{{ \sf  k}_{2j}}=\cdots=
\frac{\pi_n}{{ \sf  k}_{nj}}\\
\frac{\tilde\pi_1}{{ \sf  k}_{j1}}=
\frac{\tilde\pi_2}{{ \sf  k}_{j2}}=\cdots=
\frac{\tilde\pi_n}{{ \sf  k}_{jn}}, \end{eqnarray*}
where ${ \sf  k}_{ij}$ is the algebraic complements of an element
$i,j$ of the matrix ${ \sf  k}$.\end{lemma}
\begin{proof} Follows from the Kramer's rule.\end{proof}

Now, applying the Lemma, we obtain, with obvious identification
of minors:
\begin{equation}
\frac{\pi_1\tilde\pi_1}{ H_{11}}=\cdots
=\frac{\pi_4\tilde\pi_1}{ H_{41}}= \text{ const}.
\end{equation}
The constant in the very right hand side of this equation equals
  $\frac{1}{4}$, so, we obtain
\begin{equation}
\pi_k\tilde\pi_j=\frac{1}{4}H_{kj}.
\end{equation}

Expanding the determinant in the right hand
side of the (\ref{bakergen}), we see, that
the theorem follows.
\end{proof}

In the case of genus $2$,  when the vector $  \boldsymbol{
  \pi}_{21}= (-  \wp_{222},   \wp_{221}, -  \wp_{211},
\wp_{111}) ^T$ exhausts
all the possible $  \wp_{ijk}$--functions,  the relation
  \eqref{bakergen} was thoroughly studied by Baker   \cite{ba07}
in connection with geometry of Kummer and Weddle surfaces.
\index{extended cubic relation!for genus 2}

\index{Kummer!variety!hyperelliptic}
\section{Matrix realization of hyperelliptic Kummer
varieties} Here we present the explicit matrix realization  (see
\cite{bel96}) of hyperelliptic Jacobians $  \mathrm{ Jac} (V) $
and Kummer varieties $  \mathrm{ Kum} (V) $ of the curves $V$ with
the fixed branching point $e_{2g+2}=a=  \infty$.  Our approach is
based on the results of Section   \ref{hyper-Kum}.

Let us consider the space $  \mathsf{ H}$ of complex symmetric
$ (g+2)   \times (g+2) $-matrices  $  \mathrm{  H}=  \{  \mathrm{
h}_{k, s}  \}$,   with $  \mathrm{  h}_{g+2, g+2}=0$ and $  \mathrm{
h}_{g+1, g+2}=2$.  Let us put in correspondence  to $  \mathrm{
H}  \in   \mathsf{ H}$  a symmetric $g  \times g$--matrix $
\mathrm{ A} (  \mathrm{  H}) $,  with entries $a_{k, s}=  \det
\mathrm{  H} \left[{}_{s, g+1, g+2}^{k, g+1, g+2}  \right]$.

From  the Sylvester's identity \index{Sylvester's identity}  \eqref{Sylvester} follows
that rank of the matrix $  \mathrm{  H}  \in   \mathsf{ H}$ does
not exceed $3$ if and only if rank of the  matrix $  \mathrm{  A}
(  \mathrm{  H}) $ does not exceed $1$.

Let us put  $\mathsf{KH}=   \left  \{  \mathrm{  H}  \in
\mathsf{ H}:  \mathrm{ rank}  \mathrm{  H}  \leq 3  \right  \} $.
  For each complex symmetric $g  \times g$--matrix $  \mathrm{
A}=  \{a_{k, s}  \}$ of rank not greater $1$,  there exists,
defined up to sign, a $g$--dimensional column vector $ \mathbf{
  z}=  \mathbf{  z} (  \mathrm{  A}) $,  such that $  \mathrm{
A}=-4  \mathbf{  z}  \cdot   \mathbf{  z}^T$.

Let us introduce vectors
$  \mathbf{  h}_k=  \{  \mathrm{  h}_{k, s};  \;s=1,   \ldots, g
  \}  \,   \in   \mathbb{
C}^g$.

  \begin{lemma}  \label{geom-2}
  Map
  \begin{eqnarray*}
&  \gamma: \mathsf{KH}   \to  (  \mathbb{ C}^g/  \pm)
  \times  \mathbb{ C}^g  \times  \mathbb{
C}^g  \times  \mathbb{ C}^1  \\
&  \gamma (  \mathrm{  H}) =
-  \left (  \mathbf{  z}  \left (  \mathrm{  A} (  \mathrm{  H})   \right) ,   \mathbf{
h}_{g+1},    \mathbf{  h}_{g+2},   \mathrm{ h}_{g+1, g+1}  \right)
  \end{eqnarray*}
 is a homeomorphism.
  \end{lemma}

  \begin{proof} follows from the relation:
  \[
4   \hat{  \mathrm{  H}}=4   \mathbf{  z}  \cdot   \mathbf{  z}^T+2
  \left (  \mathbf{  h}_{g+2}  \mathbf{  h}_{g+1}^T+  \mathbf{  h}_{g+1}
  \mathbf{  h}_{g+2}^T  \right) -  \mathrm{ h}_{g+1, g+1}
 \mathbf{  h}_{g+2}  \mathbf{  h}_{g+2}^T
  \]
where $  \hat{  \mathrm{  H}}$ is the matrix composed of the column
vectors $  \mathbf{  h}_k,    \,  k=1,   \ldots, g$,
and $  \mathbf{  z}= (  \mathbf{
z}  \left (  \mathrm{  A} (  \mathrm{  H})   \right) $.
  \end{proof}

Let us introduce  the $2$--sheeted ramified
 covering
$  \pi:\mathsf{JH}  \to \mathsf{KH}$,   which the
covering $  \mathbb{ C}^g  \to (  \mathbb{ C}^g/  \pm) $
induces by the map
 $  \gamma$.
\begin{cor}
 $  \hat   \gamma: \mathsf{JH}   \cong   \mathbb{ C}^{3g+1}$.
  \end{cor}
Now let us consider the  universal space
$W_g$ of $g$--th
symmetric powers of hyperelliptic curves
  \[V=
  \left  \{  (y, x)    \in   \mathbb{ C}^2:   \, y^2=4 x^{2g+1} +
  \sum_{k=0}^{2g}  \lambda_{2g-k}x^{2g-k}  \right  \}   \]  as an algebraic
subvariety in $ (  \mathbb{ C}^2) ^g  \times
  \mathbb{ C}^{2g+1}$ with coordinates
  \[  \left  \{
  \left ( (y_1, x_1) ,   \ldots,  (y_g, x_g)   \right) ,   \;  \lambda_{2g},   \ldots,
  \lambda_0   \right  \},   \] where $ (  \mathbb{ C}^2) ^g$ is  $g$--th
symmetric power of the space $  \mathbb{ C}^2$.

Let us define the map   \[  \lambda: \mathsf{JH}  \cong
 \mathbb{ C}^{3g+1}  \to
 (  \mathbb{ C}^2) ^g  \times   \mathbb{ C}^{2g+1}  \] in the following way:
  \begin{itemize}
  \item for
$  \boldsymbol{ G}= (  \mathbf{  z},   \mathbf{  h}_{g+1},   \mathbf{
h}_{g+2},   \mathrm{ h}_{g+1, g+1})
   \in   \mathbb{ C}^{3g+1}$ construct by Lemma
  \ref{geom-2} the matrix $  \pi (  \boldsymbol{ G}) =  \mathrm{
H}=  \{  \mathrm{  h}_{k, s}  \}  \in \mathsf{KH}$
  \item  put
  \[  \lambda (  \boldsymbol{ G}) =  \{ (y_k, x_k) ,
  \lambda_r;  \;k=1,   \ldots, g,   \,  r=0,   \ldots,  2g,
  \}  \]
where $  \{x_1,    \ldots,  x_g  \}$ is the set of roots of the equation
$2 x^g+   \mathbf{  h}_{g+2}^T   \mathbf{ X}=0$,  and
$y_k=  \mathbf{  z}^T   \mathbf{ X}_k$,  and
$  \lambda_r=  \sum_{i+j=r+2}   \mathrm{ h}_{i, j}$.
  \end{itemize}
Here
$  \mathbf{ X}_k= (1, x_k,   \ldots, x_k^{g-1}) ^T$.

  \begin{theorem}. Map  $  \lambda$ induces map
$\mathsf{JH}  \cong   \mathbb{ C}^{3g+1}  \to W_g$ .  \end{theorem}

  \begin{proof} Direct check shows,  that  the identity is valid
  \[
  \mathbf{ X}_k^T   \mathrm{  A}
  \mathbf{ X}_s+4  \sum_{i, j=1}^{g+2}
  \mathrm{ h}_{i, j}x_{k}^{i-1}x_s^{j-1}=0,    \]
where
  $  \mathrm{A}=  \mathrm{  A} (  \mathrm{  H}) $  and $  \mathrm{
  H}= \pi (  \boldsymbol{ G}). $ Putting $k=s$ and using $
\mathrm{  A}=4   \mathbf{  z} \cdot \mathbf{  z}^T$,  we have
$$y_k^2=4 x_k^{2g+1}+ \sum_{s=0}^{2g}  \lambda_{2g-s}x_k^{2g-s}$$.
\end{proof}

Now it is all ready to give the description of our realization
of varieties $T^g=  \mathrm{ Jac} (V) $ and
$K^g=  \mathrm{ Kum} (V) $ of the hyperelliptic curves.

For each nonsingular curve            \index{curve!nonsingular}
$V=\left  \{
 (y, x) ,  y^2=4
x^{2g+1} +
\sum_{s=0}^{2g}  \lambda_{2g-s}x^{2g-s}  \right  \} $
define the map
  \[  \gamma:  \;T^g  \backslash (\sigma)    \to   \mathsf{H}:
\gamma (\boldsymbol{u}) =  \mathrm{  H}=  \{  \mathrm{ h}_{k, s}
\}, \] where \begin{multline*}
\mathrm{ h}_{k, s}=4  \wp_{k-1, s-1}-2 (  \wp_{s,
k-2}+ \wp_{s-2, k}) \\+  \frac12 [  \delta_{ks} (  \lambda_{2s-2}+
\lambda_{2k-2}) +  \delta_{k+1, s} \lambda_{2k-1}+  \delta_{k,
  s+1}  \lambda_{2s-1}].\end{multline*}

\begin{theorem} The map  $  \gamma$  induces map
$T^g  \backslash (  \sigma)   \to \mathsf{KH}$,  such that
$  \wp_{ggk}  \wp_{ggs}=  \dfrac14 a_{ks} (  \gamma (u) ) $,   i.e
$  \gamma$  is lifted to
  \[
{  \tilde   \gamma}:T^g   \backslash (  \sigma)   \to \mathsf{JH}
\cong   \mathbb{ C}^{3g+1}   \;   \text{with }  \;   \mathbf{  z}=
(  \wp_{gg1}, \ldots,   \wp_{ggg} ) ^T.    \] Composition of maps
$  \lambda{  \tilde \gamma}:  \;T^g  \backslash (  \sigma)   \to
  W_g$  defines the inversion of the Abel map \index{Abel map} $
\mathfrak{ A}:  \;  (V) ^g  \to T^g$ and, therefore,  the map $
\tilde \gamma$ is an embedding.  \label{geom-3} \end{theorem}

So we have obtained the explicit realization of the Kummer variety
$K^g=T^g  \backslash (  \sigma) /  \pm$  of the Riemann surface of
a hyperelliptic curve  $V$ of genus $g$ as a subvariety in the
variety of matrices $\mathsf{KH}$.

\section{Universal space of hyperelliptic Jacobians }

\subsection{Dubrovin-Novikov theorem}\label{dntheorem}
As a consequence of the  Theorem
  \ref{geom-3},  particularly,  follows a new proof of the theorem
by B.A. Dubrovin and S.P. Novikov  about rationality of the
universal space of the Jacobians of Riemann surfaces of
hyperelliptic curves $V$ of genus  $g$ with the fixed branching
point $e_{2g+2}=\infty$ \cite{dn74}.

\subsection{Solving of polynomial dynamical system}
\index{space!universal}
In this section we give explicit description of a dynamical system
defined on the universal space of the
Jacobians of the Riemann surfaces of the canonical
hyperelliptic curves of genus $g$, this space being congruent to
$\mathbb{C}^{3g}$. The dynamical system \index{dynamical system!
on $sl(2,\mathbb{C})$}
constructed below has the property of interest for us: the
trajectories of its evolution completely lay in the fibers of the
universal bundle of canonical hyperelliptic Jacobians. Next we
will construct a mapping of $\mathbb{C}^{3g}$ to a $g$-dimensional
$sl(2,\mathbb{C})$-module $\mathcal{V}_g$, such that this mapping
takes the dynamical system just discussed to an integrable
dynamical system on $\mathcal{V}_g$.

Let $\mathcal{D}$ be some differentiation, $L$ and $M$ be some
matrices, then a dynamical system defined by an equation \[
\mathcal{D}L=[L,M] \] possesses a set of invariants, which are the
coefficients of the polynomial on $\lambda$ \[
f(\lambda)=\det(L-\lambda \cdot I),\quad
\mathcal{D}f(\lambda)\equiv 0\quad\forall \lambda \in \mathbb{C}.
\] This important fact is one of the corner stones of the
contemporary theory of integrable systems. We will need here a
particular case of this statement, namely the case when $L$ and
$M$ are traceless $2\times 2$ matrices.

Let
$\mathcal{D}=\mathcal{D}_s=\sum_{i=1}^g s^{i-1}\partial_i$, where
$s$ is an arbitrary parameter.
Put \begin{align*} L=L(t)=&
\begin{pmatrix} v(t)&u(t)
\\w(t)&-v(t)
\end{pmatrix},\\
M=M(s,t)=&\frac{1}{2}\begin{pmatrix}
\beta(s,t)&\alpha(s,t)\\
\gamma(s,t)&-\beta(s,t)
\end{pmatrix},
\end{align*}
where $t$ and $s$ are some algebraically independent
commuting parameters.  Functions $u(t),v(t),w(t)$ and
$\alpha(s,t),\beta(s,t);\Gamma(s,t)$ are assumed to be regular
(analytic) functions of parameters $s$ and $t$.

We are going to define a dynamical system by the following
equations of motion:
\begin{align}
\mathcal{D}_s
x(t)&=\alpha(s,t)v(t)-\beta(s,t)u(t),\notag\\
2 \mathcal{D}_s
y(t)&=\gamma(s,t)u(t)-\alpha(s,t)w(t),\label{LM_eq}\\
\mathcal{D}_s
z(t)&=\beta(s,t)w(t)-\gamma(s,t)v(t).\notag
\end{align}
Then for  the function \begin{equation*}
\mathcal{F}(t)=v(t)^2+u(t)w(t)
\end{equation*}
we have
\[
\mathcal{D}_s\mathcal{F}(t)=0,\quad \forall s,t.
\]

\begin{prop} Let
$(\mathbf{X},\mathbf{Y},\mathbf{Z})=(x_1,\dots,x_g,y_1,\dots,y_g,
z_1,\dots,z_g)\in\mathbb{C}^{3g}$,
$\partial_i=\frac{\partial}{\partial u_i}$, $i=1,\dots,g$.
Dynamical system defined by the system of equations
\begin{align}
&\partial_{i}x_{k} -\partial_{k}x_{i}=0\notag\\ &\partial_{i}y_{k}
-\partial_{k}y_{i}=0\notag\\ &\partial_{i}y_{k}
-\partial_{k}y_{i}=0\notag\\ &\partial_{i-1}x_{k}
-\partial_{k-1}x_{i}= (\delta_{i,g+1}-x_{i})y_k
-(\delta_{k,g+1}-x_{k})y_i\label{dyn}\\ &\partial_{i-1}y_{k}
-\partial_{k-1}y_{i}= (\delta_{i,g+1}-x_{i})z_k
-(\delta_{k,g+1}-x_{k})z_i\notag\\ &\partial_{i-1}z_{k}
-\partial_{k-1}z_{i}=y_k z_i-y_i z_k+\notag\\
&\quad+4\big[(\delta_{i,g+1}-x_{i})(2y_kx_g+y_{k-1})
-(\delta_{k,g+1}-x_{k})(2y_i x_g+y_{i-1}) \big]\notag\\
&\qquad i,k=1,\dots,g+1,\notag
\end{align}
where $\partial_{g+1}=\partial_{0}=0$,
$x_0 = x_g = y_0 = y_g = z_0 = z_g $ possesses $2g$ conserved
quantities $f_L, \, L=1,\dots,2g$ defined by \begin{align}
f_L(\mathbf{X},\mathbf{Y},\mathbf{Z})=&\sum_{i+k=L} y_i y_k
-4x_g(\delta_{k,g+1}-x_{k})(\delta_{i,g+1}-x_{i})\notag
+\\&(\delta_{k,g+1}-x_{k})(z_i+6
x_g(\delta_{i,g+1}-x_{i})+2(\delta_{i-1,g+1}-x_{i-1}))+\label{f_in}\\
&(\delta_{i,g+1}-x_{i})(z_k+6
x_g(\delta_{k,g+1}-x_{k})+2(\delta_{k-1,g+1}-x_{k-1})).\notag
\end{align}
\end{prop}
\begin{proof}
Let us multiply each of the equations \eqref{dyn} by $s^{i-1}t^{k-1}$
and carry out summation over $i$ and $k$ from $1$ to
$g+1$.
Introducing functions \begin{align*}
x(t)&=2\big(t^g-\sum_{i=1}^g t^{i-1}x_i\big),\quad
y(t)=\sum_{i=1}^g t^{i-1}y_i,\\
z(t)&=\sum_{i=1}^g t^{i-1}z_i.
\end{align*}
and vector field $\mathcal{D}_s=\sum_{i=1}^g s^{i-1}\partial_i$,
we have for these equations the form:
\begin{align*}
\mathcal{D}_s x(t)&-\mathcal{D}_t x(s)=0\\
\mathcal{D}_s y(t)&-\mathcal{D}_t y(s)=0\\
\mathcal{D}_s z(t)&-\mathcal{D}_t z(s)=0\\
s\mathcal{D}_s x(t)&-t\mathcal{D}_t x(s)=-x(s)y(t)+x(t)y(s)\\
s\mathcal{D}_s y(t)&-t\mathcal{D}_t y(s)=x(s)z(t)-x(t)z(s)\\
s\mathcal{D}_s z(t)&-t\mathcal{D}_t z(s)=z(s)y(t)-z(t)y(s)\\
&+4\big[
x(s)y(t)(t+2x_g)-x(t)y(s)(s+2x_g)
\big].
\end{align*}
Hence we have the latter three equations rewritten as
\begin{align*}
\mathcal{D}_s x(t)&=-\frac{x(s)y(t)-x(t)y(s)}{s-t}\\
\mathcal{D}_s y(t)&=\frac{x(s)z(t)-x(t)z(s)}{s-t}\\
\mathcal{D}_s z(t)&=\frac{z(s)y(t)-z(t)y(s)}{s-t}\\
&+\frac{4}{s-t}\big[ x(s)y(t)(t+2x_g)-x(t)y(s)(s+2x_g) \big]
\end{align*}
and also $\mathcal{D}_s x_g=y(s)$.

The equations obtained may be written in the form \eqref{LM_eq} if
we put
\begin{equation*} u(t)=2x(t),\quad v(t)=y(t),\quad
w(t)=z(t)+2x(t)(t+2 x_g) \end{equation*}
and
\begin{equation*}
\alpha(s,t)=\frac{2x(s)}{t-s},\quad
\beta(s,t)=\frac{y(s)}{t-s},\quad \gamma(s,t)=\frac{z(s)+2
x(s)(t+2x_g)}{t-s}.
\end{equation*}
Corresponding integral has the form
\begin{equation*}
\mathcal{F}(t)=y(t)^2+2x(t)\big(z(t)+2x(t)(t+2x_g)\big),
\end{equation*}
and the conserved quantities $f_L$ in \eqref{f_in} are the
coefficients at the powers $t^L$ of this integral.
\end{proof}

We can integrate the system \eqref{dyn} in terms of Kleinian
functions.
\begin{cor} Let $\{u_i\}_{i=1,\dots,g}$ be the canonical
coordinates on the Jacobian of the curve $V$ defined by
\eqref{curve} with $\lambda_{2g+2}=0$, $\lambda_{2g+1}=4$ and
$\lambda_{2g}=0$, then vector function
$\mathrm{Jac}(V)\to\mathbb{C}^{3g}:\boldsymbol{u}
\mapsto\big(\wp(\boldsymbol{u}),\wp'(\boldsymbol{u}),
\wp''(\boldsymbol{u})\big)$ is a solution of the system
\eqref{dyn}.
\end{cor}
This Corollary is verified by \eqref{wp3} and \eqref{i-1,k-1}.

The system \eqref{dyn} preserves the structure of the universal
space $\mathcal{HT}'_g\cong\mathbb{C}^{3g}$ of the
hyperelliptic Jacobians of canonical curves of genus $g$.
\begin{prop} Each point $$(\mathbf{X},\mathbf{Y},\mathbf{Z})=(x_1,\dots,x_g,y_1,\dots,y_g,
z_1,\dots,z_g)\in\mathbb{C}^{3g}$$ defines uniquely the Jacobian
of the Riemann surface of the canonical hyperelliptic curve
$$
V(\mu,\nu)=\big\{
(\mu,\nu)\in \mathbb{C}^2\mid
\mu^2-4\nu^{2g+1}-\sum_{L=0}^{2g-1}\nu^L\lambda_{L}=0
\big\},
$$
the constants $\lambda_L$ being defined by formula
$$
\lambda_L=f_L(\mathbf{X},\mathbf{Y},\mathbf{Z})
$$
and  functions $f_L$ are as given by \eqref{f_in}.  \end{prop}

Consider as an example the family  of elliptic curves:
\begin{equation}
\label{ellipt}
\big\{
(x,y)\in \mathbb{C}^2:\, y^2=4 x^3 -g_2 x-g_3;\, (g_2,g_3)\in
\mathbb{C}^2 \big\}. \end{equation} The universal space of the
 curves \eqref{ellipt} is $W_1\cong \mathbb{C}^3$. Let us
introduce functions \begin{align*} f_1(x,y,z)&=12 x^2-2 z\\
f_2(x,y,z)&=8 x^3+y^2-2x z.
\end{align*}
If we put:
\begin{equation*}
f_1(x,y,z)=g_2
\quad\text{and}\quad
f_0(x,y,z)=g_3,\quad g_2,g_3\in \mathbb{C}
\end{equation*}
then eliminating $z$ we obtain the curve \eqref{ellipt} with given
$g_2$ and $g_3$. Using the Weierstrass parametrization of the
cubic \eqref{ellipt}
\begin{equation*}
\wp'(u)^2=4 \wp(u)^3-g_2\wp(u)-g_3,\quad
\wp'(u)=6\wp(u)^2-\frac{1}{2}g_2
\end{equation*}
and the mapping
$T^1/(\sigma)\to\mathbb{C}^3:
u\to\big(\wp(u),\wp'(u),\wp''(u)\big)$ we can integrate  a
dynamical system:
\begin{equation}
\begin{cases}
\dot x=y\\
\dot y=z\\
\dot z=12 x y \label{DS-simple}
\end{cases}
\end{equation}
and $f_0$ and $f_1$ are its integrals of motion.

\index{Weierstrass!cubic}
Let us consider the Lie algebra $sl(2,\mathbb{C})$ of the
                           $2\times2$ traceless
non-degenerate matrices
\begin{equation*} \begin{pmatrix} \beta&\alpha\\ \gamma&-\beta
\end{pmatrix}
\end{equation*}
and the following homeomorphism:
\begin{equation*}
j_1:\mathbb{C}^3\to sl(2,\mathbb{C}):\, j(x,y,z)=\begin{pmatrix}
\beta&\alpha\\
\gamma&-\beta
\end{pmatrix}=M_0,
\end{equation*}
where $\alpha=-x$, $\beta=-\frac{1}{2}y$ and $\gamma=2
x^2-\frac{1}{2}z$. Then we can rewrite \eqref{DS-simple} as
\begin{equation*}
\dot M_0=[M_1,M_0],
\end{equation*}
where
\begin{equation*}
M_1=\phi(M_0)=
\begin{pmatrix}
0&1\\
-2\alpha&0
\end{pmatrix}.
\end{equation*}
We have obtained a dynamical system on $sl(2,\mathbb{C})$.
The integrals of motion are defined by
\begin{equation*}
F_k(M_0)=f_{k-1}\big(j_1^{-1}(M_0)\big) ,\,k=1,2
\end{equation*}
and we have
\begin{align*}
F_1(M_0)&=4(\alpha^2+\gamma)\\
F_2(M_0)&=4(\beta^2+\alpha\gamma)
\end{align*}
For the mapping  $F:sl(2,\mathbb{C})\to\mathbb{C}^2:M_0\to
(F_1,F_2)$ we find
\begin{equation*}
\nabla F=\frac{\partial(F_1,F_2)}{\partial(\alpha,\beta,\gamma)}=
4\begin{pmatrix}
2 \alpha&0&1\\
\gamma&2\beta&\alpha
\end{pmatrix},
\end{equation*}
and $\mathrm{rank}(\nabla F)<2$ if and only if $\beta=0$ and
$\gamma=2 \alpha^2$.

We give  the generalization of the above elliptic
construction to the case of the universal space of hyperelliptic
curves of arbitrary genus $g$.

Consider the universal space $W^g$ of the hyperelliptic curves
\index{space!universal}
\begin{equation}
\label{hellipt}
\big\{
(x,y)\in \mathbb{C}^2:\, y^2=4 x^{2g+1}
-\sum_{i=0}^{2g-1}\lambda_i x^i;\,(\lambda_0,\dots,\lambda_{2g-1})
\in \mathbb{C}^{2g} \big\}
\end{equation}
(we have put $\lambda_{2g}=0$) is isomorphic to
$\mathbb{C}^{3g}$. Using mapping
$T^g/(\sigma)\to\mathbb{C}^{3g}:
\boldsymbol{u}\to\big(\boldsymbol{\wp}(\boldsymbol{u}),
\boldsymbol{\wp}'(\boldsymbol{u}),\boldsymbol{\wp}''(\boldsymbol{u})\big)$

Let us consider the space
$\mathcal{V}_g=\underset{g}{\underbrace{sl(2,\mathbb{C})\times
\dots\times sl(2,\mathbb{C})}}$. A point in $\mathcal{V}_g$ is the
set $\big(M_0,M_1,\dots,M_{g-1}\big)$ with $M_k=\bigl(
\begin{smallmatrix}
\beta_{k+1}&\alpha_{k+1}\\
\gamma_{k+1}&-\beta_{k+1}
\end{smallmatrix}
\bigr)\in sl(2,\mathbb{C})$. Put $M_g=\phi(M_{g-1})=
\bigl(
\begin{smallmatrix}
0&1\\
-2\alpha_{g}&0
\end{smallmatrix}
\bigr)$ and
$W_k=[M_k,E]$, where $E=\bigl(
\begin{smallmatrix}
0&0\\
-1&0
\end{smallmatrix}
\bigr)$ . That is
$W_k=\bigl(
\begin{smallmatrix}
-\alpha_{k+1}&0\\
2 \beta_{k+1}&\alpha_{k+1}
\end{smallmatrix}
\bigr)$.

We introduce the dynamical system on $\mathcal{V}_g$ as follows:
\begin{enumerate}
\item
\begin{equation}
\partial_k M_i-\partial_i M_k=[M_k,M_i],\quad
i,k=0,\dots,g\label{dyn_1} \end{equation}
we assume $\partial_0
M_k\equiv 0$ \item
\begin{equation} \partial_{k+1}
M_i-\partial_{i+1} M_k=[W_k,W_i],\quad i,k=0,\dots,g-1
\label{dyn_2} \end{equation} \end{enumerate} The first set of
equations gives $\frac{g(g+1)}{2}$ nontrivial equations, and the
second $\frac{g(g-1)}{2}$; so totally we obtain the system of
$g^2$ equations. For example for small values of $g$ we have:
\begin{enumerate}
\item{$g=1$; coordinates: $(M_0)$}
\begin{equation*}
\partial_1 M_0=[M_1,M_0];
\end{equation*}
\item{$g=2$; coordinates: $(M_0,M_1)$}
\begin{equation*}
\begin{cases}
\partial_{1} M_0=[M_1,M_0]\\
\partial_{2} M_0=[M_2,M_0]
\end{cases},
\begin{cases}
\partial_{1} M_1=\partial_{2} M_0+[W_1,W_0]\\
\partial_{2} M_1=\partial_{1} M_2+[M_2,M_1]
\end{cases} ;
\end{equation*}
\item{$g=3$; coordinates: $(M_0,M_1,M_2)$}
\begin{equation*}
\begin{gathered}
\begin{cases}
\partial_{1} M_0=[M_1,M_0]\\
\partial_{2} M_0=[M_2,M_0]\\
\partial_{3} M_0=[M_3,M_0]
\end{cases},
\begin{cases}
\partial_{1} M_1=\partial_{2} M_0+[W_1,W_0]\\
\partial_{2} M_1=\partial_{1} M_2+[M_2,M_1]\\
\partial_{3} M_1=\partial_{1} M_3+[M_3,M_1]
\end{cases},\\
\begin{cases}
\partial_{1} M_2=\partial_{3} M_0+[W_2,W_0]\\
\partial_{2} M_2=\partial_{3} M_1+[W_2,W_1]\\
\partial_{3} M_2=\partial_{2} M_3+[M_3,M_2]
\end{cases}.
\end{gathered}
\end{equation*}
\end{enumerate}
The integrals of motion are given by
\begin{equation*}
F_k(M_0,\dots M_{g-1})=f_{k-1}\big(j_g^{-1}(M_0,\dots
M_{g-1})\big) ,\,k=1,\dots 2g. \end{equation*} This result leads
to the following
\begin{cor}
The mapping $j_g:\mathbb{C}^{3g}\to \mathcal{V}_g$ takes an
integrable dynamical system of the form \eqref{dyn} to an
integrable system of the form \eqref{dyn_1}-\eqref{dyn_2}.
\end{cor}

\chapter[Addition theorems]{Addition theorems for
hyperelliptic functions}\label{chap:add}
\index{Kleinian function!hyperelliptic!addition theorems for}

\section{Introduction}
During the last 30 years the addition laws of elliptic functions
stay in the focus of the studies in the nonlinear equations of
Mathematical Physics. A large part of the interest was drawn to the
subject by the works of F.~Calogero~\cite{FC75,FC75a,FC76}, where
several important problems were reduced to the elliptic  addition
laws.  The term ``Calogero-Moser model'' being widely used in
literature, the papers caused a large series of publications, where
on one hand more advanced problems were posed and on the other hand
some advances were made in the theory of functional equations. The
``addition theorems'' for Weierstrass elliptic functions:
\begin{gather}
\frac{\sigma(u+v)\sigma(u-v)}{\sigma(u)^2\sigma(v)^2}=\wp(v)-\wp(u),\label{si}\\
(\zeta(u)+\zeta(v)+\zeta(w))^2=\wp(u)+\wp(v)+\wp(w),\quad
u+v+w=0,\label{ze}
\end{gather}
played the key r\^ole in the works of that period. At the same time,
the development of the algebro-geometric methods of solution of
integrable systems \cite{dmn76,dkn01} employed in an essential way
the addition formulas for theta functions of several variables. The
same addition formulas were needed in applications of Hirota method.
In \cite{bk93,bk96} the addition theorems  for vector Baker-Akhiezer
functions of several variables are obtained and a program is put
forward to apply the addition theorems to problems of the theory of
integrable systems, in particular, to multidimensional analogs of
Calogero-Moser type systems.

The fundamental fact of the elliptic functions theory is that any
elliptic function can be represented as a rational function of
Weierstrass functions $\wp$ and $\wp'$. The corresponding result
 in the theory of hyperelliptic Abelian
functions is formulated as follows: any hyperelliptic function can
be represented as a rational function of vector functions
$\bwp=(\wp_{g1},\dots,\wp_{gg} )$ and
$\bwp'=(\wp_{gg1},\dots,\wp_{ggg})$, where $g$ is the genus of the
hyperelliptic curve on the Jacobi variety of which the field of
Abelian functions is built.

In the present paper we find the explicit formulas for the addition
law of the vector functions $\bwp$ and $\bwp'$. As an application
the higher genus analogs of the Frobenius-Stickelberger formula
\eqref{ze} are obtained. In particular, for the genus $2$
sigma-function we obtain the following trilinear differential
addition theorem
\begin{equation*}
\big[2(\partial_{u_1}+\partial_{v_1}+\partial_{w_1})+
(\partial_{u_2}+\partial_{v_2}+\partial_{w_2})^3\big]\sigma(u)\sigma(v)\sigma(w)\big|_{u+v+w=0}=0.
\end{equation*}

Our approach is based on the explicit construction of the groupoid
structure that is adequate to describe the algebraic structure of
the space of $g$-th symmetric powers of hyperelliptic curves.

\section{Groupoids}

\subsection{Topological groupoids}
\begin{definition} Take a topological space $\mathsf{Y}$.

A space $\mathsf{X}$ together with a mapping
$p_{\mathsf{X}}:\mathsf{X}\to Y$ is called \emph{a space over
$\mathsf{Y}$}. The mapping $p_{\mathsf{X}}$ is called ``an anchor''
in Differential Geometry.

Let two  spaces  $\mathsf{X}_1$ and $\mathsf{X}_2$ over $\mathsf{Y}$
be given. A mapping  $f:\mathsf{X}_1\to \mathsf{X}_2$ is called
\emph{a mapping over $\mathsf{Y}$}, if $p_{\mathsf{X}_2}\circ
f(x)=p_{\mathsf{X}_1}(x)$ for any point $x\in \mathsf{X}_1$.

By \emph{the direct product over $\mathsf{Y}$} of the spaces
 $\mathsf{X}_1$ and $\mathsf{X}_2$
 over $\mathsf{Y}$ we call the space $ \mathsf{X}_1\times_{\mathsf{Y}}\mathsf{X}_2=\{(x_1,x_2)\in
\mathsf{X}_1\times \mathsf{X}_2\mid
p_{\mathsf{X}_1}(x_1)=p_{\mathsf{X}_2}(x_2)\}$ together with the
mapping
$p_{\mathsf{X}_1\times_{\mathsf{Y}}\mathsf{X}_2}(x_1,x_2)=p_{\mathsf{X}_1}(x_1)$.
\end{definition}
The space  $\mathsf{Y}$ together with the identity mapping
$p_\mathsf{Y}$ is considered as the space over itself.

\begin{definition} A space $\mathsf{X}$ together with a mapping
$p_{\mathsf{X}}:\mathsf{X}\to \mathsf{Y}$ is called \emph{ a
groupoid over $\mathsf{Y}$}, if  there are defined the structure
mappings over $\mathsf{Y}$
\begin{equation*}
\mu: \mathsf{X}\times_\mathsf{Y} \mathsf{X}\to
\mathsf{X}\quad\text{and} \quad\inv: \mathsf{X}\to \mathsf{X}
\end{equation*}
that satisfy the axioms
\begin{enumerate}
\item $\mu(\mu(x_1,x_2),x_3)=
\mu(x_1,\mu(x_2,x_3)),$ provided $
p_{\mathsf{X}}(x_1)=p_{\mathsf{X}}(x_2)=p_{\mathsf{X}}(x_3).$
\item $\mu(\mu(x_1,x_2),\inv(x_2))=x_1$,  provided $
p_{\mathsf{X}}(x_1)=p_{\mathsf{X}}(x_2)$.
\end{enumerate}
The mapping  $\mu$ may not be defined for all pairs $x_1$ and $x_2$
from $\mathsf{X}$.
\end{definition}
\begin{definition} A groupoid structure on $\mathsf{X}$
over the space  $\mathsf{Y}$ is called commutative, if
 $\mu(x_1,x_2)=\mu(x_2,x_1)$, provided $
p_{\mathsf{X}}(x_1)=p_{\mathsf{X}}(x_2)$.
\end{definition}
\begin{definition}
A groupoid structure on the algebraic variety $\mathsf{X}$ over the
algebraic variety $\mathsf{Y}$ is called \emph{algebraic}, if the
mapping $p_\mathsf{X}$ as well as  the structure mappings
 $\mu$ and  $\inv$ are algebraic.
\end{definition}

\subsection{Algebraic groupoids defined by plane curves}
We take as $\mathsf{Y}$ the space  $\mathbb{C}^{N}$ with coordinates
$\Lambda=(\lambda_{i})$, $i=1,\dots,N$. Let $f(x,y,\Lambda),$ where
$(x,y)\in\mathbb{C}^2$, be a polynomial in $x$ and $y$. Define the
 family of plane curves
\begin{equation*}
V=\{(x,y,\Lambda)\in\mathbb{C}^2\times\mathbb{C}^N\mid f(x,y,
\Lambda)=0\}.
\end{equation*}
We assume that at a generic value of $\Lambda$, genus of the curve
from $V$ has fixed value $g$.

Let us take as $\mathsf{X}$  the universal fiber-bundle of  $g$-th
symmetric powers of the algebraic curves from $V$. A point in
$\mathsf{X}$ is represented  by the collection of an unordered set
of $g$ pairs $(x_i,y_i)\in \mathbb{C}^2$ and  an $N$-dimensional
vector $\Lambda$ that are related by $f(x_i,y_i,\Lambda)=0$,
$i=1,\dots,g$.

The mapping $p_{\mathsf{X}}$ takes the collection to the point
$\Lambda\in \mathbb{C}^{N}.$

Let  $\phi(x,y)$ be an entire rational function on the curve  $V$
with the parameters $\Lambda$. A zero of the function $\phi(x,y)$ on
the curve $V$ is  the point  $(\xi,\eta)\in\mathbb{C}^2$, such that
$\{ f(\xi,\eta,\Lambda)=0,\phi(\xi,\eta)=0\}$. The total number of
zeros of the function $\phi(x,y)$ is called the order of
$\phi(x,y)$.

The further construction is based on the following fact.
\begin{lemma}Let  $\phi(x,y)$  be an order  $2g+k,$ $k\geqslant0$,
entire rational function on the curve
 $V$. Then the function  $\phi(x,y)$ is completely defined (up to a
constant with respect to $(x,y)$ factor) by any collection of  $g+k$
its zeros.
\end{lemma}
This fact is a consequence of Weierstrass gap theorem
(L{\"u}kensatz). In particular, an ordinary univariate polynomial is
an entire rational function on the curve of genus  $g=0$ and is
completely defined by the collection of all its zeros.

Let us construct the mapping $\inv$.

Let a point $U_{1}=\{[(x^{(1)}_i,y^{(1)}_i)],\Lambda\}\in
\mathsf{X}$ be given. Let $R^{(1)}_{2g}(x,y)$ be the entire rational
function
 of order $2g$ on the curve $V$ defined by the vector $\Lambda$,
 such that $R^{(1)}_{2g}(x,y)$  is zero in $U_1$, that is
$R^{(1)}_{2g}(x^{(1)}_i,y^{(1)}_i)=0$, $i=1,\dots,g$. Denote by
$[(x^{(2)}_i,y^{(2)}_i)]$ the complement of
$[(x^{(1)}_i,y^{(1)}_i)]$ in the set of zeros of
$R^{(1)}_{2g}(x,y)$. Denote by  $U_2$ the point in $\mathsf{X}$ thus
obtained and set $\inv(U_1)=U_2$.

So, the set of zeros of the function $R^{(1)}_{2g}(x,y)$, which
defines the mapping $\inv$, is the pair of points
$\{U_1,\inv(U_1)\}$  from $\mathsf{X}$ and
$p_{\mathsf{X}}(U_1)=p_{\mathsf{X}}(U_2)$.

\begin{lemma} The mapping $\inv$ is an involution, that is
$\inv\circ\inv(U_1)=U_1$.
\end{lemma}

Let us construct the mapping $\mu$.

Let two points $U_1=\{[(x^{(1)}_i,y^{(1)}_i)],\Lambda\}$ and
$U_2=\{[(x^{(2)}_i,y^{(2)}_i)],\Lambda\}$ from $\mathsf{X}$ be
given. Let $R^{(1,2)}_{3g}(x,y)$  be the entire rational function of
order $3g$ on the curve $V$ defined by the vector $\Lambda$, such
that $R^{(1,2)}_{3g}(x,y)$ is zero in  $\inv(U_1)$ and $\inv(U_2)$.
Denote by  $[(x^{(3)}_i,y^{(3)}_i)]$ the complementary $g$ zeros of
$R^{(1,2)}_{3g}(x,y)$ on the curve $V$. Denote by  $U_3$ the point
in $\mathsf{X}$ thus obtained and set $\mu(U_1,U_2)=U_3$.

So, the set of zeros of the function $R^{(1,2)}_{3g}(x,y)$, which
defines the mapping $\mu$, is the triple   of points
$\{\inv(U_1),\inv(U_2), \mu(U_1,U_2)\}$ from $\mathsf{X}$ and
$p_{\mathsf{X}}(U_1)=p_{\mathsf{X}}(U_2)=p_{\mathsf{X}}(\mu(U_1,U_2))$.
\begin{theorem}\label{universal-groupoid} The above mappings  $\mu$ and $\inv$ define the structure of
the commutative algebraic groupoid over $\mathsf{Y}=\mathbb{C}^{N}$
on the universal fiber-bundle $\mathsf{X}$ of
 $g$-th symmetric powers of the plane algebraic curves from the family
 $V$.
\end{theorem}
\begin{proof}
The mapping $\mu$ is symmetric with respect to $U_1$ and $U_2$, and
thus defines the commutative operation. By the construction the
mappings $p_\mathsf{X}$, $\mu$ and $\inv$ are algebraic.
\begin{lemma}\label{associatif} The mapping $\mu$ is associative.
\end{lemma}
\begin{proof}Let three points  $U_1, U_2$ and $U_3$ be given. Let us
assign
\begin{equation*}
U_{4}=\mu(U_1,U_2),\quad U_5=\mu(U_4,U_3),\quad
U_{6}=\mu(U_2,U_3),\quad U_7=\mu(U_1,U_6).\end{equation*} We have to
show that $U_5=U_7.$

Let $R_{3g}^{(i,j)}(x,y)$ be the function defining the point
$\mu(U_i,U_j)$, and  $R_{2g}^{(i)}(x,y)$ be the function defining
the point $\inv(U_i)$.

Consider the product $R_{3g}^{(1,2)}(x,y)R_{3g}^{(4,3)}(x,y)$. It is
a function of order  $6g$ with zeros at  \[\{\inv (U_1),\inv
(U_2),U_4,\inv(U_4),\inv(U_3),U_5\}.\]  Therefore, the function
\begin{equation*}
Q_1(x,y)=\frac{R_{3g}^{(1,2)}(x,y)R_{3g}^{(4,3)}(x,y)}{R_{2g}^{(4)}(x,y)}\end{equation*}
is an entire function of order  $4g$ with the zeros $\{\inv
(U_1),\inv (U_2),\inv(U_3),U_5\}$.

Similarly, the product $R_{3g}^{(2,3)}(x,y)R_{3g}^{(1,6)}(x,y)$ is a
function of order $6g$ with zeros at \[\{\inv (U_2),\inv
(U_3),U_6,\inv(U_6),\inv(U_1),U_7\}.\] Hence we find that
\begin{equation*}
Q_2(x,y)=\frac{R_{3g}^{(2,3)}(x,y)R_{3g}^{(1,6)}(x,y)}{R_{2g}^{(6)}(x,y)}\end{equation*}
is an entire function of order  $4g$ with the zeros $\{\inv
(U_1),\inv (U_2),\inv(U_3),U_7\}$.

The functions $Q_1(x,y)$ and  $Q_2(x,y)$ have order $4g$ and both
vanish at the points \[\{\inv (U_1),\inv (U_2),\inv(U_3)\}.\] Thus
by Weierstrass gap theorem $Q_1(x,y)=Q_2(x,y)$ and, therefore,
$U_5=U_7$.
\end{proof}

\begin{lemma}\label{weak_inverse}
The mappings  $\mu$ and $\inv$  satisfy the axiom \textup{2}.
\end{lemma}
\begin{proof}Let two points $U_1$ and $U_2$ be given. Assign
\begin{equation*}U_{3}=\mu(U_1,U_2),\quad U_4=\inv(U_2),\quad U_5=\mu(U_3,U_4).\end{equation*}
We have to show that  $U_5=U_1.$

Consider the product  $R_{3g}^{(1,2)}(x,y)R_{3g}^{(3,4)}(x,y)$,
which is the function of order $6g$ with the zeros \[\{\inv(U_1),
\inv(U_2), U_3,\inv(U_3),\inv(U_4),U_5\}.\] Since
\[\inv(U_4)=\inv\circ\inv(U_2)=U_2,\] the function
\begin{equation*}
Q(x,y)=\frac{R_{3g}^{(1,2)}(x,y)R_{3g}^{(3,4)}(x,y)}{R_{2g}^{(2)}(x,y)R_{2g}^{(3)}(x,y)}
\end{equation*}
is the entire function of order $2g$ with zeros at
$\{\inv(U_1),U_5\}$, that is  $Q(x,y)=R_{2g}^{(5)}(x,y)$. Hence it
follows that $U_5=\inv\circ\inv(U_1)=U_1.$
\end{proof}
The Theorem is proved.
\end{proof}
The Lemma below is useful for constructing the addition laws on our
groupoids.
\begin{lemma}\label{even_part}
Given $U_1,U_2\in \mathsf{X}$, let us assign
 $U_{3}=\mu(U_1,U_2)$ and $U_{i+3}=\inv(U_i)$,
$i=1,2.$ Then
\begin{equation*}
R^{(3)}_{2g}(x,y)=
\frac{R^{(1,2)}_{3g}(x,y)R^{(4,5)}_{3g}(x,y)}{R^{(1)}_{2g}(x,y)R^{(2)}_{2g}(x,y)}.
\end{equation*}
\end{lemma}
The formula of Lemma \ref{even_part} is important because its left
hand side depends formally  on $U_3$ only, while the right hand side
is completely defined by the pair $U_1, U_2$.

The above general construction becomes  effective once we fix a
model of the family of curves, that is once the polynomial
$f(x,y,\Lambda)$ is given. We are especially interested in the
models of the form, cf. for instance \cite{bel99a,bl02,bl04},
\begin{equation*}
f(x,y,\Lambda)=y^n-x^s-\sum \lambda_{ns-in-js}x^i y^j,
\end{equation*}
where $\gcd(n,s)=1$ and the summation is carried out over the range
$0<i<s-1$, $0<j<n-1$ under the condition $ns-in-js\geqslant 0$. It
is important that a model of the kind (possibly with singular
points) exists for an arbitrary curve. At the generic values of
$\Lambda$ a curve in such a family has genus $g=(n-1)(s-1)/2.$ In
this paper we consider in detail the case $(n,s)=(2,2g+1)$, that is
the families of hyperelliptic curves.

\section{Hyperelliptic groupoid on $\mathbb{C}^{3g}$}
A hyperelliptic curve  $V$ of genus $g$ is usually defined by a
polynomial of the form
\begin{equation*}
f(x,y,\lambda_{0},\lambda_{2},\dots)=y^2-4x^{2g+1}-\sum_{i=0}^{2g-1}\lambda_{i}x^i.
\end{equation*}
In this paper we apply the change of variables
\begin{equation*}(x,y,\lambda_{2g-1},\lambda_{2g-2},\dots,\lambda_{0})\to
(x, 2 y, 4\lambda_4,
4\lambda_6,\dots,4\lambda_{4g+2}),\end{equation*} in order to
simplify the formulas in the sequel. Below we study the
constructions related to the hyperelliptic curves defined by the
polynomials of the form
\begin{equation}
\label{hypp} f(x,y,\lambda_{4g+2},\lambda_{4g},\dots)=
y^2-x^{2g+1}-\sum_{i=0}^{2g-1}\lambda_{4g+2-2i}x^i.
\end{equation}
Let us introduce the grading by assigning $\deg x=2,$ $\deg y=2g+1$
and $\deg\lambda_k=k$. Then the polynomial $f(x,y,\Lambda)$ becomes
a homogeneous polynomial of the weight $4g+2.$

An entire function on  $V$ has  a unique representation as the
polynomial $R(x,y)=r_0(x)+r_1(x)y,$ where
$r_0(x),r_1(x)\in\mathbb{C}[x].$ In such representation we have not
more than one monomial $x^i y^j$ of each weight, by definition $\deg
x^i y^j=2i+j(2g+1)$. The order of a function $R(x,y)$ is equal to
the maximum of weights of the monomials that occur in $R(x,y)$. In
fact, on the set of zeros of the polynomial $R(x,y)$ we have
$y=-r_0(x)/r_1(x)$. Therefore, the zeros of $R(x,y)$ that lie on the
curve are defined by the roots  $x_1,\dots,x_m$ of the equation
$r_0(x)^2-r_1(x)^2(x^{2g+1}+\lambda_4 x^{2g-1}+\dots)=0.$ The total
number of the roots is equal  $m=\max(2\deg_x r_0(x),2g+1+2\deg_x
r_1(x))$, where $\deg_{x}r_j(x)$ denotes the degree of the
polynomial $r_j(x)$ in $x$, which is exactly the highest weight of
the monomials in $R(x,y)$.

In this case Weierstrass gap theorem  asserts that the ordered
sequence of nonnegative integers $\{\deg x^i y^j\}$, $j=0,1$,
$i=0,1,\dots$, has precisely  $g$ ``gaps'' in comparison to the
sequence of all nonnegative integers. All of the gaps are less than
$2g$.
\begin{lemma}
For a given point $U_1=\{[(x^{(1)}_i,y^{(1)}_i)],\Lambda\}\in
\mathsf{X}$ the entire function  $R^{(1)}_{2g}(x,y)$ defining the
mapping  $\inv$ has the form
\begin{equation*}R^{(1)}_{2g}(x,y)=(x-x^{(1)}_1)\dotsc(x-x^{(1)}_g).\end{equation*}
\end{lemma}\begin{proof}
In fact, as $\deg y>2g$, any entire function of order $2g$ does not
depend on $y$.
\end{proof}
The function  $R^{(1)}_{2g}(x,y)$  defines the unique point
\begin{equation*}\inv(U_1)=\{[(x^{(1)}_i,-y^{(1)}_i)],\Lambda\},\end{equation*} which, obviously,
also belongs to  $\mathsf{X}$.

Let us construct the functions $R_{3g}^{(i,j)}(x,y)$ that have the
properties required in Lemmas \ref{associatif} and
\ref{weak_inverse}.
\begin{lemma}
Define the $(2g+1)$-dimensional row-vector
\begin{equation*}
m(x,y)=(1,x,\dots,x^{2g-1-\rho},y, y x,\dots, y x^{\rho}),\quad
\rho=\Big[\frac{g-1}{2}\Big],
\end{equation*}
which is  composed of all monomials $x^iy^j$ of weight not higher
than $3g$ (the restriction $j=0,1$ applies). Then, up to a factor
constant in $(x,y)$, the function $R_{3g}^{(1,2)}(x,y)$ is equal to
the determinant of the matrix composed of $2g+1$ rows $m(x,y)$,
 $m(x_i^{(1)},-y_i^{(1)})$ and
 $m(x_i^{(2)},-y_i^{(2)})$, $i=1,\dots, g.$
\end{lemma}\begin{proof}
By the construction the function
 $R_{3g}^{(1,2)}(x,y)$ vanishes at the points $\inv(U_1)$ and $\inv(U_2)$,
and is uniquely defined by this property. As
$\max(4g-2-2\rho,2g+1+2\rho)=3g$,  at fixed $\Lambda$ the function
$R_{3g}^{(1,2)}(x,y)$ has  $2g$  zeros at the given points of the
curve and the collection of zeros at $[(x^{(3)}_i,y^{(3)}_i)],$
$i=1,\dots,g$, which defines the unique point $U_3$ in $\mathsf{X}$.
\end{proof}

Let $\Sym^n(\mathsf{M})$ denote the  $n$-th symmetric power of the
space $\mathsf{M}$. A point of the space $\Sym^n(\mathsf{M})$ is an
unordered collection $[m_1,\dots,m_n],$ $m_i\in \mathsf{M}$.

Consider the space $\mathsf{S}=\Sym^{g}(\mathbb{C}^2)\times
\mathbb{C}^g$ and let us define the mapping
$p_{\mathsf{S}}:\mathsf{S}\to \mathsf{Y}=\mathbb{C}^{2g}.$ Take a
point $T=\{[(\xi_i,\eta_i)],Z\}\in \mathsf{S}$. Denote by
$\mathcal{V}$ the Vandermonde matrix, composed of $g$ rows $(1,
\xi_{i},\dots,\xi_{i}^{g-1})$, denote by $\mathcal{X}$ the diagonal
matrix $\diag (\xi_1^g,\dots,\xi_g^g)$ and by  $\mathcal{Y}$ the
vector $(\eta_1^2-\xi_1^{2g+1},\dots,\eta_g^2-\xi_g^{2g+1})^T$. Set
\begin{equation*}p_S(T)=(Z_1,Z_2),\quad\text{where}\quad
Z_1=\mathcal{V}^{-1}\mathcal{Y}-(\mathcal{V}^{-1}\mathcal{X}\mathcal{V})
Z  \quad\text{and}\quad   Z_2=Z.\end{equation*} It is clear that the
domain of definition of the mapping $p_\mathsf{S}$ is the open and
everywhere dense subset  $\mathsf{S}_0$ in $\mathsf{S}$ consisting
of the points $\{[(\xi_i,\eta_i)],Z\}$ such that the determinant
$\mathcal{V}$ does not vanish.

We define the mappings  $\gamma:\mathsf{X}\to \mathsf{S}$ and
$\delta: \mathsf{S}\to \mathsf{X}$ by the following formulas: let
$U\in \mathsf{X}$ and $T\in \mathsf{S}$, then
\[
\gamma(U)=\gamma(\{[(x_i,y_i)],\Lambda\})=\{[(x_i,y_i)],\Lambda_2\},
\]
where $ \Lambda_2=(\lambda_{2(g-i)+4}),$ $i=1,\dots,g,$
\[
\delta(T)=\delta(\{[(\xi_i,\eta_i)],Z\})=\{[(\xi_i,\eta_i)],P_S(T)\}.
\]
By the construction, the mappings are mappings over $\mathsf{Y}$.
The domain of definition of $\delta$ coincides with the domain
$\mathsf{S}_0\subset \mathsf{S}$  of definition of the mapping
$p_\mathsf{S}$. Let $T\in \mathsf{S}_0$, then $\gamma\circ
\delta(T)=T$. Let $\gamma(U)\in \mathsf{S}_0$, then $\delta\circ
\gamma(U)=U$. Thus we have
\begin{lemma} \label{birazio} The mappings $\gamma$ and
$\delta$ establish the birational equivalence of the spaces
$\mathsf{X}$ and $\mathsf{S}$ over $\mathsf{Y}=\mathbb{C}^{2g}.$
\end{lemma}
The assertion of Lemma \ref{birazio} helps to transfer onto the
space $\mathsf{S}$ the groupoid over $\mathsf{Y}$ structure, which
is introduced by Theorem \ref{universal-groupoid} on the space
$\mathsf{X}$. Let $T_1, T_2\in \mathsf{S}$. The birational
equivalence induces the mappings  $\mu_{*}$ and $\inv_{*}$ that are
defined by the formulas
\begin{equation*}
\mu_{*}(T_1,T_2)=\gamma\circ\mu(\delta(T_1),\delta(T_2)),\quad
\inv_{*}(T_1)=\gamma\circ\inv\circ\,\delta(T_1).
\end{equation*}

\begin{theorem}
The mappings  $\mu_{*}$ and $\inv_{*}$ define the structure of
commutative algebraic groupoid over the space
$\mathsf{Y}=\mathbb{C}^{2g}$ on the space $\mathsf{S}$.
\end{theorem}

Let us proceed to constructing the structure of algebraic groupoid
over  $\mathbb{C}^{2g}$ on the space $\mathbb{C}^{3g}$. The
classical Vi{\`e}te mapping is the homeomorphism of spaces
$\Sym^g(\mathbb{C})\to \mathbb{C}^g$. Let us use Vi{\`e}te mapping
to construct a birational equivalence
$\varphi:\Sym^{g}(\mathbb{C}^2)\to \mathbb{C}^{2g}$.

Let $[(\xi_j,\eta_j)]\in \Sym^{g}(\mathbb{C}^2)$ and
$(p_{2g+1},p_{2g},\dots,p_2)\in \mathbb{C}^{2g}$. Let us assign
$P_{\odd}=(p_{2g+1},p_{2g-1},\dots,p_3)^{t}$,
$P_{\even}=(p_{2g},p_{2g-2},\dots,p_2)^{t}$ and
$X=(1,x,\dots,x^{g-1})^T$. Define the mapping  $\varphi$ and its
inverse $\psi$ with the help of the relations
\begin{gather*}
x^g-\sum_{i=1}^g p_{2i}x^{g-i}=x^g-X^TP_{\even}=\prod_{j=1}^g(x-\xi_j),\\
\eta_i=P_{\odd}^T X|_{x=\xi_i},\quad i=1,\dots,g.
\end{gather*}
Note, that $\varphi$ is a rational mapping, while $\psi$ is a
nonsingular algebraic mapping.

Using the mapping $\varphi$, we obtain the mapping
\begin{equation*}\varphi_1=\varphi\times\id:\mathsf{S}=
\Sym^g(\mathbb{C}^2)\times\mathbb{C}^g\to\mathbb{C}^{2g}\times\mathbb{C}^g\cong
\mathbb{C}^{3g}\end{equation*} and its inverse
$\psi_1=\psi\times\id$.

\emph{The companion matrix  of a polynomial $x^g-X^TP_{\even}$} is
the matrix
\begin{equation*}C=\sum_{i=1}^ge_i(e_{i-1}+p_{2(g-i+1)}e_{g})^T,\end{equation*}
where $e_i$ is the $i$-th basis vector in $\mathbb{C}^g$. Its
characteristic polynomial $|x\cdot 1_g-C|$ is $x^g-X^TP_{\even}.$

\begin{example}The companion matrices $V$ of the polynomials $x^g-X^TP_{\even}$
for $g=1,2,3,4$, have the form
\begin{equation*}
p_2,\quad
\begin{pmatrix}
0&p_4\\
1&p_2
\end{pmatrix},
\quad
\begin{pmatrix}
0&0&p_6\\
1&0&p_4\\
0&1&p_2
\end{pmatrix}
,\quad
\begin{pmatrix}
0&0&0&p_8\\
1&0&0&p_6\\
0&1&0&p_4\\
0&0&1&p_2
\end{pmatrix}.
\end{equation*}
Note, that the companion matrix for $g=k$ is included in the
companion matrix for $g>k$ as the lower right $k\times k$ submatrix.
\end{example}
We make use of the following property of a companion matrix.
\begin{lemma} Let the polynomial
$p(x)=x^g-\sum_{i=1}^g p_{2i}x^{g-i}$ and one of its roots  $\xi$ be
given. Set
 $\Upsilon=(1,\xi,\dots,\xi^{g-1})^T$. Then
the relations
\begin{equation*}
\xi^k \Upsilon^TA=\Upsilon^TC^kA,\quad k=0,1,2,\dots,
\end{equation*}
hold for an arbitrary vector $A\in \mathbb{C}^g$. \label{companion}
\end{lemma}
\begin{proof}Let $A=(a_1,\dots,a_g),$ then
\begin{gather*}
\xi \Upsilon^T A=\xi\sum_{i=1}^g
a_i\xi^{i-1}=\\
\qquad a_g\xi^g+\sum_{i=2}^g a_{i-1}\xi^{i-1}
=\sum_{i=1}^g((1-\delta_{i,1})a_{i-1}+a_gp_{2(g-i+1)})\xi^{i-1}.
\end{gather*}
Thus the Lemma holds for $k=1$. One can complete the proof by
induction.
\end{proof}
The mapping $p_{\mathbb{C}^{3g}}:\mathbb{C}^{3g}\to\mathbb{C}^{2g},$
with respect to which $\varphi_1$ is a mapping over
$\mathbb{C}^{2g}$, is given by the formula
\begin{gather}\label{p_C^3g}
p_{\mathbb{C}^{3g}}(P,Z)=(Z_1,Z_2),\\
Z_1=\Big(\sum_{i=1}^g p_{2i+1}C^{g-i}\Big)P_{\odd}-C^{g}(C
P_{\even}+Z),\quad Z_2=Z. \notag
\end{gather}
Using Lemma \ref{companion} one can directly verify the ``over''
property, that is, that
$p_{\mathbb{C}^{3g}}\circ\varphi_1(T)=p_\mathsf{S}(T)$ for any $T\in
\mathsf{S}_0$.

Let $A_1, A_2\in \mathbb{C}^{3g}$. The birational equivalence
 $\varphi_1$ induces the mappings
$\mu_{**}$ and $\inv_{**}$ defined by formulas
\begin{equation*}
\mu_{**}(A_1,A_2)=\varphi_1\circ\mu_{*}(\psi_1(A_1),\psi_1(A_2)),\quad
\inv_{**}(A_1)=\varphi_1\circ\inv_{*}\circ\,\psi_1(A_1).
\end{equation*}
\begin{theorem}
The mappings  $\mu_{**}$ and $\inv_{**}$  define the structure of
commutative algebraic groupoid over the space
$\mathsf{Y}=\mathbb{C}^{2g}$ on the space $\mathbb{C}^{3g}$.
\end{theorem}
\subsection{The addition law}
In what follows we use the shorthand notation
\begin{equation*}
\overline{A}=\inv_{**}(A)\quad\text{and}\quad A_1\star
A_2=\mu_{**}(A_1,A_2).
\end{equation*}
\begin{lemma}
Let $A=(P_{\even},P_{\odd},Z)\in \mathbb{C}^{3g}.$ Then
\begin{equation*}
\overline{A}=(P_{\even},-P_{\odd},Z).
\end{equation*}
\end{lemma}
Introduce the $(g\times \infty)$-matrix
\begin{equation*}
K(A)=\big(Y,C Y, C^2Y,\dots\big),
\end{equation*}
that is composed of the $(g\times 2)$-matrix
$Y=(P_{\even},P_{\odd})$ with the help of the companion matrix $V$
of the polynomial$x^g-X^TP_{\even}$. Denote by $L(A)$ the matrix
composed of the first  $g$ columns of $K(A)$, and denote by
$\ell(A)$ the  $(g+1)$-st column of $K(A)$.
\begin{theorem} Let $A_1,A_2\in \mathbb{C}^{3g}$ and let
$A_3={A}_1\star A_2$, then
\begin{equation*}
\mathrm{rank}
\begin{pmatrix}
1_g&L(\overline{A}_1)&\ell(\overline{A}_1)\\
1_g&L(\overline{A}_2)&\ell(\overline{A}_2)\\
1_g&L(A_3)&\ell(A_3)\\
\end{pmatrix}<2g+1.
\end{equation*}\label{additto_1}
\end{theorem}
\begin{proof} Suppose  the points  $U_i=\delta\circ\psi_1(A_i),$
$i=1,2$, are defined. We rewrite the function $R^{(1,2)}_{3g}(x,y)$
as a linear combination of  monomials
\begin{equation*}
R^{(1,2)}_{3g}(x,y)=\sum_{i,j,w(i,j)\geqslant 0} h_{w(i,j)}x^iy^j
=r_1(x)y+x^gr_2(x)+r_3(x),\end{equation*} where
$w(i,j)=3g-(2g+1)j-2i$,
\begin{equation*}\begin{gathered}
r_1(x)=\sum_{i=0}^{\rho}h_{g-2i-1}x^i,\\
r_2(x)=\sum_{i=0}^{g-\rho-1}h_{g-2i}x^i,\\
r_3(x)=\sum_{i=0}^{g-1}h_{3g-2i}x^i,
\end{gathered} \qquad
\rho=\Big[\frac{g-1}{2}\Big].
\end{equation*}
Let us set $h_0=1.$

We assign weights to the parameters $h_k$ by the formula $\deg
h_k=k$. Then $\deg R^{(1,2)}_{3g}(x,y)=3g$.

Let $A=(P_{\even},P_{\odd},Z)\in\mathbb{C}^{3g}$ be the point
defining  any of the collections of $g$ zeros of the function
$R_{3g}^{(1,2)}(x,y).$ Consider the function
$Q(x)=R^{(1,2)}_{3g}(x,X^TP_{\odd})$. By the construction
$Q(\xi)=0,$ if $(x^g-X^TP_{\even})|_{x=\xi}=0.$ Let us apply Lemma
\ref{companion}. We obtain
\begin{equation}
Q(\xi)= \Upsilon^T\big(
r_1(C)P_{\odd}+r_2(C)P_{\even}+H_1\big),\label{quan}
\end{equation}
where $H_{1}=(h_{3g},h_{3g-2},\dots, h_{g+2})$. Using the above
notation, we come to the relation
\begin{equation*}
Q(\xi)=\Upsilon^T\big( H_1+L(A)H_2+\ell(A)\big),
\end{equation*}
where $H_2=(h_g,h_{g-1},\dots,h_1)$. Suppose the polynomial
$x^g-X^TP_{\even}$ has no multiple roots, then from the equalities
$Q(\xi_j)=0,$ $j=1,\dots,g$ one can conclude that
\begin{equation}
H_1+L(A)H_2+\ell(A)=0. \label{juan}\end{equation}

By substituting the points $\overline{A}_1, \overline{A}_2$ and
$A_3$ to \eqref{juan}, we obtain the system of $3g$ linear
equations, which is satisfied by the coefficients $H_1,H_2$ of the
entire function $R^{(1,2)}_{3g}(x,y)$. The assertion of the Theorem
is the compatibility condition of the system of linear equations
obtained.
\end{proof}
\begin{cor}The vectors $H_1,H_2$ of coefficients of the entire function
 $R^{(1,2)}_{3g}(x,y)$ are expressed by the formulas
\begin{gather*}
H_2(A_1,A_2)=-\big[L(\overline{A}_1)-
L(\overline{A}_2)\big]^{-1}(\ell(\overline{A}_1)-\ell(\overline{A}_2)),\\
H_1(A_1,A_2)=-\frac{1}{2}\big[(\ell(\overline{A}_1)+
\ell(\overline{A}_2))-(L(\overline{A}_1)+L(\overline{A}_2))H_2(A_1,A_2)\big].
\end{gather*}
as vector functions of the points $A_1$ and $A_2$
 from $\mathbb{C}^{3g}$.
\end{cor}
Now, we know the coefficients $H_1,H_2$ of $R^{(1,2)}_{3g}(x,y)$ and
we can give the expression of $P^{(3)}_{\odd}$ as a function of
$A_1, A_2$ and $P^{(3)}_{\even}$. It follows from \eqref{quan} that
the following assertion holds.
\begin{lemma}
\begin{equation}\label{p3_odd}
P_{\odd}^{(3)}=-\big[r_1(C^{(3)})\big]^{-1}(H_1+r_2(C^{(3)})P_{\even}^{(3)}),
\end{equation}
where  $C^{(3)}$ is the companion matrix of the polynomial
$x^g-X^TP_{\even}^{(3)}$.
\end{lemma}
Let us find the explicit formula for the function
$R_{3g}^{(1,2)}(x,y)$ as a function of the points $A_1$ and $A_2$.
We introduce the $((2g+1)\times\infty)$-matrix
\begin{equation*}
F(x,y;A_1,A_2)=
\begin{pmatrix}
X^T&\mathcal{K}(x,y)\\
1_g&K(A_1)\\
1_g&K(A_2)
\end{pmatrix},
\end{equation*}
where $\mathcal{K}(x,y)=(x^g,y,\dots,x^{g+k},y x^k,\dots).$

Denote by $G(x,y;A_1,A_2)$ the matrix composed of the first $2g+1$
columns of the matrix $F(x,y;A_1,A_2).$
\begin{theorem}The entire rational function
$R_{3g}^{(1,2)}(x,y)$ defining the operation $A_1\star A_2$ has the
form
\begin{equation}\label{R_explicit}
R_{3g}^{(1,2)}(x,y)=\frac{|G(x,y;\overline{A}_1,
\overline{A}_2)|}{|L(\overline{A}_2)-L(\overline{A}_1)|}.
\end{equation}
\end{theorem}
By a use of the formula \eqref{R_explicit} and Lemma \ref{even_part}
we can find  $P_{\even}^{(3)}.$ Similar to the condition of Lemma
\ref{even_part} denote $A_4=\overline{A}_1$ and
$A_5=\overline{A}_2$. One can easily show that
$R_{3g}^{(4,5)}(x,y)=R_{3g}^{(1,2)}(x,-y)$. Thus, the product
 $R_{3g}^{(1,2)}(x,y)\,R_{3g}^{(4,5)}(x,y)$ is an even function in
 $y$. Set
\begin{gather*}
\Phi(x,y^2)=R_{3g}^{(1,2)}(x,y)\,R_{3g}^{(4,5)}(x,y)\\ \qquad=
(-1)^g\frac{|G(x,y;A_1,A_2)|}{|L(A_2)-L(A_1)|}\frac{|G(x,-y;A_1,A_2)|}{|L(A_2)-L(A_1)|}.
\end{gather*}
Therefore, $\Phi(x,y^2)$, as a function on the curve $V$, is the
polynomial in  $x$ and the parameters $Z_1$ and $Z_2$. The values of
$Z_1$ and $Z_2$ are defined by the mapping $p_{\mathbb{C}^{3g}}$
according to \eqref{p_C^3g}, and
$p_{\mathbb{C}^{3g}}(A_1)=p_{\mathbb{C}^{3g}}(A_2)=(Z_1,Z_2)$.
Namely, we have
\begin{equation*}
R_{3g}^{(1,2)}(x,y)\,R_{3g}^{(4,5)}(x,y)=
\Phi(x,x^{2g+1}+x^g X^TZ_2+X^TZ_1).\end{equation*}
Lemma \ref{even_part} asserts that dividing the polynomial
$\Phi(x,x^{2g+1}+x^g X^TZ_2+X^TZ_1)$ by
$(x^g-X^TP^{(1)}_{\even})(x^g-X^TP^{(2)}_{\even})$ gives the zero
remainder and the quotient equal $x^g-X^TP^{(3)}_{\even}$. Thus, the
calculation is reduced to the classical algorithm of polynomial
division.
\begin{theorem}\label{zakon} Consider the space
$\mathbb{C}^{3g}$ together with the mapping $p_{\mathbb{C}^{3g}}$
defined by \eqref{p_C^3g} as a groupoid over $\mathbb{C}^{2g}$. Let
$A_1=(P^{(1)}_{\even},P^{(1)}_{\odd}, Z)$ and
$A_2=(P^{(2)}_{\even},P^{(2)}_{\odd}, Z)$ be the points from
$\mathbb{C}^{3g}$ such that
$p_{\mathbb{C}^{3g}}(A_1)=p_{\mathbb{C}^{3g}}(A_2)=(Z_1,Z_2)\in\mathbb{C}^{2g}$.

Then the addition law has the form $ A_1\star A_2=A_3,$ where the
coordinates of the point $A_3=(P^{(3)}_{\even},P^{(3)}_{\odd}, Z)$
are given by the formulas
\begin{gather*}x^g-X^TP_{\even}^{(3)}= \frac{\Phi(x,x^{2g+1}+x^g
X^TZ_2+X^TZ_1)}{(x^g-X^TP^{(1)}_{\even})(x^g-X^TP^{(2)}_{\even})},
\\
P_{\odd}^{(3)}=-\big[r_1(C^{(3)})\big]^{-1}(H_1+r_2(C^{(3)})P_{\even}^{(3)}).
\end{gather*}
\end{theorem}
\begin{example}\label{genus1.1}
Let $g=1.$ The family of curves $V$ is defined by the polynomial
\begin{equation*}
f(x,y,\Lambda)=y^2-x^3-\lambda_4x-\lambda_6.
\end{equation*} In the coordinates $(\lambda_6,\lambda_4)$ on
$\mathbb{C}^2$ and $(p_2,p_3,z_4)$ on $\mathbb{C}^3$ the mapping
 $p_{\mathbb{C}^3}$ is given by the formula
\begin{equation*}
(\lambda_6,\lambda_4)=(p_3^2-p_2(p_2^2+z_4),z_4)
\end{equation*}
Let us write down the addition formulas for the points on the
groupoid $\mathbb{C}^3$ over $\mathbb{C}^2$. Set
$A_1=(u_2,u_3,\lambda_4)$, $A_2=(v_2,v_3,\lambda_4)$ and suppose
$p_{\mathbb{C}^3}(A_1)=p_{\mathbb{C}^3}(A_2)=(\lambda_6,\lambda_4)$.

Let $A_1\star A_2=A_3=(w_2,w_3,\lambda_4)$.

We have: $R_2^{(1)}(x,y)=x-u_2,$ $
 L(A_1)=u_2$, $\ell(A_1)=u_3,$ and so on.
\begin{equation*}F(x,y,A_1,A_2)=
\begin{pmatrix}
1&x&y&x^2&yx&\dots\\
1&u_2&u_3&u_2^2&\hdotsfor{2}\\
1&v_2&v_3&\hdotsfor{3}
\end{pmatrix}.
\end{equation*}
Thus, the function defining the operation  $A_1\star A_2$ has the
expression
\begin{equation*}
R_{3}^{(1,2)}(x,y)=y+\frac{v_3-u_3}{v_2-u_2}x-\frac{u_2v_3-u_3v_2}{v_2-u_2}.
\end{equation*}
Whence, we find $r_1(x)=1,$ $r_2(x)=\dfrac{v_3-u_3}{v_2-u_2},$
$H_1=-\dfrac{u_2v_3-u_3v_2}{v_2-u_2}$ and, by \eqref{p3_odd},
\begin{equation*}
w_3=\frac{u_2v_3-u_3v_2}{v_2-u_2}-\frac{v_3-u_3}{v_2-u_2}w_2.
\end{equation*}
Further, $\Phi(x, x^3+\lambda_4x+\lambda_6)=x^3+\lambda_4
x+\lambda_6-\Big(x
\dfrac{v_3-u_3}{v_2-u_2}-\dfrac{u_2v_3-u_3v_2}{v_2-u_2}\Big)^2.$
Upon dividing the polynomial $\Phi(x, x^3+\lambda_4x+\lambda_6)$ by
the polynomial $(x-u_2)(x-v_2)$ we find
\begin{gather*}
\Phi(x,x^3+\lambda_4x+\lambda_6)\\ \qquad
=\Big(x+u_2+v_2-\Big(\frac{v_3-u_3}{v_2-u_2}\Big)^2\Big)(x^2-(u_2+v_2)x+v_2u_2)+\dots.
\end{gather*}
And, finally, we obtain the addition law of the elliptic groupoid in
the following form
\begin{gather*}
w_2=-(u_2+v_2)+h^2,\\
w_3=-\frac{1}{2}(u_3+v_3)+\frac{3}{2}(u_2+v_2)h-h^3,
\quad\text{where}\quad h=\Big(\frac{v_3-u_3}{v_2-u_2}\Big).
\end{gather*}

One may check directly that
$p_{\mathbb{C}^3}(A_3)=(\lambda_6,\lambda_4).$

\medskip Let $g=2.$ The family of curves $V$ is defined by the
polynomial
\begin{equation*}
f(x,y,\Lambda)=y^2-x^5-\lambda_4x^3-\lambda_6x^2-\lambda_8x-\lambda_{10}
\end{equation*}
In the coordinates
$(\Lambda_1,\Lambda_2)=((\lambda_{10},\lambda_8)^T,(\lambda_6,\lambda_4)^T)$
on $\mathbb{C}^4$ and $(P_{\even},P_{\odd},Z)$ on $\mathbb{C}^6,$
where $P_{\even}=(p_4,p_2)^T,$ $P_{\odd}=(p_5,p_3)^T$, and
$Z=(z_6,z_4)^T$, the mapping $p_{\mathbb{C}^6}$ is given by the
formula
\begin{gather*}(\Lambda_1,\Lambda_2)
=\bigg( \!\begin{pmatrix}p_5^2+p_3^2p_4-p_2p_4(p_2^2+p_4+z_4)-p_4(p_2p_4+z_6)\\
2p_3p_5+p_2p_3^2-(p_2^2+p_4)(p_2^2+p_4+z_4)-p_2(p_2p_4+z_6)
\end{pmatrix},Z\!\bigg)\end{gather*}
Let us write down the addition formulas for the points on the
groupoid  $\mathbb{C}^6$ over $\mathbb{C}^4$. Set
$A_1=((u_4,u_2)^T,(u_5,u_3)^T,(\lambda_4),\lambda_6)^T)$,
$A_2=((v_4,v_2)^T,(v_5,v_3)^T,(\lambda_4,\lambda_6)^T)$ and suppose
$p_{\mathbb{C}^6}(A_1)=p_{\mathbb{C}^6}(A_2)=((\lambda_{10},\lambda_8)^T,(\lambda_6,\lambda_4)^T)$.

Let  $A_3=A_1\star A_2,$
$A_3=((w_4,w_2)^T,(w_5,w_3)^T,(\lambda_6,\lambda_4)^T).$

We omit the calculation, which is carried out by the same scheme as
for $g=1$, and pass to the result. Set $h=h_1.$ We have
\begin{equation*}
h=-\frac{
\begin{vmatrix}
v_4-u_4&v_2v_4-u_2u_4\\
v_2-u_2&v_4+v_2^2-(u_4+u_2^2)
\end{vmatrix}}{
\begin{vmatrix}
v_4-u_4&v_5-u_5\\
v_2-u_2&v_3-u_3
\end{vmatrix}}.
\end{equation*}
To shorten the formulas it is convenient to employ the linear
differential operator
\begin{equation*}
\mathcal{L}=\frac{1}{2}\{(u_3-v_3)(\partial_{u_2}-\partial_{v_2})+
(u_5-v_5)(\partial_{u_4}-\partial_{v_4})\},
\end{equation*}
It is important to note that $\mathcal{L}$ adds unity to the weight,
$\deg\mathcal{L}=1,$ and that it is  \emph{tangent} to the singular
set where the addition is not defined:
\begin{equation*}
\mathcal{L}\{(u_2-v_2)(u_5-v_5)-(u_3-v_3)(u_4-v_4)\}=0.
\end{equation*}
Let  $h'=\mathcal{L}(h)$ and $h''=\mathcal{L}(h')$. Note, that
$\mathcal{L}(h'')=0$. Using this notation the addition formulas are
written down as follows
\begin{gather*}
w_2=\frac{1}{2}(u_2+v_2)+2h'+h^2,\\
w_3=\frac{1}{2}(u_3+v_3)+\frac{5}{4}(u_2+v_2)h+2h''+3h'h+h^3,\\
w_4=-\frac{1}{2}(u_4+v_4)-u_2v_2+
\frac{1}{8}(u_2+v_2)^2+(u_3+v_3)h\\ \qquad-\frac{1}{2}(u_2+v_2)(h'-h^2)-2h h'',\\
w_5=-\frac{1}{2}(u_5+v_5)-\frac{1}{2}(u_2u_3+v_2v_3)\\ \qquad-
   \big\{\frac{1}{8}(u_2+v_2)^2+v_2u_2+\frac{1}{2}(u_4+v_4)\big\}h
 +(u_3+v_3)(h'+h^2)  \\ \qquad-\frac{1}{2}(u_2+v_2)(h''-h h'-2h^3)-2(h'+h^2)h''.
\end{gather*}
\end{example}

\section{Addition theorems}
For each curve $V$ from the family \eqref{hypp} consider the Jacobi
variety  $\Jac(V)$. The set of all the Jacobi varieties is
\emph{the universal space $\mathsf{U}$ of the Jacobi varieties of
the genus $g$ hyperelliptic curves}.  The points of $\mathsf{U}$ are
pairs $(u,\Lambda),$ where the vector
 $u=(u_1,\dots,u_g)$ belongs to the Jacobi variety of the curve
 with parameters $\Lambda$. The mapping
${p}_{\mathsf{U}}:\mathsf{U}\to\mathbb{C}^{2g}$ that acts as
${p}_{\mathsf{U}}(u,\Lambda)=\Lambda$ makes $\mathsf{U}$  the space
over $\mathbb{C}^{2g}.$ There is a natural groupoid over
$\mathbb{C}^{2g}$ structure on  $\mathsf{U}$. Evidently, the
mappings
 $\mu((u,\Lambda),(v,\Lambda))=(u+v,\Lambda)$ and
$\inv(u,\Lambda)=(-u,\Lambda)$  satisfy the groupoid over
$\mathbb{C}^{2g}$ axioms.

\subsection{Case of $\wp$-functions}
Let us define the mapping
$\pi:\mathsf{U}\to\mathbb{C}^{3g}$ over $\mathbb{C}^{2g}$ by putting
into correspondence a point $(u,\Lambda)\in \mathsf{U}$  and the
point $(\bwp(u),\bwp'(u)/2,\Lambda_2)\in\mathbb{C}^{3g}$, where
\[
\bwp(u)=(\wp_{g,j}(u))^T, \quad\bwp'(u)=(\wp_{g,g,j}(u))^T,\quad
\Lambda_{2}=(\lambda_{2(g-i+2)}),\;\;  i=1,\dots,g.\] Here
\[\wp_{i,j}(u)=-\partial_{u_i}\partial_{u_j}\log\sigma(u)\quad\text{and}\quad
\wp_{i,j,k}(u)=-\partial_{u_i}\partial_{u_j}\log\sigma(u) \] and
$\sigma(u)$ is the hyperelliptic sigma-function
\cite{ba97,ba07,bel97a,bel97b}.
\begin{theorem} The mapping  $\pi:\mathsf{U}\to\mathbb{C}^{3g}$
over $\mathbb{C}^{2g}$ is a birational isomorphism of groupoids:
\begin{equation*}
\pi(u+v,\Lambda)=\pi(u,\Lambda)\star\pi(v,\Lambda),\quad
\pi(-u,\Lambda)=\overline{\pi(u,\Lambda)}.
\end{equation*}\label{additio_2}
\end{theorem}
\begin{proof} First, by Abel theorem any triple of points
 $(u,v,w)\in (\Jac(V))^{3}$ that satisfies the condition
$u+v+w=0$  corresponds to the set of zeros $(x_i,y_i)$,
$i=1,\dots,3g$, of an entire rational function of order $3g$ on the
curve $V$. Namely, Let $X=(1,x,\dots,x^{g-1})^T$, then
\begin{equation}\label{x2uvw}
u=\sum_{i=1}^g\int_{\infty}^{x_i}X\,\frac{\mathrm{d}x}{2y},\quad
v=\sum_{i=1}^g\int_{\infty}^{x_{i+g}}X\,\frac{\mathrm{d}x}{2y},\quad
w=\sum_{i=1}^g\int_{\infty}^{x_{i+2g}}X\,\frac{\mathrm{d}x}{2y}.
\end{equation}
(For shortness, instead of indicating the end point of integration
explicitly, we give only the first coordinate.)

Second, for  the given value $u\in\Jac(V)$ the system of $g$
equations
\begin{equation*}
u-\sum_{i=1}^{g}\int_{\infty}^{x_i}X\frac{\mathrm{d}x}{2y}=0
\end{equation*}
with respect to the unknowns $(x_i,y_i)\in V$ is equivalent to the
system of algebraic equations
\begin{equation*}
x^g-\sum_{k=1}^g\wp_{g,k}(u)x^{k-1}=0,\quad
2y-\sum_{k=1}^g\wp_{g,g,k}(u)x^{k-1}=0,
\end{equation*}
the roots of which are the required points $(x_i,y_i)\in V$.

The combination of the two facts implies that the construction of
the preceding  sections provides the isomorphism.
\end{proof}
Above all note that $2g$ hyperelliptic functions\[
\bwp(u)=(\wp_{g,1}(u),\dots,\wp_{g,g}(u))^T\quad\text{and}\quad
\bwp'(u)=(\wp_{g,g,1}(u),\dots,\wp_{g,g,g}(u))^T\] form a basis of
the field of hyperelliptic Abelian functions, i.e., any function of
the field can be expressed as a rational function in $\bwp(u)$ and
$\bwp'(u)$. The assertion of Theorem \ref{additio_2} written down in
the coordinates of  $\mathbb{C}^{3g}$ takes the form of the addition
theorem for the basis functions $\bwp(u)$ and $\bwp'(u)$.
\begin{cor}\label{dura}
The basis hyperelliptic Abelian functions
\begin{equation*}\bwp(u)=(\wp_{g,1}(u),\dots, \wp_{g,g}(u))^T\quad\text{and}\quad
\bwp'(u)=(\wp_{g,g,1}(u),\dots,\wp_{g,g,g}(u))^T\end{equation*}
respect the addition law
\begin{equation*}
(\bwp(u+v),\bwp'(u+v)/2,\Lambda_2)=
(\bwp(u),\bwp'(u)/2,\Lambda_2)\star(\bwp(v),\bwp'(v)/2,\Lambda_2),
\end{equation*}
the formula of which is given in Theorem \ref{zakon}.
\end{cor}
Thus we have obtained a solution the problem to construct  an
explicit and effectively computable formula of the addition law in
the fields  of hyperelliptic Abelian functions.

\subsection{Case of   $\zeta$-functions}
One has  $g$ functions $\zeta_{i}(u)=\partial_{u_i}\log\sigma(u)$
and  the functions are not Abelian. However, by an application of
Abel theorem for the second kind integrals (see \cite{ba97}) one
obtains the addition theorems for $\zeta$-functions as well. On one
hand, any $\zeta$-function can be represented as the sum of $g$
second kind integrals and an Abelian function. On the other hand, an
Abelian sum of the second kind integrals with the end points at the
set of zeros of an entire rational function $R(x,y)$ is expressed
rationally in terms of the coefficients of $R(x,y)$. We employ the
function \eqref{R_explicit} computed in the variables indicated in
Corollary \ref{dura}.
\begin{theorem}\label{zeta_g} Let $(u,v,w)\in(\Jac(V))^{3}$ and $u+v+w=0.$
Then
\begin{gather*}
\zeta_g(u)+\zeta_g(v)+\zeta_g(w)=-h_1,
\end{gather*}
where $h_1$ is the rational function in $\bwp(u),\bwp'(u)$ and
$\bwp(v),\bwp'(v)$ equal to the coefficient of the monomial of the
weight $3g-1$ in the function \eqref{R_explicit} computed in the
variables indicated in Corollary \ref{dura}.
\end{theorem}
\begin{proof}
We have the identity (see \cite{ba97},\cite[p.~41]{bel97b})
\begin{gather*}\zeta_g(u)+\sum_{i=1}^g\int_{\infty}^{x_i}x^g\,\frac{\mathrm{d}x}{2y}=0,\quad
\zeta_g(v)+\sum_{i=1}^g\int_{\infty}^{x_{i+g}}x^g\,\frac{\mathrm{d}x}{2y}=0,\\\qquad
\zeta_g(w)+\sum_{i=1}^g\int_{\infty}^{x_{i+2g}}x^g\,\frac{\mathrm{d}x}{2y}=0.
\end{gather*}
Suppose that the closed path $\gamma$ encloses all zeros
$(x_1,y_1),\dots,(x_{3g},y_{3g})$ of the function $R_{3g}(x,y)$.
Then we have
\begin{gather*}
\sum_{k=1}^{3g}\int_{\infty}^{x_k}x^g\,\frac{\mathrm{d}x}{2y}=\frac{1}{2\pi\imath}
\oint_{\gamma}\mathrm{d}\big(\log
R_{3g}(x,y)\big)\int_{\infty}^{x}x^g\,\frac{\mathrm{d}x}{2y}.
\end{gather*}
 Because $\mathrm{d}\log
R_{3g}(x,y)/\mathrm{d} x$ is a rational function on the curve and,
hence, a uniform function, the total residue of $\mathrm{d}\big(\log
R_{3g}(x,y)\big)\int_{\infty}^{x}x^g\mathrm{d}x/(2y)$ on the Riemann
surface of the curve $V$ is zero. To write down this fact explicitly
consider the parametrization
\begin{equation*}
(x(\xi),y(\xi))=(\xi^{-2},\xi^{-2g-1}\rho(\xi)), \qquad
\rho(\xi)=1+\frac{\lambda_4}{2}\xi^4+\frac{\lambda_6}{2}\xi^6+O(\xi^8),
\end{equation*}
of the curve $V$ near the point at infinity and denote
$R_{3g}(\xi)=R_{3g}(x(\xi),y(\xi))$. We obtain
\begin{equation*}-
\mathrm{Res}_{\xi}\bigg[\frac{
R_{3g}'(\xi)}{R_{3g}(\xi)}\int_{\infty}^{x(\xi)}x^g\,\frac{\mathrm{d}x}{2y}\bigg]+\sum_{i=1}^{3g}
\mathrm{Res}_{x=x_i}\!\bigg[\mathrm{d}\big(\log
R_{3g}(x,y)\big)\int_{\infty}^{x}x^g\,\frac{\mathrm{d}x}{2y}\bigg]\!=0,
\end{equation*}
which is in fact a particular case of Abel theorem. Thus, the final
expression is
\[
\zeta_g(u)+\zeta_g(v)+\zeta_g(w)=-\mathrm{Res}_{\xi}\bigg[\frac{
R_{3g}'(\xi)}{R_{3g}(\xi)}\int_{\infty}^{x(\xi)}x^g\,\frac{\mathrm{d}x}{2y}\bigg].
\]
It remains  to use the expansions
\begin{gather*}
\int_{\infty}^{x(\xi)}x^g\,\frac{\mathrm{d}x}{2y}=\frac{1}{\xi}
+\frac{\lambda_4}{6}\xi^3+O(\xi^5),\\ R_{3g}(\xi)=
\xi^{-3g}(1+h_1\xi+h_2\xi^2+h_3\xi^3+O(\xi^4))
\end{gather*}
to compute the residue.
\end{proof}
A similar argument leads from the identity (see
\cite{ba97},\cite[p.~41]{bel97b})
\begin{equation*}\zeta_{g-1}(u)+\sum_{i=1}^g\int_{\infty}^{x_i}
(3x^{g+1}+\lambda_4
x^{g-1})\dfrac{\mathrm{d}x}{2y}=\frac{1}{2}\wp_{g,g,g}(u),
\end{equation*} to the following assertion.
\begin{theorem}In the conditions of Theorem \ref{zeta_g}  we have
\begin{gather*}
\zeta_{g-1}(u)+\zeta_{g-1}(v)+\zeta_{g-1}(w)\\\qquad-\frac{1}{2}(
\wp_{g,g,g}(u)+\wp_{g,g,g}(v)+\wp_{g,g,g}(w))=-h_1^3+3h_1h_2-3h_3,
\end{gather*}
\label{zeta_g-1} where $h_2$ and $h_3$ are the  coefficients of the
monomials of weight $3g-2$ and $3g-3$ in the function indicated in
Theorem \ref{zeta_g}.
\end{theorem}
\begin{example}\label{ex:g1} Let $g=1$. The function $R_3(x,y)$ has the form $y+h_1
x+h_3$, where $2h_1=(\wp'(u)-\wp'(v))/(\wp(u)-\wp(v))$,  cf. Example
\ref{genus1.1}. Thus, Theorem \ref{zeta_g} gives the classic formula
\begin{equation}
\zeta(u)+\zeta(v)-\zeta(u+v)=-\frac{1}{2}\bigg(\frac{\wp'(u)-\wp'(v)}{\wp(u)-\wp(v)}\bigg).\label{FS_1}
\end{equation}
As $h_2=0$ and $2h_3=(\wp'(v)\wp(u)-\wp'(u)\wp(v))/(\wp(u)-\wp(v))$,
cf. Example \ref{genus1.1}, Theorem \ref{zeta_g-1}  yields the
relation
\begin{equation*}
-\wp'(u)-\wp'(v)+\wp'(u+v)=-\frac{1}{4}\bigg(\frac{\wp'(u)
-\wp'(v)}{\wp(u)-\wp(v)}\bigg)^3-3\frac{\wp'(v)\wp(u)-\wp'(u)\wp(v)}{\wp(u)-\wp(v)},
\end{equation*}
which is the addition formula for Weierstrass $\wp'$-function.
\end{example}
The fact below follows directly from Lemma \ref{even_part}.
\begin{lemma}\label{p_gg} $
\wp_{g,g}(u)+\wp_{g,g}(v)+\wp_{g,g}(u+v)=h_1^2-2h_2.$
\end{lemma}
Combining Lemma \ref{p_gg} with Theorem \ref{zeta_g} we find
\begin{equation}\label{FS_g}
\big(\zeta_g(u)+\zeta_g(v)-\zeta_g(u+v)\big)^2=
\wp_{g,g}(u)+\wp_{g,g}(v)+\wp_{g,g}(u+v)+2h_2
\end{equation}
In the case $g=1$  due to the fact that $h_2=0$  formula
\eqref{FS_g} gives the famous relation
 \begin{equation*}
(\zeta(u)+\zeta(v)-\zeta(u+v))^2=\wp(u)+\wp(v)+\wp(u+v).
 \end{equation*}
discovered by Frobenius and Stickelberger.

\begin{example} Let us pass to the case $g=2$, we have $R_6(x,y)=x^2+h_1 y+h_2 x^2+h_4
 x+h_6$. Note that $h_3=0$.  The coefficient  $h_1$ is expressed as
follows, cf. Example \ref{genus1.1},
\begin{equation*}
h_1=-2\frac{
\begin{vmatrix}
\wp_{2,1}(v)-\wp_{2,1}(u)&\wp_{2,2}(u)\wp_{2,1}(v)-\wp_{2,2}(v)\wp_{2,1}(u)\\
\wp_{2,2}(v)-\wp_{2,2}(u)&\wp_{2,1}(v)-\wp_{2,1}(u)
\end{vmatrix}}{
\begin{vmatrix}
\wp_{2,1}(v)-\wp_{2,1}(u)&\wp_{2,2,1}(v)-\wp_{2,2,1}(u)\\
\wp_{2,2}(v)-\wp_{2,2}(u)&\wp_{2,2,2}(v)-\wp_{2,2,2}(u)
\end{vmatrix}}.
 \end{equation*}
And the coefficient $h_2$, respectively,
\begin{equation*}
h_2=\frac{
\begin{vmatrix}
\wp_{2,2}(v)\wp_{2,1}(v)-\wp_{2,2}(u)\wp_{2,1}(u)&\wp_{2,2,1}(v)-\wp_{2,2,1}(u)\\
\wp_{2,1}(v)+\wp_{2,2}(v)^2-\wp_{2,1}(u)-\wp_{2,2}(u)^2&\wp_{2,2,2}(v)-\wp_{2,2,2}(u)&
\end{vmatrix}}{
\begin{vmatrix}
\wp_{2,1}(v)-\wp_{2,1}(u)&\wp_{2,2,1}(v)-\wp_{2,2,1}(u)\\
\wp_{2,2}(v)-\wp_{2,2}(u)&\wp_{2,2,2}(v)-\wp_{2,2,2}(u)
\end{vmatrix}}.
 \end{equation*}
We come to the relations
\begin{gather*}
\zeta_2(u)+\zeta_2(v)-\zeta_2(u+v)=-h_1,\\
\wp_{2,2}(u)+\wp_{2,2}(v)+\wp_{2,2}(u+v)=h_1^2-2h_2,\\
\zeta_{1}(u)+\zeta_{1}(v)-\zeta_{1}(u+v)\\
\qquad-\frac{1}{2}(
\wp_{2,2,2}(u)+\wp_{2,2,2}(v)-\wp_{2,2,2}(u+v))=-h_1^3+3h_1h_2.
\end{gather*}
Hence, by eliminating $h_1$ and $h_2$, we obtain the identity
\begin{equation}
2\mathfrak{z}_{1}-\mathfrak{p}_{2,2,2}-3\mathfrak{p}_{2,2}\mathfrak{z}_2+\mathfrak{z}_2^3=0,\label{z-p}
\end{equation}
where $\mathfrak{z}_{i}=\zeta_i(u)+\zeta_i(v)+\zeta_i(w)$ and
$\mathfrak{p}_{i,j,\dots}=\wp_{i,j,\dots}(u)+\wp_{i,j,\dots}(v)+\wp_{i,j,\dots}(w)$,
provided $u+v+w=0.$
\end{example}
\subsection{Case of $\sigma$-functions}
Formula \eqref{z-p} leads to an important corollary.
\begin{theorem} The genus $2$ sigma-function respects the trilinear addition law
\begin{gather*}
\big[2D_1+ D_2^3\big]\sigma(u)\sigma(v)\sigma(w)\big|_{u+v+w=0}=0,
\end{gather*}
where
$D_j=\partial_{u_j}+\partial_{v_j}+\partial_{w_j}$.\label{trilinea}
\end{theorem}
\begin{proof}
Let us multiply the left hand side of \eqref{z-p} by the product
$\sigma(u)\sigma(v)\sigma(w)$, then \eqref{z-p} becomes the
trilinear relation
\begin{gather*}
\big[2(\partial_{u_1}+\partial_{v_1}+\partial_{w_1})+
(\partial_{u_2}+\partial_{v_2}+\partial_{w_2})^3\big]\sigma(u)\sigma(v)\sigma(w)\big|_{u+v+w=0}=0,
\end{gather*}
which is satisfied by the genus $2$ sigma-function.
\end{proof}
It is important to notice that the elliptic identity \eqref{FS_1} is
equivalent to the trilinear addition law
\begin{gather*}
\big[(\partial_{u}+\partial_{v}+\partial_{w})^2\big]\sigma(u)\sigma(v)\sigma(w)\big|_{u+v+w=0}=0,
\end{gather*}
which is satisfied by Weierstrass sigma-function.
 Let us denote
$D=(\partial_{u}+\partial_{v}+\partial_{w})$ and
$\psi=\sigma(u)\sigma(v)\sigma(w)$. The functions
\[
(D+h_1)\psi,\quad (D^3+6h_3)\psi, \quad (D^4-6\lambda_4)\psi,  \quad
(D^5+18\lambda_4 D)\psi,\quad (D^6-6^3\lambda_6)\psi,
\]
where $h_1$ and $h_2$ are given in Example \ref{ex:g1}, vanish on
the plane $u+v+w=0$. Moreover, one can show that for any $k>3$ there
exist unique polynomials $q_0, q_1, q_3\in
\mathbb{Q}[\lambda_4,\lambda_6]$ such that
\[
(D^k+q_3D^3+q_1 D+q_0)\psi\big|_{u+v+w=0}=0,
\]
and at least one of the polynomials $q_0,q_1,q_3$ is nontrivial.

For the hyperelliptic sigma-function of an arbitrary genus $g$ we
propose the following hypothesis. Let
$\mathscr{P}=\mathbb{Q}[\Lambda]$. Consider the ring
$\mathscr{Q}=\mathscr{P}[D_1,\dots,D_g]$ as a graded ring of linear
differential operators. We conjecture that there exists a collection
of $3g$ linear operators $Q_i\in\mathscr{Q},$ $\deg Q_i=i,$ where
$i=1,\dots, 3g$, such that
\[
\Big\{\sum_{i=0}^{3g}Q_i\xi^{3g-i}+R_{3g}(\xi^2,\xi^{2g+1})\Big\}\sigma(u)\sigma(v)\sigma(w)\big|_{u+v+w=0}=0,
\]
where $Q_0=1$ and $R_{3g}(x,y)$ is the function \eqref{R_explicit}
computed in the variables indicated in Corollary \ref{dura}. Thus,
$g$ operators $Q_{g+2i-1},$ $i=1,\dots,g$ define the trilinear
relations
\[
Q_{g+2i-1}\sigma(u)\sigma(v)\sigma(w)\big|_{u+v+w=0}=0,\quad
i=1,\dots,g.
\]
Note, that the assertions of Theorem \ref{zeta_g}, Lemma \ref{p_gg},
and Theorem \ref{zeta_g-1} imply the relations
\begin{gather*}
(D_g+h_1)\sigma(u)\sigma(v)\sigma(w)\big|_{u+v+w=0}=0,\\
(D_g^2-2h_2)\sigma(u)\sigma(v)\sigma(w)\big|_{u+v+w=0}=0,\\
(2D_{g-1}+D_g^3+6h_3)\sigma(u)\sigma(v)\sigma(w)\big|_{u+v+w=0}=0.
\end{gather*}

We shall return to the problem of explicit description of the
trilinear addition theorems for the hyperelliptic sigma-function in
our future publications.

\section{Pfaffian addition law}
In the Weierstrass' theory of elliptic functions and its
applications the relation \[
\frac{\sigma(u+v)\sigma(u-v)}{\sigma(u)^2\sigma(v)^2}=\wp(v)-\wp(u),
\] plays the key part.
Baker \cite{ba98} has shown, starting with the K{\"o}nigsberger
formula
\index{addition theorem!of K{\"o}nigsberger}

\[
\sigma(\boldsymbol{u}+\boldsymbol{v})\sigma(\boldsymbol{u}
-\boldsymbol{v}) =
\sum_{k=1}^{2^g}\sigma[\epsilon_R
-\epsilon_i](\boldsymbol{u})^2\sigma[\epsilon_i](\boldsymbol{v})^2,
\]
where $\epsilon_R$ is the characteristic of vector of Riemann
constants, and the characteristics
 $\epsilon_i$ belong to the subgroup $A_{2^g}$ of the group of
half-integer characteristics, generated by $g$ shifts to the
half-periods of the type $\mathfrak{a}$, that in the case, when
$\sigma$ is a hyperelliptic $\sigma$-function, the ratio
$\frac{\sigma(\boldsymbol{u}+\boldsymbol{v})\sigma(\boldsymbol{u}-
\boldsymbol{v})}{\sigma(\boldsymbol{u})^2\sigma(\boldsymbol{v})^2}$
may be expressed as a polynomial on $\wp_{i,j}(\boldsymbol{u})$
and $\wp_{i,j}(\boldsymbol{v})$.
Baker described an algorithm of construction of these polynomials,
based on the reexpression of the ratios of squares of
$\sigma[\epsilon]$-functions
in terms of functions  $\wp_{i,j}$ and derived
the polynomials for genera $2$ and $3$.
In case of higher genera  this
algorithm remains ineffective even with the use of modern
means of symbolic computation.

Below we give some of our recent results which allow to
give explicit expression for the polynomials discussed for
arbitrary genera. We briefly discuss application of this result
to the addition theorems for Kleinian functions.
\subsection{Notations}
Let us introduce the following functions
$m_{i,k}:\mathrm{Jac}(V)\times\mathrm{Jac}(V)\to \mathbb{C}$:
\begin{equation}
m_{i,k}(\boldsymbol{u},\boldsymbol{v})=
\boldsymbol{r}^T_{i,k}(\boldsymbol{u}) N
\boldsymbol{r}_{i,k}(\boldsymbol{v}),\label{new:m}
\end{equation}
where
\begin{align*}
\boldsymbol{r}^T_{i,k}(\boldsymbol{v})&=(\wp_{i,k+1}(\boldsymbol{v})-
\wp_{i+1,k}(\boldsymbol{v}),\wp_{g,k+1}(\boldsymbol{v}),
\wp_{g,i+1}(\boldsymbol{v}),1),\\
\intertext{and}
N&=
\begin{pmatrix}
0&0&0&-1\\
0&0&-1&0\\
0&1&0&0\\
1&0&0&0
\end{pmatrix}
\end{align*}
As is seen from \eqref{new:m}
\begin{equation*}
m_{i,k}(\boldsymbol{u},\boldsymbol{v})=-
m_{k,i}(\boldsymbol{u},\boldsymbol{v})=
m_{k,i}(\boldsymbol{v},\boldsymbol{u})=-
m_{i,k}(\boldsymbol{v},\boldsymbol{u}).
\end{equation*}
In expanded form:
\begin{align*}
m_{i,k}(\boldsymbol{u},\boldsymbol{v})&=\wp_{g,i+1}(\boldsymbol{u})
\wp_{g,k+1}(\boldsymbol{v})-
\wp_{g,i+1}(\boldsymbol{v})\wp_{g,k+1}(\boldsymbol{u})\\
&+
\wp_{i,k+1}(\boldsymbol{v})-\wp_{i,k+1}(\boldsymbol{u})-
\wp_{i+1,k}(\boldsymbol{v})+\wp_{i+1,k}(\boldsymbol{u}) .
\end{align*}
Note, that  $m_{i,k}(\boldsymbol{u},\boldsymbol{v})$ depends not
only on its indices $i, k$ but also on the genus $g$. For
instance,  function $m_{1,2}$ evaluated in genus $2$ is not the
same as in genus $1$.
\index{matrix!skew-symmetric}
Next, given a number $k\in \mathbb{N}$, we construct a
skew-symmetric $2[\frac{k+1}{2}]\times 2[\frac{k+1}{2}]$--matrix
\begin{equation*}
M_k(\boldsymbol{u},\boldsymbol{v})=
\{m_{i,j}(\boldsymbol{u},\boldsymbol{v})\}\qquad{i,j\, \in\,
g-k\,,\ldots,\,g-\frac{1+(-1)^k}{2}}\quad,
\end{equation*} so that both for $k=2n+1$ and $k=2n+2$ we have
$(2n+2)\times(2n+2)$--matrices, but with different ranges of
indices.

Let us consider some examples to make this notation clear:
\begin{align*}
&k=1\quad M_1(\boldsymbol{u},\boldsymbol{v})=
\begin{pmatrix}
0&m_{g,g-1}(\boldsymbol{u},\boldsymbol{v})\\
-m_{g,g-1}(\boldsymbol{u},\boldsymbol{v})&0
\end{pmatrix}
\\
&k=2\quad M_2(\boldsymbol{u},\boldsymbol{v})=
\begin{pmatrix}
0&m_{g-1,g-2}(\boldsymbol{u},\boldsymbol{v})\\
-m_{g-1,g-2}(\boldsymbol{u},\boldsymbol{v})&0
\end{pmatrix}
\\
&k=3\\
&M_3(\boldsymbol{u},\boldsymbol{v})=\\&
\begin{pmatrix}
0&m_ {g-2,g-3}(\boldsymbol{u},\boldsymbol{v})
&m_{g-1,g-3}(\boldsymbol{u},\boldsymbol{v})&
m_{g,g-3}(\boldsymbol{u},\boldsymbol{v})\\
-m_{g-2,g-3}(\boldsymbol{u},\boldsymbol{v})&0&
 m_{g-1,g-2}(\boldsymbol{u},\boldsymbol{v})&
m_{g,g-2}(\boldsymbol{u},\boldsymbol{v})\\
-m_{g-1,g-3}(\boldsymbol{u},\boldsymbol{v})
&-m_{g-1,g-2}(\boldsymbol{u},\boldsymbol{v})&0&
 m_{g,g-1}(\boldsymbol{u},\boldsymbol{v})\\
- m_{g,g-3}(\boldsymbol{u},\boldsymbol{v})
&-m_{g,g-2}(\boldsymbol{u},\boldsymbol{v})
&-m_{g,g-1  }(\boldsymbol{u},\boldsymbol{v})&0
\end{pmatrix}
\\
&k=4\\
&M_4(\boldsymbol{u},\boldsymbol{v})=\\&
\begin{pmatrix}
0&m_ {g-3,g-4}(\boldsymbol{u},\boldsymbol{v})
&m_{g-2,g-4}(\boldsymbol{u},\boldsymbol{v})&
m_{g-1,g-4}(\boldsymbol{u},\boldsymbol{v})\\
-m_{g-3,g-4}(\boldsymbol{u},\boldsymbol{v})&0&
 m_{g-2,g-3}(\boldsymbol{u},\boldsymbol{v})&
m_{g-1,g-3}(\boldsymbol{u},\boldsymbol{v})\\
-m_{g-2,g-4}(\boldsymbol{u},\boldsymbol{v})
&-m_{g-2,g-3}(\boldsymbol{u},\boldsymbol{v})&0&
 m_{g-1,g-2}(\boldsymbol{u},\boldsymbol{v})\\
- m_{g-1,g-4}(\boldsymbol{u},\boldsymbol{v})
&-m_{g-1,g-3}(\boldsymbol{u},\boldsymbol{v})
&-m_{g-1,g-2  }(\boldsymbol{u},\boldsymbol{v})&0
\end{pmatrix}
\end{align*}

We shall denote the Pfaffian of a matrix \index{Pfaffian}
$M_k(\boldsymbol{u},\boldsymbol{v})$ by
$\mathcal{F}_k(\boldsymbol{u},\boldsymbol{v})$:
\begin{equation*}
\mathcal{F}_k(\boldsymbol{u},\boldsymbol{v})=
\mathrm{proof}M_k(\boldsymbol{u},\boldsymbol{v})=\sqrt{\det
M_k(\boldsymbol{u},\boldsymbol{v})},
\end{equation*}
and the sign of the square root is to be chosen so, that
$\mathrm{proof}\big(
\begin{smallmatrix}
0&-1_n\\1_n&0
\end{smallmatrix}
\big)=1$.

\subsection{Bilinear addition formula}
Our approach is based on the following Theorem, which is proved
using the theory of the Kleinian functions developed above.
\begin{theorem} The functions
$\mathcal{F}_k(\boldsymbol{u},\boldsymbol{v})$ satisfy the
recursive relation \label{new:d_eq} \begin{equation*} \left[
\left(\frac{\partial^2}{\partial u_{g}^2}-
\frac{\partial^2}{\partial v_{g}^2}\right)\mathrm{ln}\mathcal{
F}_{k-1}(\boldsymbol{u},\boldsymbol{v})+ 2\mathcal{
F}_1(\boldsymbol{u},\boldsymbol{v}) \right]
\mathcal{F}_{k-1}(\boldsymbol{u},\boldsymbol{v})^2-
4 \mathcal{F}_k(\boldsymbol{u},\boldsymbol{v})
\mathcal{F}_{k-2}(\boldsymbol{u},\boldsymbol{v})=0
\end{equation*}
for any genus $g$.
\end{theorem}

By definition, $\mathcal{F}_k(\boldsymbol{u},\boldsymbol{v})$ is
identically zero, when $k>g$.  So the function
$\phi=\mathrm{ln} \big(\mathcal{F}_g(\boldsymbol{u},\boldsymbol{v})\big)$
solves the  linear differential equation
\[
\left(\frac{\partial^2}{\partial u_{g}^2}-
\frac{\partial^2}{\partial v_{g}^2}\right)\phi+ 2\mathcal{
F}_1(\boldsymbol{u},\boldsymbol{v})=0.
\]
This equation is also solved by the function
$\widehat{\phi}=\mathrm{ln}\dfrac{\sigma(\boldsymbol{u}+\boldsymbol{v})
\sigma(\boldsymbol{u}-\boldsymbol{v})}{
\sigma(\boldsymbol{u})^2\sigma(\boldsymbol{v})^2}$. It is
possible to show that $\widehat{\phi}=\phi$, and even more
\begin{prop} For any genus $g$ we have:
\begin{equation} \frac{\sigma(\boldsymbol{u}+\boldsymbol{v})
\sigma(\boldsymbol{u}-\boldsymbol{v})}{
\sigma(\boldsymbol{u})^2\sigma(\boldsymbol{v})^2}=
\mathcal{F}_g(\boldsymbol{u},\boldsymbol{v}).\label{new:main}
\end{equation}
\end{prop}
Note, that the formula \eqref{new:main} has a remarkable property,
that the form of expression
$\mathcal{F}_g(\boldsymbol{u},\boldsymbol{v})$ actually is not
apparently dependent on the moduli of an underlying curve. The
formula  depends on the moduli   only in a ``hidden'' manner
through the functions $\wp_{i,j}$ which may be considered to be
merely coordinates in $\mathbb{C}^{n(n-1)/2}$ (cf.  Section
\ref{hyper-Jac}).  When we impose the algebraic relations
\eqref{Kijkl} on the functions $\wp_{i,j}$, the relation
\eqref{new:main} is restricted to the Kummer variety of the
Riemann surface of the corresponding curve.

Let us consider examples of small genera:

Let genus $g=1$, we have the classical formula of the
elliptic theory \cite{ba55}  \begin{equation*} \frac{\sigma(u+v)
\sigma(u-v)}{ \sigma(u)^2\sigma(v)^2}=\wp(v)-\wp(u).
\end{equation*}

For genera $2$ and $3$ we restore the results by Baker
\cite{ba98}, in genus $g=2$ we have:
\begin{align}
&\frac{\sigma(\boldsymbol{u}+\boldsymbol{v})
\sigma(\boldsymbol{u}-\boldsymbol{v})}{
\sigma(\boldsymbol{u})^2\sigma(\boldsymbol{v})^2}\label{badd2}\\&=
\wp_{22}(\boldsymbol{u})\wp_{21}(\boldsymbol{v})-
\wp_{21}(\boldsymbol{u})\wp_{22}(\boldsymbol{v})-
\wp_{11}(\boldsymbol{u})+\wp_{11}(\boldsymbol{v})
;
\end{align}
\index{addition theorem!for $\wp$--functions at $g=3$}
and in genus $g=3$:
\begin{align*}
&\frac{\sigma(\boldsymbol{u}+\boldsymbol{v})
\sigma(\boldsymbol{u}-\boldsymbol{v})}{
\sigma(\boldsymbol{u})^2\sigma(\boldsymbol{v})^2}=\\
&(\wp_{31}(\boldsymbol{u})-\wp_{31}(\boldsymbol{v}))\cdot\\
&(
\wp_{33}(\boldsymbol{v})\wp_{32}(\boldsymbol{u})
-\wp_{33}(\boldsymbol{u})\wp_{32}(\boldsymbol{v})
+\wp_{31}(\boldsymbol{v})-\wp_{31}(\boldsymbol{u})
-\wp_{22}(\boldsymbol{v})+\wp_{22}(\boldsymbol{u}))-\\
&(\wp_{32}(\boldsymbol{u})-\wp_{32}(\boldsymbol{v}))\cdot\\
&(
\wp_{33}(\boldsymbol{v})\wp_{31}(\boldsymbol{u})
-\wp_{33}(\boldsymbol{u})\wp_{31}(\boldsymbol{v})
-\wp_{21}(\boldsymbol{v})+\wp_{21}(\boldsymbol{u})
)+\\
&(\wp_{33}(\boldsymbol{u})-\wp_{33}(\boldsymbol{v}))\cdot\\
&(\wp_{32}(\boldsymbol{v})\wp_{31}(\boldsymbol{u})
-\wp_{32}(\boldsymbol{u})\wp_{31}(\boldsymbol{v})
+\wp_{11}(\boldsymbol{u})-\wp_{11}(\boldsymbol{v}))
=\\
&(\wp_{31}(\boldsymbol{v})-\wp_{31}(\boldsymbol{u})
(\wp_{22}(\boldsymbol{v})-\wp_{22}(\boldsymbol{u}))-
(\wp_{31}(\boldsymbol{v})-\wp_{31}(\boldsymbol{u}))^2-
\\
&(\wp_{32}(\boldsymbol{v})-\wp_{32}(\boldsymbol{u}))
(\wp_{21}(\boldsymbol{v})-\wp_{21}(\boldsymbol{u}))+
(\wp_{33}(\boldsymbol{v})-\wp_{33}(\boldsymbol{u}))
(\wp_{11}(\boldsymbol{v})-\wp_{11}(\boldsymbol{u})).
\end{align*}

\index{addition theorem!for $\wp$--functions at $g=4$}
The formula for genus $g=4$,
\begin{align*}
&\frac{\sigma(\boldsymbol{u}+\boldsymbol{v})
\sigma(\boldsymbol{u}-\boldsymbol{v})}{
\sigma(\boldsymbol{u})^2\sigma(\boldsymbol{v})^2}=\\
&(\wp_{42}(\boldsymbol{v})\wp_{41}(\boldsymbol{u})
-\wp_{42}(\boldsymbol{u})\wp_{41}(\boldsymbol{v})
-\wp_{11}(\boldsymbol{v})
+\wp_{11}(\boldsymbol{u}))\cdot\\
&(
\wp_{44}(\boldsymbol{v})\wp_{43}(\boldsymbol{u})
-\wp_{44}(\boldsymbol{u})\wp_{43}(\boldsymbol{v})
+\wp_{42}(\boldsymbol{v})
-\wp_{42}(\boldsymbol{u})
-\wp_{33}(\boldsymbol{v})
+\wp_{33}(\boldsymbol{u}))
-\\
&(
\wp_{43}(\boldsymbol{v})\wp_{41}(\boldsymbol{u})
-\wp_{43}(\boldsymbol{u})\wp_{41}(\boldsymbol{v})
-\wp_{21}(\boldsymbol{v})
+\wp_{21}(\boldsymbol{u}))\cdot\\
&(
\wp_{44}(\boldsymbol{v})\wp_{42}(\boldsymbol{u})
-\wp_{44}(\boldsymbol{u})\wp_{42}(\boldsymbol{v})
+\wp_{41}(\boldsymbol{v})-\wp_{41}(\boldsymbol{u})
-\wp_{32}(\boldsymbol{v})
+\wp_{32}(\boldsymbol{u}))
+\\
&(
\wp_{44}(\boldsymbol{v})\wp_{41}(\boldsymbol{u})
-\wp_{44}(\boldsymbol{u})\wp_{41}(\boldsymbol{v})
+\wp_{31}(\boldsymbol{u})
-\wp_{31}(\boldsymbol{v}))\cdot\\
&(
\wp_{43}(\boldsymbol{v})\wp_{42}(\boldsymbol{u})
-\wp_{43}(\boldsymbol{u})\wp_{42}(\boldsymbol{v})
+\wp_{22}(\boldsymbol{u})-\wp_{31}(\boldsymbol{u})
-\wp_{22}(\boldsymbol{v})+\wp_{31}(\boldsymbol{v})),
\end{align*}
and formulas for higher genera described here are, to the knowledge
of authors, new.

\chapter[Reduction of Abelian function]{Reduction of Abelian functions}\label{chap:red}
We are considering here reduction of Abelian functions constructed by an algebraic curve to Abelian functions of lower genera. Nowadays that's the well developed theory that grows from known cases of reduction of holomorphic integrals of genust two curve to elliptic integrals and associated $\theta$-functions to Jacobian $\theta$-functions. We will start with two examples of such reduction.

\section{Two examples of reduction of Abelian integrals}
\begin{gather*}
\int\frac{d z} {\sqrt{ (z^2-a)(8z^3-6az-b)} }=
 \frac13\int \frac{d \xi} {\sqrt{(2a\xi-b) (\xi^2-a) } }  \\
 \qquad\qquad \xi=\frac{1}{a}(4z^3-3az) \\
 \int\frac{z \, dz}{\sqrt{(z^2-a)(8z^3-6az-b)} }=
 \frac{1}{2\sqrt{3} } \int \frac{\mathrm{d}\eta} {\sqrt{\eta^3-3 a\eta+b} } \\
 \qquad\qquad \eta=\frac{2z^3-b}{3(z^2-a)}
 \end{gather*}
The elliptic curves
\[\mathcal{E}_1:\qquad \zeta^2=(2a\xi-b)(\xi^2-a)\]
and
\[\mathcal{E}_2:\qquad\upsilon^2 =\eta^3-3a\eta +b\]
are different:
\begin{align*}
j_{\mathcal{E}_1}=\frac{16(b^2+12a^3)^3}{a^3(4a^3-b^2)},\quad
j_{\mathcal{E}_2}=\frac{6912a^3}{4a^3-b^2}
\end{align*}
Branch points of the first cover are: $\pm\sqrt{a}/2,\pm\sqrt{a}$,  because the Puiseux expansions near these points are
\[  z=t\pm \frac{\sqrt{a}}{2},\quad \xi=\mp\sqrt{a}\pm\frac{6t^2}{\sqrt{a}}+O(t^2)  \]
In all other points puiseux series are different, for example near $z=a$ we have
\[  z=a+t,\quad \xi=4a^2-3a+(12a-3)t+\ldots \]
i.e. branch number is zero. Moreover one can check that branch points of the underlying elliptic curve are lifted to branch points of the covering curve with zero branch number.

We conclude that  branch number of the first cover  $B=2$  and degree of cover $n=3$.

The second cover has the same values for $B$ and $n$

Let $\pi$ be the cover
$$\pi: X\rightarrow X'$$
and the genera of the curves $X,X'$ are correspondingly $g\ge g'\geq 0$. The curve $X$ is called a {\em covering curve} and $X'$ is called {\em underlying curve}.
The following {\bf Riemann-Hurwitz formula} is valid\[ g=\frac{B}{2}+ng'-n+1 \]
In the case of covering over the Riemann sphere, an extended complex plane
\[  g= \left[ \frac{n-1}{2} \right] \]
where $[\cdot]$ is entire part of the number.


The simplest reduction case was wound by Jacobi
\begin{align*}
&\int\frac{\mathrm{d}\xi}{\sqrt{\xi(1-\xi)(1-k^2_\pm\xi)}}
=-\sqrt{(1-\alpha)(1-\beta)}\\ &\times\int\frac{(z\pm\sqrt{\alpha\beta})\mathrm{d}z
}{\sqrt{z(z-1)(z-\alpha)(z-\beta)(z-\alpha\beta)}}.\end{align*}
with
\[\xi=\frac{(1-\alpha)(1-\beta)z}{(z-\alpha)(z-\beta)}
\]
\[k^2_{\pm}=-\frac{(\sqrt\alpha\pm\sqrt\beta)^2}
{ (1-\alpha)(1-\beta)},\]
Therefore the curve $$ X:\; w^2=z(z-1)(z-\alpha)(z-\beta )(z-\alpha\beta )$$ covers two
elliptic curves $$\mathcal{E}_{\pm}:\; \eta^2=\xi(1-\xi)(1-k_{\pm}^2\xi)$$ i.e.
\[   \pi_{\pm}: \; X  \rightarrow      \mathcal{E}_{\pm}  \]

\[\text{Let}\quad   \pi_{\pm}: \; X  \rightarrow      \mathcal{E}_{\pm}  \]

\section{Weierstrass-Poincar\'e theorem on complete reducibility}

Like in the Jacobi reduction example non-normalized holomorphic differential $\mathrm{d}\psi$ on $\mathcal{E}_+$ can be always expanded in
normalized holomorphic differentials of genus two curve,
\[  \oint_{\mathfrak{a_k}} \mathrm{d}v_i=\delta_{i,k}, \quad  \oint_{\mathfrak{b_k}} \mathrm{d}v_i=\tau_{i,k}, \]
namely
\[
\mathrm{d}\psi=\frac{\mathrm{d}\xi}{\eta}=c_1\mathrm{d}v_1+c_2\mathrm{d}v_2,\label{exp }
\]
where $c_1,c_2\in \mathbb{C}$ are constants. Evaluate periods
$\mathfrak{a},\mathfrak{b}$
\begin{align}\begin{split}
&c_1=2m_{11}\omega+2m_{12}\omega',\\
&c_2=2m_{21}\omega+2m_{22}\omega',\\
&c_1\tau_{11}+c_2\tau_{12}=2m_{31}\omega+2m_{32}\omega',\\
&c_1\tau_{12}+c_2\tau_{22}=2m_{41}\omega+2m_{42}\omega'.\end{split}\label{biremb}
\end{align}
Introduce Riemann period matrices
\[ \Pi=\left(\begin{array}{cc} 1&0\\0&1\\ \tau_{11}&\tau_{12}\\
\tau_{12}&\tau_{22}  \end{array}  \right),
\qquad \Pi'=\left(\begin{array}{c} 2\omega \\ 2\omega' \end{array} \right)
\]
Then the above condition can be rewritten as
\[ \Pi \lambda=M\Pi' \]
\[ \lambda=\left(\begin{array}{c}c_1\\ c_2 \end{array}\right), \quad
M^T=\left( \begin{array}{cccc}
m_{11}&m_{21}&m_{31}&m_{41}\\
m_{12}&m_{22}&m_{32}&m_{42}
\end{array}\right) \]

This statement  is a particular case of the {\em Weierstrass-Poincar\'e theorem (WP-theorem) on complete reducibility.}

{\bf WP-theorem} {\em A $2g\times g$ Riemann matrix $\Pi=\left( \begin{array}{c}  \mathcal{A}\\ \mathcal{B}\end{array} \right)$ admits a reduction if there exists a $g\times g_1$ matrix of complex numbers of maximal rank, a $2g_1\times g_1$ matrix of complex numbers $\Pi_1$  and a $2g\times 2 g_1$ matrix of integers $M$ also of maximal rank such that
\[
\Pi\lambda=M\Pi_1 \label{poincare1}
\]
where $1\leq g_1 <g$. When a Riemann matrix admits reduction the corresponding period matrix may be put in the quasi-block-diagonal form
\[
\tau =\left( \begin{array}{cc}  \tau_1&Q\\Q^T&\tau'   \end{array}  \right)
\]
where $Q$ is a $g_1\times (g-g_1)$ matrix with rational entries and the matrices $\tau_1$ and $\tau'$ have the properties of period matrices.}

Since $Q$ here has rational entries, there exists a diagonal
$(g-g_1)\times (g-g_1)$ matrix $D=(d_1,\ldots,d_{g-g_1})$ with positive entries for
which $(QD)_{jk}\in \mathbb{Z}$.

With $$(\boldsymbol{z},\boldsymbol{w})=\mathrm{Diag}(z_1,\ldots,z_{g_1},w_1,\ldots,w_{g-g_1})$$
the $\theta$-function associated with $\tau$ may be then expressed in terms of lower
dimensional theta functions as
\begin{align*}
\theta( (\boldsymbol{z},\boldsymbol{w});\tau  )&=\sum_{\begin{array}{c}  \boldsymbol{m}=(m_1,\ldots,m_{g-g_1})\\ 0\leq m_i \leq d_i-1 \end{array} }  \theta(\boldsymbol{z}+Q\boldsymbol{m};\tau_1 )
\theta \left( \begin{array}{c}  D^{-1} \boldsymbol{m} \\ \boldsymbol{0} \end{array} \right)(D\boldsymbol{w}; D\tau'D)
\end{align*}

\section{Humbert varieties}
Consider again relations (\ref{biremb}).  A compatibility condition
with respect to variables
$c_1,c_1,2\omega_1,2\omega_2$ is of the form
\begin{equation}\det\; \left(\begin{array}{cccc}1&0&m_{11}&m_{12}\\
0&1&m_{21}&m_{22}\\
\tau_{11}&\tau_{12}&m_{31}&m_{32}\\
\tau_{12}&\tau_{22}&m_{41}&m_{42}
\end{array}\right) = 0,\label{compatibility}\end{equation}
Denote $2\times 4$ matrix with integer entries
\[M=\left(\begin{array}{cccc}
m_{11}&m_{21}&m_{31}&m_{41}\\
m_{12}&m_{22}&m_{32}&m_{42}
\end{array}\right).\]
Write the compatibility condition (\ref{compatibility}) in the form
\[\alpha \tau_{11}+  \beta \tau_{12}+  \gamma
\tau_{22}+ \delta (\tau_{12}^2-\tau_{11}\tau_{22})+\epsilon=0,
\]
where
\begin{align*}
\alpha&=m_{41}m_{12}-m_{11}m_{42},\quad \gamma=m_{21}m_{32}-m_{31}m_{22},\\
\delta&=m_{21}m_{12}-m_{11}m_{22},\quad \epsilon=m_{31}m_{42}-m_{41}m_{32},\\
\beta&=m_{11}m_{32}-m_{31}m_{12}- (m_{21}m_{42}-m_{41}m_{22}).
\end{align*}

Set
\[N= |m_{11}m_{32}-m_{31}m_{12}+
m_{21}m_{42}-m_{41}m_{22}| \]
Calculating from the first four equations
\begin{align*}
m_{21}&=\Delta(m_{11}\gamma+m_{31}\delta),
m_{22}=\Delta(m_{12}\gamma+m_{32}\delta),\\
m_{41}&=\Delta(m_{11}\epsilon+m_{31}\alpha),
m_{42}=\Delta(m_{12}\epsilon+m_{32}\alpha)
\end{align*}
where $\Delta^{-1}={m_{11}m_{32}-m_{31}m_{12}} $, we get the equality
\[
m_{21}m_{42}-m_{41}m_{22}=-\frac{\epsilon\delta+\alpha\gamma}
{m_{11}m_{32}-m_{31}m_{12}},
\]
which helps to transform the fifth equation to the quadratic one
with the following roots:
\begin{align*}
&m_{11}m_{32}-m_{31}m_{12}&=&\frac12\left(
\beta\pm\sqrt{\beta^2-4(\epsilon\delta+\alpha\gamma)}
\right),\\
&m_{21}m_{42}-m_{41}m_{22}&=&\frac12\left(-
\beta\pm\sqrt{\beta^2-4(\epsilon\delta+\alpha\gamma)}
\right).
\end{align*}
Taking in account the definition of $N$ we get the condition
\[
N^2=\beta^2-4(\epsilon\delta+\alpha\gamma)\label{biremb3}
\]

\begin{defn} The Humbert variety $H_{\Delta}$ with an invariant with respect to symplectic transformation
$\Delta$ is the subset in the 3-dimensional  space with coordinates $\tau_{11},\tau_{12},\tau_{22}$ given by the conditions
\[\alpha \tau_{11}+  \beta \tau_{12}+  \gamma
\tau_{22}+ \delta (\tau_{12}^2-\tau_{11}\tau_{22})+\epsilon=0,
\]
with integer $\alpha,\beta,\gamma,\delta \in \mathbb{Z}$ satisfying
\[
N^2=\beta^2-4(\epsilon\delta+\alpha\gamma)
\]
The invariant $\Delta$ equals
\[ \Delta=N^2\]
\end{defn}
That's possible to show that there exist such an element
$T\in\mathrm{Sp}(4,{\mathbb Z})$ that
\[M\circ T=\left(\begin{array}{cccc}N&0&0&1\\0&0&1&0\end{array}\right) \]
Transformed period matrix gets
a form
\begin{equation}\widetilde{\tau}= T\circ \tau=
\left(\begin{array}{cc}\widetilde{\tau}_{11}&\frac{1}{N}\\
\frac{1}{N}&\widetilde{\tau}_{22}\end{array}\right).\label{ttt} \end{equation}
That means that $V$ is a covering of $N$-th degree over elliptic curves $\mathcal{E}_{\pm}$. In the Jacobi reduction example $N=2$.

The period matrix $\widetilde{\tau}$ represents a component of the Humbert variety $H_{\Delta}$, other components results the action of the group $\mathrm{Sp}(4,\mathbb{Z})$.

\subsection{Krazer transformations of the period matrix}
We will explain here how to construct transformation (\ref{ttt}) in the case of
reduction being induced by the cover over an elliptic curve, i.e. $g_1=1$ by
following to Krazer monograph, p.474-475
In this case the integer Matrix $M$ is of the form
\[
M^T=\left(\begin{array}{cccc}m_{11}&m_{21}&\ldots&m_{g,1}  \\
m_{12}&m_{22}&\ldots&m_{g,2}
\end{array}\right)
\]
To find necessary transformation we introduce matrices
$A_i,B_i,C_{i,j},D_{i,j}\in \mathrm{Sp}_{2g}(\mathbb{Z})$,
$i\neq j=1,\ldots, g $,
\begin{align*}
A_i&=\left(\delta_{k,l}+\delta_{k,i}\delta_{k+g,l}\right))_{k,l=1,\ldots,2g}\\
B_i&=\left(\delta_{k,l}(1-\delta_{k,g+i})(1-\delta_{l,i})+\delta_{k,i}
\delta_{l,g+i}
-\delta_{l,i}\delta_{k,g+i}\right)_{k,l=1,\ldots,2g}\\
C_{ij}&=\left(\delta_{k,l}+\delta_{l,j}\delta_{k,i}
-\delta_{l,g+i}\delta_{k,g+j} \right)_{k,l=1,\ldots,2g}\\
D_{ij}&=\left(  \delta_{k,i}\delta_{l,j}
                +\delta_{k,j}\delta_{l,i}
                 +\delta_{k,g+i}\delta_{l,g+j}
                  +\delta_{l,g+i}\delta_{k,g+j}\right.\\
&\quad\qquad\left.+\delta_{k,l}(1-\delta_{k,i})(1-\delta_{k,j})
                    (1-\delta_{l,g+i}) (1-\delta_{l,g+j})
\right)_{k,l=1,\ldots,2g}\end{align*}

The action of $A_i$ results
the adding of $i$-th column to the $g+i$-th. $B_i$ interchanges $i$-th and $g+i$-th columns with multiplication of the last by $-1$. $C_{i,k}$ adds $i$-th
column to the $k$-th and subtracts $g+k$-th column from $g+i$-th.The matrix
$D_{i,j}$ interchanges $i$ and $j$-th and $g+i$ and $g+j$-th columns.
\[\text{At}\; g=2,\;\;
M^T=\left(\begin{array}{cccc}m_{11}&m_{21}&\ldots&m_{g,1}  \\
m_{12}&m_{22}&\ldots&m_{g,2}
\end{array}\right)
\]
\[M^TA_1=\left(\begin{array}{cccc}m_{11}&m_{21}&m_{11}+m_{31}&m_{41}\\
m_{12}&m_{22}&m_{12}+m_{32}&m_{42}\end{array}\right)\]
\[ M^TB_1=\left(\begin{array}{cccc}-m_{31}&m_{21}&m_{11}&m_{41}\\
-m_{32}&m_{22}&m_{12}&m_{42}\end{array}\right)\]
\[ M^TB_1=\left(\begin{array}{cccc}-m_{31}&m_{21}&m_{11}&m_{41}\\
-m_{32}&m_{22}&m_{12}&m_{42}\end{array}\right)\]
\[ M^TC_{12}=\left(\begin{array}{cccc}m_{11}&m_{11}+m_{21}&m_{31}-m_{41}&m_{41}\\
m_{12}&m_{12}+m_{22}&m_{32}-m_{42}&m_{42}\end{array}\right)\]

At the first step of the algorithm transform using $A_i$ and $B_i$-matrices
the matrix $M^T$ to the form
\[
\begin{array}{c}
\left(
\begin{array}{ccc}*&\ldots &*\\
                      0&\ldots&0\end{array}
\begin{array}{ccc}
                   *&\ldots&*\\
                    *&\ldots&*\end{array} \right)
\\
\underbrace{\phantom{aaaaaaa}}_g\;\;\underbrace{\phantom{aaaaaaa}}_g
\end{array} \]
At the second step use $C_{i,j}$ and $D_{i,j}$ to transform the above matrix to the form
\[
\begin{array}{c}
\left(  \begin{array}{cccc}\pm k&*&\ldots&*\\
                      0&0&\ldots&0\end{array}
               \begin{array}{cccc}*&*&\ldots&*\\
                      \pm1&0&\ldots&0\end{array}\right)
\\
\underbrace{\phantom{aaaaaaaaaaa}}_g\;\;\;\underbrace{\phantom{aaaaaaaaaaaa}}_g
\end{array}\]

That's possible to turn $\pm$ signs in this expression by using operation $B_1^2$.
At the next step the system is reduced to the form
\[
\begin{array}{c}
\left(  \begin{array}{cccc} c&0&\ldots&0\\
                      0&0&\ldots&0\end{array}
               \begin{array}{cccc}a&b&\ldots&0\\
                      1&0&\ldots&0\end{array}\right)
\\
\underbrace{\phantom{aaaaaaaaaaa}}_g\;\;\;\underbrace{\phantom{aaaaaaaaaaaa}}_g
\end{array}\]
Finally apply operators $B_1^3$, $B_2^3$ to transform $M$ to the form
\[
\begin{array}{c}
\left(  \begin{array}{cccc} N&0&\ldots&0\\
                      0&0&\ldots&0\end{array}
               \begin{array}{ccccc}0&1&0&\ldots&0\\
                      1&0&0&\ldots&0\end{array}\right)
\\
\underbrace{\phantom{aaaaaaaaaaa}}_g\;\;\;\;\;\underbrace{\phantom{aaaaaaaaaaaaaa}}_g
\end{array}\]

\subsection{Variety $H_4$}
Following to   Shaska and V\"olklein  \cite{sv04} write
 the general genus two curve which covers two-sheetedly two elliptic curves
can be written in the form
\[ Y^2=X^6-s_1X^4+s_2X^2 -1 \]
with the discriminant $$27-18s_1s_2-s_1^2s_2^2+4s_1^3+4s_2^3\neq 0$$.
Introduce coordinates
\[ u=s_1s_2, \quad v=s_1^3+s_2^3 \]
The relative invariants are given as
\begin{align*}
J_2&=240+16u\\
J_4&=48v+4u^2-504u+1620\\
J_6&=24u^3+160uv-424u^2-20664u +96v+119880 \\
J_{10}&=64(27-18u-u^2+4v)^2
\end{align*}
Eliminating $u,v$ from these equations we get modular form $\chi_{30}$ of weight 30 that vanishing provides condition $\tau\in H_4$. That is

\begin{align*}
\chi_{30}&= -125971200000J_{10}^3+41472J_{10}J_4^5+159J_4^6J_2^3-80J_4^7J_2-236196J_{10}^2J_2^5\\&
+972J_{10}J_2^6J_4^2+8748J_{10}J_2^4J_{6}^2-1728J_4^5J_2^2J_{6}+6048J_4^4J_2J_{6}^2\\&
+1332J_4^4J_2^4J_{6}-8910J_4^3J_2^3J_{6}^2-592272J_{10}J_4^4J_2^2+77436J_{10}J_4^3J_2^4\\&
+31104J_{6}^5+29376J_2^2J_4^2J_{6}^3-47952J_2J_4J_{6}^4+12J_2^6J_4^3J_{6}-54J_2^5J_4^2J_{6}^2\\
&+108J_2^4J_4J_{6}^3-2099520000J_{10}^2J_4J_{6}-9331200J_{10}J_4^2J_{6}^2+104976000J_{10}^2J_2^2J_{6}\\
&-3499200J_{10}J_2J_{6}^3-78J_4^5J_2^5+384J_4^6J_{6}-6912J_4^3J_{6}^3J_2^7J_4^4-81J_2^3J_{6}^4\\
&+507384000J_{10}^2J_4^2J_2-19245600J_{10}^2J_4J_2^3-5832J_{10}J_2^5J_4J_{6}\\
&+4743360J_{10}J_4^3J_2J_{6}-870912J_{10}J_4^2J_2^3J_{6}+3090960J_{10}J_4J_2^2J_6^2
\end{align*}

Let the curve 
\[  Y^2=X(X-1)(X-\lambda_1)(X-\lambda_2)(X-\lambda_3) \]
Then direct calculation leads to the decomposition of $\chi_{30}$
\begin{align*}
\chi_{30}&=(\lambda_1\lambda_2-\lambda_2-\lambda_3\lambda_2+\lambda_3)^2\\
&\times (\lambda_1\lambda_2-\lambda_1+\lambda_3\lambda_1+\lambda_3\lambda_2)^2\\
&\times (\lambda_1\lambda_2-\lambda_3\lambda_1+\lambda_3\lambda_2+\lambda_3)^2\\
&\times(\lambda_1\lambda_3-\lambda_1-\lambda_3\lambda_2+\lambda_3)^2\\
&\times (\lambda_1\lambda_2+\lambda_1-\lambda_3\lambda_1+\lambda_2)^2\\
&\times (\lambda_1\lambda_2-\lambda_1-\lambda_3\lambda_1+\lambda_3)^2\\
&\quad\cdots \quad \cdots \\
&\times(\lambda_1\lambda_2-\lambda_3)^2\\&\times(\lambda_1-\lambda_2\lambda_3)^2\\
&\times(\lambda_1\lambda_3-\lambda_2)^2
\end{align*}

We refer here to the result by Pringsheim \cite{pri875}.  Let the curve $V$ and its Riemann period
matrix are given in the {\em Richelot form}
\begin{align*} y^2=x(1-x)(1-\lambda^2 x)(1-\mu^2 x)(1-\kappa^2 x),\\
\tau = \left(\begin{array}{cc} \tau_{11}&\tau_{12}\\
\tau_{12}&\tau_{22}
\end{array}\right) \end{align*}
The aforementioned 15 components of the variety $H_4$
Pringsheim distributed in 4 groups
\begin{center} {\bf I} \end{center}
\[2\tau_{12}+\tau_{11}\tau_{22}-\tau_{12}^2=0
\quad \Leftrightarrow\quad  \kappa^2=\lambda^2\mu^2\]
\begin{center} {\bf II} \end{center}
\begin{gather*}
\tau_{11}+2\tau_{12}\tau_{22}=0 \;  \Leftrightarrow\quad  \kappa^2-\lambda^2=\mu^2(1-\lambda^2) \\
\tau_{11}+2\tau_{12}-(\tau_{11}\tau_{22}-\tau_{12}^2)=0
\;  \Leftrightarrow\quad  \kappa^2(1-\mu^2)=\lambda^2(\kappa^2-\mu^2)
\end{gather*}


\begin{center} {\bf III} \end{center}
\begin{align*}&2\tau_{12}-\tau_{22}=0\\
&\hskip 4cm  \Leftrightarrow\quad  \mu^2=\kappa^2\lambda^2\\
&2\tau_{12}-\tau_{22}+\tau_{11}\tau_{22}-\tau_{12}^2=0\\
&\hskip 4cm  \Leftrightarrow\quad  \lambda^2=\kappa^2\mu^2\\
&\tau_{11}-\tau_{22}=0\\
&\hskip 4cm  \Leftrightarrow\quad \kappa^2- \lambda^2=\lambda^2(\kappa^2-\mu^2)\\
&\tau_{11}-\tau_{22}+\tau_{11}\tau_{22}-\tau_{12}^2=0\\
&\hskip 4cm  \Leftrightarrow\quad \kappa^2- \mu^2=\mu^2(\kappa^2-\lambda^2)
\end{align*}
\begin{center} {\bf IV} \end{center}
\begin{align*}&2\tau_{12}=1\\
&\hskip 4cm  \Leftrightarrow\quad \lambda^2- \mu^2=-\kappa^2(1-\lambda^2)\\
&\vdots
\end{align*}

According to the Bierman-Humbert theorem the associated Riemann period matrix is of the form
$$\tau=\left( \begin{array}{cc}  \tau_{1}&\frac12 \\
\frac12&\tau_{2} \end{array}\right)$$
Denote
\[  \theta_k=\vartheta_k(0,2\tau_1),\quad \Theta_k=
\vartheta_k(0,2\tau_2), \quad k=2,3,4.   \]
Then the following decomposition formula are valid for $\theta$-constants
\begin{align*}
&\theta\left[{}_0^1{}_0^0\right]=\theta\left[{}_0^1{}_1^0\right]=
\sqrt{2\theta_2\theta_3\Theta_3\Theta_4},\quad
\theta\left[{}_1^0{}_0^1\right]=\theta\left[{}_0^0{}_0^1\right]=
\sqrt{2\theta_3\theta_4\Theta_2\Theta_3},\\
&\theta\left[{}_0^1{}_0^1\right]=-\imath \theta\left[{}_1^1{}_1^1\right]=
\sqrt{2\theta_2\theta_4\Theta_2\Theta_4},\\
&\theta\left[{}_0^0{}_0^0\right]=
\sqrt{\theta_3^2\Theta_3^2+\theta_2\Theta_4^2+\theta_4^2\Theta_2^2},\quad
\theta\left[{}_1^0{}_1^0\right]=
\sqrt{\theta_3^2\Theta_3^2-\theta_2\Theta_4^2-\theta_4^2\Theta_2^2}\\
&\theta\left[{}_1^0{}_0^0\right]=
\sqrt{\theta_3^2\Theta_3^2-\theta_2\Theta_4^2+\theta_4^2\Theta_2^2},\quad
\theta\left[{}_0^0{}_1^0\right]=
\sqrt{\theta_3^2\Theta_3^2+\theta_2\Theta_4^2-\theta_4^2\Theta_2^2}
\end{align*}

Plugging above $\theta$-decomposition formulae to the formulae
\begin{align*}
\lambda_1&=\frac{ \theta^2\left[ {}_0^0{}_0^0 \right] }{ \theta^2\left[ {}_0^1{}_0^0 \right] }\frac{ \theta^2\left[ {}_0^0{}_0^1 \right] }{ \theta^2\left[ {}_0^1{}_0^1 \right] }\\
\lambda_2&= \frac{ \theta^2\left[ {}_0^0{}_0^1 \right] }{ \theta^2\left[ {}_0^1{}_0^1 \right] } \frac{ \theta^2\left[ {}_0^0{}_1^0 \right] }{ \theta^2\left[ {}_0^1{}_1^0 \right] }\\
\lambda_3&=\frac{ \theta^2\left[ {}_0^0{}_0^0 \right] }{ \theta^2\left[ {}_0^1{}_0^0 \right] }\frac{ \theta^2\left[ {}_0^0{}_1^0 \right] }{ \theta^2\left[ {}_0^1{}_1^0 \right] }
\end{align*}
we get the following
\[  \tau_{12}=\frac12  \quad \Leftarrow\Rightarrow (1-\lambda_1-\lambda_2)\lambda_3+\lambda_1\lambda_2=0  \]
what accords to the corresponding Pringsheim component moduli $\lambda,\mu,\kappa$ are redefined as
\[\lambda^2=1/\lambda_3,\quad \mu^2=1/\lambda_2,\quad \kappa^2=1/\lambda_1\]

\subsection{$\wp$-functions on $H_4$}
Consider again the curve.
$$w^2=z(z-1)(z-\alpha)(z-\beta)(z-\alpha\beta).$$
This curve is a 2-fold cover of two tori (elliptic curves)
$\pi_{\pm}: X = (w,z) \rightarrow E_{\pm} =
(\eta_{\pm}, \xi_{\pm})$,
\[
\eta_{\pm}^2=\xi_{\pm}(1-\xi_{\pm})(1-k_{\pm}^2\xi_{\pm})
\]
with the Jacobi moduli
$$
k_{\pm}^2=-\frac{(\sqrt{\alpha}\mp\sqrt{\beta})^2}{(1-\alpha)(1-\beta)}.
$$
Consider the Abel--Jacobi mapping
\begin{align*}
&\int_{x_0}^{x_1} \frac{\mathrm{d}z}{w} +\int_{x_0}^{x_2}
\frac{\mathrm{d}z}{w}=u_1,\\ &\int_{x_0}^{x_1}
\frac{z\mathrm{d}z}{w} +\int_{x_0}^{x_2}
\frac{z\mathrm{d}z}{w}=u_2.  \end{align*}
Again, we want to express the points $(x_i, y_i)$ as functions of
$\boldsymbol{u}=(u_1,u_2)^T$.

It is convenient to write  down the integrals in the Abel mapping in
the following form:
\begin{align*} \int_{x_0}^{x_1}
\frac{z-\sqrt{\alpha\beta}}{w}\mathrm{d}z +\int_{x_0}^{x_2}
\frac{z-\sqrt{\alpha\beta}}{w}\mathrm{d}z=u_+\label{j11}\\
\int_{x_0}^{x_1} \frac{z+\sqrt{\alpha\beta}}{w}\mathrm{d}z
+\int_{x_0}^{x_2}
\frac{z+\sqrt{\alpha\beta}}{w}\mathrm{d}z=u_-,\end{align*}
where
\[
u_{\pm}=-\sqrt{(1-\alpha)(1-\beta)}(u_2\mp\sqrt{\alpha\beta}u_1)
\]
Denote the Jacobi elliptic functions:
$${\rm sn}(u_{\pm},k_{\pm}),{\rm
cn}(u_{\pm},k_{\pm}), {\rm dn}(u_{\pm},k_{\pm})$$

Define the Darboux coordinates
\begin{align*} X_1&= {\rm sn}(u_{+},k_{+}){\rm
sn}(u_{-},k_{-}),\\ X_2&= {\rm cn}(u_{+},k_{+}){\rm
cn}(u_{-},k_{-}),\\ X_3&= {\rm dn}(u_{+},k_{+}){\rm
dn}(u_{-},k_{-}). \end{align*}


Reduce the hyperelliptic integrals in the last Abel mapping to elliptic ones,
then, using an addition theorem for the Jacobi elliptic functions, we get
\begin{align*}
X_1&=-\frac{(1-\alpha)(1-\beta)(\alpha\beta+\wp_{12})}
{(\alpha+\beta)(\wp_{12}-\alpha\beta)+\alpha\beta\wp_{22}+\wp_{11}} ,
\\
X_2&=-\frac{(1+\alpha\beta)(\alpha\beta-\wp_{12})-\alpha\beta\wp_{22}
-\wp_{11}}
{(\alpha+\beta)(\wp_{12}-\alpha\beta)+\alpha\beta\wp_{22}+\wp_{11}},
\\
X_3&=-\frac{\alpha\beta\wp_{22}-\wp_{11}}
{(\alpha+\beta)(\wp_{12}-\alpha\beta)+\alpha\beta\wp_{22}+\wp_{11}}.
\end{align*}
Here $\wp_{ij}=\wp_{ij}(u_1, u_2)$ are second derivatives of $\log\sigma(u_1, u_2)$.

These formulae can be inverted:
\begin{align*}
\wp_{11}&=(B-1)\frac{A(X_2+X_3)-B(X_3+1)}
{X_1+X_2-1},\\
\wp_{12}&=(B-1)\frac{1+X_1-X_2}{X_1+X_2-1},\\
\wp_{22}&=\frac{A(X_2-X_3)+B(X_3-1)}
{X_1+X_2-1},\label{wp22a}\end{align*}
where $A=\alpha+\beta$, $B=1+\alpha\beta$.

\section{Applications}

\subsection{2-gap elliptic potentials of the Schr\"odinger equation}
We already shown that $2\wp_{22}(\boldsymbol{u})$-function with $u_2=x$,
$u_1=c=\mathrm{const}$ represents a 2-gap quasi-periodic potential of the equation.
\medskip

{\bf Lemma} Let a genus two curve $X$ is an $N$-fold cover of an elliptic curve $\mathcal{E}$.
 According to the Weierstrass-Poincar\'e theorem exist  such homology basis that Riemann period matrix is of the form
\[  \tau =\omega^{-1}\omega'= \left( \begin{array}{cc}   T&    \frac{1}{N} \\   \frac{1}{N}& \widetilde{T}  \end{array}   \right) \]
where $\omega^{-1}=(\boldsymbol{U},\boldsymbol{V})$ is the matrix of
$\mathfrak{a}$-periods.
Then the condition $U_2=0$ is sufficient for the $\wp_{22}( c,x )$ to be an
elliptic function in $x$

Consider the case $N=4$. Using Rosenhain representation of the winding vector we conclude that condition $U_2=0$ requires vanishing of the derivative $\theta$-constant
$\theta_1\left[{}_1^1{}_1^0\right]= 0$.

Decomposing genus two theta-constants to elliptic $\theta$-constants by means of second order addition theorems, we reduce the above condition to the form
\begin{align*}
\widetilde{k}=k'(1-4k^2)
\end{align*}
where $k$ and $\widetilde{k}$ are Jacobian moduli corresponding to the moduli $T$ and $\widetilde{T}$ respectively.
Using the obtained condition we get
\begin{align}
\wp_{22}(c,x)=6\wp(x)+2\wp(x+\omega)
\end{align}
That is the Darboux--Treibich--Verdier elliptic potential associated to a 4-sheeted cover of elliptic curve.

\chapter[Rational analogs ]
{Rational analogs of Abelian functions}\label{chap:rat}

\section{Introduction}
In the theory of Abelian functions on Jacobians, the
key role is played by \textit{entire functions} that satisfy the Riemann
vanishing theorem (see, for instance, \cite{mu83}). Here we introduce
\textit{polynomials} that satisfy an analog of this theorem
and show that these polynomials are completely characterized by this
property. By rational analogs of Abelian functions we mean logarithmic
derivatives of orders $\geq 2$ of these polynomials.

We call the polynomials thus obtained the
\textit{Schur--Weierstrass polynomials}
because they are constructed from classical Schur polynomials,
which, however, correspond to special partitions related to
Weierstrass sequences. Recently, in connection with the problem to
construct rational solutions of nonlinear integrable equations
\cite{am78,kr79}, a special attention was payed to
Schur polynomials \cite{dg86,kac93}.
Since a Schur polynomial corresponding to an arbitrary partition leads to a
rational solution of the Kadomtsev--Petviashvili hierarchy,
the problem of connecting the above solutions with those defined
in terms of Abelian functions on Jacobians naturally arose.
Our results open the way to solve this problem
on the basis of the Riemann vanishing theorem.

We show our approach by the example of Weierstrass sequences defined
by a pair of coprime numbers $n$ and~$s$. Each of these
sequences generates a
class of plane curves of genus $g=(n-1)(s-1)/2$ defined by equations
of the form
\begin{equation}\notag
Y^n-X^s-\sum_{\alpha, \beta}
\lambda_{\alpha n+\beta s}X^{\alpha}Y^{\beta} =0,
\end{equation}
where $0\leq\alpha<s-1$, $0\leq\beta<n-1$, and $\alpha n+\beta s< ns$.
The dimension of moduli space of this class is
$d_{n,s} =(n+1)(s+1)/2-[s/n]-3$. We constructed \cite{bel99} entire
functions $\sigma(\boldsymbol{u};\boldsymbol{\lambda})$,
where $\boldsymbol{u}\in\mathbb{C}^g$ and
$\boldsymbol{\lambda}\in\mathbb{C}^{d_{n,s}}$, on the universal space
of the Jacobians of such curves. Similarly to the elliptic Weierstrass
\text{$\sigma$-function}, in a neighborhood of the point
$\boldsymbol{0}\in\mathbb{C}^g$, the function
$\sigma(\boldsymbol{u};\boldsymbol{\lambda})$ has
a power series expansion in
$u_1,\dots,u_g$ whose coefficients are polynomials
with rational coefficients
in $\boldsymbol{\lambda}$. The limit
$\lim_{\boldsymbol{\lambda}\to\boldsymbol{0}}
\sigma(\boldsymbol{u};\boldsymbol{\lambda})$ is defined and leads to a
polynomial, which satisfies the analog of the Riemann
vanishing theorem. As an application of our result we prove that
$\sigma(\boldsymbol{u};\boldsymbol{0})$ is
equal to the corresponding Schur--Weierstrass polynomial
up to a constant factor.


\section{Weierstrass sequences and partitions}

Let $n$ and $s$ be
a pair of coprime integers such that
$s>n \ge 2$. Consider the
set of all nonnegative integers of the form \begin{equation} an+bs,\quad
\text{where $a$, $b$ are nonnegative integers.} \label{n-gap} \end{equation}

\begin{definition} The positive integers that
are not representable in the form \eqref{n-gap} constitute
a \textit{Weierstrass sequence}. The number of these
integers is called the \textit{length} of the sequence.
\end{definition}

The above Weierstrass sequence in ascending order is denoted by
$\bW_{\!n,s}$. To prove the main properties of Weierstrass sequences, the
following elementary observation is useful.

\begin{lemma}\label{fir} The elements $w$
of the Weierstrass sequence $\bW_{\!n,s}$ are representable in the form
\begin{equation}
w=-\alpha n+\beta s,\label{WS_elt}
\end{equation}
where $\alpha$ and $\beta$ are integers such that $\alpha>0$
and $n>\beta>0$;
the numbers $\alpha$ and $\beta$ in the representation
\eqref{WS_elt} are defined uniquely.
\end{lemma}

The next lemma gives a list of the
properties of Weierstrass sequences that we need below.

\begin{lemma}\label{w-props} A Weierstrass sequence
$\bW_{\!n,s}=\{w_1,w_2,\dots\}$ has the following properties\phantom{,}:
\begin{enumerate}

\item[\phantom{,}{(1)}] its length $g$
is equal $(n-1)(s-1)/2$\phantom{,};

\item[\phantom{,}{(2)}] its maximal element $w_{g}$ is equal
$2g-1$\phantom{,};

\item[\phantom{,}{(3)}] if $w\in\bW_{\!n,s}$, then
$w_g-w\notin\bW_{\!n,s}$\phantom{,};

\item[\phantom{,}{(4)}] if $w>\widetilde{w}$, where $w \in \bW_{\!n,s}$ and
$\widetilde{w}\notin \bW_{\!n,s}$, then
$w-\widetilde{w}\in\bW_{\!n,s}$\phantom{,};

\item[\phantom{,}{(5)}] $i\le w_i\le 2i-1$ for all $i=1,\dots,g$.
\end{enumerate}
\end{lemma}

\begin{proof}
(1) Let us find the number of integers of the form
\eqref{WS_elt}. For a chosen value $\beta=\ell\in\{1,\dots,n-1\}$,
the factor $\alpha$ ranges over the values $\{1,\dots,[\ell s/n]\}$;
thus (see \cite[viii.~17, \text{viii.~18}]{PS78}),
\begin{equation}\notag g=\smash[t]{\sum_{\ell=1}^{n-1}\bigg[\frac{\ell
s}{n}\bigg] =\frac{(n-1)(s-1)}{2}\,.} \end{equation}

(2) Note that $2g-1=-n+s(n-1)$. Let there be a number $w$
of the form \eqref{WS_elt} that exceeds $2g-1$. Since
$n$ and $s$ are coprime, it follows that in the
representation \eqref{WS_elt}
of $w$ we have either $\alpha\le 1$ or $\beta>n$. The contradiction
thus obtained proves assertion (2) of the lemma.

(3) For an arbitrary number $w \in \bW_{\!n,s}$,
consider the difference $w_g-w$. According to Lemma \ref{fir}, we have
\begin{equation}\notag
w_g-w=-n+s(n-1)-(-\alpha n+\beta s)=(\alpha-1)n+(n-1-\beta)s,
\end{equation}
and the last representation is of the form \eqref{n-gap}.

(4) Let $\widetilde{w} \notin\bW_{\!n,s}$ and $w \in\bW_{\!n,s}$. Then
the representations
\begin{equation}\notag
\widetilde{w} =a n+bs\quad \text{and}\quad w=-\alpha n+\beta s
\end{equation}
hold, and hence
\begin{equation}\notag
w-\widetilde{w} =-\alpha n+\beta s-(an+bs)=-(\alpha+a)n+(\beta-b)s>0
\end{equation}
by assumption; therefore,
$w-\widetilde{w}\in\bW_{\!n,s}$ by Lemma \ref{fir}.

(5) Consider the set
$\boldsymbol{\widetilde{W}}_{n,s}=\{2g-1-w_{g+1-i}\}_{i=1,\dots,g}$.
According to assertions~(2) and (3), we have
$\boldsymbol{\widetilde{W}}_{n,s}\cap\bW_{\!n,s}=\{\varnothing\}$,
and hence $\boldsymbol{\widetilde{W}}_{n,s}\cup\bW_{\!n,s}$
consists of $2g$
nonnegative integers each of which is
less than $2g-1$, i.e., this is a permutation of the set
$\{0,1,2,\dots,2g-1\}$.

Now let $N(L)$ be the number of integers of the form
\eqref{n-gap} that are less than $L$. Since
$\{\widetilde{w}_1,\dots,\widetilde{w}_{N(w_i)}\}\subset
\boldsymbol{\widetilde{W}}_{n,s}$ and $\{w_1,\dots,w_i\}\subset
\bW_{\!n,s}$, it follows that the intersection
$\{\widetilde{w}_1,\dots,\widetilde{w}_{N(w_i)}\}\cap\{w_1,\dots,w_i\}$
is empty, and the union
$\{\widetilde{w}_1,\dots,\widetilde{w}_{N(w_i)}\}
\cup\{w_1,\dots,w_i\}$ is a permutation of the set $\{0,1,2,\dots,w_i\}$
for any $w_i\in\bW_{\!n,s}$, which yields $w_i=i+N(w_i)-1$.

Note that $\widetilde{w}_0=0<w_i$, i.e., $N(w_i)\ge 1$, and hence
the inequality $w_i\ge i$ holds for all $w_i\in\bW_{\!n,s}$.

On the other hand, by assertion~(4) we have the inclusion
$\{w_i-\widetilde{w}_1,\dots,
w_i-\widetilde{w}_{N(w_i)}\}\subset
\{w_1,\dots,w_i\}$, which yields the inequality
$N(w_i)\le i$. Thus, $w_i=i+N(w_i)-1\le 2i-1$.
\end{proof}

We denote by $\mathrm{WS}_m$ the set of all Weierstrass sequences
$\boldsymbol{w}=(w_1,\dots,w_m)$ of length $m$.

In what follows, we use the agreements
and notation adopted in \cite{Mac85}.

\begin{definition} A \textit{partition}
$\boldsymbol{\pi}$ of length $m$ is a nonincreasing
set of $m$ positive integers $\pi_i$. Denote by $\mathrm{Par}_m$
the set of all partitions of length $m$.
\end{definition}

On the set of partitions, a conjugation operation is defined
as follows:
\begin{equation}\notag
\boldsymbol{\pi}'=\{\pi_1,\dots,\pi_{m}\}'=\{\pi'_1,\dots,\pi'_{m'}\},
\quad\text{where }\,
\pi'_i=\mathrm{Card}\{j:\pi_j\ge i\}.
\end{equation}
In particular, the length $m'$ of the conjugate partition
$\boldsymbol{\pi}'$ is equal to $\pi_1$. Obviously,
$\boldsymbol{\pi}''=\boldsymbol{\pi}$.

\begin{lemma}
The formula
\begin{equation}
\varkappa(\boldsymbol{w})=\boldsymbol{\pi},\quad
\text{where }\,\pi_{k}=w_{g-k+1}+k-g,
\end{equation}
defines an embedding $\varkappa\colon\mathrm{WS}_g\to\mathrm{Par}_g$.
\end{lemma}

\begin{proof}
We must show that
$\varkappa(\bW_{\!n,s})$ is a partition. Let us show first that
$\pi_j=w_{g-j+1}+j-g>0$ for all $j=1,\dots,g$. By assertion~(5) of Lemma
\ref{w-props} we have $w_i-i\ge 0$, which  implies the desired
inequality for $i=g+1-j$.

It remains to show that
$\pi_k-\pi_{k+1}\ge 0$ for $\pi_k=w_{g-k+1}+k-g$ and
$\pi_{k+1}=w_{g-k}+k-g+1$. Indeed, the difference
\begin{equation}\notag
\pi_k-\pi_{k+1}=w_{g-k+1}-w_{g-k}-1
\end{equation}
is nonnegative because $\bW_{\!n,s}$
is a strictly increasing sequence by definition.
\end{proof}

\begin{definition} A partition $\boldsymbol{\pi}$ that is the image
of a Weierstrass sequence under the mapping $\varkappa$ is called a
\textit{Weierstrass partition}.
Let us introduce the following notation for Weierstrass partitions:
\begin{equation}\notag
\boldsymbol{\pi}_{n,s}=\varkappa(\bW_{\!n,s}).
\end{equation}
\end{definition}

\begin{lemma} The Weierstrass partitions have the following properties\phantom{,}:
\begin{enumerate}

\item[\phantom{,}{(1)}] $\boldsymbol{\pi}_{n,s}=\boldsymbol{\pi}'_{n,s}$\phantom{,};

\item[\phantom{,}{(2)}] $\boldsymbol{\pi}_{n,s}\subset\{g,g-1,\dots,1\}$,
where $g=(n-1)(s-1)/2$.
\end{enumerate}
\end{lemma}

\begin{proof}
It is known \cite[p. 3--4]{Mac85} that, for any partition
$\boldsymbol{\mu}\in
\mathrm{Par}_n$ such that $\mu_1=m$, the set consisting of
$m+n$ numbers \begin{equation}\notag \mu_i+n-i\quad(1\le i\le n),\qquad
n-1+j-\mu'_j\quad(1\le j\le m)
\end{equation}
is a permutation of the set $\{0,1,2,\dots,m+n-1\}$.

Let us apply this result to the Weierstrass partition
$\boldsymbol{\pi}_{n,s}$. Its length is equal to
$g$, and the element $\pi_1$ is equal to $w_g+1-g=g$. Thus,
the set formed by $2g$ numbers
\begin{alignat}2
\pi_i+g-i&=w_{g+1-i}&\qquad(i&=1,\dots,g),\notag\\
g-1+j-\pi'_j&=2g-1-(\pi'_j+g-j)&\qquad(j&=1,\dots, g)\notag
\end{alignat}
is a permutation of the set $\{0,1,2,\dots,2g-1\}$. That is, the set
$\{2g-1-(\pi'_j+g-j)\}_{j=1,\dots, g}$ is the complement of the sequence
$\bW_{\!n,s}$ in the set
$\{0,1,2,\dots,2g-1\}$. Since the complement of the sequence
$\bW_{\!n,s}$ in the set $\{0,1,2,\dots,2g-1\}$ is of the form
$\{2g-1-w_{g+1-j}\}_{ j=1,\dots, g}$ and $\{\pi'_1,\dots,\pi'_g\}$ is a
partition, it follows that
\begin{equation}\notag \pi'_j=w_{g+1-j}+j-g=\pi_j
\qquad(1\le j\le g).
\end{equation}

The condition $\boldsymbol{\pi}_{n,s}\subset\{g,g-1,\dots,1\}$
means that, for the elements of the sequence $\bW_{\!n,s}$, the
inequalities $w_i\le 2i-1$ hold for all $i$, $1\le i\le g$,
according to assertion~(5) of Lemma \ref{w-props}.
\end{proof}

\section{Schur-Weierstrass polynomials}

We present the notation and the results following~\cite{Mac85} as above. Let
$\Lambda_m=\bigoplus_{k\ge 0}\Lambda_m^k$ be the graded ring of symmetric
polynomials with integer coefficients in variables
$x_1,\dots,x_m$, where $\Lambda_m^k$ consists of the homogeneous symmetric
polynomials of degree $k$. Here the number $m$ is assumed to be
large enough as usual and,
for this reason, below we denote the ring $\Lambda_m$ by $\Lambda$.

Let us introduce the elementary symmetric functions $e_r$ by means of the
generating function $E(t)=\sum_{r\ge 0}e_r t^r=\prod_{i \ge
0}(1+x_i t)$. As is known, $\Lambda= \mathbb{Z}[e_1,e_2,\dots]$.
Let us introduce the elementary Newton polynomials $p_r$ by means of the
generating function $P(t)=\sum_{r\ge 1}p_r t^{r-1}=\sum_{i\ge1}x_i/(1-x_it)$.
The generating functions $P(t)$ and $E(t)$ are related by  the formula
$P(-t)E(t)=E'(t)$. The function $e_k$ is expressed in terms of
$\{p_1,\dots,p_k\}$ in the form of the determinant,
\begin{equation}\label{ptoe}
e_k=\frac{1}{k!}
\begin{vmatrix}
p_1&1&0&\hdots&0\\
p_2&p_1&2&\hdots&0\\
\hdotsfor{5}\\
p_{k-1}&p_{k-2}&p_{k-3}&\hdots&k-1\\
p_k&p_{k-1}&p_{k-2}&\hdots&p_1
\end{vmatrix}.
\end{equation}

The functions $p_r$ are algebraically independent over the field
$\mathbb{Q}$ of rational numbers, and $ \Lambda_{\mathbb{Q}}=\Lambda
\otimes_{\mathbb{Z}} \mathbb{Q}=\mathbb{Q}\kern1pt[p_1,p_2,\dots]$. Set
$p_{\boldsymbol{\pi}}=p_{\pi_1} p_{\pi_2} p_{\pi_3}\dots$ for all partitions
$\boldsymbol{\pi}=\{\pi_1,\pi_2,\dots\}$. Thus,
the elements $p_{\boldsymbol{\pi}}$ form an additive basis in
$\Lambda_{\mathbb{Q}}$. Using this fact, we endow the ring
$\Lambda$ with the inner product given by
\begin{equation}
\langle p_{\boldsymbol{\pi}},
p_{\boldsymbol{\rho}}\rangle=\delta_{{\boldsymbol{\pi}},
{\boldsymbol{\rho}}}z_{\boldsymbol{\rho}},\label{scal}
\end{equation}
where $z_{\boldsymbol{\rho}}=\prod_{k\ge 1} k^{m_k}m_k!$
and $m_k=m_k({\boldsymbol{\rho}})$ is the multiplicity of a number
$k$ in the partition ${\boldsymbol{\rho}}$ (see \cite[p.~24,~64]{Mac85}).

To an operator of multiplication by an arbitrary symmetric polynomial
$f$, the conjugate linear operator $\mathcal{D}(f)$ with respect to the inner
product \eqref{scal} corresponds,
which is completely determined by the formula $\langle
\mathcal{D}(f) u,v\rangle=\langle u, f v\rangle$ for all $u,v\in\Lambda$.
For instance, we have the relation
\begin{equation}
\mathcal{D}(p_k)=(-1)^{k-1}\sum_{r\ge0}e_r\, \frac{\partial}{\partial e_{r+k}},
\label{dual-p}
\end{equation}
where it is assumed that the symmetric
polynomials are expressed as functions of~$e_r$. Representing
a function $f\in
\Lambda$ in the basis $p_{\boldsymbol{\mu}}$ as a polynomial
$\phi(p_1,p_2,\dots)$ with rational coefficients,
we obtain (see \cite[p.~75--76]{Mac85})
\begin{equation}
\mathcal{D}(f)=\phi\bigg(\frac{\partial}{\partial p_1},\dots,
n\,\frac{\partial}{\partial p_n},\dots\bigg).
\label{dual-f}
\end{equation}

The classical Schur polynomial $s_{\boldsymbol{\pi}}$
corresponding to an arbitrary partition
$\boldsymbol{\pi}$ has the following representation in the form of a
determinant (see
\cite[p.~41]{Mac85}):
\begin{equation}
s_{\boldsymbol{\pi}}=\det(e_{\pi'_i-i+j})_{1\le i,j \le m},\quad
\text{where $m$ is the length of the partition $\boldsymbol{\pi}$}.
\label{schur}
\end{equation}
We denote by $S_{n,s}$ the
Schur polynomial \eqref{schur} corresponding to a Weierstrass
partition $\boldsymbol{\pi}_{n,s}$. By \eqref{schur} we have
\begin{equation}\notag
S_{n,s}=S_{n,s}(e_1,\dots,e_{2g-1}),\quad\text{where $g=(n-1)(s-1)/2$}.
\end{equation}

\begin{theorem} \label{S-W-P}
In the representation via the elementary Newton polynomials,
any $S_{n,s}$ is a polynomial in
$g$ variables $\{p_{w_1},\dots,p_{w_g}\}$ only, where
$\{w_1,\dots,w_g\}=\bW_{\!n,s}$.
\end{theorem}

\begin{proof}
Note that, for any symmetric function $u$,
it follows from the relation $\langle u,p_k f \rangle\equiv 0$
for all $f\in \Lambda$ that
$u$ does not depend on $p_k$ because
the inner product \eqref{scal} is nondegenerate. By
\eqref{dual-p}, the latter condition is equivalent to the relation
$\mathcal{D}(p_k)u=(-1)^{k-1}\sum_{r\ge0}e_r\,\partial u/\partial e_{r+k}\equiv 0$,
where the function $u$ is regarded as a polynomial in $e_r$.

For some $q\in\{1,2,\dots,2g-1\}$,
consider the action of the linear operator
$\mathcal{D}(p_q)$ on the polynomial $S_{n,s}$ corresponding to the
Weierstrass partition
$\boldsymbol{\pi}_{n,s}=\{\pi_1,\dots,\pi_g\}$.
By differentiating the determinant given by formula
\eqref{schur}, we obtain
\begin{equation}
\mathcal{D}(p_q)S_{n,s}(e_1,e_2,\dots,e_{2g-1})= (-1)^{q-1}
\smash[t]{\sum_{i=1}^g\sum_{j=1}^g}e_{\pi_i-i-q+j} M_{i,j}, \label{d-sum}
\end{equation}
where we performed the change of the summation index by the formula
$r=\pi_i-i-q+j$, and $M_{i,j}$ stands for the
$(i,j)$-cofactor of the matrix
$(e_{\pi_i-i+j})_{1\le i,j\le g}$.

We need an auxiliary assertion.

\begin{lemma} If the equation
\begin{equation}\label{null}
\pi_{\ell}-\ell-q=\pi_{\ell'}-\ell'
\end{equation}
is solvable with respect to $\ell'$ for a given
$q\in\mathbb{N}$ and for all $\ell \in \{1,\dots,g\}$ such that
$ \pi_{\ell}-\ell-q+g>0$, then $S_{n,s}$ does not
depend on $p_q$.\label{le-null} \end{lemma}

\begin{proof}
Note that, under the conditions of the lemma, the
sum $\sum_{j=1}^g e_{\pi_{\ell}-\ell-q+j} M_{\ell,j}$ is equal
to $S_{n,s}\delta_{\ell,\ell'}$. Then the double sum \eqref{d-sum}
is equal to $S_{n,s}\sum_{i=1}^g\delta_{i,i'}$ and vanishes
for a nonzero $q$, which proves that $S_{n,s}$ does not depend on
$p_q$.
\end{proof}

Let us return to proof of the theorem.

Let us rewrite Eq.~\eqref{null} in terms
of the Weierstrass sequence~$\bW_{\!n,s}$. We obtain
$w_{j'}=w_{j}-q$, where $j=g+1-\ell$ and $j'=g+1-\ell'$.
By assertion~(3) of Lemma~\ref{w-props}
we have $w_g-w_i\notin\bW_{\!n,s}$, and hence
the conditions of Lemma \ref{le-null} do not hold
for any $q=w_i\in\bW_{\!n,s}$. On the other hand,
by assertion~(4) of Lemma~\ref{w-props},
the conditions of Lemma~\ref{le-null} hold
for any $q\notin\bW_{\!n,s}$, and hence
we have $\mathcal{D}(p_q)S_{n,s}\equiv 0$
for any $p_q$ with such a subscript $q$ .
\end{proof}

Let $\bW_{\!n,s}=\{w_1,\dots,w_g\}$ be a
Weierstrass sequence as above.

\begin{definition}
A polynomial $\sigma_{n,s}(z_1,\dots,z_g)$ in $g$ variables that
exists by Theorem \ref{S-W-P} and is given by the identity
\begin{equation}
\sigma_{n,s}(p_{w_1},\dots,p_{w_g})\equiv S_{n,s},
\end{equation}
is called a \textit{Schur--Weierstrass polynomial}.
\end{definition}
Any Schur--Weierstrass polynomial is homogeneous
with respect to the natural grading of the ring $\Lambda$ in which
$\deg_{\Lambda}(e_r)=\deg_{\Lambda}(p_r)=r$ for $r>0$ and
$\deg_{\Lambda}(e_r)=\deg_{\Lambda}(p_r)=0$ for $r\le 0$.
In this grading we have $\deg_{\Lambda}(z_{\ell})=w_{\ell}$ and,
as follows from formula \eqref{schur}, the weight
of the polynomial $\sigma_{n,s}(z_1,\dots,z_g)$ is equal to the
sum of the elements of the partition
$\boldsymbol{\pi}_{n,s}$. We have
\begin{align*}\notag
\deg_{\Lambda}(\sigma_{n,s})&=\sum_{\ell=1}^{g}\pi_{\ell}
=\sum_{k=1}^g(w_{k}-k+1)\\&=-\frac{g(g-1)}{2}+
\sum_{i=1}^{n-1}\sum_{j=1}^{[is/n]}(is-jn)=
\frac{(n^2-1)(s^2-1)}{24}\,.
\end{align*}

\begin{lemma} The weight of the Schur--Weierstrass polynomial
$\sigma_{n,s}(z_1,\dots,z_g)$ is equal to its degree with respect to the
variable $z_1$, $\deg_{\Lambda}(\sigma_{n,s})=\deg_{z_1}(\sigma_{n,s})$.
\end{lemma}

\begin{proof} It follows from relation \eqref{ptoe} that
$e_r=p_1^r/r!+\dots$.
Further, by formula \eqref{schur} we have
$S_{n,s}=p_1^{\gamma}\det(\chi(w_{i}-j+1))_{1\le i,j\le g} +\dots$,
where $\gamma={\sum_{i=1}^{g}\pi_i}=\deg_{\Lambda}(\sigma_{n,s})$
and $\chi(n)=1/n!$ for $n\ge 0$ and $\chi(n)=0$ for $n<0$.

It remains to prove that the determinant
$\det(\chi(w_{i}-j+1))_{1\le i,j\le g}$ is nonzero.
Consider the Wronskian
$\mathcal{W}_t(t^{w_1},t^{w_2},\dots,t^{w_g})$ of the system of
monomials $\{t^{w_1},t^{w_2},\dots,t^{w_g}\}$, which
is obviously nondegenerate. The immediate calculation gives
\begin{equation}\notag
t^{(n^2-1)(s^2-1)/24}\det(\chi(w_i-j+1))_{1\le i,j \le g} =
\frac{\mathcal{W}_t(t^{w_1},\dots,t^{w_g})}{\prod_{i=1}^g{w_i!}}\,.
\end{equation}

This proves the lemma.
\end{proof}

\begin{example} Let us give the explicit expressions for the
Schur--Weierstrass polynomials with $g=1,2,3,4$.

For $g=1$ and $2$ we have
$\sigma_{2,3}=z_1$ and $\sigma_{2,5}=\frac{1}{3}(z_1^3-z_2)$,
respectively.

For $g=3$ we have
$\sigma_{2,7}=\frac{1}{45}(z_1^6-5z_1^3z_2+9z_1z_3-5z_2^2)$ and
$\sigma_{3,4}=\frac{1}{20}(z_1^5 - 5 z_1 z_2^2 + 4 z_3)$.

For $g=4$ we have $\sigma_{2,9}=\frac{1}{4725}(z_1^{10}-15 z_1^7z_2
+63 z_1^5 z_3-225 z_1^3 z_4+315 z_1^2 z_2 z_3-175 z_1 z_2^3
+9(-21 z_3^2+ 25 z_2 z_4))$ and
$\sigma_{3,5}=\frac{1}{448}(z_1^8-14 z_1^4 z_2^2+
56 z_1^2 z_2 z_3-64 z_1 z_4-\allowbreak
7( z_2^2-2 z_3)(z_2^2+2 z_3))$.
\end{example}

\begin{theorem}\label{invol}
Let $\imath\colon\mathbb{C}^g\to\mathbb{C}^g$,
$\boldsymbol{\xi}\mapsto-\boldsymbol{\xi}$, be the
canonical involution. In this case,
\begin{equation}\notag
\imath(\sigma_{n,s}(\boldsymbol{\xi}))=
\sigma_{n,s}(-\boldsymbol{\xi}) =(-1)^{(n^2-1)(s^2-1)/24}
\sigma_{n,s}(\boldsymbol{\xi}).
\end{equation}
\end{theorem}

\begin{proof} Consider the involutions $\omega$ and
$\widehat{\omega}$ on the
ring $\Lambda_{\mathbb{Q}}$ that are defined on the
multiplicative generators by the
formulas
\begin{equation}\notag
\omega(p_r)=(-1)^{r-1}p_r\quad\text{and}\quad
\widehat{\omega}(p_r)=(-1)^{r}p_r.
\end{equation}
As is known,
$\omega(s_{\boldsymbol{\rho}})=s_{\boldsymbol{\rho}'}$ for any Schur
function \cite[p.~42]{Mac85}. Since
$\widehat{\omega}(e_{r})=(-1)^re_{r}$, we have
$\widehat{\omega}(s_{\boldsymbol{\rho}})=(-1)^{\sum
\rho'_i}s_{\boldsymbol{\rho}}$ by \eqref{schur}.
The desired assertion follows from the fact that the
involution $\imath$ is the
composition $\widehat{\omega}\cdot \omega$.
\end{proof}

For an arbitrary Weierstrass sequence $\bW_{\!n,s}=\{w_1,\dots,w_g\}$,
for the corresponding Schur--Weierstrass
polynomial $\sigma_{n,s}(z_1,\dots,z_g)$, and for a chosen value of
the vector $\boldsymbol{z}=(z_1,\dots,z_g)\in\mathbb{C}^g$, we
introduce the following polynomial in $x\in \mathbb{C}$:
\begin{equation}\notag
R_{n,s}(x;\boldsymbol{z})=\sigma_{n,s}(z_1+x^{w_1},\dots,z_g+x^{w_g}).
\end{equation}

\begin{theorem}
For a given $\boldsymbol{\xi}\in \mathbb{C}^g$, the polynomial
$R_{n,s}(x;\boldsymbol{\xi})$ either does not depend on
$x$ or has at most $g$ roots.
\end{theorem}

\begin{proof}
Let us consider $\sigma_{n,s}(\xi_1,\dots,\xi_g)$ as a function
belonging to the ring
of symmetric functions $\Lambda$ in formal variables
$x_1,\dots,x_N$ with $N$ sufficiently large
(for instance, the number $2g-1$ is sufficiently large indeed because,
in this case, for any $\boldsymbol{\xi}\in \mathbb{C}^g$, the
system of equations
$\xi_{\ell}=\sum_{i=1}^{2g-1}x_{i}^{w_{\ell}}$ with respect to
$x_1,\dots,x_{2g-1}$ is solvable). Denote the function
thus defined in the basis $e_r$ by
$\mathcal{S}\,(e_1,e_2,\dots) =\sigma_{n,s}(\boldsymbol{\xi})$, and
we have expression \eqref{schur} for the function $\mathcal{S}$.
In this representation, the
polynomial $R_{n,s}(x;\boldsymbol{\xi})$ is the same function,
which is however calculated on the extended set of variables
$x,x_1,\dots,x_N$. Let us use the property of the elementary symmetric
polynomials given by $e_i(x,x_1,x_2,\dots)
=xe_{i-1}(x_1,x_2,\dots)+e_i(x_1,x_2,\dots)$ to express
$R_{n,s}$ via $\mathcal{S}$. We obtain
\begin{equation}
R_{n,s}(x;\boldsymbol{\xi})=\mathcal{S}\,(x+e_1,x e_1+e_2,\dots) =\det(x
e_{\pi_i-i+j-1}+e_{\pi_i-i+j})_{1\le i,j \le g}.
\label{shat}
\end{equation}
The determinant in \eqref{shat} is a polynomial of degree not higher
than $g$, which proves the theorem.
\end{proof}

Let us represent the polynomial $R_{n,s}(x;\boldsymbol{\xi})$
in the form
\begin{equation}\notag
R_{n,s}(x;\boldsymbol{\xi})=\sum_{\ell=0}^{g}
R_{\ell}(\boldsymbol{\xi}) x^{\ell}.
\end{equation}
Among the polynomials $R_{\ell}(\boldsymbol{\xi})$, there are
relations
\begin{equation}\notag
R_{r}(\boldsymbol{\xi})=\frac{1}{r}\sum_{q=1}^gw_q\,
\frac{\partial}{\partial \xi_{q}}\,R_{r-w_q}(\boldsymbol{\xi}),\qquad
r=1,\dots,3g-1,
\end{equation}
where we set
$R_{i}(\boldsymbol{\xi})=0$ for $i<0$ and for $i>g$. The first $g$
relations in this list lead to expressions for the polynomials
$R_{r}(\boldsymbol{\xi})$ with $r>0$ in terms of the
derivatives of the polynomial
$R_{0}(\boldsymbol{\xi})=\sigma_{n,s}(\boldsymbol{\xi})$, and
the other $2g-1$ relations form a system of
linear differential equations such that
$\sigma_{n,s}(\boldsymbol{\xi})$ is a solution of this system.

Let us introduce the special notation
$\widehat{\sigma}_{n,s}(\boldsymbol{\xi})$ for
the coefficient $R_{g}(\boldsymbol{\xi})$ of the polynomial
$R_{n,s}(x;\boldsymbol{\xi})$. As follows from the expansion
\begin{equation}\notag
\det(x e_{\pi_i-i+j-1}+e_{\pi_i-i+j})
=x^g\cdot\det(e_{\pi_i-i+j-1})+\dots,
\end{equation}
the polynomial $\widehat{\sigma}_{n,s}(\boldsymbol{\xi})$ corresponds to
the Schur
polynomial $s_{\widehat{\boldsymbol{\pi}}_{n,s}}$, where the partition
$\widehat{\boldsymbol{\pi}}_{n,s}$ is $\{\pi_{2},\dots,\pi_g\}$,
in the same sense in which the
Schur--Weierstrass polynomial $\sigma_{n,s}(\boldsymbol{\xi})$
corresponds to the Schur polynomial $s_{\boldsymbol{\pi}_{n,s}}$.
By construction, the
polynomial $\widehat{\sigma}_{n,s}(\boldsymbol{\xi})$ does not depend on
$\xi_g$. We can readily prove this fact by
using formula \eqref{ptoe}. Indeed, the
dependence on $p_{w_g}$ can be caused only by means of the function
$e_{w_g}$, which does not enter into the expression for
$\widehat{\sigma}_{n,s}(\boldsymbol{\xi})$. The weight
$\deg_{\Lambda}(\widehat{\sigma}_{n,s})$ is equal to
$\deg_{\Lambda}(\sigma_{n,s})-g$.

We also note (see the proof of Theorem
\ref{invol}) that the involution $\widehat{\omega}$ acts by the rule
$\widehat{\omega}(\widehat{\sigma}_{n,s}(\boldsymbol{\xi}))
=(-1)^{\deg_{\Lambda}(\widehat{\sigma}_{n,s})}
\widehat{\sigma}_{n,s}(\boldsymbol{\xi})$, while the involution $\omega$
can be represented as the composition
$\omega=\widehat{\omega}\cdot\imath$,
\begin{equation}
{\omega}(\widehat{\sigma}_{n,s}(\boldsymbol{\xi}))
=(-1)^{\deg_{\Lambda}(\widehat{\sigma}_{n,s})}
\widehat{\sigma}_{n,s}(-\boldsymbol{\xi}).
\label{omega}
\end{equation}
We finally obtain
\begin{equation}
R_{n,s}(x;-\boldsymbol{\xi})=
(-1)^{(n^2-1)(s^2-1)/24}{\omega}(R_{n,s}(-x;\boldsymbol{\xi})).
\label{om-R}
\end{equation}

\begin{example} Let us study the case $n=3$, $s=4$. We have
\begin{equation}
\widehat{\sigma}_{3,4}(\boldsymbol{z})=\tfrac{1}{2}(z_1^2-z_2),\quad
R_{3,4}(x;\boldsymbol{z})=
\widehat{\sigma}_{3,4}(\boldsymbol{z})(x^3+z_1
x^2+\tfrac{1}{2}(z_1^2+z_2)x)+ \sigma_{3,4}(\boldsymbol{z}).\notag
\end{equation}
On the two-dimensional subspace $(\widehat{\sigma})$ in
$\mathbb{C}^3$ given by the parametrization
$\boldsymbol{z}(\eta,\xi)=(\eta,\eta^2,\xi)$, the polynomial
$\widehat{\sigma}_{3,4}$ is identically zero and the polynomial
$R_{3,4}$ does not depend on $x$,
\begin{equation}\notag
R_{3,4}(x;\eta,\eta^2,\xi)=\tfrac{1}{5}(\xi-\eta^5).
\end{equation}
Finally, on the one-dimensional subspace
$(\widehat{\sigma})_{\mathrm{sing}}\subset
(\widehat{\sigma})\subset\mathbb{C}^3$ defined by the equation
$\boldsymbol{z}(\eta)=(\eta,\eta^2,\eta^5)$, the polynomial
$R_{3,4}(x;\boldsymbol{z}(\eta))$ is identically zero. Thus,
on the set $\mathbb{C}^3\backslash(\widehat{\sigma})$, the polynomial
$R_{3,4}(x;\boldsymbol{z}(\eta))$ is nondegenerate and has three roots.
\end{example}

\section{Inversion problem for the rational Abel mapping}

Consider the graded ring of polynomials
$\mathbb{C}[X,Y]$, where $\deg(X)=n$ and $\deg(Y)=s$. A polynomial of
the form
\begin{equation}
f(X,Y;\lambda_0,\lambda_n,\dots,\lambda_{i
n+js},\dots)=Y^n-X^s-\sum_{\substack{0\le\alpha<s-1\\
0 \le \beta<n-1}}
\lambda_{\alpha n+\beta s}X^{\alpha}Y^{\beta} \label{ns-pol}
\end{equation}
is called an \textit{$(n,s)$-polynomial} if
$\deg(f(X,Y;\boldsymbol{\lambda}))\le ns$. For
$0\le r<n s$,~the representation of a number $r$ in the form
$\alpha n+\beta s$ is unique. Therefore, an $(n,s)$-polynomial depends on
the collection $\boldsymbol{\lambda}=\{\lambda_{\alpha n+\beta s}\}$,
$0\le\alpha<s-1$, $0\le\beta<n-1$, $\alpha n+\beta s <ns$, which
consists of $d_{n,s}=ns -g-([s/n]+1)-1=(n+1)(s+1)/2-[s/n]-3$
parameters, the so-called \textit{moduli}. In the general
case we may assume that these parameters belong
to a graded ring, and if we set
$\deg(\lambda_r)=ns-r$, then the
$(n,s)$-polynomial $f(X,Y;\boldsymbol{\lambda})$
becomes homogeneous of degree $ns$ with respect to
$X$, $Y$, and $\boldsymbol{\lambda}$.

For the case in which $\boldsymbol{\lambda}$ is a set of complex numbers,
an $(n,s)$-polynomial is said to be \textit{nondegenerate} if
$$
\Delta_{X}(\Delta_{Y}(f(X,Y;\boldsymbol{\lambda})))\ne0,
$$
where $\Delta_{t}$ stands for the discriminant with respect to the $t$.

If $f(X,Y;\boldsymbol{\lambda})$ is a nondegenerate
$(n,s)$-polynomial, then the algebraic variety
\begin{equation}
V(X,Y)=\{(X,Y)\in\mathbb{C}^2:f(X,Y;\boldsymbol{\lambda})=0\}\label{cur}
\end{equation}
in $\mathbb{C}^2$ is a nonsingular affine model of a plane algebraic curve $V$ of
genus $g$ without multiple points, which realizes
an $n$-sheet covering over~$\mathbb{C}$.

On the $r$th symmetric power of the Riemann surface of the curve
$V$ we define the so-called Abel mapping
$\mathcal{A}\colon(V)^r\to\mathrm{Jac}(V)$, where
$\mathrm{Jac}(V)$ stands for the Jacobi variety of the curve $V$\!.
This mapping is defined by the holomorphic integrals
\begin{equation}
u_i=(-\alpha_i n+\beta_i s)\sum_{k=1}^{r}\int_{(X_k,Y_k)}^{\infty}
X^{\alpha_i-1}Y^{n-\beta_i-1}\,\frac{\mathrm{d}X}{f_{Y}},
\qquad i=1,\dots,g,\label{abel}
\end{equation}
where $f_{Y}=\partial f(X,Y;\boldsymbol{\lambda})/\partial Y$,
and the pair of integers $\alpha_i$ and $\beta_i$ represents the $i$th
element $-\alpha_i n+\beta_i s=w_i$ of the
Weierstrass sequence $\bW_{\!n,s}$.

Let us introduce the space $\mathcal{L}_{n,s}$ of all nondegenerate
$(n,s)$-polynomials. This space is a subspace in
$\mathbb{C}^{d_{n,s}}$, and it is the complement to the algebraic variety
formed by the points $\boldsymbol{\lambda}\in\mathbb{C}^{d_{n,s}}$ such that
$\Delta_{X}(\Delta_{Y}(f(X,Y;\boldsymbol{\lambda})))=0$.

On the space $\mathcal{L}_{n,s}$, the group $\mathbb{C}^{*}$ of nonzero complex
numbers acts by the transformations
$\tau(\boldsymbol{\lambda})=\{\lambda_k(\tau)\}=\{\tau^{ns-k}\lambda_k\}$
(the action is defined by the grading),
where $\tau\in\mathbb{C}^{*}$. Using this action, we define the rational curve
\begin{equation}\notag
V_0=\lim_{\tau\to0}\Big\{(X,Y)\in\mathbb{C}^2:\lim_{\tau\to 0}
f(X,Y;\tau(\boldsymbol{\lambda}))=Y^n-X^s=0\Big\}
\end{equation}
as the canonical limit of any nondegenerate curve
associated with an $(n,s)$-polynomial
$f(X,Y;\boldsymbol{\lambda})$ with a set of parameters
$\boldsymbol{\lambda}\in\mathcal{L}_{n,s}$.

In the canonical limit, the Abel mapping
\eqref{abel} passes to the limit mapping $\mathcal{A}_{0}$.
Let us introduce a parametrization
$\mathbb{C}\to V_0$ by the formula $t\mapsto(t^{-n},t^{-s})$.
In this parametrization, the mapping $\mathcal{A}_{0}$ reduces to the
integrals of the form
\begin{equation}
\begin{split}
\int_{0}^{\xi}
X(t)^{\alpha_i-1}Y(t)^{n-\beta_i-1}X'(t)\,\frac{\mathrm{d}t}{nY(t)^{n-1}}
&=-\int_{0}^{\xi}t^{-(\alpha_i-1)n-s(n-\beta_i-1)-(n+1)+(n-1)s}
\,\mathrm{d}t\\
&=-\int_{0}^{\xi}t^{w_i-1}\,\mathrm{d}t=-\frac{1}{w_i}\,\xi^{w_i}.
\end{split}\notag
\end{equation}

\begin{definition} For a Weierstrass sequence
$\bW_{\!n,s}\,{=}\,\{w_1,\dots,w_g\}$, we define the mapping
$A_{n,s}\colon(\mathbb{C})^r\to\mathbb{C}^g$, the so-called
\textit{rational analog of the Abel mapping}, by the formula
\begin{equation}\notag
A_{n,s}(x_1,\dots,x_r)=\bigg(\sum_{j=1}^rx_j^{w_1},\dots,
\sum_{j=1}^r x_j^{w_g}\bigg).
\end{equation}
\end{definition}

For the mapping $A_{n,s}$, the following analog of the Abel theorem holds.

\begin{theorem}
\label{abel-T}
Let $\{x_1,\dots,x_{N}\}\in(\mathbb{C})^N$\!, where $N\ge g$, be a collection
such that $x_i\ne x_j$ for all $1\le i\ne j\le N$ and
$\widehat{\sigma}_{n,s}(A_{n,s}(x_1,\dots,x_{N}))\ne0$. Then
there exists a set $\{t_1,\dots,t_{g}\}\in(\mathbb{C})^g$ such that
\begin{equation}\notag
A_{n,s}(x_1,\dots,x_{N})+A_{n,s}(t_1,\dots,t_{g})=0.
\end{equation}
\end{theorem}

\begin{proof}
Let $\boldsymbol{Q}_{N}=\{q_0,\dots,q_{N}\}$ be the
complement of the Weierstrass
sequence $\bW_{\!n,s}=\{w_1,\dots,w_g\}$ in the set $\{0,1,2,\dots,N+g\}$.
Denote by $\boldsymbol{b}(x)$ the column vector of the form
$((x^{N+g-q_j})_{j=0,\dots, N})^T$\!. Let us consider the polynomial
\begin{equation}
\varphi(x)=\det(\boldsymbol{b}(x),
\boldsymbol{b}(x_1),\boldsymbol{b}(x_2),\dots,\boldsymbol{b}(x_N))
=\varphi_0 x^{N+g}+\varphi_1
x^{N+g-1}+\dots+\varphi_{N+g-1}x+\varphi_{N+g},
\label{comp}
\end{equation}
in which the coefficients $\varphi_{w_i}$
are zero for all $i=1,\dots,g$ by construction.

Let us show first that $\varphi_0$ is nonzero.
We set $w_j=0$ for $j<1$ and $\pi_j=0$ for $j>g$; then
$N+g-q_i=N+g-(2g-1-w_{g-i})=N+\pi_{i+1}-i$ for all $i=0,\dots,N+g$.
We have $\varphi_0=|x_j^{\pi_{i+1}+N-i}|_{i,j=1,\dots,N}$.
Hence, by the direct definition of the Schur polynomials
\cite[p.~40]{Mac85}, we obtain \begin{equation}\notag
\varphi_0=\widehat{\sigma}_{n,s}(A_{n,s}(x_1,\dots,x_{N}))
\prod_{1\le i<j\le N}(x_i-x_j),
\end{equation}
and therefore $\varphi_0\ne0$ under the assumptions of the theorem.

Thus, the equation $\varphi(x)=0$ has $N+g$ roots,
$N$ of which form the set $\{x_1,\dots,x_{N}\}$;
we denote the other $g$ roots by $\{t_1,\dots,t_{g}\}$.

The polynomial
\begin{equation}\notag
\widehat{E}(x)=\frac{x^{N+g}}{\varphi_{0}}\,
\varphi\bigg(\!\!-\frac{1}{x}\bigg)= 1+\sum_{i\ge 1}(-x)^i\,
\frac{\varphi_i}{\varphi_{0}}
\end{equation}
is the generating function of the elementary symmetric functions of
$\{x_1,\dots,x_{N},t_1,\dots,t_{g}\}$.
Consider the generating function of the power sums
of these quantities,
\begin{equation}\notag
\widehat{P}(-x)=\frac{\widehat{E}'(x)}{\widehat{E}(x)}=
\sum_{i\ge 1}(-x)^{i-1} \psi_i.
\end{equation}
Let us show that the coefficients $\psi_{w_i}$ vanish for all
$i=1,\dots,g$. Indeed, as is known,
there is the following expression for
$\psi_k$ in terms of $\{\varphi_0,\dots,\varphi_k\}$ (see \cite{Mac85}):
\begin{equation}\notag
\psi_k=\frac{1}{k \varphi_0^{k}}
\begin{vmatrix}
\varphi_1&\varphi_0&0&0&\hdots&0\\
2\varphi_2&\varphi_1&\varphi_0&0&\hdots&0\\
\hdotsfor{6}\\
\hdotsfor{6}\\
(k-1)\varphi_{k-1}&\varphi_{k-2}&\hdotsfor{3}&\varphi_{0}\\
k \varphi_k&\varphi_{k-1}&\hdotsfor{3}&\varphi_{1}\\
\end{vmatrix}.
\end{equation}
Hence, $\psi_{w_i}$ is a sum with rational coefficients
of products of the form
$$
\dfrac{\varphi_{\rho_{1}}}{\varphi_{0}}\cdot
\dfrac{\varphi_{\rho_{2}}}{\varphi_{0}}
\cdots\dfrac{\varphi_{\rho_{\ell}}}{\varphi_{0}}
$$
such that $\rho_1+\rho_2+\dots+\rho_{\ell}=w_i$.

\begin{lemma} If $w\in\bW_{\!n,s}$, then any partition
$\boldsymbol{\rho}=\{\rho_1,\dots,\rho_{\ell}\}$
of the number $w$ contains at least one element of the sequence
$\bW_{\!n,s}$.
\end{lemma}

\begin{proof}
This follows from the fact that, in
the set of positive integers, the complement of a Weierstrass sequence
is a semigroup.
\end{proof}

According to the lemma, since
$\varphi_{w_1}=\varphi_{w_2}=\dots=\varphi_{w_i}=0$, we have
$\psi_{w_i}=0$, that is,
\begin{equation}
\sum_{j=1}^{N}x_j^{w_i}+
\sum_{k=1}^{g}t_k^{w_i}=0,\qquad i=1,\dots,g.
\end{equation}
\end{proof}

Let us show that the condition $x_i\ne x_j$ in the assumption of Theorem
\ref{abel-T} is not restrictive.

\begin{lemma}\label{constraint} Let $\mathcal{K}^{\;N}$ be the subspace
of vectors in $\mathbb{C}^N$ that have no equal coordinates.
Consider the composition $\mathcal{K}^{\;N}\xrightarrow{\imath}
\mathbb{C}^N \xrightarrow{\mathcal{N}}
\mathbb{C}^N\xrightarrow{\gamma_{\ell}} \mathbb{C}^{N-1}$, where $\imath$ is
the embedding,
$\mathcal{N}\,(x_1,\dots,x_N)=(z_1,\dots,z_N)$,
$z_k=\sum_{q=1}^N x_q^k$, and
$\gamma_{\ell}(z_1,\dots,z_N)=(z_1,\ldots,z_{\ell-1},z_{\ell+1},
\ldots,z_N)$. The composition $\gamma_{\ell}\circ\mathcal{N}\circ
\imath$ is onto if and only if either
$N= 0$ or $N=1\mod\ell$.
\end{lemma}

\begin{proof}
The preimage of a point $\boldsymbol{\xi}\in \mathbb{C}^{N-1}$
under the mapping $\gamma_{\ell}\circ\mathcal{N}$\, is the
set of roots of the polynomials (see, for instance, \cite{br98})
\begin{equation}\notag
P(t)=\begin{vmatrix}
\xi_1&1&0&\hdots&0&0\\
\xi_2&\xi_1&2&\hdots&0&0\\
\hdotsfor{6}\\
z_{\ell}&\xi_{\ell-1}&\xi_{\ell-2}&\hdots&0&0\\
\xi_{\ell+1}&z_{\ell}&\xi_{\ell-1}&\hdots&0&0\\
\hdotsfor{6}\\
\xi_N&\xi_{N-1}&\xi_{N-2}&\hdots&\xi_{1}&N\\
t^{N}&t^{N-1}&t^{N-2}&\hdots&t&1
\end{vmatrix}
\end{equation}
for all possible values $z_{\ell}\in \mathbb{C}$. If
$N\ne 0$, $1$ $\mathrm{mod}\,\ell$, then any point of the preimage
of the point $\boldsymbol{0}\in \mathbb{C}^{N-1}$ has a pair of equal
coordinates. If $N$ is either $0$ or $1$ $\mathrm{mod}\,\ell$, then the
discriminant of the polynomial $P(t)$ is a polynomial in $z_{\ell}$
of degree $N(N-1)/\ell$ for any value of
$\boldsymbol{\xi}\in\mathbb{C}^{N-1}$, i.e., it vanishes only for $N(N-1)/\ell$
particular values of $z_{\ell}$. \end{proof}

Consider the Jacobian of the mapping $A_{n,s}\colon(\mathbb{C})^g\to\mathbb{C}^g$:
\begin{equation}\notag
\frac{\partial (z_1,\dots,z_g)}{\partial (t_1,\dots,t_g)}=
\prod_{k=1}^g w_k\cdot
|t_j^{w_i-1}|_{1\le i,j\le g}=\prod_{k=1}^g w_k \cdot
|t_j^{(\pi_{\ell}-1)-\ell+g}|_{1\le \ell,j\le g}.
\end{equation}
It follows from the direct definition of the Schur polynomials
\cite[p.~40]{Mac85} that
\begin{equation}\notag
|t_j^{(\pi_{\ell}-1)-\ell+g}|_{1\le \ell,j\le g}
=\omega(\widehat{\sigma}_{n,s}(A_{n,s}(t_1,\dots,t_g)))
\prod_{1\le k<\ell\le g}(t_k-t_{\ell}).
\end{equation}
Hence, applying formula \eqref{omega}, we obtain the following expression
for the Jacobian:
\begin{equation}\notag
\frac{\partial (z_1,\dots,z_g)}
{\partial(t_1,\dots,t_g)}=(-1)^{\frac{(n^2-1)(s^2-1)}{24}-g}
\prod_{j=1}^gw_j\cdot\widehat{\sigma}_{n,s}
(-A_{n,s}(t_1,\dots,t_g))\prod_{1\le k<\ell\le g}(t_k-t_{\ell}).
\end{equation}

Let us show that the Schur--Weierstrass polynomial
of the form $$
\sigma_{n,s}(A_{n,s}(x)-\boldsymbol{\xi})=R_{n,s}(x;-\boldsymbol{\xi}) $$
leads to the solution of the inversion problem for the mapping
$A_{n,s}$. Set
$(\widehat{\sigma}_{n,s})=\{\boldsymbol{\xi}\in\mathbb{C}^g:
\widehat{\sigma}_{n,s}(-\boldsymbol{\xi})=0\}$
and introduce the mapping
$B_{n,s}\colon\mathbb{C}^g\backslash(\widehat{\sigma}_{n,s})\to(\mathbb{C})^g$
that sends a vector $\boldsymbol{z}\in\mathbb{C}^g
\backslash(\widehat{\sigma}_{n,s})$ into the set of roots
$\{x_1,\dots,x_g\}$ of the polynomial $R_{n,s}(x; -\boldsymbol{z})$.

\begin{theorem}\label{direct}
The mapping $B_{n,s}$ solves the inversion problem
of the mapping $A_{n,s}$, that is, for all
$\boldsymbol{\xi}\in\mathbb{C}^g\setminus(\widehat{\sigma}_{n,s})$
the relation $A_{n,s}(B_{n,s}(\boldsymbol{\xi}))=\boldsymbol{\xi}$ holds.
\end{theorem}

\begin{proof}
The representation \eqref{shat} of the polynomial
$R_{n,s}(x;\boldsymbol{\xi})$ can be reduced to the following form:
\begin{equation}\notag
R_{n,s}(x; \boldsymbol{\xi})=
\begin{vmatrix} e_{w_1}&e_{w_2}&\dots&e_{w_g}&(-x)^{g}\\
e_{w_1-1}&e_{w_2-1}&\dots&e_{w_g-1}&\quad (-x)^{g-1}\\
\hdotsfor{5}\\
\,e_{w_1+1-g}&e_{w_2+1-g}&\dots&e_{w_g+1-g}&-x\\
e_{w_1-g}&e_{w_2-g}&\dots&e_{w_g-g}&1
\end{vmatrix}.
\end{equation}
Hence, applying the transformation \eqref{om-R},
we obtain
\begin{equation}\notag
R_{n,s}(x;- \boldsymbol{\xi})=
(-1)^{(n^2-1)(s^2-1)/24}
\begin{vmatrix} h_{w_1}&h_{w_2}&\dots&h_{w_g}&x^{g}\\
h_{w_1-1}&h_{w_2-1}&\dots&h_{w_g-1}&\quad x^{g-1}\\
\hdotsfor{5}\\
\,h_{w_1+1-g}&h_{w_2+1-g}&\dots&h_{w_g+1-g}&x\\
h_{w_1-g}&h_{w_2-g}&\dots&h_{w_g-g}&1
\end{vmatrix}.
\end{equation}
The functions $h_k$ are complete symmetric functions \cite{Mac85}, which
are the images of elementary symmetric functions $e_k$ under the action
of the involution $\omega$ and satisfy the relations
$\sum_{j=0}^k(-1)^je_jh_{k-j}=0$ for any $k>0$. Note that
$2g-2=w_g-w_1\notin\bW_{\!n,s}$. By Lemma \ref{constraint}, in
the case of $N=2g-1$ and $\ell=2g-2$, we can use the
result of Theorem \ref{abel-T}. Hence,
it is sufficient to restrict ourselves to the case
$\boldsymbol{\xi}=A_{n,s}(t_1,\dots,t_g)$, i.e., we may assume that
$e_{q}=0$ for $q>g$. Thus, $\sum_{j=0}^g(-1)^je_jh_{w_i-j}=0$,
$i=1,\dots,g$, and consequently
\begin{equation}
R_{n,s}(x;-\boldsymbol{\xi})=
\widehat{\sigma}(-\boldsymbol{\xi})\sum_{j=0}^g(-1)^je_j x^j=
\widehat{\sigma}(-\boldsymbol{\xi})\prod_{i=1}^g(x-t_i).\label{zeros}
\end{equation}
This proves the theorem.
\end{proof}

Formula \eqref{zeros} immediately implies the following assertion.

\begin{cor}\label{inver} For an arbitrary point
$\{x_1,\dots,x_g\}$ of the space $(\mathbb{C})^g$ we have
either $\widehat{\sigma}_{n,s}(-A_{n,s}(x_1,\dots,\allowbreak x_g))=0$ or
$B_{n,s}(-A_{n,s}(x_1,\dots,x_g))=\{x_1,\dots,x_g\}$.
\end{cor}

\section{Rational Riemann vanishing theorem}

Let a polynomial $P(\xi_1,\dots,\xi_g)$ in
$g$ variables be given. For every $\boldsymbol{z}\in\mathbb{C}^g$,
using a Weierstrass sequence
$\bW_{\!n,s}$ of length $g$,
to this polynomial we put into correspondence a polynomial in
$x$ of the form
$$
R_{P}(x,\boldsymbol{z})=P(z_1-x^{w_1},\dots,z_g-x^{w_g})=
P(\boldsymbol{z}-A_{n,s}(x)).
$$

\begin{definition}
We say that a polynomial $P(\xi_1,\dots,\xi_g)$ satisfies the
$(n,s)$-analog of the Riemann vanishing theorem for polynomials
if the polynomial $R_{P}(x,\boldsymbol{z})$ for
$\boldsymbol{z}=A_{n,s}(x_1,\dots,x_g)$ either has exactly $g$
roots $\{t_1,\dots,t_g\}=\{x_1,\dots,x_g\}$ or is identically zero.
\end{definition}

According to Corollary \ref{inver}, the Schur--Weierstrass polynomial
$\sigma_{n,s}$ introduced above satisfies the\break
\text{$(n,s)$-analog} of the Riemann vanishing theorem for polynomials.

\begin{theorem}\label{Rie}
If a polynomial $P$ satisfies the
$(n,s)$-analog of the Riemann vanishing theorem for polynomials,
then it is equal to the Schur--Weierstrass polynomial
$\sigma_{n,s}$ up to a constant factor.
\end{theorem}

\begin{proof} Applying Theorem \ref{direct},
we see that, for any $\boldsymbol{z} \in
\mathbb{C}^g\backslash(\widehat{\sigma}_{n,s})$, the polynomial
$R_{P}(x,\boldsymbol{z})$ either has exactly $g$ roots,
which coincide with the roots of
the polynomial $R_{n,s}(x,-\boldsymbol{z})$, or is
identically zero. Thus, on an open everywhere dense subset in $\mathbb{C}^g$ we
have the relation
\begin{equation}\notag
P(\boldsymbol{z})R_{n,s}(x,-\boldsymbol{z})=R_{P}(x,\boldsymbol{z})
\sigma_{n,s}(-\boldsymbol{z}),
\end{equation}
and therefore this relation holds for all
$\boldsymbol{z}\in\mathbb{C}^g$ and $x\in \mathbb{C}$. Using the parity properties
of the polynomial $\sigma_{n,s}$, we can rewrite this relation
in the form
\begin{equation}\notag
P(\boldsymbol{z})\sigma_{n,s}(\boldsymbol{z}-A_{n,s}(x))=
P(\boldsymbol{z}-A_{n,s}(x))\sigma_{n,s}(\boldsymbol{z}).
\end{equation}
Applying the induction on $q$ to
$\boldsymbol{\xi}=\sum_{k=1}^{q}A_{n,s}(x_k)$, we can see
from the last relation that
\begin{equation}\notag
P(\boldsymbol{z})\sigma_{n,s}(\boldsymbol{z}-\boldsymbol{\xi})=
P(\boldsymbol{z}-\boldsymbol{\xi})\sigma_{n,s}(\boldsymbol{z}),
\end{equation}
which obviously proves the theorem.
\end{proof}

Theorem \ref{Rie} has an important application
in the theory of Abelian functions on the Jacobi varieties
of curves of the form \eqref{cur}
associated with nondegenerate $(n,s)$-polynomials.

We recently developed an approach in the theory of Abelian functions
that generalizes the theory of Weierstrass elliptic functions
to higher genera. The function $\sigma(\boldsymbol{u};\boldsymbol{\lambda})$
associated with a nondegenerate curve of the form \eqref{cur} is an entire
function, which has a power series expansion
\begin{equation}\notag
\sigma(\boldsymbol{u};\boldsymbol{\lambda})
=S(\boldsymbol{u})+O(\boldsymbol{\lambda})
\end{equation}
in a neighborhood of $\boldsymbol{0}\in \mathrm{Jac}(V)$;
here $S(\boldsymbol{u})$ is a polynomial in
$u_1,\dots,u_g$ of weight
\text{$(n^2-1)(s^2-1)/24$}, and $O(\boldsymbol{\lambda})$ denotes
the rest part of the power series in which the
constants ${\lambda}_r$ raised to positive powers enter as factors.

The function $\sigma(\boldsymbol{u};\boldsymbol{\lambda})$
satisfies the Riemann vanishing theorem in the following form.
Let $(X,Y)\in V$ and
let $\boldsymbol{z}$ be a point in $\mathrm{Jac}(V)$; by using the Abel
mapping \eqref{abel}, we define the function
$$
\mathcal{R}((X,Y);\boldsymbol{z})=\sigma
(\mathcal{A}\,(X,Y)-\boldsymbol{z};\boldsymbol{\lambda}).
$$
In this case, $\mathcal{R}((X,Y);\boldsymbol{z})$ is
either identically zero or has exactly
$g$ roots $\{(X_i,Y_i)\}_{i=1,\dots,g}\in(V)^g$ such that
$\mathcal{A}\,((X_1,Y_1),\dots,(X_g,Y_g))=\boldsymbol{z}$.

For any set of parameters
$\boldsymbol{\lambda}\in\mathcal{L}_{n,s}$, the above canonical limit
leads to the function
$\sigma(\boldsymbol{u};\boldsymbol{0})=S(\boldsymbol{u})$.
Since the Abel mapping
$\mathcal{A}$\, passes to a rational mapping $A_{n,s}$ under this limit process,
we obtain the following result.

\begin{theorem} The canonical limit
of the Kleinian $\sigma$-function $\sigma(\boldsymbol{u};\boldsymbol{0})$
is equal to the Schur--Weierstrass polynomial
$\sigma_{n,s}(\boldsymbol{u})$ up to a constant factor.
\end{theorem}

\begin{proof} Consider the polynomial
$\sigma(\mathcal{A}_0(X,Y)-\boldsymbol{z};\boldsymbol{0})$ at some value
$\boldsymbol{z}\in\mathbb{C}^g$. We have
$\sigma(\mathcal{A}_0(X,Y)-\boldsymbol{z};\boldsymbol{0})=
S(A_{n,s}(x)-\boldsymbol{z})$,
where we used the parametrization ($X=x^{-n}$, $Y=x^{-s}$). The last
expression shows that the polynomial $\sigma(\boldsymbol{z};\boldsymbol{0})$
satisfies the $(n,s)$-analog of the
Riemann vanishing theorem for polynomials.
\end{proof}

Note that, in the canonical limit, the
function $\mathcal{R}((X,Y);\boldsymbol{z})$ either does not depend
on the parameter $x$
or is a polynomial in $x$ and has at most $g$ roots.


\chapter{Dynamical system on the $\sigma$-divisor}\label{chap:div}

Most of known finite-dimensional integrable systems of classical mechanics and mathematical physics 
are also algebraicaly completely integrable, i.e. their invariant tori can be extended to specific
complex tori, Abelian varieties, and the complexified flows are  straight-line flow on them. In certain 
cases, like the famous Neumann system describing the motion of a point on a sphere with a quadratic
potential, or the Steklov--Lyapunov integrable case of the Kirchhoff equations, the complex tori are Jacobians of hyperelliptic curves that genera equal to the dimension of associated Liouville tori.  

On the other hand  a number of dynamical systems of physical interest admit the spectral curve whic genus    
 is bigger than the dimension of the invariant tori, and the latter are
certain non-Abelian subvarieties (strata) of Jacobians. 
Algebraic geometrical properties of such systems and types of
singularities of their complex solutions were described in \cite{Vanh995, af00, epr03, fgu07, eekl993, eekt994}.
There are know a number of examples of systems related to strata of hyperelliptic Jacobians and only recently 
a new case of dynamic over strata of Jacobian of trigonal curve was considered \cite{bef12}

\section{Restriction to the sigma-divisor}
Write Jacobi equations as $x_2\rightarrow \infty$, $(x_2,y_2)=(x,y)$, i.e.
\begin{align*} \int_{\infty}^{(x,y)} \frac{\mathrm{d}x}{y}=u_1,\qquad  \int_{\infty}^{(x,y)} \frac{x\mathrm{d}x}{y}=u_2 \end{align*}
Now variables $u_1,u_2$ are related. Because
\begin{align*} &\mathrm{Lim}_{x_2\to\infty}  (x_1+x_2)=
\wp_{22}\left(  \int_{\infty}^{(x_1,y_1)}\mathrm{d}\boldsymbol{u}+ \mathrm{Lim}_{x_2\to\infty} \int_{\infty}^{(x_2,y_2)}\mathrm{d}\boldsymbol{u} \right)
 \\
&=\frac{ \sigma_{22}\sigma-\sigma_2^2 }{\sigma^2} = \infty    \end{align*}
The aforementioned dependence is given by the equation
\[  \sigma(\boldsymbol{u}) \equiv 0  \]
That is equation of {\bf $\sigma$-divisor }, i.e.
\[    \sigma\left( \int_{\infty}^P   \mathrm{d}\boldsymbol{u} \right) \equiv 0 \quad \forall P\in X  \]


In what follows we denote $\sigma$-divisor as $(\sigma)$, that is one-dimensional sub-variety in the Jacobi variety:
\[ (\sigma) \subset \mathrm{Jac}(X) \]

Represent $x_1$ coordinate
\begin{align*}
x_1&=\mathrm{Lim}_{x_2\to\infty}\frac{x_1x_2}{x_1+x_2}=
-\mathrm{Lim}_{x_2\to\infty}\frac{\wp_{12}(\boldsymbol{u})}{\wp_{22}(\boldsymbol{u})}\\&
=-\left.\frac{\sigma_{12}\sigma-\sigma_1\sigma_2}{\sigma_{22}\sigma-\sigma_2^2}\right|_{(\sigma)}=-\frac{\sigma_1(\boldsymbol{u})}
{\sigma_2(\boldsymbol{u})}
\end{align*}
Therefore
\[  x=-\left.\frac{\sigma_1(\boldsymbol{u})}
{\sigma_2(\boldsymbol{u})}\right\vert_{(\sigma)}  \]


We also derive similar expression for the $y$ coordinate of the curve we use the formula
\[-\frac{x_1^2y_2-x_2^2y_1}{y_1-y_2}=\frac{\wp_{112}(\boldsymbol{u})}{\wp_{222}(\boldsymbol{u})}\]
and approach to $(\sigma)$ in its both sides. In this limit
\begin{align*}
x_2=\frac{1}{\xi^2}, \quad y_2=\frac{2}{\xi^5}\left(1+\frac18\lambda_4\xi^2+\left(\frac18\lambda_3-\frac{1}{128}\lambda_4^2\right)\xi^4  \right)+O(\xi)\end{align*}
and the left hand side of this equality (LHS) expands as
\[ \mathrm{LHS}=x_1^2-\frac12y_1\xi+ \frac{1}{16}y_1\lambda_4\xi^3+ O(\xi^5)\]


Let us expand the right hand side, RHS. To do that we denote
\[ v_1=\int_{\infty}^{P_1}\mathrm{d}u_1, v_2=\int_{\infty}^{P_1}\mathrm{d}u_2,  \]
Further we expand
\begin{align*}
w_1=\left.\int_{\infty}^{P_2}\mathrm{d}u_1\right|_{x_2=1/\xi^2}=-\frac{\xi^3}{3}+\frac{1}{40}\lambda_4\xi^5+O(\xi^7)\\
w_2=\left.\int_{\infty}^{P_2}\mathrm{d}u_2\right|_{x_2=1/\xi^2}=-\xi+\frac{1}{24}\lambda_4\xi^3+O(\xi^5)
\end{align*}
Plug into RHS
\[  \sigma(u_1,u_2)= \sigma(v_1+w_1,v_2+w_2) \]
and expand in $\xi$. We get
\[  RHS= x_1^2-\frac{\xi}{\sigma_2(\boldsymbol{v})} \left( x_1^2\sigma_{2,2}(\boldsymbol{v})+2x\sigma_{1,2}(\boldsymbol{v})+\sigma_{1,1}(\boldsymbol{v})   \right)+O(\xi^3) \]

Equating coefficients in of linear term of expansions in the LHS and RHS we derive necessary result. We summarize the obtained formulae

The genus two curve is locally uniformazed by the functions $x$ and  $y$ defined on $\sigma$-divisor that are given as
\begin{align}\begin{split}
 x&=-\left.\frac{\sigma_1(\boldsymbol{u})}
{\sigma_2(\boldsymbol{u})}\right\vert_{(\sigma)}\\
 y&=\left. -\frac{1}{\sigma_2(\boldsymbol{u})} \left( \sigma_{11}(\boldsymbol{u})+x^2 \sigma_{22}(\boldsymbol{u}) +2x\sigma_{12}(\boldsymbol{u})\right) \right\vert_{(\sigma)}\end{split} \label{xy}
\end{align}
Further we present equivalent representation of the $y$ coordinate,
\[ y=\left.-\frac12\frac{\sigma(2\boldsymbol{u})}{\sigma_2(\boldsymbol{u})}\right|_{(\sigma)} \]


\section{Remark on the $\sigma$-series}
$(\sigma)$ is given by condition
\[ (\sigma)= \left\{ \boldsymbol{u} \left|  \int_{\infty}^{P}\frac{\mathrm{d}x}{y}=u_1,\quad  \int_{\infty}^{P}\frac{x\mathrm{d}x}{y}=u_2\right.   \right\}  \]
Expanding these conditions near point $x=1/\xi^2$ and keeping the lower terms in $\xi$ we get equation
\[ -\frac13 \xi^3=u_1,\qquad  -\xi=u_2     \]
Eliminating $\xi$ we obtain the following expansion in the vicinity $\boldsymbol{u}\sim \boldsymbol{0}$
\[  \sigma(\boldsymbol{u})=u_1-\frac13 u_2^2 + \text{ highr order terms} \]
Further terms of $\sigma$-series can be obtained as series with coefficients defined recursively (Buchstaber-Leykin, 2005)

Let us check that pole behavior of $x$ and $y$ at infinity  in virtue of formulae (\ref{xy}):

\begin{align*}
x&= - \frac{\partial/\partial u_1 \left( u_1-\frac13 u_2^3  \right)}{\partial/\partial u_2 \left( u_1-\frac13 u_2^3  \right) }=\frac{1}{u_2^2}=\frac{1}{\xi^2}  \\
y&=-\frac{1}{u_2^2}\left( 0-\frac{1}{u_2^4} 2u_2 +0 \right)=\frac{2}{u_2^5}=\frac{2}{\xi^5}
\end{align*}

\subsection{Schur polynomials}
Polynomial $u_1-1/3 u_2^3$ has a deep sense, that's the Schur function corresponding to the partition $(2,1)$ and Young Diagram
\[  \yng(2,1)\]
Recall definitions. For any partition: $$\lambda: \alpha_1\geq \alpha_2\geq \ldots \geq \alpha_n, \qquad
|\alpha| = \sum_{i=1}^n\alpha_n $$ the {\bf Schur polynomial} of $n$ variables $x_1,\ldots, x_n$ defined as
\[ s_{\lambda} =\det ( p_{\alpha_i-i+j}(\boldsymbol{x}) ) \]
where {\bf elementary Schur polynomials } $p_m(\boldsymbol{x})$ are generated by series
\[ \sum_{m=0}^{\infty} p_m(\boldsymbol{x})t^m  =\mathrm{exp} \left\{ \sum_{n=1}^{\infty} x_n t^n  \right\} \]

Few first Schur polynomials are
\begin{align*}
&s_1=x_1  & \yng(1)\\
&s_2=x_2+\frac12 x_1^2&\yng(2)\\
&s_{1,1}=-x_2+\frac12 x_1^2&\yng(1,1)\\
&s_3=x_3+x_1x_2+\frac16x_1^3&\yng(3)\\
&s_{2,1}=-x_3+\frac13x_1^3&\yng(2,1)\\
&s_{1,1,1}=x_3-x_1x_2+\frac16x_1^3&\yng(1,1,1)\\
&\vdots
\end{align*}

\subsection{Weierstrass gap sequences}
Consider the set of all monomials
\[ x^n y^m, \quad n,m \in \mathbb{Z}_+  \]
Order $N$ of the monomial $x^ny^m$ is the order of the pole at infinity, i.e $2n+5m$. We call a number $p\in\mathbb{Z}_+$ an {\bf non-gap number} if exists monomial of order $p$ and {\bf gap number} otherwise

{\bf Weierstrass gap sequence at $\infty$}
\begin{align*}
&\overbrace{\phantom{aaaaaaaaaaaa}}^{2g=4}\\
&\overbrace{\phantom{aaaaaaaa}}^{g=2}\phantom{aaaaa}\downarrow 2g=4\\
&\overline{0},\hskip0.3cm 1,\hskip0.3cm \overline{2}, \hskip0.3cm 3, \hskip0.3cm \overline{4,\hskip0.3cm 5,\hskip0.3cm 6, \; \hskip0.3cm 7,\ldots} \\
&\uparrow \hskip1.1cm\uparrow\hskip1cm \uparrow\hskip0.5cm   \uparrow\hskip0.5cm\uparrow\hskip0.6cm\uparrow\ldots\\
&\text{const}\hskip0.5cm x \hskip1cm x^2\hskip0.4cm y\hskip0.5cm x^3\hskip0.4cm yx\ldots
\end{align*}

{\bf Weierstrass gap theorem} Exists exactly $g$ gap numbers
\[ 1= w_1<w_2<\ldots< w_g < 2g   \]

These are 1 and 3 in our case
Denote partition numbers
\[   \alpha_1 \geq \alpha_2 \geq \ldots \geq \alpha_g  \]
Then
\[   \alpha_k = w_{g-k+1}+k-g, \qquad k=1,\ldots, g  \]

\section{Second kind integrals on $(\sigma)$}
The following formulae are valid:
\begin{align*}
\int_{P_0}^{P}\mathrm{d}r_2&=-\left.\frac{\sigma_{22}(\boldsymbol{u})}{2\sigma_2(\boldsymbol{u})}\right|_{(\sigma)=0}+c_2\\
\int_{P_0}^{P}\mathrm{d}r_1&=-\left.\frac{\sigma_{12}(\boldsymbol{u})}{\sigma_2(\boldsymbol{u})}\right|_{(\sigma)=0} -x\left.\frac{\sigma_{22}(\boldsymbol{u})}{2\sigma_2(\boldsymbol{u})}\right|_{(\sigma)=0}+c_1
\end{align*}
For the proof we use alredy derived representations
\begin{align*}
-\zeta_1(\boldsymbol{u})&=\int_{P_0}^{(x_1,y_1)} \frac{ 12 x_1^3 +2\lambda_4x_1^2+\lambda_3x_1  }{4y_1} \mathrm{d}x_1 +\int_{P_0}^{(x_2,y_2)} \mathrm{d}r_1\\&-\frac12 \frac{y_1-y_2}{x_1-x_2}+C_1\\
-\zeta_2(\boldsymbol{u})&=\int_{P_0}^{(x_1,y_1)} \frac{ x_1^2 }{y_1} \mathrm{d}x_1 +\int_{P_0}^{(x_2,y_2)} \mathrm{d}r_2+C_2
\end{align*}


We have expansions
\begin{align*}
\left.\int_{\infty}^{P}\mathrm{d}u_1\right|_{x=1/\xi^2=\infty}
&=-\frac13\xi^3+\frac{\lambda_4}{40}\xi^5+O(\xi^7)\\
\left.\int_{\infty}^{P}\mathrm{d}u_2\right|_{x=1/\xi^2=\infty}&=-\xi+\frac{\lambda_4}{24}\xi^3+O(\xi^5)\\
\left.\int_{P_0\neq\infty}^{P}\mathrm{d}r_1\right|_{x=1/\xi^2=\infty}&=\frac{1}{\xi^3}+\frac{\lambda_4}{8\xi}+O(\xi)\\
\left.\int_{P_0\neq\infty}^{P}\mathrm{d}r_2\right|_{x=1/\xi^2=\infty}&=\frac{1}{\xi}+\frac{\lambda_4}{8}\xi+O(\xi^3)
\end{align*}

And also
\begin{align*}
\zeta_i\left(\int_{\infty}^{P_1}\mathrm{d}u_1+\left.\int_{\infty}^{P_2}\mathrm{d}u_2\right|_{x_2=1/\xi^2}\right)=-\left.\frac{1}{\xi}\frac{\sigma_i}{\sigma_2}\right|_{(\sigma)}+\left.\frac{\sigma_{i,2}}{\sigma_2}\right|_{(\sigma)}-\left.\frac{\sigma_i\sigma_{22}}{2\sigma_2^2}\right|_{(\sigma)}
\end{align*}
Plug these to the above representations, poles in $\xi$ will cansel; the necessary result follows.

\subsection{Restriction of 5 Baker equations to $(\sigma)$}
The third order $\sigma$-derivatives restricted to $\sigma$-divisor are expressible in lower derivatives as
\begin{align*}
\sigma_{222}&=\frac34\frac{\sigma_{22}^2}{\sigma_2}+\frac{\lambda_4}{4}\sigma_2+\sigma_1\\
\sigma_{122}&=\frac{\sigma_{22}\sigma_{12}}{\sigma_2}-\frac14\frac{\sigma_{22}^2\sigma_1}{\sigma_2}-\frac{\sigma_1^2}{\sigma_2}+\frac14\lambda_4\sigma_1\\
\sigma_{112}&=\frac{\sigma_1^3}{\sigma_2^2}-\frac{\sigma_1\sigma_{12}\sigma_{22}}{\sigma_2^2}+\frac14\frac{\sigma_1^2\sigma_{22}^2}{\sigma_2^3}-\frac14\lambda_4\frac{\sigma_1^2}{\sigma_2}+\frac12\frac{\sigma_{11}\sigma_{22}}{\sigma_2}+\frac{\sigma_{12}^2}{\sigma_2}+\frac14\lambda_3\sigma_1
\\
\sigma_{111}&=-\frac34\frac{\lambda_0\sigma_2^2}{\sigma_1}+\frac34\frac{\sigma_{11}^2}{\sigma_1}+\frac14\lambda_2\sigma_1+\frac14\sigma_2\lambda_1
\end{align*}
These formulae were given in \cite{on998}. For the proof one should expand 
$$\sigma\left(\left.\int_{\infty}^{P}
\mathrm{d}\boldsymbol{u}\right|_{x=1/\xi^2}+\int_{\infty}^{P_1}
\mathrm{d}\boldsymbol{u} \right)$$ 
at $\xi=0$ at equate to zero principal parts of poles. Various representations of three index symbols are 
compatible because of the equation of the curve to which expressions (\ref{xy}) are plugged. 

Using method of expansions one can get higher derivatives 
\begin{align*}
\sigma_{222}&=\frac34\frac{\sigma_{22}^2}{\sigma_2}+\sigma_1+\frac14\lambda_4\sigma_2\\
\sigma_{2222}&=\frac12\frac{\sigma_{22}^3}{\sigma_2^2}+\frac{2\sigma_{22}\sigma_1}{\sigma_2}+\frac12\lambda_4\sigma_{22}\\
\sigma_{22222}&=\frac{5}{16}\lambda_4\frac{\sigma_{22}^2}{\sigma_2}+\frac54\frac{\sigma_1\sigma_{22}^2}{\sigma_2^2}+\frac52\frac{\sigma_1^2}{\sigma_2}+\frac{5}{32}\frac{\sigma_{22}^4}{\sigma_2^3}+\frac14\lambda_4\sigma_1+\frac14\lambda_3\sigma_2+\frac{1}{32}\lambda_4^2\sigma_2
\end{align*}

\section{Double pendulum and $(\sigma)$}

\begin{figure}
\begin{center}
\includegraphics[width=0.4\textwidth]{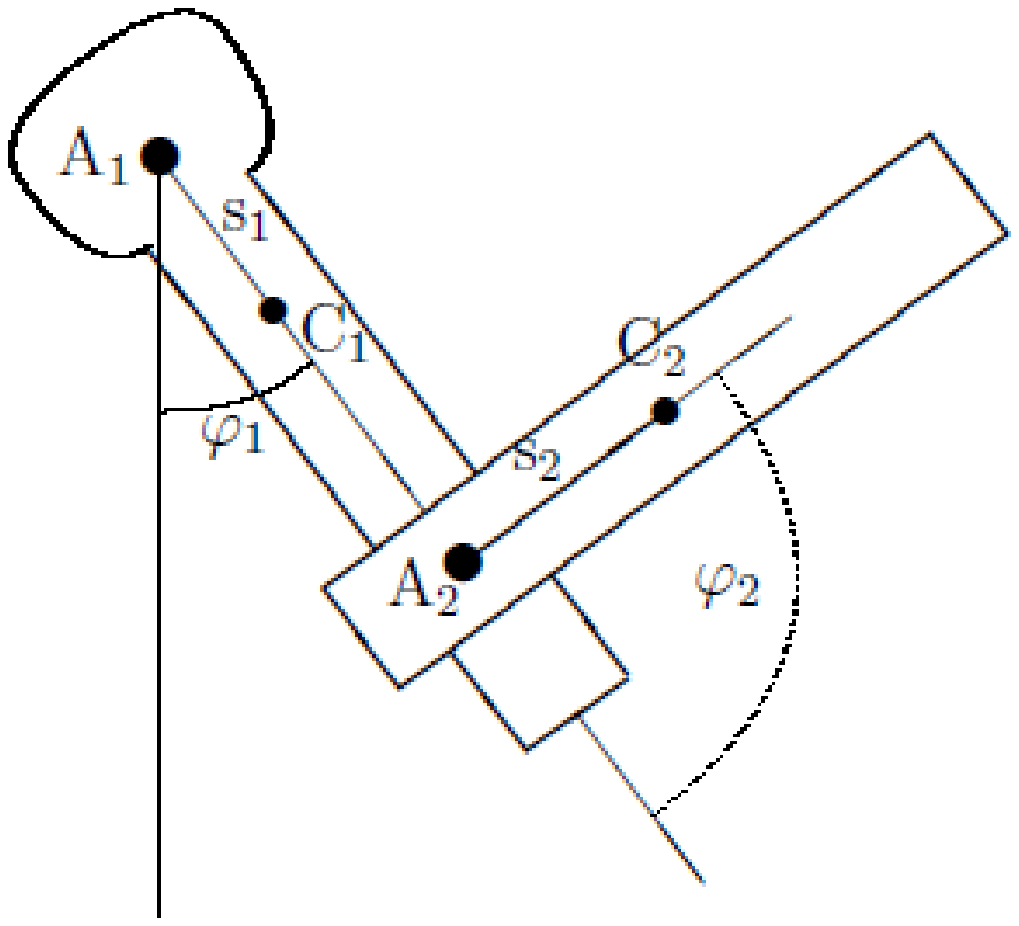}
\end{center}
\caption{Planar double pendulum}
\end{figure}


The first pendulun swings around a fixed axis $A_1$ and carries the axis $A_2$ of the second 
apart from axis; the distance between $A_1$ and $A_2$ is $a$. The configuration is determined by the two angles $\phi_1$ and $\phi_2$

\subsection{Lagrangian and Hamiltonian formalism}
Lagrangian is
\begin{align*}L= \frac12(\Theta_1+m_2a^2)\dot{\phi}_1^2+\frac12(\dot{\phi}_1+\dot{\phi}_2)^2
+m_2s_2a\dot{\phi}_1(\dot{\phi}_1+\dot{\phi}_2)\cos\,\phi_2
 \end{align*}
where $\Theta_1,\Theta_2$ are the moments of inertia of the two bodies with respect to respective suspension points, $m_2$ is the mass of the second pendulum. Scaling the energy and introducing parameters
\[  A=\frac{\Theta_1+m_2a^2}{\Theta_2},\qquad \alpha=\frac{m_2s_2a}{\Theta_2} \]
rewrite the Lagrangian in the form
\[ L=\frac12 A\dot{\phi}_1^2+\frac12(\dot{\phi}_1+\dot{\phi}_2)^2+\alpha\dot{\phi}_1(\dot{\phi}_1+\dot{\phi}_2)\cos\,\phi_2 \]

Introduce the angular momenta
\begin{align*}
p_1&=(A+1+2\alpha \cos\,\phi_2)\dot{\phi}_1+(1+\alpha \cos\,\phi_2)\dot{\phi}_2\\
p_2&=(1+\alpha \cos\,\phi_2)\dot{\phi}_1+\dot{\phi}_2
\end{align*}
The Hamiltonian becomes
\[ H=\frac12 \frac{p_1^2-2Q(\cos\,\phi_2)p_1p_2+P_1(\cos\,\phi_2)p_2^2}{P_2(\cos\,\phi_2)} \]
where
\[ Q(x)=1+\alpha x,\quad P_1(x)=A+1+2\alpha x, \quad P_2(x)=A-\alpha^2x^2 \]
Energy $H$ and the angular momentum $p_1$ are the first integrals that we will fix on levels $h$ and $l$. Hamiltonian equations are easily reduced to the inversion problem for function $z=\cos \phi_2$

\subsection{Solving the inversion problem}
\[  \frac{ (A-z^2)  \mathrm{d}z }{\sqrt{(1-z^2)(A-\alpha^2z^2)(2hA+4\alpha h z-2h-l^2)  }}=\mathrm{d} t  \]
That is inversion problem of the second kind integral
\[ \int_{P_0}^P \frac{(A-z^2)\mathrm{d}z}{y}=t+C  \]
Using the derived formulae for the second kind integrals expressed in terms of $\sigma$-derivatives restricted to $(\sigma)$ rewrite the last relation
\[ \left.\frac{\sigma_{22}(\boldsymbol{u})}{2\sigma_2(\boldsymbol{u})}\right|_{(\sigma)}+Au_1=t+C  \]

Uing this relation as well condition $\boldsymbol{u}\in (\sigma)$ we find (locally) functions
$$ u_1=u_1(t),\qquad u_2=u_2(t)$$ that should be then plug to the formula
\begin{equation}  z=cos(\phi_2)= - \frac{\sigma_1(u_1(t),u_2(t))}{\sigma_2(u_1(t),u_2(t))}  \end{equation}
That is inversion of the second kind integral by the method of restriction to $(\sigma)$-divisor.
Details and numerics are given in Enolski, Pronine, Richter (2003).

\section{On the inversion of one hyperelliptic integral}

We want to invert {\bf one} integral,
\[ t=\int_{\infty}^P \frac{ \mathrm{d} x} {y}, \quad P=(x,y)\in X  \]
The integral has 4 independent (and non-commensurable) complex periods
\begin{align*}
2\omega_{1,1}&=\oint_{\mathfrak{a}_1}\frac{ \mathrm{d} x} {y},\quad 2\omega_{1,2}=\oint_{\mathfrak{a}_2}\frac{ \mathrm{d} x} {y}\\
2\omega_{1,1}'&=\oint_{\mathfrak{b}_1}\frac{ \mathrm{d} x} {y},\quad 2\omega_{1,2}'=\oint_{\mathfrak{b}_2}\frac{ \mathrm{d} x} {y}
\end{align*}
Inverse function $x(t)$ of one complex variable should be 4-periodic, but that is not possible.


{\bf Proposition} (Fedorov-G\'omez-Ulate, 2007) Under the $B$-rule the values of the integral $t(P)$ range over the whole complex plane $\mathbb{C}$ exept for infinite number of non-intersecting identical domains $W_{i,j}$, $i,j\in\mathbb{Z}$ which called windows. Each window is a parallelogram spanned by the forbidden periods $2\omega_{1,1}$, $2\omega_{1,1}'$ while the complement to the windows is generated by the allowed periods
$2\omega_{1,2}$, $2\omega_{1,2}'$
\vskip1cm
The functions $x$ and $y$ are {\bf quasi-elliptic} i.e. they are double periodic but not defined on the whole complex plane but only on the complement to the windows.



\chapter[$\sigma$-function of $(n,s)$-curves]{$\sigma$-function of $(n,s)$-curves}
\label{chap:ns}

\section{Heat equations in a nonholonomic frame }

\subsection{Introduction}
We start to consider the following problem.

\emph{Let linear differential operators
\begin{equation}
L_i=\sum_{k=0}^{n}v_{i,k}(\la)\pa_{\la_k},\qquad i=0,\dots,n,\label{frame}
\end{equation}
define a nonholonomic frame. Find second-order linear differential operators of
the form
\begin{equation}H_{i}=\delta^{(i)}(\la)+\frac{1}{2}\sum_{a,b=1}^m
\a^{(i)}_{a,b}(\la)\pa_{u_a,u_b}+ 2\beta^{(i)}_{a,b}(\la)u_a\pa_{u_b}+
\ga^{(i)}_{a,b}(\la)u_au_b
\end{equation}
such that the system of equations
\begin{equation}
L_i(\psi)=H_i(\psi),\qquad i=0,\dots,n,\label{system}
\end{equation}
is compatible, i.e., this system has nonzero solutions
$\psi=\psi(u_1,\dots,u_m,\la_0,\dots,\la_n)$.}

We refer to system (\ref{system}) as the heat equations in the given nonholonomic frame (\ref{frame}).

We construct a transformation which assigns to any differentiable
mapping
$$\mathscr{F}\colon\mathbb{C}^{n+1}\break\to
\mathbb{C}\times\mathrm{PSp}(2m,\mathbb{C})$$
a set of operators $\{H_i\}$ compatible with the frame $\{L_i\}$ in the above
sense. Using the classical parametrization of $\mathrm{PSp}(2m,\mathbb{C})$, we
construct the operator algebra $A(\mathcal{X})$ on the space $\mathscr{R}$ of
solutions of system (\ref{system}) and also a ``primitive
solution''~$\phi_0(u,\la)$ and the primitive $\mathbb{Z}^{2m}$-invariant
solution $\phi_1(u,\la)$ obtained by averaging the solution $\phi_0$ over the
lattice contained in the algebra $A(\mathcal{X})$. One of the main our results
in this problem is that the primitive $\mathbb{Z}^{2m}$-invariant solution
coincides with the ``general sigma function''~on the bundle of Abelian tori
(see Chapt. 1).

Further the results obtained are applied to the important case in which
both the frame and the mapping $\mathscr{F}$ are defined by a family of plane
algebraic curves. We show all substantial details of the general construction
by the classical example of elliptic curves and functions which is far from
being trivial. This construction for the family of elliptic
curves enables us to present the very essence of the matter (not complicated by
technical details unavoidable for curves of higher genera).

In the case of higher genera we consider the family of plane algebraic curves
defined by the algebraic variety $\Gamma=\{(x_1,x_2,\la)\in\mathbb{C}^{n+3}\mid
f(x_1,x_2; \la)=0\}$, where
$f(x_1,x_2;\la)=x_1^{s_0}+x_2^{n_0}+\sum\la_{i}x_1^{s_i}x_2^{n_i}$. Using the
methods based on fundamental results of singularity theory
\cite{Arnold96,Zakaljukin76}, we construct vector fields $\{L_i\}$ tangent to
the discriminant variety $\Sigma=\{\la\mid\Delta(\la)=0\}$ of the family
$\Gamma$. By regarding $\Gamma$ as an algebraic bundle with the projection
$\mathsf{p}\colon\Gamma\to\mathbb{C}^{n+1}\colon(x_1,x_2,\la)\mapsto\la$, we
choose $\mathscr{F}$ in the form $(\log \Delta,\Omega)$, where $\Delta(\la)$ is
the polynomial defining the discriminant and $\Omega(\la)$ is the period matrix
for the vector of the first- and second-kind basis Abelian differentials with
respect to a symplectic basis of cycles on the fiber over a point $\la$.
Finally, using some classical \cite{ba97} and results of the Chapt.\ref{chap:rat} results, we
obtain a construction for which the difference between the left- and right-hand
sides of system (\ref{system}) gives the linear operators annihilating the sigma
function of the family of curves $\Gamma$. For the hyperelliptic sigma
functions, an explicit form of generating functions defining the frame
$\{L_i\}$ and the operators $\{H_i\}$ is indicated. In this way we derived the system of differential equations that uniquely define $\sigma$-function among solutions of this system with given initial data. Writing the solution $\sigma(\boldsymbol{u},\lambda)$ as a series in vector $\boldsymbol{u}$ we derive from this system recursive family of polynomials in $\lambda$ defining coefficients of this expansions.

Now we are ready to describe our problem formally.
Let $\la=(\la_0,\dots,\la_n)$ be a set
of commuting variables. Denote by $\CL$ the ring of differentiable functions of
$\la$. Consider a matrix $V(\la)=(v_{i,j}(\la))$, $i,j=0,1,\dots,n$, where
$v_{i,j}(\la)\in \CL$, and set $\Delta(\la)=\det V$. In what follows we assume
that $\Delta\not\equiv 0$. Introduce a moving frame $\{L_0,\dots,L_n\}$ by
setting $L_i=\sum_{q=0}^{n}v_{i,q}(\la)\pa_{\la_q}$. Let $c_{i,j}^{k}(\la)$ be
the structure functions of the frame.

Regarding $\la_i$ as the operator of multiplication by the function $\la_i$ in
the ring $\CL$, we see that the commutation relations
\begin{equation} [L_i,L_j]=\sum_{k=0}^n c_{i,j}^{k}(\la)L_k, \quad[L_i,\la_q]=v_{i,q}(\la),
\quad[\la_q,\la_r]=0\label{al}
\end{equation}
define a Lie algebra structure on the $\CL$-module with the generators
$1,L_0,\dots,L_n$.

Let $u=(u_1,\dots,u_m)$ be another set of commuting variables and let $\CU$ be
the ring of differentiable functions of $u$ and $\la$. Set
\begin{equation} Q_i=L_i+\delta^{(i)}(\la)+\frac{1}{2}\sum_{a,b=1}^m
\a^{(i)}_{a,b}(\la)\pa_{u_a,u_b}+ 2\beta^{(i)}_{a,b}(\la)u_a\pa_{u_b}+
\ga^{(i)}_{a,b}(\la)u_au_b.\label{Q:op}
\end{equation}
\begin{pro}\label{pro1} Find sufficient conditions on the data
$\bigl\{\a^{(i)}(\la),\beta^{(i)}(\la),\ga^{(i)}(\la),\delta^{(i)}(\la)\bigr\}$
for the operators \eqref{Q:op} to realize a representation of the Lie algebra
\eqref{al} in the ring of operators on $\CU$.
\end{pro}

\subsection{General solution} Note that the
relations $[Q_i,\la_q]=v_{i,q}(\la)$ automatically hold for the operators
\eqref{Q:op}. Consider the system of equations
\begin{equation}[Q_i,Q_j]=\sum_{k=0}^nc_{i,j}^k(\la)Q_k.
\label{repr}\end{equation}

Defining the operators $Q_{i}$ of the form \eqref{Q:op} is equivalent to
defining the triples
\begin{equation*}
Q_i\,\hat{}=(L_i, \;M^{(i)}(\la),\;\delta^{(i)}(\la)),
\end{equation*}
where
$M^{(i)}(\la)=\begin{pmatrix}\a^{(i)}(\la)&(\beta^{(i)}(\la))^T\\
\beta^{(i)}(\la)&\ga^{(i)}(\la)\end{pmatrix}$ is a symmetric $(2m\times2m)$
matrix. Note that the mapping $Q_i\mapsto Q_i\hat{}$\ is linear. We have
\begin{eqnarray*}
[Q_i,Q_j]\,\hat{}=([L_i,L_j],\;M^{(i)}JM^{(j)}-
M^{(j)}JM^{(i)}+L_{i}(M^{(j)})-L_{j}(M^{(i)}),\\
\frac{1}{2}\mathrm{Tr}(\a^{(i)}\gamma^{(j)}-
\a^{(j)}\gamma^{(i)})+L_{i}(\delta^{(j)})-L_{j}(\delta^{(i)})),
\end{eqnarray*}
where $J={\begin{pmatrix}0&1_m\\-1_m&0\end{pmatrix}}$. Thus, the construction
of a solution of \eqref{repr} is reduced to the solution of two systems of
equations
\begin{gather}
M^{(i)}JM^{(j)}-M^{(j)}JM^{(i)}+L_{i}(M^{(j)})-L_{j}(M^{(i)}) -\sum c_{i,j}^kM^{(k)}=0,\label{M:eq}\\
\frac{1}{2}\mathrm{Tr}(\a^{(i)}\ga^{(j)}- \a^{(j)}\ga^{(i)})
+L_{i}(\delta^{(j)})-L_{j}(\delta^{(i)}) -\sum c_{i,j}^k\delta^{(k)}=0.
\label{de:e0}
\end{gather}

In particular, it follows from \eqref{M:eq} that
\begin{equation*}
\a^{(i)}\ga^{(j)}-\a^{(j)}\ga^{(i)}-
[\beta^{(i)},\beta^{(j)}]+L_{i}(\beta^{(j)})-L_{j}(\beta^{(i)})-\sum
c_{i,j}^k\beta^{(k)}=0,
\end{equation*}
and hence we can simplify system \eqref{de:e0}. Set
$\delta^{(i)}=\chi^{(i)}+\frac{1}{2}\mathrm{Tr}\beta^{(i)}$. We obtain the equations
\begin{equation}\label{de:e1}
L_{i}(\chi^{(j)})-L_{j}(\chi^{(i)})-\sum c_{i,j}^k\chi^{(k)}=0,
\end{equation}
This system is the compatibility condition for the system
$\{L_i(\psi)-\chi^{(i)}\psi=0\}$, and therefore a set of functions $\{\chi_i\}$
is a solution of \eqref{de:e1} if and only if $\chi^{(i)}=L_i(\varphi(\la))$
for some function $\varphi(\la)\in \CL$. Thus,
\begin{equation*}
\delta^{(i)}(\la)=L_i(\varphi(\la))+\frac{1}{2}\mathrm{Tr}\beta^{(i)}(\la).
\end{equation*}

Let us proceed with system \eqref{M:eq}. Set $M^{(i)}=J A^{(i)}$. Since the
matrices $M^{(i)}$ are symmetric, it follows that $A^{(i)}J+J(A^{(i)})^T=0$,
i.e., $A^{(i)}\in\mathfrak{sp}(2m,\CL)$.

Then
\begin{equation}\label{A:e0}
[A^{(j)},A^{(i)}]+L_{i}(A^{(j)})-L_{j}(A^{(i)})-\sum c_{i,j}^kA^{(k)}=0.
\end{equation}
System \eqref{A:e0} is the compatibility condition for the system of equations
\begin{equation}\label{con:A}
L_i(\xi)+\xi A^{(i)}=0,\quad\xi=(\xi_1(\la),\dots,\xi_{2m}(\la)), \qquad
i=0,\dots,n.
\end{equation}
Denote the space of solutions of \eqref{con:A} by $\mathcal{X}$. For any
$\xi,\wt{\xi}\in\mathcal{X}$ we have
\begin{equation*}
L_i(\wt{\xi}J\xi^T)=L_i(\wt{\xi})J\xi^T+\wt{\xi}JL_i(\xi^T)= -\wt{\xi}(A^{(i)}J
+J(A^{(i)})^T)\xi^T=0\implies\wt{\xi}J\xi^T=\mathrm{const}.
\end{equation*}
Thus, the skew-symmetric bilinear function
\begin{equation*}
\mathcal{X}\otimes\mathcal{X} \to\mathbb{C}\colon\langle\wt{\xi},\xi\rangle
=\wt{\xi}J\xi^T
\end{equation*}
is defined on the linear space $\mathcal{X}$.

Let $\Omega(\la)$ be a nonsingular matrix such that $\Omega
J\Omega^T=\mathrm{const}$. Then the matrices
$\{A^{(i)}=-\Omega^{-1}L_{i}(\Omega)\}$ give a solution of \eqref{A:e0}.
Moreover, if $K$ is a nonsingular constant matrix, then the solutions defined
by matrices $K\Omega(\la)$ and $\Omega(\la)$ are equal. Let us use this fact to
fix the normalization $\Omega J\Omega^T=2\pi\imath\,J$. The choice of
normalization provides the principal polarization of the Abelian tori arising
below.

We have obtained the following solution of Problem \ref{pro1}.
\begin{theorem}\label{cons}
Let $\Omega=\Omega(\la)$ be a nonsingular $(2m\times 2m)$-matrix such that
$\Omega J\Omega^T=2\pi\imath\,J$ and $\varphi=\varphi(\la)\in\CL$. Then the
operators $\{Q_0,\dots,Q_n\}$ of the form \eqref{Q:op} that are defined by the
triples
\begin{equation*}
Q_i\,\hat{}=(L_i,\,J L_i(\Omega^{-1})\Omega,\, L_i(\varphi)+\tfrac14\mathrm{Tr}(P
L_i(\Omega^{-1})\Omega)),
\end{equation*}
where $P=\begin{pmatrix}1_m&0\\0&-1_m\end{pmatrix}$, realize a representation
of the Lie algebra \eqref{al} in the ring of operators on $\CU$.
\end{theorem}
Thus, the space of the above representations of the Lie algebras \eqref{al} is
the space of differentiable mappings $\mathscr{F}\colon\mathbb{C}^{n+1}
\to\mathbb{C}\times\mathrm{PSp}(2m,\mathbb{C})$.

For a Lie algebra \eqref{al}, denote by $\mathscr{Q}$ the set of all
representations of \eqref{al} described by Theorem~\ref{cons}. The set
$\mathscr{Q}$ is equipped with the following natural action of the group
$\mathrm{PSp}(2m,\CL)$.

\begin{lem}\label{cali} Let $T(\la)\in\mathrm{PSp}(2m,\CL)$.
The operator $T\colon\mathscr{Q}\to\mathscr{Q}$ takes a representation
$\{Q_i\}$ with $Q_i\,\hat{}=(L_i, M^{(i)}(\la),\delta^{(i)}(\la))$ to the
representation $\{Q^T_i\}$ with
\begin{equation*}(Q^T_i)\,\hat{}=
(L_i,\;-JTJM^{(i)}T^{-1}+JL_{i}(T)T^{-1},\;\delta^{(i)}
+\tfrac14\mathrm{Tr}(PL_{i}(T)T^{-1})).
\end{equation*}
\end{lem}

Let us use Theorem \ref{cons} to construct an explicit solution of Problem
\ref{pro1}.

We parametrize the set of matrices $\Omega$ satisfying the equation $\Omega
J\Omega^T=2\pi\imath\,J$ by the triples $(\omega,\tau, \varkappa)$, where
$\omega\in\mathrm{GL}(m,\CL)$, $\tau^T=\tau$, and $\varkappa^T=\varkappa$, in
the following way. Represent a matrix $\Omega$ as a $(2\times2)$ block matrix
$\Omega=\begin{pmatrix}
\Omega_{1,1}&\Omega_{1,2}\\
\Omega_{2,1}&\Omega_{2,2}\end{pmatrix}$ and set
\begin{equation}\label{param}
\Omega_{1,1}=\omega,\quad\Omega_{1,2}=\omega\varkappa,\quad\Omega_{2,1}=\tau\omega,\quad
\Omega_{2,2}=\tau\omega\varkappa +2\pi\imath (\omega^T)^{-1}.
\end{equation}
One can immediately see that the relation $\Omega J\Omega^T=2\pi\imath\,J$
holds identically. The set $\{\omega(\la)$, $\tau(\la)$,
$\varkappa(\la),\varphi(\la)\}$ is referred to as the {\it parameters\/} of the
representation of the Lie algebra \eqref{al}.

By applying Theorem \ref{cons} to the parametrized matrix $\Omega$, we obtain
the following result.

\begin{theorem}
Let $\Omega(\la)$ be a matrix parametrized by the data
$\{\omega(\la),\tau(\la), \varkappa(\la),\varphi(\la)\}$ according to
\eqref{param}. Then there is a representation of the Lie algebra \eqref{al}
realized by the operators $\{Q_i\}$ of the form \eqref{Q:op}, where
\begin{equation}\label{otk}
\begin{aligned}
\a^{(i)}&=-\tfrac{1}{2\pi\imath}\omega^T\,L_i(\tau)\,\omega,\quad
&\ga^{(i)}&=-\varkappa\,\a^{(i)}
\varkappa+\varkappa\,\beta^{(i)}+(\beta^{(i)})^T\varkappa+L_i(\varkappa),\\
(\beta^{(i)})^T&=\varkappa\,\a^{(i)}+\omega^{-1}L_i(\omega),\quad
&\delta^{(i)}&=\tfrac{1}{2}\mathrm{Tr}(\a^{(i)}\varkappa)+L_i(\varphi
+\tfrac{1}{2}\log(\det\omega)).
\end{aligned}
\end{equation}
\end{theorem}

Problem \ref{pro1} is completely solved.

\subsection{The basic solution} Fix a representation
$\{Q_i\}\in\mathscr{Q}$ and some parametrization
$\{\omega,\tau,\varkappa,\varphi\}$ of $\{Q_i\}$ according to \eqref{otk}.

Consider the system of linear differential equations
\begin{equation}
Q_i(\phi(u,\la))=0,\qquad i=0,\dots,n,\label{Q:sys}
\end{equation}
\begin{defn}
Define by $\mathscr{R}$ the set of solutions of (\ref{Q:sys}) that are entire functions with respect to $u$.
\end{defn}

\begin{lem} A function of the form
\begin{equation}
\phi(u,\la)=\mu(\la)\exp\{u^T\Phi(\la)u\},\label{anz}
\end{equation}
where $\Phi(\la)$ is a symmetric $(m\times m)$ matrix over $\CL$ and
$\mu(\la)\in\CL$, gives a solution of \eqref{Q:sys}, i.e.,
$\phi\in\mathscr{R}$, if and only if
\begin{equation*}
\Phi(\la)=-\tfrac{1}{2}\varkappa,\qquad
\mu(\la)=\mu_0(\det\omega)^{-1/2}e^{-\varphi(\la)}, \quad\mu_0\in\mathbb{C}.
\end{equation*}
\end{lem}
\begin{proof}
Let us substitute \eqref{anz} into \eqref{Q:sys}. We obtain the system of
equations
\begin{gather*}
2\Phi\a^{(i)}\Phi+\Phi\beta^{(i)}+(\beta^{(i)})^T\Phi+\ga^{(i)}+L_i(\Phi)=0,\\
L_i(\log(\mu))+\mathrm{Tr}(\a^{(i)}\Phi)+\delta^{(i)}=0
\end{gather*}
for $\Phi$ and $\mu$. By comparing the above equations with \eqref{otk}, we
immediately obtain the assertion of the lemma.\qed
\end{proof}

\begin{defn}
By the \emph{basic solution} of system \eqref{Q:sys} with the parameters
$\{\omega(\la)$, $\tau(\la)$, $\varkappa(\la)$, $\varphi(\la)\}$ we mean the
function
\begin{equation}\label{pimit}
\phi_{0}(u,\la)=\frac{\mu_0}{\sqrt{\det\omega}}
\exp\Big\{-\varphi(\la)-\frac{1}{2}u^T\varkappa u\Big\}.
\end{equation}

\end{defn}
Note that $\phi_0$ does not depend on $\tau$.

\subsection{Operator algebra} Let
$\xi=(\xi_1(\la),\dots,\xi_{2m}(\la))$. Set
\begin{equation*}
S_{\xi}=\sum_{j=1}^{m}(\xi_j(\la)\pa_{u_j}+\xi_{m+j}(\la)u_j).
\end{equation*}

\begin{lemma}\label{lino} The following conditions are equivalent

{\bf(a)} $[S_{\xi},Q_i]=0$,

{\bf (b)} $L_i(\xi)+\xi A^{(i)}=0$, i.e., $\xi\in\mathcal{X}$.
\end{lemma}

Using the nondegenerate bilinear form $\langle\cdot,\cdot\rangle$, introduce
the algebra $A(\mathcal{X})$ as the quotient algebra of the tensor algebra
$T\mathcal{X}=\mathbb{C}\cdot
1+\mathcal{X}+\mathcal{X}\otimes\mathcal{X}+\dots$ of the linear space
$\mathcal{X}$ by the relation
$\wt{\xi}\otimes\xi-\xi\otimes\wt{\xi}-\langle\wt{\xi},\xi\rangle\cdot 1=0$.

\begin{lemma}
Assigning the identity operator on $\mathscr{R}$ to the identity element $1\in
A(\mathcal{X})$, the operator $S_{\xi}$ to any element $\xi\in\mathcal{X}$, and
the composition $S_{\wt{\xi}}\circ S_{\xi}$ to any element
$\wt{\xi}\otimes\xi\in\mathcal{X}\otimes\mathcal{X}$, we obtain an action of
the algebra $A(\mathcal{X})$ on the set $\mathscr{R}$ solutions of the system(\ref{Q:sys}).
\end{lemma}

Lemma \ref{lino} implies the following assertion.

\begin{cor}\label{expo}
Let $\xi\in\mathcal{X}$. Then the operator $W_{\xi}=\exp{S_{\xi}}$ acts on
$\mathscr{R}$, i.e., it takes a solution of \eqref{Q:sys} to a solution.
\end{cor}

Set $\varrho_1=(\xi_1,\dots,\xi_m)$, $\varrho_2=(\xi_{m+1},\dots,\xi_{2m})$,
$S_1=\sum_{j=1}^{m}\xi_j\pa_{u_j}$, and $S_2=\sum_{j=1}^{m}\xi_{m+j}u_j$. Since
$[S_1,S_2]=\varrho_1\varrho_2^T$, and therefore $[S_i,[S_i,S_j]]=0$, $i,j=1,2$,
we see by the Campbell--Hausdorff formula that
\begin{equation*}
W_{\xi}=\exp{S_{\xi}}=\exp\{S_1+S_2\}=\exp\{S_2-\tfrac{1}{2}[S_2,S_1]\}\circ
\exp\{S_1\}.
\end{equation*}
Thus, the operator $W_{\xi}=W_{(\varrho_1,\varrho_2)}$ is factorized into the
composition of the shift operator $\exp\{S_1\}$ with respect to the variables
$u$ and of the operator of multiplication by the exponential of a linear
function in the variables~$u$. The action of this operator is given by the
formula
\begin{equation}\label{W}
W_{(\varrho_1,\varrho_2)}(\phi(u,\la))
=\phi(u+\varrho_1^T,\la)\exp\{\varrho_2(u+\tfrac{1}{2}\varrho_1^T)\}.
\end{equation}
Every composition of transformations \eqref{W} is a transformation of the same
form up to multiplication by a function of $\la$; under the conditions of our
construction, this holds up to multiplication by a constant.

Let us extend the algebra $A(\mathcal{X})$ by including the operators
$W_{\xi}$, $\xi\in\mathcal{X}$, into the algebra. We keep the same notation for
the extension thus obtained.

\begin{lem} Let $\xi,\wt{\xi}\in\mathcal{X}$.
The operators $W_{\wt{\xi}}$ and $W_{\xi}$ commute if and only if
\begin{equation*}
[S_{\wt{\xi}},S_{\xi}]=\wt{\xi}J\xi^T=\langle\xi, \wt{\xi}\rangle=2\pi\imath
N,\qquad N\in\mathbb{Z}.
\end{equation*}
\end{lem}

\begin{proof}{Proof} Consider the composition $W_{\xi}\circ W_{\wt{\xi}}$.
By applying the Campbell--Hausdorff formula under the assumption that
$[S_{\wt{\xi}},S_{\xi}]=\wt{\xi}J\xi^T\in\mathbb{C}$, we obtain
\begin{equation*}
\exp\{S_{\xi}+S_{\wt{\xi}}\}=W_{\xi}\circ W_{\wt{\xi}}\exp
\{-\tfrac{1}{2}[S_{\wt{\xi}},S_{\xi}]\}= W_{\wt{\xi}}\circ
W_{\xi}\exp\{\tfrac{1}{2}[S_{\wt{\xi}},S_{\xi}]\}.
\end{equation*}
The assertion of the lemma follows from the equation
\begin{equation*}
\exp\{-\tfrac{1}{2}[S_{\wt{\xi}},S_{\xi}]\}=
\exp\{\tfrac{1}{2}[S_{\wt{\xi}},S_{\xi}]\}.\eqno{\qed}
\end{equation*}
\end{proof}

Denote by $\mathscr{W}$ the Abelian group generated multiplicatively by the
commuting operators $W_{\xi}$.

By Theorem \ref{cons}, the rows of the matrix $\Omega(\la)$ form a basis of the
linear space $\mathcal{X}$. Using the normalization condition $\Omega
J\Omega^T=2\pi\imath\,J$, we obtain the following result.

\begin{cor} The group $\mathscr{W}$ is isomorphic to $\mathbb{Z}^{2m}$.
An isomorphism $H\colon\mathbb{Z}^{2m}\to\mathscr{W}$ is given by the formula
$H(\mathfrak{z})=\exp\{\pi \imath\,
\mathfrak{z}_2\mathfrak{z}_1^T\}W_{\mathfrak{z}\Omega}$, where
$\mathfrak{z}=(\mathfrak{z}_1,\mathfrak{z}_2)$ and
$\mathfrak{z}^T_1,\mathfrak{z}^T_2\in\mathbb{Z}^{m}$.

The transformation $W_{\mathfrak{z}\Omega}$ acts by formula \eqref{W} in which
the parameters $(\varrho_{1},\varrho_{2})$ are given according to the
parametrization \eqref{param} by the expressions
\begin{equation*}
\varrho_1=(\mathfrak{z}_1+\mathfrak{z}_2\tau)\omega,\qquad
\varrho_2=\varrho_1\varkappa+2\pi\imath\, \mathfrak{z}_2(\omega^T)^{-1}.
\end{equation*}
\end{cor}

\subsection{Realising of $\sigma$-function
in terms of basic solutions} Let us apply the operator
$H(\mathfrak{z})$ to the basic solution \eqref{pimit}. We obtain
\begin{equation*}
H(\mathfrak{z})(\phi_0(u,\la))
=\mu(\la)H(\mathfrak{z}_1,\mathfrak{z}_2)e^{-u^T\varkappa u/2}=
\phi_0(u,\la)\exp{\pi\imath\{\mathfrak{z}_2\tau\mathfrak{z}_2^T+
2\mathfrak{z}_2(\omega^T)^{-1}u\}}.
\end{equation*}
Write
\begin{equation*}
\phi_{1}(u,\la)= \sum_{\mathfrak{q}^T\in\mathbb{Z}^{m}}
H(0,\mathfrak{q})\phi_0(u,\la)=
\phi_0(u,\la)\sum_{\mathfrak{q}^T\in\mathbb{Z}^{m}}
\exp{\pi\imath\big\{\mathfrak{q}\tau\mathfrak{q}^T+
2\mathfrak{q}(\omega^T)^{-1}u\big\}}.
\end{equation*}

\begin{lem}\label{zin}
The function $\phi_1(u,\la)\in\mathscr{R}$ is invariant with respect to the
action of the group $\mathscr{W}\sim\mathbb{Z}^{2m}$.
\end{lem}

\begin{proof} Apply $H(\mathfrak{z}_1,\mathfrak{z}_2)$ to
$\phi_{1}$. We obtain
\begin{equation*}
\begin{split}
\hskip67pt H(\mathfrak{z}_1,\mathfrak{z}_2)(\phi_1)&=\phi_0 e^{\pi
\imath\{\mathfrak{z}_2\tau\mathfrak{z}_2^T+
2\mathfrak{z}_2(\omega^T)^{-1}u\}}\sum_{\mathfrak{q}^T\in\mathbb{Z}^{m}}
e^{\pi\imath\{\mathfrak{q}\tau\mathfrak{q}^T+
2\mathfrak{q}(\omega^T)^{-1}u+2\mathfrak{q}(\mathfrak{z}_1+\mathfrak{z}_2\tau)^T
\}}\\
&=\phi_0\sum_{\mathfrak{q}^T\in\mathbb{Z}^{m}}e^{\pi\imath\{
(\mathfrak{q}+\mathfrak{z}_2)\tau(\mathfrak{q}+\mathfrak{z}_2)^T+
2(\mathfrak{q}+\mathfrak{z}_2)(\omega^T)^{-1}u \}}=\phi_1.\hskip90pt\qed
\end{split}
\end{equation*}
\end{proof}

\begin{defn}
The function
\begin{equation}\label{pimit:Z}
\phi_{1}(u,\la)=\frac{\mu_0}{\sqrt{\det\omega}}\exp\Big\{-\varphi(\la)-\frac{1}{2}u^T\varkappa
u\Big\}\sum_{\mathfrak{q}^T\in\mathbb{Z}^{m}}\exp{\pi\imath\big\{\mathfrak{q}\tau\mathfrak{q}^T+
2\mathfrak{q}(\omega^T)^{-1}u\big\}},
\end{equation}
is called the \emph{basic $\mathbb{Z}^{2m}$-invariant solution of system
\eqref{Q:sys} with the parameters $\{\omega(\la)$, $\tau(\la)$,
$\varkappa(\la)$, $\varphi(\la)\}$}.
\end{defn}

Using definition of $\sigma$ on the universal bundle of $2m$-dimensional principally polarized Abelian tori given in Chapt. 1 we come to the results

\begin{theorem}\label{sig}
The basic $\mathbb{Z}^{2m}$-invariant solution of system \eqref{Q:sys} with
the parameters $\{\omega(\la),\linebreak[0]\tau(\la),
\varkappa(\la),\linebreak[0]\varphi(\la)\}$ coincides with the function
$\sigma$  up to notation.
\end{theorem}

\begin{proof} We need the following result.
Suppose that
\begin{equation}\label{frob}
\begin{pmatrix}
\omega_1^T&{\eta_1}^T\\
{\omega_2}^T&{\eta_2}^T
\end{pmatrix}
\begin{pmatrix}
0&1_m\\-1_m&0
\end{pmatrix}
\begin{pmatrix}
\omega_1&\omega_2\\
\eta_1&\eta_2
\end{pmatrix}
=-2\pi \imath\, \ell
\begin{pmatrix}
0&1_m\\
-1_m&0
\end{pmatrix},\qquad \ell\in\mathbb{N}.
\end{equation}

\begin{lem}\label{dime} Let $\varepsilon,
\varepsilon'\in\mathbb{C}^m$. The linear space $\mathscr{L}(\varepsilon,
\varepsilon')$ of all solutions of the functional equation
\begin{equation}\label{func-e}
\phi(u+\omega_1\mathfrak{q}+\omega_2\mathfrak{q}')=
\exp\{\tfrac{1}{2}(\eta_1\mathfrak{q}+\eta_2\mathfrak{q}')^T(2u+\omega_1\mathfrak{q}
+\omega_2\mathfrak{q}')+\varepsilon^T\omega_1\mathfrak{q}
+(\varepsilon')^T\omega_2\mathfrak{q}'\} \phi(u),
\end{equation}
where $\mathfrak{q},\mathfrak{q}'\in \mathbb{Z}^m$, in the class of entire
functions is of dimension $\ell^m$.
\end{lem}

\begin{proof}
The standard proof is to reduce the functional equation \eqref{func-e} to the
case in which
\begin{equation*}
\begin{pmatrix}
\omega_1&\omega_2\\
\eta_1&\eta_2 \end{pmatrix} =\begin{pmatrix} 1_m&\tau\\
0&-2\pi\imath\,\ell 1_m
\end{pmatrix}.
\end{equation*} Let us parametrize relations \eqref{frob} by setting
$\omega_2=\omega_1 \tau$, $\eta_1=-\varkappa \omega_1$, and
$\eta_2=-\varkappa\omega_1\tau-2\pi\imath\,\ell (\omega_1^T)^{-1}$. As usual,
$\tau^T=\tau$ and $\varkappa^T=\varkappa$. The function
\begin{equation*}
\vartheta(v)=\exp\bigg\{\bigg(\frac{1}{2}\,v^T\omega_1^T\varkappa
-\varepsilon^T\bigg) \omega_1 v\bigg\}\phi\bigg(\omega_1\bigg(v
+\frac{1}{2\pi\imath} \tau(\varepsilon'-\varepsilon)\bigg)\bigg)
\end{equation*}
satisfies the equations
\begin{equation*}
\vartheta(v+\mathsf{e}_k)=\vartheta(v) \quad\text{and}\quad
\vartheta(v+\tau\mathsf{e}_k)= \exp\{-\pi\imath\,\ell(2 v^T+
\mathsf{e}_k^T\tau)\mathsf{e}_k\} \vartheta(v),
\end{equation*}
where $\mathsf{e}_k$ is the $k$th basis vector, $k=1,\dots,n$. It follows from
the first equation that $\vartheta(v)$ admits a Fourier series expansion,
\begin{equation*}
\vartheta(v)=\sum_{\mathfrak{q}\in\mathbb{Z}^m}
c_{\mathfrak{q}}\exp\{2\pi\imath\,\mathfrak{q}^Tv\}.
\end{equation*}
By substituting the series into the other equation, we obtain the recurrence
for the coefficients $c_{\mathfrak{q}}$,
\begin{equation*}
c_{\mathfrak{q}+\ell\mathsf{e}_{k}}=c_{\mathfrak{q}}
\exp\{\pi\imath(2\mathfrak{q}^T\tau+\ell \mathsf{e}_k\tau)\mathsf{e}_k\}.
\end{equation*}
Thus, $\vartheta(v)$ is defined by the $\ell^m$ coefficients
$c_{\mathfrak{r}},\; \mathfrak{r}\in\mathbb{Z}^m/\ell\mathbb{Z}^m$.\qed
\end{proof}

It follows from the above lemma and Lemma \ref{zin} that, since the matrix
$\Omega(\la)$ satisfies \eqref{frob} with $\ell=1$, any function $\psi(u,\la)$
invariant with respect to the action of $\mathscr{W}$ given by the operators
$H(\mathfrak{z})$ can differ from $\phi_1(u,\la)$ by at most a factor depending
on $\la$.

The invariance of $\phi_1$ under the action of modular group
$\mathrm{Sp}(2m,\mathbb{Z})$ follows from the invariance of the construction of
the operators, see Theorem \ref{cons} and Lemma \ref{cali}.\qed
\end{proof}

\subsection{Case of algebraic curves}

In the above constructions, the moving frame $\{L_0,\dots,L_n\}\in
\mathrm{Der}\,\CL$ and the pair
$(\varphi,\Omega)\in\CL\times\mathrm{PSp}(2m,\CL)$ defining a representation of
this frame were assumed to be independent from each other. Here we turn to the
consideration of a construction in which the objects are defined by a family of
plane algebraic curves.

Let us begin with the general outline of the construction.

Consider a family of plane algebraic curves defined by an algebraic variety of
the form
\begin{equation*}
\Gamma=\{(x_1,x_2,\la)\in\mathbb{C}^{n+3}\mid f(x_1,x_2; \la)=0\},
\end{equation*}
where $f(x_1,x_2;\la)=f_{0}(x_1,x_2)+\sum\la_{i}x_1^{s_i}x_2^{n_i}$ is an
irreducible polynomial. The mapping
$\mathsf{p}\colon\Gamma\to\mathbb{C}^{n+1}$, where
$\mathsf{p}$$\colon(x_1,x_2,\la)\mapsto\la$, defines an algebraic fiber bundle.
Assume that the fiber over a generic base point is a nonsingular curve of genus
$g$.

By the discriminant $\Sigma$ of the polynomial $f(x_1,x_2;\la)$ we mean the
variety
\begin{equation*} \Sigma=\{\la\in
\mathbb{C}^{n+1}\mid \text{the set of solutions of }\,
(\pa_{x_1},\pa_{x_2})f(x_1,x_2;\la)=0\; \text{is not empty}\},
\end{equation*}
i.e., the fibers of the bundle $\Gamma$ over the points $\la\in\Sigma$ are
singular curves. For the moving frame $\{L_i\}$ we take a set of vector fields
tangent to $\Sigma\subset\mathbb{C}^{n+1}$. One can efficiently construct
vector fields of this kind by using methods of singularity
theory~\cite{bucley02:3}. The structure functions $v_{i,j}(\la)$ and
$c_{i,j}^k(\la)$ of the frame are polynomials. The determinant of the matrix
$V=(v_{i,j}(\la))$, i.e., the polynomial $\Delta(\la)$, defines the
discriminant $\Sigma=\{\la\mid\Delta(\la)=0\}$. With regard to this fact, we
take $\varphi(\la)=\varphi_0\log\Delta(\la)$, where $\varphi_0\in\mathbb{C};$
then the functions $\chi_i(\la)$ defining the representation of the frame at
$m=0$ are polynomials.

For the matrix $\Omega(\la)$ we take the period matrix of $g$ holomorphic basis
Abelian differentials and $g$ meromorphic ones with respect to a symplectic
basis of cycles in the fiber over a point $\la$. A basis of differentials such
that $\Omega J\Omega^T=2\pi\imath\,J$ always exists (e.g., the basis of
normalized differentials). Actually, we do not need the matrix $\Omega$. To
calculate, say, the matrix $A^{(i)}$, it suffices (1) to compose the vector
$D(x)=(\mu_i(x_1,x_2;\la))\,\mathrm{d}x_1/f_2(x_1,x_2;\la)$ of basis
differentials and (2) to apply the operator $L_i$ to $D(x)$. Then the relation
\begin{equation*}L_i(D(x))\equiv A^{(i)}D(x)
+\text{exact differential}\end{equation*} defines the matrix $A^{(i)}$
uniquely. It is of importance for our construction that one can choose basis
differentials in such a way that the entries of the matrices $\{A^{(i)}\}$ are
polynomials. One can extract the classical method of constructing the bases we
need from \cite{ba97}.

According to Theorem \ref{sig}, for an appropriate value of the constant
$\varphi_0$, we obtain the set $\{Q_i\}$ of linear differential operators (with
polynomial coefficients) annihilating the sigma function $\sigma(u,\la)$ of the
family $\Gamma$. By definition, this is the sigma function
on the bundle $\mathrm{Jac}(\Gamma)$ associated with $\Gamma$. A fiber of the
bundle $\mathrm{Jac}(\Gamma)$ over a point $\la$ is the Jacobi variety of the
algebraic curve $\mathsf{p}^{-1}(\la)$.

The value of $\varphi_0$ is chosen on the basis of the following facts. Using
the chosen basis of holomorphic differentials, we define the fiberwise
holomorphic mapping $\mathcal{A}\colon\Gamma\to\mathrm{Jac}(\Gamma)$ whose
restriction to a fiber is the classical Abel mapping. If the curve defined by
the equation $f_0(x_1,x_2)=0$ (i.e., the curve $\mathsf{p}^{-1}(0)$) is
rational, then the restriction of the mapping $\mathcal{A}$ to the fiber
$\mathsf{p}^{-1}(0)$ is a polynomial mapping. In \cite{bel99a}, polynomials
$\sigma_{0}(u)$ were constructed that satisfy an analog of the Riemann
vanishing theorem, and it was proved that this completely defines the
polynomials. It follows from Lemma \ref{zin} that $\sigma(u,\la)$ satisfies the
classical Riemann vanishing theorem. The choice $\sigma(u,0)=\sigma_{0}(u)$
defines the value of $\varphi_{0}$ uniquely.

Demonstrate the main steps of general construction for the case of elliptic curves.
To this end we consider the family of curves of
the form
\begin{equation*}\Gamma=\{(x,y,g_2,g_3)\in \mathbb{C}^4\mid 4x^3-y^2-g_2x-g_3=0\}.
\end{equation*}

{\bf (i)}  \textit{The frame and the discriminant}. The discriminant of the family
is given by the formula $\Sigma=\{(g_2,g_3)\mid g_2^3-27g_3^2=0\}$. The vector
fields tangent to $\Sigma$ are
\begin{gather*}
L_0=4g_2\pa_{g_2}+6g_3\pa_{g_3},\quad
L_2=6g_3\pa_{g_2}+\frac{1}{3}g_2^2\pa_{g_3},\quad (L_0,L_2)\Delta=(12,0)\Delta,
\end{gather*}
where $\Delta(g_2,g_3)=\frac{4}{3}(g_2^3-27g_3^2)$ is the determinant of the
coefficients of the frame $\{L_0,L_2\}$.

{\bf (ii)} \textit{The vector of canonical differentials}. Consider the
differential $D(x)^T=(-x,1)\dfrac{\mathrm{d}x}{y}$ with the period matrix
\begin{equation*}\Pi=
\begin{pmatrix}2\eta&2\eta'\\2\omega&2\omega'\end{pmatrix}=
\bigg(\oint_{a}D(x),\oint_{b}D(x)\bigg),\quad (a,b)\in
H_1({\scriptstyle\Gamma},\mathbb{Z}),\quad a\circ b=1,\quad
{\scriptstyle\Gamma}\in\Gamma,\end{equation*} which satisfies the Legendre
relation $\Pi^TJ\Pi=2\pi\imath\,J$, $J=\begin{pmatrix}0&1\\-1&0\end{pmatrix}$.

The further steps of the construction are as follows.

{\bf (iii)} \textit{Calculation of $A^{(i)}=L_{i}(\Pi)\Pi^{-1}$}. By applying $L_0$
and $L_2$ to $D(x)$, we obtain
\begin{align*}
L_0(D(x))&=\begin{pmatrix}1&0\\0&-1\end{pmatrix}D(x)
-\pa_x\bigg[\begin{pmatrix}2x&0\\0&2x\end{pmatrix}D(x)\bigg],\\
L_2(D(x))&=
\begin{pmatrix}0&\frac{1}{12}g_2\\[2pt]-1&0\end{pmatrix}D(x)
-\pa_x\bigg[\begin{pmatrix}
\frac{1}{6}g_2&-\frac{1}{2}g_3\\[2pt]-2x&-\frac{1}{3}g_2\end{pmatrix}
D(x)\bigg].
\end{align*}
Neglecting the exact differential, we obtain
\begin{equation*} A^{(0)}=\begin{pmatrix}1&0\\0&-1\end{pmatrix},\qquad
A^{(2)}=\begin{pmatrix}0&\frac{1}{12}g_2\\-1&0\end{pmatrix}
\end{equation*} which implies that
\begin{equation*} M^{(0)}=\begin{pmatrix}0&-1\\-1&0\end{pmatrix},\qquad
M^{(2)}=\begin{pmatrix}-1&0\\0&-\frac{1}{12}g_2\end{pmatrix}.\end{equation*}

{\bf (iv)}  \textit{Calculation of $\varphi_0$}. We have $\chi_i=\varphi_0
L_i(\Delta)\Delta^{-1}$. Set $\varphi_0=(\mu+1/2)/12$. Then
\begin{equation*}Q_0=-u\pa_u+4g_2\pa_{g_2}+6g_3\pa_{g_3}+\mu,\qquad
Q_2=-\frac{1}{2}\pa_u^2-\frac{1}{24}g_2u^2+6g_3\pa_{g_2}+\frac{1}{3}g_2^2\pa_{g_3}.
\end{equation*}
Make the change of variables $(g_2,g_3)\to(t g_2,t g_3)$ and pass to the limit
as $t\to 0$. We obtain the linearized operators
\begin{equation*}
Q^{\mathrm{lin}}_0=-u\pa_u+4g_2\pa_{g_2}+6g_3\pa_{g_3}+\mu,\qquad
Q^{\mathrm{lin}}_2=-\frac{1}{2}\pa_u^2+6g_3\pa_{g_2}.
\end{equation*}

Consider the system (\ref{Q:sys}). Let
$\phi(u,g_2,g_3)$ be a solution of this system in the class of entire
functions of $(u,g_2,g_3)$. Then $\phi(u,0,0)$ is a solution of the system
$Q^{\mathrm{lin}}_i(\psi)=0$, $i=0,2$. The nontrivial solutions of the latter system are
$\psi=u$ (then $\mu=1$) and $\psi=1$ (then $\mu=0$). To the nonconstant
polynomial $\phi(u,0,0)=u$, there corresponds the value $\varphi_0=1/8$.

\begin{remark}
Along with the linearization, it is useful to consider the degeneration of the
operators $Q_i$ for arbitrary values of $\mu$. Set $(g_2,g_3)=(3t^4,t^6)$. We
obtain
\begin{equation*}Q^{\mathrm{sing}}_0=-u\pa_u+t\pa_{t}+\mu,\qquad
Q^{\mathrm{sing}}_2=-\frac{1}{2}\pa_u^2-\frac{1}{8}t^4u^2+\frac{1}{2}t^3\pa_{t}.
\end{equation*}
The general solution of the system $Q^{\mathrm{sing}}_i(\psi)=0$, $i=0,2$ has the form
\begin{equation*}
\psi(u,t)=t^{-\mu}\exp\{t^2u^2\!/4\}[C_0 \cos(t u\sqrt{\mu+1/2})+C_1\sin(t
u\sqrt{\mu+1/2})].
\end{equation*}
For the integer values $\mu\leqslant 0$, the solution $\psi(u,t)$ is an entire
function for arbitrary values of the constants $C_0$ and $C_1$. For $\mu=1$, an
entire solution is obtained only if $C_0=0$. Moreover, a function $\psi(u,t)$
can be not representable as an entire function of the form $\phi(u,3t^4,t^6)$.
For instance, the representation is impossible for $\mu=-1,-2$; it is necessary
that $C_1=0$ for the even values of $\mu$ and that $C_0=0$ for the odd values
of $\mu$.
\end{remark}

{\bf (v)}  \textit{The sigma function}. The entire function satisfying the system
$Q_i(\psi)=0$ and normalized by the condition $\psi(u,0,0)=u$ is the
Weierstrass\ elliptic function $\sigma(u,g_2,g_3)$.

The relations (for their original proof, see \cite{wei82})
\begin{equation*}
\begin{split}
&Q_0(\sigma)=-u\sigma_u+4g_2\sigma_{g_2}+6g_3\sigma_{g_3}+\sigma=0,\\
&Q_2(\sigma)=-\tfrac{1}{2}\sigma_{u,u}-\tfrac{1}{24}g_2u^2\sigma
+6g_3\sigma_{g_2}+\tfrac{1}{3}g_2^2\sigma_{g_3}=0,
\end{split}
\end{equation*}
lead, in particular, to the following results.

One has the power series expansion \cite{wei82}
\begin{equation*}
\sigma(u,g_2,g_3)=u\sum_{i,j\geqslant
0}\frac{a_{i,j}}{(4i+6j+1)!}\Big(\frac{g_2u^4}{2}\Big)^i(2 g_3u^6)^j
\end{equation*}
with the \emph{integer} coefficients $a_{i,j}$ defined by the recursion
\begin{equation*}
a_{i,j}=3(i+1)a_{i+1,j-1}+\tfrac{16}{3}(j+1)a_{i-2,j+1}
-\tfrac{1}{3}(2i+3j-1)(4i+6j-1)a_{i-1,j}
\end{equation*}
with the initial conditions $\{a_{0,0}=1;\, a_{i,j}=0, \min(i,j)<0 \}$.

Let $\mathscr{E}$ be the field of elliptic functions, i.e., doubly periodic
functions of $u$ with the ``invariants'' $(g_2,g_3)$. Let us use the standard
notation $\zeta=\pa_u\log\sigma$, $\wp=-\pa_u\zeta$, and $\wp'=\pa_u\wp$. The
operators~\cite{FS:1882}
\begin{equation*}
\xi_0=-u\pa_u+L_0,\quad\xi_1=\pa_u,\quad \xi_2=-\zeta\pa_u+L_2,
\end{equation*}
are the generators of the Lie algebra $\mathrm{Der}\mathscr{E}$ of the
derivations of the field $\mathscr{E}$.

\begin{remark} We have
\begin{equation}\label{fs:comm}
[\xi_0,\xi_k]=k\xi_k,\quad[\xi_1,\xi_2]=\wp\xi_1.
\end{equation}
On the other hand, according to the formulas
\begin{align*}
[(\xi_0,\xi_1,\xi_2),g_2]=(4g_2,0,6g_3),&\qquad
[(\xi_0,\xi_1,\xi_2),g_3]=(6g_3,0,g_2^2/3),\\
[(\xi_0,\xi_1,\xi_2),\wp]=(2\wp,\wp',2\wp^2-g_2/3),&\qquad
[(\xi_0,\xi_1,\xi_2),\wp']=(3\wp',6\wp^2-g_2/2,3\wp\wp'),
\end{align*}
the operators act on the ring of polynomials $\mathbb{C}[\wp,\wp',g_2,g_3]$ as
vector fields. One can immediately show that the vector fields thus defined
satisfy relations \eqref{fs:comm} only if ${\wp'}^2=4\wp^3-g_2\wp-g_3$. Any
elliptic function $f\in\mathscr{E}$ has a unique representation of the form
$f=f_1(\wp,g_2,g_3)+f_2(\wp,g_2,g_3)\wp'$, where $f_1$ and $f_2$ are rational
functions. The mapping $\rho\colon f\mapsto(f_1,f_2)$ establishes an
isomorphism between the field $\mathscr{E}$ and the two-dimensional module $E$
over the field $\mathbb{C}(\wp,g_2,g_3)$. The ordinary multiplication of
elliptic functions defines an associative commutative bilinear mapping
$\ast\colon E\times E\to E$. Let $f,h\in\mathscr{E}$. Then
$\rho(fh)=\rho(f)\ast\rho(h)=(f_1h_1+p f_2h_2, f_1h_2+h_1f_2)$, where
$p=4\wp^3-g_2\wp-g_3$. We obtain
\begin{gather*}
\xi_0(f_1,f_2)=X_0(f_1,f_2)+(0,3f_2),\quad
\xi_1(f_1,f_2)=(0,1)\ast{X}_{-2}(f_1,f_2)+{X}_{-2}((0,1)\ast(f_1,f_2)),\\
\xi_2(f_1,f_2)=X_2(f_1,f_2)+(0,3\wp f_2),
\end{gather*}
where ${X}_{-2}=\frac{1}{2}\pa_{\wp}$, $X_0=L_0+2\wp\pa_{\wp}$, and
$X_2=L_2+\frac{1}{3}(6\wp^2-g_2)\pa_{\wp}$ are operators generating the
polynomial Lie algebra
\begin{equation*}
[X_0,X_{-2}]=-2X_{-2},\quad [X_0,X_{2}]=2X_{2},\quad [X_{-2},X_{2}]=4\wp
X_{-2}.
\end{equation*}
\end{remark}

\begin{defn}\label{ns}
Let $n$ and $s$ be a pair of coprime integers such that $s>n \ge 2$.
We will call $(n,s)$-curve an algebraic curve that belongs to the
family
$$\Gamma=\{(x_1,x_2,\la)\in\mathbb{C}^{d+2}\mid f(x_1,x_2; \la)=0\}$$ with the
polynomial
\begin{equation}\label{n-s}
f(x_1,x_2;\la)=x_1^s+x_2^n+\sum_{a, b} \la_{n s-a n-b s}x_1^{a}x_2^{b},
\end{equation}
where $0\leqslant a\leqslant s-2$, $0\leqslant b \leqslant n-2$, and $a n+b s<
ns$.
\end{defn}

We have $d =(n+1)(s+1)/2-[s/n]-3$. Set $\deg x_1=n$ and $\deg x_2=s$. Then
$f(x,\la)$ is a homogeneous polynomial of weight $ns$.

 Let us describe principal steps of the construction in the case of $(n,s)$-curves. It is convenient to introduce the following
grading. Write $\deg\la_i=i$, $\deg u_i=-i$, and $\deg L_i=i$. In contrast to
the previous sections, the number of the parameters $\la$ is denoted below by
$d$.

{{\bf (i)}}. \textit{The polynomial $\sigma_0(u)$}. The
Weierstrass sequence $(w_1,w_2,\dots)$ is the ascending set of positive
integers that are not representable in the form $an+bs$ with nonnegative
integers $a$ and $b$. Set $w(\xi)=\sum_{i}\xi^{w_i}$. We have
\begin{equation*}
w(\xi)=\frac{1}{1-\xi}-\frac{1-\xi^{ns}}{(1-\xi^n)(1-\xi^s)}\,.
\end{equation*}
This implies that the length $g=w(1)$ of the Weierstrass sequence and the sum
$G=w'(1)$ of the elements of this sequence are given by the formulas
\begin{equation*}
g=\frac{(n-1)(s-1)}{2},\qquad
G=\frac{ns(n-1)(s-1)}{4}-\frac{(n^2-1)(s^2-1)}{12}\,.
\end{equation*}
Define the Schur--Weierstrass\ polynomial $\sigma_{0}(u_{w_1},\dots,u_{w_g})$
by the formula
\begin{equation}\sigma_{0}\bigg(\frac{p_{w_1}}{w_1},\dots,\frac{p_{w_g}}{w_g}\bigg)=
c\, \frac{\det(\xi_{i}^{w_j})}{\det(\xi_{i}^{j-1})},\qquad
i,j=1,\dots,g,\label{Sch}
\end{equation}
where $p_j=\sum_{i=1}^g\xi_i^j$ is Newton's symmetric polynomial and $c
\in\mathbb{C}$. We have
\begin{equation*}
\deg \sigma_{0}(u_{w_1},\dots,u_{w_g})=-\sum_{k=1}^gw_{k}-k+1=
-G+\frac{(g-1)g}{2}=-\frac{(n^2-1)(s^2-1)}{24}\,.
\end{equation*}
Let us normalize the polynomial $\sigma_0(u)$ by the condition
\begin{equation*}
\sigma_0(u_1,0,\dots,0)=u_1^{(n^2-1)(s^2-1)/24}.
\end{equation*}

{{\bf (ii)}}. \textit{The frame and the discriminant}. For brevity, we denote below the polynomial
\eqref{n-s} simply by $f(x)$, its derivatives with respect to $x_1$ and $x_2$
by $f_1(x)$ and $f_2(x)$, and $(f_1(x),f_2(x))$ by $\pa_x f(x)$, respectively.

Let $P=\mathbb{C}[\la]$. Consider the local ring $\mathcal{R}=P[x]/\pa_x f(x)$.
The set of monomials $\mathscr{M}=\{x_1^ix_2^j\}$, $0\leqslant i\leqslant s-2$,
$0\leqslant j\leqslant n-2$, forms a basis of the ring $\mathcal{R}$. Denote by
$I=\{i_1=4g-2,i_2,\dots,i_{2g}=0\}$ the descending set of weights $\{i n+j s\}$
of the elements of $\mathscr{M}$. The ordering of $I$ is strict, and therefore
we can introduce the vector $e(x)=(e_{i_1}(x),\dots,e_{i_{2g}}(x))^T$, where
$e_{i n+j s}(x)$ stands for the monomial $x_1^ix_2^j$.

Let $\mathscr{J}$ be the ideal in $P[x,z]$ generated by the polynomials
$f_1(x),f_1(z),f_2(x),f_2(z)$. Write
\begin{equation*}
H(x,z)=\frac{1}{2}\begin{vmatrix} \dfrac{f_1(x_1,x_2)-f_1(z_1,z_2)}{x_1-z_1}&
\dfrac{f_2(x_1,x_2)-f_2(z_1,z_2)}{x_1-z_1}\\[8pt]
\dfrac{f_1(z_1,x_2)-f_1(x_1,z_2)}{x_2-z_2}&
\dfrac{f_2(z_1,x_2)-f_2(x_1,z_2)}{x_2-z_2}
\end{vmatrix}.
\end{equation*}
The polynomial $H(x,z)$ has the following properties:

\begin{enumerate}

\item[(a)] $H(x,x)=\mathrm{Hess}_xf(x)$;

\item[(b)] $H(x,z)=H(z,x)$;

\item[(c)] $H(x,z)F(x,z)\equiv H(z,x)F(z,x)\mod\mathscr{J}$ for an arbitrary
$F(x,z)\in P[x,z]$.

\end{enumerate}

The formula
\begin{equation*} \sum_{i,j\in I}^{2g}
v_{i,j}(\la)e_i(x)e_j(z)=f(x)H(x,z)\mod\mathscr{J}
\end{equation*}
defines a symmetric polynomial matrix $V(\la)=(v_{i,j})$. The matrix defines
$2g$ vector fields whose generating function is
\begin{equation} L(x)=\sum_{i\in I} e_i(x)\,
L_{2(ns-n-s)-i}=\sum_{i,j\in I} e_{i}(x)\,
v_{i,j}(\la)\,\pa_{\la_{ns-j}}.\label{reper}
\end{equation}
Note the following important fact:
$L(x)\Delta(\la)=H(x,x)\Delta(\la)=\Delta(\la) \mathrm{hess}_{x}f(x)$.

{\bf (iii)}. \textit{The vector of canonical differentials} \cite{ba97}. Introduce
the operator
\begin{equation*}
\mathcal{D}_x=f_{2}(x)\pa_{x_1}-f_{1}(x) \pa_{x_2}.
\end{equation*}
Write
\begin{align*}
\mathsf{D}(x,z)= \mathcal{D}_z\Big[\frac{f(z_1,x_2)
-f(z_1,z_2)}{(x_1-z_1)(x_2-z_2)}\Big] -\mathcal{D}_x\Big[\frac{f(x_1,z_2)
-f(x_1,x_2)}{(z_1-x_1)(z_2-x_2)}\Big].
\end{align*}
Let $\Psi$ be a $(2g\times 2g)$ matrix $\Psi=\begin{pmatrix}
\Psi_{1,1} & \Psi_{1,2} \\
\Psi_{2,1} & \Psi_{2,2} \\
\end{pmatrix}$ which is a polynomial in $\la$, where
$\Psi_{2,1}=0$, $\Psi_{2,2}=1_g$, and $\Psi_{1,2}$ is lower triangular with
zeros on the principal diagonal.
\begin{lem}\label{diff} There exists a unique matrix
$\Psi$ of the above form that satisfies the equation
\begin{equation*}
e(z)^T\Psi^TJ\Psi e(x)=\mathsf{D}(x,z).\end{equation*} Using the matrix, one
can represent the vector $D(x)$ of the canonical basis differentials as
\begin{equation*}D(x)=\Psi
e(x)\dfrac{\mathrm{d}x_1}{f_2(x)},\end{equation*} and $\deg
D(x)=(-w_1,\dots,-w_g,w_1,\dots,w_g)^T$.
\end{lem}

{{\bf iv}}. \textit{Calculation of $A^{(i)}$}. Set $R(x)=\Psi e(x)$.

\begin{lem} \label{Didea} Let $\mathfrak{D}$ be the linear space
spanned by the functions $\mathcal{D}_x\smash{\Big[\dfrac{h(x)}{f_2(x)}\Big]}$,
where $h(x)$ ranges over the ring $P[x]$. Then
\begin{equation*}
A^{(i)}R(x)\equiv L_i(R(x))-R(x)\cdot \frac{L_i(f_2(x))}{f_2(x)}\mod
\mathfrak{D}.
\end{equation*}
\end{lem}

\begin{proof} By definition, $L_i(D(x))=A^{(i)}D(x)+\mathrm{d}r(x)$,
where $r(x)$ is a $2g$-dimensional vector of functions in the field $P(x)$. On
one hand,
\begin{equation*}
L_i(D(x))=\bigg\{\dfrac{L_{i}(R(x))}{f_2(x)}+
L_i\bigg(\dfrac{1}{f_2(x)}\bigg)R(x)\bigg\}\mathrm{d} x_1.
\end{equation*}
On the other hand, the differential along the fiber $\mathsf{p}^{-1}(\la)$ of a
function $r(x)$ in $P(x)$ is represented as
$\mathrm{d}r(x)=\mathcal{D}_x[r(x)]\dfrac{\mathrm{d} x_1}{f_2(x)}$. Setting
$r(x)=\dfrac{h(x)}{f_2(x)}$, where $h(x)$ is a polynomial vector in $x$, we
obtain the assertion of the lemma.\qed
\end{proof}

{{\bf v}}. \textit{Calculation of $\varphi_0$}. The grading introduced above
enables us to evaluate $\varphi_0$.

\begin{lem} For any pair
$(n,s)$ we have $\varphi_0=1/8$.
\end{lem}

\begin{proof}
The vector field $L_0=\sum_{i\in I}(ns-i)\la_{ns-i}\pa_{\la_{ns-i}}$ is an
Euler operator. Its representative is the operator
$Q_0=L_0-\sum_{j=1}^gw_ju_{w_j}\pa_{u_{w_j}}+\delta^{(0)}$ (cf. Lemma
\ref{diff}). Consider the equation
\begin{equation*}
\chi_0+\tfrac{1}{2}\mathrm{Tr}\beta^{(0)}+\deg\sigma_0=0.
\end{equation*}
Note that $\beta^{(0)}=-\mathrm{diag}(w_1,\dots,w_g)$ and that
$\chi_0=\varphi_0\deg\Delta$ and $\deg\Delta(\la)=ns(n-1)(s-1)$ by
construction. We obtain
\begin{equation*}
\varphi_0 ns(n-1)(s-1)-\frac{G}{2}-\frac{(n^2-1)(s^2-1)}{24}=0.\eqno{\qed}
\end{equation*}
\end{proof}

{{\bf (vi)}}. \textit{The sigma function}. Summing up, we come to the following
constructive description.
\begin{theorem} Let the family of $(n,s)$-curves defined by
a polynomial of the form \eqref{n-s} be given. Take operators $Q_{i}$, $i\in
I$, of the form \eqref{Q:op} such that

\begin{enumerate}

\item[(a)] the polynomial vector fields $L_{i}$ are given by the generating
function \eqref{reper};

\item[(b)] the polynomial matrices $M^{(i)}=J A^{(i)}$ are defined by Lemma
{\ref{Didea}};

\item[(c)] the functions $\delta^{(i)}$ have the form
$\chi^{(i)}+1/2\mathrm{Tr}\beta^{(i)}$, where the generating function of the
polynomials $\chi^{(i)}$ is the Hessian of the defining polynomial multiplied
by $1/8$.

\end{enumerate}
Then the entire function satisfying the system $\{Q_i(\psi)=0\}$ and normalized
by the condition
\begin{equation*}
\psi(u_1,0,\dots,0)=u_1^{\ell},\quad\text{where}\;
\ell=\dfrac{(n^2-1)(s^2-1)}{24},\end{equation*} is the sigma function of the
family of $(n,s)$-curves.
\end{theorem}

In the case when $(n,s)$-curve is hyperelliptiv more effective description of principal steps of the construction can be presented.
The case of hyperelliptic curves is related to
$(n,s)=(2,2g+1)$; the corresponding Weierstrass\ sequence is
$(1,3,\dots,2g-1)$, and hence $G=g^2$ and $\deg \sigma_0(u)=-(g-1)g/2$.

The above desribed step leeds to more effective answers in the case of hyperelliptic curve
Let $\Gamma$ be the family of plane hyperelliptic curves of the form
\begin{equation*}
\Gamma=\{(x,y,\la)\in\mathbb{C}^{2g+1}\mid y^2-f(x)=0\},\qquad
f(x)=4x^{2g+1}+\sum_{k=0}^{2g-1}\la_{k}x^k.
\end{equation*}
We have $\deg x=2$, $\deg y=2g+1$, $\deg\la_k=2(2g+1-k)$, and $\deg
u_i=-2(g-i)-1$.

Let us use the technique of generating functions. In particular, we consider
$f(x)$ as the generating function of $\la_0,\dots,\la_{2g-1}$. Set
\begin{equation*}
L(x)=x^{2g-1}\sum_{k=0}^{2g-1}x^{-k}L_k
\end{equation*}
for the generators of the frame $L_0,\dots,L_{2g-1}$, where $\deg L_j=2j$.

\begin{theorem}\label{theorem2.5}
The structure functions of the polynomial Lie algebra generated by the frame
associated with the family $\Gamma$ of hyperelliptic curves are given by the
formulas
\begin{align*}
[L(x),L(z)]&=\mspace{-10mu}\sum_{i,j,k=0}^{2g-1}x^iz^jc_{i,j}^{k}(\la)L_k=
(\pa_x-\pa_z)\bigg[\frac{f(z)L(x)-f(x)L(z)}{x-z}\bigg]\mod(f'(x),f'(z)),\\
[L(x),f(z)]&=\sum_{i,j=0}^{2g-1}x^iz^jv_{i,j}(\la)=
\frac{f'(x)f(z)-f'(z)f(x)}{x-z}\mod(f'(x),f'(z)).
\end{align*}
\end{theorem}

\setcounter{theorem}{5}

For an arbitrary polynomial $F(x)$, write
$\mathsf{d}_{x}^{k}F(x)=(F(x)/x^{k})_{+}$, where the symbol $(\cdot)_{+}$
indicates that the terms containing negative powers of $x$ are discarded. Then
it follows from Theorem~\ref{theorem2.5} that
\begin{equation*}
L(x)=f(x) \sum_{k=0}^{2g-1}\mathsf{d}_{x}^{k+1}(f'(x))\pa_{\la_{k}}\mod f'(x).
\end{equation*}
The canonical basis differential is given by the formula
\begin{equation*}
D(x)=(R_1(x),\dots,R_{2g}(x))^T\frac{\mathrm{d}x}{y},\quad\text{where
$R_i(x)=\frac{1}{4}x^i\pa_x\mathsf{d}_{x}^{2i}f(x)$ and $R_{g+i}(x)=x^{i-1}$,
$i=1,\dots,g$.}
\end{equation*}

Our construction leads to the following result.

\begin{theorem}
Set
\begin{equation*}
h(x)=\sum_{i=1}^{g} x^{i-1}\pa_{u_i}+R_{i}(x)u_i,\qquad t(x)=\sum_{i,j=1}^g
\frac{1+\mathrm{sign}(i-j)}{2}\,j
u_j\,\mathsf{d}_{x}^{j+1}(x^{i-1}\pa_{u_i}+R_{i}(x)u_i),
\end{equation*}
and define the generating function $Q(x)$ of the second-order linear
differential operators $\{Q_{k}\}$, $k=0,\dots,2g-1$, by the formula
\begin{equation*}Q(x)=x^{2g-1}\sum_{k=0}^{2g-1}x^{-k}Q_{k}=L(x)+\frac{1}{8}f''(x)
-f(x)t(x)-h(x)\circ h(x) \mod f'(x).
\end{equation*}
Then the entire function $\psi(u,\la)$ normalized by the condition
$\psi(0,\dots,u_g,0,\dots,0)=u_g^{\ell}$, where $\ell=(g-1)g/2$, and satisfying
the equation $Q(x)\psi=0$ identically with respect to $x$ is the hyperelliptic
sigma function $\sigma(u,\la)$.
\end{theorem}



\section{Differentiation of Abelian functions on parameters}
\subsection{Statement of the problem} Recall that an \emph{Abelian function} is a meromorphic
function on a complex Abelian torus $T^g=\mathbb{C}^g\!/\Gamma$,
where $\Gamma\subset\mathbb{C}^g$ is a lattice of rank $2g$. In
other words, a meromorphic function $f$ on $\mathbb{C}^g$ is
Abelian if and only if $f(u)=f(u+\omega)$ for all
$u\in\mathbb{C}^{g}$ and $\omega\in\Gamma$. The Abelian
functions on $T^g$ form a differential field.


Let $\mathcal{F}$ be the field of Abelian functions
on the Jacobian of a genus $g$ curve.

The differential field $\mathcal{F}$ has the following
properties:

1. Let $f\in\mathcal{F}$; then
$\partial_{u_i} f\in \mathcal{F}$, $i=1,\dots,g$.

2. Let $f_1,\dots,f_{g+1}$ be any nonconstant
functions in $\mathcal{F}$; then
there exists a polynomial $P$ such that $P(f_1,\dots,
f_{g+1})(u)=0$ for all $u\in T^g$.

3. Let $f\in\mathcal{F}$ be a nonconstant
function; then any $h\in\mathcal{F}$ can be
expressed as a rational function of
$(f,\partial_{u_1}f,\dots,\partial_{u_g}f)$.

4. There exists an entire function
$\vartheta\colon\mathbb{C}^g\to\mathbb{C}$ such that
$\partial_{u_i,u_j}\log\vartheta \in \mathcal{F}$, $i,j=1,\dots,g$.

Further we need the following interpretation of the general construction from the Chapt. 1.
Let $B$ be an open dense subset of $\mathbb{C}^d$.
Consider a family $V$ of algebraic curves that
have constant genus $g$ and smoothly depend on the parameter $b\in B$. We
use the family $V$ to define a space $U$ of
Jacobians over $B$. The space $U$ has a natural structure of a smooth manifold and is the
total space of the bundle $p\colon U\to B$, where the fiber over a point $b\in B$ is the Jacobian
$J_b$ of the curve with parameter $b$.

Let $F$ be the field of
$C^{\infty}$ functions $f$ on $U$ such that the restriction of $f$ to each fiber
$J_b$ is an Abelian function.

This section deals with the following problem which is natural to divide in three parts:

(a) \textit{Find generators of the $F$-module $\Der F$ of
derivations of the field $F$.}

(b) \textit{Find commutation relations with
coefficients in $F$ for the generators of
the $F$-module $\Der F$, that is, describe the corresponding Lie
algebra structure of $\Der F$ over $F$}.

(c) \textit{Describe the action of $\Der F$ on $F$}.

\subsection{Case of the family of elliptic curves} Consider
the family
$$
V=\{(x,y,g_2,g_3)\in\mathbb{C}^2\times B\mid y^2=4x^3-g_2x-g_3\}
$$
of plane algebraic curves with parameter space
$B=\{(g_2,g_3)\in\mathbb{C}^2\mid \Delta\neq
0\}$, where $\Delta=g_2^3-27g_3^2$. The
curves in this family have constant
genus $g=1$. These curves are called Weierstrass
elliptic curves.

The fiber $T^1$ of the bundle $U$ over a point $b=(g_2,g_3)\in B$
has the form $\mathbb{C}^1\!/\Gamma$.
Here
$\Gamma=\big\{\oint_\gamma\frac{dx}{y}\big\}$, where
$\gamma$
runs over the set of cycles on the curve
$V_b=\{(x,y)\in\mathbb{C}^2\mid y^2=4x^3-g_2x-g_3\}$, is a
rank $2$ lattice.

The vector field $\ell_1=\partial_u$ is tangent to a fiber of
$U$, while the vector fields
$\ell_0=4g_2\partial_{g_2}+6g_3\partial_{g_3}$ and
$\ell_2=6g_3\partial_{g_2}+\frac{1}{3}g_2^2\partial_{g_3}$ are
tangent to the discriminant
$\{(g_2,g_3)\in\mathbb{C}^2\mid \Delta = 0\}$ of
the family (because $\ell_0\Delta=12\Delta$ and
$\ell_{2}\Delta=0$) and hence
form a basis of vector fields on the base
$B$.

The field $F$ of fiberwise Abelian functions is generated by
the coordinate functions $g_2$ and $g_3$ on
the base and by the Weierstrass elliptic functions
$\wp(u,g_2,g_3)$ and
$\wp'(u,g_2,g_3)=\partial_u\wp(u,g_2,g_3)$,
which are related by the identity $(\wp')^2 = 4\wp^3-g_2\wp-g_3$.

The operator $L_1=\ell_1 = \partial_u$ is
obviously a derivation of $F$.

To find the other two generators of the $F$-module $\Der F$, which
correspond to the basis vector
fields $\ell_0$ and $\ell_2$ on the base, we use the
following properties of the Weierstrass sigma function
$\sigma(u,g_2,g_3)$, which is an entire function of the
variables $(u,g_2,g_3)\,{\in}\,\mathbb{C}^{3}$\kern-0.5pt:

(a) $\partial_{u}^{2}\log\sigma(u,g_2,g_3)=-\wp(u,g_2,g_3)\in
F$.

(b) The function $\sigma(u,g_2,g_3)$ is a solution of the
system of linear differential equations $Q_0\sigma=0$,
$Q_2\sigma=0$, where
$$
Q_0=4g_2\partial_{g_2}+6g_3\partial_{g_3}-u\partial_u+1, \qquad
Q_2=6g_3\partial_{g_2}+\tfrac{1}{3}g_2^2\partial_{g_3}-\tfrac{1}{2}\partial_{u}^2-\tfrac{1}{24}g_2u^2.
$$
It is important to note that the coefficients of $Q_0$ and $Q_2$
are polynomial in the variables
$(u,g_2,g_3)\in\mathbb{C}^{3}$, see details in preceeding section.

Let us show how properties (a) and (b) can be used to reveal the
form of the derivations of $F$. Let us start from the equation
$Q_2\sigma=0$. We have
$Q_2=\ell_2-\frac{1}{2}\partial_{u}^2-\frac{1}{24}g_2u^2$. Let us
divide $Q_2\sigma$ by $\sigma$ and rearrange the terms with the use
of the Weierstrass functions $\zeta=\partial_u \log \sigma$ and
$\wp=-\partial_u^2 \log \sigma$. We obtain
$\ell_2\log\sigma-\frac{1}{2}\zeta^2+\frac{1}{2}\wp-\frac{1}{24}g_2u^2=0$.
Let us apply $\partial_u^2$; then, since $[\partial_u,\ell_2]=0$,
we arrive at the relation
$-\ell_2\wp+\zeta\wp'-\wp^2+\frac{1}{2}\wp''-\frac{1}{12}g_2=0$,
where $\wp''=\partial_u^2\wp$. Thus,
$(\ell_2-\zeta\partial_u)\wp=-\wp^2+\frac{1}{2}\wp''-\frac{1}{12}g_2\in
F$. Further, since
$[\ell_2-\zeta\partial_u,\partial_u]=-\wp\partial_u$, it follows
that $(\ell_2-\zeta\partial_u)\wp'\in F$. By applying
$\ell_2-\zeta\partial_u$ to a function that depends only on the
base coordinate functions $g_2$ and $g_3$ we obviously
obtain a function of $g_2$ and $g_3$. Therefore, the linear
differential operator $L_2=\ell_2-\zeta\partial_u$ takes the
generators of $F$ to elements of $F$. Consequently, $L_2$ is a
derivation of $F$.

A similar calculation for the operator $Q_0$, which also
annihilates the sigma function, yields one more derivation
$L_0=\ell_0-u\partial_u$.

The three generators
$$
L_0=-u\partial_u+4g_2\partial_{g_2}+6g_3\partial_{g_3},\quad
L_1=\partial_u,\quad
L_2=-\zeta(u,g_2,g_3)\partial_u+6g_3\partial_{g_2}+\tfrac{1}{3}g_2^2\partial_{g_3}
$$
of the $F$-module $\Der F$ were originally found by Frobenius
and Stickelberger \cite{FS:1882}.

The $F$-module $\Der F$ with generators $L_0$, $L_1$,
and $L_2$ is a Lie algebra over $F$ with
commutation relations
\begin{equation*}
[ L_0, L_k]=k L_k,\quad k=1,2,\qquad[ L_1, L_2]=\wp(u,g_2,g_3) L_1.
\end{equation*}

\subsection{Case of family of $\boldsymbol{(n,s)}$-curves}
Our aim is to single out
cases of the derivation problem that are
sufficiently general and at the same time can be
solved as efficiently as the elliptic case above descibed.
To achieve this, we

1. Choose a special family $V\to B$ of plane
algebraic curves of constant genus $g$.

2. Construct a basis $\ell_1,\dots,\ell_{d}$ of
vector fields on $B$ acting on functions $f\in F$ as linear first-order
differential operators in the coordinates of the base $B$.

3. Construct operators $H_1,\dots,H_{d}$ acting on functions
$f\in F$ as second-order differential operators along the fibers of the bundle $U\to
B$ and such that the sigma function $\sigma$ associated with the
family $V$ satisfies the system of linear differential equations $(\ell_i-H_i)\sigma=0$,
$i=1,\dots,d$.

In the previous section the most part of the program was realized
for the family of $(n,s)$-curves. In this section we suggest a certain
different approach, that permits to comlete the program.
We will come back to the description of $(n,s)$-curves within this
new approach. For the convenience of a reader we are collecting
together necessary facts and details on the family of $(n,s)$-curves,
partially repeating the material of previous chapter.


 Let $V$ be
an irreducible algebraic curve of genus $g>0$ over $\mathbb{C}$.

Let $p$ be a point on $V$. Consider the ring $M(p)$ of rational
functions on $V$ that have poles in $p$ only. Construct
a sequence $S(p)=(S_1,S_2,\dots)$ by the following rule:
$S_k=1$ if there exists a function in $M(p)$
with a pole of order $k$; otherwise, $S_k=0$, $k\in\mathbb{N}$.

\begin{df}
A point $p$ is said to be
\textit{regular} if the sequence $S(p)$ is monotone. If
$S(p)$ is nonmonotone, then $p$ is called \textit{a
Weierstrass point}.
\end{df}

{\bf Example} Let $p$ be a regular point; then
$S(p)=(0,0,\dots,0,1,1,1\dots)$. The Weierstrass points are
classified according to the number of places
where $S(p)$ fails to be monotone. A point with a single failure is
said to be \emph{normal}, and a point with the
maximum number of failures, which is
equal to the genus, is said to be
\emph{hyperelliptic}. Thus, if $p$ is a
normal Weierstrass point, then $S(p)=(0,0,\dots,0,1,0,1,1,\dots)$,
and if $p$ is a hyperelliptic Weierstrass point, then
$S(p)=(0,1,0,1,\dots,0,1,1,\dots)$.

Let $p\in V$ be a Weierstrass point. By $V_p$ we
denote the curve $V$ with puncture at $p$. Consider the
ring $\mathcal{O}$ of entire rational functions on $V_p$.

\begin{df}
Let $\phi\in\mathcal{O}$ be a nonconstant function. The total number of zeros of $\phi$ on $C_p$ is
called the \emph{order} of $\phi$ and is denoted by $\mathrm{ord}\phi$. We set
$\mathrm{ord}\phi=0$ if $\phi$ is a nonzero constant.
\end{df}

If $\phi,\varphi\in\mathcal{O}$, then $\mathrm{ord}(\phi\varphi)=
\mathrm{ord}\phi+\mathrm{ord}\varphi$.

Note that every $\phi\in\mathcal{O}$ uniquely extends to a
rational function on $V$ with a pole of multiplicity $\mathrm{ord}\phi$ at $p$.

\begin{lem}
Let $\phi,\varphi\in\mathcal{O}$ be nonconstant functions. Then
$\varphi$ is an entire algebraic function of~$\phi$.
\end{lem}

Suppose that $x,y\in\mathcal{O}$ are nonconstant functions
such that $n=\mathrm{ord}x$ is minimum possible and
$s=\mathrm{ord}y$ is minimum possible under the condition
$\gcd(n,s)=1$. By multiplying $y$ by an appropriate numerical factor, one can always ensure that
$\mathrm{ord}(y^n-x^s)<ns$. Then, since $y$ is an entire algebraic
function of $x$, the function $y^n-x^s$ can be represented as a linear combination of monomials $x^iy^j$
such that $\mathrm{ord}(x^iy^j)=ni+sj<ns$.

\begin{df}  \label{ns1} The equation
\begin{equation*}
y^n-x^s=\sum_{i,j\ge 0}^{q(i,j)>0} \alpha_{i,j}x^iy^j,
\end{equation*}
where $q(i,j)=(n-j)(s-i)-ij$
 defines a Weierstrass model of the
curve $V$ with parameters $\alpha_{i,j}$. In the case of a general Weierstrass model the parameters
$\alpha_{i,j}$ considered as algebrailcally independent with gradding $q(i,j)$
\end{df}

The best-known example of a standard Weierstrass model of elliptic curve is given by the cubic equation
$y^2=4x^3-g_2x-g_3$  ($n=2$ and $s=3$). The previous analysis leeds the following result

\begin{lem}Let $V$ be an irreducible algebraic curve that admits two non-constant functions $\phi$ and $\psi$ with coprime numbers $\mathrm{ord}\,\phi$ and $\mathrm{ord}\,\psi$ Then $V$ has a Weierstrass model. \end{lem}

Note that
For $g> 1$, the Weierstrass model \emph{is
not unique}. All Weierstrass models of a same curve are birationally equivalent.

Following an idea due to Weierstrass \cite{W1904},
instead of a ``generic curve'' we consider classes
\begin{equation*}
V=\bigg\{(x,y; \la)\in\mathbb{C}^{2+d}\biggm| y^n=x^s+\sum_{i,j\ge 0}^{q(i,j)>0}
\la_{q(i,j)}x^iy^j\bigg\}
\end{equation*}
of models of plane algebraic curves, where
$d=\frac{1}{2}(n+1)(s+1)-1$ and $q(i,j)=(n-j)(s-i)-ij$. These
classes are indexed by pairs $(n,s)$ of
integers such that $s>n>1$ and $\gcd(n,s)=1$.

The genus of a curve defined by a
Weierstrass model in an $(n,s)$-class does not exceed $g=\frac{1}{2}(n-1)(s-1)$. For
example, $(n,s)=(2,2g+1)$ in the case of
hyperelliptic curves of genus $\le
g$.

We present below necessary notions and results of
singularity theory~\cite{AVGZ2004}. In singularity theory, the
zero set of the function
\begin{equation}
f(x,y,\la)=y^n-x^s-\sum_{j=0}^{n-2}\;\sum_{i=0}^{s-2} \la_{q(i,j)}x^iy^j\label{unfold}
\end{equation}
arises as a \emph{miniversal deformation} (also known as a
semi-universal unfolding) of the so-called \emph{Pham singularity}
$y^n-x^s=0$. The miniversal deformation depends on the $(n-1)(s-1)$
coordinates of the
vector~$\la=(\la_{q(0,0)},\dots,\la_{q(s-2,n-2)})$. The number
$m=\#\{\la_k\mid k<0\}$ is called the \emph{modality} of the
singularity $y^n-x^s=0$.

The discriminant~$\Sigma\subset\mathbb{C}^{2g}$ of
the miniversal deformation $f$
is defined as follows:
\[
(\la\in\Sigma)\iff\big(\exists\, (x,y)\in\mathbb{C}^2\colon f=f_x=f_y=0\,\text{ at the point
}\,(x,y,\la)\big).
\]
Thus, if $\la\notin \Sigma$, then the genus of a curve
in the family defined by the miniversal deformation \eqref{unfold} is
\emph{not less} than $\frac{1}{2}(n-1)(s-1)$. (Recall that
the genus of a curve defined by a Weierstrass model
in the $(n,s)$-class
\textit{does not exceed}$\frac{1}{2}(n-1)(s-1)$.)

Let us impose the condition
$\la_{q(i,j)}=0$ for $q(i,j)<0$ on a
miniversal deformation, or,
equivalently, the condition
$\la_{q(s-1,j)}=\la_{q(i,n-1)}=0$ on a Weierstrass model.
Then we obtain a family of constant genus~$g=(n-1)(s-1)/2$
curves over the set~$B=\mathbb{C}^{2g-m}
\cap(\mathbb{C}^{2g}\backslash \Sigma)$.

\begin{df} \label{nsfin}
 The intersection of classes of
miniversal deformations and Weierstrass models will be called
a family of $(n,s)$-curves (compare with the Definition~\ref{ns1})
\end{df}

In what follows, we consider the family $V=\{(x,y; \la)\in\mathbb{C}^{2}\times B\mid f(x,y,\la)=0\}$ of $(n,s)$-curves defined by the polynomial
$$
f(x,y,\la)= y^n-x^s-\sum_{\substack{0\le i<s-1\\
0\le j<n-1}}^{q(i,j)>0}\la_{q(i,j)}x^iy^j,\quad\text{where
}\,q(i,j)=(n-j)(s-i)-ij.
$$
Here $\dim_\mathbb{C}B= d=2g-m$.

Note some properties of the family of
$(n,s)$-curves. The Newton polygon of $f$ with
respect to the variables $(x,y)$ is a triangle. The
polynomial $f$ is homogeneous with respect to the grading in
which $\deg x=n$, $\deg y=s$, and $\deg \la_k=k$. A generic
curve in the family $V$,
viewed as an $n$-fold covering over~$S^2$, has
$(n-1)s+1=2g+n$ branch points, of which the point
at infinity has the branching number $n-1$, while the other
$2g+n-1$ branch points
have branching numbers equal to $1$ by the
Riemann--Hurwitz formula.

 Let
$M(x,y)=(M_1(x,y),\dots, M_{2g}(x,y))$ be the set of monomials $x^iy^j$,
$i=1,\dots,s-2$, $j=1,\dots,n-2$, in ascending order of weights.

{\bf Examples} Let $(n,s)=(2,2g+1)$; then $M(x,y)=(1,x,\dots, x^{2g-1}$ . Let $(n,s)=(3,5)$; then $M(x,y)=(1,x,y,x^2,xy,x^3,y x^2,x^4)$.

For the miniversal deformation \eqref{unfold},
there is an effective method, based on Zakalyukin's theorem
\cite{Zakaljukin76}, for constructing the discriminant and its tangent vector
fields. Let $f(x,y,\la)$ be given by \eqref{unfold}. The relation\vglue-9pt
$$
M_{i}(x,y)f(x,y,\la)= \sum_{j=1}^{2g}T_{i,j}(\la)M_{j}(x,y) \mod(\partial_x,\partial_y)f(x,y,\la)
$$
uniquely determines holomorphic functions $T_{i,j}(\la)$,
$i,j=1,\dots,2g$. We denote the $2g\times2g$ matrix of
these functions by $T(\la)$.

\begin{theorem}
The discriminant $\Sigma\subset\mathbb{C}^{2g}$ of the polynomial $f(x,y,\la)$ defined
in \eqref{unfold} is the set of zeros of the holomorphic function $\Delta(\la)=\det T(\la)$, $\Sigma=\{\la \in\mathbb{C}^{2g}\mid\Delta(\la)=0\}$. Let $I=(i_1,\dots, i_{2g})$
be the set of weights $\deg \la $ arranged in ascending order. The vector fields
$\ell_j=\sum_{k=1}^{2g}\!T_{2g-j,2g-k}(\la) \,\partial/\partial\la_{i_k}$\!, $j=1,\dots, 2g$, are
tangent to $\Sigma$; that is, $\ell_j\log\Delta(\la) \in \mathbb{C}[[\la
]]$. Moreover,
$\ell_1=\sum_{k=1}^{2g}i_k\la_{i_k}\partial/\partial\la_{i_k}$ is the Euler vector field.\end{theorem}

By taking linear combinations of the fields $\ell_i$ with
coefficients holomorphic in $\la $, one can reduce the holomorphic
frame $\mathcal{L}= (\ell_1,\dots,\ell_{2g})^T$ to a special form
with symmetric coefficient matrix (see \cite{bucley02:3}). Let us keep
the notation $T(\la)$ for the symmetrized matrix. Then $T(\la)$ is
Arnold's convolution matrix for the invariants of the Pham
singularity $y^n-x^s=0$ (see \cite{Arnold96}, \cite{Givental80}).
In the sequel, we assume that the frame $\mathcal{L}$ has been
symmetrized.

Let us proceed to the family of $(n,s)$-curves. We need to
construct vector fields tangent to $\Sigma_0=\Sigma\cap H$, where $H = \{\la
\in\mathbb{C}^{2g}\mid(\la_{i_1}, \dots, \la_{i_m})=0\}$. Clearly, the tangent bundle $T\Sigma_0$
consists of the vector fields in
$T\Sigma$ that are normal to $TH$ over $H$, that is, at all points where
$(\la_{i_1},\dots,\la_{i_m})=0$. Consider the cases $m=0$ and $m=1$ in more detail.

If $m=0$ (this is the case of so-called simple singularities), then $\Sigma_0=\Sigma$, and
we can use the frame $\mathcal{L}=(\ell_1,\dots,\ell_{2g})^T$ as a basis in $TB$.
In this case, the coefficients of the frame are polynomials in $\la $.

If $m=1$, then $TH$ has the single generator
$\eta_1={\partial}/{\partial \la_{i_1}}$. Obviously, the Euler
vector field $\ell_1$ is normal to $\eta_1$ in $H$. Next, for each
pair $(j,k)$ such that $1<j<k\le2g$ the vector
field $\xi_{j,k}=T_{k,1}(\la) \ell_{j}-T_{j,1} (\la)\ell_{k}$ is
tangent to $\Sigma$ and normal to $\eta_1$ in $H$. The vector
fields $\xi_{j,k}$ satisfy the relations $T_{l,1}(\la)
\xi_{j,k}-T_{k,1}(\la)\xi_{j,l}+T_{j,1}(\la) \xi_{k,l}=0$,
$1<j<k<l\le 2g$. Therefore, for a basis in $TB$ we can take the
frame consisting of the $2g-1$ vector fields
$(\ell_1,\xi_{2,3}\dots,\xi_{2,2g})^T$. Being restricted to~$H$,
the coefficients of the frame prove to be polynomials in~$\la $.

For arbitrary $m$, a similar construction leads to a collection of $2g-m$ polynomial
vector fields. In what follows, we denote the frame in $TB$ formed by these vector fields by $\mathcal{L}=(\ell_1,\dots,\ell_{2g-m})^T$.
Let $c_{ij}^{h}(\la)$ be the structure functions of the frame
$\mathcal{L}$; that is,
\begin{equation}
[\ell_i,\ell_j]=\sum_{h=1}^{2g-m}c_{ij}^{h}(\la)\ell_h, \qquad i,j=1,\dots,2g-m.\label{struct}
\end{equation}
\subsection{Gauss--Manin connection}
In this section we present effective construction of the Gauss-Manin connection on the bundle associated with the family of punctured $(n,s)$-curves.  The equation $f(x,y,\la)=0$ in
$\mathbb{C}^{2+2g-m}$ defines the family $V$ of $(n,s)$-curves over $B=\mathbb{C}^{2g-m}\backslash
\Sigma_0$. Consider the bundle $\upo{p}\colon\upo{V}\to B$ with fiber $\upo{V}_{b}$ that is the curve $V_b$ \emph{with a puncture at infinity}.

Let $H^{1}(\upo{V}_{b},\mathbb{C})$ be the linear
$2g$-dimensional space of holomorphic $1$-forms on $\upo{V}_{b}$.

Associated with the bundle $\upo{p}:\upo{V}\to
B$ is a locally trivial bundle
$\varpi\colon\mathsf{\Omega}^1\to B$ with fiber
$H^{1}(\upo{V}_{b},\mathbb{C})$. A connection in
$\mathsf{\Omega}^1$ is called a
\textit{Gauss--Manin connection on} $\upo{V}$.

Since the point $\infty$ belongs to all curves in
$V$, we can construct a global section of $\mathsf{\Omega}^1$ by
choosing a classical basis of Abelian differentials of the
first and second kind on $V$.

Let $D(x,y,\la)=\big(D_1(x,y,\la), \dots,D_{2g}(x,y,\la)\big)$ be the vector of canonical
$1$-forms in $H^{1}({V}_{b},\mathbb{C})$. Its period matrix $\Omega$
satisfies the Legendre identity\footnote{This is special case
of the Riemann--Hodge relations.}
$$
\Omega^TJ\Omega=2\pi\imath J,\quad \text{where}\,J=\begin{pmatrix}0_g&1_g\\
-1_g&0_g\end{pmatrix}.
$$
The differential $D(x,y,\la)$ can be constructed by techniques of the
classical theory of Abelian differentials~\cite{ba97}.

For the vector field $\ell_j$, $j=1,\dots,2g-m$, on the base, the corresponding Christoffel
coefficient $\Gamma_j=(\Gamma_{j,i}^{k})$, $i,k=1,\dots,2g$,
of the Gauss--Manin connection is \emph{uniquely determined} by the fact that the
holomorphic $1$-form $\ell_jD(x,y,\la)+D(x,y,\la) \Gamma_j$ is exact along the curve. Here
the words ``along the curve'' imply that the variables $x$ and $y$ are
constrained by the equation $f(x,y,\la)=0$.

By integrating the exact form $\ell_jD(x,y,\la)+D(x,y,\la)\Gamma_j$ along the basis cycles of a
fiber, we obtain $\ell_j\Omega+ \Omega \Gamma_j=0$. It follows from the Legendre
identity that $\det\Omega\neq0$. Thus, we have

\begin{lem}\label{lem923}
$\Gamma_j=-\Omega^{-1}\ell_j\Omega$ and $J\Gamma_j+\Gamma_j^TJ=0$.
\end{lem}

This result permits us to construct the canonical differential
$D(x,y,\la)$ directly, without using the classical techniques. Put
$R(x,y,\la)=M(x,y)\,dx/\!f_{y} (x,y,\la)$. Obviously, the
differential $R(x,y,\la)$ defines a basis in $\mathsf{\Omega}^1$,
but its periods do not satisfy the Legendre identity. Let
$\chi_j(\la)$ be the Christoffel coefficient of the Gauss--Manin
connection for $R(x,y,\la)$ along $\ell_j$; that is, the $1$-form
$\ell_jR(x,y,\la)+R(x,y,\la) \chi_j $ is exact along the fibers of
the bundle $\upo{p}\colon\upo{V}\to B$. Then the differentials
$D(x,y,\la)$ and $R(x,y,\la)$ are related by
$D(x,y,\la)=R(x,y,\la)K(\la)$, where $K(\la)$ is a nondegenerate
holomorphic matrix such that $K(0)$ is a diagonal numerical matrix.
We have $\Gamma_j=K^{-1}(\chi_j K-\ell_jK)$. Hence, by the Lemma\,\ref{lem923}
we obtain the following system of $2g-m$ equations for $K$:
$$
(\chi_j K-\ell_jK)JK^T+KJ(\chi_j K-\ell_jK)^T=0, \qquad
j=1,\dots,2g-m.
$$

\subsection{Construction of $\sigma$-function} Consider the set $Sh$ of operators $W_{a,b,c}$ that act
on functions by the formula
$$
W_{a,b,c}(f(u))=f(u+a)\exp\Bigl(\frac{1}{2}\langle
2u+a,b\rangle+c\,\pi\imath\Bigr),
$$
where $u,a,b\in \mathbb{C}^g$, $c\in \mathbb{C}$,
$\imath^2=-1$, and $\langle\cdot,\cdot\rangle$
stands for the Euclidean
inner product. We refer to $W_{a,b,c}$
as the \textit{covariant shift operator}. The set
$Sh$ is closed with respect to composition,
$$
W_{a_2,b_2,c_2}W_{a_1,b_1,c_1}=
W_{a_1+a_2,\;b_1+b_2,\;c_1+c_2+\frac{1}{2\pi\imath}(\langle
b_1,a_2\rangle-\langle a_1,b_2\rangle)}.
$$

We need the following representation:
$\mathbb{Z}^{\,g}\times\mathbb{Z}^{\,g}\to Sh$,
$(n,n')\mapsto((a,b),c)=((n,n')\Omega,\phi(n,n'))$, where
$\Omega=\Omega(\la)\in \mathrm{Mat}(2g,C^{\infty}(B))$ is, as
above, the period matrix of $D(x,y,\la)$ satisfying the
Legendre identity and $\phi\colon \mathbb{Z}^{2g}\to
\mathbb{Z}_2$ is an Arf function.

\begin{df}
A function $\phi\colon\mathbb{Z}^{2g}\to
\mathbb{Z}_2$ is called an \textit{Arf function} if the \textit{Arf
identity} $\phi(q_1+q_2)=\phi(q_1)+\phi(q_2)+q_1J
q_2^T\mod 2$ holds for all $q_1,q_2\in\mathbb{Z}^{2g}$.\vglue-10pt
\end{df}

To define an Arf function $\phi$, it
suffices to choose a pair
$(\varepsilon,\varepsilon')\in\mathbb{Z}^{2g}$. Then
$$
\phi(n,n')=(\langle n+\varepsilon, n'+\varepsilon'\rangle-\langle
\varepsilon,\varepsilon'\rangle)\mod 2.
$$

Let us write $\Omega$ in block form, $\Omega=
\begin{pmatrix}
\Omega_{1,1}&\Omega_{1,2}\\
\Omega_{2,1}&\Omega_{2,2}
\end{pmatrix}$;
then we can set
$W_{\Omega}^{\varepsilon,\varepsilon'}({n,n'})= W_{a,b,c}$, where
$a=(n \Omega_{1,1}+n' \Omega_{2,1})$, $b=(n \Omega_{1,2}+n'
\Omega_{2,2})$, and $c=\langle n+\varepsilon, n'+
\varepsilon'\rangle-\langle \varepsilon,\varepsilon'\rangle$.

Since $D(x,y,\la)$ is the vector of basis differentials, it follows that $\Omega_{1,1}$ is
nondegenerate and
$\tau=\Omega_{2,1}\Omega_{1,1}^{-1}$ is symmetric; moreover, $\mathrm{Im}(\tau)$ is
positive definite by virtue of the Legendre identity.

Set
$G_{\Omega}(u)=|\Omega_{1,1}|^{-1/2}\exp(-\frac{\pi\imath}{2}\,u
\varkappa u^T)$, $\imath^2=-1$, where $\varkappa=\Omega_{1,1}^{-1}\Omega_{1,2}$ is
\textit{symmetric} by the Legendre identity. Consider the
function
\begin{equation}
\sigma(u,\Omega;\varepsilon,\varepsilon')=
\sum_{(n,n')\in\mathbb{Z}^{2g}}W_{\Omega}^{\varepsilon,
\varepsilon'}({n,n'})\,G_{\Omega}(u).\label{sigma9}
\end{equation}

\begin{theorem}\label{theorem98}
The function $\sigma(u,\Omega;\varepsilon,\varepsilon')$ given by the formula (\ref{sigma9}) is entire
in $u\in \mathbb{C}^g$.

The meromorphic functions
$\wp_{i,j}(u,\Omega;\varepsilon,\varepsilon')
=-\partial_{u_i,u_j}\log \sigma(u,\Omega;\varepsilon,\varepsilon')$
are Abelian functions with respect to the lattice generated by
the rows of the matrices $\Omega_{1,1}$ and
$\Omega_{2,1}$; that is,
$\wp_{i,j}(u,\Omega;\varepsilon,\varepsilon')=\wp_{i,j}(u+n
\Omega_{1,1}+n' \Omega_{2,1},\Omega;\varepsilon,\varepsilon')$,
$(n,n')\in\mathbb{Z}^{2g}$.\end{theorem}

\begin{proof}
By construction, $\sigma(u,\Omega;\varepsilon,\varepsilon')$ is an
eigenfunction of the covariant shift operator $W_{\Omega}^{\varepsilon,\varepsilon'}({n,n'})$
with eigenvalue $+1$ or $-1$. Therefore, the function
$\sigma(u,\Omega;\varepsilon, \varepsilon')/G_{\Omega}(u)$ admits a Fourier series expansion
with coefficients independent of $u$. The series is
convergent, since the imaginary part of $\tau$ is positive definite. On the other hand, the
difference $\log\sigma(u,\Omega;\varepsilon,\varepsilon')-\log \sigma(u+n
\Omega_{1,1}+n'\Omega_{2,1},\Omega;\varepsilon,\varepsilon')$ is a linear function of $u$.
\end{proof}

In what follows, we assume that the pair
$(\varepsilon,\varepsilon')$ is taken to be equal to the
characteristic of the vector of Riemann constants;
this characteristic can be assumed to be
the same for all curves in the family $V$.

\begin{df}
The function $\sigma(u,\la)=\Delta(\la)^{1/8} \sigma(u,\Omega(\la);\varepsilon,\varepsilon')$ is called the \emph{sigma function associated with
the family $V$ of curves}.
\end{df}

Clearly, the assertion of Theorem\,\ref{theorem98} remains valid if we replace
$\sigma(u,\Omega(\la); \varepsilon,\varepsilon')$ by $\sigma(u,\la)$.
In contrast with $\sigma(u,\Omega(\la); \varepsilon,\varepsilon')$, the
function $\sigma(u,\la)$ is holomorphic in $\la $. In particular, it has
the so-called \emph{rational limit} as $\la \to0$. The function
$\sigma(u,0)$ is a polynomial, which is completely determined by the pair
$(n,s)$ and is known as the
Schur--Weierstrass polynomial (see Chapt. \ref{chap:rat}).

Now we are ready to describe linear operators annihilated
mult-dimensional $\sigma$-function in terms of the Gauss-Manin connection.
To do that we define matrices $\alpha_j=(\alpha_j^{kl}),$
$\beta_j=(\beta_{jk}^{l})$, and $ \gamma_j=(\gamma_{jkl}),$
$k,l=1,\dots,g$, $j=1,\dots,2g-m$, by the formula
\begin{equation}
\begin{pmatrix}\alpha_j&(\beta_j)^T\\
\beta_j&\gamma_j\end{pmatrix}=-J\Gamma_j,\label{abc}
\end{equation}
where $\Gamma_j$ is the Christoffel symbol of
the Gauss--Manin connection, and
set
$$
H_j=\tfrac{1}{2}\alpha_{j}^{kl}(\la)\partial_{u_{k}}\partial_{u_{l}}+ \beta_{jk}^{l}(\la)u_{k}\partial_{u_l}+\tfrac{1}{2}\gamma_{jkl} (\la)u_{k}u_{l}+ \tfrac{1}{8}\ell_{j}(\log\Delta(\la))+\tfrac{1}{2} \beta_{jk}^{k}(\la).
$$
Here and below, summation from $1$ to $g$ over
repeated indices is assumed.

\begin{theorem}\label{theorem99}
The sigma function $\sigma(u,\la)$ satisfies the system of $2g-m$ linear differential
equations $(\ell_j-H_j)\sigma(u,\la)=0$, $j=1,\dots,2g-m$.
\end{theorem}

\begin{proof}
By using the relations
$\Gamma_j=-\Omega^{-1}\ell_j\Omega$ (cf.~Theorem{theorem98}), we find that $(\ell_j-H_j) G_{\Omega}(u)=0$ and
$[\ell_j-H_j,W_{\Omega}^{\varepsilon,\varepsilon'}({n,n'})]=0$ for
all $j=1,\dots,2g-m$ and $({n,n'})\in \mathbb{Z}^{2g}$. Details of
the proof can be found in \cite{bl04}.\qed
\end{proof}

\subsection{Operators of differentiation} Now we
intend to apply Theorems\,\ref{theorem98} and \ref{theorem99} to solve our problem for
the fields of fiberwise Abelian functions associated
with $(n,s)$-curves by the method described in details in Sect. 9.2.2
for the case in which $n=2$ and $s=3$.

Let $\mathcal{L}=(\ell_1,\dots,\ell_{2g-m})^T$ be the frame with
structure functions $c^{h}_{ij}$ (see (\ref{struct}))  and
let $\alpha_j=(\alpha_j^{kl}),$ $\beta_j=(\beta_{jk}^{l})$,
$\gamma_j=(\gamma_{jkl})$ be the matrices defined  in (\ref{abc})

\begin{theorem}
For families
of $(n,s)$-curves, the problem in
question has the following solution:

\textup{(a)} The $F$-module $\Der F$ is spanned by the $3g-m$ generators
$d_i=\partial_{u_i}$, $i=1,\dots,g$, and
$L_{j}=\ell_j-\bigl(\alpha_{j}^{kl}\zeta_{k}(u,\la)+ \beta_{jk}^{l}u_k\bigr)
\partial_{u_l}$, $j=1,\dots,2g-m$,
where $\zeta_{k}(u,\la)=\partial_{u_k} \log\sigma(u,\la)$.

\textup{(b)} The Lie algebra structure of $\Der F$ over $F$ is defined by the relations
\begin{gather*}
[d_i,d_k]=0,\qquad [L_j,d_{i}]=
-\bigl(\alpha_{j}^{kq}\wp_{iq}(u,\la)-\beta_{ji}^{k}\bigr)d_{k},\\
[L_i,L_j]=\frac{1}{2}(\alpha_{i}^{kl}\alpha_{j}^{qr}-\alpha_{j}^{kl}\alpha_{i}^{qr})
\wp_{klq}(u,\la)d_{r} +\sum_{h=1}^{2g-m}c^{h}_{ij}\,L_h.
\end{gather*}

\textup{(c)} The action of $\Der F$ on $F$ is defined by the relations
\begin{gather*}
d_i\la =0,\qquad d_i\wp_{qr}(u,\la)=\wp_{iqr}(u,\la),\\
L_j\la =\ell_j\la,\qquad L_j\wp_{qr}(u,\la)=\tfrac{1}{2}\alpha_{j}^{kl}(\wp_{klqr}-2\wp_{kq}\wp_{lr})
+\beta_{jq}^{k}\wp_{kr}+\beta_{jr}^{k}\wp_{kq}-\gamma_{jqr}.
\end{gather*}
\end{theorem}

{\bf Example,} The case $(n,s)=(2,5)$ corresponds to the family of genus~$2$ curves. We
have
$$
V=\{(x,y,\la)\in\mathbb{C}^{6}\mid f(x,y,\la)=0,\,\Delta(\la)\neq0\},
$$
where $f(x,y,\la)=y^2-(x^5+\la_4x^3+\la_6x^2+\la_8x+\la_{10})$. The polynomial $f(x,y,\la)$ is homogeneous with
respect to the grading in which $\deg x=2$, $\deg y=5$, $\deg \la_k=k$, and $m=0$. The
discriminant of $f(x,y,\la)$ is defined by the zeros of the polynomial
\begin{align*}
\Delta(\la)&=
3125\la_{10}^4\kern0.5pt{+}\kern0.5pt2(128\la_8^5\kern0.5pt{-}\kern0.5pt800\la_6\la_{10}\la_8^3\kern0.5pt{+}\kern0.5pt1000\la_4\la_{10}^2\la_8^2
\kern0.5pt{+}\kern0.5pt1125\la_6^2\la_{10}^2\la_8\kern0.5pt{-}\kern0.5pt1875\la_4\la_6\la_{10}^3)\\
&\quad+108\la_{10}\la_6^5\,{-}\,27\la_8^2\la_6^4
-630\la_4\la_8\la_{10}\la_6^3\,{+}\,144\la_4\la_8^3\la_6^2\,{+}\,825\la_4^2\la_{10}^2\la_6^2
\,{+}\,560\la_4^2\la_8^2\la_{10}\la_6\\
&\quad-128\la_4^2\la_8^4\,{-}\,900\la
_4^3\la_8\la_{10}^2\,{+}\,4\la_4^3(4\la_{10}\la_6^3
\,{-}\,\la_8^2\la_6^2\,{-}\,18\la_4\la_8\la_{10}\la_6\,{+}\,4\la_4\la_8^3\,{+}\,27
\la_4^2\la_{10}^2),
\end{align*}
which can be calculated directly by taking the resultant with respect to $x$ of the polynomials
$f(0,x,\la)$ and $\partial_{x}f(0,x,\la)$.

The vector fields $(\ell_{0},\ell_{2},\ell_{4},\ell_{6})^T= T(\la)(\partial_{\la_4},\partial_{\la_6},
\partial_{\la_8},\partial_{\la_{10}})^T$, $\deg \ell_j=j$, defined
by the matrix
$$
T=
\begin{pmatrix}
{4\la_4}&{6\la_6}&{8\la_8}&{10\la_{10}}\\[2pt]
*&8\la_8-\frac{12}5\la_4^2&10\la_{10}-\frac85\la_4\la_6&
-\frac45\la_4\la_8\\[2pt]
*&*&4\la_4\la_8-\frac{12}5\la_6^2&
6\la_4\la_{10}-\frac65\la_6\la_8\\[2pt]
*&*&*&4\la_6\la_{10}-\frac85\la_8^2
\end{pmatrix}
$$
with $\det T(\la)=\frac{16}{5}\Delta(\la)$ are tangent to the
discriminant, since
$$
(\ell_{0},\ell_{2},\ell_{4},\ell_{6})\log\Delta(\la)=(40,0,12\la_4,4\la_6).
$$
The structure functions of this frame are
polynomials,
\begin{gather*}
[\ell_0,\ell_k]=k \ell_k, \quad k=2,4,6,\qquad [\ell_2,\ell_4]=
2\ell_6-\tfrac{8}{5}\la_4
\ell_2+\tfrac{8}{5}\la_6 \ell_0,\\
[\ell_2,\ell_6]=-\tfrac{4}{5}\la_4 \ell_4+\tfrac{4}{5}\la_8
\ell_0,\qquad [\ell_4,\ell_6]=2\la_4 \ell_6-\tfrac{6}{5}\la_6
\ell_4+ \tfrac{6}{5}\la_8 \ell_2-2\la_{10} \ell_0.
\end{gather*}

In this case, $M(x,y)=(1,x,x^2,x^3)$ and
$R(x,y)=M(x,y)\dfrac{dx}{2y}$. We have
\begin{gather*}
\chi_0=\mathrm{diag}(3,1,-1,-3),\\[5pt]
\chi_2=\begin{pmatrix}
0 & -\frac{4}{5}\la_4 & 0 & \la_8 \\[2pt]
1 & 0 & -\frac{8}{5}\la_4 & 2 \la_6 \\[2pt]
0 & -1 & 0 & \frac{3}{5}\la_4 \\[2pt]
0 & 0 & -3 & 0
\end{pmatrix},\qquad
\chi_4=\begin{pmatrix}
\la_4&-\frac{6}{5}\la_6&\la_8&2\la_{10}\\[2pt]
0 & -\la_4&-\frac{2}{5}\la_6&3\la_8\\[2pt]
-1 & 0 & 0 & \frac{2}{5}\la_6\\[2pt]
0 & -3 & 0 & 0
\end{pmatrix},\\[5pt]
\chi_6=\begin{pmatrix}
0 & -\frac{3}{5}\la_8 & 2 \la _{10} & 0 \\[2pt]
-\la _4 & 0 & -\frac{1}{5}\la_8&4\la_{10}\\[2pt]
0 & 0 & 0 & \frac{1}{5}\la_8 \\[2pt]
-3 & 0 & 0 & 0
\end{pmatrix}.
\end{gather*}
The general solution of the system of equations for the matrix $K(\la)$ satisfying the holomorphy and
homogeneity conditions has the form
\[
K(\la)=\begin{pmatrix}
1 & 0 & a \la _6 & b \la _4 \\
0 & 1 & (b-1)\la _4 & 0 \\
0 & 0 & 0 & -1 \\
0 & 0 & -3 & 0
\end{pmatrix},
\]
where $(a,b)\in \mathbb{C}^2$ are free parameters. We set
$(a,b)=(0,0)$, which
corresponds to the classical construction \cite{ba97}. Then
\begin{gather*}
\Gamma_0=\mathrm{diag}(3,1,-3,-1),\quad \Gamma_2=\begin{pmatrix}
0&-\frac{4}{5}\la_4&\frac{4}{5}\la_4^2-3\la_8&0\\
1 & 0 & 0 & \frac{3}{5}\la_4\\
0 & 0 & 0 & -1\\
0 & 1 & \frac{4}{5}\la_4&0
\end{pmatrix},\\[5pt]
\Gamma_4=\begin{pmatrix}
\la_4&-\frac{6}{5}\la_6&\frac{6}{5}\la_4\la_6-6\la_{10}&-\la_8\\
0 & 0 & -\la_8& \frac{2}{5}\la_6 \\
0 & 1 & -\la_4 & 0 \\
1 & 0 & \frac{6}{5}\la_6& 0
\end{pmatrix},\quad
\Gamma_6=\begin{pmatrix}
0&-\frac{3}{5}\la_8&\frac{3}{5}\la_4\la_8&-2\la_{10}\\
0 & 0 & -2\la_{10}&\frac{1}{5}\la_8 \\
1 & 0 & 0 & 0 \\
0 & 0 & \frac{3}{5}\la_8&0
\end{pmatrix}.
\end{gather*}

Set $\deg u_i=-i$. Then the operators $H_j$
are homogeneous and $\deg H_j=j$,
\begin{align*}
H_0&={u_1\partial_{u_1}+3u_3\partial_{u_3}}-3,\\
10H_2&={5\partial_{u_1}^2+10u_1\partial_{u_3}-8\la_4u_3\partial_{u_1}}
-3\la_4u_1^2+(15\la_8-4\la_4^2)u_3^2,\\
5H_4&=5{\partial_{u_1}\partial_{u_3}\,{+}\,5\la_4u_3\partial_{u_3}\,{-}\,{6}\la_6
u_3
\partial_{u_1}}\,{-}\,\la_6u_1^2\,{+}\,5\la_8u_1u_3\,{+}\,{3}(5\la_{10}\,{-}\,\la_{4}
\la_{6})u_3^2\,{-}\,5\la_4,\\
10H_6&={5\partial_{u_3}^2-6\la_8u_3\partial_{u_1}}-
\la_8u_1^2+20\la_{10} u_1u_3 -{3}\la_{4}\la_{8}u_3^2-5\la_6.
\end{align*}
The generators of the $F$-module $\Der F$ are as follows:
\begin{gather*}
L_0=\ell_0-u_1\partial_{u_1}-3u_3\partial_{u_3},\qquad
L_1=\partial_{u_1},\\
L_2=\ell_2-\zeta_1\partial_{u_1}-u_1\partial_{u_3}-\tfrac{4}{5}\la_4u_3\partial_{u_1},\quad
L_3=\partial_{u_3},\\
L_4=\ell_4-\zeta_3
\partial_{u_1}-\zeta_1\partial_{u_3}+\la_4u_3\partial_{u_3}-\tfrac{6}{5}\la_6
u_3 \partial_{u_1},\quad
L_6=\ell_6-\zeta_3\partial_{u_3}-\tfrac{3}{5}\la_8u_3\partial_{u_1}.
\end{gather*}
The structure relations in the Lie algebra $\Der F$ over
$F$ have the form
\begin{gather*}
[L_0,L_k]=kL_k, \qquad k=1,2,3,4,6,\\
[L_1,L_2]=\wp_{1,1}L_1-L_3, \quad [L_1,L_3]=0,\quad
[L_1,L_4]=\wp_{1,3}L_1+\wp_{1,1}L_3,\\
[L_1,L_6]=\wp_{1,3}L_3,\quad
[L_2,L_4]=2L_6-\tfrac{1}{2}\wp_{1,1,1}L_3-\tfrac{8}{5}\la_4
L_2+\tfrac{1}{2}\wp_{1,1,3}L_1+\tfrac{8}{5}\la_6 L_0,\\
[L_2,L_3]=-(\wp_{1,3}-\tfrac{4}{5}\la_4)L_1,\quad
[L_2,L_6]=-\tfrac{4}{5}\la_4
L_4-\tfrac{1}{2}\wp_{1,1,3}L_3+\tfrac{1}{2}\wp_{1,3,3}L_1+
\tfrac{4}{5}\la_8 L_0,\\
[L_3,L_4]=(\wp_{1,3}+\la_4)L_3+(\wp_{3,3}-
\tfrac{6}{5}\la_6)L_1,\quad
[L_3,L_6]= \wp_{3,3}L_3-\tfrac{3}{5}\la_8L_1,\\
[L_4,L_6]=2\la_4 L_6-\tfrac{6}{5}\la_6 L_4-
\tfrac{1}{2}\wp_{1,3,3}L_3+ \tfrac{6}{5}\la_8 L_2+
\tfrac{1}{2}\wp_{3,3,3}L_1-2\la_{10} L_0.
\end{gather*}
Let us write out the formulas that define the action of $L_j$
on $F$. The action on the coordinate functions of the base is
determined by the relation $(L_0,L_1,L_2,L_3, L_4,L_6)\la
=(\ell_0,0,\ell_2,0,\ell_4,\ell_6)\la $. According to items 2 and 3
in Section~9.2.1, it suffices to write out the formulas for
$L_j\wp_{1,1}$:
\begin{gather*}
L_0\wp_{1,1}=2\wp_{1,1},\quad L_1\wp_{1,1}=\wp_{1,1,1}, \quad
L_2\wp_{1,1}=\tfrac{1}{2}\wp_{1,1,1,1}-\wp_{1,1}^2+\wp_{1,3}+
\tfrac{3}{5}\la_4,\\
\quad L_3\wp_{1,1}=\wp_{1,1,3},\quad L_4\wp_{1,1}=\wp_{1,1,1,3}-
2\wp_{1,1}\wp_{1,3}+\tfrac{2}{5}\la_6,\\
L_6\wp_{1,1}=
\tfrac{1}{2}\wp_{1,1,3,3}-\wp_{1,3}^2+\tfrac{1}{5}\la_8.
\end{gather*}

All genus $2$ curves are hyperelliptic. The universal bundle
of Jacobi varieties of genus $g$ hyperelliptic curves is a rational
variety. This is Dubrovin-Novikov theorem, see  Sect.~\ref{dntheorem}
where a fiber of the universal bundle is
viewed as a level surface of the integrals of motion
for the $g$th stationary flow of the KdV system, that
 is defined by a system of
$2g$ algebraic equations in $\mathbb{C}^{3g}$
whose degree increases with
genus. In the Sect.~\ref{dntheorem}  the coordinates were introduced such
that a fiber is defined by $2g$ equations of degree~$\le 3$.

The Dubrovin--Novikov coordinates and the coordinates from  Sect.~\ref{dntheorem}
 are the same for the
universal space of genus $1$ curves. Namely, in $\mathbb{C}^{3}$
with coordinates $(x_2,x_3,x_4)$, $\deg x_i=i$, the fiber over
$b=(g_2,g_3)\in B$ is determined by the equations
$g_2=12 x_2^2-2 x_4$, $g_3=-8 x_2^3+2 x_4 x_2-x_3^2$ and is
parametrized by the elliptic functions
$(x_2,x_3,x_4)=(\wp(u,b),\wp'(u,b),\wp''(u,b))$, where $u$ ranges
over the Jacobian $J_b$, the fiber of the bundle $U\to B$.
 Further, in these coordinates one has
$\Delta=-432x_2^3x_3^2-27x_3^4+108x_2x_3^2x_4+36x_2^2x_4^2-8x_4^3$,
and the derivations $L_0, L_1$ and $L_2$ of the field $F$
are represented by the vector fields
\begin{gather*}
L_0=2x_2\partial_{x_2}+3x_3\partial_{x_3}+4x_4\partial_{x_4},\qquad
L_1=x_3\partial_{x_2}+x_4\partial_{x_3}+12x_2x_3\partial_{x_4},\\
L_2=\tfrac{2}{3}(x_4-3x_2^2)\partial_{x_2}+3x_2x_3\partial_{x_3}
+(3x_3^2+2x_2x_4)\partial_{x_4},
\end{gather*}
which are tangent to the zero set of $\Delta$
and have polynomial coefficients.

For the universal space of genus $2$ curves the fiber over a point
$b=(\la_4,\la_6,\la_8,\la_{10})\in B$ is defined in
$\mathbb{C}^{6}$ with coordinates $(x_2,x_3,x_4,z_4,z_5,z_6)$,
$\deg x_i=i$, $\deg z_j=j$, by the equations
\begin{gather*}
6 x_2^2-x_4+4z_4+2\la_4=0,\quad
8x_2^3-2x_2(x_4+4z_4)+x_3^2+2z_6-4\la_6=0,\\
8z_4x_2^2-x_2z_6+2z_4^2-x_4z_4+x_3z_5-2\la_8=0,\quad 8 x_2 z_4^2-2
z_4z_6+z_5^2-4\la_{10}=0
\end{gather*}
and is parametrized by the hyperelliptic Abelian
functions
$$
(x_1,x_2,x_4)=(1,L_1,L_1^2)\wp_{1,1}(u,b),\qquad
(z_4,z_5,z_6)=(1,L_1,L_1^2)\wp_{1,3}(u,b),
$$
where $u=(u_1,u_3)$ ranges over the Jacobian $J_b$, the fiber of
the bundle $U\to B$. Note that, by analogy
with the elliptic case, the derivations of $F$ found above
become polynomial vector
fields tangent to the ``discriminant.''

The Dubrovin--Novikov coordinates $(U_2, U_3,\dots, U_7)$, where $\deg U_i=i$, are
expressed via $(x_2,x_3,\allowbreak x_4,z_4,z_5,z_6)$ by the formulas
\begin{gather*}
U_2=x_2,\quad U_3=x_3,\quad U_4=x_4,\quad U_5=4 (3 x_2 x_3+z_5),\\
U_6=4(3 x_3^2+3 x_2 x_4+z_6),\quad U_7=4(36 x_3 x_2^2+20z_5x_2+9
x_3 x_4+4 x_3 z_4).
\end{gather*}
Accordingly, the fiber over $b\in B$ is
parametrized as follows:
$$
(U_2,
U_3,U_4,U_5,U_6,U_7)=(1,L_1,L_1^2,L_1^3,L_1^4,L_1^{5})\wp_{1,1}(u,b).
$$


\chapter[Algebro-geometric tau function]{Algebro-geometric tau function}\label{chap:tau}

\section{Introduction}
In the mid-1970s, the results of Its, Matveev, Dubrovin and Novikov (see
\cite{dmn76}) led to the discovery of a remarkable $\theta$-functional
formula to solve the KdV equation $u_t=6uu_x-u_{xxx}$. This solution
was given as a second logarithmic derivative of a Riemann
theta-function:
\begin{equation}
u(x,t)=-\frac{\partial^2}{\partial x^2 }  \,\mathrm{ln} \,
\theta(\boldsymbol{U}x+\boldsymbol{V}t+ \boldsymbol{W})+C,\qquad
\boldsymbol{U},\boldsymbol{V},\boldsymbol{W}=\mathrm{const}\in\mathbb{C}^g
\end{equation}
This theta-function was constructed from a hyperelliptic curve $X_g$
of genus $g$, while the ``winding vectors'' $\boldsymbol{U},
\boldsymbol{V}$ are periods of Abelian differentials of the second
kind on $X_g$. Further, this formula is in a sense universal; it was
generalized by Krichever \cite{kr77} to other integrable hierarchies -
these solutions were associated with other algebraic curves.  In this
paper we consider the converse problem:

{\em Given an algebraic curve $X_g$, its Riemann period matrix $\tau$,
  and its Jacobi variety $\mathrm{Jac}(X_g)=\mathbb{C}/(1_g\oplus
  \tau)$, we may construct $\theta$-functions
  $\theta(\boldsymbol{z};M)$, $\boldsymbol{z}\in \mathrm{Jac}(X_g) $;
  then the fundamental Abelian functions on $\mathrm{Jac}(X_g)$ may be
  realized as the second logarithmic derivatives of
  $\theta(\boldsymbol{u},M)$, $\wp_{ij} = - \frac{\partial^2\ln
    \theta(\boldsymbol{u};\tau) }{\partial u_i \partial
    u_j}+C_{ij}(\tau)$.  We wish to construct all differential
  relations between these Abelian functions on $X_g$.}

In the simplest case, the Weierstrass cubic, $y^2=4x^3-g_2x-g_3$, that is
an algebraic curve of genus one. This is uniformized by the Weierstrass
elliptic functions, $x=\wp(u), y=\wp'(u)$, and these differential
relations read:
\begin{equation}
\wp''=6\wp^2-\frac{g_2}{2},\quad {\wp'}^2=4\wp^3-g_2\wp-g_3.
\label{weierstrass}
\end{equation}
In the case of higher genera, $g>1$, the derivation of analogous
equations becomes much more complicated.  In particular, the
fundamental Abelian functions are now {\em partial} derivatives of a
function of $g$ variables.  The different approaches to this problem
form the main content of the paper.  We restrict our analysis to the
case of $(n,s)$-curves introduced and investigated in this context by
Buchstaber, Enolski, and Leykin \cite{bel97b,bel99}
\begin{equation}
y^n=x^s+\sum_{si+nj<ns} \lambda_{ij} x^iy^j.
\end{equation}
These represent a natural generalization of elliptic curves to higher
genera, and include the general hyperelliptic curve ($n=2$).

To any such curve we may associate an object which is fundamental to
all our treatments of this problem, the fundamental
bi-differential, that is, the unique
symmetric meromorphic $2$-form on $X_g\times X_g$,
whose only second order pole
lies on the diagonal $Q=S$, and which satisfies:
\begin{align*}
\omega(Q,S)-\frac{\mathrm{d}\xi(Q)\mathrm{d}\xi(S)}{(\xi(Q)-\xi(S))^2 }
=f(S,Q)\mathrm{d}\xi(Q)\mathrm{d}\xi(S),
\end{align*}
where $f(S,Q)$ is holomorphic, and $\xi(Q),\xi(S)$ are local
coordinates in the vicinity of a base point $P$, $\xi(P)=0$. Usually
$\omega(Q,S)$ is realized as the second logarithmic derivative of the
prime-form or theta-function \cite{fa73}.  But in our development we
use an alternative representation of $\omega(Q,S)$ in the {\em
  algebraic form} that goes back to Weierstrass, Klein and which was
well documented by Baker \cite{ba97}
\begin{equation}
  \omega(Q,S)=\frac{\mathcal{F}(Q,S)}{f_y(Q) f_w(S)(x-z)^2}
  \mathrm{d}x\mathrm{d}z+
  \mathrm{d}\boldsymbol{u}(Q)^T\varkappa \mathrm{d}
  \boldsymbol{u}(S), \label{omegaQS}
\end{equation}
where $Q=(x,y)$, $S=(z,w)$, and the function
$\mathcal{F}(Q,S)=\mathcal{F}((x,y),(z,w))$ is a polynomial of its
arguments with coefficients depending on the parameters of the curve
$V$. $\mathcal{F}(Q,S)$ is sometimes called the {\em Kleinian polar}.
Finally, $\varkappa$ is a symmetric matrix expressed in terms of the
first and second period matrices, $2\omega$, $2\eta$ respectively, as
$\varkappa= \omega^{-1}\eta$ and providing normalization of
$\omega(Q,S)$.  We will refer to the first term on the right of
(\ref{omegaQS}), which involves the polynomial $\mathcal{F}(Q,S)$, as
the algebraic part, so that
$$
\omega(Q,S)=\omega^{\rm{alg}}(Q,S)+ \mathrm{d}
\boldsymbol{u}(Q)^T\varkappa \mathrm{d} \boldsymbol{u}(S).
$$
This representation was recently by discussed also by Nakayashiki
\cite{nakaya08}.

The algebraic representation of the fundamental differential, as
described above, lies behind the definition of the multivariate
sigma function in terms of the theta-function. This differs from $\theta$
by an exponential factor and a modular factor:
\begin{equation}
  \sigma(\boldsymbol{u}) = C(M)\mathrm{exp} \left\{
    \tfrac12 \boldsymbol{u}^T \omega^{-1}\eta \boldsymbol{u}  \right\}
  \theta\left(\tfrac12 \omega^{-1}\boldsymbol{u};M \right). \label{sigma}
\end{equation}
Here the $g\times g$ matrices $2\omega, 2\eta$ are the first and
second period matrices, and $M=\omega^{-1}\omega'$.  The modular
constant $C(M)$ is known explicitly for hyperelliptic curves and a
number of other cases, but its explicit form is not necessary here, for
the fundamental Abelian functions are independent of $C(M)$.
These modifications make $\sigma(\boldsymbol{u})$ invariant with
respect to the action of
the symplectic group, so that for any $\gamma\in{\mathrm Sp}(2g,
\mathbb{Z})$, we have:
\begin{equation}
\sigma(\boldsymbol{u};\gamma M)=\sigma(\boldsymbol{u};M).
\end{equation}
The multivariate sigma-function is the natural generalization of the
Weierstrass sigma function to algebraic curves of higher genera,
i.e. $(n,s)$-curves in this context. In his lectures \cite{w-lect},
Weierstrass started by defining the sigma-function in terms of series
with coefficients given recursively, which was the key point of the
Weierstrass theory of elliptic functions. A generalization of this
result to the genus two curve was started by Baker \cite{ba07} and
recently completed by Buchstaber and Leykin \cite{bl05}, who obtained
recurrence relations between coefficients of the sigma-series in
closed form.  In addition, Buchstaber and Leykin recently found an
operator algebra that annihilates the sigma-function of a higher
genera $(n,s)$-curve \cite{bl08}. The recursive definition of the
higher genera sigma-functions remains a challenging problem to solve,
with \cite{bl08} providing a definite step. We believe that the future
theory of the sigma and corresponding Abelian functions can be formulated
on the basis of sigma expansions that will complete the extension of
the Weierstrass theory to curves of higher genera.

In this chapter we study the interrelation of the multivariate sigma
and Sato tau functions. The $\tau$-function was introduced by Sato
\cite{sato80,sato81} in the much more general context of
integrable hierarchies. Here we deal with the `algebro-geometric
$\tau$-function' (AGT) associated with an algebraic curve.  The AGT of
the genus $g$ curve $X_g$ is defined, following Fay,
\cite{fay83,fay89}, as a function of the `times'
$\boldsymbol{t}=(t_1,\ldots,t_g,t_{g+1},\ldots)$, a point
$\boldsymbol{u}\in\mathrm{Jac}(X_g)$, as well as a point $P\in X_g$;
it is given by the formula
\begin{align*}
\tau(\boldsymbol{t};\boldsymbol{u},P) = \theta\left( \sum_{k=1}^{\infty}
\boldsymbol{U}_k(P) t_k+\boldsymbol{u} \right) \mathrm{exp}
\left\{ \frac12\sum_{m,n\geq 1} \omega_{mn}(P) t_m t_n \right\}.
\end{align*}
Here the ``winding vectors'' $\boldsymbol{U}_k(P)$ appear in the
expansion of the normalized holomorphic integral $\boldsymbol{v}$,
the quantities $\omega_{mn}(P)$ define the holomorphic part of the
expansion of the fundamental second kind differential $\omega(Q,S)$
near the point $P$.
We then introduce the $\tau$-function by the
formula:
\begin{align}
  \frac{ \tau(\boldsymbol{t};\boldsymbol{u},P)}
  {\tau(\boldsymbol{0};\boldsymbol{u},P)}= \frac{
    \sigma\left(\sum_{k=1}^{\infty} (\mathcal{A})^{-1}\boldsymbol{U}_k(P) t_k+
      \boldsymbol{u}\right)} {\sigma(\boldsymbol{u})}\mathrm{exp}
  \left\{ \frac12 \sum_{k,l=0}^{\infty}
    \omega^{\mathrm{alg}}_{k,l}(P)t_k t_l \right\}.
\label{tausigma}\end{align}

This representation of $\tau$ in terms of $\sigma$ was used by Enolski
and Harnad \cite{enhar11} to analyze the Schur function expansion of
$\tau$ for the case of algebraic curves. Recently A. Nakayashiki
\cite{nakayashiki09} has independently suggested a similar expression
for the AGT in terms of multivariate $\sigma$-functions and studied
properties of the sigma-series. In this paper we concentrate on the
application of this representation to the derivation of the differential
relations between Abelian functions of the $(n,s)$-curve, continuing
and developing the work of \cite{enhar11}.

Developing a further analogy with the Weierstrass theory of elliptic
functions, we represent the Abelian
functions, that is, $2g$-periodic
functions on $\mathrm{Jac}(X_g)$,
$f(\boldsymbol{u}+2\boldsymbol{n}\omega+2\boldsymbol{n}'\omega')
=f(\boldsymbol{u})$ $\forall \boldsymbol{n}, \boldsymbol{n}' \in
\mathbb{N}$, as second and higher logarithmic derivatives,
\begin{align}\zeta_i(\boldsymbol{u})&=\frac{\partial}{\partial u_i}
\mathrm{ln}\,\sigma(\boldsymbol{u}),\label{zeta} \\
\wp_{ij}(\boldsymbol{u})&=-\frac{\partial^2}{\partial u_i \partial u_j}
\mathrm{ln}\,\sigma(\boldsymbol{u}), \quad \wp_{ijk}(\boldsymbol{u})=
-\frac{\partial^3}{\partial u_i \partial u_j \partial u_k }
\mathrm{ln}\,\sigma(\boldsymbol{u}),\quad \rm{etc} \label{Kleinwp}
\end{align}
where $i,j,k$ etc. = $1\ldots,g$. We should remark that the $\zeta_i(\boldsymbol{u})$
are not Abelian functions.  In this notation, the genus 1
Weierstrass equations (\ref{weierstrass}) become
 \[
 \wp_{1111}=6\wp_{11}^2-\frac{g_2}{2},\quad {\wp_{111}}^2 =
 4\wp_{11}^3-g_2\wp_{11}-g_3
\]
$\wp_{ij}, \wp_{ijk}, \ldots$ are called Kleinian $\wp$-functions. They
are convenient coordinates to represent the dependent variables in
the hierarchy of integrable systems.

We compare and contrast here three approaches to obtain the
partial differential relations for the Abelian functions associated
with the $(n,s)$-curve $X_g$, using Kleinian $\wp$-functions as
coordinates.

The first of these, and the best known, is the classical approach of
comparing two different expansions of the fundamental bi-differential;
this yields first the solution of the Jacobi inversion problem for the
curve, and in higher orders, a sequence
of differential relations involving the $\wp_{ij}$.

The $\tau$-function approach to the derivation of completely
integrable systems of KP type has led to two different ways
\cite{djkm83} to obtain relations between Taylor coefficients of the
$\tau$ function expansion.  The first of these specifically exploits
the fact that these Taylor coefficients are determinants, i.e.
Pl\"ucker coordinates in the Grassmannian, and they hence satisfy the
Pl\"ucker relations \cite{fay83}. The second is based on the Bilinear
Identity, which leads to the Residue Formula \cite{fay89} involving
differential polynomials in $\tau$.

We consider and compare these two techniques, based on the
$\tau$-function method, which give a derivation of
the required differential relations; specializing to a particular
algebraic curve,
we consider its algebro-geometric $\tau$-function. The special feature
of our development is that we define this $\tau$-function in terms of
the multidimensional $\sigma$-function of the curve, leading to
coordinates that are explicitly written in terms of Kleinian
$\wp$-functions.  The differential relations we find between these
functions can be understood as arising from special solutions of
integrable hierarchies of KP type, associated with the given curve
(see for example \cite{ba97,bel97b,nakaya08}).  We will describe the
correspondence between individual differential equations for
$\wp$-functions with Young tableaux defining Pl\"ucker relations.  We
illustrate these approaches by considering two particular examples:
the genus 2 hyperelliptic curves \cite{ba07}, and the genus 3 trigonal
curve (which can be found in different places
\cite{bel97b,bel00,eemop07}) The general approach based on Pl\"ucker
coordinates for
deriving KP-flows in terms of Kleinian $\sigma$-functions was recently
discussed by Harnad and Enolski \cite{enhar11}. Here we develop this
work and consider some non-trivial examples to clarify the
interrelation of the $\tau$-functional formulation of integrable
hierarchies and $\sigma$-functional approach.  We will also consider
the relationship between of this derivation based on the Pl\"ucker
relations and that based on the Residue Formula.  Both give a
systematic way of generating the required relations, but the
differences between the two approaches are instructive.

\section{Models algebraic curves and their $\theta$,$\sigma$,$\tau$-functions}
Let $V$ be a genus $g\geq 1$ algebraic curve given by the polynomial
equation
\begin{equation}
f(x,y)=0,\quad f(x,y)=y^n+y^{n-1}a_1(x)+\ldots+a_0(x).\label{gencurve}
\end{equation}
We shall consider in what follows two relatively simple curves of the
class (\ref{gencurve}),

{\sf  Example I}:  the hyperelliptic genus two curve
\begin{equation}
y^2=4x^5+\alpha_4 x^4+\ldots+\alpha_0\label{heperel}
\end{equation}
and

{\sf  Example II}:   the cyclic trigonal genus three curve

\begin{equation}
f(x,y)=y^3 -(x^4 +\mu_3 x^3 +\mu_6 x^2 +\mu_9 x +\mu_{12}).\label{trig}
\end{equation}

We equip $V$ with a canonical basis of cycles
$(\mathfrak{a}_1,\ldots,\mathfrak{a}_g;\mathfrak{b}_1,
\ldots,\mathfrak{b}_g)\in H_1(X,\mathbb{Z})$.  We denote by
$\mathrm{d}\boldsymbol{u}=(\mathrm{d} u_1,\ldots,\mathrm{d}u_g)^T$ the
vector whose entries are independent holomorphic differentials of the
curve $V$ as well as their $\mathfrak{a}$ and $\mathfrak{b}$-periods,
\begin{equation}
  2\omega=\left(\int_{\mathfrak{a}_j} \mathrm{d}u_i\right)_{i,j=1,\ldots,g},
  \quad  2\omega'=\left(\int_{\mathfrak{b}_j}
    \mathrm{d}u_i\right)_{i,j=1,\ldots,g}
\end{equation}
The period matrix $(2\omega,2\omega')$ is the first period matrix, and the
matrix $\tau=\omega^{-1}\omega'$ belongs to the upper Siegel
half-space, $\mathfrak{S}: \tau^T=\tau,\; \mathrm{Im}\, \tau >0 $.

The $\theta$-function $\theta[\alpha](\boldsymbol{u};\tau)$ with
characteristics $ [\alpha]=\left[ \begin{array}{c}
    \boldsymbol{\alpha'}^T\\\boldsymbol{\alpha''}^T
  \end{array} \right] $ , $[2\alpha]\in \mathbb{Z}^g\times\mathbb{Z}^g $
of the algebraic curve $V$ of genus $g$ is defined through its Fourier series
\begin{equation}
  \theta[\alpha](\boldsymbol{u};\tau)
  =\sum_{\boldsymbol{n}\in\mathbb{Z}^g}
  \mathrm{exp}\left\{ \imath\pi \left( \boldsymbol{n}+
      \boldsymbol{\alpha}'\right)^T\tau \left( \boldsymbol{n}+
      \boldsymbol{\alpha}'\right)+ 2\imath\pi\left( \boldsymbol{n}
      +\boldsymbol{\alpha}'\right)^T\left( \boldsymbol{u}
      +\boldsymbol{\alpha}''\right)
  \right\}
\end{equation}

We introduce the associated meromorphic differentials
$\mathrm{d}\boldsymbol{r}=(\mathrm{d} r_1,\ldots,\mathrm{d}r_g)^T$ and
their periods
\begin{equation}
  2\eta=-\left(\int_{\mathfrak{a}_j} \mathrm{d}r_i\right)_{i,j=1,\ldots,g},
\quad  2\eta'=-\left(\int_{\mathfrak{b}_j} \mathrm{d}r_i\right)_{i,j=1,\ldots,g}
\end{equation}
form the second period matrix  $(2\eta,2\eta')$.
The period matrices satisfy the condition
\begin{equation}
  \left( \begin{array}{cc} \omega&\omega'\\ \eta&\eta' \end{array} \right)
  \left( \begin{array}{cc}0&1_g\\-1_g&0   \end{array}\right)
  \left( \begin{array}{cc} \omega&\omega'\\ \eta&\eta' \end{array} \right)^T
  =-\frac{\imath\pi}{2} \left( \begin{array}{cc}0&1_g\\-1_g&0
    \end{array}\right)
\end{equation}
Here we denote the half-periods of the holomorphic and meromorphic
differentials by $(\omega,\omega')$ and $(\eta,\eta')$ in order to
emphasize resemblance to the
Weierstrass theory. We will also use the notation
$\mathcal{A}=2\omega$ and $\mathcal{B}=2\omega'$ for the periods of
holomorphic differentials. Further we denote
\begin{equation}
\mathrm{d}\boldsymbol{v}(Q)=(\mathrm{d} v_1(Q),\ldots,\mathrm{d}
v_g(Q))=\mathcal{A}^{-1} \mathrm{d}\boldsymbol{u}(Q)
\end{equation}
as the vector of normalized holomorphic differentials.

The explicit calculation of canonical holomorphic differentials and
the meromorphic differentials conjugate to them is well understood;
in particular we have
\begin{align*}
&\mbox{\qquad{\sf Example I}}
\\
&\mathrm{d}u_1=\frac{x\,\mathrm{d}x}{y},\qquad &\mathrm{d}u_2=\frac{\mathrm{d}x}{y},&\\
\\
&\mathrm{d}r_1=\frac{x^2\,\mathrm{d}x}{y},\qquad&\mathrm{d}r_2=\frac{x(\alpha_3+2 \alpha_4 x+12 x^2)\,\mathrm{d}x}{4y};\\
\\
&\mbox{\qquad{\sf Example II}}
\end{align*}
\begin{align*}
&\mathrm{d}u_1=\frac{\mathrm{d}x}{3y},\qquad
&\mathrm{d}u_2=\frac{x\mathrm{d}x}{3y^2},\qquad
&\mathrm{d}u_3=\frac{\mathrm{d}x}{3y^2},\qquad&\\
\\
&\mathrm{d}r_1=\frac{x^2\mathrm{d}x}{3y^2},\qquad
&\mathrm{d}r_2=-\frac{2xy\mathrm{d}x}{3y^2},\qquad
&\mathrm{d}r_3=-\frac{(5x^2+3\mu_3x+\mu_6)y}{3y^2}.\qquad&
\end{align*}

\begin{remark} Note that our labeling of the differentials
is the reverse of \cite{eemop07}, with the interchange $1
\leftrightarrow 2$ in example 1, and $(1,2,3) \leftrightarrow (3,2,1)$ in
example II.
\end{remark}

We introduce the fundamental bi-differential $\omega(Q,S)$ on $X\times X$
which is uniquely defined by the conditions:

{\bf(i)}  it is symmetric:
\begin{equation}
  \omega(Q,S)=\omega(S,Q)\label{omegasym}
\end{equation}

{\bf(ii)} it has its only pole along
the diagonal $P=Q$, in which neighborhood it is expanded in a power
series according to
\begin{align}
\omega(Q,S)-\frac{\mathrm{d}\xi(Q)\mathrm{d}\xi(S)}{(\xi(Q)-\xi(S))^2 }
=\sum_{m,n\geq 1} \omega_{mn}(P) \xi(Q)^{m-1}\xi(S)^{n-1}\mathrm{d}\xi(Q)
\mathrm{d}\xi(S)
\label{omegapole}
\end{align}

{\bf(iii)}  it is normalized such that:
\begin{equation}
\oint_{\mathfrak{a}_j}\omega(Q,S)=0,\qquad j =1, \ldots, g\label{omeganorm}
\end{equation}
The well known realization of the differential $\omega(Q,S)$ involves
the Schottky-Klein prime form $E(Q,S)$, which is a $(-1/2,-1/2)$-differential
defined for arbitrary points $Q,S\in X$
\begin{equation}
E(Q,S)=\frac{ \theta[\alpha]\left( \int_{Q}^S \mathrm{d} \boldsymbol{u}
\right) }{ h_{\alpha}(Q)h_{\alpha}(S) }
\end{equation}
where $\theta[\alpha](\boldsymbol{u})$ is a $\theta$-function with
non-singular odd characteristics $[\alpha]$ and
\[ h_{\alpha}(Q)^2=\sum_{k=1}^g \frac{\partial}{\partial u_k}
\theta[\alpha](\boldsymbol{0})\,\mathrm{d} v_k(Q) \]
The bi-differential $\omega(Q,S)$ is then given by
\begin{equation}
\omega(Q,S)=\mathrm{d}_Q\mathrm{d}_S \,\mathrm{ln}\, E(Q,S)
\end{equation}

We emphasize that in this paper, we will instead rely on alternative
``algebraic'' constructions of the differential $\omega(Q,S)$, as
described in \cite{fa73}.  By following classical works such as
\cite{kl86,kl88}, together with results documented in
\cite{ba97} we realize the differential $\omega(Q,S)$ in the form\footnote{The normalizing matrix $\varkappa$ is chosen here twice bigger then $\varkappa$ used earlier}
\begin{equation}
  \omega(Q,S)=\frac{\mathcal{F}(Q,S)}{f_y(Q) f_w(S)(x-z)^2}
  \mathrm{d}x\mathrm{d}z+
  \mathrm{d}\boldsymbol{u}(Q)^T\varkappa \mathrm{d}
  \boldsymbol{u}(S)
\end{equation}
where $Q=(x,y)$, $S=(z,w)$, and the function
$\mathcal{F}(Q,S)=\mathcal{F}((x,y),(z,w))$ is a polynomial of its
arguments, with coefficients depending on the moduli of the curve
$V$.  Finally, $\varkappa$ is a symmetric matrix $\varkappa^T=\varkappa$
that is chosen to provide a normalization of $\omega(Q,S)$; it is
expressible in terms of the first and second period matrices
$\varkappa= \omega^{-1}\eta$. We will refer to the term of
$\omega(Q,S)$ including the polynomial $\mathcal{F}(Q,S)$ as its
algebraic part,
\[
\omega(Q,S)=\omega^{\rm{alg}}(Q,S)+
\mathrm{d}\boldsymbol{u}(Q)^T\varkappa \mathrm{d}\boldsymbol{u}(S)
\]
In the vicinity of a point $P$, where points $Q$ and $S$ are represented
by local
coordinates $\xi(Q)$ and $\xi(S)$ respectively, the holomorphic part
of $\omega^{\rm{alg}}(Q,S)$ is expanded in the series
\begin{align*}
  &\left.\frac{\mathcal{F}(Q,S)\mathrm{d}z \mathrm{d}x }{f_y(Q) f_w(S)
      (x-z)^2} \right|_{ x=x(Q),z=x(S)} - \frac{\mathrm{d} \xi(Q)
    \mathrm{d} \xi(S)}{(\xi(Q)-\xi(S))^2 }
  \\&=\sum_{k,l=0}^{\infty}\omega_{k,l}^{\mathrm{alg}}(P) \xi(Q)^k
  \xi(S)^l \mathrm{d} \xi(Q) \mathrm{d} \xi(S)
\end{align*}
where the set of holomorphic functions
$\omega_{k,l}^{\mathrm{alg}}(P)$ defines a projective connection.

An algorithm to construct the polynomial $\mathcal{F}(Q,S)$ is known,
see e.g.\ \cite{ba97} and therefore functions such as
$\omega_{k,l}^{\mathrm{alg}}(P)$ that are important for the
construction are considered as known.  We shall present below some
explicit expressions for $\mathcal{F}$ as well as the few first terms
of the expansions $\omega^{\rm{alg}}$ in the simplest cases:

{\sf Example I}:

\begin{equation}\label{bi-diff}
\omega^{\rm{alg}}(P,Q)=\frac{F(x,z)+2yw}{4(x-z)^2}
\frac{\mathrm{d}x}{y}\frac{\mathrm{d}z}{w},\quad P=(x,y), Q=(z,w),
\end{equation}
where
\begin{equation} \label{hypbidiff}
F(x,z)= 4 x^2 z^2 (x+z) + 2 \alpha_4 x^2 z^2 +\alpha_3
x z(x+z) + 2 \alpha_2 xz +\alpha_1(x+z) +2 \alpha_0.
\end{equation}
Expanding this gives:
\begin{align*}
\omega_{0,0}^{\rm{alg}}&=-\frac{\alpha_4}{8},\\
\omega_{0,1}^{\rm{alg}}&=\omega_{1,0}^{\rm{alg}}=0,\\
\omega_{0,2}^{\rm{alg}}&=\omega_{2,0}^{\rm{alg}}=-\frac{16\alpha_3-3\alpha_4^2}{128},\qquad
\omega_{1,1}^{\rm{alg}}=0,\\
&\ldots
\end{align*}
\begin{remark}
For hyperelliptic curves, the coefficients $\omega_{i,j}^{\rm{alg}}$
vanish if either of $i$ or $j$ is odd.
\end{remark}

{\sf Example II}:

 \begin{equation}
   \omega^{\rm{alg}}((x,y),(z,w))=\frac{\mathcal{F}((x,y),(z,w))dxdz}{(x-z)^2
     f_y(x,y)f_w(z,w)}\label{omegatrig}
\end{equation}
with the polynomial $\mathcal{F}((x,y);(z,w))$ given by the formula
\begin{align}\begin{split}
\mathcal{F}\big(&(x,y),(z,w)\big)=w^2 y^2\\
&+w\left(w \left[\frac{f(x,y)}{y}\right]_y +T(x,z)\right)
 +y\left(y \left[\frac{f(z,w)}{w}\right]_w +T(z,x)\right)\end{split}
 \label{omegatrig1}
\end{align}
and
\begin{align}
    T(x,z)&=3\mu_{12}+ ( z+2x ) \mu_9+x ( x+2\,z )\mu_6\\
    &+3\mu_3 x^2 z + x^2 z^{2}+2\,x^3z.\label{omegatrig2}
\end{align}

Expanding this about $(\infty,\infty)$ gives:
\begin{eqnarray*}
\omega_{0,0}^{\rm{alg}}&=0,\\
\omega_{0,1}^{\rm{alg}}=\omega_{1,0}^{\rm{alg}}&=-\frac{2}{3}\mu_3,\\
\omega_{0,4}^{\rm{alg}}=\omega_{4,0}^{\rm{alg}}&=-\frac{2}{3}\mu_6+\frac{5}{9}\mu_3^2,\\
\omega_{1,3}^{\rm{alg}}=\omega_{3,1}^{\rm{alg}}&=-\frac{2}{3}\mu_6+\frac{4}{9}\mu_3^2,\\
\omega_{2,2}^{\rm{alg}}&=0,\\
\ldots
\end{eqnarray*}
\begin{remark} $\omega_{i,j}^{\rm{alg}}=0$ unless $(i+j)+2\equiv 0\, \mathrm{mod}\, 3$. This is a consequence of the cyclic symmetry of the curve. \end{remark}

We restrict ourselves to the case of algebraic curves with a branch
point at infinity, and take $P=(\infty,\infty)$ to be the base point
where we expand all our functions.

If $\mathcal{A}$ is the matrix of $\mathfrak{a}$-periods of canonical
holomorphic differentials, $ u_1,\ldots, u_g$,
then we define a set of `winding vectors' as follows:
\begin{equation}
\boldsymbol{U}_k(\infty)\equiv\boldsymbol{U}_k=\mathcal{A}^{-1}
\boldsymbol{R}_k,\quad k=1,\ldots,g, \label{winvec}
\end{equation}
where $\boldsymbol{R}_1,\boldsymbol{R}_2,\ldots$ are residues of
canonical holomorphic integrals multiplied by differentials of the
second kind with poles of order $k$ at infinity, giving:
\[
\boldsymbol{R}_k= \frac{1}{k}\left. \frac{\mathrm{d}^{k-1}}{ \mathrm{d}
\xi(Q)^{k-1}}\mathrm{d} \boldsymbol{u}(Q) \frac{1}{\mathrm{d}\xi(Q) }
\right|_{Q=\infty}.
\]

Let $\boldsymbol{v}\in \mathrm{Jac}(X) $ and $\theta(\boldsymbol{v})$
be a canonical $\theta$-function, that is, a $\theta$-function with zero
characteristics:
\[
\theta(\boldsymbol{v})=\sum_{\boldsymbol{m}\in \mathbb{Z}^g} \mathrm{exp}
\left\{ \imath\pi \boldsymbol{m}^T\tau \boldsymbol{m} +2\imath \pi
\boldsymbol{v}^T \boldsymbol{m} \right\}.
\]
The principal point of this paper is the transition from $\theta$ to
$\sigma$-functions. For any point $\boldsymbol{u}\in \mathrm{Jac}(X)$
we define:
\begin{equation}
\sigma(\boldsymbol{u})=C(\tau)\theta(\mathcal{A}^{-1} \boldsymbol{u})
\mathrm{exp}\left\{\frac12  \boldsymbol{u}^T \varkappa \boldsymbol{u}
 \right\},
\end{equation}
where $\theta(\boldsymbol{v})$ is the canonical $\theta$-function, and
$C(\tau)$ is a certain modular constant that we do not need for the
results that follow.  We note that this $\sigma$-function differs from
the ``fundamental $\sigma$ function'' of the publications mentioned in
the introduction by the absence of a shift of the $\theta$-argument by
the vector of Riemann constants. Thus $\sigma(\boldsymbol{0})\neq 0$.

We introduce the Kleinian multi-variable $\zeta$ and $\wp$-functions as above
(\ref{Kleinwp}). These functions are suitable coordinates to describe
Abelian functions and the KP-type hierarchies of differential relations
between them.

In higher genera, the relations between Abelian functions become
somewhat lengthy. These relations can often be summarized concisely by
developing a matrix formulation of the theory, as has been done \cite{bel97b}
in the hyperelliptic case.
However to avoid confusion we should point out that there is at present
no known relation between the Pl\"ucker method and such a matrix
summary of the results.

The vector of normalized holomorphic differentials
$\mathrm{d}\boldsymbol{v}=(\mathrm{d} v_1,\ldots,\mathrm{d}v_g)^T$
is given by
\begin{equation}
  \mathrm{d}\boldsymbol{v}=\mathcal{A}^{-1}\mathrm{d}\boldsymbol{u}, \qquad
  \mathcal{A}=\left( \oint_{\mathfrak{a}_j}\mathrm{d} u_i \right)
\end{equation}

The Sato-Fay algebro-geometric $\tau$-function of the genus $g$ curve
$V$ of arguments $\boldsymbol{t}=(t_1,\ldots,t_g,t_{g+1},\ldots )^T$,
$\boldsymbol{u}\in \mathrm{Jac}(X)$, $P\in X$ is defined as
\begin{align}
\tau(\boldsymbol{t};\boldsymbol{u})=\theta\left( \sum_{k=1}^{\infty}
\boldsymbol{U}_k(P) t_k+\boldsymbol{u} \right) \mathrm{exp}
\left\{ \frac12\sum_{m,n\geq 1} \omega_{mn}(P) t_mt_n \right\}\label{tau}
\end{align}
Here the winding vectors defined in (\ref{winvec}),
$\boldsymbol{U}_k(P)$, appear in the expansion
of the normalized holomorphic integral $\boldsymbol{v}$ in the
vicinity of the given point $P\in X$,
\[
\int\limits_{P_0}^Q \mathrm{d}\boldsymbol{v}(Q')
=\int_{P_0}^P\mathrm{d}\boldsymbol{v}(Q')+
\sum_{k=1}^{\infty} \boldsymbol{U}_k(P) \xi(Q)^k
\]
with $\xi(Q)$ being the local coordinate of the point $Q$ in the
vicinity of the given point $P$, so that $\xi(P)=0$.

The quantities $\omega_{mn}(P)$ define the holomorphic part of the
expansion of the fundamental second kind differential $\omega(Q,S)$
near the point $P$ according to (\ref{omegapole}). Using the above definitions, we can see that the algebro-geometric
$\tau$-function is given by the formula, equivalent to (1.7):
\begin{align}
\frac{ \tau(\boldsymbol{t};\boldsymbol{u})}
{\tau(\boldsymbol{0};\boldsymbol{u})}=
\frac{ \sigma\left(\sum_{k=1}^{\infty} \boldsymbol{R}_k t_k+
\boldsymbol{u}\right)}
{\sigma(\boldsymbol{u})}\,\mathrm{exp} \left\{ \frac12
\sum_{k,l=0}^{\infty} \omega_{k,l}^{\mathrm{alg}}t_kt_l  \right\}.
\label{tausigma2}\end{align}
Here $\omega_{k,l}^{\mathrm{alg}}$ is the \emph{algebraic
part} of the holomorphic part of the expansion of the bi-differential,
as defined in Section 2:
\begin{align*}
&\left.\frac{\mathcal{F}(Q,S)\mathrm{d}z \mathrm{d}x }
{f_y(Q) f_w(S) (x-z)^2} \right|_{ x=\xi(Q),z=\xi(S)}
 - \frac{\mathrm{d} \xi(Q) \mathrm{d} \xi(S)}{(\xi(Q)-\xi(S))^2 } \\
&=\sum_{k,l=0}^{\infty}\omega_{k,l}^{\mathrm{alg}} \xi(Q)^k \xi(S)^l
\mathrm{d} \xi(Q) \mathrm{d} \xi(S) ,
\end{align*}
where $\mathcal{F}$ is a polynomial of its variables whose construction
is classically known \cite{ba97}, see also the recent exposition
\cite{na08}.  One can see that the non-algebraic part, i.e.\ the
normalizing bi-linear form, is absorbed into the
$\sigma$-function.

Once we have an expression involving derivatives of the sigma
function, we need to convert this to an expression involving
derivatives of the $\wp$-function.  To do this we start with the
definition of the $\zeta$ function (\ref{zeta}), which we write in the form
\[
\sigma_i(\mathbf{u})=\zeta_i(\mathbf{u})\sigma,
\]
then repeated differentiation gives us a ladder of relations which
enable us to recursively express any derivative of sigma in terms of
$\wp_{ij\dots k}$, $\zeta_i$ and $\sigma$, all evaluated at
$\mathbf{u}$.
\begin{align*}
  \sigma_{ij}&=\sigma_j\zeta_i-\sigma \wp_{ij}\\
  \sigma_{ijk}&=\sigma_{jk}\zeta_i-\sigma \wp_{ijk}-\sigma_k \wp_{ij}
  -\sigma_j \wp_{ik}
  \\
  \vdots
\end{align*}

In the following sections, we will look at different methods for constructing
such relations between the derivatives of $\sigma$, and hence between the
Abelian functions associated with the curve.

\section{Solution of KP hierarchy }
The starting point for this approach is the Klein formula, which compares
two different expressions for the fundamental bidifferential:
\begin{theorem}
  Let the canonical holomorphic differentials of the curve $f(x,y)=0$ be
  represented in the form
\[
u_k(x,y)=\frac{\mathcal{U}_k(x,y)}{ f_y(x,y) } \mathrm{d}x,\quad
k=1,\ldots,g;
\]
then we have:
\begin{align}
\label{klein}\begin{split}
  &\sum_{i,j=1}^g \wp_{ij}\left( \int_{P}^{(x,y)}\mathrm{d}
    \boldsymbol{u}-\sum_{k=1}^g \int_{P}^{(x_k,y_k)}\mathrm{d}
    \boldsymbol{u} +\boldsymbol{K}_P \right)
  \mathcal{U}_i(x,y)\mathcal{U}_j(x_k,y_k)\\&=\omega^{\rm{alg}}
  (x,y;x_k,y_k)\frac{f_y(x,y)}{\mathrm{d}x}\frac{f_{y_k}(x_k,y_k)}
  {\mathrm{d}x_k},\quad k=1,\ldots,g
\end{split}
\end{align}
\end{theorem}

We note that the left hand side is singular where $(x,y)$ equals any of the
$(x_k,y_k)$. We now take  the Laurent expansion of the two sides of (\ref{klein})
as  $(x,y)\rightarrow P$. Equating corresponding terms on either side will
give a polynomial equation in $(x_k,y_k)$, whose coefficients are differential
expressions in the $\wp_{ij}$. The leading term in this series leads to
the Jacobi inversion formula for the curve; higher terms lead to other polynomials
which must have the same roots, and differential relations between the $\wp_{ij}$
then follow.

\subsection{Case of genus two}
The $\sigma$-functional realization of hyperelliptic functions of a
genus two curve has already been discussed in many places, see
e.g. \cite{ba97,bel97b}. But we shall briefly describe here
the principal points of the construction to convey the structure of
the theory that we wish to develop for higher genera curves.

We consider the genus two hyperelliptic curve
\begin{equation}
y^2=4x^5+\alpha_4 x^4+\ldots+\alpha_0
\end{equation}
The algebraic part of the fundamental bi-differential is given as
above in eqn.  (\ref{bi-diff}) \cite{ba97}. The first term in the expansion is independent of $y$:
\[
\wp_{12}+x \wp_{11}-x^2=0,
\]
which is the $x$-part of the Jacobi inversion formula for this curve.
This equation gives us a way of reducing quadratic and higher terms in $x$
to linear terms.  The third term is also independent of $y$, and after
elimination of higher powers of $x$ gives a relation linear in $x$.
\begin{align*}
  &(\tfrac12 \wp_{1111} -3 \wp_{11}^2-\tfrac12 \lambda_4 \wp_{11} - 2
  \wp_{12} - \tfrac14 \lambda_3 ) x\\ &\qquad+ \tfrac12 \wp_{1112} - \tfrac12
  \lambda_4 \wp_{12} - 3 \wp_{11} \wp_{12} + \wp_{22}=0.
\end{align*}
Equating the coefficients of different powers of $x$ separately to
zero, we can solve for $\wp_{1111}$ and $\wp_{1112}$ to find:
\begin{align}
 \wp_{1111} &= 6 \wp_{11}^2 + \lambda_4 \wp_{11} + 4 \wp_{12} +
  \tfrac12 \lambda_3,\label{R1111}\\
 \wp_{1112} &= 6\wp_{11} \wp_{12} +\lambda_4 \wp_{12}  -2\wp_{22},\label{R1112}
\end{align}
to get the first two 4-index relations in the genus 2 case.  Higher
order terms follow in a similar way from higher order terms in the
expansion.  The second term in the expansion, the $y$-part of the Jacobi
inversion formula, is linear in $y$, so can
be used to eliminate any $y$-dependent terms which arise at higher order.

At every stage we need to substitute for higher derivative terms (such as
$\wp_{11111}$, for example) by using derivatives of previously derived
relations.
In addition, multiplication by a 3-index $\wp_{ijk}$ is
sometimes useful, followed by substitution of known relations which
are quadratic in the $\wp_{ijk}$. The first such relation is found to be
$$ \mathrm{Jac}_{6}:\quad \wp_{111}^2  = 4 \wp_{11}^3+\alpha_{3}
  \wp_{11}+\alpha_{4} \wp_{11}^2+4 \wp_{12} \wp_{11}+\alpha_{2}+4
  \wp_{22}.$$

It is often possible to obtain all or almost all of the required
relations using only terms from the expansion of (\ref{klein}) and
differentiation or algebraic means, see \cite{eemop07} for the case of
the genus 3 trigonal curve.  The two examples discussed in this paper,
the genus 2 hyperelliptic and the genus 3 cyclic trigonal can be
studied in this manner, although a small number of the high weight quadratic
$\wp_{ijk}$ relations in the trigonal case have not yet been derived
in this way.  A complication at higher weight is that equations from
the Kummer variety (see below) appear in addition, and it is often
difficult to separate these out.  A similar complication occurs in the
methods described below.

A variation to the classical method described above, which avoids the
problems associated with ideals containing the Kummer relation, is to
build up a series expansion of the $\sigma$-function in parallel to
deriving the PDEs in the $\wp$ variables.  This approach was followed
in \cite{balgib04,eemop07}.  The first equations from the expansion of
(\ref{klein}) are used, together with various results about the
vanishing of the sigma function on certain $\Theta$-strata, and a
knowledge of the weight of the sigma expansion, to derive the first
few terms in the sigma expansion. These can then be used to derive
some PDEs in the $\wp$, which are then used in a bootstrap fashion to
derive further terms in the sigma expansion, etc.  In the case of the
genus $6$, $(4,5)$ curve, \cite{ee09}, this method is used exclusively to
derive the required equations.

\subsection{Solving in Pl\"ucker coordinates}
 The key to this approach is Sato's formula, Theorem (5.1) below.
 This gives an expansion
 of a ratio of $\tau$-functions, $\frac{\tau(\boldsymbol{t};\boldsymbol{u})}
{\tau(\boldsymbol{0};\boldsymbol{u})}$, as a series of differential expressions
which are rational in the $\tau$-functions. The $\boldsymbol{t}$-dependence
is given in terms of Schur polynomials $s_\lambda(\boldsymbol{t})$
in the times $t_i$. The coefficients of these polynomials are determinants
of differential expressions of $\tau$, which are Pl\"ucker coordinates on
a Grassmannian. Such coordinates satisfy the Pl\"ucker relations - each partition $\lambda$ can be expanded in hooks, and the corresponding Pl\"ucker coordinates
are expressible, analogously to Giambelli's formula, as determinants of
single hook partitions. These relations give the differential relations
for the Abelian functions which we seek.

For any partition $\lambda: \alpha_1\geq \alpha_2\geq \ldots \geq
\alpha_n$ of $|\lambda|=\sum_{i=1}^n\alpha_i$, the Schur polynomial of
$n$ variables $x_1,\ldots,x_n$ is defined by
\[
s_{\lambda}(\boldsymbol{x})=\mathrm{det}\left( p_{\alpha_i-i+j}(\boldsymbol{x})
\right)_{i,j=1,\ldots,n}
\]
where the elementary Schur functions $p_m(\boldsymbol{x})$ are
generated by the series,
\[
\sum_{m=0}^{\infty} p_m(\boldsymbol{x}) t^m =\mathrm{exp}
\left\{ \sum_{n=1}^{\infty} x_n t^n\right\}
\]
The first few Schur polynomials are
\begin{align*}
  &s_{1}(\boldsymbol{x})=x_1, \\
  & s_{2}(\boldsymbol{x}) =x_2+\tfrac12
  x_1^2,\quad
  s_{1,1}(\boldsymbol{x})=-x_2+\tfrac12 x_1^2,\\
  &s_{3}(\boldsymbol{x})=x_3+x_1x_2+\tfrac16 x_1^3, \quad
  s_{2,1}(\boldsymbol{x})=-x_3+\tfrac13 x_1^3,\quad
  &s_{1,1,1}(\boldsymbol{x})=x_3-x_1x_2+\tfrac16 x_1^3\\
  &s_{4}(\boldsymbol{x})=x_4+x_1x_3+\tfrac12 x_2^2+\tfrac12 x_1^2x_2
  +\tfrac{1}{24}x_1^4, \quad \text{etc.}
\end{align*}

The Cauchy-Littlewood formula
\[
\mathrm{exp}\left\{ \sum_{n=1}^{\infty} n x_n y_n \right\}
=\sum_{\lambda}s_{\lambda}(\boldsymbol{x})
s_{\lambda}(\boldsymbol{y}),
  \]
  where $s_{\lambda}(\boldsymbol{x})$ is the Schur function of the partition
  $\lambda:\alpha_1\geq \alpha_2\geq \ldots \geq \alpha_n$, leads to
  the Taylor expansion
\begin{align*}
  f(\boldsymbol{x})=\left.\mathrm{exp}\left\{
      \sum_{n=1}^{\infty}x_n\frac{\partial}{\partial y_n} \right\}
    f(\boldsymbol{y}) \right|_{\boldsymbol{y}=0}
  =\left.\sum_{\lambda}s_{\lambda}(\boldsymbol{x})
    s_{\lambda}\left(\frac{1}{n} \frac{\partial}{\partial y_n} \right)
    f(\boldsymbol{y})\right|_{\boldsymbol{y}=0}
\end{align*}
Giambelli's formula shows how to expand a Schur function $s_{\lambda}$
in hooks,
\[
s_{\lambda}(\boldsymbol{x})=\det \left(
s_{\alpha_i+1,1^{\beta_j}}(\boldsymbol{x}) \right)_{1\leq i,j\leq r}
\]
where hook notations, $(n,1^m)$ are used. In what follows we shall
also use the Frobenius notation to describe the decomposition of a partition
into finitely many hooks:
\[
(\lambda)=(\alpha_1,\ldots,\alpha_r\vert \beta_1,\ldots,\beta_r ).
\]
In particular, a single hook has the Frobenius notation:
\[(n,1^m) = (n-1 \vert m),\]
with $n\ge 1$ and $m\ge0$;
while a partition which decomposes into two hooks is notated as:
\[(n,m, 2^k, 1^l) = (n-1, m-2 \vert k+l+1,k),\]
where $n>m>1$, $k\ge 0$ and $l\ge 0$.

\begin{theorem}[\bf Sato formula ] Let
  $\tau(\boldsymbol{t};\boldsymbol{u})$ be any function of vector
  arguments
\[
\boldsymbol{U}_1t_1,\boldsymbol{U}_2t_2,\ldots,\quad
\boldsymbol{t}=(t_1,t_2,\ldots)\in\mathbb{C}^{\infty}
\]
where $\boldsymbol{U}_1,\boldsymbol{U}_2\ldots $ is an infinite set of
constant complex vectors from $\mathbb{C}^{g}$ and
$\boldsymbol{u}\in\mathbb{C}^g$ is a parameter. Suppose that
$\tau(\boldsymbol{0};\boldsymbol{u})\neq 0$. Then for any partition
$(\lambda)=(\alpha_1,\ldots,\alpha_r\vert \beta_1,\ldots,\beta_r) $
and any $\boldsymbol{u}\in\mathbb{C}^g$
\begin{equation}
\frac{\tau(\boldsymbol{t};\boldsymbol{u})}
{\tau(\boldsymbol{0};\boldsymbol{u})}=
\sum_{\lambda} s_{\lambda}(\boldsymbol{t})\det \left( (-1)^{\beta_j+1}
A_{(\alpha_i|\beta_j)} (\boldsymbol{u}) \right)
\end{equation}
where the $A_{(m|n)}(\boldsymbol{u})$ with  $m\geq 0$, $n\geq 0$ form a
linear basis of the Grassmannian:
\begin{align}\begin{split}
    A_{(m|n)}(\boldsymbol{u})&=-A_{(n|m)}(-\boldsymbol{u})\\
    &=\left.(-1)^{n+1}s_{m+1,1^n}
      (\partial_{\boldsymbol{t}})\tau(\boldsymbol{t};\boldsymbol{u})
    \right|_{\boldsymbol{t}=0}\tau(\boldsymbol{0};\boldsymbol{u})^{-1}\\
    &=\left.\sum_{\alpha=0}^m p_{n+\alpha+1}(-\partial
      _{\boldsymbol{t}}) p_{m-\alpha}(\partial_{\boldsymbol{t}})
      \tau(\boldsymbol{t}; \boldsymbol{u})\right|_{\boldsymbol{t}=0}
    \tau(\boldsymbol{0}; \boldsymbol{u})^{-1}
\end{split}\label{Amn}
\end{align}
and
\[
\partial _{\boldsymbol{t}} =\left( \frac{\partial}{\partial t_1},
  \frac12 \frac{\partial}{\partial t_2}, \frac13
  \frac{\partial}{\partial t_3},\ldots, \right)
\]
\end{theorem}

\begin{theorem}[\bf Pl\"ucker coordinates] For any partition
  $(\lambda)=(\alpha_1,\ldots,\alpha_r\vert \beta_1,\ldots,\beta_r )$
  and any $\boldsymbol{u}\in\mathbb{C}^g$, the $\tau$-function satisfies:
\begin{equation}
  \left.\tau(\boldsymbol{0};\boldsymbol{u})^{r-1}s_{\lambda}
    (\partial_{\boldsymbol{t}})\tau(\boldsymbol{t};\boldsymbol{u})
  \right|_{\boldsymbol{t}=0}=\left.\det \left( s_{\alpha_i+1,1^{\beta_j}}
      (\partial_{\boldsymbol{t}}) \tau(\boldsymbol{t};\boldsymbol{u})
    \right|_{\boldsymbol{t}=0}  \right).\label{pr}
\end{equation}
\end{theorem}
In particular, for any curve $V$, its Sato algebro-geometric $\tau$-function
has the expansion
\begin{align}
\begin{split}
  \frac{\tau(\boldsymbol{t};\boldsymbol{u})}
  {\tau(\boldsymbol{0};\boldsymbol{u})}& =1+ A_{(0|0)}(\boldsymbol{u})
  {s_1} (\boldsymbol{t}) +A_{(1|0)}(\boldsymbol{u}){s_{2}}(\boldsymbol{t})\\&+A_{(0|1)}(\boldsymbol{u})
  {s_{1,1}}(\boldsymbol{t})+\ldots\\
  &=1+ A_{(0|0)} t_1 + A_{(1|0)} (t_2 +\tfrac12 t_1^2) +A_{(0|1)}(\boldsymbol{u})( -t_2+\tfrac12 t_1^2)+\ldots
\end{split}
\end{align}
where $A_{(m|n)}(\boldsymbol{u})$ as defined in (\ref{Amn}) are the elements
of the
semi-infinite matrix $A$,
\[
A=\left(\begin{array}{cccc} A_{(0|0)}(\boldsymbol{u})
&A_{(0|1)}(\boldsymbol{u})&A_{(0|2)}(\boldsymbol{u})
&\ldots\ldots\\
    A_{(1|0)}(\boldsymbol{u})&A_{(1|1)}(\boldsymbol{u})
&A_{(1|2)}(\boldsymbol{u})&\ldots\ldots\\
    A_{(2|0)}(\boldsymbol{u})&A_{(2|1)}(\boldsymbol{u})
&A_{(2|2)}(\boldsymbol{u})&\ldots\ldots\\
    \vdots&\vdots&\vdots&\ldots\ldots
  \end{array} \right)  = (\boldsymbol{A}_0,
\boldsymbol{A}_1,\ldots\ldots)
\]
with infinite vectors $\boldsymbol{A}_i$, $i=1,2,\ldots$. To more
complicated partitions such as
$(m_1,\ldots,m_k|n_1,\ldots,n_k)$ we associate according to the Giambelli formula certain minors of $A$,
 \[
 A_{(m_1,\ldots,m_k|n_1,\ldots,n_k)}(\boldsymbol{u})=\mathrm{det}
\left( A_{(m_i|n_j  ) (\boldsymbol{u})} \right)_{i,j=1\ldots,k}.
 \]
 Note that  these $A$'s are not independent -- they satisfy the
 Pl\"ucker relations. These are a family of differential equations
 satisfied by $\tau$, that represent  a completely integrable hierarchy
 of KP-type.

 The definitions and relations given above are valid for any
 multivariate function $\tau(\boldsymbol{t};\boldsymbol{u})$. Below
 we consider functions $\tau$-functions constructed on the Jacobi
 varieties of algebraic curves.

To each symbol  we shall put in correspondence its weight
$$
\wp_{\underbrace{\scriptstyle 1,\ldots, 1}_{k_1},
  \underbrace{\scriptstyle 2,\ldots, 2}_{k_2}, \ldots,
  \underbrace{\scriptstyle g,\ldots, g}_{k_g}}\quad \Leftrightarrow
\quad \sum_{j=1}^gk_jw_j
$$
where $w_i$ is the order of vanishing of the holomorphic integral
$\int\mathrm{d}u_i$ at infinity. In other words, if the curve has
a Weierstrass point at infinity, then $w_i=1<w_2<\ldots<w_g$ is the
Weierstrass gap sequence at infinity.

For a given weight $W$ consider all Young tableaux which
decompose into only two hooks, and write
corresponding relations between multi-index symbols
$\wp_{i_1,\ldots,i_n}$. The first non-trivial Young tableau
corresponds to the partition $\lambda=(2,2)$. There is only one
multi-index symbol of weight 4, that is $\wp_{1111}(\boldsymbol{u})$
and its corresponding Pl\"ucker relation of any curve is of the form
\begin{equation}
\wp_{1111}(\boldsymbol{u})=\text{ polynomial of even symbols} \;
\wp_{ij}(\boldsymbol{u})
\end{equation}
For higher weights a larger number of multi-index functions can be
constructed, and this number grows rapidly with increasing weight. To
find them in terms in the form of polynomials of two-index functions,
we shall write Pl\"ucker relations corresponding to independent
tableaux of the same weight, and solve the corresponding linear
systems.  The technique is best illustrated by examples.

We consider the genus two hyperelliptic curve, with the differentials and
the bidifferential chosen as in Example 1 above.
The $\sigma$-functional realization of hyperelliptic functions of such a
genus two curve has already been discussed in many places, see
e.g. \cite{ba97,bel97b}. But we shall briefly describe here
the principal points of the construction to convey the structure of
the theory that we wish to develop for higher genera curves.

The algebraic part of the fundamental bi-differential is given as
above in eqn. \ (\ref{bi-diff}) \cite{ba97}.
$$ \omega^{\rm{alg}}(P,Q)=\frac{F(x,z)+2yw}{4(x-z)^2}
\frac{\mathrm{d}x}{y}\frac{\mathrm{d}z}{w},\quad P=(x,y), Q=(z,w).$$

The simplest class of non-trivial Pl\"ucker relations is found from the class
of Young diagrams which may be decomposed into two hooks, that is, those
of the form $(2+m,2+n,2^k,1^l)$, with $m\ge n \ge 0, k \ge 0, l \ge 0$. In
Frobenius  `hook' notation these are $(m+1,n|k+l+1,l)$.
The corresponding Pl\"ucker relations read
\begin{equation}
A_{(m+1,n|k+l+1,l)} = \left|\begin{array}{cc} A_{(m+1,k+l+1)}& A_{(m+1,l)}\\
                                              A_{(n,k+l+1)}& A_{(n,l)}
\end{array}\right|.
\label{two-hook}
\end{equation}
These equations are all bilinear partial differential equations in the
$\tau$-function.  They may be expanded in terms of the Kleinian
$\wp_{ij}$ and $\zeta_i$ functions.

The first Young tableau leading to a non-trivial Pl\"ucker relation
corresponds to the partition $\lambda=(2,2)$, with Young diagram
$$ \yng(2,2).$$
Writing (\ref{pr}) in
this case we obtain after simplification
\begin{equation}
\mathrm{KdV}_{4}: \qquad \wp_{1111}(\boldsymbol{u})=
6\wp_{11}^2(\boldsymbol{u})+4\wp_{12}(\boldsymbol{u})
+\alpha_4\wp_{11}(\boldsymbol{u})+\tfrac12\alpha_3. \label{p1111}
\end{equation}
This is of course the same equation as (\ref{R1111}).
The weight of the tableau is 4 in this case, which is the same as the
weight of the equation, defined as the weighted sum of indices in
which the index ``1'' has weight 1 and index ``2'' has weight 3.  We
will use this weight correspondence in other cases too, and will
denote the weight of the object by subscript $i+3 j$ where $i$ and $j$
are respectively the numbers of 1's and 2's in the multi-index
relation.
The relation (\ref{p1111}) is the analogue of Weierstrass' equation for $\wp''$
in the genus 1 case.

The next group of tableaux are of weight 5, and correspond to the
partitions $\lambda=(3,2)$ and $\lambda=(2,2,1)$, with their
transposes, which give the same equations, in the hyperelliptic case.
Both these Pl\"ucker relations lead to the equation
\[
\left( \zeta_1(\boldsymbol{u}) + \frac{\partial}{\partial u_1}\right)
\mathrm{KdV}_{4}=0
\]
i.e. the Young tableaux of weight 5 give no new equations and we conclude
that the following correspondence is valid
\[
\left\{ \begin{array}{c}  \lambda=(2,2)\\
    \lambda=(3,3),\quad \lambda=(2,2,1) \end{array}\right\}
\Longleftrightarrow \mathrm{KdV}_{4}
\]
At weight 6 we have three independent Young tableaux with $(2,2)$ centres
\[
\yng(4,2),\hskip1cm \yng(3,3)\,,\hskip1cm \text{and} \hskip1cm \yng(3,2,1).
\]
The other two tableaux of weight 6
\[
\yng(2,2,2) \hskip1cm \text{and} \hskip1cm\yng(2,2,1,1)
\]
again give the same results  as their transposes.
\begin {remark}The Pl\"ucker relations associated
with any Young diagram and its transpose are always the same for a
hyperelliptic curve; however for general curves, this is no longer true.
\end{remark}

These three independent tableaux give an overdetermined system of three
equations.  After substituting for higher derivatives of the known
relation (\ref{p1111}), we can solve for the two unknowns $\wp_{1112}$
and $\wp_{111}^2$ to get
\begin{align}
  \mathrm{KdV}_{6}:\quad \wp_{1112} &= 6 \wp_{12} \wp_{11}-2 \wp_{22}
  + \alpha_{4} \wp_{12}\label{p1112}\\
  \mathrm{Jac}_{6}:\quad \wp_{111}^2 & = 4 \wp_{11}^3+\alpha_{3}
  \wp_{11}+\alpha_{4} \wp_{11}^2+4 \wp_{12} \wp_{11}+\alpha_{2}+4
  \wp_{22}\label{p111s}
\end{align}
The relation (\ref{p1112}) is another generalization of the equation
for $\wp''$ in the genus one case, and is identical to (\ref{R1112}) found
using the classical method. The relation (\ref{p111s})
corresponds to Weierstrass' equation for $(\wp')^2$ in the genus one
case.  In what follows we will always assume that higher derivatives of the
known relations at a lower weight have been eliminated.

At weight 7 we have 5 independent Young tableaux with (2,2) centres.
The resulting overdetermined system of equations, of weight 7, contain
$\zeta_1$ multiplied by linear combinations of the two weight 6
relations (\ref{p111s},\ref{p1112}).  In addition we get a new
relation at weight 7, namely
\[
\wp_{12} \wp_{111}-\wp_{122}-\wp_{112} \wp_{11}=0 \label{L5}
\]
We refer to relations of this type, linear in the three-index $\wp_{ijk}$,
as quasilinear.
These have no counterpart in the genus 1 theory.
They can also be derived by cross-differentiation between the two
relations (\ref{p1111},\ref{p1112}), since
\[
\frac{\partial}{\partial u_2} \wp_{1111} = \frac{\partial}{\partial
  u_1} \wp_{1112}.
\]
Equations $KdV_4$ and $KdV_6$ describe the genus 2 solutions of the
KdV hierarchy associated with the given curve. To describe the Jacobi
variety and the Kummer variety we must consider tableaux of higher
weights.

For the tableaux of weight 8 we find a set of eight overdetermined
equations with solution given by
\begin{eqnarray*}
\mathrm{ Jac}_8:\;\wp_{112}^2&=&\alpha_0-4\wp_{22}\wp_{12}+\alpha_4\wp_{12}^2
+4\wp_{11}\wp_{12}^2\\
\mathrm{KdV}_{8}:\;\wp_{1122}&=&2\wp_{11}\wp_{22}+4\wp_{12}^2+
\tfrac12\alpha_3\wp_{12},
\end{eqnarray*}
And at weight 9 the only new relation is the quasilinear relation
\[
8\wp_{122}\wp_{11}-4\wp_{12}\wp_{112}-4\wp_{222}+2\alpha_{4}\wp_{122}
-\alpha_{3}\wp_{112}-4\wp_{111}\wp_{22} = 0
\]
At weight 10 we have 18 overdetermined system of equations, which can
be solved for the 3 functions of weight 10, $\wp_{1222}, \wp_{112}^3,
\wp_{111}\wp_{122} $, giving
\begin{align*}
  \mathrm{KdV}_{10}:\;\wp_{1222}&=6\wp_{12}\wp_{22}+\alpha_2\wp_{12}
  -\tfrac12\alpha_1\wp_{11}-\alpha_0\\
  \mathrm{ Jac}_{10}^{(1)}:\;\wp_{111}\wp_{122}&=
  -\tfrac12\alpha_1\wp_{11} +2\wp_{22}\wp_{11}^2
  +2\wp_{11}\wp_{12}^2+\alpha_2\wp_{12}+4\wp_{22}\wp_{12}
  +\tfrac12\alpha_3\wp_{12}\wp_{11}\\
  \mathrm{ Jac}_{10}^{(2)}:\;\wp_{112}^2&=\alpha_0-4\wp_{22}\wp_{12}
  +\alpha_4\wp_{12}^2  +4\wp_{11}\wp_{12}^2
\end{align*}
The equations
$\mathrm{Jac}_8,\mathrm{Jac}_{10}^{(1)},\mathrm{Jac}_{10}^{(2)}$
represent an embedding of the Jacobi variety as a 3-dimensional
algebraic variety into the complex space $\mathbb{C}^5$ whose
coordinates are $\wp_{11},\wp_{12},\wp_{22},\wp_{111}, \wp_{112}$

At weight 12 we get the final 4-index relation for $\wp_{2222}$ and two
quadratic 3-index relations for $\wp_{112}\wp_{122}$ and
$\wp_{111}\wp_{222}$. We can continue in this manner at weight 14 to
get more quadratic 3-index relations \cite{bel97b}.  At the odd weights
11, 13, 15, we get quasilinear relations which can also be found by
cross-differentiation.  As a practical point we note that the
equations we derive can often contain ideals generated by the lower
weight relations, and some work is required to identify genuinely new relations.

At weight 16 we have 117 independent relations giving an
overdetermined system of equations (we have checked only a selection of
these). At this weight a new feature occurs.  As well as the
equations expressing the quadratic 3-index term $\wp_{122}\wp_{222}$
in terms of cubics in the $\wp_{ij}$, we have terms which are quartic
in the $\wp_{ij}$.  We can pick one of these quartic terms, say
$\wp_{12}^4$, and solve for this and for $\wp_{122}\wp_{222}$ to give
us two relations.  The relation involving a quartic in the $\wp_{ij}$
is just the Kummer variety of the curve, which can also be found from
the identity
\[
(\wp_{111}^2)(\wp_{112}^2)-(\wp_{111}\wp_{112})^2=0.
\]
This is the quotient of the Jacobi variety,
$\mathrm{Kum}(X)=\mathrm{Jac}(X)/\boldsymbol{u}\rightarrow -
\boldsymbol{u}$. In the case $g=2$, the Kummer variety is a surface
in $\mathbb{C}^3$ which is given analytically by a quartic equation.
The same quartic also appears, multiplied by $\wp_{11}$, etc., at weights
18 and 20.

\subsection{Case of trigonal curve of genus three}
As before we consider our $(3,4)$ example, (2.3) whose holomorphic differentials
are given in (2.x).
 \cite{eel00,bel00,eemop07}
(here we generally follow the notation of \cite{eemop07}). The fundamental second kind differential is (\ref{omegatrig}), (\ref{omegatrig1}),(\ref{omegatrig2}).

The first Young tableau leading to
a non-trivial Pl\"ucker relation, as in the genus 2 case, corresponds
to the partition $\lambda=(2,2)$.
Writing (\ref{pr}) in this case we obtain, after simplification,
\[
\wp_{1111}= 6\, \wp_{11}^{2}-3\,\wp_{22}.
\]
The weight of the tableau is again 4 in this case, which is the same
as the weight of the equation, defined as the weighted sum of indices
in which the index ``1'' has weight 1, index ``2'' has weight 2, and
index ``3'' has weight 5.  We will use this weight correspondence in
other cases too, and will denote the weight of the object by subscript
$i+2 j+5k$ where $i,j$ and $k$ are respectively the numbers of 1's,
2's and 3's in the multi-index relation. Note also that our notation
is reversed with respect to \cite{eemop07}, with the interchange $1
\leftrightarrow 3$.

In the trigonal case we no longer longer have the symmetry about the
diagonal of the tableau that we have in the genus 2 case, but we can
restrict ourselves to equations of even degree by taking the symmetric
combination of the two tableaux related by transposition.  In the
weight 5 case we have the symmetric combination
\[
\yng(3,2)\quad + \quad \yng(2,2,1),
\]
which gives the weight 5 trigonal PDE
\[
\wp_{1112}=6\,\wp_{11}\wp_{12}+3\,\mu_{3}\wp_{11}.
\]
At weight 6 we have the three symmetric tableaux
\begin{align*}
  \yng(4,2)\quad + \quad\yng(2,2,1,1), \quad \yng(3,3)\quad
  +\quad\yng(2,2,2)\,, \quad \yng(3,2,1),
\end{align*}
which give a set of three overdetermined equations with the unique
solution
\begin{align*}
  \wp_{111}^{2} & =4 \wp_{11} ^{3}+
  \wp_{12}^{2}+4\,\wp_{13}-4\,\wp_{11}\wp_{22},\\
  \wp_{1122}& =4\,\wp_{13}+2\,\mu_{6}+4\wp_{12}^{2}+
  2\,\wp_{11}\wp_{22}+3\,\mu_{3}\wp_{12}.
\end{align*}
Continuing in this way we recover the strictly trigonal versions of
the full set of equations given in \cite{eemop07}.

\subsection{Solving integrable hierarchies via residues}
The this method of deriving integrable equations on the
Jacobians of algebraic curves is based on the following relation due to Sato,
\cite{sato80}, \cite{sato81}, as applied to the algebro-geometric context by Fay,
\cite{fay83}.
\begin{theorem} [\bf Bilinear Identities] If $S$ is a small circle
  about the point $P=(z,w)\in X$, and $Q=(x,y)\in X$ is an arbitrary
  point - then
\begin{align}\begin{split}
\oint_{S^1}
\left\{\theta\left( \int_{P}^Q\boldsymbol{v}-\boldsymbol{v}
(\boldsymbol{t})-\boldsymbol{u} \right)\theta\left( \int_{P}^Q
\boldsymbol{v}-\boldsymbol{v}(\boldsymbol{T})+\boldsymbol{u}
\right)\right.\\\left.
\times\mathrm{exp}\left\{ \int^Q\Omega(\boldsymbol{t}
+\boldsymbol{T})\right\}
E(P,Q)^{-2}\right\}\mathrm{d}x=0. \end{split}\label{bilinear}
\end{align}
where
\[ \int^Q\omega(\boldsymbol{t}) =\sum_{n=1}^{\infty}t_n x^{-n}+\sum_{m,n=1}^{\infty} t_n\sigma_{mn} \frac{x^m}{m}    \]
and quantities $\sigma_{mn}$ are coefficients of the expansion of the holomorphic part of fundamental bi-differential (\ref{omegapole}).
This becomes in terms of $\tau$-function:
\begin{align}\begin{split}
\mathrm{Res}_{x=0}\frac{1}{x^2}\left[\mathrm{exp}\left\{
\sum_{1}^{\infty} t_n x^{-n}\right\}\mathrm{exp}\left\{
-\sum_{1}^{\infty}\frac{x^n}{n}\frac{\partial}{\partial t_n}
\right\} \tau(\boldsymbol{t};\boldsymbol{u})\right.\\
\times \left.\mathrm{exp}\left\{
\sum_{1}^{\infty} T_n x^{-n}\right\}\mathrm{exp}\left\{
-\sum_{1}^{\infty}\frac{x^n}{n}\frac{\partial}{\partial T_n}
\right\} \tau(\boldsymbol{T};-\boldsymbol{u})\right]=0
\end{split}\label{fay1}\end{align}
where
$\tau(\boldsymbol{t};\boldsymbol{u})$ is the algebro-geometric
$\tau$-function defined above.
\end{theorem}

\begin{proof}
To prove the equivalence  of  (fay1) and (bilinear) note first that \begin{equation}\Omega(\boldsymbol{t}+\boldsymbol{t}')
=\Omega(\boldsymbol{t})+\Omega(\boldsymbol{t}').\end{equation}
The factor in (\ref{bilinear})

\begin{equation}
\theta\left(\mathcal{A}(q)-\mathcal{A}(p)-\sum_{i=1}^{\infty}
\boldsymbol{U}_i t_i-\boldsymbol{u}\right)\mathrm{\exp}
\left\{\int_{P_0}^P\Omega(\boldsymbol{t})\right\}\end{equation}
can be expressed as
\begin{equation}\mathrm{exp}\left\{ \sum_{n=1}^{\infty} t_n \zeta^{-n} \right\}
\theta\left(\mathcal{A}(q)-\mathcal{A}(p)-\sum_{i=1}^{\infty}
\boldsymbol{U}_i t_i-\boldsymbol{u}\right)\mathrm{\exp}
\left\{\sum_{m,n=1}^{\infty}
Q_{mn} t_n \frac{\zeta^m}{n} \right\}.\end{equation}
Taking into account the form (\ref{tau}) for $\tau(\boldsymbol{t};\boldsymbol{u})$,
the last two factors in the above formula may be expressed
\begin{equation}
\theta\left(\mathcal{A}(q)-\mathcal{A}(p)-\sum_{i=1}^{\infty}
\boldsymbol{U}_i t_i-\boldsymbol{u}\right)\mathrm{\exp}
\left\{\sum_{m,n=1}^{\infty}Q_{mn} t_n \frac{\zeta^m}{n} \right\}
=\mathrm{exp}\left\{-\sum_{n=1}^{\infty} \frac{\zeta^n}{n} \frac{\partial^n}{\partial x_n} \right\}\tau(\boldsymbol{t};\boldsymbol{u}),\end{equation}
which completes the proof of the equivalence  of (\ref{fay1}) and (\ref{bilinear}).
\end{proof}

\begin{remark} Equations of this form were first written down
in terms of `vertex operators'
$$\mathrm{exp}\left\{\sum_{1}^{\infty} t_n x^{-n}\right\}
\mathrm{exp}\left\{-\sum_{1}^{\infty}\frac{x^n}{n}
\frac{\partial}{\partial t_n}\right\}$$
for general
$\tau$-functions in the works of Sato; they may be understood
as generating functions for integrable hierarchies of Hirota bilinear equations.
\end{remark}

Using this theorem we may obtain partial differential equations relating
the Kleinian symbols, $\wp_{ij}$, $\wp_{ijk}$ etc. To do that
we first insert the expression for the $\tau$-function
(\ref{tausigma})
and compute the residue.   Then we equate to zero all
coefficients of monomials in $\boldsymbol{t}$, obtaining partial
differential equations for the $\sigma$-function at argument
$\boldsymbol{u}$ or $\boldsymbol{-u}$. Next we let
$\boldsymbol{u}\rightarrow 0$, and so
$\sigma(\boldsymbol{u}),\sigma(\boldsymbol{-u})\rightarrow \sigma(0)$;
the 'time' derivatives act on $\sigma$ via its $\boldsymbol{u}$-dependence,
though, so
$\sigma_i(\boldsymbol{u})=-\sigma_i(\boldsymbol{-u})\rightarrow \pm
\sigma_i(0)$, etc.
We then replace the derivatives of the $\sigma$-function
by derivatives of the $\wp$ function using the recursive
relations described earlier.

{\sf Example I}

In the genus 2 hyperelliptic case, the first nonvanishing relations
occurs at $t$-weight 3 in the $t_i$, i.e. the coefficients of $t_1^3$ or
$t_3$.  Both these give the first 4-index relation
\[
\wp_{1111} = 6\wp_{11}^2+\alpha_4 \wp_{11}+4\wp_{12}+\tfrac12\alpha_3.
\]
The next non-zero term is at weight 5, where we recover the relation
for $\wp_{1112}$, etc.  In contrast with the previous section, this
approach only gives the even derivative relations at odd $t$-weights.

The calculations by this approach, because they involve bilinear
products of $\tau$-functions, soon become very computer-intensive,
and we have not pursued them very far.

{\sf Example II} within this method we are able to recover all quadratic relations for four indexed symbols $\wp_{ijkl}$, $1\geq i,j,k,l\leq 3$ and cubic relations for three indexed symbols $\wp_{ijk}^2$. We report here the {\em quartic} relation between even variables, that should be one from relations defining Kummer variety of the (3,4)-curve,
\begin{align*}
  \wp_{12}^{4}&- \wp_{22}^{3}-2\
  \wp_{11}\wp_{33}+2\wp_{13}^{2}+4\
  \wp_{12}^{2}\wp_{13}+\wp_{1113}\wp_{22}-6\wp_{11}
  \wp_{13}\wp_{22}+4\wp_{11}\wp_{12}\wp_{23}\\ & + \wp_{11}^{2}
  \wp_{22}^{2}-2\wp_{11}\wp_{12}^{2}\wp_{22}-\tfrac43\wp_{1113}
  \wp_{11}^{2}+8\wp_{11}^{3}\wp_{13}+2\mu_{12}+4\mu_{6}
  \wp_{11}^{3}-4\mu_{6}
  \wp_{11}\wp_{22}\\ & +3\mu_{3}\wp_{12}\wp_{13}+3\mu_{3}
  \wp_{11}\wp_{23}+\mu_{3} \wp_{12}^{3}+2\
  \mu_{6}\wp_{13}+\mu_{6} \wp_{12}^{2}+
  \mu_{9}\wp_{12}-{\mu_{3}}^{2}
  \wp_{11}^{3}\\ & -3\mu_{3}\wp_{11}\wp_{12}\wp_{22}=0.
\end{align*}

The detailed structure of the Kummer surface in the
(3,4) case is receiving further investigation and will be reported on elsewhere.

\chapter[Applications of two-dimensional $\sigma$-functions]{Applications of two-dimensional $\sigma$-functions}\label{chap:abbs}

\section{Baker function for multi-dimensional $\sigma$-function}
\index{hyperelliptic $\Phi$--functions! on Jacobian}

\index{Baker function! on Jacobian}
Let $\sigma(\boldsymbol{u})=\sigma(\boldsymbol{u};V)$ is multi-dimensional $\sigma$-function of the curve $V$. Following Baker \cite[page 421]{ba97}
 we introduce the function  on $ \mathbb{C}^g  \times \mathrm{ Jac} (V) $.
\begin{definition} Baker function
$\Phi:   \mathbb{C}^g   \times \mathrm{ Jac} V   \to
\mathbb{ C}$ is given by the formula
  \begin{eqnarray}
 &  \Phi ( \boldsymbol{ u}; \boldsymbol{\alpha};V ) =
\dfrac{  \sigma ( \boldsymbol{ \alpha}-  \boldsymbol{ u}) }{
\sigma  ( \boldsymbol{ \alpha})   \sigma  (  \boldsymbol{ u}) }
\exp (\boldsymbol{ \zeta}^T (  \boldsymbol{
  \alpha})   \boldsymbol{ u}) ,\label{bakerdef}
  \end{eqnarray}
where $  \boldsymbol{
  \zeta}^T (  \boldsymbol{
  \alpha}) = (
  \zeta_1 (  \boldsymbol{
  \alpha}) ,   \ldots,
  \zeta_g (  \boldsymbol{
  \alpha}) ) $\end{definition}

\index{Bloch!function}
Periodicity property: for any point $\boldsymbol{\Omega}\in \Gamma_V$
under the shift at a period ${\boldsymbol \Omega}_i$
\begin{equation}
\Phi({\boldsymbol u}+{\boldsymbol
\Omega}_i;{\boldsymbol \alpha})
=\xi_i({\boldsymbol \alpha})  \Phi({\boldsymbol
u}; \boldsymbol{\alpha}),\quad
 \Phi({\boldsymbol u};{\boldsymbol
 \alpha}+{\boldsymbol \Omega}_i) = \Phi({\boldsymbol
u}; \boldsymbol{\alpha}),\label{blochprop} \end{equation}
where $\xi_i({\boldsymbol
\alpha})$, $i=1,\ldots, 2g$ are functions on $ \mathrm{ Jac} (V) $
with essential singularities,
\[
\xi_i(\boldsymbol{\alpha})=\mathrm{exp}(\boldsymbol{\zeta}^T(\alpha)
\boldsymbol{\Omega}_i)-\boldsymbol{E}_i^T\boldsymbol{\alpha}.\]

From the definition follows that
\[ \nabla_{\boldsymbol{u}} \Phi( \boldsymbol{ u}; \boldsymbol{\alpha};V )=-\boldsymbol{\zeta}(\boldsymbol{\alpha}-\boldsymbol{u})
-\boldsymbol{\zeta}(\boldsymbol{u})+\boldsymbol{\zeta}(\boldsymbol{\alpha}) \]
be the vector-function on $  \mathrm{ Jac} (V) \times \mathrm{ Jac} (V) $.

In the case $g=1$ the Baker function turns to
\begin{equation}
\Phi({u};{\alpha})=\frac{\sigma ({\alpha}-u)}{
\sigma ({u})\sigma ({\alpha})}\mathrm{ exp}(\zeta({\alpha})u).
\label{lame1}\end{equation} with the standard Weierstrass $\sigma $
and $\zeta$-functions. (\ref{lame1}) represents a solution of
Lam\'e equation, \index{Lam\'e equation!for elliptic curve}
\begin{equation}
\left\{\frac{\mathrm{d}^2}{\mathrm{d}
u^2}-2\wp(u)-\wp(\alpha)\right\}\Phi({u};{\alpha}) =0.\label{lame11}
\end{equation}


\section{Second order differential equation for Baker function}
Set
\begin{equation}
L=\sum_{1\geq i \leq j \leq g } a_{i,j} \left(
L_{i,j} -2\wp_{i,j}(\boldsymbol{u})\right)-E,\label{diffeq}
\end{equation}
where $a_{i,j}$ and $E$ do not depend in $\boldsymbol{u}$. Natural to put the problem:

{\em  For constant $E$ find a symmetric matrix $(a_{i,j})_{i,j=1,\ldots,g}$ and vector $\boldsymbol{k}$ independent in $\boldsymbol{u}$ such that
\begin{equation}
L\Psi(\boldsymbol{u},\boldsymbol{\alpha},\boldsymbol{k})=0,\quad \text{where}
\quad  \Psi(\boldsymbol{u},\boldsymbol{\alpha},\boldsymbol{k})
=\Phi(\boldsymbol{u},\boldsymbol{\alpha};V)
\mathrm{exp}\{-\boldsymbol{u}^T\boldsymbol{k}\}
\end{equation}
}

\begin{df}
The subvariety ${\mathcal M}(\boldsymbol{\alpha},
\boldsymbol{k},E)$ in $ \mathrm{Jac}(V)\times \mathbb C^g\times \mathbb C$
we shall call the variety of solutions of the equation
\begin{equation} L(E)\Psi(\boldsymbol{u},\boldsymbol{\alpha},
\boldsymbol{k})=0,\label{diffeq}
\end{equation}
if for each its point  $(\boldsymbol{\alpha},\boldsymbol{k},E)$,
the equation (\ref{diffeq}) is satisfied identically in $\boldsymbol{u}$.
\end{df}

Further we will present solution of the formulated problem in the case $g=2$.
\section{Solving the problem at the case $g=2$}

\subsection{General case}

Let $V$ is the curve of genus two,
\begin{equation}
V:\quad \nu^2-(4\mu^5+\lambda_4\mu^4+\lambda_3\mu^3
+\lambda_2\mu^2+\lambda_1\mu+\lambda_0)
\end{equation}
$\boldsymbol{u}\in \mathbb{C}^2$
For a given function $\phi(\boldsymbol{u})$ put
$\phi_i(\boldsymbol{u})=\partial_i\phi(\boldsymbol{u}),
\phi_{i,j}(\boldsymbol{u})=\partial_{i,j}\phi(\boldsymbol{u})  $  etc.
$i,j =1,\ldots,2$

The function $\phi(\boldsymbol{u})$ on $\mathbb{C}^2$ belongs to the class of functions
${\mathcal B}(\sigma(\boldsymbol{u}),q)$, if
$\sigma(\boldsymbol{u})^q\phi(\boldsymbol{u})$ be entire function.

\begin{df}
The function $\phi(\boldsymbol{u})$ is called Bloch function associated with the curve $V$ with the Bloch factor $\xi$ if
\begin{equation}
\phi(\boldsymbol{u}+2\boldsymbol{\Omega}(\boldsymbol{n},\boldsymbol{n}'))=
\xi(\boldsymbol{n},\boldsymbol{n}',{\mathfrak M})
\phi(\boldsymbol{u})
,\end{equation}
and $\xi$ independent in $\boldsymbol{u}$ where $\boldsymbol{\Omega}(\boldsymbol{n},\boldsymbol{n}')\in \Gamma_V$ and $\mathfrak{M}$ is the period matrix
\[ \mathfrak{M}=\left( \begin{array}{cc}  \omega&\omega'\\  \eta&\eta'  \end{array}\right)  \]
\end{df}

Introduce sub-class  ${\mathcal B}(\sigma(\boldsymbol{u}),q,\xi)$ in
${\mathcal B}(\sigma(\boldsymbol{u}),q)$ of the Bloch functions with
given Bloch factor $\xi$. Baker function
\begin{equation}
\Phi(\boldsymbol{u},\boldsymbol{\alpha})
=\frac{\sigma(\boldsymbol{u}
+\boldsymbol{\alpha})}{\sigma(\boldsymbol{u})\sigma(\boldsymbol{\alpha})}
\exp\{- \boldsymbol{u}^T\zeta(\boldsymbol{\alpha})\}\label{blochfunction}
\end{equation}
belongs to the class  ${\mathcal B}(\sigma(\boldsymbol{u}),1,\xi)$
where
\begin{equation}\xi=\mathrm{exp}\left\{2
\boldsymbol{\zeta}^T(\alpha)\boldsymbol{\Omega}(\boldsymbol{n},\boldsymbol{n}')-
\boldsymbol{E}^T(\boldsymbol{n},\boldsymbol{n}')\boldsymbol{\alpha}
\right\}
\label{blochfactor}.\end{equation}
and at the fixed parameter $\boldsymbol{\alpha}$, vectors $\boldsymbol{n},\boldsymbol{n}'$ and matrix of half periods $\mathfrak M$  is defined uniquely by the initial condition $$\sigma(\boldsymbol{u})\Phi(\boldsymbol{u},
\boldsymbol{\alpha})|_{\boldsymbol{u}=0}=1
$$

Set
\[
F_{i,j}(\boldsymbol{u};\boldsymbol{\alpha})=\Phi_{i,j}(\boldsymbol{u};
\boldsymbol{\alpha}) -
2\wp_{i,j}(\boldsymbol{u})\Phi(\boldsymbol{u};\boldsymbol{\alpha}) ,\quad
i,j=1,2.
 \]

Introduce the vector
\begin{equation}
W(\boldsymbol{u};\boldsymbol{\alpha})^T
=\left(  \begin{array}{c}
\Phi(\boldsymbol{u};
\boldsymbol{\alpha})\\
\Phi_1(\boldsymbol{u};
\boldsymbol{\alpha})\\
\Phi_2(\boldsymbol{u};
\boldsymbol{\alpha})\\
F_{1,1}(\boldsymbol{u};\boldsymbol{\alpha})\\
F_{1,2}(\boldsymbol{u};\boldsymbol{\alpha})\\
F_{2,2}(\boldsymbol{u};\boldsymbol{\alpha})\end{array}   \right).\end{equation}

The direct checking shows that components  of
$\sigma(\boldsymbol{u})^2W(\boldsymbol{u};\boldsymbol{\alpha})$ belongs to
 ${\mathcal B}(\sigma(\boldsymbol{u}),2,\xi(\boldsymbol{\alpha}))$, where $\xi(\boldsymbol{\alpha}))$ is given by (\ref{blochfactor}). Below we shall consider Bloch functions with the Bloch factor (\ref{blochfactor}).

Put into correspondence with each entire function
$\phi(\boldsymbol{u})$ the vector
$$\boldsymbol{p}(\phi)= (\phi(\boldsymbol{0}),
\phi_1(\boldsymbol{0}),\phi_2(\boldsymbol{0}),
\phi_{2,2}(\boldsymbol{0}))^T\in\mathbb C^4$$. The following lemma
will play important role in what follows.

\begin{lemma}\label{lmatrix6x4}
\begin{equation}
\boldsymbol{p}(\sigma(\boldsymbol{u})^2\boldsymbol{W})
=\left(\begin{array}{cccccc}
0&-1&0&0&0&0\\
 1&0&0&\wp_{11}(\boldsymbol{\alpha})+\frac{\lambda_2}{4}&0
&-\wp_{22}(\boldsymbol{\alpha})\\   0&0&0&2\wp_{12}(\boldsymbol{\alpha})&
 \wp_{22}(\boldsymbol{\alpha})&-2 \\
0&\wp_{22}(\boldsymbol{\alpha})&2&2\wp_{122}(\boldsymbol{\alpha})&
\wp_{222}(\boldsymbol{\alpha})&0   \end{array}\right)
\end{equation}

\end{lemma}
\begin{proof}
.
\end{proof}

In the basis of the construction of the variety of solutions lies the
following result

\begin{theorem} \label{th1}
Let $\phi(\boldsymbol{u})$ is a entire function such that
$\phi(\boldsymbol{u})/\sigma(\boldsymbol{u})^2\in {\mathcal B}(\sigma(\boldsymbol{u})^2,2,\xi(\boldsymbol{\alpha}))$.
Then $\phi(\boldsymbol{u})\equiv 0$ if  $\boldsymbol{p}(\phi)
=\boldsymbol{0}$.
\end{theorem}

\begin{proof} Let us note that all six components of the vector
$\sigma(\boldsymbol{u})^2\boldsymbol{W}$ as well the functions
\begin{align*}
\phi^{(1)}(\boldsymbol{u};\boldsymbol{\alpha})&=
\sigma(\boldsymbol{u})^2\Phi(\boldsymbol{u};
\boldsymbol{\alpha}),\\
\phi^{(2)}(\boldsymbol{u};\boldsymbol{\alpha})&=\sigma(\boldsymbol{u})^2
\left[\Phi_1(\boldsymbol{u};\boldsymbol{\alpha})-
\frac12\wp_{22}(\boldsymbol{\alpha})
\Phi_2(\boldsymbol{u};\boldsymbol{\alpha})\right],\\
\phi^{(3)}(\boldsymbol{u};\boldsymbol{\alpha})&=\frac12\sigma(\boldsymbol{u})^2
\Phi_2(\boldsymbol{u};\boldsymbol{\alpha}),\\
\phi^{(4)}(\boldsymbol{u};\boldsymbol{\alpha})&=\frac12\sigma(\boldsymbol{u})^2
\left[
F_{22}(\boldsymbol{u};\boldsymbol{\alpha})+\wp_{22}(\boldsymbol{\alpha})
\Phi(\boldsymbol{u};\boldsymbol{\alpha})\right].
\end{align*}
satify to the conditions of the theorem \ref{th1}.

Put $\boldsymbol{p}^{(k)}=\boldsymbol{p}(\phi^{(k)})$,
$k=1,\ldots,4$. Then
\begin{equation}
\left( \begin{array}{c}  \boldsymbol{p}^{(1)} \\
\boldsymbol{p}^{(2)}\\
\boldsymbol{p}^{(3)}\\
\boldsymbol{p}^{(4)}\end{array}   \right)
 =\left(   \begin{array}{cccc}      0&1&0&0\\
 -1&0&0&0\\
 0&0&0&1\\
 0&0&-1&0 \end{array}
 \right)
\end{equation}

Then we obtain from the lemma \ref{lmatrix6x4}.
\end{proof}

\begin{prop}

The function
\begin{equation}  \phi(\boldsymbol{u})=\sigma(\boldsymbol{u})^2
\left( \exp(\boldsymbol{u}^T\boldsymbol{k}) L
\exp(-\boldsymbol{u}^T\boldsymbol{k})  \right)
\Phi(\boldsymbol{u};\boldsymbol{\alpha})\label{function}
\end{equation}
satisfies to the conditions of the theorem \ref{th1} and therefore
equation (\ref{diffeq}) is valid if and only if $\boldsymbol{p}(\phi)=
\boldsymbol{0}$.
\end{prop}

\begin{proof}
Direct computation leads to the formula
\begin{align}
&\phi(\boldsymbol{u})
=\sigma(\boldsymbol{u})^2 \\&\times \left\{
\sum a_{i,j}\left[F_{i,j}(\boldsymbol{u};\boldsymbol{\alpha})
-k_j \Phi_i(\boldsymbol{u};\boldsymbol{\alpha})
-k_i \Phi_j(\boldsymbol{u};\boldsymbol{\alpha})+k_ik_j
\Phi(\boldsymbol{u};\boldsymbol{\alpha})
\right]  +E\Phi(\boldsymbol{u};\boldsymbol{\alpha})
\right\}.\notag
\end{align}

It is easy to check that the function
$\phi(\boldsymbol{u})$ is entire. To complete to proof is sufficient to
notice that that after dividing by  $\sigma(\boldsymbol{u})^2$ we obtain
the Bloch function with the Bloch factor (\ref{blochfactor}).
\end{proof}

\begin{theorem}\label{mainthm}

The subvariety ${\mathcal M}(\boldsymbol{\lambda},
\boldsymbol{k},E)$ in $V_{\boldsymbol{\lambda}}\times \mathbb C^2\times \mathbb C$
is described by the following system of equations

\begin{align*}
&a_{1,1}k_1+a_{1,2}k_2=0,\\
&a_{1,1}\wp_{11}(\boldsymbol{\alpha})-a_{2,2}\wp_{22}(\boldsymbol{\alpha})
+(a_{11}k_1^2+2a_{12}k_1k_2+a_{22}k_2^2)+\frac{1}{4}\lambda_2+E=0,\\
&a_{1,1}\wp_{12}(\boldsymbol{\alpha})+a_{1,2}\wp_{22}(\boldsymbol{\alpha})
-a_{2,2}=0,\\
&a_{1,1}\wp_{122}(\boldsymbol{\alpha})+a_{1,2}\wp_{222}(\boldsymbol{\alpha})
+2a_{1,2}k_1+2a_{2,2}k_2=0.
\end{align*}
\end{theorem}

\begin{cor}\label{maincor}The variety of solutions of the equation (\ref{diffeq}) is
defined by the variety
\begin{equation}{\mathcal M}(\mathrm{Jac}(V))=\{\boldsymbol{\alpha}\in \mathrm{Jac}(V)|
 a_{1,1}\wp_{12}(\boldsymbol{\alpha})+a_{1,2}\wp_{22}(\boldsymbol{\alpha})
-a_{2,2}=0 \}  \end{equation}
and the following functions on ${\mathcal M}(\mathrm{Jac}(V))$
\begin{align}
k_i(\boldsymbol{\alpha})
&=\frac{a_{1,1}}{2\delta}(a_{1,i}\wp_{122}(\boldsymbol{\alpha})
                           +a_{1,2}\wp_{222}(\boldsymbol{\alpha})),\quad i=1,2\\
E &= -\frac{1}{4\delta}a_{1,1}\left(a_{1,1}
\wp_{2,2,1}(\boldsymbol{\alpha})
+a_{1,2}\wp_{2,2,2}(\boldsymbol{\alpha})
\right)^2+a_{2,2}\wp_{2,2}(\boldsymbol{\alpha})\label{energyalpha}
\end{align}
\end{cor}

The components of the quasimomenta  $\boldsymbol{Q}$ are  then given as
\begin{equation} Q_i(\boldsymbol{\alpha})
=-\zeta_1(\boldsymbol{\alpha})-k_i(\boldsymbol{\alpha}), \quad i=1,2.
\end{equation}

Using genus two formulae in the Appendix 2  we  obtain effective description of the variety $\mathcal{ M}(\mathrm{Jac}(V_{\boldsymbol{\lambda}}))$.

The matrix $(a_{i,j})_{i,j=1,2}$ defines the involution on $\mathbb C$
\begin{equation}
\varkappa: \mathbb C\rightarrow \mathbb C: \; \varkappa(\mu)\longrightarrow   \frac{a_{1,2}\mu-a_{2,2}}{a_{1,1}\mu-a_{1,2}}
\end{equation}
with the fixed points $\mu_{1,2}^{\ast}$ defined as roots of the equation
$$ a_{1,1}\mu^2+2a_{1,2}\mu+a_{2,2}=0$$

Introduce the variety

\begin{equation}
{\mathcal M}(\mathrm{Sym}^2(V_{\boldsymbol{\lambda}}))=\{ (\mu_1,\nu_1),
 (\mu_2,\nu_2) \in \mathrm{Sym}^2(V_{\boldsymbol{\lambda}}): \mu_2=\varkappa(\mu_1).\}
\end{equation}

Direct checking leads to the following result

\begin{lemma}\label{mainthm1}
The Abel map $$\mathrm{Sym}^2(V)\longrightarrow
\mathrm{Jac}(V_{\boldsymbol{\lambda}})$$
is setting the equivalence between ${\mathcal M}(\mathrm{Sym}^2(V_{\boldsymbol{\lambda}}))$ and
${\mathcal M}(\mathrm{Jac}(V_{\boldsymbol{\lambda}})$
\end{lemma}

It is remarkable that by using the Corollary \label{maincor} and
Lemma (\ref{mainthm1}) we find that the energy value is expressible in terms
of Kleinian polar $F(P,Q)$  being computed in the two fixed points  $\mu_{1,2}^{\ast}$ of the
involution $\varkappa$, namely
\begin{theorem} The restriction of the function
(\ref{energyalpha}) on the variety $\mathcal M(\mathrm{Jac}(V_{\boldsymbol{\lambda}}))$ is independent in
$\boldsymbol{\alpha} $ and equals

\begin{align}
E&=\frac{1}{2}\delta a_{1,1}^3 F(\mu_1^{\ast},\mu_2^{\ast})\label{energy}\\
&=\frac{1}{4}(a_{1,2}a_{1,1}^2\lambda_1+4a_{1,2}a_{2,2}^2
+a_{1,2}a_{1,1}\lambda_3 a_{2,2}+a_{1,1}^3\lambda_0
+a_{1,1}^2\lambda_2a_{2,2}+a_{1,1}a_{2,2}^2\lambda_4)\notag
\end{align}

\end{theorem}

Using results from the Appendix 2 we find description of $\mathrm{Jac}(V_{\boldsymbol{\lambda}})$. Denote

$$(z_1,z_2)=(\wp_{222},\wp_{122}),\quad (v_1,v_2,v_3)=(\wp_{22},\wp_{12},\wp_{11})$$,
Then from the formulae given in the Appendix 2 we get equations
\begin{align}
&z_1^2-4v_3+\lambda_3v_1+4v_1^3+4v_2v_1+\lambda_4v_1^2+\lambda_2=0,\label{eq1}\\
&z_1z_2-\left(\frac{\lambda_1}{2}+2v_2^2-2v_1v_3+\frac{\lambda_3v_2}{2}
+4v_1^2v_2+\lambda_4v_1v_2  \right)=0,\label{eq2}\\
&z_2^2-(\lambda_0-4v_2v_3+\lambda_4v_2^2+4v_1v_2^2)=0.\label{eq3}
\end{align}

Direct calculations leads to the result

\begin{theorem}
The variety $\mathrm{Jac}(V_{\boldsymbol{\lambda}})$
is described in the space $\mathbb C^3$ with coordinates $v_1,v_2,z$
by the following equation
\begin{eqnarray}
A(v_1,v_2)z^4+B(v_1,v_2)z^2+C(v_1,v_2)^2=0\label{JAC}
\end{eqnarray}
where
\begin{align}
A(v_1,v_2)&=v_2+\frac{1}{4}v_1^2\label{A}\\
B(v_1,v_2)&=-6v_2^3v_1-v_2^2\lambda_2-v_2\lambda_0
-v_2^3\lambda_4-2v_1^3v_2^2\label{B}\\
&+\frac{1}{2}
v_2\lambda_1v_1-\frac{1}{2}v_2^2
\lambda_3v_1-\frac12\lambda_4v_1^2v_2^2-\frac12
v_1^2\lambda_0\notag\\
C(v_1,v_2)&=\frac{1}{2}(4v_2^3-v_1\lambda_0+4v_1^2v_2^2+v_2
\lambda_1+v_2^2\lambda_3+v_1\lambda_4v_2^2)
\end{align}
\end{theorem}

Therefore we obtained the explicit description of the variety $\mathrm{Jac}(V_{\boldsymbol{\lambda}})$ in the terms of initial curve as 4-sheeted branch covering over the plane $\mathbb C^2$ with coordinates $v_1,v_2$. According to the Corollary \label{maincor} we obtain now the explicit description of ${\mathcal M}(\mathrm{Jac}(V_{\boldsymbol{\lambda}}))$ as
a plane curve.
\begin{eqnarray}
a(v)z^4+b(v)z^2+c(v)^2=0,\label{JAC}
\end{eqnarray}

\begin{theorem} \label{planecurve} To complete the description of $\mathrm{Jac}(V_{\boldsymbol{\lambda}})$
 it is sufficient to consider the following cases
\begin{itemize}
\item
If $a_{1,1}\neq 0$ then the variety ${\mathcal M}(\mathrm{Jac}(V))$
is described as algebraic variety in $\mathbb C^2$  with coordinates $(v,z)$
by the following equation (\ref{JAC}), where $a,b,c$ are obtained from $A(v_1,v_2),B(v_1,v_2),C(v_1,v_2) $ by
the substitution
\[  v_1=v, \quad v_2=\frac{1}{a_{1,1}}(a_{2,2}-a_{1,2}v)  \]
\item  It $a_{1,1}=0, a_{1,2}\neq 0$ then
the variety ${\mathcal M}(\mathrm{Jac}(V))$
is described as algebraic variety in $\mathbb C^2$  with coordinates $(v,z)$
by the  equation (\ref{JAC})
\[ v_1=\frac{a_{2,2}}{a_{1,2}}, \quad v_2=v \]
\item  $ a_{1,1}=0, a_{1,2}=0, a_{2,2}\neq 0 $ the variety ${\mathcal M}(\mathrm{Jac}(V))$ is empty.
\end{itemize}
\end{theorem}

\section{Variety of solutions in the rational limit}
Tending parameters $\lambda_k\rightarrow 0$ of the curve $V$ we get the equation

\begin{equation}
9u_2^2a_{1,1}+(-18u_{2}u_{1}-3u_{2}^4)a_{1,2}+(-6u_{1}u_{2}^3+9u_{1}^2
+u_{2}^6)a_{2,2}=0\label{ratcurve}
\end{equation}

The curve (\ref{ratcurve}) can be uniformized as
\begin{align}
\alpha_1(t)=&\frac{1}{3}
\sqrt{-\frac{\delta}{a_{2,2}^3a_{1,2}^3  }  }
 \sinh(t)\label{ratuniform}\\
\times&\left( \delta  \sinh(t)^2 +
3a_{2,2}a_{1,2}\sqrt{-\frac{\delta}{a_{2,2}^2}}\cosh(t)-a_{1,2}^2 ,
 \right)\notag\\
\alpha_2(t)=&
\sqrt{-\frac{\delta}{a_{2,2}^3a_{1,2}^3  }  }
 \sinh(t),\notag
\end{align}
where $\delta=a_{1,1}a_{2,2}-a_{1,2}^2$.

The potential and the Bloch function are of the form
\begin{align}
{\mathcal U}=6\frac{3a_{1,1}-3a_{1,2}u_{2}^2+6a_{2,2}u_{2}u_{1}+a_{2,2}u_{2}^4}
{(3u_1-u_2^3)^2}
\label{ratpotential}
\end{align}
\begin{align}\Psi(\boldsymbol{u};\boldsymbol{\alpha})&=
\frac{u_1+\alpha_1(t)-\frac{1}{3}  (u_2+\alpha_2(t))^3}
{(u_1-\frac13u_2^3)(u_1(t)-\frac13 u_2^3(t))}\notag\\
\times&
\mathrm{exp}   \left\{    3\frac{-u_1+\alpha_2^2(t)u_2}  {3\alpha_1(t)-\alpha_2^3(t)} -k_1(t)u_1-k_2(t)u_2 \right\}   \label{ratbloch}\end{align}
where
\begin{align}
k_i(t)&=\frac{a_{1,i}}{2\delta(3\alpha_1-\alpha_2^3)^3}\notag\\
\times&\left(
-18a_{1,1}\alpha_2(2\alpha_2^3+3\alpha_1)+6 a_{1,2}(9\alpha_1^2+21\alpha_1\alpha_2^3+\alpha_2^6)
\right)
\end{align}
and the uniformizing functions $\alpha_i(\alpha)$ are given in (\ref{ratuniform})

The energy in the rational limit is given as $E=\delta a_{1,2}a_{2,2}^2$

\index{cover}\index{covering!two-sheeted}

\section{Reduction to elliptic functions}
Let us consider the case $N=2$.
The corresponding algebraic curve has the form,
\begin{equation}
y^2=4z(z-\beta_1)(z-\beta_2)(z-\gamma_1)(z-\gamma_2),\label{roscurve}
\end{equation}
where $\beta_1\beta_2=\gamma_1\gamma_2$. Let us set
\[
a=\left(\frac{\sqrt{\gamma_1}-\sqrt{\gamma_2}}{
\sqrt{\gamma_1}+\sqrt{\gamma_2}} \right)^2
,\quad
b=\left(\frac{\sqrt{\beta_1}-\sqrt{\beta_2}}{
\sqrt{\beta_1}+\sqrt{\beta_2} }\right)^2.\]
Then the holomorphic differentials on the Riemann
surface of (\ref{roscurve}) reduce to elliptic ones,
\begin{equation} \frac{z\pm\sqrt{\beta_1\beta_2}}{ y}dz=\mp
\frac{1}{4}\sqrt{\frac{(1-a^{\pm1})(1-b^{\pm1})}{
\sqrt{\beta_1\beta_2}}} \frac{d\mu_{\pm}}{
\nu_{\pm}},\label{red}\end{equation} where
$V_{\pm}=(\mu_{\pm},\nu_{\pm})$  are elliptic curves with
$\nu_{\pm}^2=\pm(1-\mu_{\pm})(a^{\pm1}-\mu_{\pm})(b^{\pm1}-\mu_{\pm})$
and Jacobi moduli $k_{\pm}^2=(1-a^{\pm1})/( 1-b^{\pm1})$

If we put $c_1=\beta_1+\beta_2-\gamma_1-\gamma_2$ and
\begin{eqnarray}
&&\xi_1=\frac{\beta_2P_{\beta_1}-\beta_1P_{\beta_2}}{
\beta_2-\beta_1}, \quad
\xi_2=P_{\beta_1\beta_2}-\beta_1\beta_2c_1\label{xi2}\\
&&\xi_3=P_{\beta_1\beta_2}+\beta_1\beta_2c_1
-c_1\frac{\beta_2P_{\beta_1}-\beta_1P_{\beta_2}}{
\beta_2-\beta_1} \label{xi3}\\
&&\xi_4=P_{\beta_1\beta_2}-\beta_1\beta_2c_1
-c_1(\beta_2P_{\beta_1}+\beta_1P_{\beta_2}),\label{xi4}\end{eqnarray}
where
\begin{eqnarray*}
P_{\beta}&=&\wp_{12}(\boldsymbol{u})+\beta\wp_{22}(\boldsymbol{u})
-\beta^2,\cr
P_{\beta\gamma}&=&\beta\gamma\wp_{22}(\boldsymbol{
u})+(\beta+\gamma)\wp_{12}(\boldsymbol{u})+\wp_{11}(\boldsymbol{u})-
\frac{F(\beta,\gamma)}{ 4(\beta-\gamma)^2},\end{eqnarray*}
with the function $F$ given in (\ref{fx1x2}).

Then the associated with the considered reduction case singular
Kummer surface or {\em Pl\"ucker surface} is parametrised by
\index{Pl\"ucker surface}
\index{reduction!of Kummer surface}
Jacobi elliptic functions
\index{Jacobi! elliptic functions} with the moduli $k_{\pm}$,
\begin{eqnarray} &&c_4c_3^2c_1^2\xi_1^4+c_3^2\xi_3^4+c_1^2\xi_4
+c_4\xi_2^4+2(c_2c_1-c_3^2)(\xi_2^2\xi_3^2+c_1^2\xi_1^2\xi_4^2)
\nonumber\\
&&+2c_1(c_2-c_1)(\xi_2^2\xi_4^2+c_3^2\xi_1^2\xi_3^2)
-2c_1c_2(\xi_3^2\xi_4^2+c_4\xi_1^2\xi_3^2)=0,\label{kumel}
 \end{eqnarray}
where $c_1=\beta_1+\beta_2-\gamma_1-\gamma_2$,
$c_2=\beta_1+\beta_2$, $c_3=\beta_1-\beta_2$,
$c_4=c_1^2+c_3^2-2c_2c_1$ and ratios $f_i=\xi_i/\xi_4$, $i=1,2,3$
are expressed in terms of Jacobi elliptic functions as follows
\begin{eqnarray}
&&f_1(\boldsymbol{u})=(\beta_1-\beta_2)\mathrm{ cn}\;(u_+;k_+)\;{\rm
cn}\;(u_-;k_-),\nonumber\\
&&f_2(\boldsymbol{u})=(\gamma_1-\gamma_2)\mathrm{ dn}\;(u_+;k_+)\;{\rm
dn}\;(u_-;k_-),\nonumber\\
&&f_3(\boldsymbol{u})=(\beta_1-\beta_2)\mathrm{ sn}\;(u_+;k_+)\;{\rm
sn}\;(u_-;k_-),
\label{sn}\end{eqnarray}
where $\mathrm{ sn}^2\;(u_{\pm};k_{\rm})=(1-\mu_{\pm})/(1-a^{\pm1})$.

Moreover, one can invert (\ref{xi2}--\ref{xi4}) to find the
expressions of the coordinates of the Kummer surface
\index{Kummer!surface} in terms of elliptic functions. Namely, we
have \begin{eqnarray}
\wp_{11}&=&\beta_1\beta_2\frac{A(f_2(\boldsymbol{u})
+f_3(\boldsymbol{u}))-B(f_1(\boldsymbol{
u})+f_3(\boldsymbol{u})}{
A-B+f_2(\boldsymbol{u})-f_1(\boldsymbol{u})}, \label{wp11}\\
\wp_{12}&=&\beta_1\beta_2\frac{A-B+f_1(\boldsymbol{
u})-f_2(\boldsymbol{u})}{
A-B+f_2(\boldsymbol{u})-f_1(\boldsymbol{u})}, \label{wp12}\\
\wp_{22}&=&\frac{B(f_3(\boldsymbol{u})-f_1(\boldsymbol{u}))-A(f_3(\boldsymbol{
u})-f_2(\boldsymbol{u}))}{
A-B+f_2(\boldsymbol{u})-f_1(\boldsymbol{u})}, \label{wp22}
\end{eqnarray} where  $A=\beta_1+\beta_2$, $B=\gamma_1+\gamma_2$

\begin{theorem}
The spectral problem   \index{spectral problem}
\begin{eqnarray}&&\left\{
\left(1-\frac{C}{p}\right)\frac{\partial^2}{ \partial w_1^2}+
\left(1+\frac{C}{ p}\right)\frac{\partial^2}{ \partial w_2^2}-
\mathcal{U}(\boldsymbol{w})\right\}\Psi(\boldsymbol{w};\boldsymbol{\alpha})
\cr&&=\lambda(\boldsymbol{\alpha})
\Psi(\boldsymbol{w};\boldsymbol{\alpha})\label{elsp} \end{eqnarray}
being associated with the curve (\ref{roscurve}), with $p=\beta_1\beta_2=\gamma_1\gamma_2$, where the potential has the form
\begin{eqnarray}
 \mathcal{U}(\boldsymbol{w})
=\frac{f_1(\boldsymbol{w})(B-C)+f_2(\boldsymbol{w})(-A+C)+C(B-A)}{
f_1(\boldsymbol{w})-f_2(\boldsymbol{w})+B-A}, \label{rospot}
\end{eqnarray}
with $A=\beta_1+\beta_2$, $B=\gamma_1+\gamma_2$, $C$ be a constant, is
solved on the Bloch variety
$$\mathcal{M}=\left\{
(\boldsymbol{\alpha})\vert f_1(\boldsymbol{\alpha})(-2p^2+CB)+
f_2(\boldsymbol{\alpha})(2p^2-CA)+
f_3(\boldsymbol{\alpha})C(A-B)=0 \right\}$$
at the fixed energy level
\begin{equation}
\lambda(\boldsymbol\alpha)=-A-B+\frac{C}{2}\frac{AB-2C(A+B)+4p^2}
{p^2-C^2}\label{lambda}
\end{equation}
\end{theorem}

\begin{proof} It follows from the reduction formulae,
\begin{equation}
u_2- p u_1=-\frac{w_1}{ \sqrt{\gamma_1}+\sqrt{\gamma_2}},
\quad u_2+ p u_1=-\frac{w_2}{ \sqrt{\gamma_1}+\sqrt{\gamma_2}},
\end{equation}
Therefore the operator
$\sum_{i,j}\Lambda_{ij}\frac{\partial^2}{ \partial u_i\partial
u_j}$ turns to \[
 \left(2p^2\Lambda_{11}-2p\Lambda_{12}\right)\frac{\partial^2}{
 \partial w_1^2} + \left(2p^2\Lambda_{11}
+2p\Lambda_{12}\right)\frac{\partial^2}{ \partial w_2^2}
\]
under the condition $\Lambda_{22}=\Lambda_{11}p^2=\Lambda$. Let us denote
$C=\frac{\Lambda_{12}}{ \Lambda}$; then the potential
$\mathcal{U}= \sum_{i,j}\Lambda_{ij}\wp_(\boldsymbol{u}) $ takes the form
(\ref{rospot}). To calculate the spectral value
$\lambda(\boldsymbol{\alpha})$
we use  the condition
\begin{equation}
\Lambda_{22}=\Lambda_{11}\wp_{12}(\boldsymbol{\alpha})+
\Lambda_{12}\wp_{22}(\boldsymbol{\alpha})
\label{blochvar}\end{equation}

\end{proof}

\begin{remark}The Pl\"ucker surfaces parametrized by the Kleinian $\sigma $--functions
were used in \cite{es996} to  construct the Lax representation for
the two--particle systems, the  dynamics of which is split to
two tori.\end{remark}

\section{ Two-dimensional Schr\"odinger equation with Abelian potential }
In this section we describe the variety of solutions of the equation
\begin{equation}\label{eq}L(E)\psi(u;\alpha,\beta,k)=0\end{equation}
where
\begin{gather*}
L(E)=\partial_{1,2}-6\wp_{1,2}(u)+E\\ \intertext{and}
\psi(u;\alpha,\beta,k)=
\frac{\sigma(u+\alpha)\sigma(u+\beta)}{\sigma(u)^2\sigma(\alpha)\sigma(\beta)}
\exp\{-u^T(\zeta(\alpha)+\zeta(\beta)-k)\}, \end{gather*}  where
$\zeta(v)=(\zeta_1(v),\zeta_2(v))^T$; here
$\alpha,\beta,k\in\mathbb{C}^2$ and $E$ do not depend on $u$ and
are the parameters.
\bigskip

Consider the universal fiber-bundle  $\mathcal{W}$ of  the
symmetric squares $\Sym^2(\Jac(V_{\lambda}))$ of Jacobians.
Introduce a vector-function $p([\alpha,\beta],\lambda)=(x;y;z)$ on
$\mathcal{W}$, where
\begin{align*}
&x=(x_1,x_2,x_3)\quad\text{and}\quad
x_{i+j-1}=\wp_{i,j}(\alpha)+\wp_{i,j}(\beta);\\
&y=(y_1,\dots,y_4)\quad\text{and}\quad
y_{i+j+k-2}=\wp_{i,j,k}(\alpha)+\wp_{i,j,k}(\beta);\\
&z=(z_1,\dots,z_5)\quad\text{and}\quad
z_{i+j+k+\ell-3}=\wp_{i,j,k,\ell}(\alpha)+\wp_{i,j,k,\ell}(\beta);
\end{align*}
the brackets $[\cdot,\cdot]$ are used to denote an unordered pair.

Denote by
$\mathcal{M}=\mathcal{M}([\alpha,\beta],\lambda;k_1,k_2,E)$ the
variety of solutions of equation \eqref{eq}. \medskip
\begin{theorem}
$\mathcal{M}=\mathcal{M}_1 \cup\mathcal{M}_2$,
where:\\
$\mathcal{M}_1= \mathcal{M}([\alpha,\beta],\lambda;k_1,0,0)$  is a
subvariety in $\mathcal{W}$ defined by the system of equations
\begin{equation}\label{zeroE}
\begin{cases}
x_2=0,\;x_3=0,\;y_3=0,\;y_4=-2k_1;\\
 \lambda_0=-\frac{1}{8}(3k_1
y_2+z_1),\;\lambda_1 =0,\;\lambda_2=k_1^2-2x_1,\;\lambda_3=z_5;
\end{cases}
\end{equation}
$\mathcal{M}_2=\mathcal{M}([\alpha,\beta],\lambda;0,0,E)$ is a
subvariety in $\mathcal{W}$ defined by the system of equations
\begin{equation}\label{zerok}
\begin{cases}
x_2=-\frac{E}{2},\;x_3=0,\;y_3=0,\;y_4=0,\\
\lambda_0=-\frac{1}{32}(E(8 x_1+z_4) +4 z_2),\;\lambda_1
=\frac{E^2}{4},\;\lambda_2=\frac{1}{2}(z_4-4
x_1),\;\lambda_3=2E+z_5.
\end{cases}
\end{equation}
\end{theorem}
The theorem   admits of a direct proof based on the fact that the
entire function $\phi(u)=\sigma(u)^3 L\psi$, by construction,
belongs to a linear space of dimension $9$. Both equations
\eqref{zeroE} and  equations \eqref{zerok} impose the conditions
that a germ at origin of the corresponding function  $\phi(u)$ is
zero modulo monomials of order $>3$, while the coefficients of the
decomposition of the function $\phi(u)$  with respect to a basis
are completely restored from the germ. The direct deduction of the
equations \eqref{zeroE} and \eqref{zerok} uses the following
segment of $\sigma$-function series expansion:
\begin{equation*}\begin{split}
\sigma(u,\lambda)&= u_1+\tfrac{1}{24}(\lambda_2 u_1^3-8u_2^3)+\\&
\tfrac{1}{48}(\tfrac{1}{40}(\lambda_2^2+2\lambda_1\lambda_3)u_1^5
-4\lambda_0u_1^4u_2-2\lambda_1u_1^3u_2^2-2\lambda_2u_1^2u_2^3
-\lambda_3u_1u_2^4)+O(u^7).\end{split}
\end{equation*}
\bigskip

Complement the vector-function $p$ with rank $1$ matrices of
functions
$$
T=\{T_{i+j-1,k+\ell-1}=
(\wp_{i,j}(\alpha)-\wp_{i,j}(\beta))(\wp_{k,\ell}(\alpha)-\wp_{k,\ell}(\beta))\}$$
and $$S=\{S_{i,k}=(\wp_{i,2,2}(\alpha)-\wp_{i,2,2}(\beta))
(\wp_{k,2,2}(\alpha)-\wp_{k,2,2}(\beta))\},$$ and  define a
mapping $w:\mathcal{W}\to
\mathbb{C}^{21}:([\alpha,\beta],\lambda)\mapsto
((x;y;z),T,S)$.\medskip

\begin{theorem} \label{obraz} The mapping $w$ is a meromorphic embedding.
Under it the varieties $\mathcal{M}_1$ and  $\mathcal{M}_2$ are
taken to families of $3$--dimensional algebraic subvarieties
$\mathcal{N}_1(k_1)$ and  $\mathcal{N}_2(E)$ in $\mathbb{C}^{21}$.
\end{theorem}

Using formulae from Appendix 2 the proof follows.

Define a family of mappings
$\gamma(\cdot,\cdot\,;k_1,E):\mathbb{C}^2\to\Sym^2(\Sym^2(\mathbb{C}^2)):$
$$(q,r)\mapsto\big(\big[[(\mu_1^{+},\nu_1^+),(\mu_2^{+},\nu_2^+)],
[(\mu_1^{-},\nu_1^-),(\mu_2^{-},\nu_2^-)]\big]\big),$$ where
$(\mu^\pm,\nu^\pm)$ are the solutions of systems
\begin{equation}\label{symsym}
\begin{cases}
\mu^2-\frac{E}{2}+q \sqrt{r}=0\\
\nu=-2k_1\mu+(q^2+\mu)\sqrt{r}
\end{cases}\quad\text{and}\qquad
\begin{cases}
\mu^2-\frac{E}{2}-q \sqrt{r}=0\\
\nu=-2k_1\mu-(q^2+\mu)\sqrt{r}
\end{cases}.
\end{equation}
We also need a family of mappings $
\delta(\cdot,\cdot,\cdot\,;k_1,E):\mathbb{C}^3\to\mathbb{C}^4:$
$$
(q,r,s)\mapsto\big(\tfrac{1}{12}r[E-2 k_1q +4(E
s-q^2)^2],\;\tfrac{1}{4}E^2,\; k_1^2+\tfrac{3}{4}r,\;2E+4sr\big).
$$
Let $\jmath:\Jac(V_{\lambda})\to\Sym^2(V_{\lambda})
\subset\Sym^2(\mathbb{C}^2)$ be Jacoby mapping, inverse to Abel
mapping. The mapping $\jmath$ induces mapping
$$\chi:\mathcal{W}\to\Sym^2(\Sym^2(\mathbb{C}^2))\times\mathbb{C}^4:\;
\chi([\alpha,\beta],\lambda)=\big([\jmath(\alpha),\jmath(\beta)],\lambda\big).$$
\medskip
\begin{theorem}
There exist the following families of parameterizing mappings
$R_1:\mathbb{C}^3\to\mathcal{M}_1$ and
$R_2:\mathbb{C}^3\to\mathcal{M}_2$, that are uniquely defined by
the relations \begin{gather*}
\chi(R_1(q,r,s))=(\gamma(q,r;k_1,0),\delta(q,r,s;k_1,0))\intertext{and}
\chi(R_2(q,r,s))=(\gamma(q,r;0,E),\delta(q,r,s;0,E)).
\end{gather*}\label{para}
\end{theorem}
Proof uses essentially an explicit description of the varieties
$\mathcal{N}_1$ and $\mathcal{N}_2$ mentioned in Theorem
\ref{obraz} and the canonical mapping $\mathbb{C}^{21}\to
\Sym^2(\Sym^2(\mathbb{C}^2))\times\mathbb{C}^4$ that is
constructed on the ground of explicit expression of the values of
basis Abel functions at a point $\xi\in\Jac(V_{\lambda})$ in terms
of rational functions on $\jmath(\xi)$ (see Appendix 2).
\bigskip

Theorem \ref{para}  suggests the following algorithm of
construction of solutions of equation \eqref{eq}.

Let a collection  $(q,r,s;k_1)$  be given. Then the equation of
the curve  $V_{\lambda}$ is given by
\[
\nu^2=4\mu^5+4rs\mu^3+(k_1^2+\frac{3}{4}r)\mu^2 -\frac{q
r}{6}(k_1-2q^3).
\]
Using the roots of system \eqref{symsym} at $E=0$, we find
\begin{align}\label{abel}
\notag\alpha_{1}=
\int_{\infty}^{(\mu_1^+,\nu_1^+)}+\int_{\infty}^{(\mu_2^+,\nu_2^+)}
\frac{\mathrm{d}\mu}{\nu},&\quad& \beta_{1}=
\int_{\infty}^{(\mu_1^-,\nu_1^-)}+\int_{\infty}^{(\mu_2^-,\nu_2^-)}
\frac{\mathrm{d}\mu}{\nu},\\
\alpha_{2}=
\int_{\infty}^{(\mu_1^+,\nu_1^+)}+\int_{\infty}^{(\mu_2^+,\nu_2^+)}
\frac{\mu\mathrm{d}\mu}{\nu}, &\quad& \beta_{2}=
\int_{\infty}^{(\mu_1^-,\nu_1^-)}+\int_{\infty}^{(\mu_2^-,\nu_2^-)}
\frac{\mu\mathrm{d}\mu}{\nu}.
\end{align}
A subvariety in $\mathcal{M}_1$ that is singled out by fixing the
values of parameters
$\{\lambda_0,\lambda_1=0,\lambda_2,\lambda_3\}$ is a plane curve
$$
\Gamma=
\{(k_1,q)\in\mathbb{C}^2\,\mid\;2q(2q^3-k_1)(\lambda_2-k_1^2)-9\lambda_0=0\},
$$
with involution $(q,k_1)\mapsto(-q,-k_1)$, while the parameters
 $s=\frac{3\lambda_3}{16(\lambda_2-k_1^2)}$ and $
r=\frac{4}{3}(\lambda_2-k_1^2)$ become functions on it. The set of
roots of \eqref{symsym} through calculation of the integrals
\eqref{abel} defines the pair $[\alpha,\beta]$ as a function on
$\Gamma$.\medskip

Let a collection $(q,r,s;E)$ be given. Then the equation of the
curve $V_{\lambda}$ is given by
$$
\nu^2=4\mu^5+2(E+2rs)\mu^3+\frac{3}{4}r\mu^2+\frac{1}{4}E^2\mu
+\frac{1}{12}r(E+4(Es-q^2)^2).
$$
The pair $[\alpha,\beta]$ is found  by formulas \eqref{abel}
 using the roots of system \eqref{symsym} at $k_1=0$.
A subvariety in $\mathcal{M}_2$ that is singled out by fixing the
values of parameters $\{\lambda_0,\lambda_1,\lambda_2,\lambda_3\}$
consists  of  \emph{eight points with multiplicities}.
\bigskip

The above results stay under various degenerations.

For instance, at $\lambda=0$ we obtain $E=0$, and the equation
$\alpha_1^2-\alpha_1 \beta_1+\beta_1^2=0$ describes the variety of
solutions
$\mathcal{M}([(\alpha_1,0),(\beta_1,0)],0;-3/(\alpha_1+\beta_1),0,0)$.
This example played an important role in our study. The operator
$\partial_{11}+\partial_{22}-6(\wp_{11}+\wp_{22})$ in contrast to
operator $\partial_{12}-6\wp_{12}$ has no eigenfunctions of the
form $\psi(u;\alpha,\beta,k)$ at $\lambda=0$, and therefore its
variety of solutions is empty in general case, due to the fact
that  any construction with $\sigma(u;\lambda)$ admits of a
continuous passing over the limit $\lambda\to 0$.

\chapter[Baker-Akhiezer functions]
{Baker-Akhiezer functions}\label{chap:baf}

\section{Universal Abelian coverings of curves}\label{BAfunction1}
\subsection{Construction of basic functions}
Let $V$ is non-degenerate algebraic curveof genus $g$  and $\mathrm{Jac}(V)$ is its Jacobian.

Denote by $W =\{((x, y), [\gamma])\}$
the space of
pairs $((x, y), [\gamma])$, where $(x, y)\in V$ and $\gamma$ is a path on $V$ that starts at the point $\infty$, ends at the
point $(x, y)$, and belongs to the equivalence class $[\gamma]$. The paths $\gamma'$ and $\gamma''$   are considered to
be Abelian equivalent if the closed path composed of the path from $\infty$ to $(x, y)$ along $\gamma'$ and the
return path to the initial point along $\gamma''$ is homologous to zero.
 Recall that  the values of the integrals  along the path do not depend on the choice of the representative of the class $[\gamma]$. Therefore we get single-valued functions on $W$,
$$u_i(x, y)=\int_{[\gamma]}\mathrm{d}u_i \quad
\text{and}\quad r_i(x, y)=\int_{[\gamma]}\mathrm{d}r_i ,\quad i = 1, . . . , g,$$

\begin{defn}The space $W$ called
the universal Abelian
covering of the curve $V$
\end{defn}

The space $W$ is a  smooth two-dimensional variety. It covers the curve by the map $W\rightarrow V$:   $(x,y),[\gamma])\rightarrow (x,y)$. The map $A:W\rightarrow \mathbb{C}^g$ defined by the formulae $A((x,y),[\gamma])=(u_1(x,y),\ldots, u_g(x,y),[\gamma] )$ is holomorphic embedding that cover the Abel map, $A: V\rightarrow \mathrm{Jac}(V)=\mathbb{C}^g/2\omega\oplus 2\omega'$ . Denote by $W^o$ the variety $W$ without the points $(\infty,[\gamma])$ and by $V^o$ curve $V$ without infinite point. The map $A^*:W^o\rightarrow \mathbb{C}^g$ defined by the formulae $A^*((x,y),[\gamma])=(r_1(x,y),\ldots, r_g(x,y),[\gamma] )$ is holomorphic embedding that covers the Abel map, $A^*: V^o\rightarrow \mathrm{Jac}^*(V)=\mathbb{C}^g/2\eta\oplus2\eta'$ .  Note that for every $1\leq k\leq g$ the limit of product
\[   u_k(x,y,[\gamma])  r_k(x,y,[\gamma])  \]
at $(x,y)\to \infty$  exists, finite and non-zero.

\subsection{Master function }
\index{hyperelliptic $\Phi$--functions! on curve}
Here we construct systems of linear
differential operators for which the universal Abelian covering $W$ of hyperelliptic curve $V  $ is their common
spectral variety.
\begin{definition}
For a fixed point $((x_0,y_0),[\gamma_0])$ where $(x_0,y_0)  \neq \infty$  introduce the function
  \begin{eqnarray*}
&  \Phi:  \mathbb{C}  \times  \mathbb{C}^g  \times W
\to \mathbb{C}  \\ &  \Phi ( \boldsymbol{ u}; w ) =
\dfrac{  \sigma (  \boldsymbol{ \alpha}-  \boldsymbol{ u}) }{
   \sigma (  \boldsymbol{ u}) }
  \exp \left( \boldsymbol{ u}^T \int_{(x_0,y_0)}^{(x,y)}\mathrm{d}{\mathbf
r}\right) , \end{eqnarray*} where $ (x,y)   \in
  V$,  $w=((x,y),[\gamma])$,  $ \boldsymbol{ u}\in \mathbb{C}^g$,
 and $  \boldsymbol{ \alpha}=  \boldsymbol{A}(w)$  and integrsl is computed over path $\gamma'$ from the point $(x_0,y_0)$ to $(x,y)$ such that $\gamma_0+\gamma'\in[\gamma]$
\end{definition}

The function $\Phi$ is called \emph{   master function} because it allow to obtain the foregoing results


\begin{theorem}  \label{schroedinger}  The function $  \Phi=  \Phi ( \boldsymbol{
u}; w ) $ solves the Schr\"odinger equation with the
  potential $2\wp_{gg}(\boldsymbol{u})$
  \begin{equation} (  \partial_g^2-2
 \wp_{gg}(\boldsymbol{u})) \Phi= (x+  \frac{  \lambda_{2g}}{4})   \Phi,
 \label{Hill} \end{equation} with respect to $u_g$  for all $ w  \in W$.  \end{theorem}

  \begin{proof} From   \eqref{stickelberger}
  \[
  \partial_g  \Phi=  \frac{y+  \partial_g{  \mathcal{ P}}
 (x;  \boldsymbol{ u}) } {2
  \mathcal{ P} (x;  \boldsymbol{ u}) }   \Phi,
  \]
where
$  \mathcal{ P} (x;  \boldsymbol{  u}) $ is given by   \eqref{p},
hence:
  \[
  \frac{  \partial_g^2  \Phi}{  \Phi}=
 \frac{y^2- (  \partial_g  \mathcal{
P} (x;  \boldsymbol{  u}) ) ^2 +2  \mathcal{ P} (x;  \boldsymbol{
u})   \partial_g^2  \mathcal{ P} (x;  \boldsymbol{ u}) }{4  \mathcal{
P}^2 (x;  \boldsymbol{  u}) }   \]   and by   \eqref{product3} and
  \eqref{wpgggi} we obtain the theorem.  \end{proof}

From the viewpoint of classical theory differential equation the theorem \ref{schroedinger} describes solution
of the  one-dimensiona Schr\"odinger equation with almost periodic potential of the variable $u_g$.

Let us introduce the vector $  \boldsymbol{\Psi}^T= (\Phi,
\Phi_g) $, where $  \Phi_g$ stands for $\partial_g \Phi$.

\begin{theorem}
For every genus  $g \geq 1$ the vectors
$\boldsymbol{\Psi}=\boldsymbol{\Psi}(\boldsymbol{u};w)$
satisfy to the equations
\begin{equation}{\partial_k} \boldsymbol{
\Psi}= L_k(x) \boldsymbol{\Psi}, \quad k  =1, \ldots,
g,\label{com-L} \end{equation} where $L_k(x)$ are defined by
(\ref{oper-L}).  \end{theorem} In virtue of corollary \ref{zero-L}
the systems of equations (\ref{com-L}) are compatible
\begin{proof} In the accordance with the results of the preceding
section the matrix elements $L_k(x)$ have the form:
\begin{gather}L_0(x)=\begin{pmatrix}V_0&U_0 \\ W_0&-V_0
\end{pmatrix},\quad L_k(x)=D_k L_0(x)-\begin{pmatrix} 0&0
\\\wp_{gk}&0 \end{pmatrix}, \notag  \\ \intertext{where} U_0=x^g-
\sum_{i=1}^{g}x^{i-1}\wp_{gi}, \quad V_0=-\frac12 \partial_g
U_0,\\ \quad \text{and}  \quad W_0=-\frac12 \partial_g^2 U_0
+(x+2\wp_{gg}+\frac14\lambda_{2g})U_0.\notag \label{entries}
\end{gather}
Remark that  $U_0=\mathcal{P}(x;\boldsymbol{u})$.
On the other hand we have from  \eqref{stickelberger}
$$
\frac{\Phi_j}{\Phi}=\frac{(y D_j+\partial_j)\mathcal{P}
(x;\boldsymbol{u})}{2\mathcal{P} (x;\boldsymbol{u})}=
\frac{(y D_j+\partial_j)U_0}{2U_0}.
$$
Hence using the equality
$$
(\partial_j U_0)+(\partial_g U_0)(D_j U_0)-U_0(\partial_g(D_j
U_0))=0,$$ which is valid due to      \eqref{wp3},
we obtain
$$
\Phi_j=U_j \Phi_g+\Phi V_j.
$$
Differentiating the last equality by $u_g$ and expressing
$\Phi_{gg}$ after (\ref{Hill}) we obtain  $$ \Phi_{gj}=-V_j
\Phi_g+\Phi W_j, $$ what proves the theorem.  \end{proof}
\language=0 \begin{cor}
 Let the vector $\boldsymbol{u}^T=(u_1,\ldots,u_g)$ be identified
  with the vector \newline $(t_g,\ldots,t_3,t_2=t,t_1=z)$.
Then the function $  \Phi=  \Phi (u_0,   \boldsymbol{ u}; (y, x) )
$ solves the problem \[ \bigg[ \begin{pmatrix} 0&\partial
/\partial z\\ \partial /\partial t&0 \end{pmatrix}-
  \begin{pmatrix} \mathcal{U}(z,t)&0\\ -\frac14\mathcal{U}_z(z,t)&
-\frac12\mathcal{U}(z,t)+\frac18 \lambda_{2g} \end{pmatrix} \bigg]
\begin{pmatrix} \Phi\\ \Phi_z \end{pmatrix}=x \begin{pmatrix}
  \Phi\\ \Phi_z
  \end{pmatrix},
  \]
where $\mathcal{U}(z,t)=2\wp_{gg}+\frac14 \lambda_{2g}$
and subscript means differentiation.
\end{cor}

\begin{theorem}  \label{the-F} The function
$  \Phi=  \Phi (u_0,   \boldsymbol{ u}; (y, x) ) $
solves the system of
equations  \begin{gather}   \left (  \partial_k   \partial_l
-  \gamma_{kl} (x,   \boldsymbol{ u})   \partial_g +
  \beta_{kl} (x,   \boldsymbol{ u})
  \right)   \Phi=  \frac{1}{4}D_{k+l}  \big (f (x)   \big)   \Phi
  \notag  \\ \intertext{with polynomials in $x$}
\begin{aligned}
  \gamma_{kl} (x,   \boldsymbol{ u}) &=
  \frac{1}{4}   \left[
  \partial_k D_l +
  \partial_l D_k
  \right]  \sum_{i=1}^{g+2}x^{i-1}h_{g+2, i}  \quad  \text{and}  \\
  \beta_{kl} (x,   \boldsymbol{ u}) &=  \tfrac{1}{8}   \left[
 (  \partial_g  \partial_k +h_{g+2, k} )  D_l +
 (  \partial_g  \partial_l +h_{g+2, l} )  D_k
  \right]  \sum_{i=1}^{g+2}x^{i-1}h_{g+2, i}  \\
&-  \frac14   \sum  \limits_{j=k+l+2}^{2g+2}x^{j- (k+l+2) }
  \left[ \left (  \sum  \limits_{  \nu=1}^{k+1} h_{  \nu, j-  \nu}
  \right) + \left (  \sum  \limits_{  \mu=1}^{l+1} h_{j-  \mu,
  \mu}  \right)    \right] \end{aligned}  \notag \end{gather} for
all $k, l   \in 0, \ldots, g$ and arbitrary $ (y, x)    \in V$.
  \end{theorem}
The most remarkable aspect of the equations of Theorem
\ref{the-F} is the balance of powers of the polynomials $
\gamma_{kl}$,  $ \beta_{kl}$ and of the ``spectral part'', the
umbral derivative $D_{k+l} (f (x) ) $:  \begin{align*} &
\mathrm{deg}_x  \gamma_{kl} (x,   \boldsymbol{ u})   \leqslant
g-1-  \mathrm{min} (k, l) ,    \\&  \mathrm{deg}_x  \beta_{kl} (x,   \boldsymbol{
u})   \leqslant 2g- (k+l) ,   \\
&  \mathrm{deg}_x D_{k+l} (f (x) ) =2g+1- (k+l) .
  \end{align*}

Here $f (x) $ is as given in   \eqref{curve} with
$  \lambda_{2g+2}=0$ and $  \lambda_{2g+1}=4$.

Remark that the definition of the functions $\{h_{i,k}\}$ with the
help of relations \eqref{dg_H} and \eqref{wpggiwpggk} readily
yields the following formula, which reexpress the  functions
$\{h_{i,k}\}_{i,k \leq g}$ in terms of basis functions
$\{\wp_{gj},\wp_{ggj},\wp_{gggj}\}$ and of the constant
$\lambda_{2g}$:  \begin{gather*}
h_{i,k}=(8\wp_{gg}+\lambda_{2g})\wp_{gi}\wp_{gk}+2\wp_{gi}\wp_{g,k-1}+
2\wp_{g,i-1}\wp_{gk}+\\
\wp_{ggi}\wp_{ggk}-\wp_{gggi}\wp_{gk}-\wp_{gi}\wp_{gggk}.
\end{gather*}

Thus all the coefficients in the differential equations from the
theorem   \ref{the-F} are polynomials in $x$ and
in the basis functions  $\{\wp_{gj},\wp_{ggj},\wp_{gggj}\}$ .

\begin{proof}
The construction of the operators $L_k$ yields
   \[
  \Phi_{lk}=  \tfrac12
\left ( \partial_l U_k+
  \partial_k U_l  \right) {  \Phi_g}+  \left (V_lV_k +  \tfrac12
 (   \partial_l V_k+  \partial_k
V_l+U_k W_l+W_k U_l)   \right)   \Phi.
  \]
To prove the theorem, we use   \eqref{entries}  and  that   (cf.
Lemma \ref{geom-2} ) :  \begin{eqnarray*} &&D_k (V_0) D_l (V_0) +  \tfrac12
D_k (U_0) D_l (W_0) +  \tfrac12 D_l (U_0) D_k (W_0) =  \\&& -
\tfrac{1}{16}  \left (  \det H
\left[{}_{g+1}^{g+1}{}_{g+2}^{g+2}  \right] \right) (1, x,
 \ldots, x^{g+1-k}) H   \left[{}_{k}^{l}{}_{  \ldots}^{
\ldots}{}_{g+2}^{g+2}  \right] (1, x,   \ldots, x^{g+1-l}) ^T,
  \end{eqnarray*}
having in mind the relations $h_{g+2, g+2}=0$ and $h_{g+2,
g+1}=2$, we obtain the theorem from properties of the matrix $H$.
\end{proof}
Consider as an example
 of Theorem \ref{the-F} the case of genus $2$.
\begin{align} & (
\partial_2^2-2  \wp_{22}) \Phi= \tfrac{1}{4} (4 x+  \lambda_4)
\Phi,\notag   \\ & ( \partial_2 \partial_1+  \tfrac{1}{2}
\wp_{222}  \partial_2 - \wp_{22} (x+ \wp_{22}+ \tfrac{1}{4}
\lambda_4) -2  \wp_{12}) \Phi= \tfrac{1}{4} (4 x^2+  \lambda_4 x
+  \lambda_3)   \Phi, \label{ex-2}\\ & (  \partial_1^2+  \wp_{122}
  \partial_2 -2  \wp_{12} (x+ \wp_{22}+ \tfrac{1}{4}  \lambda_4) )
\Phi= \tfrac{1}{4} (4 x^3+ \lambda_4 x^2 +  \lambda_3 x+
\lambda_2)   \Phi.\notag \end{align} And the
$\Phi=\Phi(u_0,u_1,u_2;(y,x))$ of the curve $$y^2=4 x^5+ \lambda_4
x^4 +  \lambda_3 x^3+ \lambda_2x^2+ \lambda_1 x+ \lambda_0$$
  solves these equations for all $x$.

By eliminating step by step from the left hand sides of
\eqref{ex-2} the dependence in $x$ we come to the new equivalent
system of equations:
  \begin{align}
\Lambda_{22}\Phi=&( \partial_2^2-2 \wp_{22}) \Phi= \tfrac{1}{4} (4
x+ \lambda_4) \Phi,\notag \\ \Lambda_{12}\Phi=&(\partial_2
\partial_1-\wp_{22}\partial_2^2+\tfrac{1}{2}\wp_{222}\partial_2 +
\wp_{22}^2  -2  \wp_{12}) \Phi=
\tfrac{1}{4} (4 x^2+  \lambda_4 x +  \lambda_3)   \Phi, \\
\Lambda_{11}\Phi= & (
\partial_1^2
-2 \wp_{12}\partial_2^2 +\wp_{122}\partial_2+
2 \wp_{22}\wp_{12}) \Phi= \tfrac{1}{4} (4 x^3+
  \lambda_4 x^2 +  \lambda_3 x+  \lambda_2) \Phi.\notag
\label{ex-2i} \end{align} This example illustrates the general
fact that for the systems of  equations described by Theorem
\ref{the-F} one can  iteratively eliminate
dependence on $x$ in their left-hand sides.
\begin{cor}Calculating the commutators of the triple of operators
 $\Lambda_{22},\Lambda_{12},\Lambda_{11}$, we find:
\begin{gather*}
[\Lambda_{12},\Lambda_{22}]=-2 \wp_{222} A,\\
[\Lambda_{11},\Lambda_{22}]=-4 \wp_{122} A,\\
[\Lambda_{11},\Lambda_{12}]=-6 \wp_{112} A,\\
\intertext{where}
A=\partial_1-\partial_2^3+(3\wp_{22}+
\frac14\lambda_{4})\partial_2+\frac32
\wp_{222}.
\end{gather*}
The operator $A$ is the difference between $\partial_1$ and the
positive part of the formal fractional power
$\big[(L)^{3/2}\big]_+$ of the operator
$L=\partial_2^2-(2\wp_{22}+\frac16\lambda_{4})$, i.e. it is the
$A$-operator of the  $L$--$A$ pair of the classical KdV equation
with respect to the function $2\wp_{22}+\frac16\lambda_{4}$.
\end{cor} This corollary illustrates general property of operators
$\{\Lambda_{k,l}\}$ which are obtained after elimination of $x$
from the equations of Theorem \ref{the-F}:  commutation yields
operators from Lax pairs of the integrable systems.

We shall also mention here that the Theorem \ref{the-F} can be
written in the form of addition theorem of the kind "point +
divisor"
\begin{theorem}
Let $(w,z)\in V $ is arbitrary and
$$
\{(y_1,x_1),  \ldots, (y_1,x_1)\}$$ is the Abelian pre-image of
the point ${\boldsymbol u}\in \mathrm{Jac}(V)$ Then the functions
$\wp_{k,l}$ satisfy to the addition law
\begin{eqnarray}
&&\wp_{k,l}\left(\int_{a}^z\mathrm{d}{\mathbf u} +{\boldsymbol
u}\right)= \frac14 Z_k({\boldsymbol u};z,w)Z_l({\boldsymbol u};z,w)
\cr
&+&Z_g({\boldsymbol  u};z,w)\gamma_{k,l}+\beta_k({\boldsymbol
u};z)
\end{eqnarray} where polynomials in
$z$ $\beta_{kl}$ and $\gamma_{kl}$ are given in the formulation of
the Theorem \ref{the-F},
\[ Z_j(\boldsymbol{u};z,w)=\frac{(wD_j+\partial_j)
{\mathcal P}(z;\boldsymbol{u})}{2{\mathcal P}(z;\boldsymbol{u})},\]
where $D_j$ is the umbral derivative of order $j$ and $\partial_j$
is the partial derivative by the variable $u_j$.

In particular,
\begin{eqnarray}
\wp_{gg}\left(\int_{a}^z\mathrm{d}{\mathbf u} +{\boldsymbol
u}\right)& =&-(\wp_{gg}({\boldsymbol
u})+z+\frac14\lambda_{2g})\cr
&+&\frac14\left(\frac{w-\sum_{i=1}^gz^{g-i}\wp_{g-i+1,g,g}({\boldsymbol u})}
{{\mathcal P}(z;{\boldsymbol u})}
\right)^2.\label{pgg}
\end{eqnarray}
\end{theorem}

The formula (\ref{pgg}) can be found e.g. in \cite{ba97}, p.218.
\index{addition theorem!of the kind ``point+divisor"}

\section{General Baker-Akhiezer function}\label{BAfunction2}
A problem arises to characterise the master function described in the  Sect. \ref{BAfunction1}. That can be done using the theory of so called \emph{General Baker-Akhiezer function}, that was developed by Krichever \cite{kr77}. In the further exposition we will follow to the review \cite{bk06} .

 Let~$V$ be a non-singular algebraic curve of
ge\-nus~$g$ with distinguished points~$P_\alpha$ and fixed local
coordinates $k_\alpha^{-1}(Q)$ in neighbourhoods of these
distinguished points, where $k_n^{-1}(P_n)=\nobreak 0$
for $n=1,\dots,l$. Let us fix a family
${\mathbf q}=\{q_n(k)\}$ of polynomials,
\begin{equation}
\label{eq2-1}
q_n=\sum_it_{n,i}k^i.
\end{equation}
In \cite{kr77} it was shown that for any generic family of
points $v_1,\dots,v_g$ there is a function $\psi(t,Q)$,
 unique up to proportionality, such that:
\begin{itemize}
\item[(a)] {\it $\psi$ is meromorphic on $V$
outside the points $P_n$ and has at most simple poles at the
points~$v_s$ {\rm(}if they are distinct\/{\rm);}}

\item[(b)] {\it $\psi$ can be represented in a neighbourhood of $P_n$ as}
\begin{equation}
\label{eq2-2}
\psi(t,Q)=\exp\bigl(q_{n}(k_{n})\bigr)
\biggl(\sum_{s=0}^{\infty}\xi_{s,n}(t)k^{-s}\biggr), \qquad
k_{n}=k_{n}(Q).
\end{equation}
\end{itemize}
We choose a point $P_0$ and normalize the function~$\psi$ by
the condition
\begin{equation}
\label{eq2-3}
\psi(t,P_{0})=1.
\end{equation}
It should be stressed that the Baker--Akhiezer function $\psi(t,Q)$
is determined by its analytic properties with respect to the
variable~$Q$. It depends on the coefficients $t_{n,i}$ of the
polynomials~$q_n$ as on {\it external parameters}.

The {\it existence\/} of the Baker--Akhiezer function is proved by
presenting an explicit theta function formula~\cite{kr77}. We
introduce some necessary notation. First of all, we recall that
fixing a basis $\mathfrak{a}_i$,~$\mathfrak{b}_i$ of cycles on~$V$ with canonical
intersection matrix enables us to define a basis of normalized
holomorphic differentials~$\omega_i$, the matrix~$B$ of their
$b$-periods, and the corresponding theta function
$\theta(z)=\theta(z\,|\,B)$. The basis vectors $e_k$ and the
vectors $B_k=\{B_{kj}\}$ span a lattice in~$\mathbb C^g$, and the
quotient by this lattice is a $g$-dimensional torus $J(\Gamma)$,
the so-called {\it Jacobian\/} of the curve~$\Gamma$. The
map~$A\colon\Gamma\longmapsto J(\Gamma)$ given by the formula
\begin{equation}
\label{eq2-4} A_{k}(Q)=\int_{Q_{0}}^{Q}\omega_{k}
\end{equation}
is referred to as the {\it Abel map}. If a vector $Z$ is defined by
the formula
\begin{equation}
\label{eq2-5}
Z=\kappa-\sum_{s=1}^{g}A(\gamma_{s}),
\end{equation}
where $\kappa$ is a vector of Riemann constants depending on the
choice of basis cycles and on the initial point of the Abel
map~$Q_0$, then the function $\theta(A(Q)+z)$ (if it is not
identically zero) has exactly $g$ zeros on~$\Gamma$ coinciding with
the points~$\gamma_s$,
\begin{equation}
\label{eq2-6}
\theta(A(\gamma_{s})+Z)=0.
\end{equation}
We note that the function $\theta(A(Q)+Z)$ itself is multivalued
on~$\Gamma$, but the zeros of this function are well defined. Let
us introduce differentials $d\Omega_\alpha$ such that the
differential $d\Omega_{\alpha,i}$ is holomorphic
outside~$P_\alpha$, has a pole at~$P_\alpha$ of the form
$d\Omega_\alpha=d(k^i_\alpha+O(k_\alpha^{-1}))$, and is normalized
by the condition $\displaystyle\oint_{a_i}\,d\Omega_\alpha=0$. We
denote by $2\pi iU_{\alpha,i}$ the vector of its $b$-periods with
coordinates
\begin{equation}
\label{eq2-7}
2\pi iU_{\alpha,i,k}=\oint_{b_{k}}\,d\Omega_{\alpha,i}.
\end{equation}
Using the translation properties of the theta function, one can
show that the function $\psi(t,Q)$ given by the formula
\begin{equation}
\label{eq2-8}
\psi(t,Q)=\Phi(x,Q)\exp\biggl(\sum_{\alpha,i}t_{\alpha,i}
\int_{P_{0}}^{Q}\,d\Omega_{\alpha,i}\biggr),
\end{equation}
where
\begin{equation}
\label{eq2-9}
\Phi(x,Q)=\frac{\theta(A(Q)+x+Z)\,\theta(A(P_{0})+Z)}
{\theta(A(Q)+Z)\,\theta(A(P_{0})+x+Z)}
\end{equation}
and
\begin{equation}
\label{eq2-10}
x=\sum_{\alpha,i}t_{\alpha,i}U_{\alpha,i},
\end{equation}
is a {\it single-valued\/} function of the variable $Q$ on~$\Gamma$.
It follows from the definition of the differentials
$d\Omega_{\alpha,i}$ that this function has the desired form of
essential singularity at the points~$P_\alpha$, and it follows from
\eqref{eq2-6} that the function $\psi$ has poles outside these points
at the points~$\gamma_s$.

The proof of the {\it uniqueness\/} of the Baker--Akhiezer function is reduced
to the Riemann--Roch theorem, according to which a meromorphic
function on~$V$ of genus $g$ having at most $g$ generic poles is constant.
Indeed, suppose that the analytic properties of a function
$\tilde\psi$ coincide with those of~$\psi$. Then the function
$\tilde\psi\psi^{-1}$ is a meromorphic function on~$\Gamma$ whose
number of poles does not exceed~$g$. Hence, this is a constant
which is equal to~1 due to the normalization
condition~\eqref{eq2-3}.

We present another assertion which is needed in what follows
and whose proof also reduces to the Riemann--Roch theorem. For
any positive divisor $D=\sum n_iQ_i$ we consider the linear space
$L({\mathbf q},D)$ of functions having poles outside the points
$P_\alpha$ at the points $Q_i$ and of multiplicities at most $n_i$ and
having the form~\eqref{eq2-1} in a neighbourhood of $P_\alpha$. As
was proved in~\cite{kr77}, the dimension of this space is equal to
$\dim L({\mathbf q},D)=d-g+1$ for generic divisors of degree
$d=\sum n_i\ge g$.

The function $\Phi(x,Q)$ given by formula~\eqref{eq2-9} belongs to
the class of the so-called {\it factorial\/} functions~\cite{ba97}.
This is a multivalued function on~$\Gamma$. A single-valued branch of
it can be singled out if one draws cuts on~$\Gamma$ along
$a$-cycles. In what follows we use the notation~$\Gamma^*$ for the
curve~$\Gamma$ with such cuts.

\begin{lem}
\label{lem2.1} For any generic family of points
$\gamma_1,\dots,\gamma_g$ the formula~\eqref{eq2-9} with the
vector~$Z$ defined by~\eqref{eq2-5} defines a unique
function~$\Phi(x,Q)$, where $x=(x_1,\dots,x_g)$ and $Q\in\Gamma$,
having the following properties\/$:$
\begin{itemize}
\item[$1)$] $\Phi$ is a single-valued meromorphic function of~$Q$
on~$\Gamma^*$ having at most simple poles at the points~$\gamma_s$
$($if they are all distinct\/$);$

\item[$2)$] the boundary values $\Phi_j^\pm(x,Q)$, $Q\in a_j$, on
the different sides of the cuts satisfy the conditions
\begin{equation}
\label{eq2-11}
\Phi_j^+(x,Q)=e^{-2\pi ix_j}\Phi_j^-(x,Q), \qquad Q\in a_j;
\end{equation}

\item[$3)$] $\Phi$ is normalized by the condition
\begin{equation}
\label{eq2-12}
\Phi(x,P_0)=1.
\end{equation}
\end{itemize}
\end{lem}

As in the case of Baker--Akhiezer functions, for any effective
divisor $D=\sum n_iQ_i$ one can introduce the notion of associated
linear space $\mathscr L(x,D)$ as the space of meromorphic
functions on~$\Gamma^*$ having poles at the points $Q_i$ of
multiplicities not exceeding~$n_i$ and such that the boundary
values on the sides of the cuts satisfy~\eqref{eq2-11}.
It follows from the Riemann--Roch theorem
that for a generic divisor of degree $d=g$ the dimension of this
space is equal to
\begin{equation}
\label{eq2-13}
\dim\mathscr{L}(x,D)=d-g+1.
\end{equation}
According to \eqref{eq2-13}, the dimension of the space of
factorial functions with poles of multiplicity at most two at the
points $\gamma_s$ is equal to $g+1$ for any generic family of
points $\gamma_1$, \dots, $\gamma_g$. The following assertion can
be regarded as an explicit representation of a set of $g$ functions
which, together with the function~$\Phi$, define a basis in this
space.

\begin{lem}
\label{lem2.2} For any generic family of points
$\gamma_1,\dots,\gamma_g$ the function $C_k(x,Q)$ given by the
formula
\begin{equation}
\label{eq2-14}
C_k(x,Q)=\frac{\theta(A(Q)+Z+A(\gamma_k)-A(P_0))
\,\theta(A(Q)+x+Z-A(\gamma_k)+A(P_0))}
{\theta^2(A(Q)+Z)\,\theta(A(P_0)+x+Z)\,\theta(2A(\gamma_k)-A(P_0)+Z)}\,,
\end{equation}
where $Z$ is defined by~\eqref{eq2-5}, is a unique function such
that:
\begin{itemize}
\item[$1)$] $C_k$, as a function of the variable
$Q\in\Gamma^*$, is a meromorphic function with at most simple poles
at the points $\gamma_s$, $s\ne k$, and a second-order pole
of the form
\begin{equation}
\label{eq2-15}
C_k(x,Q)=\theta^{-2}(A(Q)+Z)+O(\theta^{-1}(A(Q)+Z))
\end{equation}
at the point $\gamma_k;$

\item[$2)$] the boundary values $C_{k,j}^\pm(x,Q)$, $Q\in a_j$, of
the function $C_k$ on the different sides of the cuts satisfy the
equations
\begin{equation}
\label{eq2-16}
C_{k,j}^+(x,Q)=e^{-2\pi ix_j}C_{k,j}^-(x,Q), \qquad Q\in a_j;
\end{equation}

\item[$3)$]
\begin{equation}
\label{eq2-17}
C_k(x,P_0)=0.
\end{equation}
\end{itemize}
\end{lem}

The uniqueness of a function $C_k$ having the above properties is
an immediate corollary to the Riemann--Roch theorem. The fact that
the function~$C_k$ given by the formula~\eqref{eq2-14} satisfies
these conditions can be verified immediately. Indeed, the function
$\theta(A(Q)+Z)$ vanishes at the point~$\gamma_k$, and hence the
equality~\eqref{eq2-14} means that $C_k$ has a second-order pole
with normalized leading coefficient at this point. It
follows from the equality~\eqref{eq2-5} that the first factor in
the numerator vanishes at the point~$P_0$ and at the points
$\gamma_s$, $s\ne k$. Since the first factor of the denominator has
a second-order pole at all points $\gamma_s$, the function $C_k$
has first-order poles at the points $\gamma_s$, $s\ne k$, and a
second-order pole at the point~$\gamma_k$. The difference between
the arguments of the theta functions containing $A(Q)$ in the numerator
and the denominator is equal to~$x$. This, together with the
monodromy relations  implies~\eqref{eq2-16}.

\section{Multiplication theorem}
Krichever and Novikov introduced  in~ \cite{kn987} analogues of  Laurent
bases  (KN-bases)  to solve the problem of operator quantization for closed bosonic strings.
These bases are analogues of  Laurent bases  in the case of algebraic curves of
arbitrary genus with a pair of distinguished points. In essence,
these bases $\psi_n(A)$, $A\in\Gamma$, are a special case of the
Baker--Akhiezer functions of a discrete argument
 $n\in{\mathbb Z}$. As was noted in~ \cite{kn987},
in this case the discrete Baker--Akhiezer functions $\psi_n$
satisfy the remarkable relation
\begin{equation}
\label{eq1-34}
\psi_n(A)\psi_m(A)=\sum_{k=0}^{g}c_{n,m}^k\psi_{n+m+k}(A),
\end{equation}
which was a foundation of the notions of {\it almost graded algebras}
and modules over them.

The notion of continuous analogue of KN-bases was introduced in the
papers~ \cite{gn01},\cite{gn03} of Grinevich and Novikov, where it was proved that
in a special case the corresponding Baker--Akhiezer functions $\psi(t_1,A)$ of
a continuous argument satisfy the equation
\begin{equation}
\label{eq1-35}
\psi(t_1,A)\psi(t_2,A)==\mathcal{L}\psi(t_1+t_2,A),
\end{equation}
where $\mathscr L$ is a differential operator of order~$g$ with
respect to the variable~$t_1$ and replaces the
difference operator on the right-hand side of~\eqref{eq1-34},
\begin{equation}
\mathcal{L}=\frac{\partial^g}{\partial t_3^g}-\sum_{j=1}^g a_j(t_1,t_2,t_3)
\frac{\partial^{g-j}}{\partial t_3^{g-j}}
\end{equation}

The Baker-Akhiezer technique permits to show that relation (\ref{eq1-35}) is a particular
case of the more genera relation
\begin{equation}
\Psi(u,(x,y))\Psi(v,(x,y))=\left.\mathcal{L}\Psi(w,(x,y))\right|_{w=u+v},\qquad u,v\in\mathbb{C}^g
\label{mainrel}
\end{equation}

We describe now modification of the Baker-Akhiezer function that permits to find effectively coefficients $a_j$
of the operator $\mathcal{L}$.
A special degenerate case of the Baker--Akhiezer function
namely, the basis KN-function with
parameter $u\in\mathbb C^g$, is realized in terms of the $\sigma$-function in the form
\begin{equation}
\label{eq4-1}
\Psi\bigl(u,(x,y)\bigr)=\frac{\sigma(A(x,y;[\gamma])-u)}
{\psi(x,y;[\gamma])\,\sigma(u)}
\exp\bigl\{-\langle A^{*}(x,y;[\gamma]),u\rangle\bigr\},
\end{equation}
where $\psi(x,y;[\gamma])$ is the function given by the
formula
\begin{equation}
\psi(x,y;[\gamma])=\mathrm{exp}\left\{   \int_{[\gamma]}   \langle  A^{\ast}( (x',y'),[\gamma'] ),\mathrm{d}A(x',y')  \rangle \right\}
\end{equation}
where the path $\gamma'$ is a part of the path from the base point to the point $(x',y')$,
$\langle a,b\rangle=\sum a_ib_i$ and $A^*((x,y),[\gamma])$ - is the map dual to the Abel map given by thasis seconf kind integrals with poles only in the base point, where the following expansion over local parameter is valid
$$  A^*((x(\xi),y(\xi)),[\gamma])=  \left( \xi^{-w_1}( 1 +a_1^*(\xi)),\ldots,  \xi^{-w_g}( 1 +a_g^*(\xi))  \right)^T$$
where $a_i^*(\xi)$ are series over positive powers of $\xi$ with coefficients from $\mathbb{Q}[\lambda]$.

The function $\Psi$ is single-valued on
$\mathbb C^g\times V$. As a function on~$V$, it has $g$
zeros $A^{-1}(u)\in\mathsf X$ and a unique
essential singularity $\infty\in V$, at which it has the behaviour
$\Psi\sim\xi^{-g}\exp\{p(\xi^{-1})\}(1+O(\xi))$, where $p$~is a
polynomial of degree not exceeding $2g-1$. For instance, in the
hyperelliptic case we have the following expansion with respect to the local
parameter~$\xi$ at this singular point:
\begin{equation}
\label{eq4-2}
\Psi\bigl(u,(x(\xi),y(\xi))\bigr)
=\xi^{-g}\exp\biggl\{-\rho(\xi)\biggl(
\sum_{i=1}^{g}u_i\xi^{-2i+1}\biggr)\biggr\}\bigl(1+O(\xi)\bigr),
\end{equation}
where $\rho(\xi)^2=1+\sum_{i>1}\lambda_{2i}\xi^{2i}$. Thus
 the function $\Psi$ is a
degeneration of the Baker--Akhiezer function,
corresponding to gluing an essential singularity and $g$ poles at the
base point $\infty\in V$. Due to these properties,
~$\Psi$ is an extremely convenient tool.

\chapter[Trigonal Abelian functions]{Application of the trigonal Abelian functions}\label{chap:trig}
\section{Introduction}
The present work is devoted to development of a method of
integration of nonlinear partial differential equations on the
basis of uniformization of the Jacobi varieties  of algebraic
curves with the help of $\wp$--functions of several variables,
the theory of which was founded by Klein. Our studies in this
direction (see the review \cite{bel97b}) are motivated by the
fact that under such uniformization the most important nonlinear
equations of mathematical physics appear naturally and explicitly.

Klein \cite{kl88} proposed to construct the theory of an analog
of the elliptic Weierstrass $\si$--function for the curves of
genus $>1$ as the theory of a function of several variables, which
is closely related with $\theta$--functions and has the
complementary property of \emph{modular invariance}. The
differential field of Abelian functions on the Jacobi variety of a
curve is generated by the derivatives of order $>1$ of the
logarithm of $\si$--function, that is by the functions
\[ \wp_{i_1,i_2,\dots,i_k}(\Bu,\Bla)= -\frac{\pa^k
\log\si(\Bu,\Bla)}{\pa
u_{i_1}\pa u_{i_2}\cdots\pa u_{i_k}},\quad k\ge 2.
\]

In the work of the authors \cite{bel97b} for the
hyperelliptic curves of genus $g$ the set of  $3g$ basis
 $\wp$--functions on the Jacobian is presented, for which the
fundamental algebraic relations are written down explicitly.  For
the case $g=2$ these results may be derived from the work of
Baker \cite{ba07}, while the theory  constructed by the authors
for $g>2$ requires a substantial development of the classical
methods. It is necessary to indicate that the works of classical
period lack for any comparison of the differential relations
obtained with the problems of the theory of partial differential
equations, despite the fact that from the relations for $g=2$
obtained in \cite{ba07} one can deduce solutions of a number of
differential equations, which were topical already at that time
(such as, e.g., the ``sine--Gordon'' equation).
As an exeption could serve Baker's \cite{ba03}, but even in this
article the link with known in that time important
equations of mathematical physics has not been discovered; the paper
\cite{ba03} was analysed in the context of completely integrable
equations in \cite{ma00}.

Our  approach to the theory in the general case is based on the
use of the models of curves which are called
 $(n,s)$--curves in \cite{bel99}.
The hyperelliptic curves of genus $g$ are  $(2,2g+1)$--curves in
this terminology.  Below we presents the results for
$(3,g+1)$--curves, which are the so-called \emph{trigonal} curves
of genus $g$. In the class of $(n,s)$--curves the cases $n=2$ and
$n=3$ are distinguished, as it is in these cases that the
inequalities,
 which play an important r{\^o}le in construction of the theory,
$n\le g=\frac{(n-1)(s-1)}{2}<s$ for the genus $g$ of the curve
hold.

The paper represents itself extended and improved version of the
recent paper \cite{bel00}.  \S \ref{sec:U}  contains the necessary
preliminaries and introduces the important  in the sequel notion
of the universal space $\sU_g$ of  $g$--th symmetric powers of
trigonal curves.  The main result of  \S \ref{sec:L} is Theorem
\ref{aL->U} that gives an explicit description of the space
$\sU_g$ as an algebraic subvariety in a complex linear space of
dimension  $4g$, or $4g+2$, if  $g=3\ell$, or $g=3\ell+1$. In \S
\ref{sec:K} we obtain an explicit realization of the Kummer
variety of a trigonal curve as an algebraic subvariety in
$\C^{g(g+1)/2}$ defined by a system of $\frac{(g-1)g}{2}$
polynomial equations of degree not greater than  $5$.  The
effectiveness of our approach is illustrated by the example of a
non-hyperelliptic curve of genus $3$. In \S \ref{sec:J} we single
out the collection of  $4g$ $\wp$--functions on the Jacobi variety
of a trigonal curve, which uniformize the algebraic varieties
constructed in \S \ref{sec:L}.  This section contains  the
necessary facts on $\si$--functions.  In \S \ref{sec:Eqs} we
analyze the differential relations obtained from the viewpoint of
the theory of partial differential equations. Particularly,  by
comparing with the analogous results for the hyperelliptic case,
we show, that the theory of $\wp$--functions is an effective tool
of the characterization of the Jacobians by analytic means.

The authors are grateful to  S.~P. Novikov, who
greatly influenced our understanding of the subject of the present
study.

\section{Universal
space of  $g$--th symmetric powers of trigonal curves
}\label{sec:U}
Let $\gcd(g+1,3)=1$.
The possible values of $g$ are $g=3 \ell$ and $g=3
\ell+1$, where $\ell=1,2,\dots$.
\begin{df} We call
\emph{trigonal} such a linear in parameters
$\Bla=\{\lambda_0,\lambda_3,\dots\}$ polynomial $f_{g}(x,y;\Bla)$ with the
coefficients at the highest powers of  $x$ and $y$ respectively
equal to $-1$ and $1$ that  it is homogeneous of weight $\deg
f_{g}=3(g+1)$ with respect to the grading $\deg x=3$, $\deg
y=g+1$ and $\deg\lambda_i=3g+3-i>0$.
\end{df} It follows from the definition that a trigonal
polynomial has the form
\begin{align} f_{g}(x,y;\Bla)=y^3&-
y^2(\sum_{j=0}^{[\frac{g+1}{3}]}\lambda_{2g+2+3 j}x^j)-
y(\sum_{j=0}^{[\frac{2g+2}{3}]}\lambda_{g+1+3 j}x^j)\notag \\
&-(x^{g+1}+
\sum_{j=0}^{g}\lambda_{3 j}x^j),\label{TRF}
\end{align}
and depends on a collection $2g+3$ parameters $\Bla=
\{\lambda_0,\lambda_3,\dots,\lambda_{3g+1},\lambda_{3g+2}\}$.
The set of subscripts of the members of the collection
$\Bla$ runs over all the numbers of the form
 $3 a +(g+1) b$ with non-negative integers
$a$ and $b$ in the range from $0$ to $3g+2$, and there are
exactly $g$ missing numbers, so-called gaps,
that form the \emph{Weierstrass sequence} (for more detail
see  \cite{bel99}).

\begin{df} We call a graded space, the points of which are the
collections $\Bla$, the \emph{moduli space $\sM_{g}$  of the
trigonal polynomials of weight} $3(g+1)$, and, thus
$\sM_{g}\cong\C^{2g+3}$.  \end{df}
At numerical values of the
parameters $\{\lambda_k\}$ the set of zeros of the trigonal polynomial
$\{(x,y)\in\C^2\mid f_{g}(x,y;\Bla)=0\}$ is an affine model of the
\emph{plane trigonal curve} $V(x,y)$ of genus not greater than
$g$.  Under the condition of non-degeneracy of the polynomial
$f_{g}$, i.e.,  when the values of the parameters $\{\lambda_k\}$ are
such that the triple of polynomials
$\{f_{g}(x,y;\Bla),\frac{\pa}{\pa y}f_{g}(x,y;\Bla),\frac{\pa}{\pa
x}f_{g}(x,y;\Bla)\}$ has no common zeros, the curve $V(x,y)$ is
non-singular, and its genus is equal to $g$.

Any  \emph{entire rational function on the curve} $V(x,y)$, i.e.
a function that takes finite values at the points of the affine
model $V$, can be written down in the so-called
 \emph{normal form} $p(x,y)=p_{2}(x)
y^2+p_{1}(x)y+p_{0}(x)$, where $p_{0}(x)$, $p_{1}(x)$  and
$p_{1}(x)$ are some polynomials in $x$.
The highest weight of the monomials
$x^iy^j$ that occur in the polynomial
$p(x,y)$ is called the \emph{order} of the entire rational
function.  In the sequel we need the following facts
 (see, e.g.,
\cite{ba97,che48}):  the \emph{number of zeros} of an entire
rational function $p(x,y)$ on the curve
$V(x,y)$, i.e.  the number of pairs
$\{(\xi,\eta)\in\C^2\mid f_{g}(\xi,\eta;\Bla)=0,p(\xi,\eta)=0\}$,
is equal to its order; an entire rational function on the
curve having
$2g+n$ zeros ($n\ge  0$) is defined by a set  of $g+n$
parameters.

It is known \cite{dub81,mu83} that \emph{Abel map}  takes
$k$--th symmetric power of a curve
$V(x,y)$ into its \emph{Jacobi variety}
$\Jac(V)$.  At $k$ equal (or greater than) the genus of the
curve Abel map is  ``\emph{onto}''. Below we need the following
variant of
\emph{Abel Theorem}:  on the set of zeros of an entire rational
function on the curve Abel map takes the zero value.

Further the symbol  $(X)^{k}$  stands for the
$k$--th symmetric power of a space $X$.
\begin{df}

The space
$\sU_{g}\subset{(\C^2)}^{g}\times\sM_{g}$
of all collections $(\{(x_i,y_i)\}_{i=1,\dots,g} ; \Bla)$
such that $f_{g}(x_i,y_i;\Bla)=0$ for all $i=1,\dots,g$ is called
the \emph{universal space of
$g$--th symmetric powers of trigonal curves} $V$.

The triple $(\sU_{g},
\sM_{g},p_{\sU})$, where $p_{\sU}$ is the canonical projection,
is called the \emph{universal bundle of
$g$--th symmetric powers of trigonal curves} $V$.
\end{df}

Through the whole paper we shall illustrate the results
obtained on the examples of the trigonal curves of lower
genera:

\begin{multline*}
f_{3}(x,y;\Bla)=
y^3-(\lambda_{11}x+\lambda_{8})y^2-(\lambda_{10}x^2+\lambda_{7}
x+\lambda_{4})y
-\\
(x^4+\lambda_{9}x^3+\lambda_{6}x^2+\lambda_{3}x+\lambda_{0}).
\end{multline*}

and
\begin{multline*}
f_{3}(x,y;\Bla)=
y^3-(\lambda_{13}x+\lambda_{10})y^2-(\lambda_{14}x^3+\lambda_{11}x^2+\lambda_{8}
x+\lambda_{5})y
-\\
(x^5+\lambda_{12}x^5+\lambda_{9}x^3+\lambda_{6}x^2+\lambda_{3}x+\lambda_{0}).
\end{multline*}

\begin{example}Let us agree to over-line the non-gap
entries to the Weierstrass gap sequence.
Than at the cases of genera 3 and 4 the correspondingly
the Weierstrass sequences read
\begin{align*}
\overline{0},\,1,\,2,\overline{3,\,4},\,5,\,\overline{6,\,7,\ldots}\\
\overline{0},\,1,\,2,\overline{3,},\,4,\,\overline{5},\,
\overline{6},\,7,\overline{8,\,9,\ldots}\\
\end{align*}
 \end{example}

\section{Matrix construction}\label{sec:L}
Denote by $\BW_{g}(x,y)$ the vector $(w_1,w_2,\dots,w_{g+3})^T$
formed by homogeneous elements $w_k=x^{\a} y^{\beta}$ in
ascending order of weights $\deg w_k=3\a+\beta(g+1)$ under the
condition $ 0\le \deg w_k\le  2g+2$. Let $H=\{h_{i,j}\},$
${i,j=1,\dots,g+3}$ be a symmetric matrix, whose entries are
homogeneous elements, $\deg h_{i,j}=4g+2-\deg w_i-\deg w_j$, of a
suitable graded ring, and, so, for $H$ holds the relation
$\deg(\BW_{g}^TH\BW_{g})=4g+2$.  Let us denote the space of
symmetric matrices thus graded by $\cH$.

For an arbitrary matrix $M$ put
$r(x,y;M)=\frac{\pa}{\pa y}(0,\dots,0,1) M \BW_{g}(x,y)$.
Note that if $H\in\cH$ then $\deg r(x,y;H)=g-1$.
\begin{df} Denote by
$\cT_g\subset\cH\times\sM_{g}$ the subspace of
pairs $H\in\cH$ and $\Bla\in\sM_{g}$ such that
the relation holds
\begin{equation}
\label{whw=2rf} \BW_{g}^TH\BW_{g}=2r(x,y;H) f_{g}(x,y;\Bla).
\end{equation}
Thus the bundle $(\cT_{g},
\sM_{g},p_{\cT})$ is defined, where  $p_{\cT}$ is the canonical
projection.  \end{df}

Taking into account that a trigonal polynomial includes only
the parameters $\{\lambda_k\}$ of positive weights, we find from the
definition that matrices from $\cT_g$ are of the form
\begin{equation}\label{blok} \begin{pmatrix} \hdotsfor{4}\\
\hdots&h_{g+1,g+1}&h_{g+1,g+2}&1\\ \hdots&h_{g+1,g+2}&-2&0\\
\hdots&1&0&0 \end{pmatrix},\text{ or } \begin{pmatrix}
\hdotsfor{4}\\ \hdots&-2h_{g,g+3}&0&1\\ \hdots&0&-2&0\\
\hdots&1&0&0 \end{pmatrix}, \end{equation} at $g=3\ell+1$, or
$g=3\ell$, respectively. Denote by  $p_1$ and $p_2$ projections of
the product $\cH\times\sM_{g}$ to the first and second factors.
\begin{lemma}\label{proj}
The composition of the embedding  $\cT_g
\subset\cH\times\sM_{g}$ with the projection $p_1$
is an embedding, and the  composition with the projection
$p_2$ coincides with $p_{\cT}$ and is ``onto''.
\end{lemma}
\begin{proof} 1. We have to show that if
for a given matrix $H$ there exists a trigonal
polynomial that satisfies \eqref{whw=2rf},  then this
polynomial is unique.
Indeed, otherwise we would come to a relation $r(x,y;H)(f(x,y;\Bla_1)
-f(x,y;\Bla_2))=0,\quad \forall (x,y)\in\C^2, $ which
cannot hold unless $\Bla_1=\Bla_2$.

2. We have to show that for a given $\Bla\in \sM_g$ there exists
at least one solution  $H$ of the equation \eqref{whw=2rf}. Let
$q$ be a homogeneous polynomial in $x$ and $y$ of weight $g-1$
with coefficients of non-negative weights. As $\deg y=g+1>g-1$,
such polynomial does not contain $y$.  Let $M(q,\Bla)$ be a
solution of the equation $\BW_{g}^TM\BW_{g}-2q(x)
f_{g}(x,y;\Bla)=0$ in the class of symmetric matrices with
entries of non-negative weights. Evidently, the set of the
solutions is not empty. By taking the coefficient at the highest
power of $y$  we find that $r(x,y;M(q,\Bla))=q(x)$.
\end{proof}

Let us consider the matrix subspace
$\cH_{4}=\{H\in \cH\mid \rank
H\le  4\}$.
\begin{lemma}\label{SYLV}
Suppose that the lower-right $k\times k$-submatrix $\Dt$ of a
symmetric $(N+k)\times(N+k)$-matrix $M=\{m_{i,j}\}$ is
nondegenerate.  Rank of  $M$  does not exceed
$k+1$ iff there exists a vector
$\Bz=(z_1,\dots,z_N)^T $ such that
\begin{equation}\label{sylm}
m_{i,j}=-\frac{1}{\det\Dt}\det \begin{pmatrix}
-\frac{1}{\det\Delta}z_i z_j &\Bmu_i
\\\\
 \Bmu_j^T& \Dt
\end{pmatrix} , \quad 1\le  i,j \le  N,
\end{equation}
where $\Bmu_r=(m_{r,N+1},m_{r,N+2},\dots,m_{r,N+k})$.
\end{lemma}
The above Lemma is a direct consequence of the Sylvester identity
for compound determinants \cite{ga67}.  Note that the vector
$\Bz$ in Lemma  \ref{SYLV} is defined up to the sign.  Denote by
$\Sym^{*}(k,\C )$ the set of nondegenerate symmetric  $k\times
k$-matrices and by  $\Sym_{r}(k,\C)$ the set of symmetric
$k\times k$-matrices of rank not greater than $r$ with
nondegenerate lower-right  $(r-1)\times(r-1)$-submatrix. We have
\begin{cor}\label{cover} The mapping
\[\phi:\C^{N}\times\C^{Nk}\times
\Sym^{*}(k,\C)\to\Sym_{k+1}(N+k,\C)
\]
defined by:
\begin{equation}\label{sylm1}
\phi(\Bz,(\Bnu_j)_{j=1,\dots,k},\Dt
)= \begin{pmatrix}
\left(m_{q,r}\right)_{q,r=1,\ldots,N}&
\left(\Bnu_1,\ldots, \Bnu_k\right)
\\
\left(\Bnu_1,\ldots, \Bnu_k\right)^T
&\Dt \end{pmatrix},
\end{equation}
where $m_{q,r}$ are calculated according to the
formula \eqref{sylm} with $\Bmu_r= (\nu_{r,j})_{j=1,\dots,k}$, is
a double covering ramified along the subvariety $\Bz=0$.
\end{cor} Let $\Dt=\Dt_{g}$ be equal to the corresponding block
from \eqref{blok} and fix the embedding $\C^{g}\times\C^{3g}\times
\C^{\dt}\to\C^{g}\times\C^{3g}\times \Sym^{*}(3,\C)$, where
${\dt}=2$ at $g=3\ell+1$ and  ${\dt}=0$ at $g=3\ell$.  We obtain
the parametrisation $\ph:\C^{g}\times\C^{3g}\times
\C^{\dt}\to\cH_{4}$.  The matrix $H=\ph(\Bz,(\Bga_1,\Bga_2,
\Bga_3),\Dt_{g})$ is calculated by \eqref{sylm1} with $k=3$ and
$N=g$.  Note that  the mapping $\ph$ agrees with the grading of
the matrix entries $h_{i,j}$ assumed above, if we accept the
following convention: $\deg z_i=2g+1-\deg w_i$ and $\deg
\ga_{i,r}=2g+3-r-\deg w_i$.

For a given collection
$(\Bz,(\Bga_1,\Bga_2,
\Bga_3),\Dt_{g})$ define four polynomials
$\chi(x,y)$, $\rho^{\pm}(x,y)$ and  $\tau(x,y)$ by the
formula:  \begin{equation} \label{prt}
(\chi(x,y),\rho^{\pm}(x,y),\tau(x,y))^T=
((\Bga_1,\Bga_2\pm\Bz,
\Bga_3)^T,\Dt_{g})\BW_{g}(x,y).
\end{equation}
By the construction the higher monomials of the polynomials
 $\chi(x,y)$, $\rho^{\pm}(x,y)$
and $\tau(x,y)$ have the weights $2g+2$,
$2g+1$ and $2g$ respectively.  With that the variable
$y$ enters the polynomials  $\tau(x,y)$ and
$\rho^{\pm}(x,y)$ linearly. From the definition of the mapping
$\ph$ and \eqref{prt} follows \begin{lemma}
For  $H=\ph(\Bz,( \Bga_1, \Bga_2,\Bga_3),\Dt_{g} )$ the
relation holds \begin{multline}\label{whw}
\BW^T_{g}H\BW_{g}=\\ -\frac{1}{2}\rho^{+}\rho^{-}+\tau(2\chi
+\frac{1}{2}h_{g+1,g+2}(\rho^{+}+\rho^{-})-(h_{g+1,g+1}+\frac{1}{2}
h^2_{g+1,g+2}
)\tau),
\end{multline}
and also  $r(x,y;H)=\frac{\pa}{\pa y}\tau(x,y)$  and  does not
depend on  $y$.  \end{lemma} Let $\BZ=(\Bz,( \Bga_1,
\Bga_2,\Bga_3),\Dt_{g} )$ be a point in
$\C^g\times\C^{3g}\times\C^{\dt}$. Let us rewrite the right hand
side of \eqref{whw} in the form \[ \BW^T_{g}\ph(\BZ)
\BW_{g}=Q(x,y;\BZ)\frac{\pa}{\pa y}\tau(x,y)+R(x,y;\BZ)\] so that
the degree of the polynomial  $R$ in $x$ is not greater than
$[\frac{g-1}{3}]-1$, and its degree in $y$ is not greater than
$2$.  Let us introduce the subspace
$\cL\subset\C^g\times\C^{3g}\times \C^{\dt}$ defined by
$3[\frac{g-1}{3}]$ conditions $R(x,y;\BZ)\equiv 0$. Then form
comparison with  \eqref{whw=2rf} we obtain the relation
\begin{equation}\label{Q-la} Q(x,y;\BZ)-2f_g(x,y,\Bla)=0
\end{equation}
that enables us to define the bundle $(\cL,\sM_{g},p_{\cL})$. The
projection $p_{\cL}$ maps a point   $\BZ\in\cL$ to the set of the
positive weight coefficients of the polynomial
$-\frac{1}{2}Q(x,y;\BZ)$ in ascending order of weights. Note that
the projection $p_{\cL}$ is a rational mapping.
\begin{lemma}\label{g-roots}
Let the polynomials $ \rho^{\pm}(x,y)$ and
$\tau(x,y)$ be coprime in the ring
$\C[x,y]$.  Then they have  $g$ common zeros
$\{(\xi_i^{\pm},\eta_i^{\pm})\},$ ${i=1,\dots,g}$.

Moreover,
$f_g(\xi_i^{\pm},
\eta_i^{\pm};p_{\cL}(\BZ))=0$ for all
$i=1,\dots,g$. \end{lemma}
\begin{proof} The polynomials
$ \rho^{\pm}(x,y)$ and  $\tau(x,y)$ are linear in $y$.  Their
coefficients at  $y^1$ are the polynomials in  $x$ of degrees
$[\frac{g}{3}]$ and $[\frac{g-1}{3}]$ respectively, and the
coefficients at  $y^0$ have degrees $[\frac{2g+1}{3}]$ and
$[\frac{2g}{3}]$.  By eliminating $y$, we obtain a polynomial in
$x$ of degree $\max([\frac{2g+1}{3}]+[\frac{g-1}{3}],
[\frac{2g}{3}]+[\frac{g}{3}] )$, which is equal to $g$ under the
condition $\gcd(g+1,3)=1$. By assumption the polynomial does not
vanish identically.

Next, it follows from \eqref{whw} that $
\BW^T_{g}(\xi_i^{\pm},\eta_i^{\pm})\ph(\BZ)
\BW_{g}(\xi_i^{\pm},\eta_i^{\pm})=0$ for all $i=1,\dots,g$.
Suppose that $f_g(\xi,
\eta;p_{\cL}(\BZ))\neq 0$ for some point
 $(\xi,\eta)$ form the set of common zeros.
Then, as $\frac{\pa}{\pa y}\tau(x,y)$ does not depend on $y$, we
conclude that  $ \BW^T_{g}(\xi,y)\ph(\BZ) \BW_{g}(\xi,y)=0$ for
arbitrary $y$. The last is only possible if the polynomials  $
\rho^{\pm}(x,y)$ and  $\tau(x,y)$ have a common factor
$(x-\xi)$.  \end{proof} Now we are ready to formulate the main
result of this section.
\begin{theorem}\label{aL->U} The mapping  $\a$ from the space
$\cL$ to the space $\sU_{g}$ of  $g$-th symmetric powers of
trigonal curves, which  puts  a
collection $\BZ=(\Bz,(\Bga_1, \Bga_2,
\Bga_3),\Dt_g)\in\cL$ into correspondence with
the vector  $\Bla=p_{\cL}(\BZ)$ and the set
of common zeros of the coprime polynomials
$\tau(x,y)$ and $\rho^{+}(x,y)$, defines
an identical on
the common base   $\sM_{g}$ fiberwise birational equivalence of
the bundles $\a:(\cL,\sM_{g},p_{\cL}) \to (\sU_{g},
\sM_{g},p_{\sU})$.
\end{theorem}
\begin{proof} Let us constuct the mapping
$\a^{-1}$.

Let a point
$(\{(\xi_i,\eta_i)\}_{i=1,\dots,g},\Bla)\in
\sU_{g}$ be given. According to \eqref{prt}
 to solve our problem it is sufficient to find polynomials
$\tau(x,y)$,
$\chi(x,y)$, $\rho^{+}(x,y)$ and $\rho^{-}(x,y)$.

Let $g=3\ell+1$. Denote by
$\BU(x,y)$ the vector formed by the first
$g$ components of the vector $\BW_{g}(x,y)$ and
put \begin{gather*} \Bt(x,y)=(\BU(x,y)^T,y x^{\ell})^T,\quad
\Br(x,y)=(\BU(x,y)^T,-\lambda_{3g+2}
y x^{\ell}
-2 x^{2\ell+1})^T\\
\text{and}\quad
\Bp(x,y)=(\BU(x,y)^T,y x^{\ell},
2y^2)^T.
\end{gather*}

Suppose that $\det(\BU(\xi_1,\eta_1),\dots,
\BU(\xi_g,\eta_g))\neq0$. Consider the entire rational
function \[
T(x,y)=\frac{\det(\Bt(x,y),
\Bt(\xi_1,\eta_1),\dots,\Bt(\xi_g,\eta_g))}
{\det(\BU(\xi_1,\eta_1),\dots,
\BU(\xi_g,\eta_g))}=y x^{\ell}+\dots
\]
on the curve $V$ defined by  numerical parameters $\Bla$.
The order of the function $T(x,y)$ is equal to
$\deg(y x^{\ell})=g+1+3\ell=2g$, therefore, beside
the zeros $\{(\xi_i,\eta_i)\}_{i=1,\dots,g}$  it has  $g$
more zeros $ \{(\wt{\xi}_i,\wt{\eta}_i)\}_{i=1,\dots,g}$ on the
curve $V$.  With the help of these two collections of zeros we
define a pair of entire rational functions on the
curve $V$ of order $2g+1$:  \begin{align*}
R_1(x,y)&=\frac{\det(\Br(x,y),
\Br(\xi_1,\eta_1),\dots,\Br(\xi_g,\eta_g))}
{\det(\BU(\xi_1,\eta_1),\dots,
\BU(\xi_g,\eta_g))}=-2x^{2\ell+1}-\lambda_{3g+2}y
x^{\ell}+\dots,\\ R_2(x,y)&=\frac{\det(\Br(x,y),
\Br(\wt{\xi}_1,\wt{\eta}_1),\dots,
\Br(\wt{\xi}_g,\wt{\eta}_g))}
{\det(\BU(\wt{\xi}_1,\wt{\eta}_1),
\dots,\BU(\wt{\xi}_g,\wt{\eta}_g))}
=-2x^{2\ell+1}-\lambda_{3g+2}y x^{\ell}+\dots.
\end{align*}
The function  $R_1(x,y)$ has
 $g+1$ complementary zeros
$\{(\varrho_i,\varsigma_i)\}_{i=1,\dots,g+1}$ on the curve, and
the function $R_2(x,y)$ has complementary zeros
$\{(\wt{\varrho}_i,\wt{\varsigma}_i)\}_{i=1,\dots,g+1}$.

The ratio $\frac{1}{2}(R_1(x,y)R_2(x,y))/{T(x,y)}$
represents an  \emph{entire} (all the zeros of the denominator
cancel out) rational function
$P(x,y)$ on the
curve $V$ of order  $2g+2$ and we know all of its zeros.
Let us study the behaviour of the function
$P(x,y)$ near the point $(\infty,\infty)$.
Let  $t$ be a local parameter, we have
$x=t^{-3}$ and $y=t^{-(g+1)}(1+\frac{1}{3}\lambda_{3g+2}t+O(t^2))$.
We obtain
$P(x,y)=t^{-(2g+2)}(2+\frac{4}{3}\lambda_{3g+2}t+O(t^2))$.  From the
above expressions one finds that
the function $P(x,y)$ allows a polynomial representation
of the form
$2y^2+P_{2g+1} x^{2\ell+1}+\dots$ with the coefficient
 $P_{2g+1}=0$, and so, in order to obtain an explicit expression
it is sufficient to prescribe any  $g+1$ of its zeros.
With the use of the collection
$\{(\varrho_i,\varsigma_i)\}_{i=1,\dots,g+1}$ we obtain
\[P(x,y)=\frac{\det(\Bp(x,y), \Bp(\varrho_1,\varsigma_1),
\dots,\Bp(\varrho_{g+1},\varsigma_{g+1}))}
{\det(\Bt(\varrho_{1},\varsigma_{1}),\dots,
\Bt(\varrho_{g+1},\varsigma_{g+1}))}.
\]

Consider the polynomial \[
C(x,y)=P(x,y)T(x,y)-\frac{1}{2}R_1(x,y)R_2(x,y)=
2x^{\ell}y^3-2x^{\ell}x^{g+1}+\dots
\] as a function on the curve $V$. This function has
$4g+2$ zeros.  Let us reduce the function
$C(x,y)$ to the normal form, i.e.  by adding multiples of
the trigonal polynomial
$f_g(x,y;\Bla)$ get rid of the terms that contain powers of $y$
greater than $2$.  The coefficient at  $y^3$ is equal  to
$2\frac{\pa}{\pa y}T(x,y)$, the higher powers of
 $y$ are absent, thus the difference
$C(x,y)-2f_g(x,y;\Bla)\frac{\pa}{\pa y}T(x,y)$ already has the
normal form. Note, however, that  because of the cancellation
of the higher powers of $x$ the order
of this difference is $<4g+2$.  So, the number of zeros of
the function
$C(x,y)-2f_g(x,y;\Bla)\frac{\pa}{\pa y}T(x,y)$ is greater
than its order, and, therefore, it vanishes identically.

By comparing the identity
\[
P(x,y)T(x,y)-\frac{1}{2}R_1(x,y)R_2(x,y)\equiv
2f_g(x,y;\Bla)\frac{\pa}{\pa
y}T(x,y)
\]
with the relation \eqref{whw} we obtain the following
expressions for the polynomials sought (up to permutation of the
pair $\rho^{+}$ and $\rho^{-}$):
\begin{gather*} \tau(x,y)=T(x,y),\quad\rho^{+}(x,y)=R_{1}(x,y),
\quad\rho^{-}(x,y)=R_{2}(x,y),\\
\chi(x,y)=\tfrac{1}{2}(P(x,y)-
\tfrac{h_{g+1,g+2}}{2}(R_1(x,y)+R_2(x,y))+
\tfrac{2 h_{g+1,g+1}+h^2_{g+1,g+2}}{2}T(x,y))
\end{gather*}
It remains to express the matrix entries
 $h_{g+1,g+1}$ and $h_{g+1,g+2}$
in terms of $\Bla$ and the coefficients of the
 polynomial $P(x,y)$.  It follows
from \eqref{prt} that \begin{align*}
\rho^{+}(x,y)&=-2x^{2\ell+1}+h_{g+1,g+2}y x^{\ell}+\dots, \\
\chi(x,y)&=y^2+h_{g+1,g+2}x^{2\ell+1}+h_{g+1,g+1}y x^{\ell}+\dots,
\end{align*}
whence we obtain $h_{g+1,g+2}=-\lambda_{3g+2}$ and
$h_{g+1,g+1}=\frac{1}{6}(2 P_{2g}-\lambda_{3g+2}^2)$, where $P_{2g}$
is the coefficient at  $y x^{\ell}$ of the polynomial $P(x,y)$.

The case  $g=3\ell$ is considered similarly.  \end{proof}

Below we are dealing with a number of
bundles over the same base  $\sM_{g}$
as in the above theorem.
When formulating assertions about
\emph{fiberwise mappings} of such bundles,
we imply only the mappings \emph{identical on the
base} without an explicit indication of this.

\section{Universal spaces of Jacobi and Kummer
varieties}\label{sec:K} Each collection $\Bla$ defining a
trigonal curve as the set of zeros of the polynomial \eqref{TRF}
can be unambiguously put into correspondence with the lattice
$\La= \La(\Bla)$ in $\C^g$ generated by periods of  $g$ basis
differentials
\begin{equation}\label{ADiff} \D
u_i(x,y;\Bla)=\frac{w_i(x,y)}{\frac{\pa}{\pa
y}f_{g}(x,y;\Bla)}\D x,\quad i=1,\dots,g,
\end{equation}
where $w_i(x,y)$, as above, stands for the $i$--th coordinate of
the vector $\BW(x,y)$.
\begin{df} The quotient space $\C^g\times\sM_g/\sim$
with respect to the equivalence
$(\Bu,\Bla)\sim (\Bu',\Bla')\Leftrightarrow(
\Bla=\Bla',\Bu- \Bu'\in\La(\Bla))$  is called
\emph{universal space $\sJ_g$ of Jacobi varieties of
trigonal curves}.

The quotient of the space $\sJ_g$ over the canonical involution
$(\Bu,\Bla)\to (-\Bu,\Bla)$ is called
\emph{
universal space $\sK_g$  of Kummer varieties of
trigonal curves}.
\end{df}

\begin{df} We call the triple $(\sJ_g,\sM_g,p_{\sJ})$, where
$p_{\sJ}(\Bu,\Bla)=\Bla$, the
\emph{universal bundle of Jacobi varieties of
trigonal curves}.

We call the triple
$(\sK_g,\sM_g,p_{\sK})$, where
$p_{\sK}(\Bu,\Bla)=\Bla$, the
\emph{universal bundle of Kummer varieties of
trigonal curves}.
\end{df}

The fibers of these bundles over a point
$\Bla\in\sM_g$ are called the  \emph{Jacobi variety} $\Jac(V)$ and
the \emph{Kummer variety} $\Kum(V)$
of a trigonal curve $V(x,y)$ in the model defined by the zeros of
the polynomial $f_{g}(x,y;\Bla)$.

With the help of the differentials
\eqref{ADiff} we define the Abel map
$A:\sU_g\to\C^{g}\times\sM_{g}$, $
A(\{(x_1,y_1),\dots,(x_g,y_g)\};\Bla)=
(\Bu,\Bla)$,
where $\Bu=(u_1,\dots,u_g)$ and
$u_i=\sum_{k=1}^{g}\int_{(\infty,\infty)}^{(x_k,y_k)}
\D u_{i}(x,y;{\Bla})$. This map is multivalued, as a set
$(\{(x_1,y_1),\dots,(x_g,y_g)\}\in (V)^{g}$
corresponds to a countable set of values $\Bu_1,\Bu_2,\dots$ in
$\C^g$ depending on
a choice of the integration paths.
It is known, however, that the pairwise differences
$\Bu_i-\Bu_j$ of these values lie in $\La(\Bla)$.  So, this map
defines  a fiberwise birational equivalence of bundles
$A:(\sU_{g}, \sM_{g}, p_{\sU})\to (\sJ_g,\sM_{g},p_{\sJ})$.

It follows directly from the construction of the bundle $(\cL,
\sM_{g},p_{\cL})$ that the map $(\Bz,(\Bga_1,\Bga_2,
\Bga_3),\Dt_{g})\mapsto (-\Bz,(\Bga_1,\Bga_2, \Bga_3),\Dt_{g})$
induces a fiberwise involution on it. The quotient over this
involution, which we call canonical, defines the bundle $(\cK,
\sM_{g},p_{\cK})$, where $\cK=\cT_{g}\cap(\cH_{4}\times\sM_g)$.
According to Lemma \ref{proj} and Corollary \ref{cover} the
projection $p_1$ allows to identify the space  $\cK$ with the
matrix subspace $p_1(\cT_{g})\cap\cH_{4}$.
\begin{theorem}[on parametrisation]\label{aL->J} There is a
fiberwise birational equivalence of bundles $\cA:(\cL,
\sM_{g},p_{\cL})\to (\sJ_g,\sM_g,p_{\sJ})$, which
commutes with the canonical involutions and induces a
 fiberwise birational equivalence of bundles $\cB:(\cK,
\sM_{g},p_{\cK})\to (\sK_g,\sM_g,p_{\sK})$.
\end{theorem}
\begin{proof} Consider the composition
$\cA=A\circ\a:\cL\to\sJ_g$.  First, notice that
by construction
\[ p_{\cL}(-\Bz,( \Bga_1, \Bga_2,\Bga_3),\Dt_{g}
)=p_{\cL}(\Bz,( \Bga_1, \Bga_2,\Bga_3),\Dt_{g} ).\]
In order to show that the mapping $\cA$ commutes with the
involutions, it is sufficient to verify that
\[\cA(-\Bz,( \Bga_1, \Bga_2,\Bga_3),\Dt_{g} )+\cA(\Bz,(
\Bga_1, \Bga_2,\Bga_3),\Dt_{g} )\sim(\boldsymbol{0},\Bla).  \]
Indeed, in the notation of Lemma
\ref{g-roots}, the inclusion \begin{equation}\label{Abel}
\sum_{k=1}^{g}
{\textstyle\int\limits_{(\infty,\infty)}^{(\xi_k^{+},\eta_k^{+})}}
\negthickspace
\D \Bu(x,y;p_{\cL}({\BZ})) +
\sum_{k=1}^{g}
{\textstyle\int\limits_{(\infty,\infty)}^{(\xi_k^{-},\eta_k^{-})}}
\negthickspace
\D \Bu(x,y;p_{\cL}({\BZ}))
\in \La(p_{\cL}(\BZ))
\end{equation}
follows from the classical Abel Theorem in view of the fact that
the set of common zeros of the pair of polynomials $\{\tau(x,y),
f_g(x,y;p_{\cL}({\BZ}))\}$ coincides with the set of common zeros
of the pair $\{\tau(x,y),\rho^{+}(x,y)\rho^{-}(x,y)\}$.

Second, notice that the mapping  $\cA$
takes the ramification set of the covering  $\cL\to\cK$, on
which all the zeros of the pair
$\{\tau(x,y), f_g(x,y;p_{\cL}({\BZ}))\}$ are double, into
the ramification set of the covering $\sJ_g\to\sK$, i.e.  into the
subspace of half-periods
$\{(\BOm,p_{\cL}({\BZ}) )\mid 2\BOm\in \La(p_{\cL}(\BZ))\}$, and,
therefore, the induced mapping  $\cB$ is well defined.  \end{proof}
\begin{remark} Under a proper interpretation of the periods
of Abelian integrals, i.e. of the lattice $\La(p_{\cL}(\BZ))$,
 the relation \eqref{Abel} holds also for the case
of a degenerate polynomial $f_g(x,y;p_{\cL}({\BZ}))$ (see
the discussion of Abel Theorem in
\cite{ba97}).  The map $A$, which is holomorphic in a
nondegenerate case, becomes meromorphic, and its inversion
reduces to the class of so-called generalized Jacobi inversion
problems.  A solution by means of the theory of
$\theta$--functions for a particular formulation is given in
\cite{fa73}.  \end{remark}

Fibers of the universal bundles of Jacobi and Kummer varieties
are birationaly equivalent to algebraic subvarieties.
Let us formulate our result using the notations introduced
in \S \ref{sec:L}.
\begin{cor}\label{kumm} For a fixed point $\Bla\in\sM_g$ we have:

\textup{(1)} realization of the variety  $\Jac(V)$ as an
algebraic subvariety in  $\C^{4g+{\dt}}$, where
$\dt=2(g-3[\frac{g}{3}])$.  This subvariety is defined as the set
of common zeros of the system of $2g+3+3[\frac{g-1}{3}]=3g+\dt$
polynomials that are generated by the functions
\[
\begin{split}
G_1(t)&=\ga_{g,3}^{(2-\dt)/2}(Q(t^3,t^{g+1};\BZ)-2
f_g(t^3,t^{g+1};\Bla)),\\
 G_2(t)&=R(t^3,t^{g+1};\BZ).\end{split}
\]

\textup{(2)} realization of the variety $\Kum(V)$ as an algebraic
subvariety in $\C^{g(g+1)/2}$. This subvariety is defined as the
set of common zeros of a system of $\frac{(g-1)g}{2}$ polynomials
of degree not greater than $5$. The polynomials are the
$5\times5$--minors of the $(g+3)\times(g+3)$--matrix $K$, which
is defined as an ansatz of the general solution of the equation
$\BW_{g}^TK\BW_{g}=2r(x,y; K) f_g(x,y;\Bla)$ in the matrix space
$\cH$.
\end{cor}
\begin{proof} 1.
According to the definition of the space $\cL$ the image of the
mapping  $\cA^{-1}$  in $\C^{4g+\dt}$ is defined by the
condition  $R(x,y;\BZ)=0$ for all $(x,y)\in\C^2$, whence we
obtain the generating function $G_2(t,\BZ)$. At a given value of
$\Bla$ the relation \eqref{Q-la} generates the equations of
restriction on a fiber in the bundle $(\cL,\sM_g,p_{\cL})$.

2. An arbitrary matrix $K\in\cT_g$ has $\frac{g(g+1)}{2}+3g+\dt$
parameters by definition. At a given value of  $\Bla\in\C^{2g+3}$
the relation \eqref{whw=2rf} imposes $2g+3+3[\frac{g-1}{3}]$
linear conditions on the parameters (in fact, from the formal
view-point this relation defines an entire rational function of
order $3g+3+3[\frac{g-1}{3}]$ in normal form which vanishes
identically on the curve) that single out a fiber
$p_{\cT}^{-1}(\Bla) \subset\cT_g$. So, a general solution of the
equation $\BW_{g}^T K\BW_{g}=2r(x,y; K) f_g(x,y;\Bla)$ in the
matrix space $\cH$, i.e. a point $K\in p_{\cT}^{-1}(\Bla)$, has
$\frac{g(g+1)}{2}$ parameters. It follows from the identification
of $\cK$ with the intersection $ p_1(\cT_{g})\cap\cH_{4}$
mentioned above and Lemma \ref{proj} that mapping $\cB^{-1}$
takes a fiber of the bundle $(\sK_g,\sM_g,p_{\sK})$ to the matrix
space $\{K\in p_{\cT}^{-1}(\Bla)\mid \rank K\le 4\}$.
\end{proof}
\begin{example}\label{ex34} Let  $g=3$. Then
\[\BW_{3}(x,y)=(1,x,y,x^2,y
 x,y^2)^T\quad\text{and}\quad R(x,y;
 \BZ)=0.\] The trigonal curve $V(x,y)$ is defined by polynomial
 \begin{multline*}
f_{3}(x,y;\Bla)=
y^3-(\lambda_{11}x+\lambda_{8})y^2-(\lambda_{10}x^2+\lambda_{7}
x+\lambda_{4})y
-\\
(x^4+\lambda_{9}x^3+\lambda_{6}x^2+\lambda_{3}x+\lambda_{0}).
\end{multline*}
The polynomials $\chi,\rho^{\pm},\tau$ reads
\begin{eqnarray*}
\chi&=&y^2-2\ga_{3,3}x^2+\ga_{3,1}y+\ga_{2,1}x+\ga_{1,1}\\
\rho^{\pm}&=&-2xy+\ga_{1,2}\pm z_1+(\ga_{2,2}\pm z_2)x
+(\ga_{3,2}\pm z_3)y\\
\tau&=&x^2+\ga_{1,3}+\ga_{2,3}x+\ga_{3,3}y
\end{eqnarray*}

In the space $\C^{12}$ with coordinates
$(z_i,(\ga_{i,1},\ga_{i,2},\ga_{i,3})),$ $i=1,2,3$ the Jacobi
variety $\Jac(V)$ is singled out by the system of $9$ equations:

\begin{align}\begin{split}
&4\ga_{3,3}\lambda_{0}+z_1^2+ 4\ga_{1,3}^2\ga_{3,3} +
4\ga_{1,3}\ga_{1,1} - \ga_{1,2}^2=0;\\
&2\ga_{3,3}\lambda_{3}+
z_1 z_2 + 4\ga_{1,3}\ga_{2,3}
\ga_{3,3}+ 2\ga_{2,3}\ga_{1,1}+
2\ga_{1,3}\ga_{2,1}- \ga_{1,2}\ga_{2,2}=0;\\
&2\ga_{3,3}\lambda_{4}+
z_1z_3 + 4\ga_{1,3}\ga_{3,3}^2 +
2\ga_{3,3}\ga_{1,1}+2\ga_{1,3}\ga_{3,1}-
\ga_{1,2}\ga_{3,2}=0,\\
&4\ga_{3,3}\lambda_{6}+
z_2^2+4\ga_{2,3}^2\ga_{3,3}+4\ga_{1,1}+4\ga_{2,3}
\ga_{2,1}-\ga_{2,2}^2=0;\\
&2\ga_{3,3}\lambda_{7}+
z_2z_3+ 4\ga_{2,3}\ga_{3,3}^2+
2\ga_{3,3}\ga_{2,1}+
2\ga_{2,3}\ga_{3,1}+2\ga_{1,2}
-\ga_{2,2}\ga_{3,2}=0;\\
&4\ga_{3,3}\lambda_{8}+
z_3^2+4\ga_{1,3}+
4\ga_{3,3}^3+4\ga_{3,3}\ga_{3,1}-\ga_{3,2}^2=0;\\
&\ga_{3,3}\lambda_{9}+\ga_{2,1}=0;\\
&\ga_{3,3}\lambda_{10}+\ga_{2,2}+\ga_{3,1}=0;\\
&\ga_{3,3}\lambda_{11}+\ga_{2,3}+\ga_{3,2}=0.\end{split}\label{9eq}
\end{align}

In order to describe the Kummer variety
$\Kum(V)$ in the space $\C^{6}$ with coordinates
$(v_i),\,i=2,3,4,6,7,8$,
$$v_2=\ga_{3,3},\;v_3=\ga_{2,3},\;v_4=\ga_{2,2},\;v_6=\ga_{1,3},\;
v_7=\ga_{1,2},\;v_8=\ga_{1,1}.
$$
(the subscript of a coordinate is
equal to its weight in the grading assumed)
we use the following ansatz
of the general solution $K$ of the equation $\BW_{3}^T
K\BW_{3}=2r(x,y; K) f_3(x,y;\Bla)$ at the given value of the
vector $\Bla$ in the class of the
symmetric matrices $\cH$:
\begin{equation} K= \begin{pmatrix}
-2\lambda_{0}v_2&-\lambda_{3}v_2&-\lambda_{4}v_2&v_8&v_7&v_6\\
-\lambda_{3}v_2&-2v_8-2\lambda_{6}v_2&-v_7-\lambda_{7}v_2&
-\lambda_{9}v_2&v_4&v_3\\
-\lambda_{4}v_2&-v_7-\lambda_{7}v_2&-2v_6-2\lambda_{8}v_2&-v_4
-\lambda_{10}v_2&-v_3-\lambda_{11}v_2&v_2\\
v_8&-\lambda_{9}v_2&-v_4-\lambda_{10}v_2&-2v_2&0&1\\
v_7& v_4&-v_3-\lambda_{11}v_2&0&-2&0\\
v_6&v_3&v_2&1&0&0
\end{pmatrix}.\label{k34}
\end{equation}
The coordinates $v_i$ are linked with $\ga_{i,j}$ as
$$v_2=\ga_{3,3},v_3=\ga_{2,3},v_4=\ga_{2,2},v_6=\ga_{1,3},
v_7=\ga_{1,2},v_8=\ga_{1,1}
$$

The equations that single out $\Kum(V)$
are found from the condition $\rank K\le  4$,
what reduces to three equations, on vanishing of the cofactors of
the entries  $K_{1,1}, K_{1,2}$ and $K_{2,2}$.

Remark, that the the first 6 equations from the 9, which single
out Jacobi variety can be written in compact form as
\begin{equation} \label{torso34}
\frac12z_{i}z_{j}+\det K \big[\begin{smallmatrix}
i&4  &5  &6  \\ j&4  &5  &6   \end{smallmatrix}\big]=0,\quad
i,j\in\{1,\dotsc,3\}; \end{equation} where  $ K
\big[\begin{smallmatrix} i&\dots&k\\ j&\dots&l
\end{smallmatrix}\big]$ stands for the submatrix of the
matrix $K$ formed by the entries located at the intersections of
the rows $(i,\dots,k)$ and columns $(j,\dots,l)$.
\end{example}

The last observation of the example \ref{ex34} is valid in
general. Let us apply the formula \eqref{sylm} to the matrix $K$
from assertion (2) of Corollary \ref{kumm}.  We find that there
exist a vector $\Bq=(q_1, \dots,q_g)^T$ such that:
\begin{equation} \label{torso}\frac{1}{\Delta} q_{i}q_{j}+\det K
\big[\begin{smallmatrix} i&g+1&g+2&g+3\\ j&g+1&g+2&g+3
\end{smallmatrix}\big]=0,\quad i,j\in\{1,\dotsc,g\};
\end{equation} where  $ K \big[\begin{smallmatrix} i&\dots&k\\
j&\dots&l \end{smallmatrix}\big]$ stands for the submatrix of the
matrix $K$ formed by the entries located at the intersections of
the rows $(i,\dots,k)$ and columns $(j,\dots,l)$.
Evidently, one  finds such a set  $\BZ=(\Bz,(
\Bga_1, \Bga_2,\Bga_3),\Dt_{g} )$ from the assertion (1)
that $\Bq=\Bz$ satisfies these equations.
On the other hand, we have
\begin{cor}
The Jacobi variety of a trigonal curve allows a realization
in the space $\C^{g(g+3)/2}$
as  the set of common zeros of the system of
$\frac{g(g+1)}{2}$
polynomials of the form \eqref{torso} of degree not greater than
$4$.
\end{cor}

\begin{example} \label{ex35} Let  $g=4$. Then
\[\BW_{3}(x,y)=(1,x,y,x^2,xy,x^3,y^2)^T\quad\text{and}
\quad R(x,y;
 \BZ)=0.\] The trigonal curve $V(x,y)$ is defined by polynomial
\begin{multline*}
f_{3}(x,y;\Bla)=
y^3-(\lambda_{13}x+\lambda_{10})y^2-(\lambda_{14}x^3+\lambda_{11}x^2+\lambda_{8}
x+\lambda_{5})y
-\\
(x^5+\lambda_{12}x^5+\lambda_{9}x^3+\lambda_{6}x^2+\lambda_{3}x+\lambda_{0}).
\end{multline*}
The polynomials $\chi,\rho^{\pm},\tau$ reads
\begin{eqnarray*}
\chi&=&y^2-\lambda_{14}x^3+h_{5,5}xy+\ga_{4,1}x^2+\ga_{3,1}y+\ga_{2,1}x+\ga_{1,1}\\
\rho^{\pm}&=&-2x^3-\lambda_{14}xy
+\ga_{1,2}\pm z_1+(\ga_{2,2}\pm z_2)x\\
&+&(\ga_{3,2}\pm z_3)y+(\ga_{4,2}\pm z_4)x^2\\
\tau&=&xy+\ga_{1,3}+\ga_{2,3}x+\ga_{3,3}y+\ga_{4,3}x^2
\end{eqnarray*}

The matrix $K$ is $7\times7$ matrix
\begin{equation} K= \begin{pmatrix}
-2\lambda_0v_3&-\lambda_3v_3-\lambda_0&-2\lambda_5v_3&v_{12}&v_{10}&v_9&v_8\\
-\lambda_3v_3-\lambda_0&k_{2,2}&k_{2,3}&k_{2,4}&v_{7 }&v_6&v_5\\
-2\lambda_5v_3&k_{2,3}&k_{3,3}&k_{3,4}&k_{3,5}&v_4&v_3\\
v_{12 }&k_{2,4}&k_{3,4}&k_{4,4}&k_{4,5}&k_{4,6}&v_2\\
v_{10 }&v_7&k_{3,5}&k_{4,5}&-2v_2-2\lambda_{13}&-\lambda_{14}&1\\
v_{9  }&v_6&v_4    &k_{4,6}&-\lambda_{14}&-2     &0\\
v_{8  }&v_5    &v_3    &v_2   &1     &0  &0  \\
\end{pmatrix},\label{k35}
\end{equation}
where
\begin{eqnarray*}
k_{2,2}&=&-2v_{12}-2\lambda_6v_3-2\lambda_3,\quad
k_{2,3}=-v_{10}-\lambda_8v_3-\lambda_5,\quad k_{2,4}=-v_3-\lambda_9v_3-\lambda_6\\
k_{3,3}&=&-2v_8-2\lambda_{10}v_3,\quad
k_{3,4}=-v_7-\lambda_{11}v_3-\lambda_8,\quad
k_{3,5}=-v_5-\lambda_{13}-\lambda_{10}\\
k_{4,4}&=&=-2v_6-2\lambda_{12}v_3-2\lambda_{11},\quad
k_{4,5}=-v_4-\lambda_{14}v_3-\lambda_{11},\quad k_{4,6}=-v_3-\lambda_{12}.
\end{eqnarray*}

The equations that single out $\Kum(V)$
are found from the condition $\rank K\le  4$,
what reduces to three equations on vanishing of the cofactors of
the entries  $K_{1,1}, K_{1,2}$ $K_{1,3}$and
$K_{2,2}$, $K_{2,3}$, $K_{3,3}$.

In the space $\C^{16}$ with coordinates
$(z_i,(\ga_{i,1},\ga_{i,2},\ga_{i,3})),$ $i=1,\ldots,4$ the Jacobi
variety $\Jac(V)$ is singled out by the system of $12$ equations:
\begin{eqnarray}
&&\ga_{3,1}+\lambda_{13}\ga_{3,3}+\ga_{2,3}+\lambda_{10}=0;\\
&&\ga_{3,3}\lambda_{14}+\ga_{3,2}+\ga_{4,1}+\lambda_{11}=0
\end{eqnarray}
and
\begin{equation} \label{torso35}
\frac12z_{i}z_{j}+\det K \big[\begin{smallmatrix}
i&5  &6  &7  \\ j&5  &6  &7   \end{smallmatrix}\big]=0,\quad
i,j\in\{1,\dotsc,4\}; \end{equation} where  $ K$  is given in
(\ref{k35}).

\end{example}

\begin{remark} One can show that the system of
$g+\dt-3$ polynomial equations, which follow from the generating
function $G_2(t)$, has a  rational solution \footnote{This
follows directly, e.g., from the rationality of the universal
spaces of the symmetric powers of plane algebraic curves. }.  By using this solution it is possible
starting with Corollary \ref{kumm} to come to a realization of
Jacobi variety  as an algebraic subvariety in the space
$\C^{3g+3}$. However, in view of the fact that  degrees of the
polynomials delivered by the generating function $G_1(t)$ do not
exceed  $4$ while degrees of the polynomials delivered by the
generating function $G_2(t)$ grow linearly in $g$, the degrees of
the equations in such realization would grow polynomially in  $g$.
\end{remark}

\section{Uniformization of Jacobi varieties}\label{sec:J}
The models of spaces
$\sJ_{g}$ constructed above allow explicit uniformization
by means of the theory of Abelian functions.  For the sake of
simplicity we restrict ourselves with consideration of trigonal
polynomials of the form \eqref{TRF} with the zero coefficient at
$y^2$. This does not lead to a loss of generality as
one can pass to such a form from a general polynomial with the
help  of the variables change $(x,y)\mapsto
(x,y+\frac{1}{3}(\sum_{j=0}^{[\frac{g+1}{3}]}\lambda_{2g+2+3
j}x^j))$ and further renotation of parameters.

In the first chapter of the work  by authors
\cite{bel97b} a  modification of the standard
$\theta$--function is constructed that defines a mapping
 $\si:  \C^g\times\sM_g\to\C$ being
\emph{invariant} w.r.t. modular transformations, i.e.  its
values, in the contrast with the values of $\theta$--function,
does not depend on the choice of the generators of the lattice
$\La(\Bla)$.  To denote such functions  Klein \cite{kl88}
introduced the symbol $\si$, thus stressing the analogy with the
Weierstrass elliptic  $\si$--function.  As  shown in
\cite{bel97b}, the partial derivatives of order $>1$ of
the logarithm of $\si$--function, are \emph{Abelian} and,
in the same time, \emph{modular invariant} functions, i.e. the
meromorphic functions, the domain of definition of which is the
space $\sJ_g$. The following notation is adopted for the
logarithmic derivatives of $\si$--function:  \[
\wp_{i_1,i_2,\dots,i_k}(\Bu,\Bla)= -\frac{\pa^k
\log\si(\Bu,\Bla)}{\pa u_{i_1}\pa u_{i_2}\cdots\pa u_{i_k}},\quad
k\ge 2.
\]
Note the parity property of  $\wp$--functions:
$
\wp_{i_1,\dots,i_k}(-\Bu,\Bla)=(-1)^k\wp_{i_1,\dots,i_k}(\Bu,\Bla)$.

Canonical differentials of the second kind for a nonsingular
trigonal curve of genus $g$, such that $\gcd(g+1,3)=0$,
\begin{gather*}
V(x,y)=\Big(f(x,y)
\Big)\subset \mathbb{C}^2,\\f(x,y)=y^3-y\sum_{\ell=0}^{\left[\frac{2g+2}{3}\right]}\lambda_{g+1+3
\ell}x^\ell -\sum_{\ell=0}^{g+1}\lambda_{3\ell}
x^\ell=y^3-q(x)y-p(x) ; \end{gather*} are given by the formula:
\begin{equation*}
\mathrm{d}r_{i,j}=\frac{R_{i,j}(x,y)\mathrm{d}x}{f_{y}(x,y)},\quad
j=0,1 \text{ and } i=0,\dots,\left[\frac{2g-1-j(g+1)}{3}\right];
\end{equation*}
with
\begin{multline*}
R_{i,0}(x,y)=
-y^2 \partial_x \mathrm{D}_{x}^{i+1}\big(q(x)\big)+\\
yx^{[\frac{i}{2}]}\mathrm{D}_{x}^{[\frac{3i}{2}]+2}\big(
2x\partial_x p(x)-3(i+1)p(x)\big)+
x^i\mathrm{D}_{x}^{2i+2}\big(\tfrac{1}{2}x\partial_x
q(x)^2-(i+1)q(x)^2\big)
\end{multline*}
and
\begin{multline*}
R_{i,1}(x,y)=\\yx^{i+1}\mathrm{D}_{x}^{2i+3}\big(
x\partial_x q(x)-2(i+1)q(x)\big)+
x^{2i+2}\mathrm{D}_{x}^{3i+4}\big(x\partial_x
p(x)-3(i+1)p(x)\big),
\end{multline*}
where $\mathrm{D}_{t}$ is the umbral derivative w.r.t. the
variable $t$ and $\partial_x=\frac{\partial}{\partial x}$.

The set of the second kind differentials thus defined  is
associated (dual) to the set of canonical (monomial) holomorphic
differentials
$\{\mathrm{d}u_{i,j}=x^iy^j\mathrm{d}x/f_{y}(x,y)\}$.

For the Klein-Weierstrass 2-differential we obtain
\begin{equation*}
\mathrm{d}\omega(x,z)=\frac{\mathcal{F}(x,y;z,w)}{f_y(x,y)f_w(z,w)(x-z)^2}
 \mathrm{d}x\mathrm{d}z,
\end{equation*}
with
\begin{align*}
\mathcal{F}(x,y;z,w)=&(wy+Q(x,z))(w y+Q(z,x)) +\\
  &w(w\mathrm{D}_{y}f(x,y) + T(x,z))+
  y(y\mathrm{D}_{w}f(z,w)+ T(z,x)),
\end{align*}
where
\begin{equation*}
Q(x,z)=\sum_{k=0}^{\left[\frac{2g+2}{3}\right]}
\lambda_{1+g+3k}x^{k-\left[\frac{k}{2}\right]}z^{\left[\frac{k}{2}\right]},
\end{equation*}
and
\begin{equation*}
T(x,z)=\sum_{k=0}^{g+1}3\lambda_{3k}\big(
(1-\{\tfrac{k}{3}\})x^{k-[\frac{k}{3}]}z^{[\frac{k}{3}]}+
\{\tfrac{k}{3}\}x^{k-[\frac{k}{3}]-1}z^{[\frac{k}{3}]+1}\big).
\end{equation*}

\begin{example} For the curve $V_{3,4}$ of genus $g=3$ the set of
second kind differentials
reads
\begin{align*}
\mathrm{d}r_3&=\frac{x^2\mathrm{d}x}{f_y},
\quad \mathrm{d}r_2=\frac{2xy\mathrm{d}x}{f_y}\\
\mathrm{d}r_1&=\left(-\lambda_{10}y^2+y(\lambda_6+3\lambda_9x+5x^2)
+\lambda_{10}x(\lambda_{10}x+\lambda_7)\right)\frac{\mathrm{d}x}{f_y}
\end{align*}
while the polynomial $Q(x,z)$ and $T(x,z)$ are given as
\begin{align*}
Q(x,z)&=\lambda_{10}xz+\lambda_7z+\lambda_4\\
T(x,z)&=3\lambda_0+\lambda_3(2x+z)+\lambda_6x(x+2z)\\
&+3\lambda_9x^2z+x^2z(2x+z).
\end{align*}
\end{example}

\begin{example} For the curve $V_{3,5}$ of genus $g=4$
the set of  second kind differentials reads
\begin{align*}
\mathrm{d}r_4&=\frac{yx\mathrm{d}x}{f_y},
\quad \mathrm{d}r_3=(x^2(\lambda_{12}+2x)+\lambda_{14}xy)
\frac{\mathrm{d}x}{f_y}\\
\mathrm{d}r_2&=
\left(-\lambda_{14}y^2+2xy(2x+\lambda_{12})+
       \lambda_{14}x^2(\lambda_{14}x+\lambda_{11})
\right)
\frac{\mathrm{d}x}{f_y}\\
\mathrm{d}r_1&=
\left(  -y^2(2\lambda_{14}x+\lambda_{11})
+2\lambda_{14}^2x^4+3\lambda_{11}\lambda_{14}x^3\right.\\
&\left.+ y(7x^3+\lambda_{12}x^2+3\lambda_9+\lambda_6 ) +
(2\lambda_8\lambda_{14}+\lambda_{11})x^2\right.\\
&\left.+(\lambda_{5}\lambda_{14}+\lambda_8\lambda_{11})x
\right)
\frac{\mathrm{d}x}{f_y}
\end{align*}
while the polynomial $Q(x,z)$ and $T(x,z)$ are given as
\begin{align*}
Q(x,z)&=\lambda_{14}x^2z+\lambda_{11}xz
+\lambda_8x+\lambda_5\\
T(x,z)&=3\lambda_0+\lambda_3(2x+z)
+\lambda_6x(x+2z)\\
&+3\lambda_9x^2z+\lambda_{12}x^2z(2x+z)+x^3z(x+2z).
\end{align*}
\end{example}

The results given below are based on the following
relation, an analog of which  is valid for
$\theta$--functions also \cite{fa73}.  However,
the important for us explicit dependence of the
right-hand side of this relation on the parameters of the model of
a curve can be obtained only for  modular invariant left-hand
side.  \begin{equation}\label{KF}
\sum_{i,j=1}^{g} \wp_{i,j}\Big(\Bv-\mspace{-12mu}
\textstyle{\int\limits_{(\infty,\infty)}^{(\xi,\eta)}}
\mspace{-12mu}\D \Bu,
\Bla\Big)\D
u_i(\xi,\eta;\Bla)
\D u_j(x,y;\Bla)=
\dfrac{\cF((\xi,\eta),(x,y);
\Bla)\D \xi\D x}
{(\xi-x)^2\tfrac{\pa}{\pa
y}f_g(
x,y;\Bla)\tfrac{\pa}{\pa
\eta}f_g(\xi,\eta;\Bla)},
\end{equation}
where $(x,y)\in\{\{(x_1,y_1),\dots,(x_g,y_g)\}\mid
A(\{(x_1,y_1),\dots,(x_g,y_g)\};\Bla)\sim(\Bv,\Bla)\}$, and
$(\xi,\eta)$ is an arbitrary point on the curve $V$.
In the considered case of trigonal curves for $\cF((\xi,\eta),
(x,y); \Bla)$  we have the expression
\begin{multline*} \cF((\xi,\eta), (x,y); \Bla)=3 y^2
\eta^2-y^2P_1(\xi)-\eta^2P_1(x)+ y\eta \Psi_2(P_1;x,\xi) +\\ \eta
\Psi_3(P_0;x,\xi)+ y
\Psi_3(P_0;\xi,x)+\tfrac{1}{2}\Psi_2(P_1^2;x,\xi), \end{multline*}
where $$\Psi_r(q;x,\xi)=\sum_{k\ge  0} q_k x^{k-[\frac{k}{r}]-1}\xi
^{[\frac{k}{r}]}(( r[\tfrac{k}{r}]+r-k)x+(k-r[\tfrac{k}{r}])\xi)$$
for an arbitrary polynomial $q(t)=\sum_{k\ge 0}q_k t^k$, and
polynomials $$P_1(x)=\sum_{j=0}^{[\frac{2g+2}{3}]}\lambda_{g+1+3
j}x^j \quad \text{and}\quad P_0(x)=x^{g+1}+ \sum_{j=0}^{g}\lambda_{3
j}x^j$$ are the coefficients at the powers of $y$ of the trigonal
polynomial.  Take a note that $\Psi_r(q;x,x)=r q(x)$ and, so
\[\cF((x,y),(x,y); \Bla)=(\tfrac{\pa}{\pa
y}f_g(x,y;\Bla))^2-
f_g(x,y;\Bla)\tfrac{\pa^2}{\pa
y^2}f_g(x,y;\Bla).
\]
Put $\Bwp_k^T=(
\wp_{i,k}(\Bv,\Bla)),
i=1,\dots,g$, and denote $\frac{\pa }{\pa
v_g}(\dots)=(\dots)'$.
\begin{theorem}\label{pi:J->L}
Define the mapping
$\pi:\sJ_{g}\to\C^{4g+{\dt}}$
by formulas
\begin{align} \label{zgg}
&\Bz=\Bwp_g',\quad
\Bga_3=-\Bwp_g,\quad
\Bga_2=\Bwp_{g-1},\\
\label{gamma_1}
&\Bga_1=\begin{cases}
\tfrac{1}{3}(\Bwp_g''- (6
\wp_{g,g}(\Bv,\Bla)+\lambda_{3g+1})
\Bwp_g+\BL_g),& g=3\ell\\
\tfrac{1}{3}(\Bwp_g''- (6
\wp_{g,g}(\Bv,\Bla)+
\lambda_{3g+2}^2)
\Bwp_g +\lambda_{3g+2}\Bwp_{g-1}
+\BL_g),&g=3\ell+1\end{cases},
\end{align}
where
$\BL_g^T=(\lambda_{g+1+\mathrm{deg}w_i}
\dt_{\mathrm{deg}w_i,3[\mathrm{deg}w_i/3]})$
${i=1,\dots,g}$ (here $\dt_{i,j}$ is the Kronecker symbol), and \[
h_{g+1,g+1}=2\wp_{g,g}(\Bv,\Bla),\quad
h_{g+1,g+2}=\begin{cases}0,& g=3\ell \\ -\lambda_{3g+2},& g=3\ell+1
\end{cases}.
\]

Meromorphic mapping $\pi$ uniformizes the space
$\cL\subset\C^{4g+{\dt}}$ and defines a fiberwise
mapping of bundles
$(\sJ_g,\sM_g,p_{\sJ})\to (\cL, \sM_{g},p_{\cL})$ inverse
to mapping  $\cA$  from Theorem \ref{aL->J}.
\end{theorem}
\begin{proof} Let $\si(\Bv,\Bla)\neq 0$ and $\Bv$ be not a
half-period, i.e.  $2\Bv\notin\La(\Bla)$.  Then the relation
\eqref{KF} is equivalent to the following assertion: the entire
rational function of order $2g+4$ \[
\Phi^{\pm}(x,y,\Bv)=\cF((\xi,\eta), (x,y); \Bla)
-(x-\xi)^2\sum_{i,j=1}^{g}
\wp_{i,j}\Big(\Bv\mp
\mspace{-12mu}\textstyle{\int\limits_{(\infty,\infty)}^{(\xi,\eta)}}
\mspace{-12mu}\D \Bu, \Bla\Big)
w_i(\xi,\eta)w_j(x,y) \]
has on the curve $V(x,y)$ exactly $g$ zeros $
\{(x_1^{\pm},y_1^{\pm}),\dots,(x_g^{\pm},y_g^{\pm})\}$ that
do not depend on an arbitrary point  $(\xi,\eta)\in V$.

1. Consider the case $g=3\ell$.
Using the parametrisation
$\xi=t^{-3}$ and $\eta=t^{-(g+1)}(1+\frac{1}{3}\lambda_{3g+1}
t^{2}+\dots)$,
let us study the behaviour of  $\Phi^{\pm}(x,y)$ in vicinity of
$(\xi,\eta)=(\infty,\infty)$.
We have:
\begin{eqnarray*}
\Phi^{\pm}(x,y,\Bv)=t^{-2g-4}
\sum_{i\ge 0}\Phi_i^{\pm}(x,y,\Bv)t^{i},\end{eqnarray*}
\noindent at which
\begin{eqnarray*}
\Phi_0^{\pm}(x,y,\Bv)=
\Phi_0(x,y,\Bv)=w_{g+1}(x,y)-\Bwp_{g}^T\BU(x,y),\\
\Phi_1^{\pm}(x,y,\Bv)=2w_{g+2}(x,y)
-(\Bwp_{g-1}\pm\Bwp_{g}')^T\BU(x,y),  \\
\Phi_2^{\pm}(x,y,\Bv)=\tfrac{\pa}{\pa y}
f_{g}(x,y;\Bla)-\tfrac{1}{2}
(\Bwp_{g}''\pm3\Bwp_{g-1}')^T\BU(x,y)
+\tfrac{\lambda_{3g+1}}{3}\Phi_0(x,y,\Bv),
\end{eqnarray*}
and where, as above,
$w_{i}(x,y)$
stands for the $i$--th coordinate of vector
$\BW_g(x,y)$ and $\BU(x,y)=(w_1(x,y),\dots,w_g(x,y))^T$.

The functions $\{\Phi_0(x,y,\Bv),\Phi_1^{\pm}(x,y,\Bv), \dots\}$
simultaneously vanish at the sets of points
$\{(x_1^{\pm},y_1^{\pm}),\dots,(x_g^{\pm},y_g^{\pm})\}\in(V)^g$.
According to Theorems \ref{aL->U} and \ref{aL->J} collections
$(\pm\Bz,(\Bga_{1}, \Bga_{2},\Bga_{3}))=\a^{-1}(
\{(x_j^{\pm},y_j^{\pm})\},\Bla)$ belong to $\cL$. Note, that the
condition $\si(\Bv,\Bla)\neq 0$ is equivalent to nondegeneracy of
matrices $( \BU(x_i^{\pm},y_i^{\pm})),$ $i=1,\dots,g$.

By identifying the sets of zeros of the pairs of polynomials
$\Phi_0(x,y,\Bv)$ and $\Phi_1^{\pm}(x,y,\Bv)$ with the sets of
zeros $\{(\xi_j^{\pm},\eta_j^{\pm})\}$ from Lemma \ref{g-roots},
we obtain \[ \tau(x,y)=\Phi_0(x,y,\Bv)\quad\text{and}
\quad\rho^{\pm}(x,y)=-\Phi_1^{\pm}(x,y,\Bv),
\]
whence follows \eqref{zgg}. It remains to express
the vector $\Bga_1$ through $\wp$--functions and $\Bla$.
  We utilize identity  (see  \eqref{whw=2rf} and
\eqref{whw}):  \begin{multline}\label{wp->ga}
\left[2y^2-
4\wp_{g,g}(\Bv,\Bla)w_{g+1}(x,y)
+2\BU(x,y)^T \Bga_1-
2\wp_{g,g}(\Bv,\Bla)
\Phi_0(x,y,\Bv)\right] \Phi_0(x,y,\Bv)
-\\\tfrac{1}{2}\Phi_{1}^{-}(x,y,\Bv)
\Phi_{1}^{+}(x,y,\Bv)-2f_{g}(x,y;\Bla)
\tfrac{\pa}{\pa y}\Phi_0(x,y,\Bv)=0,
\end{multline}
that is valid for all $(x,y)\in\C^2$.
Let us differentiate  \eqref{wp->ga} over $v_g$  and
substitute $(x,y)=(x_k^{+},y_k^{+})$ for some
$k\in\{1,\dots,g\}$.  We obtain:  \[
\Phi_{1}^{-}(x_k^{+},y_k^{+},\Bv)
\{
(y_{k}^{+})^2
+\BU(x_k^{+},y_k^{+})^T
(
\Bga_1+2
\wp_{g,g}(\Bv,\Bla)\Bwp_g)
+\tfrac{1}{2}\tfrac{\pa}{\pa
v_g}\Phi_{1}^{+}(x_k^{+},y_k^{+},\Bv)\}=0,
\]
where we used the identity
\[
\tfrac{\pa}{\pa
v_g}\Phi_{0}(x_k^{+},y_k^{+},\Bv)=\tfrac{1}{2}(
\Phi_{1}^{+}(x_k^{+},y_k^{+},\Bv)-
\Phi_{1}^{-}(x_k^{+},y_k^{+},\Bv))=-\tfrac{1}{2}
\Phi_{1}^{-}(x_k^{+},y_k^{+},\Bv).
\]
Thus, as
$\Phi_{1}^{-}(x_k^{+},y_k^{+},\Bv)\neq 0$, if
$\Bv$ is not a half-period,
\begin{equation}
-(y_{k}^{+})^2= \BU(x_k^{+},y_k^{+})^T \{
\Bga_1-\tfrac{1}{2}\Bwp_g''
-\tfrac{1}{2}\Bwp_{g-1}'+
2
\wp_{g,g}(\Bv,\Bla)\Bwp_g\}. \label{gam1}
\end{equation}
On the other hand, from equality
$\Phi_{2}^{+}(x_k^{+},y_k^{+},\Bv)=0$,
follows
\[
-(y_{k}^{+})^2=\BU(x_k^{+},y_k^{+})^T
\{-\tfrac{1}{6}\Bwp_g''
-\tfrac{1}{2}\Bwp_{g-1}'\}+\frac{1}{3}
\sum_{j=0}^{2g/3}\lambda_{g+1+3 j}(x_k^+)^j,
\]
or
\begin{equation}
-(y_{k}^{+})^2=\BU(x_k^{+},y_k^{+})^T
\{-\tfrac{1}{6}\Bwp_g''
-\tfrac{1}{2}\Bwp_{g-1}'-\tfrac13\lambda_{3g+1}\Bwp_{g}\}+\frac{1}{3}
\sum_{j=0}^{2g/3-1}\lambda_{g+1+3 j}(x_k^+)^j. \label{gam2}
\end{equation}
Comparing (\ref{gam1}) and (\ref{gam2}) and taking into
account  nondegeneracy of the matrix $(
\BU(x_k^{+},y_k^{+}))_{k=1,\dots,g}$, we come to \eqref{gamma_1}.

2. Consider the case $g=3\ell+1$.
Using the parametrisation
$\xi=t^{-3}$ and $\eta=t^{-(g+1)}(1+\frac{1}{3}\lambda_{3g+2}t+
\dots)$,
let us study the behaviour of  $\Phi^{\pm}(x,y)$ in vicinity of
$(\xi,\eta)=(\infty,\infty)$.
We have:
\begin{eqnarray*}
\Phi_0(x,y,\Bv)&=&w_{g+1}(x,y)-\Bwp_{g}^T\BU(x,y),\\
\Phi_1^{\pm}(x,y,\Bv)&=&2w_{g+2}(x,y)+\lambda_{3g+2}w_{g+1}(x,y)+
\lambda_{3g}w_{g}(x,y)\notag\\
&-&(\Bwp_{g-1}\pm\Bwp_{g}')^T\BU(x,y),  \\
\Phi_2^{\pm}(x,y,\Bv)&=&\tfrac{\pa}{\pa y}
f_{g}(x,y;\Bla)-\tfrac{1}{2}
(\Bwp_{g}''\mp3\Bwp_{g-1}'\mp\lambda_{2g+2}\Bwp_{g}')^T\BU(x,y)
\notag\\&+&\tfrac{\lambda_{3g+2}}{3}\Phi_1^{\pm}(x,y,\Bv),
\end{eqnarray*}
and where, as above, $w_{i}(x,y)$
stands for the $i$--th coordinate of vector
$\BW_g(x,y)$ and $\BU(x,y)=(w_1(x,y),\dots,w_g(x,y))^T$.

The functions $\{\Phi_0(x,y,\Bv),\Phi_1^{\pm}(x,y,\Bv), \dots\}$
simultaneously vanish at the sets of points
$\{(x_1^{\pm},y_1^{\pm}),\dots,(x_g^{\pm},y_g^{\pm})\}\in(V)^g$.
According to Theorems \ref{aL->U} and \ref{aL->J} collections
$(\pm\Bz,(\Bga_{1}, \Bga_{2},\Bga_{3}))=\a^{-1}(
\{(x_j^{\pm},y_j^{\pm})\},\Bla)$ belong to $\cL$. Note, that the
condition $\si(\Bv,\Bla)\neq 0$ is equivalent to nondegeneracy of
matrices $( \BU(x_i^{\pm},y_i^{\pm})),$ $i=1,\dots,g$.

By identifying the sets of zeros of the pairs of polynomials
$\Phi_0(x,y,\Bv)$ and $\Phi_1^{\pm}(x,y,\Bv)$ with the sets of
zeros $\{(\xi_j^{\pm},\eta_j^{\pm})\}$ from Lemma \ref{g-roots},
we obtain \[ \tau(x,y)=\Phi_0(x,y,\Bv)\quad\text{and}
\quad\rho^{\pm}(x,y)=-\Phi_1^{\pm}(x,y,\Bv),
\]
whence follows \eqref{zgg}. It remains to express
the vector $\Bga_1$ through $\wp$--functions and $\Bla$.
  We utilize identity  (see  \eqref{whw=2rf} and
\eqref{whw}):
\begin{multline}\label{wp->ga}
\left[2y^2+(4\wp_{g,g}(\Bv,\Bla)+\lambda_{3g+2}^2)w_{g+1}(x,y)
+\BU(x,y)^T(2 \Bga_1- \lambda_{3g+2}\Bwp_{g-1})
                             \right.\notag\\
\left. +\lambda_{3g+2}\lambda_{3g}w_{g}(x,y)-(2\wp_{g,g}(\Bv,\Bla)
+\tfrac12\lambda_{3g+2}^2)
\Phi_0(x,y,\Bv)\right]
\Phi_0(x,y,\Bv)
-\\\tfrac{1}{2}\Phi_{1}^{-}(x,y,\Bv)
\Phi_{1}^{+}(x,y,\Bv)-2f_{g}(x,y;\Bla)
\tfrac{\pa}{\pa y}\Phi_0(x,y,\Bv)=0,
\end{multline}
that is valid for all $(x,y)\in\C^2$.
Let us differentiate  \eqref{wp->ga} over $v_g$  and
substitute $(x,y)=(x_k^{+},y_k^{+})$ for some
$k\in\{1,\dots,g\}$.  We obtain:  \[
\Phi_{1}^{-}(x_k^{+},y_k^{+},\Bv)
\{
(y_{k}^{+})^2
+\BU(x_k^{+},y_k^{+})^T
(
\Bga_1+2
\wp_{g,g}(\Bv,\Bla)\Bwp_g)
+\tfrac{1}{2}\tfrac{\pa}{\pa
v_g}\Phi_{1}^{+}(x_k^{+},y_k^{+},\Bv)\}=0,
\]
where we used the identity
\[
\tfrac{\pa}{\pa
v_g}\Phi_{0}(x_k^{+},y_k^{+},\Bv)=\tfrac{1}{2}(
\Phi_{1}^{+}(x_k^{+},y_k^{+},\Bv)-
\Phi_{1}^{-}(x_k^{+},y_k^{+},\Bv))=-\tfrac{1}{2}
\Phi_{1}^{-}(x_k^{+},y_k^{+},\Bv).
\]
Thus, as
$\Phi_{1}^{-}(x_k^{+},y_k^{+},\Bv)\neq 0$, if
$\Bv$ is not a half-period,
\begin{equation}
-(y_{k}^{+})^2= \BU(x_k^{+},y_k^{+})^T \{
\Bga_1-\tfrac{1}{2}\Bwp_g''
-\tfrac{1}{2}\Bwp_{g-1}'+
2
\wp_{g,g}(\Bv,\Bla)\Bwp_g\}. \label{gam1:1}
\end{equation}
On the other hand, from equality
$\Phi_{2}^{+}(x_k^{+},y_k^{+},\Bv)=0$,
follows

\begin{eqnarray}
-(y_{k}^{+})^2&=&\BU(x_k^{+},y_k^{+})^T
\{\tfrac{1}{6}\Bwp_g''
-\tfrac{1}{2}\Bwp_{g-1}'-\tfrac13\lambda_{3g+1}\Bwp_{g}\}
+\frac{1}{6}\Bwp_{g-1}\lambda_{3g+2}
\notag\\
&-&\tfrac16\lambda_{3g+2}\Bwp_g
-\tfrac13\sum_{j=0}^{2/3(g-1)}\lambda_{g+1+3
j}(x_k^+)^j+\tfrac16\lambda_{3g+2}\lambda_{3g}w_g(x,y).  \label{gam2:1}
\end{eqnarray} Comparing (\ref{gam1:1}) and (\ref{gam2:1}) and
taking into account nondegeneracy of the matrix $(
\BU(x_k^{+},y_k^{+}))_{k=1,\dots,g}$, we come to \eqref{gamma_1}.
\end{proof}

In the course of proof of the theorem the following solution of the
trigonal Jacobi inversion problem was obtained

\begin{theorem}\label{jiptrig}

Let the curve $V$ be  nondegenerate trigonal curve
of genus $g$. Let
$w_{g+1},w_{g+2}$ are the components of the vector $\boldsymbol{W}$.
Then the  Jacobi inversion problem is
solved as
\begin{equation}
w_{g+1}=\boldsymbol{\wp}^T_g\boldsymbol{U},
\quad w_{g+2}=\left(\boldsymbol{\wp}_{g-1}\pm
{\boldsymbol{\wp}'_g}\right)^T\boldsymbol{U}.
\label{jip2}
\end{equation}
\end{theorem}

\begin{example}
Let us continue the consideration of the Example
\ref{ex34} and
uniformize the coordinates of Jacobi and Kummer
surfaces in terms of $\wp$-functions.
Consider trigonal curve $V(x,y)$ being  defined by
polynomial \begin{multline*} f_{3}(x,y;\Bla)=
 y^3-(\lambda_{10}x^2+\lambda_{7} x+\lambda_{4})y -\\
(x^4+\lambda_{9}x^3+\lambda_{6}x^2+\lambda_{3}x+\lambda_{0}).
\end{multline*}
The vectors $\boldsymbol{z}$ and
$\boldsymbol{\ga}_1,\boldsymbol{\ga}_2,\boldsymbol{\ga}_3$ are
\begin{eqnarray*}
\boldsymbol{z}&=&\left(\begin{array}{c}z_1\\z_2\\z_3\end{array}\right)=
\left(\begin{array}{c}\wp_{133}\\\wp_{233}\\\wp_{333}
\end{array}\right),
\\
\boldsymbol{\ga}_3&=&\left(\begin{array}{c}\ga_{1,3}\\\ga_{2,3}\\\ga_{3,3}
\end{array}\right)=
-\left(\begin{array}{c}\wp_{13}\\\wp_{23}\\\wp_{33}
\end{array}\right),\\
\boldsymbol{\ga}_2&=&\left(\begin{array}{c}\ga_{1,2}\\\ga_{2,2}\\
\ga_{3,2}\end{array}\right)=
\left(\begin{array}{c}\wp_{12}\\\wp_{22}\\\wp_{23}
\end{array}\right),\\
\boldsymbol{\ga}_1&=&\left(\begin{array}{c}\ga_{1,1}\\\ga_{2,1}
\\
\ga_{3,1}\end{array}\right)=
\left(\begin{array}{c}\frac13\wp_{1333}-\frac13(6\wp_{33}-\lambda_{10})\wp_{13}
\\ \frac13\wp_{2333}-\frac13(6\wp_{33}-\lambda_{10})\wp_{23}\\
\frac13\wp_{3333}-\frac13(6\wp_{33}-\lambda_{10})\wp_{33}
\end{array}\right).
\end{eqnarray*}

The polynomials then $\chi,\rho^{\pm},\tau$ reads
\begin{eqnarray*}
\chi(x,y)&=&y^2+2\wp_{33}x^2+(\lambda_{10}\wp_{33}-\wp_{22})y+
(\lambda_9\wp_{33}+\lambda_7)x\cr&+&
\frac13\wp_{1333}-2\wp_{13}\wp_{33}-\frac{\lambda_{10}}{3}\wp_{13}+\lambda_4,\\
\rho^{\pm}(x,y)&=&-2xy+\wp_{1,2}\pm \wp_{133}+(\wp_{22}\pm \wp_{233})x
+(\wp_{23}\pm \wp_{333})y,\\
\tau(x,y)&=&x^2-\wp_{13}-\wp_{23}x-\wp_{3,3}y.
\end{eqnarray*}

Introduce in  the space $\C^9$ the coordinates
$\wp_{133},\wp_{233},\wp_{333},\wp_{33},\wp_{23},\wp_{13},
\wp_{12},\wp_{22},\wp_{1333}$, which we shall refere below as
{\it
basis functions}. The Jacobi variety $\Jac(V)$ is then singled out
by the system of $6$ equations:  \[ \begin{aligned}
\wp_{133}^2=&\frac43\wp_{13}\wp_{1333}-4\wp_{33}\wp_{13}^2+\wp_{12}^2
-\frac43\lambda_{10}\wp_{13}^2+4\lambda_0\wp_{33}+4\lambda_4\wp_{13},\\
\wp_{133}\wp_{233}=&\frac23\wp_{23}\wp_{1333}+\wp_{22}\wp_{12}
-\frac23\lambda_{10}\wp_{23}\wp_{13}
+2\lambda_{9}\wp_{13}\wp_{33}+2\lambda_3\wp_{33},\\
\wp_{133}\wp_{333}=&\frac23\wp_{33}\wp_{1333}-2\wp_{13}\wp_{22}
+\wp_{23}\wp_{12}+2\lambda_4\wp_{23}+\frac23\lambda_{10}\wp_{13}\wp_{33}
+2\lambda_7\wp_{13},\\
\wp_{233}^2=&4\wp_{33}\wp_{23}^2-\frac43\wp_{1333}+\wp_{22}^2
+8\wp_{13}\wp_{33}\\+
&4\lambda_9\wp_{33}\wp_{23}+\frac43\wp_{13}+
+4\lambda_6\wp_{33}+4\lambda_7\wp_{23}-4\lambda_4\\
\wp_{233}\wp_{333}=&4\wp_{33}^2\wp_{23}-\wp_{23}\wp_{22}-2\wp_{12}
+2\lambda_{10}\wp_{23}\wp_{33}+2\lambda_9\wp_{33}^2+2\lambda_7\wp_{33},\\
\wp_{333}^2=&4\wp_{33}^3+\wp_{23}^2+4\wp_{13}-4\wp_{22}\wp_{33}
+4\lambda_{10}\wp_{33}^2\\
\end{aligned}
\]
These equations represent the first 6 ones from the 9 equation
(\ref{9eq}). The three last lead to the four-index relations
\[ \begin{aligned}
\wp_{3333}=&6\wp_{33}^2-3\wp_{22}+4\lambda_{10}\wp_{33}\\
\wp_{2333}=&6\wp_{23}\wp_{33}+\lambda_{10}\wp_{23}+3\lambda_9\wp_{33}-3\lambda_7
 \end{aligned}
\]

The coordinates $(v_i),\,i=2,3,4,6,7,8$
entering to the matrix $K$, given in (\ref{k34}),  are expressed
\begin{eqnarray*}v_2&=&\wp_{3,3},\quad
v_3=\wp_{2,3},\quad v_4=\wp_{2,2},\quad v_6=\wp_{1,3},\\
v_7&=&\wp_{1,2},\quad
v_8=\frac13(\wp_{1333}+(6\wp_{33}+\lambda_{10})\wp_{13}).
\end{eqnarray*}
The equations that single out $\Kum(V)$
are found from the condition $\rank K\le  4$,
what reduces to three equations on vanishing of the cofactors of
the entries  $K_{1,1}, K_{1,2}$ and
$K_{2,2}$.

To complete the example we shall give below the  set of
expressions of three and four  index symbols in terms of even basis
functions. Namely we have
\begin{align*}
&-2\wp_{33}\wp_{123}
-\wp_{23}\wp_{133}+2\wp_{13}\wp_{233}+\wp_{12}\wp_{333}=0\\
&2\wp_{133}-2\wp_{33}\wp_{233}+\wp_{23}\wp_{233}+\wp_{22}\wp_{333}=0\\
&-\wp_{222}+\lambda_{10}\wp_{233}+2\wp_{33}\wp_{233}-2\wp_{23}\wp_{333}=0
 \end{align*}
also
\begin{align*}
\wp_{3333}& =6{\wp_{33}}^2 +{\mu_1}^2\wp_{33} -3 \wp_{22}
  + 2 \mu_1\wp_{23} - 4\mu_2 \wp_{33} -2 \mu_4,\\
\wp_{2333}& =6 \wp_{23}\wp_{33} + {\mu_1}^2\wp_{23}
  + 3\mu_3\wp_{33} - \mu_2\wp_{23} -\mu_5 -\mu_1\wp_{22},\\
\wp_{2233}& = 4{\wp_{23}}^{2}+ 2 \wp_{33}\wp_{22}
  +\mu_1\mu_3\wp_{33} - \mu_2\wp_{22} + 2 \mu_6
  + 3\mu_3\wp_{23} + \mu_1\mu_2\wp_{23} + 4 \wp_{13},\\
\wp_{2223}& =6 \wp_{22}\wp_{23} + 4 \mu_1\wp_{13}
  + \mu_1\mu_3\wp_{23} + \mu_2\mu_3\wp_{33} + 2 \mu_3\mu_4
  + {\mu_2}^2\wp_{23} + 4 \mu_4\wp_{23} + 3 \mu_3\wp_{22}\\
  &\qquad + 2\mu_1\mu_6 + \mu_2\mu_5 - 2 \mu_5\wp_{33},\\
\wp_{2222}& = 6{\wp_{22}}^2 -2 \mu_2\mu_3\wp_{23}
+ \mu_1\mu_2\mu_5 + 2 \mu_1\mu_3\mu_4 + 24\wp_{13}\wp_{33}
  + 4{\mu_1}^2\wp_{13} - 4\mu_2\wp_{13} - 4 \wp_{1333} \\
  &\qquad + 4\mu_5\wp_{23} +2 {\mu_1}^2\mu_6 - 2 \mu_2\mu_6
  + \mu_3\mu_ 5 - 3{\mu_3}^2\wp_{33} + 12\mu_6\wp_{33}
  + 4 \mu_4\wp_{22} \\
  &\qquad + {\mu_2}^2\wp_{22} +  4\mu_1\mu_3\wp_{22},\\
\wp_{1233}& = 4 \wp_{13}\wp_{23} +2 \wp_{33}\wp_{12}
  - 2\mu_1\wp_{33}\wp_{13} - \tfrac13{\mu_1}^{3}\wp_{13}
  + \tfrac13\mu_1\wp_{1333} +\tfrac13{\mu_1}^2\wp_{12}
  + 3 \mu_3\wp_{13} \\
  &\qquad +\tfrac13 \mu_1\mu_8 +\tfrac43\mu_1\mu_2\wp_{13}
  - \mu_2\wp_{12} + \mu_9,\\
\wp_{1223}& = 4 \wp_{23}\wp_{12} + 2 \wp_{13}\wp_{22}
  - 2\mu_2\wp_{33}\wp_{13} - 2\mu_8\wp_{33}
  - \tfrac23\mu_8\mu_2 +\tfrac13\mu_2\wp_{1333} + 3 \mu_3 \wp_{12}
  + 4\mu_4\wp_{13} \\
  &\qquad + \tfrac43 {\mu_2}^2\wp_{13} - 2 \wp_{11}
  - \tfrac13{\mu_1}^2\mu_2\wp_{13} +\tfrac13 \mu_1\mu_2\wp_{12}
  + \mu_1\mu_3\wp_{13},\\
\wp_{1222}& =6 \wp_{22}\wp_{12} + 6\mu_9\wp_{33} - \mu_3\wp_{1333}
  + 4 \mu_5\wp_{13} + {\mu_2}^2\wp_{12} - \mu_2\mu_9
  + 4 \mu_ 4\wp_{12} - 2 \mu_1\wp_{11}  \\
  &\qquad + 6\mu_3\wp_{33}\wp_{13} -3\mu_2\mu_3\wp_{13}
  + {\mu_1}^2\mu_3\wp_{13} + 3 \mu_1\mu_3\wp_{12} - \mu_1\mu_2\mu_8,\\
\wp_{1133}& =4{\wp_{13}}^2 +2 \wp_{33}\wp_{11} - \mu_9\wp_{23}
  + 2\mu_6\wp_{13} +\mu_8\wp_{22} - \mu_5\wp_{12}
  + \tfrac23 \mu_4\wp_{1333} + \tfrac23 \mu_4\mu_8 \\
  &\qquad +2\mu_2\mu_8\wp_{33} - 4\mu_4\wp_{13}\wp_{33}
  + \tfrac23\mu_2\mu_4\wp_{13} + \mu_1\mu_9\wp_{33} - \mu_1\mu_8\wp_{23}
  + \mu_1\mu_5\wp_{13}  \\
  &\qquad -\tfrac23{\mu_1}^2\mu_4\wp_{13} + \tfrac23 \mu_1\mu_4\wp_{12},\\
\wp_{1123}& =4 \wp_{12}\wp_{13} + 2 \wp_{23}\wp_{11}
  + 2\mu_3\mu_4\wp_{13} - \mu_3\mu_8\wp_{33} - 2\mu_5\wp_{13}\wp_{33}
  + \mu_2\mu_8\wp_{23} + \tfrac43\mu_2\mu_5\wp_{13} \\
  &\qquad -\mu_9\wp_{22} + 2\mu_6\wp_{12} + \tfrac13\mu_5\wp_{1333}
  + \tfrac13\mu_5\mu_8 +\mu_1\mu_9\wp_{23} -\tfrac13{\mu_1}^2\mu_5\wp_{13}
  +\tfrac13 \mu_1\mu_5\wp_{12},\\
\wp_{1122}& = 4{\wp_{12}}^2 +2 \wp_{11}\wp_{22} +\tfrac23
  {\mu_1}^2\mu_6\wp_{13} + \tfrac43 \mu_1\mu_6\wp_{12}
  + \mu_3\mu_9\wp_{33} + \mu_2\mu_9\wp_{23}
  + 8\mu_{12}\wp_{33}  \\
  &\qquad + 2 \mu_3\mu_4\wp_{12}-\tfrac23\mu_6\wp_{1333}
  + 4\mu_8\wp_{13} -\tfrac23\mu_6\mu_8 + 4\mu_6\wp_{33}\wp_{13}
  - \mu_3\mu_8\wp_{23} + \mu_3\mu_5\wp_{13} \\
  &\qquad - \tfrac83 \mu_2\mu_6\wp_{13} + \mu_2\mu_8\wp_{22}
  + \mu_2\mu_5\wp_{12},\\
\end{align*}
\begin{align*}
\wp_{1113}& = 6 \wp_{13}\wp_{11} + 6 \mu_2\mu_8\wp_{13}
  - 2\mu_2\mu_{12}\wp_{33} - {\mu_1}^2\mu_8\wp_{13}
  + 4\mu_1\mu_{12}\wp_{23} + \mu_1\mu_8\wp_{12} + \mu_5\mu_9\wp_{33} \\
  &\qquad + {\mu_5}^2\wp_{13} - 2 \mu_4\mu_9\wp_{23} + \mu_1\mu_9\wp_{13}
  - 6\mu_8\wp_{33}\wp_{13} - 2\mu_6\mu_8\wp_{33} + \mu_8\wp_{1333}
  - 4\mu_4\mu_{12} \\
  &\qquad + 3 \mu_9\wp_{12} - 6 \mu_{12}\wp_{22} - \mu_5\mu_8\wp_{23}
  + 4 \mu_4\mu_6\wp_{13},\\
\wp_{1112}& = 6 \wp_{12}\wp_{11} + 6\mu_3\mu_{12}\wp_{33}
  + 3 \mu_3\mu_8\wp_{13} - 2\mu_6\mu_8\wp_{23} - \mu_1{\mu_8}^2
  + 5\mu_2\mu_8\wp_{12} + 4\mu_2\mu_{12}\wp_{23}  \\
  &\qquad - 2 \mu_1\mu_{12}\wp_{22} + 4 \mu_4\mu_6\wp_{12}
  - \mu_5\mu_8\wp_{22} + {\mu_5}^2\wp_{12} + 4\mu_5\mu_{12}
  - \mu_9\wp_{1333} - 4 \mu_1\mu_{12}\mu_4 \\
  &\qquad + {\mu_ 1}^2\mu_9\wp_{13} + 3 \mu_1\mu_9\wp_{12}
  - 2 \mu_4\mu_9\wp_{22} +\mu_5\mu_9\wp_{23} -4 \mu_2\mu_9\wp_{13}
  + 6 \mu_9\wp_{13}\wp_{33} - 3 \mu_8\mu_9,\\
\wp_{1111}& =6{\wp_{11}}^2 + 4 \mu_4\mu_9\wp_{12} - 8{\mu_4}^2\mu_{12}
  - 2{\mu_2}^2\mu_4\mu_{12} -3 {\mu_8}^2\wp_{22} - 2 \mu_4{\mu_8}^2
  + {\mu_5}^2\wp_{11} -3{\mu_9}^2\wp_{33} \\
  &\qquad - 4\mu_{12}\wp_{1333} + 24 \mu_{12}\wp_{33}\wp_{13}
  + 12 \mu_5\mu_{12}\wp_{23} + \mu_2 \mu_4\mu_5\mu_9
  - 6 \mu_1\mu_3\mu_4\mu_{12} \\
  &\qquad +\mu_1\mu_2\mu_5\mu_{12} +2 {\mu_6}^2\mu_8 +2{\mu_2}^2{\mu_8}^2
  - \mu_5\mu_6\mu_9 - 2 \mu_5\mu_9\wp_{13}
  + 4\mu_4\mu_6\wp_{11} + 4\mu_6\mu_8\wp_{13} \\
  &\qquad + 8\mu_2\mu_8\wp_{11} - 6 \mu_2\mu_6\mu_{12}
  - 12 \mu_2\mu_{12}\wp_{13} + 4{\mu_1}^2\mu_{12}\wp_{13}
  + 2 {\mu_1}^2\mu_6\mu_{12} + 2\mu _8\mu_5\wp_{12} \\
  &\qquad - 6 \mu_8\mu_9\wp_{23} - 12\mu_4\mu_{12}\wp_{22} + \mu_2 {\mu_5}^2\mu_8
  + 2\mu_1\mu_4\mu_6\mu_9 +\mu_1\mu_5\mu_6\mu_8 + 12\mu_6\mu_{12}\wp_{33} \\
  &\qquad + 4\mu_1\mu_9\wp_{11} + 2\mu_3 {\mu_4}^2\mu_9
  + 9\mu_3\mu_5\mu_{12} - 2\mu_1\mu_3{\mu_8}^2 - 6\mu_3\mu_8\mu_9
  + 2\mu_1\mu_2\mu_8\mu_9 \\
  &\qquad + \mu_3\mu_4\mu_5\mu_8  + 2\mu_2\mu_4\mu_6 \mu_8
  + 2\mu_2{\mu_9}^2.\\
\wp_{1233}&=\lambda_3+\lambda_{10}\wp_{12}+4\wp_{13}\wp_{23}+2\wp_{12}\wp_{33}\\
\wp_{2233}&=2\lambda_6+4\wp_{13}+\lambda_0\wp_{22}+4\wp_{23}^2
+2\wp_{22}\wp_{33}\\
\wp_{2223}&=\lambda_7\lambda_{10}+\lambda_{10}^2\wp_{23}+6\wp_{22}\wp_{23}
+2\lambda_7\wp_{33}\\
\wp_{2222}&=2\lambda_6\lambda_{10}+4\lambda_{10}\wp_{13}+\lambda_{10}^2\wp_{22}
+6\wp_{22}^2\\
&-4\lambda_7\wp_{23}+12\lambda_6\wp_{33}+4(6\wp_{13}\wp_{33}-\wp_{1333})\\
\wp_{1222}&=\lambda_3\lambda_{10}+\lambda_{10}^2\wp_{12}-4\lambda_7\wp_{13}
+6\wp_{12}\wp_{22}+6\lambda_3\wp_{33}\\
\wp_{1133}&=\lambda_7\wp_{12}+2\lambda_6\wp_{13}+4\wp_{13}^2-\lambda_4\wp_{22}
-la_3\wp_{23}+2\lambda_4\lambda_{10}\wp_{33}+2\wp_{11}\wp_{33}\\
\wp_{1123}&=\tfrac13\lambda_4\lambda_7+2\lambda_6\wp_{12}
+\tfrac43\lambda_7\lambda_{10}\wp_{13}+4\wp_{12}\wp_{13}-\lambda_3\wp_{22}\\
&+\lambda_4\lambda_{10}\wp_{23}+2\wp_{11}\wp_{23}
+\tfrac13(6\wp_{13}\wp_{33}-\wp_{1333})\\
\wp_{1223}&=-\frac23\lambda_4\lambda_{10}-\wp_{11}+\frac43\lambda_{10}^2\wp_{13}
+2\wp_{13}\wp_{22}\\
&+4\wp_{12}\wp_{23}+2\lambda_4\wp_{33}+\frac13\lambda_{10}
(6\wp_{13}\wp_{33}-\wp_{1333})\\
\wp_{1122}&=\frac23\lambda_4\lambda_6+\lambda_7\lambda_{10}\wp_{12}+4\wp_{12}^2
-4\lambda_4\wp_{13}+\frac83\lambda_6\lambda_{10}\wp_{13}\\
&=\lambda_4\lambda_{10}\wp_{22}+2\wp_{11}\wp_{22}-\lambda_3\lambda_{10}\wp_{23}
+8\lambda_0\wp_{33} +\frac23\lambda_6(6\wp_{13}\wp_{33}-\wp_{1333})\\
\wp_{1113}&=3\lambda_3\wp_{12}+\lambda_7^2\wp_{13}+6\lambda_4\lambda_{10}\wp_{13}
+6\wp_{11}\wp_{13}-6\lambda_0\wp_{22}-\lambda_4\lambda_7\wp_{23}\\
&+2\lambda_4\lambda_6\wp_{33}-\lambda_3\lambda_7\wp_{33}+2\lambda_0\lambda_{10}\wp_{33}
+\lambda_4(6\wp_{13}\wp_{33}-\wp_{1333})
\end{align*}
\end{example}


\section{Comparison with hyperelliptic case}

Let $V$ be the hyperelliptic curve of genus $g$ realized as two-sheeted covering
of the extended complex plane,

\begin{equation}
y^2-\sum_{k=0}^{2g+2}\lambda_{k}x^k=0.
\end{equation}

The vector $\boldsymbol{W}_2$ is given as

\[ \boldsymbol{W}_2=(1,x,\ldots,x^{g-1})^T.  \]
Introduce the matrix
\begin{eqnarray}
H=\{h_{i,k}\}_{i,k=1,\ldots,g+2}, \qquad h_{ik}=4\wp_{i-1,k-1}
-2\wp_{k,i-2}-2\wp_{i,k-2}\cr
+\frac12\left(\delta_{i,k}(\lambda_{2i-2}+\lambda_{2k-2})+\delta_{k,i+1}
\lambda_{2i-1}+\delta_{i,k+1}\lambda_{2k-1}\right)
\end{eqnarray}
Denote the
minors of the matrix $H$ as follows
\[ H[{}_{j_1}^{k_1}{}^{\ldots}_{\ldots}{}_{j_m}^{k_n}]
\{h_{i_k,j_l}\}_{k=1,\ldots,h;l=1,\ldots,n}
\]
\begin{theorem}[\cite{bel97b}]
The matrix $H $ has the following
property
Let ${\boldsymbol
W}=(1,x,\ldots,x^{g+1})$ then for arbitrary vectors the equality
is valid
\begin{eqnarray*}
&&y_ry_s={\boldsymbol W}_r^TH{\boldsymbol W}_s,\quad
\text{and in particular}\\
&&{\boldsymbol W}^TH{\boldsymbol W}=\sum_{i=0}^{2g+2}\lambda_ix^i
\end{eqnarray*}
\end{theorem}

The Jacobi inversion problem is solved as

\begin{theorem}\label{jiphyp}
Let the curve $V$ be  nondegenerate hyperelliptic curve
of genus $g$ given by the eqution
$$ y^2-4x^{2g+1}-\sum_{k=0}^{2g}\lambda_kx^k=0 $$ and
$w_{g+1},w_{g+2}$ are entries to the Weiestrass gap sequence.
Then the  Jacobi inversion problem is
solved as
\begin{equation}
w_{g+1}=\boldsymbol{\wp}^T_g\boldsymbol{U},
\quad w_{g+2}={\boldsymbol{\wp}'}_g^T\boldsymbol{U}.
\label{jip2}
\end{equation}
\end{theorem}

\begin{theorem}
Let $(z_1,w_1),\ldots, (z_g,w_g)$ be the divisor, then
the vectors $\boldsymbol{Z}=(1,z_r,\ldots,z_r^{g+1})$,
$r=1,\ldots,g$ are  orthogonal to the  last collumn of the
matrix $H$ or equivalentely $z_r$ are the roots of the equation
\begin{equation}z^g-\wp_{g,g}(\boldsymbol{u})z^{g-1}-\ldots
-\wp_{1,g}(\boldsymbol{u})=0\label{jachyp1}
 \end{equation}
which yeilds the solution of the Jacobi inversion problem
where the second coordinate of the divisor is defined as follows
\begin{equation}w_k=-\sum_{j=1}^g\wp_{g,g,j}(\boldsymbol{u})z_k^{g-j}
\label{jachyp2}
\end{equation}
\end{theorem}

The meromorphic embedding of the Jacobi variety   $\mathrm{Jac}(V)$ into
${\mathbb C}^{g+\frac12 g(g+1)}$ is described by the following

\begin{theorem}
Introduce in  ${\mathbb C}^{g+\frac12 g(g+1)}$ the coordinates:
$\frac12g(g+1)$ even functions $\wp_{i,j}$ and $g$ odd $\wp_{g,g,i}$, $i=1,\ldots,g$
Then

 $\mathrm{rank}\,H=3$ in generic points and
\begin{equation}
-\frac14\wp_{i,g,g}\wp_{k,g,g}
=\det \;H\left[{}_i^k{}_{g+1}^{g+1}{}_{g+2}^{g+2}\right],\quad
\forall i,k=1,\ldots,g\label{cubics} \end{equation} The
intersection of $g(g+1)/2$ cubics defines the Jacobi variety as
algebraic variety in ${\mathbf C}^{g+\frac12g(g+1)}$.
\end{theorem}

The meromorphic embedding of the Kummer variety   $\mathrm{Kum}(V)$ into
${\mathbb C}^{\frac12 g(g+1)}$ is described by the following

\begin{theorem}
Introduce in  ${\mathbb C}^{\frac12 g(g+1)}$ the coordinates:
$\frac12g(g+1)$ even functions $\wp_{i,j}$
Then the
intersection of $g(g-1)/2$ quartics \begin{equation}
\det
\;H\left[{}_i^k{}_j^l{}_{g+1}^{g+1}{}_{g+2}^{g+2}\right]=0,\forall
i\neq j,k\neq l=1,\ldots,g\label{quartics}
\end{equation} defines the Kummer variety as algebraic variety in
${\mathbf C}^{\frac12g(g+1)}$.
\end{theorem}

\begin{theorem}
The following equality is valid
 \begin{eqnarray} \mathbf{ R}^T \boldsymbol{
  \pi}_{jl}  \boldsymbol{   \pi}_{ik}^T  \mathbf{ S}=  \frac{1}{4}
\det   \left (  \begin{array}{cc} H
  \left[{}^i_j{}^k_l{}^{g+1}_{g+1}{}^{g+2}_{g+2}  \right]
&  \mathbf{ S}  \\  \mathbf{ R}^T&0
  \end{array}  \right) ,   \label{bakergen}
  \end{eqnarray}
where $  \mathbf{ R},   \,    \mathbf{ S}  \in   \mathbf{  C}^4$ are
arbitrary vectors and
  \[  \boldsymbol{   \pi}_{ik}=  \left (
  \begin{array}{c} -  \wp_{ggk}  \\
  \wp_{ggi}  \\
  \wp_{g, i, k-1}-  \wp_{g, i-1, k}  \\
  \wp_{g-1, i, k-1}-
  \wp_{g-1, k, i-1}+
  \wp_{g, k, i-2}-
  \wp_{g, i, k-2}
  \end{array}
  \right) .  \]
\end{theorem}

\begin{example}
We shall give a number of formulas to illustrate the
content of this section in the case of genus three. We consider
the case of the curve defined by an equation
\begin{equation*}
y^2=4 x^7+\sum_{j=0}^6\lambda_jx^j.
\end{equation*}

The principal matrix $H$ is the $5\times5$ matrix of the form
\begin{eqnarray*}
H=\left(\begin{array}{ccccc}
\lambda_0&\frac12\lambda_1&-2\wp_{11}&-2\wp_{12}&-2\wp_{13}\\
\frac12\lambda_1&4\wp_{11}+\lambda_2&2\wp_{12}+\frac12\lambda_3&
4\wp_{13}-2\wp_{22}&-2\wp_{23}\\
-2\wp_{11}&2\wp_{12}+\frac12\lambda_3&4\wp_{22}-4\wp_{13}+\lambda_4&
2\wp_{23}+\frac12\lambda_5&-2\wp_{33}\\
-2\wp_{12}&4\wp_{13}-2\wp_{22}&2\wp_{23}+\frac12\lambda_5&4\wp_{33}+
\lambda_6&2\\
-2\wp_{13}&-2\wp_{23}&-2\wp_{33}&2&0\end{array}\right)
\end{eqnarray*}

We give list of the
expressions of the $\wp_{ijk}$-functions, that is the complete
list of the first derivatives of the $\wp_{i,j}$ over the
canonical fields $\partial_i$, as linear combinations if the basis
functions.  In this case the basis set consists of functions
$\wp_{333},\wp_{233},\wp_{133}$ and $\wp_{33},\wp_{23},\wp_{13}$.
The basic cubic relations are
\begin{align*}
\wp_{333}^2&=4\wp_{33}^3+\lambda_6\wp_{33}^2+4\wp_{23}\wp_{33}
+\lambda_5\wp_{33}+4\wp_{22}-4\wp_{13}+\lambda_4,\\
\wp_{233}^2&=4\wp_{23}^2\wp_{33}+\lambda_6\wp_{23}^2-4\wp_{22}\wp_{23}
+8\wp_{13}\wp_{23}+4\wp_{11}+\lambda_2,\\
\wp_{133}^2&=4\wp_{13}^2\wp_{33}+\lambda_6\wp_{13}^2-4\wp_{12}\wp_{13}
+\lambda_0, \\
\wp_{233}\wp_{333}&=4\wp_{33}^2\wp_{23}+\lambda_6\wp_{23}\wp_{33}
-4\wp_{22}\wp_{33},\\
&+4\wp_{13}\wp_{33}+2\wp_{23}^2-\lambda_5\wp_{23}+2\wp_{12}+\lambda_3,\\
\wp_{133}\wp_{233}&=4\wp_{13}\wp_{23}\wp_{33}+\lambda_6\wp_{13}\wp_{23}
-2\wp_{12}\wp_{23}-2\wp_{13}\wp_{22}+4\wp_{13}^2+\frac12\lambda_1,\\
\wp_{133}\wp_{333}&=4\wp_{13}\wp_{33}^2+\lambda_6\wp_{13}\wp_{33}
-2\wp_{12}\wp_{33}+2\wp_{13}\wp_{23}+\frac12\lambda_5\wp_{13}-2\wp_{11}.
\end{align*}

Remaining cubic relations can be derived with the aids of the
above formulae and relations

\begin{align*}
\wp_{223}&=-\wp_{333}\wp_{23}+\wp_{233}\wp_{33}+\wp_{133},\\
\wp_{123}&=-\wp_{333}\wp_{13}+\wp_{133}\wp_{33},\\
\wp_{113}&=-\wp_{233}\wp_{13}+\wp_{133}\wp_{23},\\
\wp_{222}&=
\wp_{333}\big(2\wp_{23}(\wp_{33}+\frac{\lambda_{6}}{4})+4\wp_{13}
  -\wp_{22}\big)-
\wp_{233}\big(2\wp_{33}(\wp_{33}+\frac{\lambda_{6}}{4})+\wp_{23}
  +\frac{\lambda_{5}}{4}\big)
\\&-2\wp_{133}\wp_{33},\\
\wp_{122} &=
\wp_{333}\big(2\wp_{13}(\wp_{33}+\frac{\lambda_{6}}{4})-\wp_{12}\big)+
\wp_{233}\wp_{13}-
\wp_{133}\big(2\wp_{33}(\wp_{33}+\frac{\lambda_{6}}{4})+\wp_{23}
  +\frac{\lambda_{5}}{4}\big),
\\
\wp_{112} &=
\wp_{233}\big(2\wp_{13}(\wp_{33}+\frac{\lambda_{6}}{4})-\wp_{12}\big)-
\wp_{133}\big(2\wp_{23}(\wp_{33}+\frac{\lambda_{6}}{4})+2\wp_{13}
  -\wp_{22}\big),
\\
\wp_{111}&=
\wp_{333}\big(\wp_{13}\wp_{22}-2\wp_{13}^2-\wp_{12}\wp_{23}\big)+
\wp_{233}\big(2\wp_{13}(\wp_{23}+\frac{\lambda_{5}}{4})-\wp_{12}\wp_{33}\big),
\\ &-
\wp_{133}\big(2\wp_{23}(\wp_{23}+\frac{\lambda_{5}}{4})-\wp_{33}(2\wp_{13}
  -\wp_{22})\big)
\end{align*}

Along with the above expressions for the first derivatives of the
$\wp_{i,j}$ we obtain an anologous list for the second
derivatives,
but here we give the expressions by the $\wp_{i,j}$--functions
themselves and the constants $\lambda_k$:
\begin{align*}
\wp_{3333}&= \lambda_5/2+4\wp_{23}+\lambda_6\wp_{33}+6\wp_{33}^2,
\\
\wp_{2333}&=6\wp_{13}-2\wp_{22}+\lambda_6\wp_{23}+6\wp_{23}\wp_{33}, \\
\wp_{1333}&=-2\wp_{12}+\lambda_6\wp_{13}+6\wp_{13}\wp_{33}, \\
\wp_{2233}&=-2\wp_{12}+\lambda_6\wp_{13}+\lambda_5\wp_{23}/2+4\wp_{23}^2+
2\wp_{22}\wp_{33} ,\\
\wp_{1233}&=\lambda_5\wp_{1 3}/2+4\wp_{13}\wp_{23}+2\wp_{12}\wp_{33}, \\
\wp_{1133}&=6\wp_{13}^2-2\wp_{13}\wp_{22}+2\wp_{12}\wp_{23}, \\
\wp_{2223}&=-\lambda_2-6\wp_{11}+\lambda_5\wp_{13}+\lambda_4\wp_{23}+6\wp_{22}\wp_{23}-
\lambda_3\wp_{33}/2 ,\\
\wp_{1223}&=-\lambda_1/2+\lambda_4\wp_{13}-2\wp_{13}^2+4\wp_{13}\wp_{22}+
2\wp_{12}\wp_{23}+2\wp_{11}\wp_{33}, \\
\wp_{1123}&=-\lambda_0+\lambda_3\wp_{13}/2+4\wp_{12}\wp_{13}
+2\wp_{11}\wp_{23} ,\\
\wp_{1113}&=\lambda_2\wp_{13}+6\wp_{11}\wp_{13}-\lambda_1\wp_{23}/2
+\lambda_0\wp_{33} ,\\
\wp_{2222}&=-3\lambda_1/2+\lambda_3\lambda_5/8-\lambda_2\lambda_6/2
-3\lambda_6\wp_{11}+
 \lambda_5\wp_{12}+12\wp_{13}^2+ \\
&\lambda_4\wp_{22}-12\wp_{13}\wp_{22}+
 6\wp_{22}^2+\lambda_3\wp_{23}+12\wp_{12}\wp_{23}-3\lambda_2\wp_{33}
-12\wp_{11}\wp_{33}, \\
\wp_{1222}&=-2\lambda_0-\lambda_1\lambda_6/4-\lambda_5\wp_{11}/2
+\lambda_4\wp_{12}+
 \lambda_3\wp_{13}+6\wp_{12}\wp_{22}-3\lambda_1\wp_{33}/2, \\
\wp_{1122}&=-\lambda_0\lambda_6/2+\lambda_3\wp_{12}/2+4\wp_{12}^2
+\lambda_2\wp_{13}+
  2\wp_{11}\wp_{22}-\lambda_1\wp_{23}/2-2\lambda_0\wp_{33} ,\\
\wp_{1112}&=-\lambda_0\lambda_5/4+\lambda_2\wp_{12}+6\wp_{11}\wp_{12}
+3\lambda_1\wp_{13}/2-
  \lambda_1\wp_{22}/2-2\lambda_0\wp_{23},\\
\wp_{1111}&=-\frac12\lambda_0\lambda_4+\frac18\lambda_1\lambda_3
-3\lambda_0\wp_{22}+\lambda_1\wp_{12}+\lambda_2\wp_{11}+
4\lambda_0\wp_{13} +6\wp_{11}^2.
\end{align*}
\end{example}

\section{Nonlinear differential equations
integrable in trigonal $\wp$--functions} \label{sec:Eqs}
In view of the results of Theorem \ref{pi:J->L} generating
functions  $G_1(t)$ and $G_2(t)$ introduced in the assertion  (1)
of Corollary \ref{kumm} obtain another interpretation.  Let the
vector $\BZ$ be given by formulas \eqref{zgg} and \eqref{gamma_1},
then the polynomials generated by functions $G_1(t)$ and $G_2(t)$
are \emph{differential relations}, to which $\wp$--functions
satisfy.

On the other hand,  defining the vector $\BZ$ by formulas
\begin{align*}
&\Bga_1=\tfrac{1}{3}(\Bu''-(6
u_{g}+\dt_{g,3[g/3]}\lambda_{3g+1}+\dt_{g,1+3[g/3]}
\tfrac{1}{2}\lambda_{3g+2}^2)
\Bu
+\dt_{g,1+3[g/3]}\lambda_{3g+2}\Bq
+\BL_g)\\&
\Bga_2=\Bq,\;
\Bga_3=-\Bu,\;
\Bz=\Bu',\;
h_{g+1,g+1}=2u_g,\;
h_{g+1,g+2}=-\dt_{g,1+3[g/3]}\lambda_{3g+2},
\end{align*}
we obtain with the help of generating functions
 $G_1(t)$ and  $G_2(t)$ a collection of differential
polynomials
$S_{\Bla}^{0}=S_{\Bla}(\Bq, \Bu,\Bu',\Bu'')$.  Let us
complete the collection $S_{\Bla}^{0}$ with the collection
of linear expressions \[ S^{1}=\{q_i'-
\tfrac{\pa }{\pa v_{g-1}}u_i;\quad \tfrac{\pa}{\pa v_i}u_k-
\tfrac{\pa}{\pa v_k}u_i,\quad \tfrac{\pa}{\pa v_i}q_k-
\tfrac{\pa}{\pa v_k}q_i\}\quad i,k=1,\dots,g, \; i>k \] to the
collection $\wh{S}_{\Bla}= S_{\Bla}^{0}\cup
S^{1}=\wh{S}_{\Bla}(\Bq, \Bu,\tfrac{\pa}{\pa
v_1}\Bq,\dots,\Bq'=\tfrac{\pa}{\pa
v_g}\Bq,\tfrac{\pa}{\pa
v_1}\Bu,\dots,\Bu'=\tfrac{\pa}{\pa
v_g}\Bu,
\Bu'')$.
\begin{prop}\label{set}
System $\wh{S}_{\Bla}=0$ of nonlinear differential
equations of second order
with respect to functions
$\Bq$ and $\Bu$ allows a solution in trigonal
$\wp$--functions $\Bq=\Bwp_{g-1}$ and $\Bu=\Bwp_{g}$.
\end{prop}
Introduce a grading of the independent variables
$\Bv^T=(v_1,\dots,v_g)$ according to the rule $\deg v_i=\deg
w_i-2g+1$, so that  $\deg v_g=-1$, $\deg v_{g-1}=-2$ and so on,
(note that the sequence
 $-\deg \Bv$ forms the Weierstrass sequence generated by a
pair of coprime integers $(3,g+1)$, see \cite{bel99}). For
functions $\Bq$ and  $\Bu$ we assume the weights that follow from
the convention adopted in \S \ref{sec:L} about the weights of
coordinates of vectors $\Bga_2$ and $\Bga_3$, and, so $\deg
u_g=2$, $\deg u_{g-1}=\deg q_g=3$ and so on.  Members of the
collection  $\wh{S}_{\Bla}$  are homogeneous differential
polynomials with respect to this grading.  Let us consider  the
lower weight polynomials in more detail.

Each of the generating functions $G_1(t)$ and $G_2(t)$
produce one polynomial of each weight.  At that the function
$G_2(t)$ generates polynomials of weight not less than
 $2g-3[\frac{g-1}{3}]=g+3-\dt$.  Thus the elements of the lowest
weights $3$ and $4$ in the collection
$S_{\Bla}^{0}$ are generated by the function
$G_1(t)$.  Namely, there is a single polynomial
 $\psi_3=q_g-u_{g-1}$ of weight $3$ and a single polynomial of
weight $4$:  \begin{gather*}
\psi_4=u_g''-6u_g^2+3q_{g-1}-a q_{g}-b u_{g} -c,
\end{gather*}
\noindent where
\begin{gather*}
\{a,b,c\}=\{2\dt_{g,1+3[g/3]}\lambda_{3g+2},
4\dt_{g,3[g/3]}\lambda_{3g+1}+
\dt_{g,1+3[g/3]}\lambda_{3g+2}^2,2\dt_{g,1+3[g/3]}\lambda_{3g-1}\}.
\end{gather*}
Denote $v_g=\xi$, $v_{g-1}=\eta$ and $u_g=U/2$, and partial
derivatives of  $U$ over $\xi$  and $\eta$ denote by subscripts,
for instance, $\tfrac{\pa}{\pa \xi}U=U_{\xi}$, $\tfrac{\pa^2}{\pa
\xi^2}U=U_{\xi\xi}$ etc. From the system of equations
$(\psi_3,\psi_4, S^{1})=0$ follows a differential equation
w.r.t.  $U$ of the form:
\begin{equation}
3U_{\eta\eta}+(U_{\xi\xi\xi}-6U_{\xi}U)_{\xi}- a U_{\xi\eta}-b
U_{\xi\xi}=0,\label{boo} \end{equation} that is the well-known
\emph{Boussinesq  equation}. By Proposition
\ref{set} the equation \eqref{boo} is satisfied by the function
$U=2\wp_{g,g}(u_1,\dots,\eta,\xi,\Bla)$ constructed by the model
of a trigonal curve defined by equation
\[ y^3-
y(\sum_{j=0}^{[\frac{2g+2}{3}]}\lambda_{g+1+3 j}x^j)-(x^{g+1}+
\sum_{j=0}^{g}\lambda_{3 j}x^j)=0.  \]

Let us compare the result obtained with an analogous
result from the hyperelliptic theory
\cite{bel97b}.  The hyperelliptic function
 $2\wp_{g,g}(u_1,\dots,\eta,\xi,\Bla)$ constructed by the model of
a  hyperelliptic curve
 \[ y^2- 4x^{2g+1}- \sum_{j=0}^{2g-1}\lambda_{2 j}x^j=0, \]
is a solution of \emph{Korteweg-de Vries hierarchy} and,
particularly, of the classical KdV equation:
\begin{equation}\label{KdV} 2U_{\eta}-U_{\xi\xi\xi}+6 U
U_{\xi}=0.  \end{equation}
At certain constraints imposed on the values of the parameters
$\Bla$ the hyperelliptic function
 $2\wp_{1,1}$ of weight $4g-2$ is a solution of
\emph{Kadomtsev-Petviashvili equation}:
\begin{equation}\label{KP}
4U_{\xi\zeta}-3U_{\eta\eta}-(U_{\xi\xi\xi}-6U_{\xi}U)_{\xi}=0,
\end{equation}
and with the help of the function $\wp_{1,g}$
of weight $2g-2$ a solution of
sine-Gordon equation is constructed.

In both cases \eqref{boo} and \eqref{KdV} the solution
$U=2\wp_{g,g}$ is a function of weight $2$,
the least weight possible for a
$\wp$--function, while the weights $\{-\deg\xi,-\deg\eta\}$
are the initial segments of the corresponding
Weierstrass sequences:  $\{1,2\}$ for
\eqref{boo} and  $\{1,3\}$ for \eqref{KdV}. At that the
right-hand sides of \eqref{boo} and  \eqref{KdV} are obtained
by (multiple) differentiation w.r.t. $\xi$ of the
polynomial of weight $4$ that corresponds to the relation
$\wp_{g,g,g,g}-6\wp_{g,g}^2+\dots=0$.

Elements of weights $>4$ from the family $\wh{S}_{\Bla}$ after
repeated differentiation lead to a system of nonlinear
differential equations w.r.t. the function $U$ that is to
\emph{Boussinesq hierarchy}.
 Systems
of differential equations satisfied by the higher weight
$\wp$--functions form families that do not reduce to Boussinesq hierarchy.
We are going to consider this matter in detail in our
nearest publications.
\begin{prop} For the function
$U=2\wp_{g,g}(u_1,\dots,\zeta,\eta,\xi,\Bla)$ of weight $2$
built by the $(n,s)$--model of a curve
$y^n-x^s+\dots=0$ to be a solution of equation \eqref{KP}, it is
necessary that the number of sheets of the model $n$ is not
less than $4$.
\end{prop}
\begin{proof}
Indeed, let us apply a scale transform
$(U,\xi,\eta,\zeta)\mapsto(t^2 U,t^{-\a_1}\xi,
t^{-\a_2} \eta,t^{-\a_3}\zeta)$ to equation \eqref{KP}.

From the claim of homogeneity we obtain
$\{\a_1,\a_2,\a_3\}=\{1,2,3\}$, i.e.  the initial segment
of the Weierstrass sequence contains the triple $\{1,2,3\}$, and,
therefore, $n>3$.
\end{proof}
In some cases the condition  $n>3$ is sufficient. By the
homogeneity reasons a polynomial relation of weight $4$ for
$\wp$--functions, if \emph{exists}, is always of the form
\[
\wp_{g,g,g,g}=6\wp_{g,g}^2+a_0\wp_{g,g-2}+b_0\wp_{g-1,g-1}+
a_1\wp_{g,g-1}+a_2\wp_{g,g}+a_4,
\]
where subscripts of parameters  $a$ and $b$
are equal to their weights.  At the ``rational limit''
$\Bla\to\boldsymbol{0}$ the parameters  $a_4,a_2$ and $a_1$
vanish, while the function $\si(\Bv,\Bla)$
becomes a special Schur polynomial
$\si_{n,s}(\Bv)$, so-called   \emph{Schur-Weierstrass
polynomial} (see  \cite{bel99}). As is  known
\cite{kac93}, doubled second  partial derivative of any
Schur polynomial w.r.t. the variable of weight $1$, and,
particularly, the rational limit of the function
$2\wp_{g,g}$, is a rational solution of equation
\eqref{KP}, whence follow
$a_0=-4$ and $b_0=3$.

\chapter*{Appendix I: A set of formulae for genus $2$}
Material of this appendix represents elements of a handbook on the genus two hyperelliptic (ultraelliptic) functions. We mostly collected here formulae involving hyperelliptic $\sigma$ functions that are relevant to the above presented exposition.

\section{Hyperelliptic curve of genus two}\label{sec:curvesgenus2}
We consider a hyperelliptic curve $X_2$ of genus two
\begin{align}
\begin{split}
w^2 & = 4(z-e_1)(z-e_2)(z-e_3)(z-e_4)(z-e_5)\\
& = 4 z^5 + \lambda_4 z^4 + \lambda_3 z^3 + \lambda_2 z^2 + \lambda_1 z + \lambda_0 \, .
\end{split} \label{curve2}
\end{align}
The basic holomorphic and meromorphic differentials are
\begin{align}
\mathrm{d}u_1 &=\frac{\mathrm{d}z}{w}\,, & \qquad \mathrm{d}r_1 & = \frac{12 z^3+2\lambda_4z^2+\lambda_3z}{4w}\mathrm{d}z \,, \\
\mathrm{d}u_2 &=\frac{z\mathrm{d}z}{w}\,, & \qquad \mathrm{d}r_2 & = \frac{z^2}{w}\mathrm{d}z \,.
\end{align}

For arbitrary points $P=(x,y), Q=(z,w)\in V$  fundamental non-normalized bi-differential $\mathrm{d}\widehat{\omega}(P,Q)$ is given as
\begin{align}
\mathrm{d}\widehat{\omega}(P,Q)= \frac{F(x,z)+2yw}{4(x-z)^2}\frac{\mathrm{d}x}{y} \frac{\mathrm{d}z}{w}
\end{align}
with $F(x,z)$ given by the formula
\begin{equation}
F(x,z)=\sum_{k=0}^2  z^kx^k( 2\lambda_{2k}+\lambda_{2k+1}(x+z)  )
\end{equation}
Equivalently this formula is written as
\begin{equation}
F(x,z)=  \frac{\partial}{\partial z}  \frac{w+y}{2 y (x-z)}= \sum_{k=1}^2 \mathrm{d}u_{k}(P)\mathrm{d}r_{k}(Q)
\end{equation}

For running point $P=(x,y)\in V$ and two arbitrary fixed points  $P_1=(x_1,y_1), P_2=(x_2,y_2)\in V$  the non-normalized third kind differential $\omega_{P_1,P_2}(P)$ with first order poles in
$P_1=(x_1,y_1), P_2=(x_2,y_2)$ and residues $+1$ and $-1$ is given
\begin{equation}
\omega_{P_1,P_2}(P)=\frac{w+y}{2 (x-z)} \frac{\mathrm{d}x}{y} +  \sum_{k=0}^2  \mathrm{d}u_k(P) \int_{P_1}^{P_2}r_k(P')
\end{equation}

Then the Jacobi inversion problem for the equations
\begin{align}\begin{split}
\int_{\infty}^{(z_1,w_1)}\frac{ \mathrm{d}z}{w}+\int_{\infty}^{(z_2,w_2)}\frac{ \mathrm{d}z}{w}=u_1 \,,\\
\int_{\infty}^{(z_1,w_1)}\frac{z \mathrm{d}z}{w}+\int_{\infty}^{(z_2,w_2)}\frac{z \mathrm{d}z}{w}=u_2\end{split} \label{JIP}
\end{align}
is solved in terms of $\wp$-functions in the form
\begin{align}\begin{split}
z_1+z_2&=\wp_{22}(\boldsymbol{u}), \quad z_1z_2=-\wp_{12}(\boldsymbol{u}) \,,\\
w_k&= \wp_{222}(\boldsymbol{u})z_k + \wp_{122}(\boldsymbol{u}), \quad k=1,2 \,.
\end{split} \label{SOLJIP}
\end{align}

\subsection{Characteristics in genus two}
The homology basis of the curve is fixed by defining the set of
half-periods corresponding to the branch points. The characteristics of the Abelian images of the branch points are defined as
\begin{equation}
[\boldsymbol{\mathfrak A}_i]=\left[\int_{\infty}^{(e_i,0)} \mathrm{d}\boldsymbol{u}\right] = \begin{pmatrix} \boldsymbol{\varepsilon}_i^{'^T} \\ \boldsymbol{\varepsilon}_i^{^T} \end{pmatrix} = \begin{pmatrix} \varepsilon_{i,1}' & \varepsilon_{i,2}' \\ \varepsilon_{i,1} & \varepsilon_{i,2} \end{pmatrix} \,,
\end{equation}
that can be also written as
\[\boldsymbol{\mathfrak A}_i=2\omega  \boldsymbol{\varepsilon}_i+ 2\omega' \boldsymbol{\varepsilon'}_i, \quad i=1,\ldots,6 \,. \]

\begin{figure}
\begin{center}
\unitlength 0.7mm \linethickness{0.6pt}
\begin{picture}(150.00,80.00)
\put(9.,33.){\line(1,0){12.}} \put(9.,33.){\circle*{1}}
\put(21.,33.){\circle*{1}} \put(10.,29.){\makebox(0,0)[cc]{$e_1$}}
\put(21.,29.){\makebox(0,0)[cc]{$e_2$}}
\put(15.,33.){\oval(20,30.)}
\put(8.,17.){\makebox(0,0)[cc]{$\mathfrak{ a}_1$}}
\put(15.,48.){\vector(1,0){1.0}}
\put(32.,33.){\line(1,0){9.}} \put(32.,33.){\circle*{1}}
\put(41.,33.){\circle*{1}} \put(33.,29.){\makebox(0,0)[cc]{$e_3$}}
\put(42.,29.){\makebox(0,0)[cc]{$e_4$}}
\put(37.,33.){\oval(18.,26.)}
\put(30.,19.){\makebox(0,0)[cc]{$\mathfrak{a}_2$}}
\put(36.,46.){\vector(1,0){1.0}}
\put(100.,33.00) {\line(1,0){33.}} \put(100.,33.){\circle*{1}}
\put(133.,33.){\circle*{1}}
\put(101.,29.){\makebox(0,0)[cc]{$e_{5}$}}
\put(132.,29.){\makebox(0,0)[cc]{$e_{6}=\infty$}}
\put(59.,58.){\makebox(0,0)[cc]{$\mathfrak{b}_1$}}
\put(63.,62.){\vector(1,0){1.0}}
\bezier{484}(15.,33.00)(15.,62.)(65.,62.)
\bezier{816}(65.00,62.)(119.00,62.00)(119.00,33.00)
\bezier{35}(15.,33.00)(15.,5.)(65.,5.)
\bezier{35}(65.00,5.)(119.00,5.00)(119.00,33.00)
\put(70.,44.){\makebox(0,0)[cc]{$\mathfrak{b}_2$}}
\put(74.00,48.){\vector(1,0){1.0}}
\bezier{384}(37.,33.00)(37.,48.)(76.00,48.)
\bezier{516}(76.00,48.)(111.00,48.00)(111.00,33.00)
\bezier{30}(37.,33.00)(37.,19.)(76.00,19.)
\bezier{30}(76.00,19.)(111.00,19.00)(111.00,33.00)
\end{picture}
\end{center}
\caption{Homology basis on the Riemann surface of the curve $X_2$ with real branch points $e_1 < e_2 <\ldots < e_{6}=\infty$ (upper sheet). The cuts are drawn from $e_{2i-1}$ to $e_{2i}$, $i=1,2,3$. The $\mathfrak{b}$--cycles are completed on the lower sheet (dotted lines).} \label{figure-2}
\end{figure}
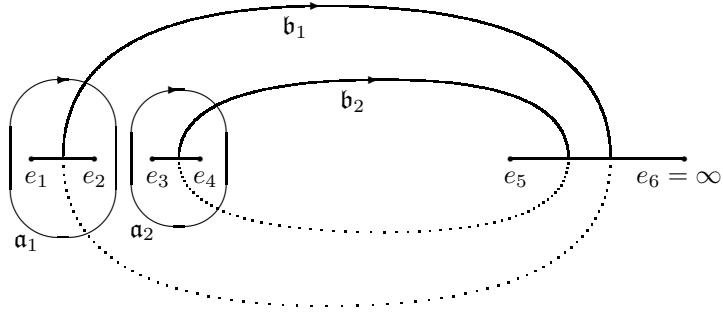

In the homology basis given in Figure \ref{figure-2} the characteristics of the branch points are
\begin{align}
[\boldsymbol{\mathfrak A}_1] & = \frac{1}{2}
\begin{pmatrix} 1 & 0 \\ 0 & 0 \end{pmatrix} \, , & \quad
[\boldsymbol{\mathfrak A}_2] & = \frac{1}{2}
\begin{pmatrix} 1 & 0 \\ 1 & 0 \end{pmatrix} \, , & \quad
[\boldsymbol{\mathfrak A}_3] & = \frac{1}{2}
\begin{pmatrix} 0 & 1 \\ 1 & 0 \end{pmatrix} \\
[\boldsymbol{\mathfrak A}_4] & = \frac{1}{2}
\begin{pmatrix} 0 & 1 \\ 1 & 1 \end{pmatrix}\, , & \quad
[\boldsymbol{\mathfrak A}_5] & = \frac{1}{2}
\begin{pmatrix} 0 & 0 \\ 1 & 1 \end{pmatrix}\, , & \quad
[\boldsymbol{\mathfrak A}_6] & = \frac{1}{2}
\begin{pmatrix} 0 & 0 \\ 0 & 0 \end{pmatrix} \, .
\label{hombas2}
\end{align}
The characteristics of the vector of Riemann constants $\boldsymbol{K}_{\infty}$ yield
\begin{equation}
[\boldsymbol{K}_{\infty}] = [\boldsymbol{\mathfrak{A}}_2]+ [\boldsymbol{\mathfrak{A}}_4] = \frac{1}{2} \begin{pmatrix} 1 & 1 \\ 0 & 1 \end{pmatrix} \, .
\end{equation}

From the above characteristics we build 16 half-periods. Denote 10 half-periods for $i\neq j = 1,\ldots, 5$ that are images of two branch points as
\begin{align}
\boldsymbol{\Omega}_{ij}=2\omega ( \boldsymbol{\varepsilon}_i+\boldsymbol{\varepsilon}_j)+
2\omega' (\boldsymbol{\varepsilon}_i'+  \boldsymbol{\varepsilon}_j') ,\quad  i = 1, \ldots, 5 \,.
\label{even}
\end{align}
Then the characteristics of the $6$ half-periods
\begin{equation}
\left[ (2\omega)^{-1} \boldsymbol{\mathfrak A}_i+\boldsymbol{K}_{\infty} \right] =: \delta_i  ,\quad i= 1, \ldots, 6
\end{equation}
are nonsingular and odd, whereas the characteristics of the $10$ half-periods
\begin{equation}
\left[(2\omega)^{-1} \boldsymbol{\Omega}_{ij}+\boldsymbol{K}_{\infty}\right] =: \varepsilon_{ij}, \quad 1 \leq i < j \leq 5
\end{equation}
are nonsingular and even.

Odd characteristics correspond to partitions $\{6\}\cup \{ 1,\ldots,5 \}$ and
$\{k\}\cup\{i_1,\ldots,i_4,6\}$ for $i_1,\ldots,i_4\neq k$. The first partition from these two corresponds to $\Theta_0$ and the second to $\Theta_1$.

The correspondence between branch points and $16=10+6$ characteristics is given (in the fixed homology basis as follows)
\begin{align*}\label{}
\boldsymbol{\Omega}_{1,2}&=[ \boldsymbol{ \mathfrak
A}_1+\boldsymbol{ \mathfrak
A}_2+\boldsymbol{K}_{\infty}]=\left[\begin{array}{cc}1&1\\
1&1\end{array}\right]\quad \equiv   \{1,2\}\\
\boldsymbol{\Omega}_{1,3}&=[ \boldsymbol{ \mathfrak
A}_1+\boldsymbol{ \mathfrak
A}_3+\boldsymbol{K}_{\infty}]=\left[\begin{array}{cc}0&0\\
1&1\end{array}\right]\quad \equiv  \{1,3\}\\
\boldsymbol{\Omega}_{1,4}&=[ \boldsymbol{ \mathfrak
A}_1+\boldsymbol{ \mathfrak
A}_4+\boldsymbol{K}_{\infty}]=\left[\begin{array}{cc}0&0\\
1&0\end{array}\right]\quad \equiv \{1,4\}\\
\boldsymbol{\Omega}_{1,5}&=[ \boldsymbol{ \mathfrak
A}_1+\boldsymbol{ \mathfrak
A}_5+\boldsymbol{K}_{\infty}]=\left[\begin{array}{cc}0&1\\
1&0\end{array}\right]\quad \equiv \{1,5\}\\
\boldsymbol{\Omega}_{2,3}&=[ \boldsymbol{ \mathfrak
A}_2+\boldsymbol{ \mathfrak
A}_3+\boldsymbol{K}_{\infty}]=\left[\begin{array}{cc}0&0\\
0&1\end{array}\right]\quad \equiv \{2,3\}\\
\boldsymbol{\Omega}_{2,4}&=[ \boldsymbol{ \mathfrak
A}_2+\boldsymbol{ \mathfrak
A}_4+\boldsymbol{K}_{\infty}]=\left[\begin{array}{cc}0&0\\
0&0\end{array}\right]\quad \equiv \{2,4\}\\
\boldsymbol{\Omega}_{2,5}&=[ \boldsymbol{ \mathfrak
A}_2+\boldsymbol{ \mathfrak
A}_5+\boldsymbol{K}_{\infty}]=\left[\begin{array}{cc}0&1\\
0&0\end{array}\right]\quad \equiv \{2,5\}\\
\boldsymbol{\Omega}_{3,4}&=[ \boldsymbol{ \mathfrak
A}_3+\boldsymbol{ \mathfrak
A}_4+\boldsymbol{K}_{\infty}]=\left[\begin{array}{cc}1&1\\
0&0\end{array}\right]\quad \equiv \{3,4\}\\
\boldsymbol{\Omega}_{3,5}&=[ \boldsymbol{ \mathfrak
A}_3+\boldsymbol{ \mathfrak
A}_5+\boldsymbol{K}_{\infty}]=\left[\begin{array}{cc}1&0\\
0&0\end{array}\right]\quad \equiv \{3,5\}\\
\boldsymbol{\Omega}_{4,5}&=[ \boldsymbol{ \mathfrak
A}_4+\boldsymbol{ \mathfrak
A}_5+\boldsymbol{K}_{\infty}]=\left[\begin{array}{cc}1&0\\
0&1\end{array}\right]\quad \equiv \{4,5\}
\end{align*}

\begin{align*}\label{}
\boldsymbol{\Omega}_{1}&=[ \boldsymbol{ \mathfrak
A}_1+\boldsymbol{K}_{\infty}]=\left[\begin{array}{cc}0&1\\
0&1\end{array}\right]\quad \equiv \{1\}\\
\boldsymbol{\Omega}_{2}&=[ \boldsymbol{ \mathfrak
A}_2+\boldsymbol{K}_{\infty}]=\left[\begin{array}{cc}0&1\\
1&1\end{array}\right]\quad \equiv \{2\}\\
\boldsymbol{\Omega}_{3}&=[ \boldsymbol{ \mathfrak
A}_3+\boldsymbol{K}_{\infty}]=\left[\begin{array}{cc}1&0\\
1&1\end{array}\right]\quad \equiv \{3\}\\
\boldsymbol{\Omega}_{4}&=[ \boldsymbol{ \mathfrak
A}_4+\boldsymbol{K}_{\infty}]=\left[\begin{array}{cc}1&0\\
1&0\end{array}\right]\quad \equiv \{4\}\\
\boldsymbol{\Omega}_{5}&=[ \boldsymbol{ \mathfrak
A}_5+\boldsymbol{K}_{\infty}]=\left[\begin{array}{cc}1&1\\
1&0\end{array}\right]\quad \equiv \{5\}\\
\boldsymbol{\Omega}_{6}&=[ \boldsymbol{ \mathfrak
A}_6+\boldsymbol{K}_{\infty}]=\left[\begin{array}{cc}1&1\\
1&0\end{array}\right]\quad \equiv \{6\}
\end{align*}

From the solution of the Jacobi inversion problem we obtain for any $i,j =1,\ldots,5$, $i\neq j$
\begin{equation}
e_i+e_j=\wp_{22}(\boldsymbol{\Omega}_{ij}),\quad -e_ie_j=\wp_{12}(\boldsymbol{\Omega}_{ij})\,.
\label{JIPE1}
\end{equation}
From the relation
\[ \wp_{11}(\boldsymbol{u}) = \frac{F(x_1,x_2)-2y_1y_2}{4(x_1-x_2)^2} \]

one can also find
\begin{equation}
e_ie_j(e_p+e_q+e_r)+e_pe_qe_r=\wp_{11}(\boldsymbol{\Omega}_{ij}) \,,
\label{JIPE2}
\end{equation}
where $i$,$j$,$p$,$q$, and $r$ are mutually different.

$\wp_{i,j}(\boldsymbol{\Omega}_i)=\infty$ for all odd
half-periods. For even half-periods we have
\begin{align*}
\wp_{22}(\boldsymbol{\Omega}_{i,j})&=e_i+e_j,
\\
\wp_{12}(\boldsymbol{\Omega}_{i,j})&=-e_ie_j,\\
\wp_{11}(\boldsymbol{\Omega}_{i,j})&=e_{i,j}\equiv
e_ie_j(e_k+e_l+e_k)+e_ke_le_m.
\end{align*}
for all $1\leq i< j\leq 5  $ and $k\neq l \neq i\neq j$.

From~\eqref{JIPE1} and~\eqref{JIPE2} we obtain an expression for the matrix $\varkappa$ that is useful for numeric calculations because it reduces the second period matrix to an expression in the first period matrix and $\theta$--constants, namely, in the case $e_i=e_1,e_j=e_2$,
\begin{equation}
\varkappa=-\frac12 \left(\begin{array}{cc} e_1e_2(e_3+e_4+e_5)+e_3e_4e_5&-e_1e_2\\
-e_1e_2&e_1+e_2 \end{array}\right)-\frac12 {(2\omega)^{-1}}^T \frac{1}{\theta[\varepsilon]} \left( \begin{array}{cc}
\theta_{11}[\varepsilon]&\theta_{12}[\varepsilon]\\
\theta_{12}[\varepsilon]&\theta_{22}[\varepsilon] \end{array} \right)(2\omega)^{-1} \,,
\label{kappa2}
\end{equation}
where the characteristic $[\varepsilon]$ in the fixed homology basis reads
\[  [\varepsilon]=[\boldsymbol{\mathfrak{A}}_1]+ [\boldsymbol{\mathfrak{A}}_2]+[\boldsymbol{K}_{\infty}]
=\left[\begin{array}{cc} \frac12& \frac12\\
\frac12&\frac12\end{array}\right].   \]
Because exist $C_{5}^2=10$ variants to choose $[\varepsilon]$ one could sum up them to find $\varkappa$ in the form
\begin{equation}
\varkappa=\frac{1}{20} \left(\begin{array}{cc} \lambda_2& \frac34\lambda_3\\
\frac34\lambda_3&\lambda_4 \end{array}\right)-\frac{1}{20} {(2\omega)^{-1}}^T  \sum_{\rm{even}\; \varepsilon } \frac{1}{\theta[\varepsilon]} \left( \begin{array}{cc}
\theta_{11}[\varepsilon]&\theta_{12}[\varepsilon]\\
\theta_{12}[\varepsilon]&\theta_{22}[\varepsilon] \end{array} \right)(2\omega)^{-1} \,,
\label{kappa2a}
\end{equation}

\subsection{Inversion of a holomorphic integral}
Taking the limit $z_2\rightarrow \infty$ in the Jacobi inversion problem (\ref{JIP}) we obtain
\begin{equation}
\int_{\infty}^{(z,w)} \frac{\mathrm{d}z}{w}=u_1,\quad \int_{\infty}^{(z,w)} \frac{z\mathrm{d}z}{w}=u_2 \,.
\end{equation}
The same limit in the ratio
\begin{equation}
\frac{\wp_{12}(\boldsymbol{u})}{\wp_{22}(\boldsymbol{u})} = - \frac{z_1z_2}{z_1+z_2}
\end{equation}
leads to the Grant-Jorgenson formula (\cite{gr91},\cite{jor92}),
\begin{equation}
z=-\left.\frac{\sigma_1(\boldsymbol{u})}{\sigma_2(\boldsymbol{u})}\right|_{(\sigma)}
\label{grjor}\end{equation}
In terms of $\theta$--functions this can be given the form
\begin{equation}
z = \left.-\frac{\partial_{\boldsymbol{U}}  \theta[\boldsymbol{K}_{\infty}]((2\omega)^{-1}\boldsymbol{u};\tau)}{\partial_{\boldsymbol{V}}  \theta[\boldsymbol{K}_{\infty}]((2\omega)^{-1}\boldsymbol{u};\tau)}\right|_{\theta((2\omega)^{-1}\boldsymbol{u};\tau)=0},
\end{equation}
where here and below $\partial_{\boldsymbol{U}} = \sum_{j=1}^g U_j \frac{\partial}{\partial z_j}$ is the derivative along the direction $\boldsymbol{U}$. Here we introduced the ``winding vectors" $\boldsymbol{U}$, $\boldsymbol{V}$ as column vectors of the inverse matrix
\begin{equation}
(2\omega)^{-1}=(\boldsymbol{U},\boldsymbol{V}) \,.
\end{equation}

For the $w$ coordinate two equivalent representations can be given
\begin{align}
\begin{split}
w&=-\left.\frac12\frac{\sigma(2\boldsymbol{u})}{\sigma_2^4(\boldsymbol{u})}\right|_{(\sigma)}\\
w&=\left.\frac{1}{\sigma_2(\boldsymbol{u})}\left(\sigma_{11}(\boldsymbol{u})+2z\sigma_{12}(\boldsymbol{u})+z^2\sigma_{22}(\boldsymbol{u})\right)\right|_{(\sigma)}
\end{split}\label{wcoordinate}
\end{align}

From~\eqref{grjor} we obtain for all finite branch points
\begin{equation}
e_i=-\frac{\sigma_1(\boldsymbol{\mathfrak A}_i ) }{\sigma_2(\boldsymbol{\mathfrak A}_i) }, \quad \text{ equivalently}, \quad e_i=-\frac{\partial_{\boldsymbol{U}}\theta[\delta_i]}{\partial_{\boldsymbol{V}}\theta[\delta_i]}, \quad i=1,\ldots,5 \label{e2a} \,.
\end{equation}
This formula was mentioned by Bolza \cite{bolza86} (see his Eq.~(6)) for the case of a genus two curve with finite branch points.

The $\zeta$-formula reads

\begin{align}\begin{split}
&-\zeta_1(\boldsymbol{u})+2\mathfrak{n}_1+\frac12 \frac{w_1-w_2}{z_1-z_2}=\int_{(e_2,0)}^{(z_1,w_1)}\mathrm{d}r_1(z,w)+
\int_{(e_4,0)}^{(z_2,w_2)}\mathrm{d}r_1(z,w),\\
&-\zeta_2(\boldsymbol{u})+2\mathfrak{n}_2=\int_{(e_2,0)}^{(z_1,w_1)}\mathrm{d}r_2(z,w)+
\int_{(e_4,0)}^{(z_2,w_2)}\mathrm{d}r_2(z,w) \ , \end{split} \label{zetaformula2}
\end{align}
where $\mathfrak{n}_j=\sum_{i=1}^{2}\eta^{\prime}_{ji}{\varepsilon}'_{i} + \eta_{ji}{\varepsilon}_{i}$. Here the characteristics ${\varepsilon}'_{i}$ and ${\varepsilon}_{i}$ of $\boldsymbol{K}_{\infty}$ are not reduced.

Choosing $(z_1,w_1)=(Z,W)$, $(z_2,w_2)=(e_4,0) $ we get from (\ref{zetaformula2})

\begin{align}\begin{split}
&-\zeta_1\left(\int_{(e_2,0)}^{(Z,W)} \mathrm{d}\boldsymbol{u} + \boldsymbol{K}_{\infty} \right)+2\mathfrak{n}_1 +\frac12\frac{W}{Z-e_4}=\int_{(e_2,0)}^{(Z,W)}\mathrm{d}r_1(z,w),\\
&-\zeta_2\left(\int_{(e_2,0)}^{(Z,W)} \mathrm{d}\boldsymbol{u} + \boldsymbol{K}_{\infty} \right)+2\mathfrak{n}_2=\int_{(e_2,0)}^{(Z,W)}\mathrm{d}r_2(z,w) \ .
\end{split} \label{zeta2}
\end{align}

The inversion formula for the integral of the third kind  is written as

 \begin{align}\begin{split}
& W\int_{P'}^P\frac{1}{x-Z}\frac{\mathrm{d}x}{y}
=-2\left(\boldsymbol{u}^T-{\boldsymbol{u}'}^T\right)
\int_{(e_2,0)}^{(Z,W)} \mathrm{d}\boldsymbol{r}
+\mathrm{ln} \frac{\sigma\left(\boldsymbol{u}-\boldsymbol{v} - \boldsymbol{K}_{\infty} \right)}{\sigma\left(\boldsymbol{u}+\boldsymbol{v} - \boldsymbol{K}_{\infty}  \right)}-\mathrm{ln} \frac{\sigma\left(\boldsymbol{u}'-\boldsymbol{v} - \boldsymbol{K}_{\infty} \right)}{\sigma\left(\boldsymbol{u}'+\boldsymbol{v} - \boldsymbol{K}_{\infty}  \right)}
\end{split} \label{thirdinv2}
\end{align}
with
\[\boldsymbol{v}=\int_{(e_2,0)}^{(Z,W)}\mathrm{d}\boldsymbol{u},\quad \boldsymbol{u}=\int_{\infty}^{P}\mathrm{d}\boldsymbol{u},\quad \boldsymbol{u}'=\int_{\infty}^{P'}\mathrm{d}\boldsymbol{u} \]
and $\boldsymbol{u}\in\Theta_1$, $\boldsymbol{u}'\in\Theta_1$. The integrals $\displaystyle{\int_{(e_2,0)}^{(Z,W)} \mathrm{d}\boldsymbol{r}}$ are given by the formula~\eqref{zeta2}.

In the case when the base point $P'$ is chosen to be a branch point, say $(e_2,0)$ then the final formula takes the form

\begin{eqnarray}
 W\int_{(e_2,0)}^{P}\frac{1}{x-Z}\frac{\mathrm{d}x}{y}
&=& 2\left(\boldsymbol{u}^T-\boldsymbol{\mathfrak{A}}_2^T\right) \left[\boldsymbol{\zeta}(\boldsymbol{v}+\boldsymbol{K}_{\infty})-2( \eta^{\prime}\boldsymbol{\varepsilon}'_{\boldsymbol{K}_{\infty}} + \eta\boldsymbol{\varepsilon}_{\boldsymbol{K}_{\infty}} )-\frac12 \boldsymbol{\mathfrak{Z}}(Z,W)\right]   \nonumber \\ & + &\mathrm{ln} \frac{\sigma\left(\boldsymbol{u}-\boldsymbol{v} - \boldsymbol{K}_{\infty} \right)}{\sigma\left(\boldsymbol{u}+\boldsymbol{v} - \boldsymbol{K}_{\infty}  \right)}-\mathrm{ln} \frac{\sigma\left(\boldsymbol{\mathfrak{A}}_2-\boldsymbol{v} - \boldsymbol{K}_{\infty} \right)}{\sigma\left(\boldsymbol{\mathfrak{A}}_2+\boldsymbol{v}  - \boldsymbol{K}_{\infty} \right)} \ .
 \label{thirdinv2final1}
\end{eqnarray}

\section{Modular equation for $\sigma$-function and recursion}
Denote the vector
\begin{equation*}
\begin{pmatrix}
-\sigma+\frac{1}{3}u_{2}\sigma_{2}+u_{1}\sigma_{1}\\ \\
\big( \frac{3}{8}\lambda_{1}-\frac{1}{40}\lambda_{3}(
\lambda_{3}u_{1}^{2}+3 u_{2}^{2})
\big)\sigma
-\frac{1}{5}\lambda_{3}u_{1}\sigma_{2}+u_{2}\sigma_{1}
+\frac{1}{2}\sigma_{2,2}\\ \\
\big(2\lambda_{0}u_{1}u_{2}-\frac{1}{10}\lambda_{1}u_{2}^2
-\frac{1}{2}\lambda_{2}-\frac{3}{40}\lambda_{3}\lambda_{1}u_{1}^2
\big)\sigma-\frac{3}{5}\lambda_{1}u_{1}\sigma_{2}+2\sigma_{1,1}\\ \\
\big(
\frac{3}{4}\lambda_{0}u_{1}^2 +
\frac{1}{4}\lambda_{1}u_{1}u_{2}-
\frac{1}{20}\lambda_{2}u_{2}^2-
\frac{3}{80}\lambda_{3} \lambda_{2}u_{1}^2
\big)
-\big(\frac{3}{10}\lambda_{2}u_{1}+\frac{1}{12}\lambda_{3}\big)\sigma_{2}
+\sigma_{1,2}
\end{pmatrix}
\end{equation*}
by $\boldsymbol{\xi}$ and the matrix
\begin{equation*}
\begin{pmatrix}
-\frac{10}{3}\lambda_{0}&-\frac{8}{3}\lambda_{1}&-2\lambda_{2}&
-\frac{4}{3}\lambda_{3}\\ \\
\frac{1}{5}\lambda_{1}\lambda_{3}&-10\lambda_{0}+
\frac{2}{5}\lambda_{2}\lambda_{3}&-8\lambda_{1}+\frac{3}{5}\lambda_{3}^{2}&
-6\lambda_{2}\\ \\
\frac{8}{5}\lambda_{1}^{2}-4\lambda_{0}\lambda_{2}&
\frac{6}{5}\lambda_{1}\lambda_{2}-6\lambda_{0}\lambda_{3}&
\frac{4}{5}\lambda_{1}\lambda_{3}&-40\lambda_{0}\\ \\
\frac{3}{10}\lambda_{1}\lambda_{2}-\frac{2}{3}\lambda_{0}\lambda_{3}&
\frac{3}{5}\lambda_{2}^{2}-\frac{1}{3}\lambda_{1}\lambda_{3}&
-10\lambda_{0}+\frac{9}{10}\lambda_{2}\lambda_{3}&
-8\lambda_{1}+\frac{1}{3}\lambda_{3}^2
\end{pmatrix}
\end{equation*}
by $m$. Then for
\begin{equation*}
\boldsymbol{\Lambda}=\bigg(\frac{\partial \sigma}{\partial \lambda_{0}},
\frac{\partial \sigma}{\partial \lambda_{1}},
\frac{\partial \sigma}{\partial \lambda_{2}},
\frac{\partial \sigma}{\partial \lambda_{3}}\bigg)^T,
\end{equation*}
we have
\begin{equation}
\boxed{\boldsymbol{\xi}+m\boldsymbol{\Lambda}=0}\label{def-1}
\end{equation}
iff $\sigma=\sigma(u_1,u_2;\lambda_{0},\lambda_{1},\lambda_{2},\lambda_{3})$ is
the fundamental $\sigma$--function associated with a genus $2$
curve
\begin{equation*}
V=\big\{(x,y)\in\mathbf{C}^2\,|\,f(x,y)=y^2-4x^5-\sum_{i=0}^{3}\lambda_{i}x^{i}=0\big\}.
\end{equation*}
Note, that $\det m$ is proportional to the discriminant
$\Delta_{z}\big(\Delta_{w}\big(f(z,w)\big)\big)$.

As usual, denote
\begin{gather*}
\zeta_1(\boldsymbol u)=\frac{\partial}{\partial u_1}\mathrm{ ln}\;
\sigma(\boldsymbol u),\quad \zeta_2(\boldsymbol
u)=\frac{\partial}{\partial u_2}\mathrm{ ln}\; \sigma(\boldsymbol
u),\quad \\  \wp_{11}(\boldsymbol u)=-\frac{\partial^2}{
\partial u_1^2}\mathrm{ ln}\; \sigma(\boldsymbol
u),\quad \wp_{12}(\boldsymbol u)=-\frac{\partial^2}{\partial
u_1\partial u_2}\mathrm{ ln}\; \sigma(\boldsymbol u),\\
\wp_{22}(\boldsymbol u)=-\frac{\partial^2}{\partial u_2^2}\mathrm{
ln}\; \sigma(\boldsymbol u). \end{gather*}

Let us discuss the structure of the equations
\eqref{def-1}.

Note first,
that the matrix $m$ is equivalent to the symmetric
matrix \begin{equation*} \mu=\begin{pmatrix} -4\lambda_{3} & -6\lambda_{2} &
-8\lambda_{1} & -10\lambda_{0} \\ \\ -6 \lambda_{2} & -8\lambda_{1} +
\frac{3}{5} \lambda_{3}^{2}& -10 \lambda_{0} +\frac{2}{5}\lambda_{2}
\lambda_{3} & \frac{1}{5}\lambda_{1} \lambda_{3} \\ \\ -8\lambda_{1} &
-10\lambda_{0}+\frac{2}{5}\lambda_{2} \lambda_{3} & \frac{3 }{5}\lambda_{2}^{2}
- \lambda_{1}\lambda_{3} & \frac{3}{10}\lambda_{1}\lambda_{2}-
\frac{3}{2}\lambda_{0}\lambda_{3} \\ \\ -10\lambda_{0} &
\frac{1}{5}\lambda_{1}\lambda_{3}&\frac{3}
{10}\lambda_{1}\lambda_{2}-\frac{3}{2}\lambda_{0}\lambda_{3}
&\frac{2}{5}\lambda_{1}^{2}-\lambda_{0}\lambda_{2}
\end{pmatrix}.
\end{equation*}
This is the reason to look for a representation of \eqref{def-1} with the help
of the symmetric matrices only.

Let $\mathcal{V}_4$ be a $4$--dimensional vector space over the
ring $\mathcal{R}$ of entire functions of  the six variables
$\{u_1,u_2;\lambda_0,\lambda_1, \lambda_2,\lambda_3\}$.
Introduce a pair of linear first-order operators
acting from
$\mathcal{R}$ to $\mathcal{V}_{4}$
\begin{equation*}\boldsymbol{\epsilon}=\begin{pmatrix}
\partial/\partial \lambda_{3}\\
\partial/\partial \lambda_{2}\\
\partial/\partial \lambda_{1}\\
\partial/\partial \lambda_{0}\end{pmatrix}\quad
\text{
and}\quad\boldsymbol{\delta}(\boldsymbol{\alpha})
=\begin{pmatrix}\alpha_{1}\partial/\partial u_1\\
\alpha_{2}\partial/\partial u_2\\
\alpha_{3} u_2\\
\alpha_{4} u_1\end{pmatrix}
\end{equation*} with $\boldsymbol{\alpha}\in\mathbf{Z}^4$.
Introduce a row-vector
$\boldsymbol{\gamma}=(\gamma_1,\gamma_2,\gamma_3,\gamma_4)$ with $\gamma_i$
being symmetric $4\times4$--matrices over the ring
$\mathbf{Q}[\lambda_3,\lambda_2,\lambda_1,\lambda_0]$:
\begin{align*}
\gamma_1&= \begin{pmatrix} 0 & 0 & 0 & \frac{1}{2}\\ \\ 0 & 0 &
 -\frac{1872}{6125} & 0 \\ \\ 0 & -\frac{1872}{6125} & 0 &
-\frac{3}{28}\lambda_{3} \\ \\
\frac{1}{2}& 0 & -\frac{3}{28}\lambda_{3}& 0
\end{pmatrix};\\ \\
\gamma_2&=
\begin{pmatrix}
0 & 0 & \frac{3132}{6125} & 0 \\ \\
0 & \frac{648}{30625} & 0 & 0 \\  \\
\frac{3132}{6125} & 0 & \frac{486}{6125}\lambda_{3}
 &\frac{162}{6125}\lambda_{2} \\  \\
0 & 0 & \frac{162}{6125}\lambda_{2} &
\frac{9}{12250}(25\lambda_{3}^2-339\lambda_{1})\end{pmatrix};
\\ \\
\gamma_3&=
\begin{pmatrix}
0 & -\frac{4428}{6125} & 0 & \frac{3}{28}\lambda_{3} \\ \\
-\frac{4428}{6125} & 0 & -\frac{486}{6125}\lambda_{3} &
-\frac{162}{6125}\lambda_{2} \\ \\
0 &-\frac{486}{6125}\lambda_{3}  & -\frac{324}{6125}\lambda_{2}
& -\frac{162}{6125}\lambda_{1} \\ \\
\frac{3}{28}\lambda_{3}
&-\frac{162}{6125}\lambda_{2}&-\frac{162}{6125}\lambda_{1} &
\frac{27}{245}(4\lambda_{2}\lambda_{3}-\lambda_{0}) \end{pmatrix};\\ \\
\gamma_4&=\begin{pmatrix}
-\frac{43}{14}& 0 & -\frac{3}{28}\lambda_{3}& 0 \\ \\
0 &\frac{216}{6125}\lambda_{3} & \frac{162}{6125}\lambda_{2} &
\frac{9}{490}( 15\lambda_{1}-\lambda_{3}^{2}) \\ \\
-\frac{3}{28}\lambda_{3} & \frac{162}{6125}\lambda_{2}&
\frac{216}{875}\lambda_{1} &
\frac{27}{980}(20\lambda_{0}-\lambda_{2}\lambda_{3}) \\ \\
0 & \frac{9}{490}(15\lambda_{1}- \lambda_{3}^{2}) &
\frac {27}{980}( 20\lambda_{0} - \lambda_{2}\lambda_{3})&
-\frac{27} {1960}\lambda_{1}\lambda_{3}
\end{pmatrix}.
\end{align*}

Then the equations $\eqref{def-1}$ are equivalent to
\begin{equation}\label{def-2}
\Big\{\big[\boldsymbol{\gamma}*\boldsymbol{
\delta}(36,175,-35,-42)\big]\,
\boldsymbol{
\delta}(36,175,-35,-42)+
6^4\mu \boldsymbol{\epsilon}\Big\}\sigma=0,
\end{equation}
where $\boldsymbol{\gamma}*\boldsymbol{\delta}=\sum\gamma_i\delta_i$.
The representation \eqref{def-2}, that is the set
$\{\boldsymbol{\gamma},\boldsymbol{\alpha}\}$, is unique.
In the case of genus $1$ ($\sigma=\sigma(u;g_2,g_3)$ is the Weierstrass
function associated with an elliptic curve defined by the equation
$y_2-4x_3+g_2x+g_3=0$) the analogue of \eqref{def-2} is
\begin{equation*}
\left\{
\left[
\left(
\begin{pmatrix}
0&\frac{1}{2}\\
\frac{1}{2}&0
\end{pmatrix},\begin{pmatrix}
-1&0\\
0&\frac{g_2}{96}
\end{pmatrix}\right)*\begin{pmatrix}
\frac{\partial}{\partial u}\\
2 u
\end{pmatrix}
\right] \begin{pmatrix}
\frac{\partial}{\partial u}\\
2 u
\end{pmatrix}+\begin{pmatrix}
4g_2 & 6 g_3 \\
6 g_3 & \frac{g_2^2}{3}
\end{pmatrix}\begin{pmatrix}
\frac{\partial}{\partial g_2}\\
\frac{\partial}{\partial g_3}
\end{pmatrix}
\right\}\sigma =0,
\end{equation*}
which is also a unique representation, if we restrict ourselves to ring
operations.

The genus $2$ $\sigma$-function is represented by a formal series
\begin{multline*}
\sigma(u_1,u_2;\lambda_{0},\lambda_{1},\lambda_{2},\lambda_{3})=\\
\sum a_{\ell,m,n,o,p}u_2^{3-3 l+4 m+6 n+8
o+10 p}u_1^{\ell}\lambda_3^m \lambda_2^n\lambda_1^o\lambda_0^p=\\
u_{2}^{3}\sum
        a_{\ell,m,n,o,p}\Big(\frac{u_1}{u_{2}^{3}}\Big)^{\ell}
(\lambda_{3} u_{2}^4)^m
(\lambda_2 u_{2}^6)^n(\lambda_1 u_{2}^8)^o(\lambda_0 u_{2}^{10})^p,
\end{multline*} where the summation is carried over non-negatives
integer $\ell,m,n,o,p$ such that $4 m+6 n+8 o+10 p+3\geqslant 3
l$. Coefficients $a_{\ell,m,n,o,p}$ are rational numbers,
particularly, \begin{equation}\label{init}
a_{0,0,0,0,0}=-\frac{1}{3}\quad\text{and}\quad a_{1,0,0,0,0}=1.
\end{equation}
The equations \eqref{def-2} are equivalent to the following three
recursion equations with respect to the coefficients
$a_{\ell,m,n,o,p}$ (the first of the equations is trivial):

\begin{align*}
&a_{\ell-2,m-2,n,o,p}
-15a_{\ell-2,m,n,o-1,p}\\&\qquad
-8(3\ell-4m-6n-8o-10p-2)a_{\ell-1,m-1,n,o,p}\\&\qquad
-24(n+1)a_{\ell,m-2,n+1,o,p}
-16(o+1)a_{\ell,m-1,n-1,o+1,p}\\&\qquad
-8(p+1)a_{\ell,m-1,n,o-1,p+1}
+3a_{\ell,m-1,n,o,p}\\&\qquad
-20(3\ell-4m-6n-8o-10p-3)(3\ell-4m-6n-8o-10p-2)a_{\ell,m,n,o,p}\\&\qquad
+400(o+1)a_{\ell,m,n,o+1,p-1}
+320(n+1)a_{\ell,m,n+1,o-1,p}\\&\qquad
+240(m+1)a_{\ell,m+1,n-1,o,p}
-40(\ell+1)a_{\ell+1,m,n,o,p}
=0;
\\ \\
&3a_{\ell-2,m-1,n-1,o,p}
-60a_{\ell-2,m,n,o,p-1}\\&\qquad
-24(3\ell-4m-6n-8o-10p)a_{\ell-1,m,n-1,o,p}\\&\qquad
-20a_{\ell-1,m,n,o-1,p}
-4(5\ell+8n-20o-30p-5)a_{\ell,m-1,n,o,p}\\&\qquad
-48(o+1)a_{\ell,m,n-2,o+1,p}
-24(p+1)a_{\ell,m,n-1,o-1,p+1}\\&\qquad
+4a_{\ell,m,n-1,o,p}+
800(n+1)a_{\ell,m,n+1,o,p-1}\\&\qquad
+640(m+1)a_{\ell,m+1,n,o-1,p}\\&\qquad
+80(\ell+1)(3\ell-4m-6n-8o-10p)a_{\ell+1,m,n,o,p}
=0;
\\ \\
&3a_{\ell-2,m-1,n,o-1,p}-
24(3\ell-4m-6n-8o-10p+2)a_{\ell-1,m,n,o-1,p}\\&\qquad
-80a_{\ell-1,m,n,o,p-1}
+240(o+1)a_{\ell,m-1,n,o+1,p-1}\\&\qquad
-32(n+1)a_{\ell,m-1,n+1,o-1,p}
-4(12o-40p-5)a_{\ell,m,n-1,o,p}\\&\qquad
-64(p+1)a_{\ell,m,n,o-2,p+1}
+4a_{\ell,m,n,o-1,p}\\&\qquad
+1600(m+1)a_{\ell,m+1,n,o,p-1}
-80(\ell+1)(\ell+2)a_{\ell+2,m,n,o,p}
=0.
\end{align*}

The above equations together with initial conditions \eqref{init}
give the complete information about the expansion of
the $\sigma$--function.
\begin{align*}
\sigma(u_1,u_2;\lambda_{0},\lambda_{1},\lambda_{2},\lambda_{3})&=
\\&
u_1 -\frac{1}{3}u_2^3
\\&
-\left(
\frac{u_1 u_2^4}{48}
+\frac{u_2^7}{5040}
\right)
\lambda_{3}
\\&
+ \left(
\frac{u_1^3}{24}
-\frac{u_1^2 u_2^3}{24}
-\frac{u_1 u_2^6}{360}
+\frac{u_2^9}{22680}
\right)
\lambda_{2}
\\&
+ \left(
-\frac{u_1^3 u_2^2}{24}
-\frac{u_1^2 u_2^5}{120}
-\frac{u_1u_2^8}{5040}
+\frac{u_2^{11}}{99792}
\right)
\lambda_{1}
\\&
+\left(
-\frac{u_1^4 u_2}{12}
-\frac{u_1^3 u_2^4}{72}
-\frac{u_1^2 u_2^7}{504}
+\frac{u_1 u_2^{10}}{22680}
+\frac{u_2^{13}}{1389960}
\right)
\lambda_{0}
\\&+\dots
\end{align*}

\section{Jacobi inversion problem}
The equations of the Jacobi inversion problem
\index{Jacobi!inversion problem!solition of!for genus two}
\begin{align*}
  u_1&=\int_{a_1}^{x_1}\mathrm{d}u_1
+\int_{a_2}^{x_2}\mathrm{d}u_1, \\
u_2&=\int_{a_1}^{x_1}\mathrm{d}u_2 +\int_{a_2}^{x_2}\mathrm{d}u_2;
\end{align*} are equivalent to an algebraic equation
\begin{equation}
\mathcal{P}(x,\boldsymbol{u})=x^2-\wp_{22}(\boldsymbol{u})x
-\wp_{12}(\boldsymbol{ u})=0,\label{bolza2}\end{equation} that is,
the sought pair $(x_1,x_2)$ is the pair of roots of
 \eqref{bolza2}. So we have
\begin{equation}\wp_{22}(\boldsymbol{u})
=x_1+x_2,\;\wp_{12}(\boldsymbol{u})=-x_1x_2.\label{b1}\end{equation}
The corresponding $y_i$ is expressed as
\index{$\wp$--function!fundamental relation!for genus two}
 \begin{equation}
y_i=\wp_{222}(\boldsymbol{u})x_i+\wp_{122}(\boldsymbol{ u}),\quad
 i=1,2. \label{y2}\end{equation} There is the following expression
for the function $\wp_{11}(\boldsymbol{u})$ in terms of $x_1,x_2$
and $y_1,y_2$:  \begin{equation} \wp_{11}(\boldsymbol{u})=
\frac{F(x_1,x_2)-2y_1y_2}{4(x_1-x_2)^2}, \label{b2}\end{equation}
where $F(x_1,x_2)$ be Kleinian $2$-polar,
\begin{equation}F(x_1,x_2)=\sum_{r=0}^2x_1^rx_2^r[2\lambda_{2r}
+\lambda_{2r+1}(x_1+x_2)].\label{fx1x2}
\end{equation}

The following formulae are valid for the derivatives
\begin{eqnarray*}
\frac{\mathrm{d}x_1}{\mathrm{d}u_1}&=&-\frac{y_1x_2}{x_1-x_2},\quad
\frac{\mathrm{d}x_2}{\mathrm{d}u_1}=\frac{y_2x_1}{x_1-x_2},\cr
\frac{\mathrm{d}x_1}{\mathrm{d}u_2}&=&\frac{y_1}{x_1-x_2},\quad
\frac{\mathrm{d}x_2}{\mathrm{d}u_2}=-\frac{y_2}{x_1-x_2}.
\end{eqnarray*}

Further we have from  \eqref{y2}
\begin{eqnarray} \wp_{222}(\boldsymbol{
u})&=&\frac{y_1-y_2}{ x_1-x_2},\quad \wp_{221}(\boldsymbol{
u})=\frac{x_1y_2-x_2y_1}{ x_1-x_2},\nonumber\\
\wp_{211}(\boldsymbol{ u})&=&-\frac{x_1^2y_2-x_2^2y_1}{
x_1-x_2}\nonumber\\
\wp_{111}(\boldsymbol{u})&=&\frac{y_2\psi(x_1,x_2)-y_1\psi(x_2,x_1)}{
4(x_1-x_2)^3}, \label{thirdder} \end{eqnarray} where
\begin{eqnarray*}
\psi(x_1,x_2) &=& 4\lambda_0 + \lambda_1(3x_1 +
x_2) + 2\lambda_2x_1(x_1 +x_2) + \lambda_3x_1^2(x_1 +3x_2)\\
&+&4\lambda_4x_1^3x_2 +4x_1^3x_2(3x_1 +x_2).\end{eqnarray*}
\index{$\zeta$--function!principal relation!for genus two}

\section{$\zeta$-functions and $\wp$-function}

The relations between $\zeta$-functions and
the differentials of the second kind  are the following
\begin{align*}
-&\zeta_1\left(
\int_{a}^x
\mathrm{d}\mathbf{u}+\int_{a_1}^{x_1}\mathrm{d}\mathbf{u}+
\int_{a_2}^{x_2}\mathrm{d}\mathbf{u} \right)
=\int_{a}^x\mathrm{d}r_1+\int_{a_1}^{x_1}\mathrm{d}r_1
+
\int_{a_2}^{x_2}\mathrm{d}r_1\cr
+&f_{1}\big({}^{x}_{y}\big|
{}^{x_1\, x_2}
_{y_1\,y_2}\big),\\
-&\zeta_2\left(
\int_{a}^x\mathrm{d}\mathbf{u}+\int_{a_1}^{x_1}\mathrm{d}\mathbf{u}+
\int_{a_2}^{x_2}\mathrm{d}\mathbf{u} \right)=
\int_{a}^x\mathrm{d}r_2 +\int_{a_1}^{x_1}\mathrm{d}r_2
+\int_{a_2}^{x_2}\mathrm{d}r_2\cr +& f_{2}\big({}^{x}_{y}\big|
{}^{x_1\, x_2}_{y_1\,y_2}\big),
\end{align*}
where
\begin{align*}
f_{1}\big({}^{x}_{y}\big|
{}^{x_1\, x_2}_{y_1\,y_2}\big)
&=-\frac{y(x-x_1-x_2)}{2(x-x_1)(x-x_2)}
-\frac{y_1(x_1-x-x_2)}{2(x_1-x)(x_1-x_2)}\\
&-\frac{y_2(x_2-x-x_1)}{2(x_2-x)(x_2-x_1)}
\cr=& -\frac{y(x-\wp_{22}({\boldsymbol u}))-x\wp_{122}({\boldsymbol
u})-\wp_{112}({\boldsymbol u})}{2\mathcal{ P}(x,{\boldsymbol
u})}-\frac12\wp_{222}({\boldsymbol u}), \\
f_{2}\big({}^{x}_{y}\big|
{}^{x_1\, x_2}_{y_1\,y_2}\big)
&=
-\frac{y}{2(x-x_1)(x-x_2)}
-\frac{y_1}{2(x_1-x)(x_1-x_2)}\\
&-\frac{y_2}{2(x_2-x)(x_2-x_1)}\\
&=-\frac{y-x\wp_{222}({\boldsymbol u})-\wp_{122}({\boldsymbol
u})}{2\mathcal{ P}(x;\boldsymbol{u})}.
\end{align*}
Taking the limit $x\to a=\infty$, we have
\begin{align*}
-\zeta_1(\boldsymbol{u})
&=
 \int_{a_1}^{x_1}\mathrm{d}r_1 +\int_{a_2}^{x_2}\mathrm{d}r_1
 -\frac12\wp_{222}({\boldsymbol
u}),\\
-\zeta_2(\boldsymbol{u})&=
\int_{a_1}^{x_1}\mathrm{d}r_2 +\int_{a_2}^{x_2}\mathrm{d}r_2.
\end{align*}
\nopagebreak
All the possible pairwise products of the $\wp_{ijk}$-functions are
expressed  as follows in terms of $\wp_{22},\wp_{12},\wp_{11}$ and
constants $\lambda_s$ :
\begin{align*}
\wp_{222}^2&=4\wp_{11}+\lambda_3\wp_{22}+4\wp_{22}^3+4\wp_{12}\wp_{22}
+\lambda_4\wp_{22}^2+\lambda_2
,\\
\wp_{222}\wp_{221}&=\frac12\lambda_1+2\wp_{12}^2-2\wp_{11}\wp_{22}
+\frac12\lambda_3\wp_{12}
\\
&+4\wp_{12}\wp_{22}^2+\lambda_4\wp_{12}\wp_{22}
, \\
\wp_{221}^2&=\lambda_0-4\wp_{11}\wp_{12}+\lambda_4\wp_{12}^2
+4\wp_{22}\wp_{12}^2
,\\
\wp_{222}\wp_{211}&= -\frac12\lambda_1\wp_{22}+2\wp_{11}\wp_{22}^2
+2\wp_{22}\wp_{12}^2+\lambda_2\wp_{12}
\\
&+4\wp_{11}\wp_{12}
+\frac12\lambda_3\wp_{12}\wp_{22}
,\\
\wp_{222}\wp_{111}&=-4\wp_{11}^2-2\wp_{12}^3-\frac18\lambda_1\lambda_3-
\frac12\lambda_1\wp_{12}-\frac14\lambda_1\lambda_4\wp_{22}
\\
&-\lambda_1\wp_{22}^2-\lambda_2\wp_{11}-\frac12\lambda_3\wp_{11}\wp_{22}+
6\wp_{11}\wp_{22}\wp_{12}
\\
&-\lambda_3\wp_{12}^2+\frac12\lambda_2\lambda_4\wp_{12}
+2\lambda_2\wp_{12}\wp_{22}
\\
&+2\lambda_4\wp_{12}\wp_{11}
-\frac18\lambda_3^2\wp_{12}
,\\
\wp_{221}\wp_{211}&=2\wp_{12}^3+\frac12\lambda_3\wp_{12}^2
+\frac12\lambda_1\wp_{12}+2\wp_{11}\wp_{22}\wp_{12}
-\lambda_0\wp_{22}
,\\
\wp_{221}\wp_{111}&= 2\wp_{11}\wp_{12}^2
+\frac14\lambda_1\lambda_4\wp_{12}
+\lambda_1\wp_{22}\wp_{12}+\frac12\lambda_3\wp_{12}\wp_{11}
\\
&-\frac14\lambda_0\lambda_3-\lambda_0\wp_{12}-2\lambda_0\wp_{22}^2
-\frac12\wp_{22}\lambda_4\lambda_0
+2\wp_{22}\wp_{11}^2-\frac12\lambda_1\wp_{11}
,\\
\wp_{211}^2&=
\lambda_0\wp_{22}^2-\lambda_1\wp_{22}\wp_{12}+\lambda_2\wp_{12}^2
+4\wp_{11}\wp_{12}^2
,\\
\wp_{211}\wp_{111}&=-\frac12\lambda_0\lambda_2-2\lambda_0\wp_{11}
-\frac14\lambda_0\lambda_3\wp_{22}+\frac18\lambda_1^2
\\
&-\lambda_0\wp_{22}\wp_{12}-\frac12\lambda_1\wp_{11}\wp_{22}
+\frac12\lambda_1\wp_{12}^2
\\
&+4\wp_{12}\wp_{11}^2+\frac18\lambda_1\lambda_3\wp_{12}
+\lambda_2\wp_{12}\wp_{11}
,\\
\wp_{111}^2&=\frac{1}{16}(\lambda_1^2\lambda_4+\lambda_0\lambda_3^2
 -4\lambda_0\lambda_3\lambda_4)+(\frac14\lambda_1^2
-\lambda_0\lambda_2)\wp_{22}
\\
&+\frac12\wp_{12}\lambda_0\lambda_3
+(\frac14\lambda_1\lambda_3-\lambda_0\lambda_4)\wp_{11}
+\lambda_0\wp_{12}^2
\\
&+\lambda_1\wp_{12}\wp_{11}+\lambda_2\wp_{11}^2
-4\lambda_0\wp_{22}\wp_{11}+4\wp_{11}^3.
\end{align*}
These expressions may be rewritten in the form of an ``extended
 cubic relation''.
For arbitrary  $l,k\in \mathbf{  C}^4$
\begin{equation}\boldsymbol{l}^T\pi\pi^T
\boldsymbol{k}=-\frac14{\det}\;\left(\begin{array}{cc}H&\boldsymbol{l}\\
\boldsymbol{k}^T&0\end{array}\right),\label{kum1}\end{equation}
where
$\pi^T=(\wp_{222},-\wp_{221},\wp_{211},-\wp_{111})$
and   $H$ is $4\times4$ matrix:
\begin{equation*}
H=
 \left (  \begin{array}{cccc}  \lambda_0&  \frac{1}{2}
 \lambda_1&-2  \wp_{11}&-2  \wp_{12}  \\
  \frac{1}{2}  \lambda_1&  \lambda_2+4  \wp_{11}&  \frac{1}{2}
\lambda_3+ 2  \wp_{12}&-2  \wp_{22}  \\-2  \wp_{11}&  \frac{1}{2}
 \lambda_3+2  \wp_{12}&  \lambda_4+4  \wp_{22}&2  \\
-2  \wp_{12}&-2  \wp_{22}&2&0  \end{array}  \right) .
 \end{equation*}
The vector $\pi$ satisfies the equation $H\pi=0$, or in
detailed form \begin{eqnarray}
&&-\wp_{12}\wp_{222}+\wp_{22}\wp_{221}+\wp_{211}=0,\label{e1}\\
&&2\wp_{11}\wp_{222}+\left(\frac12\lambda_3+2\wp_{12}\right)\wp_{221}
\notag\\&&-(\lambda_{4}+4\wp_{22})\wp_{211}+2\wp_{111}=0,\label{e2}\\
&&\frac12\lambda_1\wp_{222}-\left(\lambda_2+4\wp_{11}\right)\wp_{221}
\notag\\&&+\left(\frac12\lambda_{3}+2\wp_{12}\right)\wp_{211}+2\wp_{22}\wp_{111}=0,\label{e3}\\
&&-\lambda_0\wp_{222}+\frac12\lambda_1\wp_{221}
+2\wp_{11}\wp_{211}-2\wp_{12}\wp_{111}=0,\label{e4}
\end{eqnarray}
and so the functions $\wp_{22},\wp_{12}$ and $\wp_{11}$ are
 related by the equation
\begin{equation}\mathrm{ det}
H=0.\label{kum5} \end{equation} The equation (\ref{kum5}) defines
the quartic Kummer surface $\mathbf K$ in $\mathbf C^3$ with
coordinates $X=\wp_{22},Y=\wp_{12},Z=\wp_{11}$ \cite{hu05}.

\index{Kummer!surface}
The $\wp_{ijkl}$-functions are expressed as follows
\begin{align}
&\wp_{2222}=6\wp_{22}^2+\frac12\lambda_3+\lambda_4\wp_{22}
+4\wp_{12},\label{eeq1}\\
&\wp_{2221}=6\wp_{22}\wp_{12}+\lambda_4\wp_{12}-2\wp_{11},\label{eeq3}\\
&\wp_{2211}=2\wp_{22}\wp_{11}+4\wp_{12}^2+
\frac12\lambda_3\wp_{12},\label{eeq5}\\
&\wp_{2111}=6\wp_{12}\wp_{11}+\lambda_2\wp_{12}
-\frac12\lambda_1\wp_{22}-\lambda_0,\label{eeq4}\\
&\wp_{1111}=6\wp_{11}^2-3\lambda_0\wp_{22}
+\lambda_1\wp_{12}+\lambda_2\wp_{11}-
\frac12\lambda_0\lambda_4+\frac18\lambda_1\lambda_3.\label{eeq2}
\end{align}
All these formulas may be condensed into a single expression.
Namely, the following expression, which can be interpreted as a direct
analogue of the Hirota bilinear
relation \cite{hi72,hi80},
$$
\left\{\frac13\Delta\Delta^T+\Delta^T\!\begin{pmatrix}0&0&1\\ 0&-1/2&0\\
1&0&0\end{pmatrix}\epsilon_{\eta,\eta}
\epsilon_{\eta,\eta}\cdot\epsilon_{\eta,\eta}^T-(\xi-\eta)^4
\epsilon_{\eta,\xi}\epsilon_{\eta,\xi}^T\right\}\sigma({\boldsymbol u})
\sigma({\boldsymbol u}')\bigg|_{{\boldsymbol u}'={\boldsymbol u}}
$$
is identically $0$, where
$\Delta^T=(\Delta_1^2,2\Delta_1\Delta_2,\Delta_2^2)$ with
$\Delta_i=\partial/\partial u_i-\partial/\partial u_i'$ and also
$\epsilon^T_{\xi,\eta}=(1,\eta+\xi,\eta\xi)$. After evaluation, the
powers of parameters $\eta$ and $\xi$ are replaced according to the
rules $\eta^k,\xi^k\to\lambda_kk!(6-k)!/6!$ by the constants defining the curve.

The functions $\wp_{12}$ and $\wp_{22}$ can be reexpressed by the aids of
(\ref{eeq1}), (\ref{eeq3}) in terms of the function $\wp_{2,2}$
and its higher derivatives as follows

\begin{eqnarray*}
\wp_{12}&=&\frac14\wp_{2222}-\frac32\wp_{22}^2-\frac14\lambda_4\wp_{22}-\frac18\lambda_3\\
\wp_{11}&=&-\frac18\wp_{22222}+\left(\frac34\wp_{22}+\frac18\lambda_4\right)\wp_{2222}\\
&+&\left(\frac32\wp_{22}+\frac18\lambda_4\right)\wp_{222}-\frac92\wp_{22}^3\\
&-&\frac32\lambda_4\wp_{22}^2-\left(\frac38\lambda_3+\frac18\lambda_4^2\right)
\wp_{22}-\frac16\lambda_4\lambda_3.
\end{eqnarray*}

The substitution of these formulae to the equation of the Kummer
surface leads to the associated differential equation.

\section{Representations of the group of characteristics}
Recall that in the case of genus one shift on a half period of the
Weierstrass elliptic function is given as fractional linear
transformation of the form
\begin{equation}\label{eltrans}
  \wp(u+\omega_i)= \frac{a_1\wp(u)+a_2}{d_1\wp(u)+d_2}
\end{equation}
with $a_1=e_i, a_2=-e_i(e_k+e_l)+e_ke_l, d_1=1,d_2=-e_i$ and
indices $k,l$ complement the set $\{i\}$ up to $\{1,2,3\}$. The
transformations (\ref{eltrans}) at $i=1,2,3$
\[ T_i= \left(\begin{array}{cc}   a_1&a_2\\ d_1&d_2  \end{array}\right)
=\left(\begin{array}{cc}  e_i&-e_i(e_k+e_l)+e_ke_l\\
1&-e_i  \end{array} \right),\quad \det\,T_i=(e_i-e_k)(e_i-e_l)  \]
complemented by the identical transformation form the group, which is isomorphic to
the group of characteristics. In other words we have a
representation of the group of characteristics.

In the the case of the genus two the picture is analogous. We have

\begin{align}
\wp_{2,2}(\boldsymbol{u}+\boldsymbol{\Omega})=\frac{a_1\wp_{2,2}(\boldsymbol{u})
+a_2\wp_{1,2}(\boldsymbol{u})+a_3\wp_{1,1}(\boldsymbol{u})+a_4}{d_1\wp_{2,2}(\boldsymbol{u})
+d_2\wp_{1,2}(\boldsymbol{u})+d_3\wp_{1,1}(\boldsymbol{u})+d_4},\label{wp22trans}\\
\notag\\
\wp_{1,2}(\boldsymbol{u}+\boldsymbol{\Omega})=\frac{b_1\wp_{2,2}(\boldsymbol{u})
+b_2\wp_{1,2}(\boldsymbol{u})+b_3\wp_{1,1}(\boldsymbol{u})+b_4}{d_1\wp_{2,2}(\boldsymbol{u})
+d_2\wp_{1,2}(\boldsymbol{u})+d_3\wp_{1,1}(\boldsymbol{u})+d_4},\label{wp12trans}\\
\notag\\
\wp_{1,1}(\boldsymbol{u}+\boldsymbol{\Omega})=\frac{c_1\wp_{2,2}(\boldsymbol{u})
+c_2\wp_{1,2}(\boldsymbol{u})+c_3\wp_{1,1}(\boldsymbol{u})+c_4}{d_1\wp_{2,2}(\boldsymbol{u})
+d_2\wp_{1,2}(\boldsymbol{u})+d_3\wp_{1,1}(\boldsymbol{u})+d_4},\label{wp11trans}
\end{align}
where  $\boldsymbol{\Omega}$ is a half-period. In the case when
$\boldsymbol{\Omega}$ is even half-period,
$\boldsymbol{\Omega}=\boldsymbol{\Omega}_{i,j}$ the coefficients
of the transformation are given by

\begin{align*}
a_1&= S_3-S_1s_2\quad a_2=-s_2-S_1s_1+S_2\\
a_3&=- s_1,\quad a_4=-s_2^2+S_2s_2+S_1s_2s_1\\
b_1&=-S_3s_1+S_2s_2,\quad b_2=S_1s_2-S_3,\\
b_3&=s_2,\quad b_4= -2S_3s_2+S_3s_1^2-S_2s_2s_1\\
c_1&=-2 S_3 s_2 + S_3 s_1^2  - S_2 s_2 s_1,\quad c_2=s_2^2  - S_1
s_2 s_1 - S_2 s_2\\
c_3&=-S_3 - S_1 s_2,\quad c_4=- S_2^2  s_2 + 4 S_3 S_1 s_2 + S_2
s_2 S_1 s_1\\& + S_3 S_2 s_1 + S_2 s_2^2 - S_3 S_1 s_1^2  - S_3s_1s_2\\
d_1&=-s_2,\quad d_2=-s_1,\quad d_3=-1,\quad d_4=S_1s_2+S_3
\end{align*}
where symmetric functions $s_k=s_k[i,j]$, $k=1,2$ and
$S_l=S_l[i,j]$, $l=1,2,3$ corresponds to the partition of
branching points $\{e_i,e_j\}\cup \{ e_p,e_q,e_r \}$ and are given
as
\begin{align*}
  s_1[i,j]&=e_i+e_j,\; s_2[i,j]=e_ie_j\\
  S_1[i,j]&=e_p+e_q+e_r,\;S_2[i,j]=e_pe_q+e_pe_r+e_qq_r,\;S_3[i,j]=e_pe_ge_r
\end{align*}
are built on arbitrary permutations of indices $\{i,j,p,q,r\}$
from  $\{1,2,3,4,5 \} $

In the case of odd half period
$\boldsymbol{\Omega}=\boldsymbol{\Omega}_{i}$, $i=1,\ldots,6$.
\begin{align*}
a_1&=-e_i^2,\quad a_2=0,\quad a_3=1,\quad a_4=e_i^2T_1-e_iT_2\\
b_1&=0,\quad b_2=-e_i^2,\quad b_3=-e_i,\quad b_4=e_iT_3-T_4\\
c_1&=-e_iT_3+T_4,\quad c_2=e_i(e_iT_1-T_2),\quad c_3=e_i^2,\quad
c_4=0,\\
d_1&=-e_i,\quad d_2=-1, \quad d_3=0,\quad d_4=e_i^2,
\end{align*}
where symmetric functions $T_l=S_l[i]$, $l=1,2,3,4$ corresponds to
the partition of branching points $\{e_i\}\cup \{ e_p,e_q,e_r,e_s
\}$ and are given as
\begin{align*}
T_1[i]&=e_p+e_q+e_r+e_s,\ldots, \;T_4[i]=e_pe_ge_re_s.
\end{align*}

The matrices
\begin{equation}
  \Gamma_{i,j}=\left(\begin{array} {cccc} a_1[i,j]&a_2[i,j]&a_3[,ji]&a_4[i,j]\\
  b_1[i,j]&b_2[i,j]&b_3[i,j]&b_4[i,j] \\
  c_1[i,j]&c_2[i,j]&c_3[i,j]&c_4[i,j]\\
  d_1[i,j]&d_2[i,j]&d_3[i,j]&d_4[i,j]
  \end{array}\right), \quad i\neq j\in \{1,\ldots,5\}
\end{equation}
and
\begin{equation}
  \Gamma_{i}=\left(\begin{array} {cccc} a_1[i]&a_2[i]&a_3[i]&a_4[i]\\
  b_1[i]&b_2[i]&b_3[i]&b_4[i] \\
  c_1[i]&c_2[i]&c_3[i]&c_4[i]\\
  d_1[i]&d_2[i]&d_3[i]&d_4[i]
  \end{array}\right), \quad i\neq j\in \{1,\ldots,5\}
\end{equation}
defining the transformation (\ref{wp22trans}-\ref{wp11trans}) generates the group of characteristices.
Remark that
\begin{align*} & \det \Gamma_{i,j} = (e_i-e_p) ^2(e_i-e_q)^2 (e_i-e_r)^2 (e_j-e_p) ^2(e_j-e_q)^2 (e_j-e_r)^2\\
 &\det \Gamma_{i} = (e_i-e_p) ^2(e_i-e_q)^2 (e_i-e_r)^2
 \end{align*}

In the case when $\boldsymbol\Omega$ is even or odd half period has
the following remarkable property: the product $\Gamma J$ where
\begin{equation}\label{jmatrix}
  J=\left(\begin{array} {cccc}0&-1&0&0\\
                        1&0&0&0 \\
                        0&0&0&1\\
                        0&0&-1&0    \end{array}\right)
\end{equation}
is a symmetric matrix.

Altogether the projective transformation corresponding to even and
odd half-periods forms a group satisfying the following relations

\begin{align*}
  &[\Gamma_{i,j},\Gamma_{k,l}]=0 \quad i,j,k,l \quad  \text {are all
  different}, \\
&\{\Gamma_{i,j},\Gamma_{i,k}\}=0, \quad  k\neq j, j\neq i, k \neq
i,\\
 &[\Gamma_{i,j},\Gamma_{k}]=0,\quad i,j,k \quad  \text {are all
  different}, \\
 &\{\Gamma_{l},\Gamma_{k}\}=0,\quad k \neq l, \\
&\Gamma_{i,j}\Gamma_{k,l}=\Gamma_m (e_i-e_k)(e_i-e_l)
(e_j-e_k)(e_j-e_l)  \quad i,j,k,l,m \quad \text {are all
  different},\\
&\Gamma_{m}\Gamma_{i,j}=\Gamma_{k,l}(e_m-e_i)(e_m-e_j), \quad
i,j,k,l,m \quad \text {are all
  different},
\end{align*}
where $[\cdot,\cdot] $ means commutator and $\{\cdot,\cdot\} $
means anti-commutator.

This group is the group of symmetries of the Kummer surface.

The transformations (\ref{wp22trans},\ref{wp12trans},\ref{wp11trans} ) can be written in the vector form. Let
\[   \boldsymbol{\wp}(\boldsymbol{u})=  \left(  \begin{array}{c} \wp_{22}(\boldsymbol{u})\\
\wp_{12}(\boldsymbol{u})\\ \wp_{11}(\boldsymbol{u})\\ 1 \end{array}  \right) \]
The shifts at an even half-period $\boldsymbol{\Omega}_{i,j}$ and odd  half-period $\boldsymbol{\Omega}_{i}$
are given by the formulae
\begin{align*}
& \boldsymbol{\wp}(\boldsymbol{u}+\boldsymbol{\Omega}_{i,j})= \frac{1}{\mathcal{P}_{i,j}} \Gamma_{i,j}\boldsymbol{\wp}(\boldsymbol{u}),\quad 1\leq i<j\leq 5\\
 &\boldsymbol{\wp}(\boldsymbol{u}+\boldsymbol{\Omega}_{i})= \frac{1}{\mathcal{P}_{i}} \Gamma_{i}\boldsymbol{\wp}(\boldsymbol{u}),\quad i=1,\ldots,5
\end{align*}
where
\begin{align*}
&\mathcal{P}_{i,j}=-\wp_{2,2}(\boldsymbol{u})s_2[i,j]-\wp_{1,2}(\boldsymbol{u})s_1[i,j]-\wp_{1,1}(\boldsymbol{u})s_2+
S_1[p,q,r]s_2[i,j]+S_3[p,q,r]\\
&\mathcal{P}_{i}=e_i^2-\wp_{2,2}(\boldsymbol{u})e_i--\wp_{1,2}(\boldsymbol{u})
\end{align*}

\section{Addition theorems}
\subsection{Baker addition formula}
\begin{eqnarray}
&&\frac{\sigma ({\boldsymbol u}+{\boldsymbol
v})\sigma({\boldsymbol u}-{\boldsymbol v})} {\sigma^2({\boldsymbol
u})\sigma^2({\boldsymbol
v})}\label{bad}\\&&\qquad
=\wp_{22}({\boldsymbol
u})\wp_{12}({\boldsymbol v}) -\wp_{12}({\boldsymbol
u})\wp_{22}({\boldsymbol v})+\wp_{11}({\boldsymbol
v})-\wp_{11}({\boldsymbol u}),
\nonumber \end{eqnarray} which plays
very important role in what follows. We shall also denote below
the polynomial in $\wp_{ij}(\boldsymbol{u})$,
$\wp_{ij}(\boldsymbol{v})$ standing in the right hand side of
(\ref{bad}) as \begin{equation} B({\boldsymbol u},{\boldsymbol
v})=\wp_{22}({\boldsymbol u})\wp_{12}({\boldsymbol v})
-\wp_{12}({\boldsymbol u})\wp_{22}({\boldsymbol
v})+\wp_{11}({\boldsymbol v})-\wp_{11}({\boldsymbol u}).
\label{mmmm}\end{equation}

From here follows

\begin{equation}
\frac{\sigma(2\boldsymbol{u})}{\sigma^4(\boldsymbol{u})}= \wp_{12}(\boldsymbol{u}) \wp_{222}(\boldsymbol{u})
-\wp_{22}(\boldsymbol{u}) \wp_{122}(\boldsymbol{u})-\wp_{112}(\boldsymbol{u})
\end{equation}

Denote
\[  M(\boldsymbol{u})=  \wp_{12}(\boldsymbol{u}) \wp_{222}(\boldsymbol{u})
-\wp_{22}(\boldsymbol{u}) \wp_{122}(\boldsymbol{u})-\wp_{112}(\boldsymbol{u}) \]

Also, for any $P=(x,y)\in V$, $\boldsymbol{v}\in\mathrm{Jac}(V)$, $\boldsymbol{u}=\left( \int_{\infty}^P\mathrm{d}u_1, \int_{\infty}^P\mathrm{d}u_2 \right)^T\in (\sigma)$
\begin{align}
\frac{\sigma\left(\boldsymbol{v}
+\int_{\infty}^P\mathrm{d}\boldsymbol{u}\right)
\sigma\left(\boldsymbol{v}-\int_{\infty}^P\mathrm{d}\boldsymbol{u}\right)}  {\sigma^2(\boldsymbol{v})\sigma_2^2(\boldsymbol{u})  }
=x^2-\wp_{22}(\boldsymbol{u})x-\wp_{12}(\boldsymbol{u})
\end{align}

\subsection{Analog of the Frobenius Stickelberger formula}
The addition rule on $\mathrm{Jac}(X)\times \mathrm{Jac}(X)\times
\mathrm{Jac}(X)$, i.e. a generalization of the Frobenius Stickelberger addition formula to the genus two curves can be given as follows. Let
$$
\Delta(\boldsymbol{x},\boldsymbol{y};\boldsymbol{x}',\boldsymbol{y}';{\bf
x}'',{\bf y}'')= \frac14\left |\begin {array}{cccccc} 1 & x_1 & {x_1}^{2}
& y_1&{x_1
      }^{3}&x_1y_1\\\noalign{\medskip}1 & x_2 & {x_2}^2 & y_2 &
    {x_2}^3 & x_2y_2\\\noalign{\medskip}1 & { x'}_1 & {{ x'}_1}^{2} &
    { y'}_1&{{ x'}_1}^{3} & { x'}_1{ y'}_1\\\noalign{\medskip}1 & {
      x'}_2&{{ x'}_2}^{2} & { y'}_2 & {{ x'}_2}^{3}&{ x'}_2{ y'}_2
    \\\noalign{\medskip}1&{ x''}_1 & {{ x''}_1}^{2}&{ y''}_ 1 &
    {{x''}_1}^{3} & { x''}_1 { y''}_1 \\\noalign{\medskip}1&{ x''}_2 &
    {{ x''}_2}^{2} & { y''}_ 2&{{ x''}_2}^{3}&{ x''}_2{ y''}_2\end
  {array} \right |.
$$
Also we denote
\begin{align*}
\boldsymbol{u}&=\int^{(x_1,y_1)}_{(\infty,\infty)}
d\boldsymbol{u}+ \int^{(x_2,y_2)}_{(\infty,\infty)}
d\boldsymbol{u}\\
\boldsymbol{u}'&=\int^{(x_1',y_1')}_{(\infty,\infty)}
d\boldsymbol{u}+ \int^{(x_2',y_2')}_{(\infty,\infty)}
d\boldsymbol{u}\\
\boldsymbol{u}''&=\int^{(x_1'',y_1'')}_{(\infty,\infty)}
d\boldsymbol{u}+ \int^{(x_2'',y_2'')}_{(\infty,\infty)}
d\boldsymbol{u},
\end{align*}
Then

\begin{align}
  &\frac{\sigma(\boldsymbol{u}+\boldsymbol{u}'+\boldsymbol{u}'')
\sigma(\boldsymbol{u}-\boldsymbol{u}')\sigma(
      \boldsymbol{u}'-\boldsymbol{u}'')\sigma( \boldsymbol{u}''-
\boldsymbol{u})}
  {\sigma(\boldsymbol{u})^3\sigma(\boldsymbol{u}')^3\sigma(\boldsymbol{u}'')^3}
  =\Delta(\boldsymbol{x},\boldsymbol{y};\boldsymbol{x}',\boldsymbol{y}';
\boldsymbol{      x}'',
  \boldsymbol{y}'')\nonumber\\
  &\quad\times\frac{\Delta(\boldsymbol{x},\boldsymbol{y};\boldsymbol{x}',
\boldsymbol{      -y}')
  \Delta(\boldsymbol{x}',\boldsymbol{y}';\boldsymbol{x}'',\boldsymbol{-y}'')
  \Delta(\boldsymbol{      x}'',\boldsymbol{y}'';\boldsymbol{x},
-\boldsymbol{y})}
  {V(\boldsymbol{x},\boldsymbol{x}',\boldsymbol{      x}'')
  V^2(\boldsymbol{x})V^2(\boldsymbol{x}')V^2(\boldsymbol{x}'')})\label{formula}
\end{align}
where $V$ is the Vandermonde determinant of its arguments
\begin{eqnarray}
  V(\boldsymbol{x})&=&V(x_1,x_2)=\left |\begin {array}{cc} 1&x_{{1}}
      \\\noalign{\medskip}1&x_{{2}}\end {array}\right |,\\
  V(\boldsymbol{x},\boldsymbol{x}')&=&V(x_1,x_2,x_1',x_2')=\left |
\begin {array}{cccc}
  1&x_{{1}}&{x_{{1}}}^{2}&x_{{1}}^3
  \\\noalign{\medskip}1&x_{{2}}&{x_{{2}}}^{2}&x_{{2}}^3
  \\\noalign{\medskip}1&{\it x'}_{{1}}&{{x'}_{{1}}}^{2}&{ x'}_{{
      1}}^3\\\noalign{\medskip}1&{x'}_{{2}}&{{x'}_{{2}}}^{2}&{x'}
  _{{2}}^3\end {array}\right |,
\end{eqnarray}
thus $V(\boldsymbol{x},\boldsymbol{x}',\boldsymbol{      x}'')$ is
$6\times6$ Vandermonde determinant.

In terms of Kleinian $\wp$-functions we get
\begin{align*}
 &\frac{\sigma(\boldsymbol{u}+\boldsymbol{u}'+\boldsymbol{u}'')\,
\sigma(\boldsymbol{u}-\boldsymbol{u}')\,\sigma(\boldsymbol{u}'-\boldsymbol{u}'')\,
    \sigma(\boldsymbol{u}''-\boldsymbol{u})}
  {\sigma(\boldsymbol{u})^3\,\sigma(\boldsymbol{u}')^3\,\sigma(\boldsymbol{u}'')^3}=\\
&\frac18\,\wp_{{112}}\wp'_{{122}}\wp''_{{222}}-\frac18\,\wp_{{112}}\wp'_{{222}
}\wp''_{{122}}-\\
&\frac14 \, \bigl( -\wp''_{{12}}\wp_{{22}}+\wp'_{{12}}\wp_{{22}}-\wp'_{{22}
}\wp_{{12}}+\wp''_{{22}}\wp'_{{12}}-2\,\wp''_{{11}}+\\
&\qquad \wp''_{{22}}\wp_{{
12}}-\wp'_{{22}}\wp''_{{12}}+ 2\,\wp'_{{11}} \big) \wp_{{111}}-\\
&\frac14 \,
 \bigl( 2\,\wp''_{{22}}\wp_{{22}}\wp'_{{12}}-2\,\wp''_{{12}}\wp'_{{22}
}\wp_{{22}}-\wp''_{{22}}\wp_{{11}}+\wp'_{{22}}\wp_{{11}}+\wp_{{12}}
\wp'_{{12}}+\\
&\qquad \wp'_{{11}}\wp'_{{22}}-2\,\wp''_{{11}}\wp'_{{22}}-\wp''_{{
11}}\wp''_{{22}}-\wp_{{12}}\wp''_{{12}}+2\,\wp'_{{11}}\wp''_{{22}}-{
\wp'_{{12}}}^{2}+{\wp''_{{12}}}^{2} \big) \wp_{{112}}+\\
&\frac14 \, \bigl( -
\wp'_{{11}}\wp'_{{22}}\wp''_{{22}}+\wp''_{{22}}\wp'_{{12}}\wp_{{12}}-
\wp'_{{22}}\wp''_{{12}}\wp_{{12}}+\wp''_{{11}}\wp'_{{22}}\wp''_{{22}}-
2\,\wp''_{{11}}\wp'_{{12}}-\\
&\qquad {\wp''_{{12}}}^{2}\wp'_{{22}}+2\,\wp'_{{12}
}\wp_{{11}}-2\,\wp''_{{12}}\wp_{{11}}+{\wp'_{{12}}}^{2}\wp''_{{22}}+2
\,\wp'_{{11}}\wp''_{{12}} \big) \wp_{{122}}+\\
&\frac14 \, \big(
  \wp'_{{11}}\wp'_{{22}}\wp''_{{12}}-\wp'_{{22}}\wp''_{{12}}
\wp_{{11}}-\wp''_{{11}}
\wp''_{{22}}\wp'_{{12}}+\\
& \qquad {\wp''_{{12}}}^{2}\wp'_{{12}}-\wp''_{{12}}{
\wp'_{{12}}}^{2}+\wp''_{{22}}\wp'_{{12}}\wp_{{11}} \big) \wp_{{222}}
+\\
&+ \text{cyclic permutations of } \,\wp,\wp',\wp'',
\end{align*}
where $\wp=\wp(\boldsymbol{u})$, $\wp'=\wp(\boldsymbol{u}')$,
$\wp''=\wp(\boldsymbol{u}'')$.

\subsection{Addition on strata of the $\theta$-divisor}
Recall that the
$\theta$-divisor $(\theta)$ is a subvariety in $\mathrm{Jac}(X)$ given
by the equation
\begin{equation} \theta(\boldsymbol{u}|\tau)=0,\quad \text
{or equivalently}\quad
\sigma(\boldsymbol{u})=0.
\end{equation}
According to  Riemann's vanishing theorem, points from $(\theta)$
are represented by
\begin{equation}\boldsymbol{u}
=\int\limits_{(\infty,\infty)}^{(x,y)}d \boldsymbol{u}
-2\omega \boldsymbol{K}_{\infty}.
\end{equation}

The co-ordinates of a point of the curve $(x,y)$ can
be given in terms of the $\sigma$-functions restricted to the
$\theta$-divisor as follows (see \cite{gr91,jor92})
$$
x_i=-\left.\frac{ \sigma_1(\boldsymbol{u}_i)}{\sigma_2(
\boldsymbol{u}_i)} \right|_{(\theta)}, \quad 2y_i=-\left.\frac{
\sigma(2
      \boldsymbol{u}_i)}{\sigma_2^4(\boldsymbol{u}_i)}
      \right|_{(\theta)}.
$$

We shall use the following result, a special case of a result by
\^Onishi \cite{on02,on04,on05}.
\begin{theorem} Let $V$ be an algebraic curve of of genus 2. Then we
have
\begin{align}
  & \frac{\sigma(\boldsymbol{u}_0+\boldsymbol{u}_1+\dots+\boldsymbol{u}_n)
    \prod_{0\leq k<l\leq n}\sigma( \boldsymbol{u}_k-\boldsymbol{u}_l)
  }{\sigma^{n+1}_2(\boldsymbol{u}_0) \ldots
    \sigma^{n+1}_2(\boldsymbol{u}_n)}\notag\\
  &\qquad\qquad=\frac{1}{2^{[n/2-1]}}\left |\begin {array}{cccc}
      1&w_1(x_0,y_0)&\dots&w_n(x_0,y_0)\\\noalign{\medskip}
      1&w_1(x_1,y_1)&\dots&w_n(x_1,y_1)\\\noalign{\medskip}
      \vdots&\vdots&\vdots&\vdots\\\noalign{\medskip}
      1&w_1(x_n,y_n)&\dots&w_n(x_n,y_n)\end {array}\right
      |,\label{onishi}
\end{align}
where $[\cdot ]$ means integer part, $(x_0,y_0), \dots, (x_n,y_n)$ is a non-special divisor on
$V$,
 $\boldsymbol{ u}_i$ is the Abel image
$$
\boldsymbol{u}_i=\int^{\,(x_i,y_i)}_{\,(\infty,\infty)}d{\bf u},
$$
and $\sigma_2(\boldsymbol{u}_i)$ is the value of the $\sigma$
derivative restricted to the $\theta$-divisor, $(\theta):
\sigma(\boldsymbol{u}_i)=0$
\begin{eqnarray*}
\sigma_2(\boldsymbol{u}_i)&=&\sigma_2(u_{i_1},u_{i_2})\\
&=&\frac{\partial}{\partial u_{i_2} }\,
  \sigma(u_{i_1},u_{i_2})|_{\sigma(\boldsymbol{u}_i)=0}.
  \end{eqnarray*}
The factor $1/2^{[n/2-1]}$ arises in our version of this theorem as we
use a different normalization of the curve than \^Onishi.
\end{theorem}
In particular, for $n=1$ we have
$$
\frac{\sigma(\boldsymbol{u}_0+\boldsymbol{u}_1)\sigma(\boldsymbol{u}_0-\boldsymbol{u}_1)
}
{\sigma^2_2(\boldsymbol{u}_0)\sigma^{2}_2(\boldsymbol{u}_1)}=x_1-x_0.
$$

Remark that the Baker addition formula can be obtained from the above addition formula. To show that we introduce the notation
$$
\Delta({\bf x,y;x',y'})= \frac12 \left |\begin {array}{cccc}
    1&x_{{1}}&{x_{{1}}}^{2}&y_{{1}}
    \\\noalign{\medskip}1&x_{{2}}&{x_{{2}}}^{2}&y_{{2}}
    \\\noalign{\medskip}1&{\it x'}_{{1}}&{{\it x'}_{{1}}}^{2}&{\it
      y'}_{{ 1}}\\\noalign{\medskip}1&{\it x'}_{{2}}&{{\it
        x'}_{{2}}}^{2}&{\it y'} _{{2}}\end {array}\right |,
$$
where $\boldsymbol{x}=(x_1,x_2)^T, \boldsymbol{y}=(y_1,y_2)^T$,
etc. Then by applying the above theorem, we obtain after
simplification
$$
\frac{\sigma(\boldsymbol{u}+\boldsymbol{u}')\sigma(\boldsymbol{u}-\boldsymbol{u}')
} {\sigma^2_2(\boldsymbol{    u})\sigma^{2}_2(\boldsymbol{u}')}=
\frac{\Delta(\boldsymbol{
    x},\boldsymbol{y};\boldsymbol{x}',\boldsymbol{y}')
    \Delta(\boldsymbol{x},\boldsymbol{y};\boldsymbol{x}',-\boldsymbol{y}')}
    {V(\boldsymbol{ x},\boldsymbol{x}') V(\boldsymbol{
    x})V(\boldsymbol{x}')},
$$
where $V$ is the Vandermonde determinant. After expanding these determinants, factorizing, applying the
equation of the curve (\ref{curve}) and the expressions of symmetric functions
we arrive finally at the required formula (\ref{onishi}).
This is a simplified version of the calculation carried out
by Baker \cite [pp. 331-332]{ba97} and also derived by him using
another method in 1907 \cite[pg.  100]{ba07}.

\section{$\theta$ and $\sigma$-quotients for the case of genus two}

The formulae presented in this Section permit to
express any $\theta$--quotient with half integer characteristics
as rational function on $V^g$.  We shall give these expressions
following {\it Rosenhain} \cite{ros851} for the genus two curve
being represented in the form ({\it Richelot normal form}),
\[y^2=x(1-x)(1-\kappa^2x)(1-\lambda^2x)(1-\mu^2x) .\]
Then the moduli of the curve, $\kappa,\mu,\nu$ are expressible in terms
of $\theta$--constants as follows

\index{Rosenhain!formulae for the moduli of the curve}

\begin{eqnarray}
\kappa^2&=&\frac{\ffBB\ffBC}{\ffCB\ffCC},\quad\lambda^2=
\frac{\ffBO\ffBB}{\ffCO\ffCB}\nonumber\\
\mu^2&=&\frac{\ffBO\ffBC}{\ffCO\ffCC}\nonumber\end{eqnarray}
\begin{eqnarray}
\kkI&=&\frac{\ffOB\ffOC}{\ffCB\ffCC},\quad\llI=
\frac{\ffOO\ffOB}{\ffCO\ffCB}\nonumber\\
\mmI&=&\frac{\ffOO\ffOC}{\ffCO\ffCC}\nonumber\end{eqnarray}
\begin{eqnarray}
\llk&=&\frac{\ffBB\ffAA\ffOB}{\ffCB\ffCO\ffCC},\quad\mml=
\frac{\ffBO\ffAA\ffOO}{\ffCO\ffCB\ffCC}\nonumber\\
\mmk&=&\frac{\ffOO\ffAA\ffOC}{\ffCC\ffCB\ffCO}\nonumber\\
\end{eqnarray}

All the $\theta$--quotients with half integer characteristics are
given as follows\index{Rosenhain!functions}
\begin{eqnarray}
&&\frac{\ffAO(\boldsymbol{\mathfrak A}({\mathcal D}))}
{\ffOO(\boldsymbol{\mathfrak A}({\mathcal D}))}=\kappa\lambda\mu \;
x_1x_2,\nonumber\\
&&\frac{\ffBO(\boldsymbol{\mathfrak A}({\mathcal D}))}
{\ffOO(\boldsymbol{\mathfrak A}({\mathcal D}))}=-\frac{\kappa\lambda\mu}{
\kI\lI\mI}\;(1-x_1)(1-x_2),\nonumber\\
&&\frac{\ffCA(\boldsymbol{\mathfrak A}({\mathcal D}))}
{\ffOO(\boldsymbol{\mathfrak A}({\mathcal D}))}=-\frac{\lambda\mu}{
\kI\lk\mk}\;(1-\kappa^2x_1)(1-\kappa^2x_2),\nonumber\\
&&\frac{\ffCB(\boldsymbol{\mathfrak A}({\mathcal D}))}
{\ffOO(\boldsymbol{\mathfrak A}({\mathcal D}))}=-\frac{\kappa\mu}{
\lI\ml\lk}\;(1-\lambda^2x_1)(1-\lambda^2x_2),\nonumber\\
&&\frac{\ffCC(\boldsymbol{\mathfrak A}({\mathcal D}))}
{\ffOO(\boldsymbol{\mathfrak A}({\mathcal D}))}=-\frac{\kappa\lambda}{
\mI\mk\ml}\;(1-\mu^2x_1)(1-\mu^2x_2),\nonumber\\
&&\frac{\ffCO(\boldsymbol{\mathfrak A}({\mathcal D}))}
{\ffOO(\boldsymbol{\mathfrak A}({\mathcal D}))}=-\frac{x_1x_2(1-x_1)(1-x_2)}{
\kI\lI\mI(x_2-x_1)^2},\nonumber\\
&&\phantom{\frac{\ffCO(\boldsymbol{\mathfrak A}({\mathcal D}))}
{\ffOO(\boldsymbol{\mathfrak A}({\mathcal D}))}}\times\left(\frac{y_1}{
x_1(1-x_1)}\pm\frac{y_2}{ x_2(1-x_2)}\right)^2,
\nonumber\end{eqnarray}\begin{eqnarray}
&&\frac{\fOA(\boldsymbol{\mathfrak A}({\mathcal D}))}
{\fOO(\boldsymbol{\mathfrak A}({\mathcal D}))}=
-\frac{\lambda\mu(1-\lambda^2
x_1)(1-\lambda^2 x_2)(1-\mu^2 x_1)(1-\mu^2 x_2)}{
\lI\mI\lk\mk(x_2-x_1)^2}\nonumber\\
&&\phantom{\frac{\ffCO(\boldsymbol{\mathfrak A}({\mathcal D}))}
{\ffOO(\boldsymbol{\mathfrak A}({\mathcal D}))}}\times\left(\frac{y_1}{
(1-\lambda^2x_1)(1-\mu^2x_1)}\pm\frac{y_1}{
(1-\lambda^2x_2)(1-\mu^2x_2)} \right)^2,\nonumber\end{eqnarray}\begin{eqnarray}
&&\frac{\ffOB(\boldsymbol{\mathfrak A}({\mathcal D}))}
{\ffOO(\boldsymbol{\mathfrak A}({\mathcal D}))}=-\frac{\kappa\mu(1-\mu^2
x_1)(1-\mu^2 x_2)(1-\kappa^2 x_1)(1-\kappa^2 x_2)}{
\kI\mI\lk\ml(x_2-x_1)^2}\nonumber\\
&&\phantom{\frac{\ffCO(\boldsymbol{\mathfrak A}({\mathcal D}))}
{\ffOO(\boldsymbol{\mathfrak A}({\mathcal D}))}}\times\left(\frac{y_1}{
(1-\mu^2x_1)(1-\kappa^2x_1)}\pm\frac{y_2}{
(1-\mu^2x_2)(1-\kappa^2x_2)} \right)^2,\nonumber\end{eqnarray}\begin{eqnarray}
&&\frac{\ffOC(\boldsymbol{\mathfrak A}({\mathcal D}))}
{\ffOO(\boldsymbol{\mathfrak A}({\mathcal D}))}
=-\frac{\kappa\lambda(1-\lambda^2
x_1)(1-\lambda^2 x_2)(1-\kappa^2 x_1)(1-\kappa^2 x_2)}{
\kI\lI\mk\ml(x_2-x_1)^2}\nonumber\\
&&\phantom{\frac{\ffCO(\boldsymbol{\mathfrak A}({\mathcal D}))}
{\ffOO(\boldsymbol{\mathfrak A}({\mathcal D}))}}\times\left({y_1}{
(1-\lambda^2x_1)(1-\kappa^2x_1)}\pm{y_2}{
(1-\lambda^2x_2)(1-\kappa^2x_2)} \right)^2,\nonumber\end{eqnarray}
\begin{eqnarray}
&&\frac{\ffAA(\boldsymbol{\mathfrak A}({\mathcal D}))}
{\ffOO(\boldsymbol{\mathfrak A}({\mathcal D}))}=\frac{\kappa(1-
x_1)(1-x_2)(1-\kappa^2 x_1)(1-\kappa^2 x_2)}{
\mI\lI\lk\mk(x_2-x_1)^2}\nonumber\\
&&\phantom{\frac{\ffCO(\boldsymbol{\mathfrak A}({\mathcal D}))}
{\ffOO(\boldsymbol{\mathfrak A}({\mathcal D}))}}\times\left(\frac{y_1}{
(1-x_1)(1-\kappa^2x_1)}\pm\frac{y_2}{ (1-x_2)(1-\kappa^2x_2)}
\right)^2,\nonumber\end{eqnarray}
\begin{eqnarray}
&&\frac{\ffAB(\boldsymbol{\mathfrak A}({\mathcal D}))}
{\ffOO(\boldsymbol{\mathfrak A}({\mathcal D}))}=\frac{\lambda(1-
x_1)(1-x_2)(1-\lambda^2 x_1)(1-\lambda^2 x_2)}{
\mI\kI\ml\lk(x_2-x_1)^2}\nonumber\\
&&\phantom{\frac{\ffCO(\boldsymbol{\mathfrak A}({\mathcal D}))}
{\ffOO(\boldsymbol{\mathfrak A}({\mathcal D}))}}\times\left(\frac{y_1}{
(1-x_1)(1-\lambda^2x_1)}\pm\frac{y_2}{ (1-x_2)(1-\lambda^2x_2)}
\right)^2,\nonumber\end{eqnarray}\begin{eqnarray}
&&\frac{\ffAC(\boldsymbol{\mathfrak A}({\mathcal D}))}
{\ffOO(\boldsymbol{\mathfrak A}({\mathcal D}))}
=\frac{\mu(1- x_1)(1-x_2)(1-\mu^2
x_1)(1-\mu^2 x_2)}{ \kI\lI\mk\ml(x_2-x_1)^2}\nonumber\\
&&\phantom{\frac{\ffCO(\boldsymbol{\mathfrak A}({\mathcal D}))}
{\ffOO(\boldsymbol{\mathfrak A}({\mathcal D}))}}\times\left(\frac{y_1}{
(1-x_1)(1-\mu^2x_1)}\pm\frac{y_2}{ (1-x_2)(1-\mu^2x_2)}
\right)^2,\nonumber\end{eqnarray}\begin{eqnarray}
&&\frac{\ffBA(\boldsymbol{\mathfrak A}({\mathcal D}))}
{\ffOO(\boldsymbol{\mathfrak A}({\mathcal D}))}
=\frac{\kappa x_1x_2(1-\kappa^2
x_1)(1-\kappa^2 x_2)}{ \kI\lk\mk(x_2-x_1)^2}\nonumber\\
&&\phantom{\frac{\ffCO(\boldsymbol{\mathfrak A}({\mathcal D}))}
{\ffOO(\boldsymbol{\mathfrak A}({\mathcal D}))}}\times\left(\frac{y_1}{
x_1(1-\kappa^2x_1)}\pm\frac{y_2}{ x_2(1-\kappa^2x_2)}
\right)^2,\nonumber\end{eqnarray}\begin{eqnarray}
&&\frac{\ffBB(\boldsymbol{\mathfrak A}({\mathcal D}))}
{\ffOO(\boldsymbol{\mathfrak A}({\mathcal D}))}
=\frac{\lambda x_1x_2(1-\lambda^2
x_1)(1-\lambda^2 x_2)}{ \lI\ml\lk(x_2-x_1)^2}\nonumber\\
&&\phantom{\frac{\ffCO(\boldsymbol{\mathfrak A}({\mathcal D}))}
{\ffOO(\boldsymbol{\mathfrak A}({\mathcal D}))}}\times\left(\frac{y_1}{
x_1(1-\lambda^2x_1)}\pm\frac{y_2}{ x_2(1-\lambda^2x_2)}
\right)^2,\nonumber\end{eqnarray}\begin{eqnarray}
&&\frac{\ffBC(\boldsymbol{\mathfrak A}({\mathcal D}))}
{\ffOO(\boldsymbol{\mathfrak A}({\mathcal D}))}=\frac{\mu x_1x_2(1-\mu^2
x_1)(1-\mu^2 x_2)}{ \mI\mk\ml(x_2-x_1)^2}\nonumber\\
&&\phantom{\frac{\ffCO(\boldsymbol{\mathfrak A}({\mathcal D}))}
{\ffOO(\boldsymbol{\mathfrak A}({\mathcal D})
)}}\times\left(\frac{y_1}{
x_1(1-\mu^2x_1)}\pm\frac{y_2}{ x_2(1-\mu^2x_2)}
\right)^2.\nonumber\end{eqnarray}

The following expressions
in terms of Kleinian functions are valid in the case of genus two.
Let
\[ y^2=R(x),\quad R(x)=4\prod_{k=1}^5(x-e_k),\quad e_i\neq e_j \]
also
\[  \boldsymbol{v}=(2\omega)^{-1}\sum_{k=1}^2
\int\limits_{(\infty,\infty)}^{(x_k,y_k)}
d \boldsymbol{u}-\boldsymbol{K}_{\infty},
\]
then
\begin{align}
\frac{  \theta^2[\boldsymbol{\mathfrak A}_k](\boldsymbol{v}|\tau)     }
      {\theta^2(\boldsymbol{v}|\tau)    }&=
\frac{\mathrm{e}^ { -\frac{\imath\pi}{2} |\boldsymbol{\mathfrak A}_k|
 } }  {\sqrt{2R'(e_k)}}
\mathcal{P}_k(\boldsymbol{u}),\label{qp}\\
\frac{\theta^2[\boldsymbol{\mathfrak A}_k+{\boldsymbol{\mathfrak
A}}_l](\boldsymbol{v}|\tau)}
      {\theta^2(\boldsymbol{v}|\tau)}
&=\frac{\mathrm{e}^{  -\frac{\imath\pi}{2}\{|\boldsymbol{\mathfrak
A}_k|+|\boldsymbol{\mathfrak A}_l|\}
 } (e_k-e_l)}  {\sqrt{2R'(e_k)R'(e_l)}}
\mathcal{Q}_{k,l}(\boldsymbol{u}),\label{qq}\\
 \frac{\theta[\boldsymbol{\mathfrak
A}_k](\boldsymbol{v}|\tau) \theta[\boldsymbol{\mathfrak
A}_l](\boldsymbol{v}|\tau) \theta[\boldsymbol{\mathfrak A}_k+
\boldsymbol{\mathfrak
A}_l](\boldsymbol{v}|\tau)}
      {\theta^3(\boldsymbol{v}|\tau)}
&=\frac{\mathrm{e}^{  -\frac{\imath\pi}{2}\{|\boldsymbol{\mathfrak
A}_k|+|\boldsymbol{\mathfrak A}_l|\}
 } (e_k-e_l)}  {\sqrt{2R'(e_k)R'(e_l)}}
\mathcal{R}_{k,l}(\boldsymbol{u}),\label{qr}
\end{align}
where
\begin{align*}
{\mathcal P}_k(\boldsymbol{u})&=e_k^2
-\wp_{22}(\boldsymbol{u})e_k-\wp_{1,2}(\boldsymbol{u}),\\
{\mathcal Q}_{k,l}(\boldsymbol{u}) &=\wp_{11}(\boldsymbol{u})+
\wp_{12}(\boldsymbol{u})(e_k+e_l)+\wp_{22}(\boldsymbol{u})e_ke_l
+e_{k,l},\\
{\mathcal R}_{k,l}(\boldsymbol{u})&= \wp_{112}(\boldsymbol{u})+
(e_k+e_l)\wp_{122}(\boldsymbol{u})+e_ke_l\wp_{222}(\boldsymbol{u})
,\end{align*} $|\boldsymbol{\mathfrak
A}_k|=(-1)^{\boldsymbol{\delta}^T_k\boldsymbol{\delta}_k}$ for the
characteristic  $[\boldsymbol{\mathfrak
A}_k]=\left[{}_{\boldsymbol{\epsilon}^T}^{\boldsymbol{\delta}^T}
\right]$ and the quantities $e_{k,l}=e_ie_j(e_p+e_q+e_r)+e_pe_qe_r.$

The complete set of hyperelliptic formulae is given in the case of genus two by Forsyth \cite{for882}.

\section{Rosenhain formulae}
Rosenhain discovered in 1851 the following:

{\bf Theorem} \emph{Let $[\delta_1],\ldots,[\delta_6]$ be the six odd characteristics and $[\delta_1]$ and $[\delta_2]$ are any two from them. With the remaining four $[\delta_{i+1}]$, $i=1,\ldots,4$ and $$[\varepsilon_i]=[\delta_1]+[\delta_2]+\delta_{i+2}(\mathrm{mod}\;1)$$
there esist 15 relations. The first one reads
 \begin{align}
\label{rosder}
\begin{split} D[\delta_1,\delta_2]&=\theta_1[\delta_1]\theta_2[\delta_2]-\theta_2[\delta_1]\theta_1[\delta_2]\\
&\pm\pi^2\theta[\varepsilon_1]\theta[\varepsilon_2]\theta[\varepsilon_3]\theta[\varepsilon_4]
\end{split}
\end{align}
remaining 14 are obtained by cyclic permutations
}
The complete list of the Rosenhain formulae \cite{ros851} for the
case $g=2$

\begin{eqnarray*}
D\left(
\left[\begin{array}{ll}\scriptstyle{0}&\!\!\!\!\scriptstyle{1}\cr
\scriptstyle{0}&\!\!\!\!\scriptstyle{1}\end{array}\right],
\left[\begin{array}{ll}\scriptstyle{1}&\!\!\!\!\scriptstyle{1}\cr
\scriptstyle{0}&\!\!\!\!\scriptstyle{1}\end{array}\right]
\right)=\pi^2
\theta\ma{0}{0}{1}{0}\theta\ma{0}{0}{1}{1}\theta\ma{1}{1}{1}{1}
\theta\ma{0}{1}{1}{0},\quad\left(\ma{1}{0}{0}{0}\right);
\end{eqnarray*}
\begin{eqnarray*}
D\left(
\left[\begin{array}{ll}\scriptstyle{1}&\!\!\!\!\scriptstyle{1}\cr
\scriptstyle{1}&\!\!\!\!\scriptstyle{0}\end{array}\right],
\left[\begin{array}{ll}\scriptstyle{1}&\!\!\!\!\scriptstyle{0}\cr
\scriptstyle{1}&\!\!\!\!\scriptstyle{0}\end{array}\right]
\right)=\pi^2
\theta\ma{1}{1}{1}{1}\theta\ma{0}{0}{0}{1}\theta\ma{0}{0}{1}{1}
\theta\ma{1}{0}{0}{1},\quad\left(\ma{0}{1}{0}{0}\right);
\end{eqnarray*}
\begin{eqnarray*}
D\left(
\left[\begin{array}{ll}\scriptstyle{1}&\!\!\!\!\scriptstyle{0}\cr
\scriptstyle{1}&\!\!\!\!\scriptstyle{1}\end{array}\right],
\left[\begin{array}{ll}\scriptstyle{0}&\!\!\!\!\scriptstyle{1}\cr
\scriptstyle{1}&\!\!\!\!\scriptstyle{1}\end{array}\right]
\right)=\pi^2
\theta\ma{0}{1}{1}{0}\theta\ma{1}{0}{0}{1}\theta\ma{0}{0}{1}{0}
\theta\ma{0}{0}{0}{1},\quad\left(\ma{1}{1}{0}{0}\right);
\end{eqnarray*}
\begin{eqnarray*}
D\left(
\left[\begin{array}{ll}\scriptstyle{0}&\!\!\!\!\scriptstyle{1}\cr
\scriptstyle{1}&\!\!\!\!\scriptstyle{1}\end{array}\right],
\left[\begin{array}{ll}\scriptstyle{0}&\!\!\!\!\scriptstyle{1}\cr
\scriptstyle{0}&\!\!\!\!\scriptstyle{1}\end{array}\right]
\right)=\pi^2
\theta\ma{1}{0}{0}{0}\theta\ma{1}{0}{0}{1}\theta\ma{1}{1}{0}{0}
\theta\ma{1}{1}{1}{1},\quad\left(\ma{0}{0}{1}{0}\right);
\end{eqnarray*}

\begin{eqnarray*}
D\left(
\left[\begin{array}{ll}\scriptstyle{0}&\!\!\!\!\scriptstyle{1}\cr
\scriptstyle{1}&\!\!\!\!\scriptstyle{1}\end{array}\right],
\left[\begin{array}{ll}\scriptstyle{1}&\!\!\!\!\scriptstyle{1}\cr
\scriptstyle{0}&\!\!\!\!\scriptstyle{1}\end{array}\right]
\right)=\pi^2
\theta\ma{0}{0}{0}{0}\theta\ma{0}{0}{0}{1}\theta\ma{1}{1}{1}{1}
\theta\ma{0}{1}{0}{0},\quad\left(\ma{1}{0}{1}{0}\right);
\end{eqnarray*}
\begin{eqnarray*}
D\left(
\left[\begin{array}{ll}\scriptstyle{1}&\!\!\!\!\scriptstyle{0}\cr
\scriptstyle{1}&\!\!\!\!\scriptstyle{1}\end{array}\right],
\left[\begin{array}{ll}\scriptstyle{1}&\!\!\!\!\scriptstyle{1}\cr
\scriptstyle{0}&\!\!\!\!\scriptstyle{1}\end{array}\right]
\right)=\pi^2
\theta\ma{1}{1}{0}{0}\theta\ma{0}{0}{1}{1}\theta\ma{0}{0}{0}{1}
\theta\ma{1}{0}{0}{0},\quad\left(\ma{0}{1}{1}{0}\right);
\end{eqnarray*}
\begin{eqnarray*}
D\left(
\left[\begin{array}{ll}\scriptstyle{1}&\!\!\!\!\scriptstyle{0}\cr
\scriptstyle{1}&\!\!\!\!\scriptstyle{1}\end{array}\right],
\left[\begin{array}{ll}\scriptstyle{0}&\!\!\!\!\scriptstyle{1}\cr
\scriptstyle{0}&\!\!\!\!\scriptstyle{1}\end{array}\right]
\right)=\pi^2
\theta\ma{0}{1}{0}{0}\theta\ma{1}{0}{0}{1}\theta\ma{0}{0}{0}{0}
\theta\ma{0}{0}{1}{1},\quad\left(\ma{1}{1}{1}{0}\right);\end{eqnarray*}

\begin{eqnarray*}
D\left(
\left[\begin{array}{ll}\scriptstyle{1}&\!\!\!\!\scriptstyle{0}\cr
\scriptstyle{1}&\!\!\!\!\scriptstyle{0}\end{array}\right],
\left[\begin{array}{ll}\scriptstyle{1}&\!\!\!\!\scriptstyle{0}\cr
\scriptstyle{1}&\!\!\!\!\scriptstyle{1}\end{array}\right]
\right)=\pi^2
\theta\ma{0}{1}{0}{0}\theta\ma{0}{1}{1}{0}\theta\ma{1}{1}{1}{1}
\theta\ma{1}{1}{0}{0},\quad\left(\ma{0}{0}{0}{1}\right);
\end{eqnarray*}
\begin{eqnarray*}
D\left(
\left[\begin{array}{ll}\scriptstyle{1}&\!\!\!\!\scriptstyle{1}\cr
\scriptstyle{1}&\!\!\!\!\scriptstyle{0}\end{array}\right],
\left[\begin{array}{ll}\scriptstyle{0}&\!\!\!\!\scriptstyle{1}\cr
\scriptstyle{1}&\!\!\!\!\scriptstyle{1}\end{array}\right]
\right)=\pi^2
\theta\ma{0}{0}{1}{1}\theta\ma{0}{0}{1}{0}\theta\ma{1}{1}{0}{0}
\theta\ma{0}{1}{0}{0},\quad\left(\ma{1}{0}{0}{1}\right);\end{eqnarray*}
\begin{eqnarray*}
D\left(
\left[\begin{array}{ll}\scriptstyle{1}&\!\!\!\!\scriptstyle{1}\cr
\scriptstyle{1}&\!\!\!\!\scriptstyle{0}\end{array}\right],
\left[\begin{array}{ll}\scriptstyle{1}&\!\!\!\!\scriptstyle{0}\cr
\scriptstyle{1}&\!\!\!\!\scriptstyle{1}\end{array}\right]
\right)=\pi^2
\theta\ma{1}{1}{1}{1}\theta\ma{0}{0}{0}{0}\theta\ma{0}{0}{1}{0}
\theta\ma{1}{0}{0}{0},\quad\left(\ma{0}{1}{0}{1}\right);\end{eqnarray*}
\begin{eqnarray*}
D\left(
\left[\begin{array}{ll}\scriptstyle{1}&\!\!\!\!\scriptstyle{0}\cr
\scriptstyle{1}&\!\!\!\!\scriptstyle{0}\end{array}\right],
\left[\begin{array}{ll}\scriptstyle{0}&\!\!\!\!\scriptstyle{1}\cr
\scriptstyle{1}&\!\!\!\!\scriptstyle{1}\end{array}\right]
\right)=\pi^2
\theta\ma{0}{1}{1}{0}\theta\ma{1}{0}{0}{0}\theta\ma{0}{0}{1}{1}
\theta\ma{0}{0}{0}{0},\quad\left(\ma{1}{1}{0}{1}\right);\end{eqnarray*}

\begin{eqnarray*}
D\left(
\left[\begin{array}{ll}\scriptstyle{1}&\!\!\!\!\scriptstyle{1}\cr
\scriptstyle{1}&\!\!\!\!\scriptstyle{0}\end{array}\right],
\left[\begin{array}{ll}\scriptstyle{1}&\!\!\!\!\scriptstyle{1}\cr
\scriptstyle{0}&\!\!\!\!\scriptstyle{1}\end{array}\right]
\right)=\pi^2
\theta\ma{1}{0}{0}{1}\theta\ma{1}{0}{0}{0}\theta\ma{0}{1}{1}{0}
\theta\ma{0}{1}{0}{0},\quad\left(\ma{0}{0}{1}{1}\right);\end{eqnarray*}
\begin{eqnarray*}
D\left(
\left[\begin{array}{ll}\scriptstyle{1}&\!\!\!\!\scriptstyle{1}\cr
\scriptstyle{1}&\!\!\!\!\scriptstyle{0}\end{array}\right],
\left[\begin{array}{ll}\scriptstyle{0}&\!\!\!\!\scriptstyle{1}\cr
\scriptstyle{0}&\!\!\!\!\scriptstyle{1}\end{array}\right]
\right)=\pi^2
\theta\ma{0}{0}{0}{1}\theta\ma{0}{0}{0}{0}\theta\ma{1}{1}{0}{0}
\theta\ma{0}{1}{1}{0},\quad\left(\ma{1}{0}{1}{1}\right);\end{eqnarray*}
\begin{eqnarray*}
D\left(
\left[\begin{array}{ll}\scriptstyle{1}&\!\!\!\!\scriptstyle{0}\cr
\scriptstyle{1}&\!\!\!\!\scriptstyle{0}\end{array}\right],
\left[\begin{array}{ll}\scriptstyle{1}&\!\!\!\!\scriptstyle{1}\cr
\scriptstyle{0}&\!\!\!\!\scriptstyle{1}\end{array}\right]
\right)=\pi^2
\theta\ma{1}{1}{0}{0}\theta\ma{0}{0}{1}{0}\theta\ma{0}{0}{0}{0}
\theta\ma{1}{0}{0}{1},\quad\left(\ma{0}{1}{1}{1}\right);\end{eqnarray*}
\begin{eqnarray*}
D\left(
\left[\begin{array}{ll}\scriptstyle{1}&\!\!\!\!\scriptstyle{0}\cr
\scriptstyle{1}&\!\!\!\!\scriptstyle{0}\end{array}\right],
\left[\begin{array}{ll}\scriptstyle{0}&\!\!\!\!\scriptstyle{1}\cr
\scriptstyle{0}&\!\!\!\!\scriptstyle{1}\end{array}\right]
\right)=\pi^2
\theta\ma{0}{1}{0}{0}\theta\ma{1}{0}{0}{0}\theta\ma{0}{0}{0}{1}
\theta\ma{0}{0}{1}{0},\quad\left(\ma{1}{1}{1}{1}\right).
\end{eqnarray*}
We pointed at the right margin the characteristic, which is the
sum of characteristics of each entry to the corresponding
equality.

\subsection{Winding vectors}
For any $k \neq l \neq p,q,r$ from the set $\{1,\ldots,5\}$ (there
are 10 different possibilities) the following representation is
valid (see theorem~\ref{thomaematrix})
\begin{equation}
2\omega = \frac{T_{pqr}}{\theta\{k,l\}(e_k-e_l)^{\frac32}}
 \begin{pmatrix} \M 1&-1\\-e_l& \M e_k
 \end{pmatrix}
 \begin{pmatrix}\frac{1}{T_l}&0\\0&\frac{1}{T_k}
 \end{pmatrix}
 \begin{pmatrix}
   \theta_1\{l\}&\theta_2\{l\}\\\theta_1\{k\}&\theta_2\{k\}
   \end{pmatrix},
\end{equation}
where
\begin{equation}
\begin{split}
 T_k&=\theta\{p,k\}  \theta\{q,k\} \theta\{r,k\} ,\\
 T_l&=\theta\{p,l\}  \theta\{q,l\} \theta\{r,l\} ,\\
 T_{pqr} & =\theta\{p,q\}  \theta\{p,r\} \theta\{q,r\}.
\end{split}
\end{equation}
Remark that the modular form of the weight 5, $\chi_5$, is given in
this notations as
\[\chi_5=\prod_{\text{even}\; [\varepsilon]} \theta[\varepsilon] = T_k T_l T_{pqr}\theta\{k,l\}  \]
with $k\neq l \neq p \neq q \neq r \in \{1,2,3,4,5\}$.

Choosing $k=2$ and $l=4$, and normalizing the curve with $e_2=0$ and
$e_4=1$, we obtain the explicit result
\begin{equation}
 2\omega = \frac{T_{135}^2}{\chi_5}
 \begin{pmatrix}
  T_{135}\,\theta_1\left[{}_0^1{}_1^1\right]_2  &
  T_{135}\,\theta_2\left[{}_0^1{}_1^1\right]_2 \\
  T_4\;\;\; \theta_1\left[{}_1^0{}_1^1\right]_2  &
  T_4\;\;\; \theta_2\left[{}_1^0{}_1^1\right]_2
 \end{pmatrix} ,
\end{equation}
where
\begin{equation}
\begin{split}
 T_2&= \theta\left[{}_1^1{}_1^1\right]_2
     \theta\left[{}_0^0{}_1^0\right]_2
     \theta\left[{}_0^0{}_0^1\right]_2 ,\\
 T_4&= \theta\left[{}_1^0{}_0^0\right]_2
     \theta\left[{}_0^1{}_0^1\right]_2
     \theta\left[{}_0^1{}_1^0\right]_2 , \\
 T_{135} & = \theta\left[{}_1^0{}_1^0\right]_2
     \theta\left[{}_1^0{}_0^1\right]_2
     \theta\left[{}_0^1{}_0^0\right]_2.
\end{split}
\end{equation}
In the derivation of this result we used the equality ($i = 1,2$)
\[
 \theta_i\left[{}_1^0{}_1^1\right]\theta\left[{}_0^1{}_0^1\right]
 \theta\left[{}_1^0{}_0^0\right] \theta\left[{}_0^1{}_1^0\right] -
 \theta_i\left[{}_1^1{}_0^0\right]\theta\left[{}_0^0{}_0^1\right]
 \theta\left[{}_0^0{}_1^0\right] \theta\left[{}_1^1{}_1^1\right]
 =
 \theta_i\left[{}_0^1{}_1^1\right]\theta\left[{}_1^0{}_1^0\right]
 \theta\left[{}_1^0{}_0^1\right] \theta\left[{}_0^1{}_0^0\right],
\]
\[\text{equivalently}\quad
\theta_i\{2\}T_4 -
 \theta_i\{4\}T_2
 =
 \theta_i\{6\}  T_{135}, \]
 which can be derived from addition theorems as
given, e.\,g., on p.~342 in~ Baker(1897); a complete set of such
relations can be found in~ Forsyth (1882).

The entries to the inverse matrix $\rho=2\omega^{-1}$ are the
normalizing constants of the holomorphic differentials. Its columns
represent the so-called winding vectors in the Its-Matveev
formula~ Its\&Matveev (1975) for the genus two solution of the KdV equation.
For the case of $g=2$ the general formula~(\ref{finomega}) for
$\rho$ reduces to
\begin{equation}
 \rho =  (2\omega)^{-1} =
 \frac{\sqrt{e_k-e_l}}{\pi^2 T^2_{pqr}}
 \Bigl(\begin{array}{cc}\M\theta_2\{k\}&-\theta_2\{l\}\\
                       -\theta_1\{k\}&\M\theta_1\{l\}
       \end{array}
 \Bigr)  \Bigl(
 \begin{array}{cc}T_l&0\\0&T_k\end{array}\Bigr)
 \Bigl(\begin{array}{cc}
 e_k&1\\ e_l&1\end{array}\Bigr).
 \label{g2wind}
\end{equation}
With the normalization $e_2=0$, $e_4=1$, this coincides with the
formulae given in the Rosenhain memoir of Rosenhain (1851), page 75:
\begin{equation}
 \label{eq:Rosenhain}
 \rho = \frac{1}{\pi^2T^2_{135}}
 \begin{pmatrix}
  \M -T_4\,\theta_2\left[{}_1^0{}_1^1\right]_2  &
  T_{135}\theta_2\left[{}_0^1{}_1^1\right]_2 \\
  T_4\,\theta_1\left[{}_1^0{}_1^1\right]_2  &
  \M -T_{135}\theta_1\left[{}_0^1{}_1^1\right]_2
 \end{pmatrix} .
\end{equation}

\section{Absolute invariants of genus two curve}
Absolute invariants of the genus two hyperelliptic curves serves to check equivalence of the curves under the M\"obius transformations. In the case of genus two curve exist 3 absolute invariants that could be expressed both in terms of coefficients of the curves as well in theta-constants. We will use both representations.

Let the curve is given as sextic,
\begin{equation}  y^2=u_0x^6+u_1x^5+u_2x^4+u_3x^3+u_4x^2+u_5x+u_6. \label{sextic}  \end{equation}
Denote branch points as $e_j,j=1,\ldots,6$ and denote $(i,j)=e_i-e_j$. Then relative invariants with respect to the M\"obius transformations are given as
\begin{align*}
A(u)&=u_0^2\sum_{\rm{fifteen}}(12)^2(34)^2(56)^2\\
B(u)&=u_0^4\sum_{\rm{ten}}(12)^2(23)^2(31)^2(45)^2(56)^2(64)^2\\
C(u)&=u_0^6\sum_{\rm{sixty}}(12)^2(23)^2(31)^2(45)^2(56)^2(64)^2
(14)^2(25)^2(36)^2\\
D(u)&=u_0^{10}\prod_{j<k}(j,k)^2,
\end{align*}
where $D(u)$ is just the discriminant of the sextic. Direct computation leads to the following expressions for $A,B,C,D$ in terms of coefficients of (\ref{sextic}).

\begin{align*}
A&=2(20u_1u_5-8u_2u_4+3u_3^2-8u_1u_3-120u_6u_0+20u_0u_4 )\\
B&=4(u_2^2u_4^2-3u_1u_3u_4^2-3u_2^2u_3u_5+9u_1u_2^2u_5+u_1u_2u_4u_5
-20u_1^2u_5^2+12u_2^3u_6\\&-45u_1u_2u_3u_6+75u_1^2u_4u_6
+12u_0u_4^3-135u_0u_1u_5u_6-126u_0u_2u_4u_6+81u_0u_3^2u_6\\&+75u_0u_2u_5^2-45u_0u_3u_4u_5+405u_0^2u_6^2)\\
\end{align*}
\begin{align*}
C&=-12u_{2}^3u_{4}^3-12u_{2}^2u_{3}^3u_{5}+4u_{2}^2u_{3}^2u_{4}^2-119u_{1}u_{2}u_{3}^2u_{4}u_{5}
-12u_{1}u_{3}^3u_{4}^2+36u_{1}u_{3}^4u_{5}+88u_{1}^2u_{3}^2u_{5}^2\\
&+246u_{1}u_{2}^2u_{3}u_{4}u_{6}-930u_{1}^2u_{2}u_{3}u_{5}u_{6}+30u_{2}^3u_{3}^2u_{6}
-80u_{2}^4u_{4}u_{6}+38u_{2}^3u_{3}u_{4}u_{5}+1125u_{1}^3u_{3}u_{6}^2\\
&-450u_{1}^2u_{2}^2u_{6}^2-160u_{1}^3u_{5}^3+14u_{1}u_{2}^2u_{4}^2u_{5}+38u_{1}u_{2}u_{3}u_{4}^3
+13u_{1}u_{2}^2u_{3}u_{5}^2+32u_{1}^2u_{2}u_{4}u_{5}^2-99u_{1}u_{2}u_{3}^3u_{6}\\
&+308u_{1}u_{2}^3u_{5}u_{6}-320u_{1}^2u_{2}u_{4}^2u_{6}+800u_{1}^3u_{4}u_{5}u_{6}
+165u_{1}^2u_{3}^2u_{4}u_{6}-48u_{2}^3u_{6}^2u_{0}+81u_{3}^4u_{6}u_{0}\\
&-48u_{4}^3u_{6}u_{0}^2-5022u_{3}^2u_{6}^2u_{0}^2+30u_{3}^2u_{4}^3u_{0}-80u_{2}u_{4}^4u_{0}
-59940u_{6}^3u_{0}^3+308u_{1}u_{4}^3u_{5}u_{0}-320u_{2}^2u_{4}u_{5}^2u_{0}\\
&-99u_{3}^3u_{4}u_{5}u_{0}+800u_{1}u_{2}u_{5}^3u_{0}+165u_{2}u_{3}^2u_{5}^2u_{0}
-9300u_{1}^2u_{4}u_{6}^2u_{0}-9300u_{2}u_{5}^2u_{6}u_{0}^2+212u_{2}^2u_{4}^2u_{6}u_{0}\\
&+10332u_{2}u_{4}u_{6}^2u_{0}^2-1120u_{1}^2u_{5}^2u_{6}u_{0}+29970u_{1}u_{5}u_{6}^2u_{0}^2
-18u_{2}^4u_{5}^2+1736u_{1}u_{2}u_{4}u_{5}u_{6}u_{0}+13u_{1}^2u_{3}u_{4}^2u_{5}\\
&-18u_{1}^2u_{4}^4-450u_{4}^2u_{5}^2u_{0}^2+1125u_{3}u_{5}^3u_{0}^2-930u_{1}u_{3}u_{4}u_{5}^2u_{0}
+246u_{2}u_{3}u_{4}^2u_{5}u_{0}+1530u_{1}u_{2}u_{3}u_{6}^2u_{0}\\
&-438u_{1}u_{3}u_{4}^2u_{6}u_{0}-234u_{2}u_{3}^2u_{4}u_{6}u_{0}+909u_{1}u_{3}^2u_{5}u_{6}u_{0}
-438u_{2}^2u_{3}u_{5}u_{6}u_{0}+1530u_{3}u_{4}u_{5}u_{6}u_{0}^2
\end{align*}
These formulae could be found in \cite{sv04}.

Alternatively the relative invariants can be rewritten by means of Bolza (\ref{e2a}) and Rosenhain (\ref{rosder}) formulae in terms of the $\theta$-constants as follows
\begin{align}
J_2&=48\pi^{12}\;\frac{\sum\limits_{15\;\rm{terms}
}\prod\limits_{k=1,\ldots,6,\;\sum{[\varepsilon_k]=0}}\;
\theta^4[\varepsilon_k]}
{\left(\prod\limits_{6 \;\rm{odd}\; [\delta]}
\theta_{1}[\delta]\right)^2 }\\
J_4&=72\pi^{24}\;\frac{
\sum\limits_{10\;\rm{even}\; [\varepsilon]}\theta^8[\varepsilon_k]
\prod\limits_{10 \;\rm{even}\; [\varepsilon]}
\theta^4[\varepsilon] }
{\left(\prod\limits_{6 \;\rm{odd}\; [\delta]}
\theta_{1}[\delta]\right)^4 }\\
J_6&=12\pi^{36}\;\frac{
\sum\limits_{60\;\rm{terms}}\theta[\varepsilon]^8
\sum\limits _{6\; \rm{terms}\;
[\varepsilon_k]\neq [\varepsilon],\;
\sum[\varepsilon_k]=0    } \prod\limits_{k=1}^6\;\theta^4[\varepsilon_k]
}
{\left(\prod\limits_{6 \;\rm{odd}\; [\delta]}
\theta_{1}[\delta]\right)^6 }\\
J_{10}&=\pi^{60}\;\frac{
\prod\limits_{10 \;\rm{even}\; [\varepsilon]}
\theta^{12}[\varepsilon] }
{\left(\prod\limits_{6 \;\rm{odd}\; [\delta]}
\theta_{1}[\delta]\right)^{10} }
\end{align}
To the best knowledge of the authors this representation of relative invariants in terms of theta-constants is published to the first time.

Summarizing we get for the absolute invariants taken in the form of the \cite{sv04}

\begin{theorem}
Absolute invariants which are quotients of relative invariants of the same
order that we fix in the form
\begin{align}
i_1=144\frac{J_{4}}{J_{2}^2},\quad i_2=-1728\frac{(J_4J_2-3J_6)}{J_{2}^3},
\quad i_3=486\frac{J_{10}}{J_{2}^5}
\end{align}
are represented in terms of of coefficients of the sextic $u_0,\ldots, u_6$ or
$\theta$-constants.
\end{theorem}

The two presented representations of the absolute invariant could be implemented for the check if Riemann period matrix of a genus two curve and realization of the curve as a sextic/quitic belongs to the same curve.

\chapter*{Appendix II: A set of formulae for genus $3$}

\section{Hyperelliptic curve of genus three}\label{sec:curvesgenus3}

Consider also the hyperelliptic curve $X_3$ of genus three with seven real zeros.
Let the curve $X_3$ be given by
\begin{align}\begin{split}
w^2&=4(z-e_1)(z-e_2)(z-e_3)(z-e_4)(z-e_5)(z-e_6)(z-e_7)\\
&=4z^{7}+\lambda_6z^6+\ldots+\lambda_1z+\lambda_0 \,.
\end{split}
\label{curve3}
\end{align}

The complete set of holomorphic and meromorphic differentials with a unique pole at infinity is
\begin{align}
\mathrm{d}u_1 & = \frac{\mathrm{d}z}{w}\,, & \qquad \mathrm{d}r_1 & = z(20z^4+4\lambda_6z^3+3\lambda_5z^2+2\lambda_4z+\lambda_3)
 \frac{\mathrm{d}z}{4w} \,, \nonumber \\
\mathrm{d}u_2 &= \frac{z\mathrm{d}z}{w}\,, & \qquad \mathrm{d}r_2 & = z^2(12z^2+2\lambda_6z+\lambda_5)\frac{\mathrm{d}z}{4w}\,, \\
\mathrm{d}u_3 &= \frac{z^2\mathrm{d}z}{w}\,, & \qquad  \mathrm{d}r_3 & = \frac{z^3\mathrm{d}z}{w} \,. \nonumber
\end{align}
Again we introduce the winding vectors
\begin{equation}
(2\omega)^{-1} = \left( \boldsymbol{U}, \boldsymbol{V}, \boldsymbol{W} \right) \, .
\end{equation}

The Jacobi inversion problem for the equations
\begin{align}\begin{split}
\int_{\infty}^{(z_1,w_1)}\frac{ \mathrm{d}z}{w}
+\int_{\infty}^{(z_2,w_2)}\frac{ \mathrm{d}z}{w}
+\int_{\infty}^{(z_3,w_3)}\frac{ \mathrm{d}z}{w}
=u_1,\\
\int_{\infty}^{(z_1,w_1)}\frac{z \mathrm{d}z}{w}
+\int_{\infty}^{(z_2,w_2)}\frac{z \mathrm{d}z}{w}
+\int_{\infty}^{(z_3,w_3)}\frac{z \mathrm{d}z}{w}
=u_2,\\
\int_{\infty}^{(z_1,w_1)}\frac{z^2 \mathrm{d}z}{w}
+\int_{\infty}^{(z_2,w_2)}\frac{z^2 \mathrm{d}z}{w}
+\int_{\infty}^{(z_3,w_3)}\frac{z^2 \mathrm{d}z}{w}
=u_3\end{split} \label{JIP3}
\end{align}
is solved by
\begin{align}\begin{split}
z_1+z_2+z_3&=\wp_{33}(\boldsymbol{u}), \quad z_1z_2+z_1z_3+z_2z_3=-\wp_{23}(\boldsymbol{u}), \quad z_1z_2z_3= \wp_{13}(\boldsymbol{u})\\
w_k&=\wp_{333}(\boldsymbol{u})z_k^2+\wp_{233}(\boldsymbol{u})z_k + \wp_{133}(\boldsymbol{u}) , \quad k=1,2,3 \,.
\end{split} \label{SOLJIP3}
\end{align}

\begin{figure}
\begin{center}
\unitlength 0.7mm \linethickness{0.6pt}
\begin{picture}(150.00,80.00)
\put(9.,33.){\line(1,0){12.}} \put(9.,33.){\circle*{1}}
\put(21.,33.){\circle*{1}} \put(10.,29.){\makebox(0,0)[cc]{$e_1$}}
\put(21.,29.){\makebox(0,0)[cc]{$e_2$}}
\put(15.,33.){\oval(20,30.)}
\put(8.,17.){\makebox(0,0)[cc]{$\mathfrak{ a}_1$}}
\put(15.,48.){\vector(1,0){1.0}}
\put(32.,33.){\line(1,0){9.}} \put(32.,33.){\circle*{1}}
\put(41.,33.){\circle*{1}} \put(33.,29.){\makebox(0,0)[cc]{$e_3$}}
\put(42.,29.){\makebox(0,0)[cc]{$e_4$}}
\put(37.,33.){\oval(18.,26.)}
\put(30.,19.){\makebox(0,0)[cc]{$\mathfrak{a}_2$}}
\put(36.,46.){\vector(1,0){1.0}}
\put(57.,33.){\line(1,0){10.}} \put(57.,33.){\circle*{1}}
\put(67.,33.){\circle*{1}} \put(57.,29.){\makebox(0,0)[cc]{$e_5$}}
\put(67.,29.){\makebox(0,0)[cc]{$e_6$}}
\put(62.,33.){\oval(18.,21.)}
\put(54.,21.){\makebox(0,0)[cc]{$\mathfrak{a}_3$}}
\put(62.,43.5){\vector(1,0){1.0}}
\put(100.,33.00) {\line(1,0){33.}} \put(100.,33.){\circle*{1}}
\put(133.,33.){\circle*{1}}
\put(101.,29.){\makebox(0,0)[cc]{$e_{7}$}}
\put(132.,29.){\makebox(0,0)[cc]{$e_{8}=\infty$}}
\put(66.,63.){\makebox(0,0)[cc]{$\mathfrak{b}_1$}}
\put(70.,66.){\vector(1,0){1.0}}
\bezier{484}(15.,33.)(15.,66.)(70.,66.)
\bezier{816}(70.,66.)(120.,66.00)(120.,33.)
\bezier{35}(15.,33.)(15.,0.)(70.,0.)
\bezier{35}(70.,0.)(120.,0.)(120.,33.)
\put(70.,55.){\makebox(0,0)[cc]{$\mathfrak{b}_2$}}
\put(74.00,58.){\vector(1,0){1.0}}
\bezier{384}(37.,33.)(37.,58.)(76.,58.)
\bezier{516}(76.,58.)(115.,58.)(115.00,33.00)
\bezier{30}(37.,33.00)(37.,8.)(76.00,8.)
\bezier{30}(76.00,8.)(115.00,8.00)(115.00,33.00)
\put(82.,42.){\makebox(0,0)[cc]{$\mathfrak{b}_3$}}
\put(85,45.){\vector(1,0){1.0}}
\bezier{384}(62.,33.)(62.,45.)(85.,45.)
\bezier{516}(85.,45.)(110.,45.)(110.00,33.00)
\bezier{30}(62.,33.00)(62.,21.)(85.00,21.)
\bezier{30}(85.00,21.)(110.00,21.00)(110.00,33.00)
\end{picture}
\end{center}
\caption{Homology basis on the Riemann surface of the curve
$X_3$ with real branch points $e_1 < e_2 <\ldots <
e_{8}=\infty$ (upper sheet).  The cuts are drawn from $e_{2i-1}$
to $e_{2i}$, $i=1,2,4$.  The $\mathfrak{b}$--cycles are completed on the
lower sheet (dotted lines).} \label{figure-3}
\end{figure}

\subsection{Characteristics in genus three}

Let $\mathfrak{A}_k$ be the Abelian image of the $k$-th branch point, namely
\begin{equation}
\boldsymbol{\mathfrak{A}}_k=\int_{\infty}^{(e_k,0)} \mathrm{d}\boldsymbol{u}= 2\omega \boldsymbol{\varepsilon}_k+2\omega' \boldsymbol{\varepsilon}_k', \quad k=1,\ldots,8 \,,
\end{equation}
where $\boldsymbol{\varepsilon}_k$ and $\boldsymbol{\varepsilon}_k'$ are column vectors whose entries $\varepsilon_{k,j}$, $\varepsilon'_{k,j}$ are $\frac{1}{2}$ or $0$ for all $k=1,\ldots,8$, $j=1,2,3$.

The correspondence between the branch points and the characteristics in the fixed homology basis is given as
\begin{align}\begin{split}
[\boldsymbol{{\mathfrak A}}_1]= \frac{1}{2}
\begin{pmatrix} 1 & 0 & 0 \\ 0 & 0 & 0 \end{pmatrix}\, ,\quad
[\boldsymbol{{\mathfrak A}}_2]= \frac{1}{2}
\begin{pmatrix} 1 & 0 & 0 \\ 1 & 0 & 0 \end{pmatrix}\, ,\quad
[\boldsymbol{{\mathfrak A}}_3] = \frac{1}{2}
\begin{pmatrix} 0 & 1 & 0 \\ 1 & 0 & 0 \end{pmatrix}\, , \\
[\boldsymbol{{\mathfrak A}}_4]= \frac{1}{2}
\begin{pmatrix} 0 & 1 & 0 \\ 1 & 1 & 0 \end{pmatrix}\, ,\quad
[\boldsymbol{{\mathfrak A}}_5] = \frac{1}{2}
\begin{pmatrix} 0 & 0 & 1 \\ 1 & 1 & 0 \end{pmatrix}\, ,\quad
[\boldsymbol{{\mathfrak A}}_6]= \frac{1}{2}
\begin{pmatrix} 0 & 0 & 1 \\ 1 & 1 & 1 \end{pmatrix}\, , \\
[\boldsymbol{{\mathfrak A}}_7]= \frac{1}{2}
\begin{pmatrix} 0 & 0 & 0 \\ 1 & 1 & 1 \end{pmatrix}\, ,\quad
[\boldsymbol{{\mathfrak A}}_8]= \frac{1}{2}
\begin{pmatrix} 0 & 0 & 0 \\ 0 & 0 & 0 \end{pmatrix}\, . \end{split}
\label{hombasis_gen3}
\end{align}

The vector of Riemann constants $\boldsymbol{K}_{\infty}$ with the base point at infinity is given in the above basis by the even singular characteristics,
\begin{equation}
[\boldsymbol{K}_{\infty}]=[\boldsymbol{\mathfrak A}_2]+[\boldsymbol{\mathfrak A}_4]+[\boldsymbol{\mathfrak A}_6] = \frac12 \begin{pmatrix} 1 & 1 & 1 \\ 1 & 0 & 1 \end{pmatrix}  \, .
\end{equation}

From the above characteristics 64 half-periods can be built as follows. Start with singular even characteristics, there should be only one such characteristic that corresponds to the vector of Riemann constants $\boldsymbol{K}_{\infty}$. The corresponding partition reads $\mathcal{I}_2\cup \mathcal{J}_2 = \{ \} \cup \{1, 2, \ldots, 8\}$ and the $\theta$--function $\theta(\boldsymbol{K}_{\infty}+\boldsymbol{v})$ vanishes at the origin $\boldsymbol{v}=0$ to the order $m=2$.

The half-periods $\boldsymbol{\Delta}_1=(2\omega)^{-1}\boldsymbol{\mathfrak A}_k+\boldsymbol{K}_{\infty}\in \Theta_1$ correspond to partitions
\begin{equation}
\mathcal{I}_1\cup \mathcal{J}_1 =  \{ k,8 \} \cup \{ j_1,\ldots, j_6 \}, \quad j_1,\ldots,j_6 \notin \{8,k\}
\end{equation}
and the $\theta$--function $\theta(\boldsymbol{\Delta}_1+\boldsymbol{v})$ vanishes at the origin $\boldsymbol{v}=0$ to the order $m=1$.

Also denote the $21$ half-periods that are images of two branch points
\begin{align}
\boldsymbol{\Omega}_{ij}=2\omega ( \boldsymbol{\varepsilon}_i+\boldsymbol{\varepsilon}_j)+2\omega' (\boldsymbol{\varepsilon}_i'+  \boldsymbol{\varepsilon}_j') ,\quad  i,j = 1,\ldots, 7, i\neq j \ .
\label{odd2}
\end{align}
The half-periods $\boldsymbol{\Delta}_1=(2\omega)^{-1}\boldsymbol{\Omega}_{ij}
+\boldsymbol{K}_{\infty}\in \Theta_2$ correspond to the partitions
\begin{equation}
\mathcal{I}_1\cup \mathcal{J}_1 = \{ i,j \} \cup \{ j_1,\ldots, j_6 \}, \quad j_1,\ldots,j_6 \notin \{i,j\}
\end{equation}
and the $\theta$--function $\theta(\boldsymbol{\Delta}_1 + \boldsymbol{v})$ vanishes at the origin, $\boldsymbol{v}=0$, as before to the order $m=1$.
Therefore the characteristics of the $7$ half-periods
\begin{equation}
\left[(2\omega)^{-1} \boldsymbol{\mathfrak A}_i+\boldsymbol{K}_{\infty}\right] =: \delta_i \, ,\quad i=1,\ldots,7
\end{equation}
are nonsingular and odd as well as the characteristics of the $21$ half-periods
\begin{equation}
\left[(2\omega)^{-1} \boldsymbol{\Omega}_{ij}+\boldsymbol{K}_{\infty}\right] =: \delta_{ij}   \, ,\quad 1\leq i<j\leq 7 \,.
\end{equation}
We finally introduce the $35$ half-periods that are images of three branch points
\begin{align}
\boldsymbol{\Omega}_{ijk}=2\omega ( \boldsymbol{\varepsilon}_i+\boldsymbol{\varepsilon}_j+\boldsymbol{\varepsilon}_k)+ 2\omega' (\boldsymbol{\varepsilon}_i'+  \boldsymbol{\varepsilon}_j'+  \boldsymbol{\varepsilon}_k')\in \mathrm{Jac}(X_g) ,\quad  1\leq i<j<k\leq 7 \,.
\end{align}
The half-periods $\boldsymbol{\Delta}_2=(2\omega)^{-1}\boldsymbol{\Omega}_{ijk}
+\boldsymbol{K}_{\infty}$ correspond to the partitions
\begin{equation}
\mathcal{I}_0\cup \mathcal{J}_0 =  \{ i,j,k,8 \} \cup \{ j_1,\ldots, j_4 \}, \quad j_1,\ldots,j_4 \notin \{i,j,k,8\}\,.
\end{equation}
The $\theta$--function $\theta(\boldsymbol{\Delta}_2+\boldsymbol{v})$
does not vanish at the origin $\boldsymbol{v}=0$.

Furthermore, the $35$ characteristics
\begin{equation}
[\varepsilon_{ijk}]=  \left[ (2\omega)^{-1} \boldsymbol{\Omega}_{ijk}+\boldsymbol{K}_{\infty} \right],\quad 1\leq i<j<k\leq 7
\end{equation}
are even and nonsingular while the characteristic $[\boldsymbol{K}_{\infty}]$ is even and singular. Altogether we got all $64=4^3$ characteristics classified by the partitions of the branch points.

\subsection{Inversion of a holomorphic integral}

All three holomorphic integrals,
\begin{align}
\int_{\infty}^{(x,w)} \frac{\mathrm{d}z}{w}=u_1,\quad
\int_{\infty}^{(x,w)} \frac{z\mathrm{d}z}{w}=u_2,\quad
\int_{\infty}^{(x,w)} \frac{z^2\mathrm{d}z}{w}=u_3
\end{align}
are inverted by the same formula (\ref{onishi}). Nevertheless, there are three different cases for which one of the variables $u_1,u_2,u_3$ is considered as independent while the remaining two result from solving the divisor conditions $\sigma(\boldsymbol{u})=\sigma_3(\boldsymbol{u})=0$.

Formula (\ref{onishi}) can be rewritten in terms of $\theta$--functions as
\begin{align}
x=-\frac{\partial^2_{\boldsymbol{U},\boldsymbol{W}}
\theta[\boldsymbol{K}_{\infty}]((2\omega)^{-1}\boldsymbol{u})
+2(\partial_{\boldsymbol{U}}
\theta[\boldsymbol{K}_{\infty}]((2\omega)^{-1}\boldsymbol{u}))
\boldsymbol{e}_3^T\varkappa\boldsymbol{u}}{\partial^2_{\boldsymbol{V},\boldsymbol{W}}
\theta[\boldsymbol{K}_{\infty}]((2\omega)^{-1}\boldsymbol{u})
+2(\partial_{\boldsymbol{V}}
\theta[\boldsymbol{K}_{\infty}]((2\omega)^{-1}\boldsymbol{u}))
\boldsymbol{e}_3^T\varkappa\boldsymbol{u}}, \label{xtheta}
\end{align}
where $\boldsymbol{e}_3=(0,0,1)^T$. This represents the solution of the inversion problem.

From the solution of the Jacobi inversion problem follows for any $1\leq i<j<k \leq 7$,
\begin{align}\begin{split}
&e_i+e_j+e_k=\wp_{33}(\boldsymbol{\Omega}_{ijk}),\\
&-e_ie_j-e_ie_k-e_je_k=\wp_{23}(\boldsymbol{\Omega}_{ijk}),\\
&e_ie_je_k=\wp_{13}(\boldsymbol{\Omega}_{ijk}) \,.\end{split}
\label{JIPE31}
\end{align}
For remaining two-index symbol we find
\begin{align}\begin{split}
&\wp_{12}(\boldsymbol{\Omega}_{ijk})=-s_3S_1-S_4 \\& \wp_{11}(\boldsymbol{\Omega}_{ijk})=s_3S_2+s_1S_4 , \\ & \wp_{22}(\boldsymbol{\Omega}_{ijk})=S_3+2s_3+s_2S_1 \,,\end{split}
\label{JIPE32}
\end{align}
where $s_l$ are the elementary symmetric functions of order $l$ of the branch points $e_i,e_j,e_k$ and $S_l$ are the elementary symmetric functions of order $l$ of the remaining branch points $\{1,\ldots,7\} \setminus \{i,j,k\}$.

From (\ref{JIPE31}) and (\ref{JIPE32}) one can find the expression for the matrix $\varkappa$. To do that consider the half-period $\boldsymbol{\Omega}_{123}$
\begin{align}
\varkappa = -\frac12 \mathfrak{P}(\boldsymbol{\Omega}_{123}) -\frac12 {(2\omega)^{-1}}^T H(\boldsymbol{\Omega}_{123}) (2\omega)^{-1}
\end{align}
with
\begin{equation}
H(\boldsymbol{\Omega}_{123})=\frac{1}{\theta[\varepsilon]}\left(\begin{array}{ccc}
\theta_{11}[\varepsilon]&\theta_{12}[\varepsilon]&\theta_{13}[\varepsilon]\\
\theta_{12}[\varepsilon]&\theta_{22}[\varepsilon]&\theta_{23}[\varepsilon]\\
\theta_{13}[\varepsilon]&\theta_{23}[\varepsilon]&\theta_{33}[\varepsilon]
 \end{array}\right),\quad [\varepsilon]=\left[ \begin{array}{ccc} \frac12&0& \frac12\\  \frac12&0& \frac12  \end{array}   \right] \ .
\end{equation}

For the branch points $e_1,\ldots, e_8$ the expression
\begin{equation}
e_i=-\frac{\partial_{\boldsymbol{U}}\left[
\partial_{\boldsymbol{W}}+2\boldsymbol{\mathfrak A}^T_i\varkappa \boldsymbol{e}_3
\right]\theta[\boldsymbol{K}_{\infty}]((2\omega)^{-1} \boldsymbol{\mathfrak A}_i;\tau)}
{\partial_{\boldsymbol{V}}\left[
\partial_{\boldsymbol{W}}+2\boldsymbol{\mathfrak A}^T_i\varkappa \boldsymbol{e}_3
\right]\theta[\boldsymbol{K}_{\infty}]((2\omega)^{-1} \boldsymbol{\mathfrak A}_i;\tau)}
\label{thomae1a}
\end{equation}
is valid. Furthermore we have for $i,j=1,\ldots,8$, $i\neq j$
\begin{align}\begin{split}
e_i + e_j & = - \frac{\sigma_2(\boldsymbol{\Omega}_{ij})}
{\sigma_3(\boldsymbol{\Omega}_{ij})} \equiv
\frac{\partial_{\boldsymbol{V}} \theta[\delta_{ij}]}
{\partial_{\boldsymbol{W}} \theta[\delta_{ij}]}, \\
e_i e_j & = \frac{\sigma_1(\boldsymbol{\Omega}_{ij})}
{\sigma_3(\boldsymbol{\Omega}_{ij})} \equiv
\frac{\partial_{\boldsymbol{V}} \theta[\delta_{ij}]}
{\partial_{\boldsymbol{W}} \theta[\delta_{ij}]} \, ,
\end{split}\label{thomae2}
\end{align}
and for $i=1,\ldots,7$
\begin{equation}
e_i = -\frac{\sigma_1(\boldsymbol{\mathfrak A}_i)} {\sigma_2(\boldsymbol{\mathfrak A}_i)}
= - \frac{\partial_{\boldsymbol{U}} \theta[\delta_{i}]}
{\partial_{\boldsymbol{V}} \theta[\delta_{i}]} \, .
\end{equation}

The $\zeta$-formula reads

\begin{align}\begin{split}
&-\zeta_1(\boldsymbol{u})+2\mathfrak{n}_1+\frac12 \frac{w_1(z_1-z_2-z_3)}{(z_1-z_2)(z_1-z_3)}+\text{permutations}\\&\hskip 2cm=\int_{(e_2,0)}^{(z_1,w_1)}\mathrm{d}r_1(z,w)+
\int_{(e_4,0)}^{(z_2,w_2)}\mathrm{d}r_1(z,w)+\int_{(e_6,0)}^{(z_3,w_3)}\mathrm{d}r_1(z,w)\\
&-\zeta_2(\boldsymbol{u})+2\mathfrak{n}_2+\frac12\frac{w_1}{(z_1-z_2)(z_1-z_3)}+\text{permutations} \\&\hskip 2cm =\int_{(e_2,0)}^{(z_1,w_1)}\mathrm{d}r_2(z,w)+
\int_{(e_4,0)}^{(z_2,w_2)}\mathrm{d}r_2(z,w)+\int_{(e_6,0)}^{(z_3,w_3)}\mathrm{d}r_2(z,w)\\
&-\zeta_3(\boldsymbol{u})+2\mathfrak{n}_3=\int_{(e_2,0)}^{(z_1,w_1)}\mathrm{d}r_3(z,w)+
\int_{(e_4,0)}^{(z_2,w_2)}\mathrm{d}r_3(z,w)+\int_{(e_6,0)}^{(z_3,w_3)}\mathrm{d}r_3(z,w) \ ,
\end{split} \label{zetaformula3}
\end{align}
where $\mathfrak{n}_j=\sum_{i=1}^{3}\eta^{\prime}_{ji}{\varepsilon}'_{i} + \eta_{ji}{\varepsilon}_{i}$. Here the characteristics ${\varepsilon}'_{i}$ and ${\varepsilon}_{i}$ of $\boldsymbol{K}_{\infty}$ are not reduced.

Choosing $(z_1,w_1)=(Z,W),(z_2,w_2)=(e_4,0)$, $(z_3,w_3)=(e_6,0) $ we get from~\eqref{zetaformula3}
\begin{align}
\begin{split}
&-\zeta_1\left(\int_{(e_2,0)}^{(Z,W)} \mathrm{d}\boldsymbol{u} + \boldsymbol{K}_{\infty}\right)+2\mathfrak{n}_1+\frac12\frac{W(Z-e_4+e_6)}{(Z-e_4)(Z-e_6)}=\int_{(e_2,0)}^{(Z,W)}\mathrm{d}r_1(z,w)\\
&-\zeta_2\left(\int_{(e_2,0)}^{(Z,W)} \mathrm{d}\boldsymbol{u} + \boldsymbol{K}_{\infty}\right)+2\mathfrak{n}_2+\frac12\frac{W}{(Z-e_4)(Z-e_6)}=\int_{(e_2,0)}^{(Z,W)}\mathrm{d}r_2(z,w)\\
&-\zeta_3\left(\int_{(e_2,0)}^{(Z,W)} \mathrm{d}\boldsymbol{u} + \boldsymbol{K}_{\infty} \right)+2\mathfrak{n}_3=\int_{(e_2,0)}^{(Z,W)}\mathrm{d}r_3(z,w)
\end{split} \label{zetaformula3_2}
\end{align}

The inversion formula for the integral of the third kind is written as

\begin{align}\begin{split}
& W\int_{P'}^P\frac{1}{x-Z}\frac{\mathrm{d}x}{y}
=-2\left(\boldsymbol{u}^T-{\boldsymbol{u}'}^T\right)
\int_{(e_2,0)}^{(Z,W)} \mathrm{d}\boldsymbol{r}
+\mathrm{ln} \frac{\sigma\left(\boldsymbol{u}-\boldsymbol{v} - \boldsymbol{K}_{\infty} \right)}{\sigma\left(\boldsymbol{u}+\boldsymbol{v} - \boldsymbol{K}_{\infty}  \right)}-\mathrm{ln} \frac{\sigma\left(\boldsymbol{u}'-\boldsymbol{v} - \boldsymbol{K}_{\infty} \right)}{\sigma\left(\boldsymbol{u}'+\boldsymbol{v} - \boldsymbol{K}_{\infty}  \right)}
\end{split} \label{thirdinv3}
\end{align}
with
\[\boldsymbol{v}=\int_{(e_2,0)}^{(Z,W)}\mathrm{d}\boldsymbol{u},\quad \boldsymbol{u}=\int_{\infty}^{P}\mathrm{d}\boldsymbol{u},\quad \boldsymbol{u}'=\int_{\infty}^{P'}\mathrm{d}\boldsymbol{u} \]
and $\boldsymbol{u}\in\Theta_1$, $\boldsymbol{u}'\in\Theta_1$. The integrals $\displaystyle{\int_{(e_2,0)}^{(Z,W)} \mathrm{d}\boldsymbol{r}}$ are given by the formula~\eqref{zetaformula3_2}.
Matrix $H$ be  $5\times5$ matrix
\begin{eqnarray*}
H=\left(\begin{array}{ccccc}
\lambda_0&\frac12\lambda_1&-2\wp_{11}&-2\wp_{12}&-2\wp_{13}\\
\frac12\lambda_1&4\wp_{11}+\lambda_2&2\wp_{12}+\frac12\lambda_3&
4\wp_{13}-2\wp_{22}&-2\wp_{23}\\
-2\wp_{11}&2\wp_{12}+\frac12\lambda_3&4\wp_{22}-4\wp_{13}+\lambda_4&
2\wp_{23}+\frac12\lambda_5&-2\wp_{33}\\
-2\wp_{12}&4\wp_{13}-2\wp_{22}&2\wp_{23}+\frac12\lambda_5&4\wp_{33}+
\lambda_6&2\\
-2\wp_{13}&-2\wp_{23}&-2\wp_{33}&2&0\end{array}\right)
\end{eqnarray*}

\subsection{Complete set of expressions of $\wp_{ijk}$ and
$\wp_{ijkl}$--functions} We give list of the expressions of the
$\wp_{ijk}$-functions, that is the complete list of the first
derivatives of the $\wp_{i,j}$ over the canonical fields
$\partial_i$, as linear combinations if the basis functions.  In
this case the basis set consists of functions
$\wp_{333},\wp_{233},\wp_{133}$ and $\wp_{33},\wp_{23},\wp_{13}$.
The basic cubic relations are
\begin{align*}
\wp_{333}^2&=4\wp_{33}^3+\lambda_6\wp_{33}^2+4\wp_{23}\wp_{33}
+\lambda_5\wp_{33}+4\wp_{22}-4\wp_{13}+\lambda_4,\\
\wp_{233}^2&=4\wp_{23}^2\wp_{33}+\lambda_6\wp_{23}^2-4\wp_{22}\wp_{23}
+8\wp_{13}\wp_{23}+4\wp_{11}+\lambda_2,\\
\wp_{133}^2&=4\wp_{13}^2\wp_{33}+\lambda_6\wp_{13}^2-4\wp_{12}\wp_{13}
+\lambda_0, \\
\wp_{233}\wp_{333}&=4\wp_{33}^2\wp_{23}+\lambda_6\wp_{23}\wp_{33}
-4\wp_{22}\wp_{33},\\
&+4\wp_{13}\wp_{33}+2\wp_{23}^2-\lambda_5\wp_{23}+2\wp_{12}+\lambda_3,\\
\wp_{133}\wp_{233}&=4\wp_{13}\wp_{23}\wp_{33}+\lambda_6\wp_{13}\wp_{23}
-2\wp_{12}\wp_{23}-2\wp_{13}\wp_{22}+4\wp_{13}^2+\frac12\lambda_1,\\
\wp_{133}\wp_{333}&=4\wp_{13}\wp_{33}^2+\lambda_6\wp_{13}\wp_{33}
-2\wp_{12}\wp_{33}+2\wp_{13}\wp_{23}+\frac12\lambda_5\wp_{13}-2\wp_{11}.
\end{align*}

Remaining cubic relations can be derived with the aids of the
above formulae and relations

\begin{align*}
\wp_{223}&=-\wp_{333}\wp_{23}+\wp_{233}\wp_{33}+\wp_{133},\\
\wp_{123}&=-\wp_{333}\wp_{13}+\wp_{133}\wp_{33},\\
\wp_{113}&=-\wp_{233}\wp_{13}+\wp_{133}\wp_{23},\\
\wp_{222}&=
\wp_{333}\big(2\wp_{23}(\wp_{33}+\frac{\lambda_{6}}{4})+4\wp_{13}
  -\wp_{22}\big)-
\wp_{233}\big(2\wp_{33}(\wp_{33}+\frac{\lambda_{6}}{4})+\wp_{23}
  +\frac{\lambda_{5}}{4}\big)
\\&-2\wp_{133}\wp_{33},\\
\wp_{122} &=
\wp_{333}\big(2\wp_{13}(\wp_{33}+\frac{\lambda_{6}}{4})-\wp_{12}\big)+
\wp_{233}\wp_{13}-
\wp_{133}\big(2\wp_{33}(\wp_{33}+\frac{\lambda_{6}}{4})+\wp_{23}
  +\frac{\lambda_{5}}{4}\big),
\\
\wp_{112} &=
\wp_{233}\big(2\wp_{13}(\wp_{33}+\frac{\lambda_{6}}{4})-\wp_{12}\big)-
\wp_{133}\big(2\wp_{23}(\wp_{33}+\frac{\lambda_{6}}{4})+2\wp_{13}
  -\wp_{22}\big),
\\
\wp_{111}&=
\wp_{333}\big(\wp_{13}\wp_{22}-2\wp_{13}^2-\wp_{12}\wp_{23}\big)+
\wp_{233}\big(2\wp_{13}(\wp_{23}+\frac{\lambda_{5}}{4})
-\wp_{12}\wp_{33}\big),
\\ &-
\wp_{133}\big(2\wp_{23}(\wp_{23}+\frac{\lambda_{5}}{4})
-\wp_{33}(2\wp_{13}
  -\wp_{22})\big)
\end{align*}
These expressions are derived by successive differentiation from
the main relations \eqref{wpgggi}, \eqref{wp3},
\eqref{i-1,k-1} and \eqref{product3}.

Along with the above expressions for the first derivatives of the
$\wp_{i,j}$ we obtain an analogous list for the second
derivatives, but here we give the expressions by the
$\wp_{i,j}$--functions themselves and the constants $\lambda_k$:
\begin{align*}
\wp_{3333}&= \lambda_5/2+4\wp_{23}+\lambda_6\wp_{33}+6\wp_{33}^2,
\\
\wp_{2333}&=6\wp_{13}-2\wp_{22}+\lambda_6\wp_{23}+6\wp_{23}\wp_{33}, \\
\wp_{1333}&=-2\wp_{12}+\lambda_6\wp_{13}+6\wp_{13}\wp_{33}, \\
\wp_{2233}&=-2\wp_{12}+\lambda_6\wp_{13}+\lambda_5\wp_{23}/2
+4\wp_{23}^2+
2\wp_{22}\wp_{33} ,\\
\wp_{1233}&=\lambda_5\wp_{1 3}/2+4\wp_{13}\wp_{23}+2\wp_{12}\wp_{33}, \\
\wp_{1133}&=6\wp_{13}^2-2\wp_{13}\wp_{22}+2\wp_{12}\wp_{23}, \\
\wp_{2223}&=-\lambda_2-6\wp_{11}+\lambda_5\wp_{13}+\lambda_4\wp_{23}
+6\wp_{22}\wp_{23}-
\lambda_3\wp_{33}/2 ,\\
\wp_{1223}&=-\lambda_1/2+\lambda_4\wp_{13}-2\wp_{13}^2+4\wp_{13}\wp_{22}+
2\wp_{12}\wp_{23}+2\wp_{11}\wp_{33}, \\
\wp_{1123}&=-\lambda_0+\lambda_3\wp_{13}/2+4\wp_{12}\wp_{13}
+2\wp_{11}\wp_{23} ,\\
\wp_{1113}&=\lambda_2\wp_{13}+6\wp_{11}\wp_{13}-\lambda_1\wp_{23}/2
+\lambda_0\wp_{33} ,\\
\wp_{2222}&=-3\lambda_1/2+\lambda_3\lambda_5/8-\lambda_2\lambda_6/2
-3\lambda_6\wp_{11}+
 \lambda_5\wp_{12}+12\wp_{13}^2+ \\
&\lambda_4\wp_{22}-12\wp_{13}\wp_{22}+
 6\wp_{22}^2+\lambda_3\wp_{23}+12\wp_{12}\wp_{23}-3\lambda_2\wp_{33}
-12\wp_{11}\wp_{33}, \\
\wp_{1222}&=-2\lambda_0-\lambda_1\lambda_6/4-\lambda_5\wp_{11}/2
+\lambda_4\wp_{12}+
 \lambda_3\wp_{13}+6\wp_{12}\wp_{22}-3\lambda_1\wp_{33}/2, \\
\wp_{1122}&=-\lambda_0\lambda_6/2+\lambda_3\wp_{12}/2+4\wp_{12}^2
+\lambda_2\wp_{13}+
  2\wp_{11}\wp_{22}-\lambda_1\wp_{23}/2-2\lambda_0\wp_{33} ,\\
\wp_{1112}&=-\lambda_0\lambda_5/4+\lambda_2\wp_{12}+6\wp_{11}\wp_{12}
+3\lambda_1\wp_{13}/2-
  \lambda_1\wp_{22}/2-2\lambda_0\wp_{23},\\
\wp_{1111}&=-\frac12\lambda_0\lambda_4+\frac18\lambda_1\lambda_3
-3\lambda_0\wp_{22}+\lambda_1\wp_{12}+\lambda_2\wp_{11}+
4\lambda_0\wp_{13} +6\wp_{11}^2.
\end{align*}
It is clear, that the lists of this type may be derived for
arbitrary genus.

\subsection{Second and third terms of the
expansion of $\sigma$-function in vicinity $\boldsymbol{u}=0$}
\index{$\sigma$--function!second term of expansion} The expansion
of the $\sigma$-function of genus $3$ in the vicinity of
$\boldsymbol{u}=0$ may be derived  out of the complete set of
$\wp_{ijkl}$-functions. Indeed, the $\sigma$-function is the even
function  and its expansion is given according to Proposition
\ref{expansion_1} by formula:
\begin{equation*}
\sigma(u_1,u_2,u_3)=u_1u_3-u_2^2+
\frac{1}{24}\sum_{i,j,k,l=1}^3s_{ijkl}u_{i}u_{j}u_{k}u_{l}+\dots
\end{equation*}
with coefficients $s_{ijkl}\in \mathbb{C}$ being completely
symmetric over their indices:
\begin{equation*}
s_{ijkl}=\frac{\partial^4}{\partial u_{i}\partial u_{j}\partial
u_{k}\partial u_{l}}\sigma(u_1,u_2,u_3)\bigg|_{{\boldsymbol u}=
{\boldsymbol 0}}.
\end{equation*}

Straightforwardly, substituting this expansion to the equations
\begin{equation*}
\sigma^4(\wp_{ijkl}-\{\dots\})=0,
\end{equation*}
where $\{\dots\}$ stands for the proper expression from the above
list, we determine all the coefficients $s_{ijkl}$ and obtain:

\begin{equation}
\sigma(u_1,u_2,u_3)=u_1u_3-u_2^2+\sigma_4(u_1,u_2,u_3)
+\sigma_6(u_1,u_2,u_3)+\text{higher order terms},
\label{sigmaexpans} \end{equation}
where
\begin{align*}
24\sigma_4(u_1,u_2,u_3)
= -& 2 \lambda_{0} u_{1}^4 +2 \lambda_{1}
u_{1}^3 u_{2} +\lambda_{2} u_{1}^2 (3 u_{2}^2-u_{1}
u_{3})\nonumber\\ &+2 \lambda_{3} u_{1} u_{2}^3 +2 \lambda_{4}
u_{2}^4 +2 \lambda_{5} u_{2}^3  u_{3} +\lambda_{6} u_{3}^2 (3
u_{2}^2-u_{1} u_{3})\nonumber\\
&+8 u_{2} u_{3}^3
\end{align*}

\begin{align*}
5760\sigma_6(u_1,u_2,u_3)&=
128u_3^6 + 384{u_1}u_2^5{\lambda_0} - 960u_1^2u_2^3{u_3}
{\lambda_0} - 960u_1^3{u_2}u_3^2{\lambda_0} + 72u_2^6{\lambda_1}
\\&- 360{u_1}u_2^4{u_3}{\lambda_1} - 720u_1^2u_2^2u_3^2
     {\lambda_1} - 80u_1^3u_3^3{\lambda_1} + 8u_1^6\lambda_1^2 -
96u_2^5{u_3}{\lambda_2}
\\&- 480{u_1}u_2^3u_3^2{\lambda_2} -
240u_1^2{u_2}u_3^3{\lambda_2} - 44u_1^6{\lambda_0}{\lambda_2} -
12u_1^5{u_2}{\lambda_1}{\lambda_2}
\\&- 15u_1^4u_2^2\lambda_2^2 +
3u_1^5{u_3}\lambda_2^2 - 240u_2^4u_3^2 {\lambda_3} -
240{u_1}u_2^2u_3^3{\lambda_3} - 72u_1^5{u_2}{\lambda_0}
{\lambda_3}
\\   &- 30u_1^4u_2^2{\lambda_1}{\lambda_3} + 6u_1^5{u_3}{\lambda_1}
{\lambda_3} - 20u_1^3u_2^3{\lambda_2}{\lambda_3} -
     320u_2^3u_3^3{\lambda_4}
\\&- 120u_1^4u_2^2{\lambda_0}{\lambda_4} -
24u_1^5{u_3}{\lambda_0}{\lambda_4} - 80u_1^3u_2^3{\lambda_1}
{\lambda_4} - 60u_1^2u_2^4{\lambda_2}{\lambda_4}
\\&-
24{u_1}u_2^5{\lambda_3}{\lambda_4} - 16u_2^6\lambda_4^2 -
120u_2^2u_3^4{\lambda_5} + 24{u_1}u_3^5{\lambda_5} - 80u_1^3u_2^3{\lambda_0}
     {\lambda_5}
\\&- 120u_1^4{u_2}{u_3}{\lambda_0}{\lambda_5} -
60u_1^2u_2^4{\lambda_1} {\lambda_5} -
60u_1^3u_2^2{u_3}{\lambda_1}{\lambda_5} - 24{u_1}u_2^5{\lambda_2}
     {\lambda_5}
\\&- 60u_1^2u_2^3{u_3}{\lambda_2}{\lambda_5} -
     2u_2^6{\lambda_3}{\lambda_5} -
    30{u_1}u_2^4{u_3}{\lambda_3}{\lambda_5} - 24u_2^5{u_3}{\lambda_4}
     {\lambda_5}
\\&- 48{u_2}u_3^5{\lambda_6} -
    240u_1^3u_2^2{u_3}{\lambda_0}{\lambda_6} -
    60u_1^4u_3^2{\lambda_0}{\lambda_6} - 24{u_1}u_2^5{\lambda_1}
     {\lambda_6}
\\&- 120u_1^2u_2^3{u_3}{\lambda_1}{\lambda_6} -
    60u_1^3{u_2}u_3^2{\lambda_1} {\lambda_6} -
   12u_2^6{\lambda_2}{\lambda_6} - 60{u_1}u_2^4{u_3}{\lambda_2}
     {\lambda_6}
\\&- 90u_1^2u_2^2u_3^2{\lambda_2}{\lambda_6} +
     10u_1^3u_3^3{\lambda_2} {\lambda_6} -
     24u_2^5{u_3}{\lambda_3}{\lambda_6}
\\&- 60{u_1}u_2^3u_3^2{\lambda_3}{\lambda_6} - 60u_2^4u_3^2{\lambda_4}
     {\lambda_6} - 20u_2^3u_3^3{\lambda_5} {\lambda_6}
\\&- 15u_2^2u_3^4\lambda_6^2 + 3{u_1}u_3^5\lambda_6^2
\end{align*}

In general, the coefficients in the expansions of the fundamental
$\sigma$-functions depend only on the symmetric functions of the
branching points $e_i$ of the curve, i.e on the constants
$\lambda_j$.

\section{Restriction to strata of the $\theta$-divisor}
Suppose that the $\theta$-divisor is stratified as
\[
\Theta_0=\boldsymbol{K}_{\infty}\subset  \Theta_1=\boldsymbol{K}_{\infty}+\int_{
\infty}^P \mathrm{d}\boldsymbol{u} \subset \Theta_2=\boldsymbol{K}_{\infty}+\int_{
\infty}^{P_1} \mathrm{d}\boldsymbol{u}+\int_{\
\infty}^{P_2} \mathrm{d}\boldsymbol{u}\subset\mathrm{Jac}(V) \]
Let $\boldsymbol{u}\in \Theta_1$. Then
\begin{align*}
&\int_{a}^{x_1}\mathrm{d}r_3=-\frac{\sigma_{23} (\boldsymbol{u} )}
{\sigma_2(\boldsymbol{u}) }+c_3,\quad
\int_{a}^{x_1}\mathrm{d}r_2=-\frac12\frac{\sigma_{22}(\boldsymbol{u}
)}{\sigma_2(\boldsymbol{u} )}+c_2,\\
&\int_{a}^{x_1}\mathrm{d}r_1=\frac12
\frac{\sigma_{1}(\boldsymbol{u}) \sigma_{22}(\boldsymbol{u}) }
{\sigma_2(\boldsymbol{u})^2 }-
\frac{\sigma_{12}(\boldsymbol{u}) }{\sigma_2(\boldsymbol{u} )}+c_1,
\end{align*}
where the constants $c_i$ are fixed by requiring the
right hand side to vanish at $x_1=a$

Let $\boldsymbol{u}\in \Theta_2$. Then
\begin{align*}
\int_{a}^{x_1}\mathrm{d}r_3+\int_{a}^{x_2}\mathrm{d}r_3&=
-\frac12\frac{\sigma_{33}(\boldsymbol{u}) }{\sigma_3 (\boldsymbol{u}) }+C_1,\\
\int_{a}^{x_1}\mathrm{d}r_2+\int_{a}^{x_2}\mathrm{d}r_2&=
-\frac{\sigma_{23}(\boldsymbol{u} )}{\sigma_3}+\frac12 \frac{\sigma_{2}(\boldsymbol{u})\sigma_{33}(\boldsymbol{u})}
{\sigma_3(\boldsymbol{u})^2 }+C_2,\\
\int_{a}^{x_1}\mathrm{d}r_1+\int_{a}^{x_2}\mathrm{d}r_1&=
-\frac12 \frac{\sigma_{22}(\boldsymbol{u})}{\sigma_3(\boldsymbol{u})}
+\frac{\sigma_2(\boldsymbol{u})\sigma_{23}(\boldsymbol{u})}{\sigma_3(\boldsymbol{u})^2}
-\frac12 \frac{\sigma_{33}(\boldsymbol{u})\sigma_2(\boldsymbol{u})^2}{\sigma_3(\boldsymbol{u})^3}\\
&-\frac{\sigma_{13}(\boldsymbol{u})}{\sigma_3(\boldsymbol{u})}
+\frac12 \frac{\sigma_{33}(\boldsymbol{u})\sigma_1(\boldsymbol{u})}{\sigma_3(\boldsymbol{u})^2}
+C_3,\notag
\end{align*}
where the constants $C_i$ are chosen so that the right hand side vanishes at $x_1=x_2=a$.

\subsection{Relations on $\Theta_2$}
The complete set of the $\sigma$-relations on the second
stratum of the $\theta$-divisor
\begin{align*}
\sigma_{333}&=\sigma_2+\frac34\frac{\sigma_{33}^2}{\sigma_3}
+\frac14\lambda_6\sigma_3,\\
\sigma_{233}&=-\frac{\sigma_2^2}{\sigma_3}
-\frac14\frac{\sigma_2\sigma_{33}^2}{\sigma_3^2}
+\frac14\lambda_6\sigma_2+\frac{\sigma_{33}\sigma_{23}}{\sigma_3}
+2\sigma_1,\\
\sigma_{133}&=-\frac{\sigma_1\sigma_2}{\sigma_3}
-\frac14\frac{\sigma_1\sigma_{33}^2}{\sigma_3^2}+\frac14
\lambda_6\sigma_1
+\frac{\sigma_{33}\sigma_{13}}{\sigma_3},\\
\sigma_{223}&=-\frac{\sigma_2\sigma_{33}\sigma_{23}}{\sigma_{3}^2}+
\frac{\sigma_2^3}{\sigma_3^2}
+\frac14\frac{\sigma_2^2\sigma_{33}^2}{\sigma_3^3}
+\frac12\lambda_6\sigma_1-3\frac{\sigma_2\sigma_1}{\sigma_3}-
\frac14\lambda_6  \frac{\sigma_2^2}{\sigma_3}\\
&+\frac12\frac{\sigma_{33}\sigma_{22}}{\sigma_3}
+\frac{\sigma_{23}^2}{\sigma_3}
+\frac14\lambda_5\sigma_2,\\
\sigma_{123}&=\frac{\sigma_{23}\sigma_{13}}{\sigma_3}+
\frac{\sigma_2^2\sigma_1}{\sigma_3^2}
+\frac14\frac{\sigma_2\sigma_1\sigma_{33}^2}{\sigma_3^3}
-\frac14\lambda_6\frac{\sigma_2\sigma_1}{\sigma_3} \\
&-\frac12\frac{\sigma_2\sigma_{33}\sigma_{13}}{\sigma_3^2}
-\frac12\frac{\sigma_1\sigma_{33}\sigma_{23}}{\sigma_3^2}
-\frac{\sigma_1^2}{\sigma_3}+\frac12
\frac{\sigma_{33}\sigma_{12}}{\sigma_3}
+\frac14\lambda_5\sigma_1,\\
\sigma_{122}&=
\frac14\frac{\sigma_2^2\sigma_1\sigma_{33}}{\sigma_3^4}
+\frac{\sigma_2^3\sigma_1}{\sigma_3^3}
-\frac{\sigma_1\sigma_2\sigma_{23}\sigma_{33}}{\sigma_3^2}
-3\frac{\sigma_2\sigma_{1}^2}{\sigma_3^2}
+\frac12\frac{\sigma_1\sigma_{22}\sigma_{33}}{\sigma_3}\\
\end{align*}

\begin{align*}
&+\frac14\frac{\sigma_1\sigma_{23}^2}{\sigma_3^2}
-\frac{\sigma_{23}\sigma_1\sigma_{22}}{\sigma_2\sigma_3}
+2\frac{\sigma_1^3}{\sigma_2\sigma_3}
+\frac{\sigma_{22}\sigma_{12}}{\sigma_2}+\frac{\lambda_6}{2}\left(\frac{\sigma_1^2}{\sigma_3}
-\frac12\frac{\sigma_2^2\sigma_1}{\sigma_3^4}\right)\\
&+\frac{\lambda_5}{2}\left(\frac{\sigma_1\sigma_2}{\sigma_3}
-\frac{\sigma_1^2}{\sigma_2}\right)
+\frac{\lambda_3}{2}\frac{\sigma_1\sigma_3}{\sigma_2}
-\frac{\lambda_1}{2}\frac{\sigma_3^2}{\sigma_2},\\
\sigma_{222}&= -\frac{3\sigma_2\sigma_{11}\sigma_{22}}{2\sigma_1^2}
+\lambda_3 \sigma_3 - 3 \lambda_0 \frac{\sigma_3 \sigma_2^3}{\sigma_1^3}
-\lambda_2\frac{3\sigma_2 \sigma_3}{2\sigma_1}+\lambda_2 \frac{3\sigma_2^3}{4\sigma_1^2}
+3\frac{\sigma_2^2 \sigma_{11}\sigma_{12}}{\sigma_1^3}\\
&
+ \lambda_1\frac{9\sigma_3 \sigma_2^2}{4\sigma_1^2}
-\lambda_1 \frac{3\sigma_2^4}{4\sigma_1^3}+ \lambda_0 \frac{3\sigma_2^5}{4\sigma_1^4}
+3\sigma_2\lambda_0 \frac{\sigma_3^2}{\sigma_1^2}
-3\frac{\sigma_2\sigma_{12}^2}{\sigma_1^2}- \frac{3\sigma_2^3\sigma_{11}^2}{4\sigma_1^4}
-\lambda_3 \frac{3\sigma_2^2}{4\sigma_1}\\
& -\frac{3\sigma_3^2}{2\sigma_1}+\lambda_4\sigma_2
-\frac12 \lambda_5 \sigma_1+3\frac{\sigma_{12}\sigma_{22}}{\sigma_1},\\
\sigma_{113}&=
\frac{\sigma_1\sigma_{23}}{\sigma_3^2}-
\frac{\sigma_2\sigma_1^2
}  {\sigma_3^2}-\frac12\frac{\sigma_{22}\sigma_{12}}{\sigma_2}
-\frac32\frac{\sigma_{23}\sigma_1\sigma_{22}}{\sigma_2\sigma_3}
-\frac{\sigma_1^3}{\sigma_2\sigma_3}-\frac{\sigma_3\sigma_{12^2}}{\sigma_2}+
2\frac{\sigma_1^4}{\sigma_2^3}\\
&-\frac{\sigma_1\sigma_{33}\sigma_{12}}{\sigma-1\sigma_3}
+\frac{\sigma_{22}\sigma_{33}\sigma_162}{\sigma_2^2\sigma_3}
-\frac{\sigma_{22}\sigma_{13}\sigma_1}{\sigma_2^2}+
2\frac{\sigma_{12}\sigma_{13}}{\sigma_2}+
\frac12\frac{\sigma_{22}\sigma_{13}}{\sigma_3}-
\frac{\sigma_1\sigma_{33}\sigma_{23}}{\sigma_2\sigma_3}\\
&+\frac{\sigma_1^2\sigma_{23}^2}{\sigma_2^2\sigma_3}
-\frac{\sigma_2\sigma_{23}\sigma_{13}}{\sigma_2^2}
+\frac{\sigma_{23}\sigma_{12}}{\sigma_3}-\frac{\sigma_{23}\sigma_1^2\sigma_{22}}{\sigma_2^3}
+\frac{\sigma_3\sigma_1\sigma_{22}\sigma_{12}}{\sigma_2^3}
+\frac{\sigma_1\sigma_{33}\sigma_{13}}{\sigma_3^2}\\
&+\lambda_6\left(\frac14\frac{\sigma_1^2}{\sigma_3}+\frac12
\frac{\sigma_1^3}{\sigma_2^2}  \right)
-\lambda_5\left(\frac14\frac{\sigma_1^2}{\sigma_3}
+\frac12\frac{\sigma_3\sigma_1^3}{\sigma_2^3}
\right)+\lambda_3\left(\frac14\frac{\sigma_1\sigma_3}{\sigma_2}
+\frac12\frac{\sigma_3^2\sigma_1^3}{\sigma_2^3} \right)\\
&-\lambda_2\frac12\frac{\sigma_3^2\sigma_1}{\sigma_2^3}
+\lambda_1\left(\frac14\frac{\sigma_3^2}{\sigma_2}
-\frac12\frac{\sigma_3^3\sigma_1}{\sigma_2^3}\right)
+\lambda_0\frac{\sigma_3}{\sigma_2^2},\\
\sigma_{112}&=\frac14\frac{\sigma_2\sigma_1^2\sigma_{33}^2}{\sigma_3^4}
+3\frac{\sigma_1^3}{\sigma_3^2}
+\frac{\sigma_{23}\sigma_1^2\sigma_{22}}{\sigma_3\sigma_2^2}
-\frac{\sigma_1\sigma_{22}\sigma_{12}}{\sigma_2^2}
-\frac{\sigma_2^2\sigma_1^2}{\sigma_3^3}
-2\frac{\sigma_1\sigma_{23}\sigma_{13}}{\sigma_3^3}\\
&+2\frac{\sigma_1^2\sigma_{33}\sigma_{23}}{\sigma_3^3}
-\frac12\frac{\sigma_2\sigma_{22}\sigma_{13}}{\sigma_3^2}
+\frac{\sigma_{23}\sigma_{11}}{\sigma_1}
-\frac{\sigma_{22}\sigma_{33}\sigma_1^2}{\sigma_3^2\sigma_2}
+\frac{\sigma_{22}\sigma_{13}\sigma_1}{\sigma_3\sigma_2}\\
&-\frac12\frac{\sigma_{22}^2\sigma_1}{\sigma_3\sigma_2}+
\frac12\frac{\sigma_{22}\sigma_{12}}{\sigma_3}
-\frac{\sigma_1^2\sigma_{23}^2}{\sigma_3^2\sigma_2}
-\frac{\sigma_2\sigma_1\sigma_{23}}{\sigma_3^3}
+\frac{\sigma_{23}\sigma_{13}\sigma_2^2}{\sigma_3^2}\\
&+\frac32\frac{\sigma_{23}\sigma_1\sigma_{22}}{\sigma_3^2}
-\frac{\sigma_{23}\sigma_{12}\sigma_2}{\sigma_3^2}+\frac{\sigma_{12}^2}{\sigma_2}
-2\frac{\sigma_1^4}{\sigma_3\sigma_2^2}+\lambda_6\left(
\frac14\frac{\sigma_2\sigma_1^2}{\sigma_3^2}
-\frac12\frac{\sigma_1^3}{\sigma_2\sigma_3}
\right) \\&+\lambda_5\left(-\frac14\frac{\sigma_1^2}{\sigma_3}
+\frac12\frac{\sigma_1^3}{\sigma_2^2}\right)+
\lambda_3\left(-\frac12\frac{\sigma_1^2\sigma_3}{\sigma_2^2}+\frac14\sigma_1
\right) +\lambda_2\frac{\sigma_1\sigma_3}{\sigma_2}
+\lambda_1\left(\frac12
\frac{\sigma_3^2\sigma_1}{\sigma_2^2}-\frac14\sigma_3\right),\\
 \sigma_{111}&=3\frac{\sigma_{23}\sigma_{1}^3\sigma_{22}
}{\sigma_{3}\sigma_{2}^3  }+3\frac{\sigma_{2} \sigma_{1}^3 }{\sigma_{3}^3}
-6\frac{\sigma_{1}^5 }{\sigma_{3}\sigma_{2}^3   }
-3\frac{\sigma_{2}^2\sigma_{13}^2  }{\sigma_{3}^3 }+3\frac{\sigma_{1}^4
}{\sigma_{3}^2\sigma_{2}  } +3\frac{\sigma_{2} \sigma_{13} \sigma_{12}
}{\sigma_{3}^2 } \\
&-3\frac{\sigma_{1}^3\sigma_{23}^2
}{\sigma_{2}^2\sigma_{2}^2  } -3\frac{\sigma_{1}^2\sigma_{23}^2  }{\sigma_{3} }
-\frac32\frac{\sigma_{22}^2\sigma_1^2}{\sigma_3\sigma_2^2}
+6\frac{\sigma_{13}^2\sigma_1}{\sigma_3^22}
-3\frac{\sigma_1^2\sigma_{22}^2\sigma_{12}}{\sigma_{2}^3}+
3\frac{\sigma_1\sigma_{12}^2}{\sigma_2^2}\\
&-3\frac{\sigma_1\sigma_{23}\sigma_{12}}{\sigma_3^2}-
6\frac{\sigma_1^2\sigma_{33}\sigma_{13}}{\sigma_3^3}+
\frac92\frac{\sigma_{23}\sigma_1^2\sigma_{22}}{\sigma_3^2\sigma_2}+
3\frac{\sigma_1^2\sigma_{33}\sigma_{12}}{\sigma_3^2\sigma_2}-
3\frac{\sigma_{22}\sigma_{33}\sigma_1^3}{\sigma_3^3\sigma_2}\\
&-6\frac{\sigma_1\sigma_{12}\sigma_{13}}{\sigma_2\sigma_3}
+6\frac{\sigma_2\sigma_1\sigma_{23}\sigma_{13}}{\sigma_3^3}
+3\frac{\sigma_{22}\sigma_{13}\sigma_1^2}{\sigma_3\sigma_2^2}
-\frac92\frac{\sigma_1\sigma_{22}\sigma_{13}}{\sigma_3^2}+
\frac32\frac{\sigma_1\sigma_{22}\sigma_{12}}{\sigma_3\sigma_2}\\
&+\frac34\frac{\sigma_1^3\sigma_{33}^2}{\sigma_3^4}-\frac32
\lambda_6\left(\frac12\frac{\sigma_1^3}{\sigma_3}
+\frac{\sigma_1^4}{\sigma_3\sigma_2^2}\right)
+\frac32\lambda_5\left(\frac{\sigma_1^4}{\sigma_2^3}
+\frac12\frac{\sigma_1^3}{\sigma_3\sigma_2}\right)\\
&-\frac32\lambda_3\left(\frac{\sigma_1^3\sigma_3}{\sigma_2}
+\frac12\frac{\sigma_1^2}{\sigma_2}\right)+\lambda_2\left(\sigma_1
+\frac32\frac{\sigma_2\sigma_1^2}{\sigma_2^2}\right)\\
&+\frac12\lambda_1\left(-\sigma_2-\frac32\frac{\sigma_3\sigma_1}{\sigma_2}
+3\frac{\sigma_3^2\sigma_1^2}{\sigma_2^3}\right)+\lambda_0\left(\sigma_3
-3\frac{\sigma_3^2\sigma_1}{\sigma_2^2}\right)
\end{align*}
and also relations
\begin{eqnarray*}
-\sigma_{11}\sigma_3^2-
\sigma_1^2\sigma_{33}+\sigma_2(\sigma_2\sigma_{13}
-\sigma_{23}\sigma_2)+(\sigma_2\sigma_{12}-\sigma_{22}\sigma_1
+2\sigma_1\sigma_{12})\sigma_3=0.
\end{eqnarray*}

\subsection{Relations on $\Theta_1$}
The complete set of the $\sigma$-relations on the first
stratum of the $\theta$-divisor \cite{on998}
\begin{align*}
\sigma_{333}&=-2\sigma_2,\quad \sigma_{233}=\frac{\sigma_{23}^2}{\sigma_2}
-\sigma_1,\\
\sigma_{133}&=\frac{\sigma_{1}\sigma_{23}^2}{\sigma_2^2}
+\frac{\sigma_1^2}{\sigma_2},\quad \sigma_{223}=
\frac{\sigma_{23}\sigma_{22}}{\sigma_2}
-\frac{2\sigma_1}{\sigma_2},\\
\sigma_{113}&=\frac{\sigma_{23}\sigma_{11}}{\sigma_2},\quad
\sigma_{123}=\frac{\sigma_{23}\sigma_{12}}{\sigma_2},\\
\sigma_{111}&=\frac{\lambda_1}{4}\sigma_2
-\frac34 \frac{\lambda_0 \sigma_2^2}{\sigma_1}
+\frac34 \frac{\sigma_{11}^2}{\sigma_1}+\frac{\lambda_2}{4}\sigma_1,\\
\sigma_{112}&=\frac{\lambda_2}{4}\sigma_2
+\frac{\sigma_{12}\sigma_{11}}{\sigma_1}
-\frac{\lambda_1 \sigma_2^2}{4 \sigma_1}
+\frac{\lambda_0 \sigma_2^3}{4 \sigma_1^2}
-\frac{\sigma_2 \sigma_{11}^2}{4\sigma_1^2},\\
\sigma_{222}&=-\frac{3\sigma_2 \sigma_{11}\sigma_{22}}{2 \sigma_1^2}-
\frac{3\lambda_3\sigma_2^2}{4 \sigma_1}
-\frac{3\sigma_2 \sigma_{12}^2}
{\sigma_1^2}
-\frac{3\lambda_1\sigma_2^4}{4 \sigma_1^3}
-\frac{3\sigma_2^3 \sigma_{11}^2}{4 \sigma_1^4}
+\frac{3\lambda_2\sigma_2^3}{4 \sigma_1^2}
\notag\\&+\frac{3\sigma_2^2 \sigma_{11}\sigma_{12}}{\sigma_1^3}
+\frac{3\lambda_0\sigma_2^5}{4 \sigma_1^4}
+\lambda_4 \sigma_2-\lambda_5\sigma_1
+\frac{3 \sigma_{12}\sigma_{22}}{\sigma_1},\\
\sigma_{122}&=\frac{\lambda_3}{4}\sigma_2
+\frac{\sigma_{12}^2}{\sigma_1}
+\frac{\lambda_1\sigma_2^3}{4 \sigma_1^2}
+\frac{\sigma_{11}\sigma_{22}}{2 \sigma_1}
-\frac{\lambda_2 \sigma_2^2}{4 \sigma_1}
-\frac{\sigma_2 \sigma_{11}\sigma_{12}}{\sigma_1^2}
-\frac{\lambda_0\sigma_2^4}{4 \sigma_1^3}
+\frac{\sigma_{11}^2\sigma_2^2}{4 \sigma_1^3}.
\end{align*}

The relation:
\[ \sigma_{1}\sigma_{23}=\sigma_{2}\sigma_{13}.   \]

\chapter{Recent References }
\begin{enumerate}
\item  S.~Abenda and Yu.~Fedorov,\emph{On the Weak Kowalevski-Painlev\'e Propert for Hyperelliptically Separable Systems}, Acta Appl. Math. \textbf{60} (2000), 137-178

\item Chris Athorne. \emph{Identities for hyperelliptic $\wp$-functions of genus one, two and three in covariant form}. J.Phys.A:Math.Gen, {\bf 41} (2008) 20pp, DOI: 10.1088/1751-8113/41/41/415202

\item Chris Athorne. \emph{A generalization of Baker's quadratic formulae for hyperelliptic Weierstrass $\wp$-functions}. Phys.Lett A, {\bf 375} (2011) 2689-2693

\item Ch. Athorne, and J.C. Eilbeck, and V.Z.
Enolskii,  \emph{Identities for classical $\wp$ functions},
J. Geom. and Phys., {\bf 48}, (2003) 354-368.

\item Ch. Athorne and J.C.~ Eilbeck and V.Z.~Enolskii.
\emph{An {SL}(2) covariant theory of genus 2 hyper-elliptic functions},
Math. Proc. Camb. Phil. Soc., {\bf 136}, no.2 (2004) 269-286.
\label{aee04}

\item S.~Baldwin and J.~Gibbons,
  \emph{Hyperelliptic reduction of the {B}enney moment
  equations} J.Phys.A:Math.Gen, {\bf 36}31, (2003) 8393-8413.

\item S.~Baldwin and J.~Gibbons, \emph{Higher genus hyperelliptic reductions of the {B}enney
               equations}, J.Phys.A:Math.Gen., {\bf 37}:20 (2004)  5341-5354.

\item S.~Baldwin and J.~ C.~Eilbeck and J.~Gibbons and Y.~{\^O}nishi. \emph{Abelian functions for cyclic trigonal curve of genus four.} J.Geom.Phys., {\bf 58} (2008) 450-467.

\item E.~D.~Belokolos and  V.~Z.~Enolskii and M.~Salerno.
\emph{ Wannier functions of elliptic one-gap potentials}, J.Phys.:Math.Gen.
{\bf 37} (2004) 9685-9704.

\item E.D.~Belokolos and V.Z.~Enolskii
\emph{Reduction of {A}belian {F}unctions and
              {A}lgebraically {I}ntegrable {S}ystems,}
Journal of Mathematical Sciences, Part I: 106:6 (2001) 3395-3486;
Part II: 108:3 (2002) 295-374.

\item E.~D.~Belokolos and  V.~Z.~Enolskii and M.~Salerno
\emph{ Wannier functions for quasiperiodic finite-gap
potentials.}  Teor. Mat. Fiz. {\bf 144}: 2 (2005) 1081-1099.

\item H.~Braden and V.~Enolskii and A.~E.~Hone.
\emph{Bilinear recurrences and addition
formulae for hyperelliptic sigma functions}, J.Nonlin.Math.Phys, {\bf 12}, Supplement 2 (2005) 46-62.

\item
V.~M.~Buchstaber and E.~Yu.~Bunkova
\emph{The heat equation and Families of
Two-Dimensional Sigma Functions.}, Proceedings of the Steklov
Institute of Mathematics, \textbf{266} (2009), 1--28.

\item
V.M.~Buchstaber  and J.~C.~Eilbeck and V.Z.~Enolskii and D.V.~Leykin and
M.~Salerno,
\emph{$2D$ and $3D$ Schr{\"o}dinger equation with
Abelian potential}
J.  Math.  Phys., {\bf 43} (2002) 2858-2881.

\item
V.~M.~Buchstaber, \, \emph{Heat equations and sigma functions.},
accepted to American Institute of Physics, 2009.

\item
Buchstaber V.M., Enolskii V.Z. and Leykin D.V.
\emph{Trigonal Abelian functions and completely
integrable equations} Funct. Anal. Appl. {\bf 34}:3 (2000) 1-15.

\item
V.M.~Buchstaber and V.Z.~Enolskii and D.~V.~Leykin \emph{
 Rational analogues of Abelian functions}
Funk. Anal. Appl., {\bf 33}:2 (1999) 1-15.

\item
Buchstaber V.M., Enolskii V.Z. and Leykin D.V.
\emph{$\sigma$-functions of $(n,s)$-curves}
Uspekhi Matem. Nauk, {\bf 54}:3 (1999) 155-156.

\item V.~M.~Buchstaber, V.~Z.~Enolskii, \emph{Abelian Bloch solutions of
the two-dimentional Schr\"odinger equation.}, Russian Math. Surveys,
v.~50, Issue~1, 1995, 195--197.

\item
V.~M.~Buchstaber, I.~M.~Krichever, \emph{Multidimensional vector
addition theorems and the Riemann theta functions.}, Int. Math. Res.
Notices, N~10, 1996, 505--513.

\item
V.~M.~Buchstaber, V.~Z.~Enolskii, \emph{Explicit algebraic
description of hyperelliptic Jacobians based on Klein's
sigma-function.}, Functional Anal. Appl., 30:1, 1996, 44--47.

\item
V.~M.~Buchstaber, D.~V.~Leykin, V.Z.Enolskii \emph{A matrix realization of
hyperelliptic Kummer varieties.}, Russian Math. Surveys, v.~51,
Issue~2, 1996, 319--320.

\item
V.~M.~Buchstaber, V.~Z.~Enolskii, D.~V.~Leykin, \emph{Hyperelliptic
Kleinian functions and applications.}, Solitons, Geometry and
Topology: On the Crossroad, V.~M.~Buchstaber, S.~P.~Novikov Editors,
AMS Trans., ser.~2, v.~179, 1997, 1--33.

\item
V.~M.~Buchstaber, V.~Z.~Enolskii, D.~V.~Leykin, \emph{Kleinian
functions, hyperelliptic Jacobians and applications.}, Reviews in
Mathematics and Math. Physics, I.~M.~Krichever, S.~P.~ Novikov
Editors, v.~10, part~2, Gordon and Breach, London, 1997, 3--120.

\item
V.~M.~Buchstaber, V.~Z.~Enolskii, D.~V.~Leykin, \emph{A recursive
family of differential polynomials generated by the Sylvester
identity, and addition theorems for hyperelliptic Klein functions.},
in Functional Anal. Appl., 31:4, 1997, 240--251.

\item
V.~M.~Buchstaber, H.W.Braden, \emph{The general analytic solution of
a functional equation of addition type.}, SIAM, J. Math. Anal.,
v.~28, N~4, 1997, 903--923.

\item
V.M.~Buchstaber, H.~W.~Braden, \emph{Integrable systems with
pairwise interactions and functional equations.}, Reviews in
Mathematics and Math. Physics, I.~M.~Krichever, S.~P.~Novikov
Editors, v.~10, part~2, Gordon and Breach, London, 1997, 121--166.

\item
V.M.~Buchstaber and J.C.~Eilbeck and V.Z.~Enolskii and D.V.~Leykin,
and M.~Salerno,
\emph{$2D$ and $3D$ Schr{\"o}dinger equation with
Abelian potential} J.  Math.  Phys., {\bf 43} (2002) 2858-2881.
 \label{beels02}

\item
V.M.~Buchstaber, V.~Z.~Enolskii, D.~V.~Leykin,
\emph{Sigma-functions of (n,s)-curves.}, Russian Math. Surveys,
v.~54, Issue~3, 1999, 628--629.

\item
V.M.~Buchstaber, V.~Z.~Enolskii, D.~V.~Leykin, \emph{Uniformization
of Jacobi manifolds of trigonal curves, and nonlinear differential
equations.}, Funct. Anal. Appl., 34:3, 2000, 159--171.

\item
V.M.~Buchstaber, D.~V.~Leykin, \emph{The manifold of solutions of
the two-dimensional analogue of the two-zone Lame equation.},
Russ. Math. Surv., v.~56, Issue~6, 2001, 1155--1157.

\item
V.M.~Buchstaber, D.~V.~Leykin, \emph{Lie algebras associated with
sigma-functions and versal deformations.}, in Russ. Math. Surv.,
v.~57, Issue~3, 2002, 584--586.

\item
V.M.~Buchstaber, D.~V.~Leykin, \emph{Graded Lie algebras defining
hyperelliptic sigma-functions.}, Doklady Math. Sci., \textbf{66},  4:2 (2003)
87--90.

\item
V.M.~Buchstaber, D.~V.~Leykin, \emph{Polynomial Lie algebras.},
Funct. Anal. Appl., 36:4, 2002, 267--280.

\item
V.M.~Buchstaber, D.~V.~Leykin, M.~V.~Pavlov, \emph{Egorov's
hydrodynamic chains, Chazy equation and group $SL(2,\mathbb{C})$.},
Funct. Anal. Appl., \textbf{37}:4 (2003) 251--262.

\item
V.M.~Buchstaber, D.~V.~Leykin, \emph{Heat Equations in a
Nonholonomic Frame.}, Funct. Anal. Appl., \textbf{38}:2, 2004, 88--101.

\item
V.M.~Buchstaber, D.~V.~Leykin, \emph{Hyperelliptic Addition Law.},
J. Nonlin. Math. Phys., \textbf{12}, suppl.~1, (2005) 106--123.

\item
V.M.~Buchstaber, D.~V.~Leykin, \emph{Trilinear functional
equations.}, Russ. Math. Surv., \textbf{60}: (2005) 341--343.

\item
V.M.~Buchstaber, D.~V.~Leykin, \emph{Addition laws on Jacobian
varieties of plane algebraic curves.}, Proceedings of the Steklov
Math. Inst., \textbf{251}:4 (2005) 49--120.

\item
V.M.~Buchstaber, D.~V.~Leykin, \emph{Differential and
differential-functional equations for sigma-functions.}, Fourth
International Conference on Differential and Functional Differential
Equations, Moscow, August 14-21, 2005, Book of abstracts.

\item
V.M.~Buchstaber, I.~M.~Krichever, \emph{Integrable equations,
addition theorems, and the Riemann-Schottky problem.}, Russ. Math.
Surv., \textbf{61}:1  (2006) 19--78.

\item
V.M.~Buchstaber, D.~V.~Leykin, \emph{Functional equations defining
multiplication in a continuous Krichever-Novikov basis.}, Russ.
Math. Surv., \textbf{61}:1 (2006) 165--167.

\item
V.M.~Buchstaber, D.~V.~Leykin, \emph{Differentiation of Abelian
functions over its parameters.}, Russ. Math. Surv., \textbf{62}:4 (2007) 787--789.

\item
V.M.~Buchstaber, D.~V.~Leykin, \emph{Solution of the problem of
differentiation of Abelian functions over parameters for families of
$\boldsymbol{(n,s)}$-curves.}, Funct. Anal. Appl., \textbf{42}:4 (2008)
268--278.

\item
V.M.~Buchstaber, S.~Yu.~Shorina, \emph{Commuting multidimensional
third-order differential operators defining a KdV hierarchy.},
Russ. Math. Surv., \textbf{58}:3, 2003, 610--612.

\item
V.M.~Buchstaber, S.~Yu.~Shorina, \emph{The $w$-Function of a
Solution the $g$-th Stationary KdV Equation.}, Russian Math.
Surveys, \textbf{58}:4 (2003) 780--781.

\item
V.M.~Buchstaber, S.~Yu.~Shorina, \emph{$w$-Function of the KdV
Hierarchy.}, Geometry, Topology, and Mathematical Physics,
S.P.Novikov's seminar: 2002-2003, V.M.Buchstaber, I.M.Krichever
Editors, Providence, RI, Amer. Math. Soc. Transl., \textbf{212}, ser.~2,
(2004) 41--46.

\item J.~ W.~ S.~ Cassels  and E.~V.~Flynn \emph{Prologomena to a Middlebrow Arithmetic of Curves of Genus 2,} London Mathematical Society Lecture Note Series \textbf{230}, Cambridge, Cambridge University Press

\item K.~Cho and A.~Nakayashiki, \emph{Differential structure of Abelian functions} Int.J.Math. \textbf{61} (2011) 961-985

\item M.~ England and J.~ C.~ Eilbeck. \emph{Abelian functions associated with a cyclic tetragonal curve of genus six}. J.Phys.A: Math.Theor. {\bf 42} (2009)
    Article Number: 095210   DOI: 10.1088/1751-8113/42/9/095210

\item M.~ England. \emph{Deriving bases for Abelian functions} Computational Methods and Function Theory {\it to be published} arXiv:1103.0468

\item M.~ England and J.~Gibbons. \emph{A genus six cyclic tetragonal reduction of the Benney equations}
    J.Phys.A: Math.Theor. {\bf 42} (2009) Article Number: 375202   DOI: 10.1088/1751-8113/42/37/375202

\item Eilbeck Chris and Enolskii Victor,
and Matsutani Shigeki and \^Onishi Yoshihiro, and Previato Emma.\emph{
Addition formulae over the Jacobian pre-image of hyperelliptic Wirtinger varieties}.  J. f\"ur reine angew. Math.
{\bf 619}, (2008) 37-48. \label{eemop08}

\item Eilbeck Chris and Enolski Victor and Gibbons John. \emph{Sigma, Tau and Abelian functions of algebraic curves,}  J.Phys.A:Math.Theor.  {\bf 43}: 45, (2010)  455216(20p)

\item J.C.~Eilbeck and V.Z.~Enolski and E.~Previato \emph{
Spectral Curves of Operators with Elliptic Coefficients,}
Symmetry, Integrability and Geometry: Methods and Applications,
SIGMA {\bf 3} (2007) 045, 17p. \label{eep07}

\item Eilbeck Chris and  Enolskii Victor,
and Matsutani Shigeki and \^Onishi Yoshihiro, and Previato Emma.
\emph{Abelian Functions for Trigonal Curve of Genus Three,} Int. Math.
Res. Notices, {\bf 2007}, (2007): rnm 140-38.

\item J.C.~Eilbeck and V.Z.~Enolskii. \emph{Bilinear operators and the power
series for the
 Weierstrass $\sigma$-function }
J.Phys.A: Math.Gen. ,{\bf 33} (2000) 791-794.

\item
J.C.~Eilbeck and V.Z.~Enolskii and H.~Holden
\emph{The hyperelliptic $\zeta$- functions and the integrable
massive {T}hirring system,} Proc. London Math. Soc. A, {\bf
459} (2003) 1581-1610.

\item J.C.~Eilbeck and V.Z.~Enolskii and E.~Previato \emph{
On a generalized Frobenius-Stickelberger addition theorem to higher
genera.} Lett.Math.Phys., {\bf 63} (2003) 5-17.

\item  J.C.~Eilbeck and S.~Matsutani and Y.~Onishi. \emph{Addition formulae for Abelian functions associated with specialized curves}, Phil.Trans. Roy.Soc. A-Math.Phys.Engin.Sci. \textbf{369}   Issue: 1939  (2011) 1245-1263

\item J.C.~Eilbeck and M. England and Y.~ Onishi  
  \emph{Abelian functions associated with genus three algebraic curves},
    LMS J. Copmput.Math. {\bf 14}, (2011) 291-326;ArXiv:1008.0289.

\item Enolskii Victor, Matsutani Shigeki  and \^Onishi Yoshihiro.
\emph{The Addition Law attached to a stratification of a hyperelliptic
Jacobian variety}, Tokyo J. Math. {\bf 31}: 1 (2008) 27-38.

\item V.Z.Enolski and E.~Previato. \emph{
Ultra-elliptic solitons.}
Uspekhi Matem. Nauk, {\bf 62}:4 (2007) 173-174. Translation in: Russian Math. Surveys 62 (2007), no. 4, 796--798 \label{ep07}

\item V.Z.~Enolskii and M.~Pronine and P.~Richter,
\emph{Double pendulum and $\theta$-divisor}, J.Nonlin.Sci., {\bf 13} (2003)
157-174.

\item V.Z.~Enolskii and M.Salerno, \emph{
  Lax representation for two particle dynamics split on two
                tori} J.Phys.A.:Math.Gen., {\bf 29}:17 (1996) L425-431.

\item
J.C.~Eilbeck and V.Z.~Enolskii, and D.V.~Leykin.
\emph{On the {K}leinian construction of {A}belian functions of canonical
  algebraic curve.}
In: Proceedings of the Conference SIDE III: Symmetries of
  Integrable Differences Equations ,
  Saubadia, May 1998,
  CRM Proceedings and Lecture Notes, pp 121-138, 2000.

\item Enolski Victor and Harnad John, \emph{Schur function expansions of KP tau functions associated to algebraic curves.} Russ. Math. Surv. \textbf{66}:4 (2011) 137-178.

\item Enolski Victor and Hackmann Eva and Kagramanova Valeria and Kunz Jutta and L{\"a}m\-mer\-zahl Claus.\emph{
Inversion of hyperelliptic integrals of arbitrary genus with application to
geodesic equations in higher di\-men\-si\-ons, }  J.Geom.Phys.  \textbf{61} (2011) 899-921,

\item Enolski Victor and Hartmann Betti and Kagramanova Valeria and Kunz Jutta and L{\"a}mmerzahl Claus and Parinya Sirimachan.  \emph{Inversion of a general hyperelliptic integral and particle motion
in Ho\v{r}ava-Lifshitz black hole space-times}. J.Math.Phys., {\bf 53},  (2012), 012504;
Preprint 2011: arXiv:1106.2408v1 [gr-qc] \label{ehkklp11}

\item Enolski Victor and Hartmann Betti and Kagramanova Valeria and Kunz Jutta and L{\"a}mmerzahl Claus and Parinya Sirimachan.  \emph{Particle motion
in Ho\v{r}ava-Lifshitz black hole space-times} Preprint 1011: arXiv: 1106.4913 [gr-qc] \label{ehkklp11a}

\bibitem[EEKL993]{eekl993} J.~C. Eilbeck, and V.~Z. Enolskii, and V.~ Z.~ Kuznetsov, and D.~V. Leykin.
 \emph{Linear $r$-matrix algebra for systems separable in parabolic coordinates},
 Phys. Lett. A \textbf{ 180} (1993), no.~3, 208--214.

\bibitem[EEO11]{eeo11} J.~C. Eilbeck, and M.~England, and  Yo.~\^Onishi,
 \emph{Abelian functions associated with genus three algebraic curves},
 LMS J. Comput.Math. \textbf{ 14}, (2011) 291-326.

\bibitem [EEKT994]{eekt994}J.~C.~Eilbeck, and V.~Z.~ Enolskii V.Z., and V.~B.~Kuznetsov, and A.~V.~Tsiganov A.V.,
\emph{Linear $r$-matrix algebra for classical separable systems},  J.Phys. A: Math. Gen., \textbf{ 27}, (1994)  567-578.

\item D. Grant, \emph{Formal groups in
genus two}, J. reine ang. Math., {\bf 411} (1990)
96--121.

\item Eva Hackmann and Claus L\"ammerzahl,\emph{{Complete Analytic Solution of the Geodesic Equation in Schwarzschild-(Anti-)de Sitter Spacetimes,}} Phys.Rev.Lett., {\bf 100} (2008)
171101:1-4.

\item Eva Hackmann and Claus L\"ammerzahl,\emph{Geodesic equation in Schwarzschild-(anti-)de Sitter space--times: Analytical solutions and applications,} Phys. Rev., \textbf{D 78} (2008) 024035.

\item Eva Hackmann and Valeria Kagramanova and Jutta Kunz and Claus L\"ammerzahl, \emph{Analytic solutions of the geodesic equations in higher dimensional static spherically symmetric spacetimes,} Phys. Rev., \textbf{D 78} (2008) 124018.

\item Eva Hackmann and Valeria Kagramanova and Jutta Kunz and Claus L\"ammerzahl, \emph{Analytic solutions of the geodesic equation in axially symmetric space--times,} Europhys. Lett., \textbf{88} (2009) 30008.

\item Eva Hackmann and Valeria Kagramanova and Jutta Kunz and Claus L\"ammerzahl, \emph{Analytical solution of the geodesic equation in {Kerr-(anti-) de Sitter} space-time,} Phys. Rev., \textbf{D 81} (2010) 044020.

\item Eva Hackmann and Betti Hartmann and Claus L\"ammerzahl and Parinya Sirimachan, \emph{The complete set of solutions of the geodesic equations in the space-time of a Schwarzschild black hole pierced by a cosmic string,} Phys. Rev., \textbf{D 81} (2010) 064016.

\item Eva Hackmann and Betti Hartmann and Claus L\"ammerzahl and Parinya Sirimachan, \emph{Test particle motion in the space-time of a Kerr black hole pierced by a cosmic string,} Phys. Rev., \textbf{D 82} (2010) 044024.

\item J.~D. ~Edelstein and M. ~G{\'o}mez-Reino
and M.~Mari{\~n}o, \emph{Blowup formulae in {D}onaldson-{W}itten theory and integrable
hierarchies,} Adv. Theor. Math. Phys.,
{\bf 4} (2000) 503--543.

\item Yuri N.~Fedorov and David G\'omez-Ullate, \emph{Dynamical systems on infinitely sheeted Riemann surfaces}, Physica D, \textbf{227} (2007), 120-134

\item J.~Jorgenson
, \emph{On directional derivatives of the theta function along its divisor,} Israel J.Math., {\bf 77} (1992) 274-284.

\item A.~N.~W.~Hone. \emph{Elliptic curves and quadratic recurrence sequences.}
Bulletin of the London Mathematical Society {\bf 37} (2005) 161-171;
Corrigendum, ibid. {\bf 38} (2006) 741-742.

\item A.~N.~W.~Hone, \emph{Sigma function solution of the initial value problem for Somos 5
sequences.} Transactions of the American Mathematical Society {\bf 359} (2007) 5019-5034.

\item A.~N.~W.~Hone, \emph{Laurent Polynomials and Superintegrable Maps.}
Symmetry, Integrability and Geometry: Methods and Applications {\bf 3} (2007) 022, 18 pp.

 \item A.~N.~W.~Hone, \emph{Analytic solution and integrability for a bilinear recurrence of
order six.}
Applicable Analysis \textbf{89}:4 (2010) 473-492

\item Andrew~N.~W.~Hone and Christine Swart. \emph{Integrality and the Laurent phenomenon for Somos 4 and Somos 5 sequences} Math. Proc. Cambridge Phil. Soc.  \textbf{145} (2008)  65-85

\item A.~N.~W.~Hone. \emph{Sigma function solution of the initial value problem for somos 5 sequences} Transac. Amer.Math.Soc. \textbf{359}:10, 5019-5034

\item Valeria Kagramanova and Jutta Kunz and Eva Hackmann and Claus L\"ammerzahl, \emph{Analytic treatment of complete and incomplete geodesics in Taub-NUT space-times,} Phys. Rev., \textbf{D 81} (2010) 124044.

\item Yu.Kodama and S.~Matsutani and E.~Previato, \emph{Quasi-periodic and periodic solutions of the Toda lattice via the hyperelliptic sigma functions} Preprint: arXiv:1008.0509

\item D.Korotkin and V.Shramchenko, \emph{On higher genus Weierstrass sigma-functions.}
Physica D, 2012

\item S. ~Matsutani,\emph{
         Closed loop solitons and sigma functions:
          classical and quantized elasticas with genera one and two},
          J.~Geom.~Phys., {\bf 39} (2001) 50-61.

\item S. ~Matsutani,\emph{
         Hyperelliptic solutions of KdV and KP equations:
          reevaluation of Baker's study on hyperelliptic sigma functions},
          J.~Phys.~A: Math.~\& Gen., {\bf 34} (2001) 4721-4732.

\item S. ~Matsutani,\emph{
       Hyperelliptic loop solitons with genus $g$:
       investigation of a quantized elastica},
      J.~Geom.~Phys., {\bf 43} (2002)  146-162.

\item S.~ Matsutani,
       \emph{Elliptic and hyperelliptic solutions of discrete Painlev\'e I
       and its Extensions to third difference equation},
        Phys.~Lett.~A, {\bf 300} (2002)  233-242.

\item S.~ Matsutani, \emph{
       Hyperelliptic solutions of modified Kortweg-de Vries equation
       of genus $g$:  essentials of Miura transformation},
       J.~Phys.~A:~Math.~\& Gen., {\bf 35} (2002)
          4321-4333.

\item S. ~Matsutani,\emph{
       Toda equations and $\sigma$ function},
        J.~Non.~Math.~Phys. {\bf 10} (2003)  1-15.

\item S. ~Matsutani,\emph{
       Recursion relation of hyperelliptic psi-functions of genus two},
       Integral Trans.~Spec. Function. {\bf 14} (2003) 517-527.

\item S. ~Matsutani,\emph{Hyperelliptic al Functions of sine-Gordon equation} In: New development in mathematical physics research, 2004, Nova Science, edited by V.Benton: invited paper pp. 177-200

  \item S. ~Matsutani,\emph{Relations of al Functions over Subjarieties in a Hyperelliptic Jacobian} CUBO A Math. J \textbf{7} (2005) 75-85

 \item S. ~Matsutani,\emph{Reality conditions of loop solitons genus $g$ hyperelliptic al functions} Electron.J.Diff.Eqns. \textbf{2007} (2007) 1-12

 \item S. ~Matsutani,\emph{Neumann system and hyperelliptic al functions} Surveys Math.Appl. \textbf{2008} (2008) 13-25

\item S. ~Matsutani, \emph{
Relations in a quantized elastica},
J. Phys. A: Math. Theor. {\bf 41} (2008)  {075201}(12pp).

\item S.~Matsutani and Y. Onishi, \emph
      {On the moduli of a quantized elastica in $\mathbb{P}$ and
      KdV flows:
       study of hyperelliptic curves as an extension of Euler's perspective
       of elastica I},
       Rev.~Math.~Phys. {\bf 15}:6 (2003)  559-628.

\item S.~Matsutani, and E.~Previato. \emph{
 Jacobi inversion on strata of the Jacobian of the $C_{rs}$ curve
 $y^r = f(x)$},
 J. Math. Soc. Jpn. {\bf 60} (2008) 1009-1044.

\item S. ~Matsutani and E.~Previato \emph{A generalized Kiepert formula for $C_{ab}$ curves} Israel J.Math. \textbf{171} (2009) 305-323

\item  S.~ Matsutani, and Emma Previato,\emph{
A class of solutions of the dispersionless KP equation},
Phys. Lett. A.  {\bf 373} (2009)
3001-3004

\item S. ~Matsutani and E.~Previato \emph{Jacobi inversion on strata of the Jacobian of the $C_{rs}$ curve  $y^r=f(x)\; II$ } Preprint (2011) arXiv: 1006.1090 [math.AG]

\item S. ~Matsutani Sigma functions for a space curve (3, 4, 5) type with an appendix by J. Komeda, arXiv:1112.4137 [math-ph]

\item A.~Nakayashiki,\emph{
{S}igma {F}unction as {a} {T}au {F}unction}.
Int. Math. Res. Notices, 2010-3 (2010) 373-394

\item A.~Nakayashiki, \emph{{A}lgebraic {E}xpression of {S}igma {F}unctions of
  $(n,s)$ {C}urves}, Asian J.Math. \textbf{14}:2 (2010), 174-211;
   "arXiv:0803.2083", 2008

\item  A. Nakayashiki,\emph{
On Hyperelliptic Abelian Functions of Genus 3}:
J.Geom.Phys \textbf{61} (2011) 961-985

\item
F.W.~Nijhoff and V.Z.~Enolskii
\emph{Integrable mappings of KdV type and hyperelliptic
addition theorems,}
\newblock Symmetries and Integrabilty of Difference
          Equations (SIDE II),
    Lecture Note Series,
\newblock Canterbery, July 1996,\newblock Edited by P Clarkson
and F. W. Nijhoff , 64-78,
\newblock "Cambridge University Press", 1999.

\item Y.~ {\^O}nishi, \emph{Complex
multiplication formulae for hyperelliptic curve of genus three},
Tokyo J. Math., {\bf 21:2} (1998)
381--431. A list of corrections is available at {\em http://www.ccn.yamanashi.ac.jp/~yonishi/$\#$publications}

\item Y. ~{\^O}nishi, \emph{On the
{G}alois group corresponding to the formula of {G}rant,}
Comment. Math. Univ. St. Paul., {\bf 42} (1991)
37--48.

\item Y. ~{\^O}nishi,   \emph{ Determinant
expressions for some {A}belian functions in genus two},
Glasgow Math. J. , {\bf 44}:3 (2002)
353-364.

\item Y. ~{\^O}nishi,
 Determinantal expressions for hyperelliptic functions in genus three,
{\em Tokyo J. Math.}, {\bf 27} (2004) 299-312.
A list of corrections is available at {\em http://www.ccn.yamanashi.ac.jp/~yonishi/$\#$publications}

\item Y.~ {\^O}nishi, \emph{Determinant
{E}xpressions for {H}yperelliptic
            {A}belian {F}unctions (with an {Appendix} by
            {S}higeki {M}atsutani:  {C}onnection of the formula
             of {C}antor and {B}rioshi-{K}iepert type)},
Proc. Edinburgh Math. Soc., {\bf 48} (2005) 705-742.

\item Y. ~{\^O}nishi, \emph{
Abelian functions for trigonal curves of degree four and determinantal formulae in purely trigonal case.}
International Journal of Mathematics, {\bf 20}:4 (2009) 427-44.

\end{enumerate}

\newcommand{\etalchar}[1]{$^{#1}$}
\providecommand{\bysame}{\leavevmode\hbox to3em{\hrulefill}\thinspace}
\providecommand{\MR}{\relax\ifhmode\unskip\space\fi MR }
\providecommand{\MRhref}[2]{%
  \href{http://www.ams.org/mathscinet-getitem?mr=#1}{#2}
}
\providecommand{\href}[2]{#2}


\begin{thebibliography}{EGRM00}


\bibitem[AM978]{am78}
M. Adler and J. Moser, \emph{On a class of
polynomials connected with the Korteweg--de Vries equation}, Comm.
Math. Phys., \textbf{61}, 1--30 (1978).

\bibitem[AF00]{af00}
S.~Abenda and Yu.~Fedorov,\emph{ On the weak Kowalevski-Painlev\'e property for hyperelliptically separable systems},
Acta Appl.Math. \textbf{60} (2000), 137-178.

\bibitem[Akh965]{ak65}
N.~I. Akhiezer, \emph{The classical moment problem and some related questions
  in analysis}, Oliver and Boyd, London, New York, (1965).

\bibitem[Arn96]{Arnold96}
V.~I.~Arnold, Singularities of Caustics and Wave Fronts, Kluwer Academic
Publishers, Dordrecht, 1996.

\bibitem[AGZV85,88]{AVGZ2004}
V. ~I. ~Arnold, S. ~M.~ Gusein-Zade, and A. ~N.~ Varchenko
Singularities of Differentiable Maps I, II
Birkhauser, Boston--Basel--Stuttgart, 1985, 1988

\bibitem[BG04]{balgib04}
S.~Baldwin and J.~Gibbons, \emph{Higher genus hyperelliptic reductions of the
  {B}enney equations.}, J.Phys.A:Math.Gen. \textbf{37} (2004), no.~20,
  5341--5354.

\bibitem[Bak897]{ba97}
H.~F. Baker, \emph{Abel's theorem and the allied theory of theta functions},
  Cambridge Univ. Press, Cambridge, (1897), Reprinted in 1995.

\bibitem[Bak898]{ba98}
\bysame, \emph{On the hyperelliptic sigma functions}, Amer. Journ. Math.
  \textbf{20} (1898), 301--384.

\bibitem[Bak903]{ba03}
\bysame, \emph{On a system of differential equations leading to periodic
  functions}, Acta Math. \textbf{27} (1903), 135--156.

\bibitem[Bak907]{ba07}
\bysame, \emph{Multiply {P}eriodic {F}unctions}, Cambridge Univ. Press,
  Cambridge, 1907.

\bibitem[BE955]{ba55}
H.~Bateman and A.~Erdelyi, \emph{Higher {T}ranscendental {F}unctions}, vol.~2,
  McGraw-Hill, New York, 1955.

\bibitem[Bea87]{bea87} A.Beavville, \emph{Le probleme de Schottky et la conjecture  de Novikov} Sem. N.Bourbaki, \textbf{1986-1987}, exp.  $n^o675$
101-112

\bibitem[BBE{\etalchar{+}}994]{bbeim94}
E.~D. Belokolos, A.~I. Bobenko, V.~Z. Enolskii, A.~R. Its, and V.~B. Matveev,
  \emph{Algebro {G}eometric {A}pproach to {N}onlinear {I}ntegrable
  {E}quations}, Springer, Berlin, 1994.

 \bibitem[BE01]{be01}Belokolos E.D. and Enolskii V.Z
\emph{Reduction of {A}belian {F}unctions and
              {A}lgebraically {I}ntegrable {S}ystems,}
Journal of Mathematical Sciences, Part I: 106:6 (2001) 3395-3486;
Part II: 108:3 (2002) 295-374.

\bibitem[Bol886]{bolza86} O.~Bolza,
\emph{Ueber die {R}eduction hyperelliptischer {I}ntegrale erster {O}rdnung und erster {G}attung auf elliptische durch eine {T}ransformation vierten {G}rades}, Math. Ann., \textbf{XXVIII} (1886), 447.

\bibitem[Bol895]{bo95}
O.~Bolza, \emph{On the first and second logarithmic derivatives of
  hyperelliptic $\sigma$--functions}, Amer. Journ. Math. \textbf{17} (1895),
  11--36.

\bibitem[Bol899]{bo99}
O.~Bolza, \emph{The partial differential equations for the hyperelliptic $\theta$
and $\sigma$-functions,} Amer. J. Math., \textbf{21}, 107--125 (1899).

\bibitem[Bol900]{bo00}
O.~Bolza, \emph{Remarks concerning the expansions of the hyperelliptic
$\sigma$-functions,} Amer. J. Math., \textbf{22}, 101--112 (1900).


\bibitem[BEF12]{bef12}
H. W. Braden and V.Z. Enolski and Yu.N.Fedorov
\emph{Dynamics on strata of trigonal Jacobians in some integrable problems of rigid body motion}, 
Preprint, 2012

\bibitem[Buc990]{bu90}
V.~M. Buchstaber, \emph{Functional equations associated with addition theorems
  for elliptic functions and two-valued algebraic groups}, Russ. Math. Sur.
  \textbf{45} (1990), 231--214.

\bibitem[BE996]{be96b}
V.~M. Buchstaber and V.~Z. Enolskii, \emph{Explicit algebraic description of
  hyperelliptic {J}acobian on the background of {K}leinian
  $\sigma$--functions}, Func. Anal. Appl. \textbf{30} (1996), no.~1, 57--60.



\bibitem[BK993]{bk93} Buchstaber V M, Krichever I M, \emph{Vector addition
theorems and Baker-Akhiezer functions}, Teor. Mat. Fiz.
\textbf{94} (1993) 200--212.

\bibitem[BK996]{bk96} Buchstaber V M, Krichever I M, \emph{Multidimesional
vector addition theorems and the Riemann theta function}
Internat. Math. Res. Notices \textbf{10} (1996) 505--513.


\bibitem[BK06]{bk06} Buchstaber V M, Krichever I M, \emph{Integrable equations, addition theorems and the Riemann-Schottky problem}
Uspekhi Matem.Nauk \textbf{61}:1 (2006) 25-84. Engl.transl.
Russ.Math.Surv.  \textbf{61}:1 (2006), 19-78


\bibitem[BEL999]{bel99}
Buchstaber V.M., Enolskii V.Z. and Leykin D.V.
\emph{$\sigma$-functions of $(n,s)$-curves}
Uspekhi Matem. Nauk, {\bf 54}:3 (1999) 155-156.

\bibitem[BL02]{bl02} Buchstaber V M,  Leykin D V, \emph{Lie algebras associated with
$\sigma$-functions and versal deformations}, Uspekhi Math.
Nauk, \textbf{57} No.~3 (2002) 145--146.

\bibitem[BL04]{bl04} Buchstaber V M,  Leykin D V, \emph{Heat equations in a nonholonomic frame,
} Funkts. Anal. Prilozhen, \textbf{38} No.~2 (2004) 12--27.

\bibitem[BL05]{bl05}
V.~M. Buchstaber and D.~V. Leykin, \emph{Addition {L}aws on {J}acobian
  {V}ariety of {P}lane {A}lgebraic {C}}, Proceedings of the Steklov Institute
  of Mathematics \textbf{251} (2005), 1--72.

\bibitem[BL08]{bl08} V.~M. Buchstaber and D.~V. Leykin,
\emph{Solution of the problem of differentiation of {A}belian
  functions over parameters for families of $(n,s)$ curves}, Functional. Anal.
  and its Appl. \textbf{42} (2008), no.~4, 268--278, Translation of:
  Funksional. Anal. i Prilozhen, 42(2008) 24-36.


\bibitem[BL07]{bule9}
V. M. Buchstaber and D. V. Leykin
\emph{Differentiation of
Abelian functions with respect to parameters} Uspekhi Mat. Nauk
\textbf{62}, No.4 (2007) 153--154. English transl.:
Russian Math. Surveys \textbf{ 62}, No. 4 (2007) 787--789


\bibitem[BEL996]{bel96}
V.~M. Buchstaber, V.~Z. Enolskii, and D.~V. Leykin, \emph{Matrix realization of
  hyperelliptic {K}ummer varieties}, Uspekhi Matem. Nauk \textbf{51} (1996),
  no.~2, 147--148.

\bibitem[BEL997a]{bel97a} V.~M. Buchstaber, V.~Z. Enolskii, and
 D.~V. Leykin, \emph{Hyperelliptic {K}leinian functions and applications.}
American Mathematical Society Translations, Advances in Mathematical Sciences
\newblock {\em Solitons Geometry and Topology: On Crossroad. AMS
Translations}, Editors: V.M.Buchstaber and S.P.Novikov \textbf{179}, Ser. 2, pp. 1-34, 1997.
Publisher American Mathematical Society

\bibitem[BEL997b]{bel97b} V.~M. Buchstaber, V.~Z. Enolskii, and
 D.~V. Leykin, \emph{{K}leinian functions, hyperelliptic {J}acobians and
  applications}, Reviews in Mathematics and Mathematical Physics (London)
  (S.~P. Novikov and I.~M. Krichever, eds.), vol. 10:2, Gordon and Breach,
  1997, pp.~1--125.

\bibitem[BEL997b]{bel97c}
V.~M. Buchstaber, V.~Z. Enolskii, and D.~V. Leykin, \emph{A recursive family of differential polynomials generated by
  {S}ylvester's identity and addition theorems for hyperelliptic {K}leinian
  functions}, Func. Anal. Appl. \textbf{31} (1997), no.~4, 240--251.

\bibitem[BEL999]{bel99} V.~M. Buchstaber, V.~Z. Enolskii, and D.~V. Leykin, \emph{$\sigma$-functions of $(n,s)$-curves}, Uspekhi Matem. Nauk
  \textbf{54} (1999), no.~3, 155--156.

\bibitem[BEL00]{bel00}
V.~M. Buchstaber, V.~Z. Enolskii, and D.~V. Leykin,
\emph{Uniformisation of {J}acobi varieties of trigonal curves and
  nonlinear differential equations}, Func. Anal. Appl. \textbf{34} (2000),
  no.~3, 159--171.

\bibitem[BEL999]{bel99a}
V.~M.~Bukhshtaber, D.~V.~Leikin, and V.~Z.~Enol'skii, \emph{Rational analogues of
Abelian functions,} Funkts. Anal. Prilozhen., \textbf{33}, No.~2, 1--15
(1999).

\bibitem[BL02]{bucley02:1}
V.~M.~Bukhshtaber and D.~V.~Leikin, \emph{Lie algebras associated with
$\sigma$-functions and versal deformations,} Usp. Mat. Nauk, \textbf{57},
No.~3, 145--146 (2002).

\bibitem[BL02]{bucley02:2}
V.~M.~Bukhshtaber and D.~V.~Leikin, \emph{Graded Lie algebras that define
hyperelliptic sigma functions,} Dokl. RAS, \textbf{385}, No.~5, 583--586
(2002).

\bibitem[BL02a]{bucley02:3} V.~M.~Bukhshtaber and D.~V.~Leikin,
\emph{Polynomial Lie algebras} Funkts. Anal. Prilozhen., \textbf{36}, No.~4,
2002, 18--34 (2002). English transl.:
Funct. Anal. Appl. \textbf{36} No.4, 267--280, (2002)

\bibitem[BL04]{bl04}
V. M. Buchstaber and D. V. Leykin
\emph{Heat equations in a nonholonomic frame} Funkts. Anal. Prilozhen.
\textbf{38}, No. 2 (2004) 12--27, English transl.: Funct. Anal. Appl.
\textbf{38}:2 (2004) 88--101

\bibitem[BR998]{br98}
V.~M. Buchstaber and E.~G. Rees,
\emph{Multivalued groups, $n$-Hopf algebras and $n$-ring
homomorphisms,} In: Lie groups and Lie algebras, Kluwer,
1998, pp.~85--107.

\bibitem[BC928]{bc28a}
J.~L. Burchnall and T.~W. Chaundi, \emph{Commutative ordinary differential
  operators. {II}}, Proc. London Math. Soc \textbf{118} (1928), 557--583.

\bibitem[Bur888]{bur88}
H.~Burkhardt, \emph{Beitr{\"a}ge zur {T}heorie der hyperelliptische
  {S}igmafunktionen}, Math. Ann. \textbf{32} (1888), 381--442.

\bibitem[Cal975]{FC75} Calogero F, Exactly solvable one-dimensional many-body
problems, \emph{Lett. Nuovo Cimento} \textbf{13} (1975) 411--416.

\bibitem[Cal975a]{FC75a} Calogero F, \emph{One-dimensional many-body problems with
pair interactions whose ground-state wavefunction is of product
type}, Lett. Nuovo Cimento, \textbf{13} (1975) 507--511.

\bibitem[Cal976]{FC76} Calogero F, \emph{On a functional equation connected with integrable
many-body problems}, Lett. Nuovo Cimento, \textbf{16} (1976)
77--80.

\bibitem[CEEK00]{ceek00}
P.~L. Christiansen, J.~C. Eilbeck, V.~Z. Enolskii, and N.~A. Kostov,
  \emph{Quasi periodic solutions of {M}anakov type coupled nonlinear
  {S}chr{\"o}dinger equations}, Proc. R. Soc. Lond. A \textbf{456} (2000),
  2263--2281.

\bibitem[Con956]{co56}
F.~Conforto, \emph{Abelsche {F}unktionen und algebraische {G}eometry},
  Springer, Berlin, G{\"o}ttingen, Heiselberg, 1956.

\bibitem[CV990]{cv90}
O.~A. Chalykh and A.~P. Veselov, \emph{Commutative rings of partial
  differential operators and {L}ie algebras}, Commun. Math. Phys \textbf{126}
  (1990), 597--611.

\bibitem[Che948]{che48}
N.~G. Chebotarev, \emph{The theory of algebraic functions}, Gostekchisdat,
  Moscow, 1948.

\bibitem[DJKM983]{djkm83}
E.~Date, M.~Jimbo, M.~Kashiwara, and T.~Miwa, \emph{Transformation groups for
  soliton equations}, Proc. {RIMS} {S}ymposium on {N}onlinear {I}ntegrable
  {S}ystems -- {C}lassical and {Q}uantum {F}ield {T}heory (M.~Jimbo and
  T.~Miwa, eds.), World Scientific, 1983.

\bibitem[Dub981]{dub81}
B. ~A.~ Dubrovin, \emph{Theta functions and nonlinear equations},
Russ. Math. Surveys, \textbf{36} (1981) 11--80.

\bibitem[Dub994]{D1994}
B. A. Dubrovin
\emph{Geometry of 2D topological field theories}
Lecture Notes in Math. \textbf{1620}, (1994) 120--348

\bibitem[DMN976]{dmn76}
B.~A. Dubrovin, V.~B. Matveev, and S.~P. Novikov, \emph{Nonlinear equations of
  the {K}d{V} type, finite gap linear operators and Abelian varieties}, Uspekhi
  Matem. Nauk \textbf{31} (1976), no.~1, 55--135.

\bibitem[DKN976]{dkn76}
B.~A. Dubrovin, I.~M. Krichever, and S.~P. Novikov, \emph{The {S}chr{\"o}dinger
  equation in a periodic field on {R}iemann surface}, Dokl. AN SSR \textbf{229}
  (1976), no.~1, 15--18.

\bibitem[DMN976]{dmn76}
B.~A. Dubrovin, V.~B. Matveev, and S.~P. Novikov, \emph{Nonlinear equations of
  the {K}d{V} type, finite gap linear operators and Abelian varieties}, Uspekhi
  Matem. Nauk \textbf{31} (1976), no.~1, 55--135.

\bibitem[DN974]{dn74}
B.~A. Dubrovin and S.~P. Novikov, \emph{Periodic and quasiperiodic analogues of
  multisoliton solutions of {K}orteweg-de {V}ries equation}, Soviet JETP
  \textbf{67} (1974), no.~6, 2131--2143.

\bibitem[DN974]{DN1974}
B. A. Dubrovin and S. P. Novikov
\emph{A periodic problem for the Korteweg--de Vries and
Sturm--Liouville equations. Their connection with
algebraic geometry} Dokl. Akad. Nauk SSSR
\textbf{219} No. 3 (1974) 531--534. English transl.:
Soviet Math. Dokl. \textbf{15}(1974) 1597--1601

\bibitem[DKN01]{dkn01} Dubrovin B A, Krichever I M, Novikov S P,
\emph{Integrable systems.I}, In Encyclopaedia of Mathematical Sciences vol
4, Editors: Arnold V I, Novikov S P, Springer Verlag, Berlin, 2001,
177--332.

\bibitem[DG986]{dg86}
J.~J. Duistermaat and F.~A. Gruenbaum,
\emph{Differential equations in spectral parameter}, Comm. Math. Phys.,
\textbf{103}, 177--204 (1986).

\bibitem[EGRM00]{eg-rm00}
J.~D. Edelstein, M.~G{\'o}mez-Reino, and M.~Mari$\tilde{\rm n}$o, \emph{Blowup
  formulae in {D}onaldson-{W}itten theory and integrable hierarchies},
  Adv.Theor.Math.Phys. \textbf{4} (2000), no.~3, 503--543, hep-th/0006113.

\bibitem[EH11]{enhar11}
V.~Enolski and J.~Harnad, \emph{Schur function expansions of {KP} tau functions
  associated to algebraic curves},  Russ. Math. Surv.\textbf{66} no. 4, 137-178, (2011)
  {\em in Russian}
    English translation: Russ.Math.Surv. \textbf{66} no. 4, 767-807, 2011. Preprint: 2010,
     arXiv:1012.3152 [math-ph]

\bibitem[EEK00]{eek00}
J.~C. Eilbeck, V.~Z. Enolskii, and N.~A. Kostov, \emph{Quasi periodic solutions
  for {V}ector nonlinear {S}chr{\"o}dinger equations}, J. Math. Phys.
  \textbf{41} (2000), 8236--8248.



\bibitem[EEL00]{eel00}
J.~C. Eilbeck, V.~Z. Enolskii, and D.~V. Leykin, \emph{On the {K}leinian
  construction of {A}belian functions of canonical algebraic curves},
  Proceedings of the Conference SIDE III: Symmetries of Integrable Differences
  Equations , Saubadia, May 1998, CRM Proceedings and Lecture Notes 25, 2000,
  pp.~121--138.

\bibitem[EEM{\etalchar{+}}07]{eemop07}
J.~C. Eilbeck, V.~Z. Enolski, S.~Matsutani, Y.~\^Onishi, and E.~Previato,
  \emph{Abelian functions for trigonal curves of genus three}, Int. Math. Res.
  Notices \textbf{2007} (2007), rnm 140--38.





\bibitem[EEKL993]{eekl93} J.~C. Eilbeck, and V.~Z. Enolskii, and V.~ Z.~ Kuznetsov, and D.~V. Leykin.
 \emph{Linear $r$-matrix algebra for systems separable in parabolic coordinates},
 Phys. Lett. A \textbf{ 180} (1993), no.~3, 208--214.

\bibitem[EEO11]{eeo11} J.~C. Eilbeck, and M.~England, and  Yo.~\^Onishi,
 \emph{Abelian functions associated with genus three algebraic curves},
 LMS J. Comput.Math. \textbf{ 14}, (2011) 291-326.

\bibitem [EEKT994]{eekt94}J.~C.~Eilbeck, and V.~Z.~ Enolskii V.Z., and V.~B.~Kuznetsov, and A.~V.~Tsiganov A.V.,
\emph{Linear $r$-matrix algebra for classical separable systems},  J.Phys. A: Math. Gen., \textbf{ 27}, (1994)  567-578.

bibitem[EE09]{ee09}
M.~England and J.~C. Eilbeck, \emph{Abelian functions associated with a cyclic
  tetragonal curve of genus six}, J. Phys. A \textbf{42} (2009), 095210.


\bibitem[ES96]{es996}
V.~Z. Enolskii and M.~Salerno, \emph{Lax reprentation for two particle dynamics
  splitted on two tori}, J. Phys. A.: Math. Gen. \textbf{29} (1996), no.~17,
  L425--431.


\bibitem[EPR03]{epr03} V.Z.~Enolskii and M.~Pronine and P.~Richter,
\emph{Double pendulum and $\theta$-divisor}, J.Nonlin.Sci., {\bf 13} (2003)
157-174.

\bibitem[EHKKLP12]{ehkklp12} V.Enolski and B. Hartmann Betti and
V. Kagramanova and J.Kunz Jutta and C. L{\"a}mmerzahl and P. Sirimachan.
\emph{Inversion of a general hyperelliptic integral and particle motion
in Ho\v{r}ava-Lifshitz black hole space-times}. J.Math.Phys., \textbf{53},  (2012),  012504;
arXiv:1106.2408v1 [gr-qc]


\bibitem[Fay973]{fa73}
J.~D. Fay, \emph{Theta functions on {R}iemann surfaces}, Lectures Notes in
  Mathematics (Berlin), vol. 352, Springer, 1973.

\bibitem[Fay979]{fa79}
\bysame, \emph{On the {R}iemann-{J}acobi formula}, Nachrichten der {A}kadedemie
  der {W}issenschaften in {G}{\"o}ttingen. {I}{I}.
  {M}athematisch-{P}hysikalische {K}lasse \textbf{5} (1979), 61--73.

\bibitem[Fay983]{fay83}
\bysame, \emph{Bilinear identities for theta functions}, Mathematics {R}eport
  83-168, 1983.

\bibitem[Fay989]{fay89}
John Fay, \emph{Schottky {R}elations on $\frac12({C}-{C})$}, Proceedings of
  {S}ymposia in {P}ure {M}athematics, vol. 49:1, 1989, Theta {Functions},
  {B}owdoin 1987, pp.~485--501.

\bibitem[FK980]{fk80}
H.~M. Farkas and I.~Kra, \emph{Riemann {S}urfaces}, Springer, New York, 1980.

\bibitem[FG07]{fgu07}
Yu.N.Fedorov and D.~G\'omes-Ulate
\emph{Dynamical systems on infinitely sheeted {Riemann} surfaces}, Physica D \textbf{ 227} (2007) 120-134.

\bibitem[FKT92]{fkt992}
J.~Feldman, H.~Kn{\"o}rrer, and E.~Trubowitz, \emph{There is no two-dimensional
  analogue of {L}am\'e equation}, Math. Ann. \textbf{294} (1992), 295--324.

\bibitem[For882]{for882}
A.~R.~Forsyth, \emph{Memoir on the Theta-Functions, Particularly Those of Two Variables}
Phil.Trans.Roy.Soc.London \textbf{173} (1882), 783-862.

\bibitem[FS880]{fs80}
G.~Frobenius and L.~Stickelberger, \emph{Ueber die {A}ddition und
  {M}ultiplication der elliptischen {F}unctionen}, J. reine angew. Math.
  \textbf{86} (1880), 146--184.

\bibitem[FS882]{FS:1882}
F.~Frobenius und L.~Stickelberger, \emph{\"Uber die Differentiation der
elliptischen Functionen nach den Perioden und Invarianten,} J. Reine Angew.
Math., \textbf{92}, 311--327 (1882).

\bibitem[Gan967]{ga67}
F.~R. Gantmaxer, \emph{The theory of matrices}, Gostekchisdat, Moscow, 1967.


\bibitem[Giv980]{Givental80}
A. B. Givental
\emph{Displacement of invariants of groups that are generated by
reflections and are connected with simple singularities of functions}
Funkts. Anal. Prilozhen. \textbf{14} No. 2 (1980) 4--14
English transl.: Funct. Anal. Appl. \textbf{14}, No. 2(1980), 81--89.


\bibitem[Van995]{Vanh}
P.~Vanhaecke \emph{
Stratification of hyperelliptic Jacobians and the Sato Grassmannian}
 Acta Appl. Math., \textbf{ 40} (1995), 143--172.


\bibitem[GH978]{gh78}
P.~Griffitth and J.~Harris, \emph{Principles of {A}lgebraic {G}eometry}, Wiley,
  New York, 1978.

\bibitem[GN01]{gn01}P.G.~Grinevich and S.P.Novokov, \emph{
Real finite-zone solutions of the sine–Gordon
equation: a formula for the topological charge}, UMN (Russ. Math. Surv) \textbf{56}, no. 5  (2001), 980-981


\bibitem[GN03]{gn03}P.G.~Grinevich and S.P.Novokov, \emph{Topological Charge of the real finite-gap periodic Sine-Gordon solutions.} Comm. Pure Appl. Math, \textbf{ LVI}, dedicated to the memory of Juergen Moser,
,( 2003), 956-978,  arXiv, math-ph/0111039.

\bibitem[G{\"o}p847]{go47}
G.~G{\"o}ppel, \emph{Theoriae transcendentium {A}belianarum primi ordinis
  adumbrato levis}, J. reine angew. {M}ath. \textbf{35} (1847), 277.

\bibitem[Gra991]{gr91}
D.~Grant, \emph{A generalization of a formula of {E}isenstein}, Proc. London
  Math. Soc. \textbf{62} (1991), 121--132.

\bibitem[GH03]{gh03} F.~Gesztesy and H.~Holden,
\emph{{S}oliton {E}quations and {T}heir
{A}lgebro-Geometric Solutions. $(1+1)$-{D}imensional
Continuous Models}, Cambridge University Press, Cambridge, U.K., 2003

\bibitem[GW995a]{gw95a}
F~Gesztesy and R~Weikard, \emph{On {P}icard potentials}, Differential Integral
  Equations \textbf{8} (1995), 1453--1476.

\bibitem[GW995b]{gw95}
\bysame, \emph{Treibich-{V}erdier potentials and the stationary (m){K}d{V}
  hierarchy}, Math. Z. \textbf{219} (1995), 451--476.

\bibitem[GW996]{gw96}
\bysame, \emph{Picard potentials and {H}ill's equation on a torus}, Acta Math.
  \textbf{176} (1996), 73--107.

\bibitem[Hir972]{hi72}
R.~Hirota, \emph{Exact solutions of the {K}orteweg -- de {V}ries equation for
  multiple collisions of solitons}, Phys. Rev. Lett. \textbf{27} (1972),
  1192--1194.

\bibitem[Hir980]{hi80}
\bysame, \emph{Direct methods in soliton theory}, Topics in {C}urrent {P}hysics
  (New-York) (R.~Bullough and P.~Caudrey, eds.), vol.~17, Springer, 1980,
  pp.~157--175.

\bibitem[HJ986]{hj86}
R.~A. Horn and C.~R. Johnson, \emph{Matrix analysis}, Cambridge University
  Press, Cambridge, England, 1986.

\bibitem[Hud905]{hu05}
R.~W. H.~T. Hudson, \emph{{K}ummer's quartic surface}, Cambridge university
  press, Cambridge, 1990, First published 1905.

\bibitem[Igu972]{ig72}
J.~Igusa, \emph{Arithmetic variety of moduli for curves of genus 2}, Ann. of
  Math. (2) \textbf{72} (1972), 612--649.


\bibitem[Igu972]{igusa72}
J.~Igusa, \emph{Theta {F}unctions, {G}rund. {M}ath. {W}iss.}, vol. 194,
  Springer, Berlin, 1972.

\bibitem[Igu982]{ig82}
\bysame, \emph{Problems on {A}belian functions at the time of {P}oincar{\'e}
  and some at present}, Bull. of the AMS \textbf{6} (1982), 161--174.

\bibitem[IM975]{im75}
A.~R. Its and V.~B. Matveev, \emph{Hill's operators with a finite number of
  lacunae and multisoliton solutions of the {K}orteweg-de {V}ries equation},
  Teor. Mat. Fiz. \textbf{23} (1975), 51--67.


\bibitem[Jor992]{jor92}
J. Jorgenson, \emph{On directional derivatives of the theta function along its divisor}, Israel J.Math., {\bf 77}, (1992) 274-284.

\bibitem[Kac93]{kac93}
V. G. Kac, Infinite Dimensional Lie Algebras, Birkh\"auser,
1983.

\bibitem[Kle886]{kl86}
F.~Klein, \emph{{\"U}ber hyperelliptische {S}igmafunctionen}, Math. Ann.
  \textbf{27} (1886), 431--464.

\bibitem[Kle888]{kl88}
\bysame, \emph{{\"U}ber hyperelliptische {S}igmafunctionen}, Math. Ann.
  \textbf{32} (1888), 351--380.

\bibitem[Kle890]{kl90}
\bysame, \emph{Zur Theorie der {A}bel'schen {F}unctionen}, Math. Ann.
  \textbf{36} (1890).

\bibitem[Kle923]{kl923}
\bysame, \emph{Vorbemerkungen zu den Arbeiten \"uber hyperelliptische und {A}belsche {F}unktionen}, Gesammele Mathematische Abhandlungen
 Vol. \textbf{3}, S. 317-322 (1923).

\bibitem[Kle923a]{kl23}
F.~Klein, \emph{{\"U}ber hyperelliptische Sigmafunktionen,} Gesammelte
{M}athematische {A}bhandlungen, Vol.~3, Teubner, Berlin, 1923, pp.~323--387.

\bibitem[KSh12]{ksh12}
 D.~Korotkin and V.~Shramchenko, \emph{On higher genus Weierstrass sigma-functions.}
Physica D, (2012)


\bibitem[Kra889]{K:90}
A.~Krazer, \emph{Zur Bildung allgemeiner $\sigma$-Functionen,} Math. Ann.,
\textbf{33}, 591--599 (1889).

\bibitem[Kra903]{kr03}
A.~Krazer, \emph{Lehrbuch der {T}hetafunktionen}, Teubner, Leipzig, 1903,
  reprinted by AMS Chelsea Publishing, 1998.

\bibitem[KW915]{kw15}
A.~Krazer and W.~Wirtinger, \emph{Abelsche {F}unktionen und allgemeine
  {T}hetafunktionen}, pp.~603--882, Teubner, 1915.

\bibitem[Kri977]{kr77}
I.~M. Krichever, \emph{The method of algebraic geometry in the theory of
  nonlinear equations}, Russian. Math. Surveys \textbf{32} (1977), 180--208.

\bibitem[Kri979]{kr79}
I.~M. Krichever, \emph{On the rational solutions of the
Zaharov--Shabat equations and
completely integrable systems of $N$ particles on a line,}
Zap. Nauchn. Sem. LOMI, \textbf{84}, No.~1, 117--130 (1979).

\bibitem[Kri980]{kr80}
\bysame, \emph{Elliptic solutions of {K}adomtsev-{P}etviashvili equation and
  integrable particle systems}, Funct. Anal. Appl. \textbf{14} (1980), 45--54.

\bibitem[KN987]{kn987}
I.~M. Krichever and S.~P.~Novikov
\emph{ Algebras of Virasoro type, Riemann surfaces
and the structures of soliton theory},  Funkts. Anal. i ego Pril.,
, \textbf{21}, No. 2, (1987) 46–63; English translation: Funct. Anal. Appl.,
\textbf{ 21}, No 2 (1987)



\bibitem[Ley995]{le95}
D.~V. Leykin, \emph{On {W}eierstrass cubic for hyperelliptic functions},
  Uspekhi Matem. Nauk \textbf{50} (1995), no.~6, 191--192.


\bibitem[Mac995]{Mac85}
I. G. Macdonald, \emph{Symmetric Functions and Hall Polynomials}, Clarendon
Press, Oxford, 1995.


\bibitem[Mar979]{ma79}
A.~I. Marcushevich, \emph{Introduction into the classical theory of Abelian
  functions}, Nauka, Moscow, 1979.

\bibitem[Mat01]{mat01}
S.~Matsutani,\emph{
         Closed loop solitons and sigma functions:
          classical and quantized elasticas with genera one and two},
          J.~Geom.~Phys., {\bf 39} (2001) 50-61.

\bibitem[Mat00]{ma00}
S.~Matsutani, \emph{Hyperelliptic solutions of {K}d{V} and {KP} equations:
  {R}eevaluation of {B}aker's {S}tudy on {H}yper{Elliptic} {S}igma
  {F}unctions}, arXiv: nlin.SI/00070001, 2000.

\bibitem[Mui928]{mu28}
T.~Muir, \emph{A {T}reatise on the {T}heory of {D}eterminants}, Dover
  publications, Inc., New York, 1928.

\bibitem[Mum975]{mu75}
D.~Mumford, \emph{Curves and their {J}acobians}, University of {M}ichigan
  press, Ann Arbor, 1975.

\bibitem[Mum984]{mu83}
\bysame, \emph{Tata lectures on theta, vol.1,\,vol.2}, Birkh\"auser, Boston,
  1983, 1984.

\bibitem[Nak08a]{nakaya08}
A.~Nakayashiki, \emph{{A}lgebraic {E}xpression of {S}igma {F}unctions of
  $(n,s)$ {C}urves}, Asian J.Math. \textbf{14}:2 (2010), 174-211; arXiv:0803.2083, 2008.

\bibitem[Nak08b]{na08}
\bysame, \emph{Tau and {S}igma}, Private communication, Iwate, 2008.

\bibitem[Nak09]{nakayashiki09}
\bysame,
\emph{
{S}igma {F}unction as {A} {T}au {F}unction}
Int. Math. Res. Notices, 2009, doi:10.1093/imrn/rnp135:
arXiv:0904.0846.


\bibitem[Nov974]{no74}
S.~P. Novikov, \emph{Periodic problem for the {K}orteweg de {V}ries equation},
  Funk. Anal. Appl. \textbf{74} (1974), 54--66.

\bibitem[Nov983]{no83}
\bysame, \emph{Two dimensional {S}chr{\"o}dinger operators in periodic fields},
  Itogi Nauki i Techniki, VINITI \textbf{23} (1983), 3--35.

\bibitem[NRS10]{nrs10} D.P.Novikov and P.K.Romanovski and S.G.Sadovnichuk. \emph{Some new methods of finite gap integration of solitonic equations}   (2010) Preprint: Omsk 2010

\bibitem[{\^O}ni998]{on998}
Y.~{\^O}nishi, \emph{Complex multiplication formulae for hyperelliptic curve of
  genus three}, Tokyo J. Math. \textbf{21} (1998), no.~2, 381--431.

\bibitem[{\^O}ni02]{on02} ~Y.{\^O}nishi,   \emph{ Determinant
expressions for some {A}belian functions in genus two},
Glasgow Math. J. , {\bf 44}:3 (2002)
353-364.


\bibitem[{\^O}ni04]{on04} ~Y. {\^O}nishi,\emph{
 Determinantal expressions for hyperelliptic functions in genus three,}
Tokyo J. Math., {\bf 27} (2004) 299-312.
A list of corrections is available at {\em http://www.ccn.yamanashi.ac.jp/~yonishi/$\#$publications}


\bibitem[{\^O}ni05]{on05} Y.~ {\^O}nishi, \emph{Determinant
{E}xpressions for {H}yperelliptic
            {A}belian {F}unctions (with an {Appendix} by
            {S}higeki {M}atsutani:  {C}onnection of the formula
             of {C}antor and {B}rioshi-{K}iepert type)},
Proc. Edinburgh Math. Soc., {\bf 48} (2005) 705-742.

\bibitem[PS964]{PS78}
G. Polya and G. Szeg\"o, Aufgaben und Lehrs\"atze aus der Analysis,
V.~2, Springer-Verlag, 1964.

\bibitem[Pri875]{pri875}
A.~Pringsheim.~\emph{ Zur Transformation zweiten Grades der hyperelliptischen Functionen
erster Ordnung} Math.Ann., \textbf{9}(1875), 445-475

\bibitem[Rom984]{ro84}
S.~M. Roman, \emph{The umbral calculus}, Academic Press, New York, 1984.

\bibitem[Ros851]{ros851}
G.~Rosenhain, \emph{Abhandlung {\"u}ber die {F}unktionen zweier {V}ariabler mit
  vier {P}erioden}, M{\'e}m. pr{\'e}s. l'Acad. de Sci. de France des savants
  \textbf{XI} (1851), 361--455, The paper is dated 1846. German Translation: H.
  Weber (Ed.), Engelmann-Verlag, Leipzig 1895.


\bibitem[SM980]{sato80}
M.~Sato and Y.Sato (Mori), \emph{On {H}irota's bilinear equations {I}}, {RIMS}
  {K}okyuroku \textbf{338} (1980), 183.

\bibitem[SM981]{sato81}
\bysame, \emph{On {H}irota's bilinear equations {II}}, {RIMS} {K}okyuroku
  \textbf{414} (1981), 181.


\bibitem[SV04]{sv04} Tanush Shaska and Helmut V\"olklein. \emph{Elliptic subfields and automorphisms of genus 2 function fields } in: Algebra, arithmetic and geometry with applications (West Lafayette, IN, 2000) 703-723, Springer, Berlin, 2004

\bibitem[Sh86]{shi86}T.Shiota, \emph{Characterization of Jacobian varieties in terms of soliton equations}, Invent. math. \text{83} (1986), 333-382

\bibitem[Tho870]{tho870}
J.~Thomae, \emph{Beitrag zur {B}estimmung von $\vartheta(0,0,\ldots,0)$ durch
  die {K}lassenmoduln algebraischer {F}unctionen}, J. reine angew. Math.
  \textbf{71} (1870), 201--222.

\bibitem[Van995]{Vanh995}
P.~Vanhaecke \emph{
Stratification of hyperelliptic Jacobians and the Sato Grassmannian}
 Acta Appl. Math., \textbf{ 40} (1995), 143--172

\bibitem[VN984a]{vn84a}
A.~P. Veselov and S.~P. Novikov, \emph{Finite zone two dimensional periodic
  {S}chr{\"o}dinger operators: case of potential}, Dokl. AN. SSSR \textbf{279}
  (1984), no.~4, 784--788.

\bibitem[VN984b]{vn84}
\bysame, \emph{Finite zone two dimensional periodic {S}chr{\"o}dinger
  operators: explicit formulae and evolution equations}, Dokl. AN. SSSR
  \textbf{279} (1984), no.~1, 20--24.

\bibitem[Wil888]{wi88}
E.~Wiltheiss, \emph{Ueber die {P}otenzreihen der hyperelliptischen
  {T}hetafunktionen}, Math. Ann. \textbf{31} (1888), 410--423.

\bibitem[Wil888a]{wi88a}
E.~Wiltheiss, \emph{Partielle Differentialgleichungen der hyperelliptischen
Thetafunctionen und der Perioden derselben,} Math. Ann., \textbf{31}, 134--155
(1888).

\bibitem[WW973]{ww73}
E.~T. Whittaker and G.~N. Watson, \emph{A course of modern analysis}, CUP,
  Cambridge, 1973.

\bibitem[Wei849]{w49}
K.~Weierstrass, \emph{Beitrag zur {T}heorie der {A}bel'schen {I}ntegrale},
  Jahresber. K{\"o}nigl. {K}atholischen Gymnasiums zu Braunsberg in dem
  Schuljahre 1848/49 (1849), 3--23.

\bibitem[Wei854]{w54}
\bysame, \emph{Zur {T}heorie der {A}belschen {F}unctionen}, J. reine angew.
  Math. \textbf{47} (1854), 289--306.

\bibitem[Wei893]{w-lect}
K.~Weierstrass, \emph{Formeln und {L}ehrs{\"a}tze zum {G}ebrauche der
  elliptischen {F}unctionen}, Springer, 1893, bearbeitet und herausgegeben von
  H. A. Schwarz.

\bibitem[Wei894]{wei82}
K.~Weierstrass, \emph{Zur {T}heorie der elliptischen {F}unctionen,} Mathematische
{W}erke, Vol.~2, Teubner, Berlin, 1894, pp.~245--255.

\bibitem[Wei894]{wei1894}
K. Weierstrass
\emph{Zur Theorie der elliptischen Funktionen}
Mathematische Werke, Bd. 2
Berlin, Teubner 1894 245--255

\bibitem[Wei904]{W1904}
K. Weierstrass
Abelsche Funktionen Gesammelte Werke, Bd. 4,  1904


\bibitem[Zak76]{Zakaljukin76}
V.~M.~Zakalyukin, \emph{Rearrangements of wave fronts that depend on a certain
parameter,} Funkts. Anal. Prilozhen., \textbf{10}, No.~2, 69--70 (1976).
English transl.: Funct. Anal. Appl. \textbf{10}
No.2 (1976) 139--140

\end{thebibliography}
\end{document}